\documentclass[a4paper,11pt, twoside, openany]{book}
\usepackage[english]{babel}
\usepackage[utf8]{inputenc}
\usepackage{graphicx}
\usepackage{amsmath}
\usepackage{amsthm}
\usepackage{amssymb}
\usepackage{mathtools}
\usepackage{physics}
\usepackage{nccmath}
\usepackage{stackengine,amsmath}
\usepackage{float}
\usepackage{latexsym}
\usepackage{scalerel}
\usepackage{fancyhdr}
\usepackage{booktabs}
\usepackage{listings}
\usepackage{braket}
\usepackage{titlesec}
\usepackage{tensor}
\usepackage{cancel} 
\usepackage{setspace}
\usepackage{lscape}
\usepackage{bm}
\usepackage{comment}
\usepackage{url}
\usepackage{caption}
\usepackage{indentfirst}
\usepackage{subcaption}
\usepackage[left=2.8cm, right=2.8cm, top=3.2cm, bottom=2.8cm]{geometry}
\usepackage{amsfonts}
\usepackage{bbm}
\usepackage{verbatim}
\usepackage{dsfont}
\usepackage{booktabs}
\usepackage{slashed}
\usepackage{epsfig}
\usepackage{color}
\usepackage[table]{xcolor}
\usepackage[]{caption}
\usepackage{overpic}
\usepackage{enumerate}
\usepackage{hhline}
\usepackage{multirow}
\usepackage{xspace}
\usepackage{setspace}
\usepackage[title]{appendix}
\usepackage[T1]{fontenc}
\usepackage{titlesec}
\usepackage{xcolor}
\usepackage{blindtext}
\usepackage[OT1]{fontenc}
\usepackage{bibentry}
\usepackage[square,numbers,compress]{natbib}
\usepackage{transparent}
\usepackage[hang,flushmargin]{footmisc}
\usepackage{blkarray}
\usepackage{empheq}
\usepackage[indent]{parskip}
\usepackage[skins,theorems,most]{tcolorbox}
\usepackage{svg}

\usepackage{hhline}
\usepackage{diagbox}

\usepackage{xr-hyper}
\usepackage[colorlinks,linkcolor=blue,citecolor=blue]{hyperref}
\hypersetup{
    colorlinks=true,
    linkcolor=blue,
    urlcolor=cyan,
    citecolor=cyan,}

\graphicspath{{Images/}}

\newcommand{\chapnumfont}{
  \usefont{T1}{pnc}{b}{n}
  \fontsize{80}{80}
  \selectfont
}
\colorlet{chapnumcol}{gray!75}  

\definecolor{Berry}{rgb}{0.60,0.06,0.34}
\definecolor{uam}{rgb}{0.32, 0.5 , 0.18}
\definecolor{csic}{rgb}{0.82, 0.03 , 0.15}
\colorlet{Teja}{csic!90!black}

\titleformat{\chapter}[display]
{\bfseries}
{\vspace{-120pt}\filleft\chapnumfont\textcolor{chapnumcol}{\thechapter}\\\vspace{-2pt}\color{chapnumcol}\titlerule[.1pt]%
\vspace{2pt}%
\titlerule \vspace{-12pt}}
{-90pt}
{
\huge\filleft\usefont{OT1}{lmss}{b}{n}}
[\vspace{8pt}\color{chapnumcol}\titlerule
\vspace{2pt}%
\titlerule \vspace{-10pt}
]

\newtheoremstyle{indentedupright}
  {15pt}
  {15pt}
  {} 
  {\parindent}
  {\bfseries}
  {.}
  {.5em}
  {}
  
\newtheoremstyle{indenteditalic}
  {15pt}
  {15pt}
  {\itshape} 
  {\parindent}
  {\bfseries}
  {.}
  {.5em}
  {}

\newtheoremstyle{indentedremark}
  {}
  {}
  {} 
  {\parindent}
  {\itshape}
  {.}
  {.5em}
  {}
  
\theoremstyle{indenteditalic}

\theoremstyle{indentedremark}

\theoremstyle{indentedupright}

\newcommand{\bmat}{\left(\begin{array}}
\newcommand{\emat}{\end{array}\right)}

\newcommand{\pr}{\mathbbm{R}}

\def\bZ{\mathbb{Z}}
\def\Z{\mathbb{Z}}
\def\R{\mathbb{R}}
\def\C{\mathbb{C}}

\def\P{\mathbb{P}}
\def\CK {{\cal K}}
\def\a {\alpha}
\def\b {\beta}

\def\ov{\overline}

\def\Tr{\text{Tr}}
\def\IM{\text{Im}\,}
\def\RE{\text{Re}\,}
\def\ov{\overline}
\def\1{{\bf 1}}
\def\2{{\bf 2}}
\def\3{{\bf 3}}
\def\4{{\bf 4}}
\def\6{{\bf 6}}

\def\half{\frac{1}{2}}

\def\targ#1#2{\genfrac{[}{]}{0pt}{}{#1}{#2}}
\def\targ2#1#2{\genfrac{}{}{0pt}{}{#1}{#2}}
\def\half{{\textstyle\frac{1}{2}}}

\definecolor{mygr}{rgb}{0,0.6,0}
\definecolor{mygrey}{rgb}{0,0.1,0.2}
\definecolor{myblue}{rgb}{0,0.5,0.9}
\definecolor{myblue2}{rgb}{0,0.5,0.5}
\definecolor{myblue3}{rgb}{0,0.7,0.9}
\definecolor{myblue4}{rgb}{0,0.6,0.6}
\definecolor{myorange}{rgb}{1,0.5,0}
\definecolor{mypurple}{rgb}{0.6,0,1}
\definecolor{mygolden}{rgb}{1,0.8,0.2}
\definecolor{mycyan}{rgb}{0,1,1}
\definecolor{mymagenta}{rgb}{1,0,1}
\definecolor{mykiwi}{rgb}{0.8,1,0.5}
\definecolor{mybrown}{cmyk}{0.14, 0.42, 0.56, 0.2}
\definecolor{myturq}{cmyk}{0.99, 0, 0.2, 0.4}
\definecolor{myaubergine2}{cmyk}{0.4, 0.5, 0, 0.1}
\definecolor{myaubergine}{cmyk}{0.6,0.85,0,0}
\definecolor{CycleGreen}{cmyk}{0.52,0,1,0}
\definecolor{CycleBrown}{cmyk}{0, 0.4, 0.9, 0.2}

\DeclareFontFamily{U}{rcjhbltx}{}
\DeclareFontShape{U}{rcjhbltx}{m}{n}{<->rcjhbltx}{}
\DeclareSymbolFont{hebrewletters}{U}{rcjhbltx}{m}{n}

\DeclareMathSymbol{\lamed}{\mathord}{hebrewletters}{108}
\DeclareMathSymbol{\mem}{\mathord}{hebrewletters}{109}
\DeclareMathSymbol{\ayin}{\mathord}{hebrewletters}{96}
\DeclareMathSymbol{\tsadi}{\mathord}{hebrewletters}{118}
\DeclareMathSymbol{\qof}{\mathord}{hebrewletters}{113}
\DeclareMathSymbol{\resh}{\mathord}{hebrewletters}{114}
\DeclareMathSymbol{\pe}{\mathord}{hebrewletters}{112}
\DeclareMathSymbol{\pesofit}{\mathord}{hebrewletters}{80}
\DeclareMathSymbol{\samekh}{\mathord}{hebrewletters}{115}
\DeclareMathSymbol{\tav}{\mathord}{hebrewletters}{116}
\DeclareMathSymbol{\vav}{\mathord}{hebrewletters}{119}
\DeclareMathSymbol{\het}{\mathord}{hebrewletters}{120}
\DeclareMathSymbol{\yod}{\mathord}{hebrewletters}{121}
\DeclareMathSymbol{\zayin}{\mathord}{hebrewletters}{122}
\DeclareMathSymbol{\alephdot}{\mathord}{hebrewletters}{128}
\DeclareMathSymbol{\tsadisofit}{\mathord}{hebrewletters}{90}
\DeclareMathSymbol{\shin}{\mathord}{hebrewletters}{152}

\def\CN {{\cal N}}

\def\sig{{\sigma}}
\def\d{{\delta}}
\def\be{\begin{equation}}
\def\ee{\end{equation}}
\def\bea{\begin{eqnarray}}
\def\eea{\end{eqnarray}}
\def\bes{\begin{subequations}}
\def\ees{\end{subequations}}
\def\raw{\rightarrow}

\def\eps{{\epsilon}}
\def\oh{\frac{1}{2}}

\def\re{\mbox{Re}\, }
\def\im{\mbox{Im}\, }
\def\tr{\mbox{Tr}}

\def\cy {{\text{CY}}}
\def\IZ{\mathbb{Z}}
\def\om{\omega}
\def\Om{\Omega}
\usepackage{multicol}
\usepackage{float}
\usepackage{caption}
\def\mk {{\mathcal K}}

\def\tr {{\tilde{\rho}}}
\def\p {{\partial}}

\def\g {{\gamma}}

\def\CO {{\cal O}}

\newcommand{\cF}{\mathcal{F}}

\newcommand{\cK}{\mathcal{K}}
\newcommand{\cM}{\mathcal{M}}
\newcommand{\cN}{\mathcal{N}}
\newcommand{\cO}{\mathcal{O}}
\newcommand{\cA}{\mathcal{A}}
\newcommand{\cB}{\mathcal{B}}
\newcommand{\cT}{\mathcal{T}}

\newcommand{\II}{\mathbb{I}}

\newcommand*\widefbox[1]{\fbox{\hspace{2em}#1\hspace{2em}}}

\newenvironment{eqn}{\begin{equation}\begin{aligned}}{\end{aligned}\end{equation}\noindent}
\newenvironment{eqn*}{\begin{equation*}\begin{aligned}}{\end{aligned}\end{equation*}\noindent}

\newcommand{\F}{{\cal F}}

\newcommand{\V}{{\cal V}}
\renewcommand{\S}{{\cal S}}
\newcommand{\cH}{{\cal H}}

\renewcommand{\Re}{{\rm Re}\,}
\renewcommand{\Im}{{\rm Im}\,}

\newcommand{\ie}{{\em i.e.} }

\renewcommand{\and}{\mbox{and}}

\usepackage{xcolor}

\newcommand{\Nf}{N_{\rm flux}}

\renewcommand{\bm}{\boldmath}

\usepackage{subfiles} 

\begin{document}

\onehalfspacing

\pagestyle{fancy}
\frontmatter 
\fancyhf{}
\fancyhf[EFC,OFC]{\thepage}
\pagenumbering{Roman}

\begin{titlepage}
\vspace*{-0.25in}
\begin{center}

\noindent\makebox[\linewidth]{\rule{\textwidth}{2pt}}\vspace{0.15cm} \\
{ \LARGE \bfseries Moduli Stabilization and Stability\\ in Type II/F-theory flux compactifications\\[0.3cm] }
\noindent\makebox[\linewidth]{\rule{\textwidth}{2pt}}

\vspace{1cm}
\normalsize{Memoria de Tesis Doctoral realizada por
\vspace{0.15cm}\\
\textbf{David Prieto Rodríguez\vspace{0.15cm}}\\
presentada ante el Departamento de Física Teórica\\ de la Universidad Autónoma de Madrid\\ para optar al Título de Doctor en Física Teórica\\}
\vspace{1cm}
\normalsize{Tesis Doctoral dirigida por}\vspace{0.15cm}\\
\normalsize{\textbf{Fernando G. Marchesano Buznego}\vspace{0.15cm}},\\
\normalsize{investigador científico del CSIC}\\
\vspace{1cm}
\normalsize{Departamento de Física Teórica\\ Universidad Autónoma de Madrid}\\
\vspace{0.35cm}
\normalsize{Instituto de Física Teórica UAM-CSIC}\\
\vspace{0.3cm}
\begin{figure}[H]
    \centering\hspace{1.25cm}
    \begin{subfigure}[c]{0.3\textwidth}
    \centering
     \includegraphics[width=0.9\textwidth]{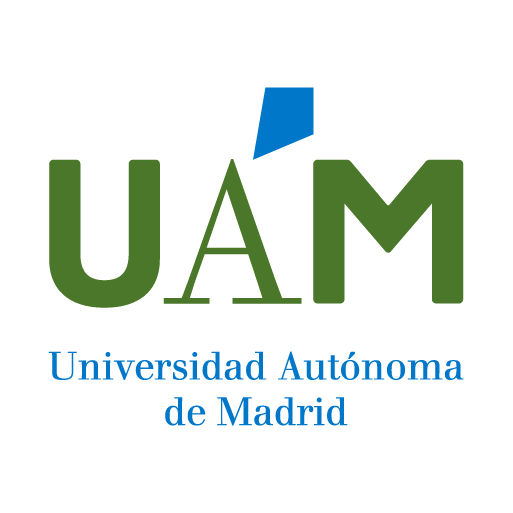}\hfill
    \end{subfigure}\hspace{0.5cm}
    \begin{subfigure}[c]{0.5\textwidth}
    \centering
     \includegraphics[width=0.9\textwidth]{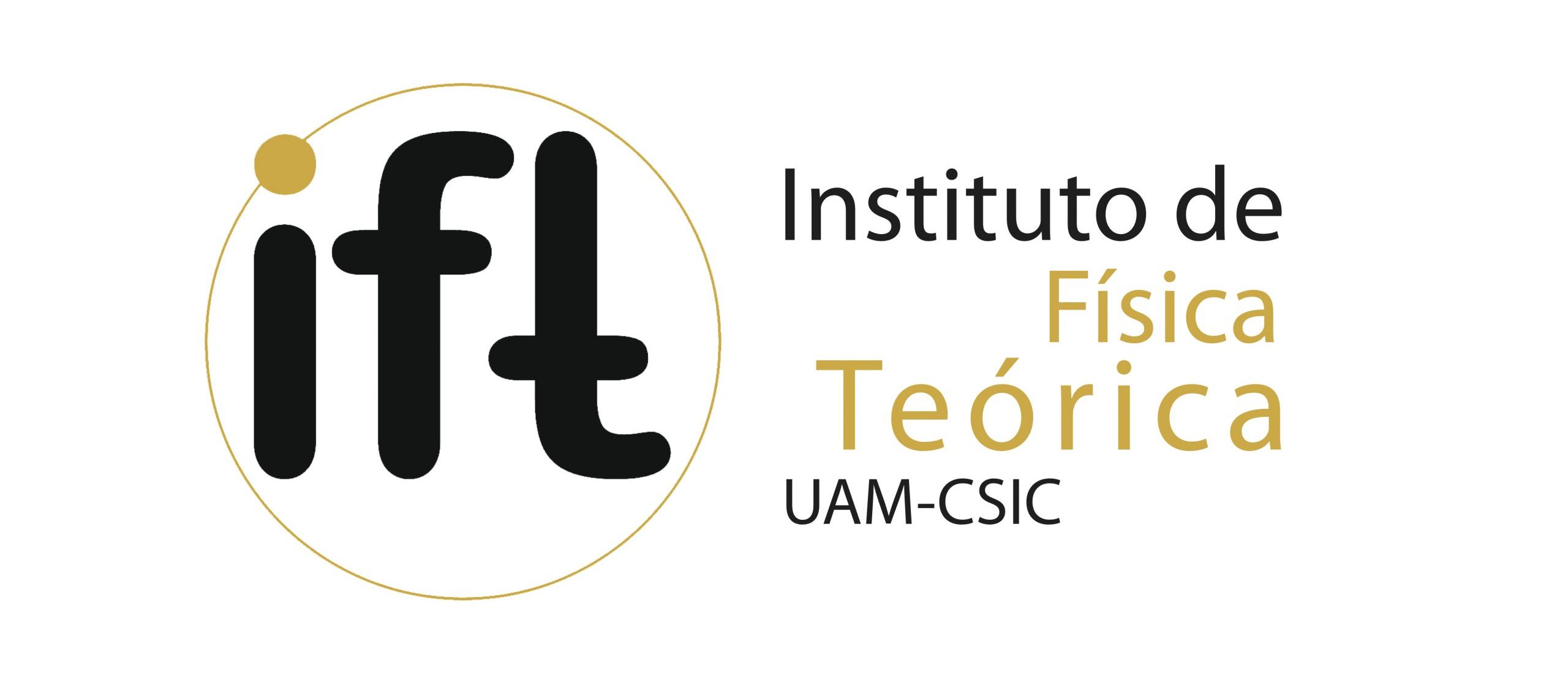}\hfill
    \end{subfigure}\vspace{-0.75cm}\\
    \hspace{0.6cm}\begin{subfigure}[b]{0.5\textwidth}
    \centering
     \includegraphics[width=0.9\textwidth]{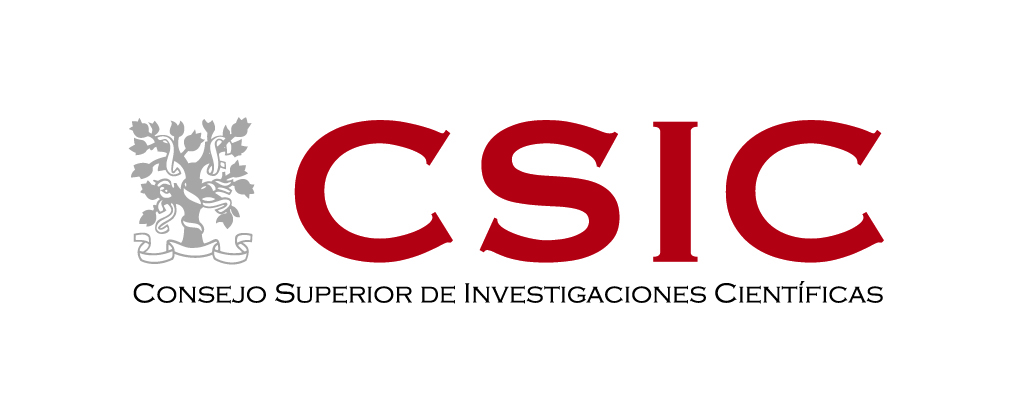}\hfill
    \end{subfigure}\\
\end{figure}
\vspace{0.25cm}
\normalsize Octubre, 2023
\end{center}
\end{titlepage}

\onehalfspacing


\thispagestyle{empty}
\phantom{a}\vspace{5cm} 
\begin{flushright} 
\textit{A mis padres.}
\end{flushright}


\newpage
\thispagestyle{empty}
 \vspace*{1.5cm}
\textbf{Esta tesis doctoral está basada en los siguientes artículos:}
 \vspace*{0.5cm}

\noindent [1]\, \textit{Systematics of Type IIA moduli stabilisation,}\\
\phantom{[1]\, }F. Marchesano, \textbf{D. Prieto}, J. Quirant, P. Shukla\\
\phantom{[1]\, }\href{https://doi.org/10.1007/JHEP11(2020)113}{JHEP 11 (2020) 113}  \href{https://arxiv.org/abs/2007.00672}{[2007.00672]}

\noindent [2]\, \textit{F-theory flux vacua at large complex structure,}\\
\phantom{[2]\, }F. Marchesano, \textbf{D. Prieto}, M. Wiesner\\
\phantom{[2]\, }\href{https://doi.org/10.1007/JHEP08(2021)077}{JHEP 08 (2021) 077}  \href{https://arxiv.org/abs/2105.09326}{[2105.09326]} 

\noindent [3]\, \textit{BIonic membranes and AdS instabilities,}\\
\phantom{[3]\, }F. Marchesano,\textbf{ D. Prieto}, J. Quirant\\
\phantom{[3]\, }\href{https://doi.org/10.1007/JHEP07(2022)118}{JHEP 07 (2022) 118}  \href{https://arxiv.org/abs/2110.11370}{[2110.11370]} 

\noindent [4]\, \textit{Membranes in AdS4 orientifold vacua and their
Weak Gravity Conjecture,}\\
\phantom{[4]\, }G. F. Casas, F. Marchesano, \textbf{D. Prieto}\\
\phantom{[4]\, }\href{https://doi.org/10.1007/JHEP09(2022)034}{JHEP 09 (2022) 034}  \href{https://arxiv.org/abs/2204.11892}{[2204.11892]]} 

\noindent [5]\, \textit{Analytics of type IIB flux vacua
and their mass spectra,}\\
\phantom{[5]\, }T. Coudarchet, F. Marchesano, \textbf{D. Prieto}, M. A. Urkiola\\
\phantom{[5]\, }\href{https://doi.org/10.1007/JHEP01(2023)152}{JHEP 01 (2023) 152}  \href{https://arxiv.org/abs/2212.02533}{[2212.02533]} 

\phantom{\cite{Marchesano:2020uqz,Marchesano:2021gyv,Marchesano:2021ycx,Casas:2022mnz, Coudarchet:2022fcl,Coudarchet:2023mmm, systematicsfollowup}}

Otros artículos escritos durante la realización del doctorado y que no se incluyen en esta tesis son

\noindent [6]\, \textit{Symmetric fluxes and small
tadpoles,}\\
\phantom{[6]\, }T. Coudarchet, F. Marchesano, \textbf{D. Prieto}, M. A. Urkiola\\
\phantom{[6]\, }\href{https://doi.org/10.1007/JHEP08(2023)016}{JHEP 08 (2023) 016}  \href{https://arxiv.org/abs/2304.04789}{[2304.04789]} 

\noindent [7]\, \textit{Symmetric fluxes and small
tadpoles,}\\
\phantom{[7]\, }R. Carrasco, T. Coudarchet, F. Marchesano, \textbf{D. Prieto}\\
\phantom{[7]\, }\href{https://doi.org/10.1007/JHEP11(2023)094}{JHEP 11 (2023) 094}  \href{https://arxiv.org/abs/2309.00043}{[2309.00043]}


\newpage
\chapter*{ Acknowledgments}

Es un deber y un placer comenzar agradeciendo a Fernando todo su apoyo durante los últimos cuatro años. Desde que me acogió en el IFT cuando estaba todavía un poco perdido al terminar el máster, siempre ha estado disponible para ofrecer su consejo, resolver cualquier duda y compartir su increíble intuición física y matemática. Gracias por abrirme la puerta al mundo de la investigación y por guiarme, incluso a distancia durante la pandemia, de forma paciente, cercana y flexible por el exótico reino de la Teoría de Cuerdas. ¡Ha sido un placer aprender y trabajar contigo!

I would also like to express my gratitude towards all the people I have worked with and whose involvement has been crucial in the development of the results of this thesis: thank you Pramod, Thibaut and Mikel for instructive and exciting collaborations. Special thanks to Max for helping me understand the geometry of F-theory during my first year of PhD, while facing social distancing and teleworking. Muchas gracias también a Joan, que ha sido mi faro entre la niebla de la burocracia y junto a quien he luchado para construir unos convenios de signos (espero) consistentes, y a Gonzalo, mi hermanito académico cuyas preguntas me han servido para replantearme y aprender muchas cosas. ¡Espero que podamos seguir colaborando en el futuro!  

En siguiente lugar me gustaría dar las gracias a Luis, Ángel y más recientemente Miguel, que con sus explicaciones, preguntas y agudos comentarios durante los SPLEs me han impulsado a desarrollar una actitud más crítica y a aprender mucha física. Gracias de nuevo a Ángel por su curso de Teoría de Cuerdas que me ayudó a aclarar unas cuantas dudas y a organizar las ideas. También quiero agradecer a Irene por darme la oportunidad de pasar tres meses en el CERN. Ser partícipe de la actividad de un grupo tan dinámico ha sido muy enriquecedor. Gracias también a José y Bernardo por su cálida bienvenida y hacer la estancia mucho más entretenida.  And thank you Thomas for giving me the chance to continue learning and doing research!

No puedo olvidarme de mis compañeros de despacho Alessandro, Ginevra, Gabriel, Marienza y especialmente Roberta y Sergio, mi inseparable compañero de fatigas desde que llegué al IFT. Gracias a todos por crear un entorno de trabajo acogedor, colaborativo y amistoso. 

Una mención especial para Alberto por los tres meses en el CERN que pasamos “together!” compartiendo casa y despacho. Hiciste que la experiencia fuese mucho más grata y las discusiones que tuvimos sobre los distintos proyectos me han enseñado mucho.

Para no alargarme  e ir agrupando, quiero dar las gracias a todos los demás miembros, antiguos y nuevos, del grupo de teoría de cuerdas que aún no he nombrado. Matilda, Matteo, Luca, Jesús, Lorenzo y el pletórico Nacho, gracias por llenar de energía todas las actividades del IFT y hacer los viajes de las conferencias mucho más divertidos.

El buen ambiente en el IFT no se restringe al grupo de cuerdas. Entre otros me gustaría destacar a Jorge, Fran y Llorenç, que animaron muchas comidas en el CBM, Mattia, ha sido un placer dar clase de Electrodinámica contigo, y Rafa, seguro que pronto conseguimos acabar el proyecto que tenemos entre manos. Finalmente, quiero agradecer también a la gente de secretaría del IFT por su constante ayuda y consejo con todos los papeleos.

A mis amigos de la infancia, Nacho, Guille, Sergio y Javi, gracias por hacerme compañía en la distancia siempre que lo necesitaba.  Y a mis amigos de la carrera, Rafa, Kike, Marcos, Diego, Rosa y todos los demás, espero que mantengamos las reuniones tradicionales.

Por último quiero expresar mi más sincera gratitud a toda mi familia. A mis tíos, especialmente a Ana por ayudarme a encontrar camino en la carrera académica. A Claudia y a Filip por apuntarse a cualquier aventura y otorgarme el gran honor de ser la cobaya cero en sus experimentos lingüísticos. A mi abuela y sobretodo a mis padres. Gracias por estar ahí siempre que os necesito, por apoyarme y acompañarme en cada uno de mis pasos. Gracias por hacerme quien soy.


\newpage

\chapter*{\vspace{-3.5cm}Abstract}
\singlespacing
\vspace{-1 cm}
In this thesis we study String Theory compactifications to four dimensions focusing on the moduli stabilization process and the associated vacua structure in various frameworks, from Type IIA to F-theory. We interpret the results in the context of the Swampland Program.

We start with a basic introduction to  String Theory and the Swampland conjectures to lay out all the ingredients used throughout the thesis. We also summarize the geometrical aspects of Calabi-Yau orientifolds and their role in massive Type IIA compactifications. We end the review with a discussion on the current state of the field, presenting the approximated 10d solutions to the equations of motion with fluxes and the bilinear formalism of the 4d effective potential created by the RR and NSNS flux quanta.

Having introduced all the key concepts and background results, we generalize the bilinear formalism of the scalar potential to include the contributions of geometric and non-geometric fluxes, which is later used to perform a systematic search of vacua. Using an Ansatz motivated by the goal of achieving stable de Sitter vacua, we study the equations of motion of Type IIA with metric fluxes. We obtain only AdS vacua, both SUSY and non-SUSY, checking their stability and generalizing several results from the literature. We try to find scale separation but fail to do so in the studied solutions. 

We also consider the 10d uplift of  $AdS_4$ vacua arising from the 4d massive Type IIA effective theory with only RR and NSNS fluxes. Using the language of $SU(3)\times SU(3)$ structures and performing an expansion around the smearing approximation in powers of the string coupling, we study the stability of the supersymmetric solution and its non-supersymmetric partner (associated with the former by a change of sign in the RR 4-form field strength  flux). We contrast the results with the Weak Gravity Conjecture and the AdS instability conjecture in several toroidal orbifold examples and find that some non-supersymmetric cases are in tension with the predictions of those conjectures, hinting at the existence of additional corrections that have not been taken into account.

After briefly introducing F-theory and Type IIB compactifications, we study moduli stabilization in the complex structure sector of F-theory compactifications over elliptically fibered Calabi-Yau 4-folds in the limit of Large Complex Structure. Using homological mirror symmetry, we are able to replicate the analysis for the Type IIA case and give a bilinear expression for the scalar potential, allowing for a simpler and more detailed study of the vacua structure. In the process, we find two distinct families of flux configurations compatible with the tadpole constraints that allow for full moduli stabilization. The first one requires polynomial corrections to fix all the moduli and the flux contribution to the tadpole scales with the dimension of the moduli space. In contrast, in the second family, polynomial corrections are not needed and only a pair of fluxes enters the tadpole independently of the number of moduli. We thoroughly examine the former in the Type IIB limit, where the superpotential is also quadratic and polynomial corrections can be considered at all orders. We argue that vacua fall into three classes depending on the choice of flux quanta. In particular, we provide analytic expressions for the vacuum expectation values and flux-induced masses of the axio-dilaton and complex structure fields in a large subclass of vacua, independently of the Calabi-Yau and the number of moduli. Finally, we show that at this level of approximation supersymmetric vacua always contain flat directions.

\newpage
\chapter*{\vspace{-3.5cm}Resumen}
\singlespacing
\vspace{-1 cm}

En esta tesis estudiamos las compactificaciones de Teoría de Cuerdas a cuatro dimensiones centrándonos en el proceso de estabilización de módulos y su estructura de vacíos asociada en varios escenarios, desde la teoría Tipo IIA a la teoría F. Los resultados obtenidos son interpretados en el contexto del Programa de la Ciénaga.

Comenzamos con una introducción básica a la Teoría de Cuerdas y las conjeturas de la Ciénaga para presentar todas las piezas utilizadas a lo largo de la tesis. También resumimos los aspectos geométricos de los \textit{orientifolds} de variedades Calabi-Yau y su papel en las compactificaciones de Tipo IIA masiva. Terminamos el repaso con una discusión sobre el estado actual del campo, presentando las soluciones aproximadas a las ecuaciones de movimiento con flujos en 10d y el formalismo bilineal del potencial efectivo en 4d creado por los cuantos de flujo RR y NSNS.

Una vez introducidos todos los conceptos clave y los resultados previos, generalizamos el formalismo bilineal del potencial escalar para incluir las contribuciones de los flujos geométricos y no geométricos. Este formalismo es utilizado posteriormente para realizar una búsqueda sistemática de vacíos. Utilizando un Ansatz motivado por el objetivo de conseguir vacíos de Sitter estables, estudiamos las ecuaciones de movimiento del Tipo IIA con flujos métricos. Hallamos sólo vacíos AdS, tanto SUSY como no-SUSY, comprobando su estabilidad y generalizando varios resultados de la literatura. Intentamos encontrar separación de escalas pero no lo logramos para soluciones estudiadas. 

También consideramos la extensión a 10d de los vacíos $AdS_4$ que surgen en la teoría efectiva  de Tipo IIA masiva en 4d al activar únicamente  flujos RR y NSNS. Utilizando el lenguaje de estructuras $SU(3)\times SU(3)$ y realizando una expansión en torno a la aproximación \textit{smearing} en términos del acoplamiento de cuerdas, estudiamos la estabilidad de la solución supersimétrica y de su pareja no supersimétrica (asociada a la primera por un cambio de signo en el flujo de la 4-forma RR). Contrastamos los resultados con la Conjetura de la Gravedad Débil y la Conjetura de inestabilidad de AdS en varios ejemplos de \textit{orbifolds} toroidales y encontramos que algunos casos no supersimétricos están en tensión con las predicciones de dichas conjeturas. Esto apunta a la existencia de correcciones adicionales que no se han tenido en cuenta.

Tras una breve introducción sobre las compactificaciones en la teoría F y en la teoría tipo IIB, pasamos a estudiar la estabilización de los módulos en el sector de estructura compleja para las compactificaciones de teoría F en variedades Calabi-Yau de 8 dimensiones fibradas elípticamente en el límite de gran estructura compleja. Utilizando  simetría especular homológica, somos capaces de replicar el análisis para el caso de la teoría Tipo IIA y dar una expresión bilineal para el potencial escalar, permitiendo un estudio más simple y detallado de la estructura de los vacíos. En el proceso, encontramos dos familias distintas de configu-raciones de flujos compatibles con las restricciones \textit{tadpole} que permiten la estabilización completa de los módulos. La primera requiere correcciones polinómicas para fijar todos los módulos y la contribución de los flujos al \textit{tadpole} escala con la dimensión del espacio de módulos. En la segunda familia, en cambio, no se necesitan correcciones polinómicas y sólo una pareja de flujos entra en el \textit{tadpole } independientemente del número de módulos. Examinamos en detalle la primera de estas familias en el límite de Tipo IIB, donde el superpotencial también es cuadrático y las correcciones polinómicas pueden ser tratadas a todos los órdenes. Argumentamos que los vacíos en este caso se dividen en tres clases dependiendo de la elección de los cuantos de flujo. En particular, proporcionamos expresiones analíticas para los valores de vacío esperados y las masas  del axiodilatón y de los campos de estructura compleja inducidas por flujos en una gran subclase de vacíos, independientemente de la variedad Calabi-Yau considerada y del número de módulos. Finalmente, mostramos que a este nivel de aproximación los vacíos supersimétricos siempre contienen direcciones planas.

\let\cleardoublepage=\clearpage

\singlespacing
\renewcommand{\contentsname}{Contents}
\tableofcontents
\onehalfspacing


\mainmatter


\part[\textcolor{Teja}{ Preliminaries}]{\scshape \textcolor{Teja}{\huge Preliminaries}}
\label{part: preliminaries}


\fancyhf{}
\renewcommand{\chaptermark}[1]{\markboth{#1}{}}
\renewcommand{\sectionmark}[1]{\markright{#1}}
\fancyhf[EHL]{\textit{\thechapter. \nouppercase{\leftmark}}}
\fancyhf[OHR]{}
\fancyhf[EFC,OFC]{\thepage}

\chapter{Introduction}

It is the goal of any theoretical physicist to delve ever deeper into the most fundamental laws that shape our world, driven by the wonder at the mesmerizing power of mathematical language to describe and predict natural phenomena, in a utopian attempt to find a simple and elegant framework that provides the building blocks from which empirical reality can be reconstructed. Such an underlying theory may not exist or may be unattainable, but its search helps to push forward the frontiers of knowledge in both Physics and Mathematics, and each new advance reveals many more tempting questions to be answered.

At the end of the nineteenth century, with the great successes of classical mechanics, thermodynamics and electromagnetism, one might have thought that the work of the physicist was essentially finished. However, several observations did not  fit well with the established theories, such as the precession of the perihelion of Mercury, the black body radiation and the photoelectric effect. These cracks became the threads that led to the development of our modern understanding of physics. In the course of the twentieth century, research split into two distinct paths: the study of the very large, i.e. stars, galaxies and even the universe as a whole, and the study of the very small, the tiniest constituents of such universe. 

On the one hand, general relativity, proposed by Albert Einstein in 1915 \cite{Einstein:1916vd}, is able to predict with great precision the dynamics that govern the motion of objects at very large distances. From the orbit of satellites around the Earth to the study of black holes and even the expansion of the universe, general relativity (GR) has been successfully tested. The most recent achievement is the astonishing detection of gravitational waves by the LIGO collaboration \cite{LIGOScientific:2016aoc} one hundred years after their prediction \cite{Einstein:1916cc}. 

On the other hand, Quantum Field Theory (QFT) provides an excellent framework for explaining phenomena at the level of elementary particles. The most important QFT, the Standard Model (SM), is able to describe three of the four fundamental forces and their associated particles. The Standard Model has also experienced a massive breakthrough in recent years with the experimental detection of the last missing piece in 2012: the Higgs boson \cite{CMS:2012qbp,ATLAS:2012yve}. Today, the Standard Model stands as one of the most successful scientific theories, experimentally supported by the LHC up to energies of 10 TeV and providing extremely accurate measurements (see for instance \cite{Parker:2018vye}).  

Despite the achievements of the last decade, and in a way reminiscent of the situation a century earlier, there are several open questions that neither branch is able to answer. Among these, the strong CP problem, the origin of neutrino masses or the nature of dark matter and dark energy stand out. The most appealing solution to at least some of these problems is to unify the two paths mentioned above - General Relativity and Quantum Field Theory - into a single theory: Quantum Gravity. 

Unification has been both a guideline and a trend throughout the history of Physics. From Newton's work uniting the dynamics of the Earth and the skies, to Special Relativity combining classical mechanics and electromagnetism, and finally Quantum Field Theory merging Special Relativity and Quantum Mechanics, our understanding of the natural world has grown through the construction of theories that explain seemingly independent phenomena.  Nevertheless, combining gravity with the other three interactions contained in the Standard Model has proved to be an extremely challenging task, due to the divergences that arise when gravity is treated as a QFT \cite{Feynman:1963ax,tHooft:1974toh,Goroff:1985sz}. Furthermore, Quantum Physics and General Relativity give rise to discrepancies when considering the cosmological constant, i.e., the vacuum energy 
of our universe. Estimates using QFT methods exceed the value derived from astronomical observations using a GR model by more than 100 orders of magnitude. Nevertheless, there is some reason to believe that there must be a common framework in which these two scales coexist and merge. This is the case for black holes, classical GR solutions that have an entropy associated with a purely quantum radiation process. Therefore, a different approach is needed to meet the challenge. 

In this context, String Theory is currently the best proposal for a theory of Quantum Gravity, offering a way to unify Quantum Field Theory and General Relativity in a single framework capable of describing all interactions. It does this by replacing the previous paradigm of particles with extended objects along one dimension: strings. Particles  in QFT constructions are interpreted as different vibrational modes of the more fundamental string.  This seemingly simple change resolves the divergences found in the QFT approach and provides a topological understanding of the UV-IR dependence that characterizes quantum gravity, leading to a UV-finite theory.

The history of String Theory is marked by revolutions. After its initial introduction as an alternative way to describe the strong interaction, its popularity hastily grew in the 1970s, when it was
realized that the massless excitation spectrum of closed strings contains a spin-2 field, i.e., a graviton. This means that a theory containing closed string (that is, any theory of strings) always gravitates. Additionally, the introduction of supersymmetry allowed to describe both bosonic and fermionic states in a unified framework. 

Another fascinating property of String Theory is that, contrary to general QFTs, it only has one external parameter: the length of the string. All other quantities are either completely fixed or determined as dynamical objects' expectation values. That even applies to the number of dimensions of the theory, which is fixed to be ten due to consistency requirements. In this 10-dimensional space, five different consistent superstring theories were found. 

The second revolution arrived in the 1990s, headed by two major discoveries. One was the observation of different dualities between the five aforementioned superstring theories, which could then be interpreted as limits of a unique theory in eleven dimensions, named M-theory \cite{Witten:1995ex}. The other was the realization that String Theory also predicts the existence of higher dimensional objects, D-branes, as non-perturbative states \cite{Dai:1989ua, Polchinski:1995mt}. These provide the tools to build far richer constructions capable of describing all interactions.

Until the arrival of String Theory, our understanding of nature has been framed (consciously or not) in terms of effective theories valid up to a certain  energy scale called cut-off. This is particularly apparent in Quantum Field Theories, that generically have a cut-off over which they are ill-defined. Depending on whether the QFT is renormalizable or not, a finite or infinite number of parameters is required to define the theory below the cut-off. In contrast, String Theory is UV-complete and its only external parameter, the string length, could be an artifact of the perturbative description that becomes dynamical in the bulk of M-theory. These observations hint at the fact that String Theory (in its broadest interpretation) could truly be the final fundamental theory.

However enticing as it may be, String Theory is not exempt of difficulties. The root of many of them lies in the fact that, as we mentioned before, the formulation of the theory requires the existence of six extra dimensions. Consequently,  to give an acceptable description of the observed universe, an explanation must be provided to justify why they are not detected . This is commonly achieved through the process of compactification, whose core idea is the assumption that the additional dimensions are so small that they are unreachable by the current experiments.  The particular characteristics of the compactified 4-dimensional effective theory depend on the geometry of the compact space and the configuration of fluxes (vacuum expectation values for the internal field strengths) and branes that populate it. Since we do not have direct observations of these characteristics, many candidates are allowed, which generates a colossal set of 4-dimensional EFTs (up to the order $10^{272000}$ according to recent estimations \cite{Taylor:2015xtz}) often referred to as String Landscape.

In the process of compactification, many geometrical quantities that characterize the internal 6-dimensional space become massless scalars in the effective theory: the moduli. Given that we do not observe these massless particles in our daily life, we need a mechanism capable of granting them mass. Such process is known as moduli stabilization. Our current picture of the string Landscape is tightly connected to the different mechanisms for moduli stabilization. This is because a simple procedure to generate an ensemble of vacua is to consider an EFT with a perturbative multi-dimensional moduli space, and implement one or several moduli-fixing mechanisms that select a discrete set of points in that space. Such philosophy is usually realized by means of background fluxes threading the compact space.

Given the vast amount of possible vacua and the lack of a selection procedure to find the one that describes our Universe, String Theory has often been criticized for its lack of predictive power. After all, with so many options, what does stop us from choosing the most exotic effective theory we can imagine?  This concern has been addressed in the Swampland Program, which was originally proposed in \cite{Vafa:2005ui} and has become an highly active field of research in recent years. It is centered around the idea that not every 4-dimensional EFT can be uplifted to a complete ultraviolet theory of Quantum Gravity.  In fact, the Swampland Program expects that those EFTs that can, constitute a set of zero-measure with respect to the full set of EFTs. Therefore, compactifications coming from String Theory, as a theory of quantum gravity, are actually a very selective ensemble. 

Consequently, this has introduced a change of paradigm. Instead
of searching for particular models of our universe, the focus has shifted to the study of the generic properties that any EFT needs to satisfy in order to be embedded in Quantum Gravity. A theory failing to do so is then said to live in the Swamp. These required properties are formulated in terms of conjectures,
which aim to set boundaries in the space of EFTs that clearly separate the Landscape from the Swamp. A great effort is currently being taken by the community in order to upgrade these conjectures to full results. In the process, we are obtaining a very valuable insight on the nature of Quantum Gravity.

\textbf{Plan of this Thesis}

The thesis aims to advance one step in this direction by developing tools that allow us to systematically explore the vacua structure of EFTs arising from flux compactifications in Type II and F-theory. The information we extract from the said analysis will help us to better understand their properties and test the predictions derived from Swampland conjectures. The thesis is structured in five parts. The main results derived from the research undertaken during the PhD are contained in chapters \ref{ch: systematics}, \ref{ch: bionic}, \ref{ch: membranes}, \ref{ch: Ftheory} and \ref{ch: typeIIB}, while chapters \ref{ch: basics}, \ref{ch: calabi-yau} and \ref{ch: Fintro} offer a review of the subjects addressed in our work and of recent progress on the field. 

\begin{itemize}
    \item In the remainder of part \ref{part: preliminaries}, we will present a review of the basic concepts of String Theory in order to provide a background knowledge that contextualizes the different elements and techniques employed in the following chapters. With the aim of making the thesis as self-contained as possible, we will briefly introduce the bosonic string and its main properties. Then, we will motivate the use of supersymmetry and take a quick tour through the most important aspects of superstring theories, with special focus on Type IIA and Type IIB theories. We will follow with an explanation of the role that non-perturbative states and dualities play in the current understanding of String Theory, and we will end by  providing a summary of the philosophy of the Swampland Program and some of its most important conjectures.
    \item Part \ref{part: type IIA} will be focused on massive Type IIA flux compactifications. In chapter \ref{ch: calabi-yau}, we will discuss the necessary internal geometric requirements for standard compactifications and the corresponding properties of the moduli space. We will also revisit the 10-dimensional equations of motion, the techniques employed to obtain approximated $AdS_4$ solutions to these equations, and the current results regarding the structure of the 4-dimensional vacua with RR and NSNS fluxes. In the following chapters, we push forward the boundaries of our knowledge on both fronts. In chapter \ref{ch: systematics} we study from a 4-dimensional perspective the vacua structure of  compactifications  adding geometric and non-geometric fluxes and test if they are capable of providing de Sitter vacua and scale separation. In chapter \ref{ch: bionic} we consider the 10-dimensional uplift of two families of 4-dimensional solutions presented in \cite{Marchesano:2019hfb} (one supersymmetric and one non-supersymmetric) and see how their properties match with the predictions of the Weak Gravity Conjecture and the AdS instability conjecture. Such behaviour is further tested in chapter \ref{ch: membranes} considering explicit setups using toroidal orbifolds. 
    \item In part \ref{part: F and B}, we turn our attention to type IIB and F-theory compactifications. Chapter \ref{ch: Fintro} offers a review of the main elements of these compactifications building on top of the common elements introduced in chapter \ref{ch: calabi-yau} and emphasizing the connections between the different theories due to mirror symmetry. In chapter \ref{ch: Ftheory} we study the complex structure moduli stabilization process of F-theory compactifications  on elliptically fibered Calabi-Yau four-folds in the large complex structure limit. We find out that there are  two generic families of flux quanta that allow for full moduli stabilization while satisfying the tadpole cancellation conditions. In one of them, polynomial corrections are required to stabilize all moduli and the saxionic vacuum expectation values are bounded by both the D3-brane tadpole and the contribution of the polynomial corrections. In the other family, the saxionic vacuum expectation values are unbounded and the flux contribution to the tadpole is just a function of a single pair of flux quanta independently of the actual size of the moduli space. We particularize the first of these two families to the Type IIB limit case in chapter \ref{ch: typeIIB}, where we analyze the equations of motion in greater detail and provide a particular Ansatz for which we find solutions exact in the polynomial correction and compute their associated mass spectra.
    \item We close the thesis in part \ref{part: conclusions} summarizing the main results and including some remarks for future research directions. Finally, part \ref{part: Appendix} contains several appendices with technical material and some long computations that complement the main text.
\end{itemize}


\fancyhf{}
\renewcommand{\chaptermark}[1]{\markboth{#1}{}}
\renewcommand{\sectionmark}[1]{\markright{#1}}
\fancyhf[EHL]{\textit{\thechapter. \nouppercase{\leftmark}}}
\fancyhf[OHR]{\textit{\thesection. \nouppercase{\rightmark}}}
\fancyhf[EFC,OFC]{\thepage}

\graphicspath{{Images/Basics}}

\ifSubfilesClassLoaded{%
\tableofcontents
}{}

\setcounter{chapter}{1}
\chapter{Unraveling the basics}
\label{ch: basics}

In this chapter we will review some of the most fundamental concepts of String Theory and the Swampland Program that will be required and provide context to understand the work presented in the thesis.

In section \ref{basic-sec: bosonic string} we present a broad introduction to the first and simplest String Theory, the bosonic string, emphasizing the main characteristics that distinguish it from the dynamics of standard particles and  the limitations that require the inclusion of supersymmetry. In section \ref{basic-sec: superstrings} we review the basic aspects of supersymmetric String Theories, putting special focus on Type IIA and Type IIB, which will be extensively studied in the later chapters of this thesis. We then move on to consider non-perturbative states in section \ref{basic-sec: non perturbative and dualities}, motivating their existence, their action as well as their behaviour in different limits, which leads us to discuss the fascinating web of dualities that connect all superstring theories. Finally, once the String Theory background is well established, we review the core philosophy of the Swampland Program and some of their most essential conjectures in section \ref{basic-sec: swampland program}, highlighting those aspects that will become relevant when studying compactifications in the following chapters. 

This quick tour over the foundations of String Theory aims to build a framework as self-contained as possible to develop our results. However, it is far from a comprehensive description. For an in-depth general treatment of String Theory, we refer the reader to the books and reviews \cite{Ibanez:2012zz,Blumenhagen:2013fgp,Becker:2006dvp,Polchinski:1998rq, Polchinski:1998rr, Green:2012oqa, Green:2012pqa, Tong:2009np, Kiritsis:1997hj}. Regarding the more recent topic of the Swampland program there are as well excellent specific reviews \cite{Brennan:2017rbf,Palti:2019pca,vanBeest:2021lhn,Grana:2021zvf, Agmon:2022thq}.

\section{First steps: Bosonic  Strings}
\label{basic-sec: bosonic string}

\subsection{Actions and Symmetries}

The core idea behind String Theory is to replace the point-like particle, essential element of most physical theories, with a small object that extends along one spatial dimension: the string. This apparently innocuous change has dramatic propagating consequences that give rise to fascinating new properties. While all particles are topologically identical, two different topologies are allowed when moving up to one-dimensional compact objects: the circle and the segment. Hence, two distinct objects, the close and the open string, can be used as the elemental pieces of the theory. In contrast with particles, both types of strings share the capability to vibrate along their internal structure. 

The first step to describe the dynamics of the new object is to define its action in a relativistic framework. This is achieved by generalizing the point-like particle action to account for the extended nature of the string.

The action of a free particle of mass $m$ moving in a $d$-dimensional Minkowski spacetime $\mathbb{M}_d$ is proportional to the invariant length of its worldline , i.e., a one-dimensional subspace of  $\ell\subset \mathbb{M}_d$ usually parametrized by the proper time of the particle ($\tau$) and with embedding function $X^M(\tau):\ell\rightarrow \mathbb{M}_d$. As a straightforward extension, the string action is proportional to the area of the two-dimensional surface $\Sigma\subset\mathbb{M}_d$ spawned by the string as it propagates. This surface, known as the worldsheet, is described by two parameters $\sigma^a=(\tau, \sigma)$ and has embedding functions $X^M(\tau,\sigma): \Sigma \rightarrow \mathbb{M}_d$. The resulting action is called the Nambu-Goto action
\begin{equation}
    S_{\textrm{particle}}=-m\int_\ell d\tau \sqrt{\eta_{MN}\dot{X}^M \dot{X}^N}\Rightarrow S_{\textrm{NG}}=-T_s\int_\Sigma d\tau d\sigma \sqrt{-\textrm{det} h}\, ,
\end{equation}
where $h$ is the two-dimensional worldsheet metric induced from the spacetime geometry and $T_s$ is the tension of the string. In natural units ($\hbar=c=1$) the tension has dimensions of mass squared and is related to the string scale through $M_s^2=2\pi T_s$. It will also be convenient to introduce another related quantity: the universal Regge slope, given by $\alpha'=1/2\pi T_s$, which has dimensions of length squared and satisfies $\ell_s=2\pi \sqrt{\alpha'}$ with $\ell_s$ the length of the string. 

Strings can be classified according to the topology of the worldsheet, so we can distinguish between closed and open strings and oriented and unoriented strings. Closed strings are associated with worldsheets without boundaries and do not have endpoints, whereas open strings have two distinct endpoints and their propagation produces worldsheets with boundaries. Similarly, it is possible to distinguish between oriented and unoriented strings depending on whether the worldsheet is an orientable manifold or not.

The Nambu-Goto action is the simplest action that can be built. However, the presence of the square root  makes quantization difficult. In order to prevent this issue, a new auxiliary field called $g_{ab}$ is added. This field serves as a metric for the worldsheet, which is now considered an independent space rather than an embedding on the original spacetime. The resulting action is named after Polyakov \cite{Polyakov:1981rd} and takes the form
\begin{equation}
    S_P=-\frac{T_s}{2}\int_\Sigma d\sigma d\tau \sqrt{-\textrm{det}g}\ g^{ab}(\tau,\sigma)\partial_a X^M \partial_b X^N \eta_{MN}\,.
    \label{basic-eq: polyakov action}
\end{equation}
Note that in the Polyakov action we have two different metrics for the worldsheet, the one induced from the spacetime in which the string is propagating ($h$) and an intrinsic one ($g$), in principle unrelated to the first. From this point of view, the new action describes a two-dimensional field theory coupled to two-dimensional gravity (non-dynamical), independently of  the number of spacetime dimensions. Satisfying the equations of motion derived from the Polyakov action requires $h_{ab}\propto g_{ab}$ and both metrics need to be conformally related. It is then easy to see that Nambu-Goto and Polyakov actions are equivalent at the classical level. 

Let us now discuss the symmetries of the actions. Both of them have $d$-dimensional Poincaré invariance as a global symmetry from the worldsheet viewpoint and a two-dimensional diffeomorfism invariance (invariance under redefinitions of the parameters $\tau,\sigma$) as a gauge symmetry of the worldsheet. However, they are not interchangeable in this regard. One of the main advantages of Polyakov action is the existence of an additional symmetry: Weyl invariance. This is a two-dimensional transformation that leaves the same embedding functions but modifies the worldsheet metric by $g_{ab}'=\Omega(\tau, \sigma)g_{ab}$. A consequence of this additional symmetry is that the classical string is conformally invariant. Indeed,  Polyakov action illuminates a fundamental property of string theory: conformal invariance. Its preservation will put many constraints on the kind of geometries and allowed interactions. It will also permit to solve the equations of motion and provide great insight into the nature of gravity.

The aforementioned symmetries can be used to remove the degrees of freedom of the intrinsic metric $g_{ab}$. In particular, it is possible to restrict the worldsheet geometry to be flat without loss of generality. The resulting gauge, known  as the conformal gauge, is then given by $g_{ab}=\eta_{ab}$. In the new coordinates ($\xi$), the Polyakov action becomes
\begin{equation}
    S_P=-\frac{T_s}{2}\int_\Sigma d^2\xi \eta^{ab}\partial_a X^M\partial_b XN \eta_{MN}\,.
\end{equation}

At this point, it is straightforward to derive the equations of motion from variational analysis, resulting in a set of $d$ 2-dimensional independent wave equations
\begin{equation}
    (\partial^2_t-\partial_\sigma^2)X^M=0\,.
\end{equation}

In order to find a physical string solution for the classical theory, additional constraints will be needed. The first one is the tracelessness of the energy-momentum tensor of the worldsheet,  which arises from the conformal invariance. It provides a set of equations known as Virasoro constraints. The second, inherent to any partial differential equation, are the boundary conditions and depend on the nature of the string considered. Periodicity conditions are imposed for closed strings, while Dirichlet or Neumann conditions are required for open strings.
\begin{align}
    \textrm{Closed:} & \qquad X^M(t,\sigma+l)=X^M(t,\sigma)\,,\\
    \textrm{Open:}& \qquad \begin{cases}
    \partial_\sigma X^{M}|_{\sigma=0,l}=0\,, \qquad &  (\textrm{Neumann})\\
    \delta X^{M}|_{\sigma=0,l}=0\,. \quad &(\textrm{Dirichlet})\footnotemark
    \end{cases}
    \label{basic-eq: boundary conditions bosonic string}
\end{align}
\footnotetext{Dirichlet boundary conditions break Poincaré invariance. As it will be discussed when addressing superstring theories, Dirichlet conditions are associated with lower-dimensional objects known as D-branes.}
Under the periodic boundary conditions, the classical solution for the closed string is a combination of left and right-moving waves:
\begin{equation}
    X^M(\sigma,t)=X^M_R(\tau-\sigma)+X^M_L(\tau+\sigma)\,,
    \label{basic-eq: LR closed string}
\end{equation}
with
\begin{align}
\begin{split}
    X^M_R(\tau-\sigma)= & \frac{1}{2}x^M+\frac{1}{2}\alpha'p^M \cdot(\tau-\sigma) +i\sqrt{\frac{\alpha'}{2}}\sum_{n\neq 0}\frac{1}{n}\alpha_{n}^M e^{-\frac{2\pi}{l}in(\tau-\sigma)}\,,\\
    X^M_L(\tau+\sigma)= & \frac{1}{2}x^M+\frac{1}{2}\alpha'p^M \cdot(\tau+\sigma) +i\sqrt{\frac{\alpha'}{2}}\sum_{n\neq 0}\frac{1}{n}\tilde{\alpha}_{n}^M e^{-\frac{2\pi}{l}in(\tau+\sigma)}\,,
\end{split}
\label{basic-eq: closed string oscillation expansion}
\end{align}
where $x^M$ are the spacetime coordinates for the center of mass at $\tau=0$, $p^M$ its momentum and the coefficients $\alpha_n^M,\tilde{\alpha}_n^M$ represent the amplitudes of the $n$-mode momentum for left and right movers respectively. Similar expressions can be derived for the open string and the two boundary conditions available on its two borders. In contrast with the closed string, only one set of oscillator modes will be present, since the boundary reflects right modes into left modes and vice versa. We will study its properties in more detail when we address the supersymmetric theories.

\subsection{Quantization of the Closed String}

There are three ways in which the classical string theory can be quantized: the canonical quantization, the path-integral quantization and the light-cone quantization. They mainly differ in their implementation of the Virasoro constraints, each with its own advantages and disadvantages. The first two focus on the Conformal Field Theory defined over the worldsheet and preserve manifest Lorentz invariance, but are populated by ghost states.\footnote{Unphysical states required to keep gauge symmetries. For a theory to be well defined, they must decouple from the physical Hilbert space.} The light-cone quantization fixes all remaining gauge freedom  and solves the Virasoro constraints explicitly before quantizing, but loses manifest Lorentz invariance in the process. We will focus on this last method.

In the previous section, the symmetries of the action were employed to fix the conformal gauge, characterized by a  flat intrinsic worldsheet metric. Such condition still leaves some redundancies due to residual gauge freedom that can be removed with a suitable Weyl rescaling. This allows to introduce the spacetime light-cone coordinates
\begin{equation}
    X^{\pm}=\frac{1}{\sqrt{2}}(X^0\pm X^1)\,, \qquad \textrm{with} \quad X^+(\tau,\sigma)=\tau\,.
\end{equation}
The remaining coordinates $X^i$ with $i=2,\dots, d$ are kept the same. Let $\xi_L,\xi_R$ be the new parameters of the worldsheet after the gauge fixing. The Virasoro constraints are then given by the following linear relations in $X^-$
\begin{equation}
    \partial_{\xi_L}X^-_L=\frac{1}{2}\left(\partial_{\xi_L}X^i_L\right)^2\,, \quad \partial_{\xi_R}X^-_R=\frac{1}{2}\left(\partial_{\xi_R}X^i_R\right)^2\,,
\end{equation}
which fixed $X^-$ in terms of $X^i$, leaving the center of mass term as the only degree of freedom of $X^-$. Studying the Lagrangian associated to the Polyakov action in the light-cone gauge, it is easy to check that the center of mass momenta for $X^+$ and $X^-$ satisfy
\begin{equation}
    p_-=-p^+=-\frac{l}{2\pi\alpha'}\,.
\end{equation}
As a consequence of this analysis, there are actually only $D-2$ oscillation modes for the closed string, whose expressions can be derived from \eqref{basic-eq: LR closed string} and \eqref{basic-eq: closed string oscillation expansion}
\begin{equation}
    X^i=x^i+\frac{p^i}{p^+}\tau+i\sqrt{\frac{\alpha'}{2}}\sum_{n\neq0}\left[\frac{\alpha_n^i}{n}e^{-2\pi in(\tau+\sigma) l}+\frac{\tilde{\alpha}^i_n}{n}e^{-2\pi i n(\tau-\sigma)l} \right]\,.
    \label{basic-eq: boson oscillatory expansion light cone}
\end{equation}
Quantization is achieved by promoting the worldsheet degrees of freedom to operators subjected to canonical commutation relations.
\begin{equation}
\begin{gathered}
        [x^-,p^+]=-i\,, \quad [X^i,p_j]=i\delta^i_j\,,  \quad [\alpha_m^i,\tilde{\alpha}_n^j]=0\,,\\
        [\alpha_m^i,\alpha_n^j]=[\tilde{\alpha}^i_m,\alpha^j_n]=m\delta^i_j\delta_{m,-n}\,.
\end{gathered}
\end{equation}
The resulting Hamiltonian is given by the combination of the free center of mass motion plus a sum of two infinite sets of simple harmonic oscillators, each with a different frequency.
\begin{equation}
    H=\sum_{i=2}^{d-1}\frac{p_i^2}{2p^+}+\frac{1}{\alpha'p^+}\left[\sum_i\sum_{n>0}(\alpha^i_{-n}\alpha^i_n+\tilde{\alpha}^i_{-n}\tilde{\alpha}^i_n)+E_0+\tilde{E}_0\right]\,,
\end{equation}
where $E_0$ and $\tilde{E}_0$ are the zero point energies associated to the left and right-moving modes respectively.

Following the standard techniques of Quantum Field Theory, the Fock space  can be  built defining the vacuum state $\ket{0}$ as the common kernel of the annihilation operators $\alpha^i_n$ and $\tilde{\alpha}^i_n$ for $n>0$. All other states are built by acting on $\ket{0}$ with the creation operators given by $n<0$. The infinite set of two dimensional quantum harmonic oscillation states describe the spectrum of spacetime particles of the theory.

It is worth noting the existence of one last degree of freedom in the worldsheet parametrization: the selection of the reference line $\sigma=0$ for the periodic spatial coordinate. Physical states should not depend on this arbitrary choice, which imposes that the number operators of left moving ($N$) and right-moving ($\tilde{N}$) systems are the same. This property, known as the level-matching constraint, is the only link between both sectors.

From the spacetime perspective, each oscillation state corresponds to a particle of mass
\begin{equation}
    M^2=-p^2=2p^+H-(p_i)^2=\frac{2}{\alpha'}(N+\tilde{N}+E_0+\tilde{E}_0)\,.
    \label{basic-eq: closed boson string mass}
\end{equation}
Hence, the mass of the state grows with the number of oscillation modes that are turned on. Using the commutation relations, the Virasoro constraints and normal ordering requirements, it is possible to deduce that the zero-point energy corresponds to the following divergent sum
\begin{equation}
    E_0=\tilde{E}_0=\sum_{i=2}^d \frac{1}{2}\sum_{n=0}^\infty n\,,
\end{equation}
which after a convenient regularization using the Riemann Zeta function gives
\begin{equation}
    E_0=-\frac{d-2}{24}\,.
\end{equation}
Therefore the lightest states of the theory have the following masses
\begin{align}
  \ket{0}: &\hspace{-2cm} & N=\tilde{N}=0 & \Rightarrow M^2=-\frac{2}{\alpha'}\frac{d-2}{12}\,,\\
  \alpha_{-1}^i\tilde{\alpha}_{-1}^j\ket{0}: &\hspace{-2cm}  &N=\tilde{N}=1 &\Rightarrow M^2=\frac{2}{\alpha'}\left(2-\frac{d-2}{12}\right)\,.
  \label{basic-eq: light states bosonic string}
\end{align}

We conclude that the lightest states of the closed string are a scalar with negative squared mass (tachyon) for $d>2$ and a two index tensor field that can be split in its traceless symmetric part ($G_{MN}$), its antisymmetric part ($B_{MN}$) and its trace ($\phi$). They correspond to the d-dimensional graviton, a d-dimensional 2-form and a d-dimensional scalar, known as the dilaton.

\subsection{Anomalies, critical dimension and tachyons}

Now that the string has been quantized, it is essential to consider whether the classical symmetries of the theory have survived the process or, on the contrary, some anomalies have appeared. In particular,  Lorentz invariance is no longer manifest under the light cone quantization, only the $SO(d-2)$ subgroup associated to the coordinates $X^i$. In order to find the conditions under which the full Lorentz invariance is recovered, it is helpful to consider Wigner classification of the representations of the Poincaré group. This result splits the information that characterizes the representation into two parts: the momentum of the states and the representation of little group (stabilizer) associated to that momentum ($SO(d-1)$ for massive particles and $SO(d-2)$ for massless ones). In \eqref{basic-eq: light states bosonic string}, the first excited states are incompatible with an $SO(d-1)$ representation but fit nicely in the tensor representation of $SO(d-2)$. Therefore, to restore Lorentz invariance those states must be massless, which fixes the allowed spacetime dimension of our theory to $d=26$.

The striking requirement of 22 extra spatial dimensions that have been so far invisible to any experiment might raise some suspicions regarding the validity of the Zeta function regularization. This initially dubious step is legitimized by the results obtained from the other two quantization procedures, which also demand the critical dimension $d=26$ to decouple the ghosts from the physical states. 

All three approaches are joined together from the perspective of the conformal symmetry of classical Polyakov action. This symmetry is a key element that allows to solve the quantum String Theory completely. Consequently, the quantization must be performed in a way that avoids any conformal anomaly, which fixes the dimension the theory can live in. Such anomalies manifest as ghosts in the canonical and path integral quantization and as loss of manifest Lorentz invariance in the light-cone quantization. 

It is important to note that imposing $d=26$ has a negative implication: the ground state has $M^2= -4/\alpha'$ and hence becomes tachyonic, rendering the theory unstable. This problem will be addressed in section \ref{basic-sec: superstrings}.

\subsection{Curved spacetime and background fields}

As it was discussed in \eqref{basic-eq: light states bosonic string}, the light spectrum of the closed strings generates the graviton, a 2-form and a scalar (dilaton). The graviton represents a perturbation in the spacetime metric, which we took to be flat. Since deformations of the metric are inherent to the theory, it is reasonable to consider a string theory defined on curved backgrounds and replace $\eta_{MN}$ in \eqref{basic-eq: polyakov action} by a general metric $G_{MN}[X]$ that depends on the coordinates $X^M(\tau,\sigma)$.\footnote{Because of its similarities with some actions describing strong interactions, the resulting action is historically known as a 2d non-linear sigma model.} Such metric can be understood as the summation of the effects of all gravitons present in the background through which the studied string propagates. Then the updated action is
\begin{equation}
    S_P^G=-\frac{T_s}{2}\int_\Sigma d\sigma d\tau \sqrt{-\textrm{det}g}g^{ab}(\tau,\sigma)\partial_a X^M \partial_b X^N G_{MN}[X]\,.
\end{equation}
The other two massless states (antisymmetric tensor field $B_2$ and dilaton $\phi$) can also generate a background with which the string can interact. This effect is described by the following action, which must be added to the generalization of \eqref{basic-eq: polyakov action} to curved backgrounds:
\begin{equation}
    S_B=\frac{1}{2\pi\alpha'}\int_{\Sigma}B_2\,, \quad S_\phi=\frac{1}{4\pi}\int_\Sigma \sqrt{-g}R[g]\phi\,,
\end{equation}
with $R[g]$ the curvature scalar of the worldsheet metric $g$. 

The action $S_\phi$ plays an important role when describing interacting strings. Since 2d gravity is not dynamical, the integral will not depend on the metric. In the case of constant dilaton, the result is a topological invariant known as the Euler Characteristic ($\chi)$ 
\begin{equation}
    S_\phi=\chi\phi  =(2-2n_g-n_b-n_c)\phi\,,
\end{equation}
where $n_g$ is the number of handles of the worldsheet manifold, $n_b$ the number of boundaries and $n_c$ the number of crosscaps (only present in non-orientable worldsheets).

\subsection{Interactions and perturbative expansions}
\label{basic-susbsec: interaction and expansion}

Strings can have non-trivial interactions in the spacetime theory, as befits a candidate to substitute the notion of particle as the fundamental object in Physics. Considering for simplicity a flat background ($G_{MN}=\eta_{MN}$, $B_{MN}=0$), such interactions manifest as different topologies in the (non-interacting) worldsheet. From the perspective of the conformal field theory of the worldsheet, the path integral formulation can be used to compute string scattering amplitudes. Instead of summing over all paths connecting the initial and final states, now a sum over all possible worldsheet (WS) geometries must be performed, as shown in figure \ref{basic-fig: genus expansion}. The schematic formula for the scattering amplitude is
\begin{equation}
    \braket{\textrm{out}|\textrm{evolution}|\textrm{in}}\sim \frac{1}{Z}\sum_{\textrm{WS topologies}}[\mathcal{D}X]e^{-S_P[X]-S_\phi}\mathcal{O}_{\rm in}[X]\mathcal{O}_{\rm out}[X]\,,
\end{equation}
where $\mathcal{O}[X]$ are the vertex operators associated to the initial and final states and $Z$ is the partition function, which takes the form
\begin{equation}
Z=\sum_{\textrm{WS topologies}}[\mathcal{D}X]e^{-S_P[X]-S_\phi}\,.   
\label{basic-eq: partition function}
\end{equation}

\begin{figure}[htbp]
    \centering
    \includegraphics{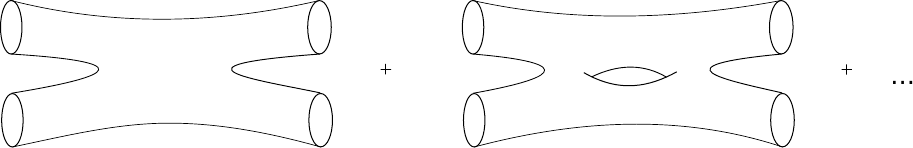}
    \caption{String generalization of the scattering process of four point-particles. Each component of the sum contributes with a power of the string coupling $g_s$ that depends on the genus of the respective worldsheet.}
    \label{basic-fig: genus expansion}
\end{figure}

Due to the abundant symmetries of the action, the sum over worldsheet topologies must be taken modulo diffeomorphisms and Weyl transformations. Then, describing the string interactions requires classifying the topologically distinct two-dimensional surfaces.  It is clear that the interaction between closed strings can only be mediated through surfaces without endpoint boundaries.\footnote{Note that there is another type of boundaries, known as source boundaries, corresponding to the initial and final string configurations of the system. They are present in interactions between closed strings as the one described in figure \ref{basic-fig: genus expansion}.} The set of topologically inequivalent oriented dimensional surfaces without boundaries is completely characterized by the number of handles (genus). Therefore, the sum over oriented topologies can be ordered as a sum over manifolds with different values of the genus $n_g$. The weight of each term is controlled by the contribution of $e^{-S_\phi}=e^{-\chi \phi}$. Hence, the constant background value of the dilaton plays the role of the string coupling of the theory. Defining
\begin{equation}
    g_s=e^{\phi}\,,
\end{equation}
a worldsheet with genus $n_g$ is  weighted by a factor $e^{-(2-2n_g)\phi}=g_s^{-(2-2n_g)}$. This expansion can be easily generalized to open strings by including worldsheets with boundaries. Denoting by $n_b$ the number of borders, each distinct topology is weighted by $g_s^{-(2-2n_g-n_b)}$. Similarly, unoriented strings can be added accounting for the number of crosscaps $n_c$ of the surface. We conclude that the perturbative expansion is ordered by powers of the string coupling depending on the Euler characteristic of the worldsheet that mediates the interaction:
\begin{equation}
    g_s^{-\chi}\,, \quad \chi=2 -2n_g-n_b-n_c\,.
\end{equation}
This shows that the string coupling is not an external parameter but the vacuum expectation value of one of its fields. In fact, the absence of external parameters is a general feature of string theory that sets it apart from quantum field theories. 

A fundamental result of the extended nature of strings is the smearing of the interaction vertices along a region of typical size $L_s\approx 1/M_s$ (as can be seen in figure \ref{basic-fig: genus expansion}). This eliminates the ultraviolet divergences of the quantum  field theory, which then becomes an effective low energy limit of string theory with a cut-off $M_s$. Moreover, the malleable structure of the worldsheet allows to deform it, transforming ultraviolet regimes into infrared ones. The duality that arises between low and high energy systems is another consequence of diffeomorphism and conformal invariance.

\begin{tcolorbox}[breakable, enhanced,  colback=uam!10!white, colframe=uam!85!black, title=Modular Invariance]
\begin{small}

Let us consider the effects of conformal invariance in more detail for the first non-trivial entry of the genus expansion: the closed oriented string one-loop vacuum amplitude. To do so, we will focus on the entry associated to $n_g=1$, $n_b=n_c=0$ in the partition function expansion \eqref{basic-eq: partition function}. The worldsheet geometry corresponds  to a torus, which describes a closed string propagating and closing back to itself. A sum must then be performed over all possible inequivalent worldsheet geometries with the topology of a two-torus. The two-torus can be described as the complex manifold resulting from modding out the complex plane by translation vectors in a two-dimensional lattice. Hence, letting $z=\sigma+i\tau$ be the complex coordinate, the torus is built through the identification $z\sim z+l$ and $z\sim z+ul$, with $u=u_1+iu_2$ and $u_1,u_2,l\in \mathbb{R}$. The displacement $l$ is simply the length of the string and the factor $u$, known as the complex structure of the torus, distinguishes between different worldsheet geometries. It is important to note, however, that this identification is not one-to-one: several values of $u$ can describe the same lattice and hence the same torus. Such changes in $u$ amount to global diffeomorphisms on the torus that are not smoothly connected with the identity and are generated by the transformations $u\rightarrow u +1$ and $u\rightarrow -\frac{1}{u}$. Together they generate the modular group of the torus, whose general element acts as
\begin{equation}
    u\rightarrow \frac{au+b}{cu+d}\,,\quad  \rm{with\ } a,b,c,d\in \mathbb{Z} \quad \rm{and\ } ad-bc=1\,, 
    \label{basic-eq: modular group}
\end{equation}
which is just a generic element of $SL(2,\mathbb{Z})$.

The partition function should be invariant under modular transformations, since they just reflect the arbitrariness in the choice of the worldsheet parametrization. The identification of geometries by the modular group and in particular under the action $u\rightarrow -1/u$ provides the map between UV and IR strings propagating regimes described before. 

\end{small}
\end{tcolorbox}

The analysis of the interactions discussed above is still valid in the presence of non-trivial backgrounds but adds additional challenges. When considering fluctuations of the metric, 2-form $B_2$ and dilaton, the worldsheet action $S_P^G+S_B+S_\phi$ becomes interacting and the theory is generally not exactly solvable. The common approach then is to perform a perturbative expansion around the free theory using $\alpha'/R^2$ (with $R$ the curvature radius) as the expansion parameter. Therefore, String Theory is described through a double expansion. The first one is the genus expansion, which sums over different topologies weighted by the string coupling $g_s$ and is equivalent to the spacetime loop expansion in QFT. The second one is the $\alpha'$ expansion, which for each fixed worldsheet topology describes that worldsheet loop expansion weighted by $\alpha'/R^2$.

\section{Superstrings}
\label{basic-sec: superstrings}

\subsection{Worldsheet supersymmetry and Type II theories}
\label{basic-subsec: Type II theories}

The theory that has been considered until this point is known as the bosonic string, since all its degrees of freedom are scalars and tensors (integer spin). It has provided an excellent framework to develop the basic notions of the new paradigm but is hindered by two challenges: the presence of a tachyon and the absence of spacetime fermions in the spectrum of both open and closed strings. The nature and consequences of the tachyon are still a matter of active research but the lack of fermions entails an insurmountable obstacle for a theory aiming to describe our Universe.  The simplest way to solve such problem in a mathematically consistent manner is to modify the worldsheet field content introducing supersymmetry.

In the bosonic string, the worldsheet description consists of $d$ scalar fields $X^M$ coupled to two-dimensional gravity. Each of these scalars is now associated with a new fermionic two-dimensional spinor field $\psi^M$. Both sets of fields are coupled to $N=1$ supergravity, whose multiplet contains the worldsheet metric $g_{ab}$ and the gravitino $\chi_a$. The resulting action, analogous to Polyakov's, is
\begin{equation}
    S_P=\frac{1}{4\pi\alpha'}\int \sqrt{\textrm{det}g}\ \eta_{MN}\left[g^{ab}\partial_aX^M \partial_b X^N +\frac{i}{2}\psi^M\slashed\partial \psi^N+\frac{i}{2}\chi_a\gamma^a\gamma^b\psi^M\left(\partial_bX^N-\frac{i}{4}\chi_b\psi^N\right)\right]\,.
\end{equation}
The bosonic fields accept the same expansion and quantization as in the previous theory. Introducing again the light-cone gauge, only $d-2$ bosonic fields $X^i$ are dynamical. Consistent with supersymmetry, the generalization of Virasoro constraints leaves only the corresponding transverse fermionic fields $\psi^i$ as independent degrees of freedom. In regards to the fermionic fields, new boundary conditions are also demanded. Focusing on closed strings, the periodicity requirement has an additional sign choice freedom, since fermionic fields always appear quadratically on physical observers. Hence, it is possible to distinguish between antiperiodic and periodic boundary conditions, commonly known as Neveu-Schwarz and Ramond conditions respectively \cite{Neveu:1971rx,Ramond:1971gb}.
\begin{align}
    \textrm{Neveu-Schwarz (NS)}:& \quad  \psi^i(\tau,\sigma+l)=-\psi^i(\tau,\sigma)\,,\\
     \textrm{Ramond (R)}:& \quad  \psi^i(\tau,\sigma+l)=\psi^i(\tau,\sigma)\,.
\end{align}

As was the case for the bosonic degrees of freedom, performing a split between the left and right sectors of the fermionic fields is also possible.  Boundary conditions can be assigned independently to both sectors, but must be kept the same for all values of $M$  to preserve Lorentz invariance. Therefore, there are four families of closed strings depending on this choice: NS-NS, NS-R, R-NS and R-R.

Setting some subtleties aside, the oscillatory expansion and quantization can be performed in an analogous manner to the bosonic fields, imposing anticommutation instead of commutation relations. The light spectrum of the  resulting full theory is dominated by the fermionic creation operators (the bosonic operators give rise to heavier states), which has several important consequences.

The first consequence is that the requirements for cancelling the conformal anomaly are modified. The critical dimension for the superstring theory is $d=10$. Hence, six additional dimensions must be dealt with to achieve an effective description of our Universe. Particular aspects of such process will be the main focus of this thesis. From now on we will assume $d=10$ unless stated otherwise.

The second consequence is that at low energies we can focus only on the fermionic sector of the worldsheet in order to build a supergravity effective action. To construct it, one must take into account that the oscillator operators differ depending on the choice of boundary conditions. For antiperiodic (NS) conditions, half-integer modes are required ($\psi^i_{r+1/2}$), while periodic conditions (R) demand integer modes ($\psi^i_r$). These effects propagate to the spectrum of light states, including the ground state.

The NS ground state $\ket{0}_{NS}$ is defined by the relations $\psi^i_{r+1/2}\ket{0}=0$, $\forall r>0$. It is very similar in nature to the bosonic string theory ground state. In contrast, the construction of the R ground state is more involved, since the operators $\psi_0^i$ do not increase the energy of the string state, leading to a degeneracy. Therefore, in addition to the element $\ket{0}_R$ given by   $\psi^i_{r}\ket{0}=0$, $\forall r>0$, one must consider the result of the different combinations of operators $\psi_0^i$ acting over $\ket{0}$. The consequence is that the ground state of the R sector behaves like a 16-component spinor representation of $SO(8)$, which can be decomposed into two irreducible spinor representations of opposite chirality. The light spectrum is summarized in table \ref{basic-table: R and NS spectrum},  together with the associated masses and representation. There are two important aspects to highlight. Firstly, the NS sector generates a tachyonic state which, as in the bosonic string, leads to instabilities. Secondly, the Ramond sector generates spacetime fermions, achieving the goal that motivated the introduction of supersymmetry. This non-trivial result (the NS sector does not have fermions) highlights again the deep connection between the worldsheet and the spacetime: requiring supersymmetry on the worldsheet provides fermionic states on the spacetime.  
\begin{table}[htbp]
\centering
\def\arraystretch{1.5}
\begin{tabular}{c|ccc}
\multicolumn{1}{l|}{} & State                       & $\frac{\alpha'}{2}M^2$ & $SO(8)$        \\ \hline
\multirow{2}{*}{NS}   & $\ket{0}_{NS}$              & $-\frac{1}{2}$         & $\mathbf{1}$   \\
                      & $\psi^i_{-1/2}\ket{0}_{NS}$ & $0$                    & $\mathbf{8}_V$ \\ \hline
\multirow{2}{*}{R}    & $\ket{\mathbf{8}_C}$        & $0$                    & $\mathbf{8}_C$ \\
                      & $\ket{\mathbf{8}_S}$        & $0$                    & $\mathbf{8}_S$
\end{tabular}
\caption{Light spectrum of the NS and R sectors.}
\label{basic-table: R and NS spectrum}
\end{table}

The full theory will combine the right and left moving sectors. Each one allows independent R or NS boundary conditions, but  must satisfy the level-matching constraint and hence $M_L^2=M_R^2$. To obtain a consistent construction, the gluing of both sectors must be carefully performed, since not all possible states are allowed in the final theory. In particular we demand that the theory is tachyon free and that the one-loop vacuum amplitude preserves modular invariance. The appropriate selection is obtained through the Gliozzi–Scherk–Olive (GSO) projection \cite{Gliozzi:1976qd} and it is implemented with the operator $(-1)^F$, which anticommutes with every fermionic oscillator. When acting over the NS sector, it removes the even number fermionic oscillators, including the tachyonic state. When acting over the R sector, two options are available: it either selects the chiral spectrum $\mathbf{8}_S$ and removes $\mathbf{8}_C$ or vice versa. This final choice has to be taken independently in the left and right-moving sectors. Due to parity relations, there are  only two different theories based on this election. If the same choice is made on both sectors we obtain a chiral theory known as type IIB. If we instead choose different GSO projections for the R sector of left and right-moving modes, the result is a non-chiral theory known as type IIA.  It is worth noting that the two theories are spacetime supersymmetric, displaying once again the powerful and useful constraint provided by the nature of worldsheet (modular invariance), and all the surviving low-energy states are massless. We summarize the field content in tables \ref{basic-table: type IIB spectrum} and \ref{basic-table: type IIA spectrum}.

\begin{table}[htbp]
\centering
\def\arraystretch{1.5}
\begin{tabular}{cccc}
Sector & $\ket{\ }_L\otimes \ket{\ }_R$        & $SO(8)$                      & Field content                         \\ \hline
NS-NS & $\mathbf{8}_V\otimes\mathbf{8}_V$ & $\mathbf{1}+\mathbf{28}_V+\mathbf{35}_V$ & $\phi, B_{MN},G_{MN}$           \\
NS-R   & $\mathbf{8}_V\otimes\mathbf{8}_C$ & $\mathbf{8}_S+\mathbf{56}_S$ & $\lambda_\alpha^1,\psi^1_{M\alpha}$ \\
R-NS   & $\mathbf{8}_C\otimes\mathbf{8}_V$ & $\mathbf{8}_S+\mathbf{56}_S$ & $\lambda_\alpha^2,\psi^2_{M\alpha}$ \\
R-R   & $\mathbf{8}_C\otimes\mathbf{8}_C$ & $\mathbf{1}+\mathbf{28}_C+\mathbf{35}_C$ & $a,C_{MN},C_{MNLK}$
\end{tabular}
\caption{Type IIB massless spectrum.}
\label{basic-table: type IIB spectrum}
\end{table}

\begin{table}[htbp]
\centering
\def\arraystretch{1.5}
\begin{tabular}{cccc}
Sector & $\ket{\ }_L\otimes \ket{\ }_R$        & $SO(8)$                      & Field content                         \\ \hline
NS-NS & $\mathbf{8}_V\otimes\mathbf{8}_V$ & $\mathbf{1}+\mathbf{28}_V+\mathbf{35}_V$ & $\phi, B_{MN},G_{MN}$           \\
NS-R   & $\mathbf{8}_V\otimes\mathbf{8}_S$ & $\mathbf{8}_C+\mathbf{56}_C$ & $\lambda_\alpha^1,\psi^1_{M\alpha}$ \\
R-NS   & $\mathbf{8}_C\otimes\mathbf{8}_V$ & $\mathbf{8}_S+\mathbf{56}_S$ & $\lambda_{\dot{\alpha}}^2,\psi^2_{M\dot{\alpha}}$ \\
R-R   & $\mathbf{8}_C\otimes\mathbf{8}_C$ & $\mathbf{8}_V+\mathbf{56}_V$ & $C_{M},C_{MNK}$
\end{tabular}
\caption{Type IIA massless spectrum.}
\label{basic-table: type IIA spectrum}
\end{table}

Both theories share the same NS-NS sector, containing a dilaton $\phi$, a 2-form $B_2$ and a graviton $G_{MN}$. The NS-R and R-NS sectors contain the fermionic degrees of freedom: two Rarita-Schinger fields $\psi_{M\alpha}$ (gravitinos of the spacetime supersymetry) and two spinors $\lambda_\alpha$ (known as dilatinos). These pair of fermionic families share the same chirality in Type IIB theory and are of opposite chirality in Type IIA. Finally, the R-R sector provides additional forms. Type IIB contains a scalar $a=C_0$ (usually called axion), a 2-form $C_2$ and a 4-form $C_4$. Meanwhile, type IIA has odd forms $C_1$ and $C_3$. In both cases, the fields $C_i$ play the role of generalized gauge potentials and the theories display local supersymmetry with $32$ supercharges. Their spectra match the gravity multiplet of chiral (type IIB) and non-chiral (type IIA) 10 $\mathcal{N}=2$ superalgebra.

The 10d low energy effective action for the bosonic sector of type IIB is 
\begin{equation}
\begin{aligned}
    S_{IIB}=&\frac{1}{2\kappa^2_{10}}\int_{\mathcal{M}_{10}} d^{10}x \sqrt{-G}\left[e^{-2\phi}(R+4\partial_{M}\phi\partial^M \phi-\frac{1}{2}|H_3|^2)-\frac{1}{2}|F_1|^2-\frac{1}{2}|\tilde{F}_3|^2-\frac{1}{2}|\tilde{F}_5|^2\right]\\
    &-\frac{1}{4\kappa_{10}^2}\int_{\mathcal{M}_{10}}  C_4\wedge H_3\wedge F_3\,,
\end{aligned}
\label{basic-eq: type IIB action}
\end{equation}
where $\mathcal{M}_{10}$ is the 10-dimensional space (not necessarily Minkowski), the norm $|F_p|^2$ is defined in appendix \ref{ch: ap conventions}, $2\kappa_{10}^2=(2\pi)^7\alpha'^4$ is the 10d gravitational strength and $H_3$ and $F_p$ are the field strengths of the p-forms of type IIB, that is, $H_3=dB_2$ and
\begin{equation}
    F_1=dC_0\,,\quad \tilde{F}_3=F_3-C_0H_3\,, \quad \tilde{F}_5=F_5-\frac{1}{2}C_2\wedge H_3 +\frac{1}{2}B_2\wedge F_3\,.
    \label{basic-eq: Type IIB form flux definitions}
\end{equation}
Strictly speaking, the expression above is not an action but a pseudoaction, since it needs to be complemented with an external self-duality constraint for the 5-form field strength.
\begin{equation}
    \star_{10}\tilde{F}_5=\tilde{F}_5
\end{equation}

The 10d effective type IIA action is built in a similar manner, obtaining
\begin{equation}
    \begin{aligned}
            S_{IIA}=&\frac{1}{2\kappa^2_{10}}\int_{\mathcal{M}_{10}} d^{10}x \sqrt{-G}\left[e^{-2\phi}(R+4\partial_{M}\phi\partial^M \phi-\frac{1}{2}|H_3|^2)-\frac{1}{2}|F_2|^2-\frac{1}{2}|\tilde{F}_4|^2\right]\\
    &-\frac{1}{4\kappa_{10}^2}\int_{\mathcal{M}_{10}}  B_2\wedge F_4\wedge F_4\,,
    \end{aligned}
\end{equation}
where all common elements with the type IIB action share the same definition and
\begin{equation} 
    F_2=dC_1\,, \quad \tilde{F}_4=dC_3-C_1\wedge H_3\,.
\end{equation}
It is important to mention that Type IIA supergravity effective action admits a deformation by a mass parameter $m$, called Romans mass, that plays the role of a background field strength $F_0$. The resulting massive type IIA action modifies the higher form field strengths 
\begin{equation}
    F_2=dC_1+mB_2\,,\qquad \tilde{F}_4=dC_3-C_1\wedge H_3+\frac{1}{2} B_2\wedge B_2\,,
\end{equation}
and includes a kinetic and Chern-Simons term for the new field
\begin{equation}
    S_{IIA {\rm mass}}=\tilde{S}_{\rm IIA}-\frac{1}{4\kappa^2_{10}}\int \sqrt{-G}m^2+\frac{1}{2\kappa_{10}^2}\int mF_{10}\,.
    \label{basic-eq: massive type IIA action}
\end{equation}
It is worth noting that $F_2$ is no longer closed in massive type IIA, instead it satisfies $dF_2=mH$. Therefore, it is convenient to define a twisted exterior derivative $d_H\equiv d-H\wedge$, which is consistent as long as $dH=0$. 

From all this discussion, we conclude that through the use of worldsheet supersymmetry and the GSO projection we have been able to build two consistent quantum theories that describe the dynamics of a graviton together with several generalized gauge fields. To further enrich the content of our theory  with the final goal of describing the field content of the observed Universe, other objects (such as open strings and non-perturbative states) shall be added.

\subsection{Open String}

\subsubsection*{Generalities}

So far, the discussion has been mainly focused on closed strings. One may wonder how open strings arise in superstring theories and what are their properties. Some of the more basic characteristics of open strings were already introduced alongside the closed bosonic string in the previous section. The symmetries of their worldsheets, their quantization and interactions are analogous to its closed version. The two main differences  are the boundary conditions and the fact that open strings cannot exist in isolation. Contrary to closed strings, which constitute a consistent theory on their own, a theory of open strings always requires closed strings. The reason behind this is simple: two open strings interact by joining their end points and locality makes the interaction of the two endpoints of one string indistinguishable from the two endpoints of two different strings. Consequently, an open string can always interact with itself, creating a closed string, as shown in figure \ref{basic-fig: open string interaction}.
\begin{figure}[htbp]
\centering
 \hspace{0.5cm}\begin{subfigure}[t]{0.34\textwidth}
    \centering
    \includegraphics[width=\textwidth]{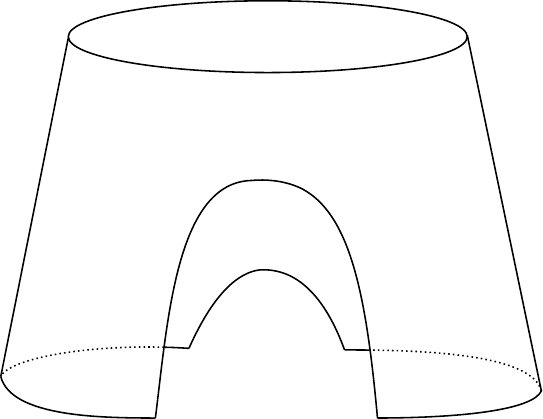}
    \caption{Open string interacting to generate closed strings.}
    \label{basic-fig: open string interaction}
\end{subfigure}\hfill
\begin{subfigure}[t]{0.46\textwidth}
    \centering
    \includegraphics[width=\textwidth]{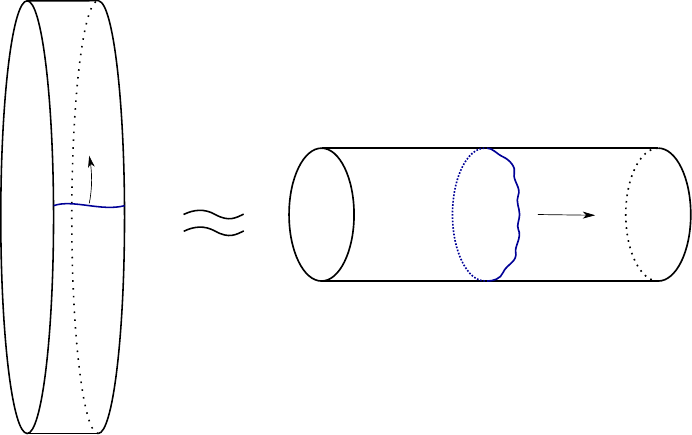}
    \caption{Open-closed duality.}
    \label{basic-fig: open closed duality}
\end{subfigure} \hspace{0.5cm}
\caption{Properties of open string worldsheets.}
\end{figure}

Another interesting property of open strings is the open-closed duality, which relates the one-loop open string amplitude with the propagation of a closed string. The resulting annulus diagram (figure \ref{basic-fig: open closed duality}) can be regarded both as an open string gluing into itself or as a tree-level diagram where a closed string appears from the vacuum and then disappears again. Due to this duality, the requirements of modular invariance for closed strings imposed by the GSO projection extend to open strings. Consequently, the GSO projection must also be applied to open strings in order to guarantee the consistency of the theory.

The final new feature of open strings is that they can have discrete degrees of freedom associated to their endpoints, called Chan-Paton indices. They are non-dynamical and propagate unchanged, providing a way to label each boundary of the worldsheet. The number of allowed choices for the indices, $N$, defines different theories.

\subsubsection*{Boundary conditions}

Similar to the closed string, the open superstring also has two sets of degrees of freedom. The one associated to the spacetime coordinates, split in left and right sectors  ($X^i_L,X^i_R$) and their fermionic partners ($\psi^i_L,\psi^i_R$). Their boundary conditions determine their oscillatory expansion. The bosonic conditions were already discussed in \eqref{basic-eq: boundary conditions bosonic string}, they can be either Dirichlet (D) or Neumann (N).

In addition to the choice of Dirichlet vs Neumann, for the fermionic degrees of freedom the situation is analogous to the closed string: there are two possible options differing in a sign, called Neveu-Schwarz (NS) and Ramond (R). Both pairs of choices combine non-trivially and impose the following relations:
\begin{equation}
\begin{aligned}
        {\rm NN}: \quad & & \psi^i_L&=\psi_R^i \quad \textrm{at $\sigma=0$}\,, &\qquad \rm{DD} \quad & & \psi^i_L&=-\psi^i_R \quad \textrm{at $\sigma=0$}\,,\\
        & & \psi^i_L&=\eta \psi^i_R \quad \textrm{at $\sigma=l$}\,, & & & \psi^i_L&=-\eta\psi^i_R \quad \textrm{at $\sigma=l$}\,,\\
        {\rm ND}: \quad & & \psi^i_L&=\psi_R^i \quad \textrm{at $\sigma=0$}\,, &\qquad \rm{DN} \quad & & \psi^i_L&=-\psi^i_R \quad \textrm{at $\sigma=0$}\,,\\
        & & \psi^i_L&=-\eta \psi^i_R \quad \textrm{at $\sigma=l$}\,, & & & \psi^i_L&=+\eta\psi^i_R \quad \textrm{at $\sigma=l$}\,,       
\end{aligned}
\label{basic-eq: fermionic boundary conditions}
\end{equation}
where $\eta=+1$ in the Ramond sector and $\eta=-1$ in the Neveu-Schwarz sector.

It is important to note that contrary to the closed string, the left and right sectors of the open fermionic oscillators are coupled. Consequently, the degrees of freedom are reduced by half and the dynamics are fully described by just one of the sectors. 

\subsubsection*{NN open string spectrum}

Since Dirichlet conditions break Poincaré invariance, the most natural option is to demand Neumann conditions in all coordinates at both endpoints. Hence, we will impose these boundary conditions unless stated otherwise. 

Quantization of the fermionic degrees of freedom together with the GSO projection provides the light string spectrum of the open superstring, summarized in table \ref{basic-table: massless opens superstring spectrum}. It consists of a 10-dimensional $U(1)$ gauge boson and its supersymmetric partner, the gaugino. Together they constitute a 10d $\mathcal{N}=1$ vector multiplet. Given that type II string theories display $\mathcal{N}=2$ supersymmetry, the inclusion of open strings partially breaks supersymmetry.\footnote{This is closely related to the notion of BPS states and the link between open strings and branes. Extended supersymmetry with non-trivial central charges gives rise to a set of constraints between the mass of state and central charges known as Bogomol’nyi–Prasad–Sommerfield (BPS) bounds. Extended objects (branes) that saturate this bound generally break half of the supersymmetry of the system.} Furthermore, the relation between left and right oscillator of open strings prohibits the coupling of open strings to closed string theories whose left and right content differs. Therefore, open strings only have the potential to couple to type IIB theory. 
\begin{table}[htbp]
\centering
\def\arraystretch{1.5}
\begin{tabular}{ccccc}
 Sector  & State                       & $\alpha'M^2$ & $SO(8)$        & Field content            \\ \hline
NS & $\psi^i_{-1/2}\ket{0}_{NS}$ & 0            & $\mathbf{8}_V$ & $A_M$                  \\
R  & $\ket{8_C}$                 & 0            & $\mathbf{8}_C$ & $\lambda_{\dot{\alpha}}$
\end{tabular}
\caption{Massless spectrum of the open superstring with Neumann boundary conditions.}
\label{basic-table: massless opens superstring spectrum}
\end{table}

The gauge boson $A_M$ provides a new background field to which the worldsheet can couple, adding a term to the action of the form
\begin{equation}
    S_{\partial\Sigma}=\int_{\partial \Sigma} A\,,
\end{equation}
with $\partial \Sigma$ the boundary of the worldsheet $\Sigma$.

The spectrum of \ref{basic-table: massless opens superstring spectrum} assumes trivial Chan-Paton indices (i.e. only one index). Allowing for an arbitrary number $N$ of distinct indices means introducing $N^2$ different generators. This generates $N^2$ $U(1)$ gauge bosons $A_M^{ab}$ and the same number of gauginos. It is possible to show that the bosonic degrees of freedom can be reassembled into a $U(N)$ enhancement of the gauge symmetry. Then, an open string with Chan-Paton indices $ab$ charge $(+1,-1)$ under the gauge factors $U(1)_a$ and  $U(1)_b$ respectively.

\subsubsection*{RR tadpole}

There is an additional consistency condition that a theory with open strings needs to satisfy: the R-R tadpole cancellation condition. The origin of this requirement is the existence of tadpole interactions arising from disk diagrams describing a closed string emitted from the vacuum. Such terms arise from contributions to the effective action of the form
\begin{equation}
    Q\int dX^{10}\varphi(X)\,,
\end{equation}
where $Q$ is the coefficient of the disk tadpole and $\varphi$ is the closed string field. Poincaré invariance greatly limits the kind of fields that can enter the tadpole, the only one allowed being the 10-form $C_{10}$. Since the spacetime of superstring theories has ten dimensions, $C_{10}$ is not dynamical ($dC_{10}=0$). The associated equation of motion becomes a constraint $Q=0$.

Even though the previous reasoning is framed from the perspective of closed strings, the open-closed duality propagates this condition to the open sector. In fact, the tadpole can be regarded as a particular limit of the annulus diagram (open oriented string one-loop amplitude). Requiring that the coefficient of the tadpole vanishes demands that the Chan-Paton indices verify $N=0$. Consequently, the theory cannot have open oriented strings. 

We conclude that it is not consistent to couple a 10d Poincaré invariant open oriented string to any of the two Type II superstring theories. Adding open strings will  require further modifications of the theory such as adding objects that break Poincaré invariance or introducing unoriented strings. 

\subsubsection{DD open string spectrum and Branes}

Given the previous results, an interesting avenue to keep exploring type II theories is to consider a mixture of Neumann and Dirichlet boundary conditions for open strings. This will partially break Poincaré invariance, evading the problems that arose in the original constructions.

Consider then an open string whose fermionic worldsheet field content is given by $\psi^M$ with NN conditions for  $M=\mu=2,\cdots, p$ and DD conditions for $M=i=p+1,\cdots, 9$ (recall that $M=0,1$ are fixed in the light-cone gauge). For each DD boundary condition there is a sign flip in the Ramond sector between the left and right-moving components, as  seen in \eqref{basic-eq: fermionic boundary conditions}. Demanding compatibility of this relation with the GSO projection over the left- and right-moving sectors selects the possible values of p for type IIA (even) and type IIB (odd) theories.

Regarding the massless spectrum, the field content is very similar to the pure NN open superstring sector, albeit restricted to a p-dimensional space. The oscillator modes associated with the Dirichlet conditions contribute with additional scalar fields. The full spectrum, described in table \ref{basic-table: massless opens superstring spectrum DD+NN}, can be grouped into a $p+1$-dimensional $U(1)$ vector supermultiplet with $16$ supercharges.

\begin{table}[htbp]
\centering
\def\arraystretch{1.5}
\begin{tabular}{c|cccc}
Sector   & State                       & $\alpha'M^2$ & $SO(p-1)$        & Field content            \\ \hline
\multirow{2}{*}{NS}  & $\psi^\mu_{-1/2}\ket{0}_{NS}$ & 0            & Vector & $A_\mu$                  \\
 & $\psi^i_{-1/2}\ket{0}_{NS}$ & 0            & Scalar & $\phi^i$                  \\
R  & $\ket{8_C}$                 & 0            & Spinor & $\lambda_{\dot{\alpha}}$
\end{tabular}
\caption{Massless spectrum of the open superstring with Neumann boundary conditions in coordinates $i=1,\cdots p$ and Dirichlet conditions in coordinates $m=p+1,\cdots, 9$.}
\label{basic-table: massless opens superstring spectrum DD+NN}
\end{table}

The fact that strings are fixed at some coordinates $X^m$ can seem initially unphysical. As it will be detailed later, they are actually attached to non-perturbative objects known as D-branes (``D'' standing for Dirichlet conditions). These new objects are hypersurfaces spanning along the coordinates $X^i$ and localized on the rest. Open strings can move their endpoints inside the hypersurface but cannot break free from it. From this perspective, open string modes can be understood as elements describing the dynamics of the brane: the scalar fields $\phi^i$ parametrize the embedding of Dp-brane worldvolume\footnote{Generalization of the notion of worldline and worldsheet to higher dimensional objects.} on spacetime through its relation with the transverse space $\mathbb{R}^{9-p}$, while the gauge field $A^\mu$ generates a worldvolume flux that lives inside the brane.

\subsection{Compactifications}
\label{basic-subsec: compactifications}

Up until this point we have developed the concepts and framework to describe a self-consistent theory with  fermions, gauge fields and a graviton. However, one problem overtly challenges our state of the art observations: it requires nine spatial dimensions. A justification is thus required to address why we do not see the extra dimensions. The very nature of gravity can beautifully explain this as a curvature of spacetime. In fact, the canonical procedure to obtain dimensional reduction, known as Kaluza-Klein compactification, was developed in the context of general relativity long before the emergence of String Theory. The basic idea is to assume that the theory exits on a curved background which factorizes as 
\begin{equation}
    M_{10}=M_{4}\times X_6\,,
\end{equation}
where $M_4$ is the space we live in and $X_6$ is a compact manifold (the internal space) encompassing the six additional dimensions. Compact manifolds are bounded and thus described by a specific typical size. If this size is small enough in comparison with the energy scale at which we can experimentally operate, we will not be able to interact with extra dimensions and excite their degrees of freedom, which will therefore remain frozen and hidden. However, as we will see later on, some geometrical properties of the internal space can permeate through and provide meaningful predictions for the 4-dimensional effective theory. 

Even though dimensional reduction from 10 to 4 dimensions is the most phenomenologically important,  compactifications can be performed to any number of dimensions. Each one provides valuable insight into the process and the structure of String Theory. Thus, let us illustrate how the mechanism works for the simplest case: type II compactified on a circle $S^1$ from 10 to 9 dimensions.

In this example, the spacetime coordinate associated to the ninth space dimension is periodic. The local dynamics are identical to the uncompactified space. The difference between both arises as a global effect due to the identification of the periodic coordinate $x^9\sim x^9+2\pi R$, with $R$ the $S^1$ radius. The relation directly propagates to the worldsheet embedding coordinates
\begin{equation}
    X^9(\tau,\sigma+l)=X^{9}(\tau, \sigma)+2\pi R w\,,\quad w\in \mathbb{Z}\,.
\end{equation}
This means that it is possible to have strings occupying the same space but differing in the number of times they wrap the $S^1$ direction, as shown in figure \ref{basic-fig: winding strings}. That information is encoded in $w$, which is therefore known as winding number.

\vspace{0.4cm}
\begin{figure}[htbp]
    \centering
    \includegraphics[width=0.75\textwidth]{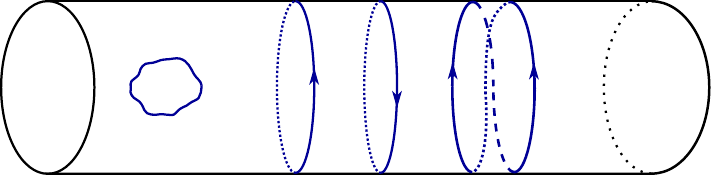}
    \caption{Closed strings in a $S^1$ compactification with different winding numbers. From left to right these are $w=0,1,-1,2$.}
    \label{basic-fig: winding strings}
\end{figure}
 
Both the full fermionic sector and the bosonic degrees of freedom of the uncompactified directions remain the same as in the original theory. The only thing that needs to be modified is the oscillatory expansion of the periodic direction degree of freedom. In the light cone gauge, \eqref{basic-eq: boson oscillatory expansion light cone} becomes
\begin{equation}
    X^{9}=x^{9}_{CM}+\frac{p^{25}}{p^+}\tau +\frac{2\pi Rw}{l}\sigma +{\rm osc.}=x^{9}_{CM}+\frac{k/R}{p^+}\tau +\frac{2\pi Rw}{l}\sigma +{\rm osc.} \,,
\end{equation}
where $k,w\in \mathbb{Z}$, $x^9_{CM}$ is the center of mass coordinate and we have omitted the oscillatory part since it is not modified. Note that in the last step we have introduced the quantization of the momentum imposed by the periodicity conditions along direction 9. 

Including the changes in the Hamiltonian, the spacetime mass formula can be derived:
\begin{equation}
    M_L^2=\frac{2}{\alpha'}\left(\frac{p_L^2}{2}+N_b+N_F+E_0\right)\,,\qquad  M_R^2=\frac{2}{\alpha'}\left(\frac{p_R^2}{2}+\tilde{N}_B+\tilde{N}_F+\tilde{E}_0\right)\, ,
    \label{basic-eq: mass spectrum compactified}
\end{equation}
where $N_F$, $N_B$, $\tilde{N}_F$ and $\tilde{N}_B$ are the fermionic and bosonic numbers of the left and right sector respectively, $E_0$ and $\tilde{E}_0$ are their Casimir energies and 
\begin{equation}
    p_L=\sqrt{\frac{\alpha'}{2}}\left(\frac{k}{R}+\frac{wR}{\alpha'}\right)\,, \qquad p_R=\sqrt{\frac{\alpha'}{2}}\left(\frac{k}{R}-\frac{wR}{\alpha'}\right)\,.
\end{equation}
Consequently, the 10-dimensional compactified theory has the mass spectrum of a 9-dimensional uncompactified theory, but with the addition of two sets of infinite states that have uniformly spaced masses. The first tower, labelled by $k$ is  a general feature of dimensional reduction associated to the quantization of the momentum in the compact dimension and it is known as Kaluza-Klein tower. The second tower, labelled by $w$ and known as the winding tower, is a purely string theory effect that arises due to the possibility of strings to wrap the compactified dimension.

\begin{tcolorbox}[breakable, enhanced,  colback=uam!10!white, colframe=uam!85!black, title=T-duality]
\begin{small}
It is worth pointing out that the mass spectrum is invariant under the transformation
\begin{equation}
    R\leftrightarrow\frac{\alpha'}{R}\,, \qquad k\leftrightarrow w\,.
\end{equation}
In the large volume limit, $R^2/\alpha'\gg 1$, the winding states become very heavy (unreachable to observations and thus effectively $w=0$), while the Kaluza-Klein tower collapses to zero mass and the momentum forms a continuum as we approach the decompactification limit. Meanwhile, in the limit $R^2/\alpha'\ll 1$, moving along the cycle requires too much energy but wrapping it with a string is much less costly and thus the winding modes start to form a continuum of their own, hinting at the existence of another dimension that is being decompactified in the dual theory. The equivalence of this relation, known as T-duality, extends to the full conformal field theory and the string interactions.

The new direction corresponds to the other choice of building the full oscillator from the left and right sectors:
\begin{equation}
    X'^9=X^9_L-X^9_R\,.
\end{equation}
By worldsheet supersymmetry, this transformation propagates to the fermionic sector. Therefore, T-duality is effectively a spacetime parity operation over the right-moving degrees of freedom. This changes the chirality of one of the spinorial groundstates of the Ramond right sector, mapping Type IIB to Type IIA and vice versa.
\end{small}
\end{tcolorbox}

Assuming $R$ is large enough to ignore winding effects but small enough so the light mass spectrum of the theory can be truncated to $k=0$,  the  field content amounts to decompose the $SO(8)$ (Little group of the Lorentz symmetry in 10d) representations of the original 10-dimensional fields into representations of SO(7) group in 9 dimensions. The vector representation of $SO(8)$ splits into the vector representation of $SO(7)$  plus and scalar ($\mathbf{8}_V\rightarrow \mathbf{7
}_V+\mathbf{1}$) while the two spinor representations of different chirality of $SO(8)$ collapse to the unique spinor representation of $SO(7)$ (there is no notion of  chirality in odd dimensions). The NSNS sector, common to type IIA and type IIB theories becomes
\begin{equation}
\begin{aligned}
    \mathbf{8}_V\otimes\mathbf{8}_V&= & \underbrace{\mathbf{7}_V\otimes \mathbf{7}_V} \hspace{0.3cm} & & + & & \underbrace{\mathbf{7}_V\otimes\mathbf{1}+\mathbf{1}\otimes\mathbf{7}_V}& & + & & \underbrace{\mathbf{1}\otimes \mathbf{1}}\,,\\
     G_{MN},B_{MN},\phi & \Rightarrow &  G_{\mu\nu},B_{\mu\nu},\phi & & & &  G_{9\mu}, B_{9\mu} \hspace{0.6cm} & & & & G_{99}\hspace{0.3cm}
\end{aligned}
\end{equation}
where $M,N$ label the 10d coordinates while $\mu,\nu$ label the 9d coordinates. Therefore, the 9-dimensional graviton, 2-form and dilaton arise as a result of the compactification together with two gauge fields $G_{9m}$, $B_{9m}$ and a scalar $G_{99}$. The scalar is particularly interesting, since its vacuum expectation value parametrizes the radius of the internal circle. Thus, as it was the case with the dilaton and the string coupling, we observe again that there are no external parameters in string theory. All of them are vacuum expectation values of dynamical scalar fields. In the current setup the new scalar $G_{99}$ has no potential, which means that any radius can be chosen for the compactification. A field of this kind, associated with a flat direction of the potential, is massless and is known as modulus. 

Similar decompositions can be performed for the other sectors obtaining the 9-dimensional theory description. Generalizing the above procedure to compactifications to lower dimensions does not introduce any conceptual difficulty. Nevertheless, while there is only a single 1-dimensional compact manifold ($\mathbf{S}^1$), the number of 6-dimensional compact manifolds is astronomical, which poses problems regarding the predictive nature of the theory in 4 dimensions.

\subsection{Other theories} 

Up until this point we have considered Type II superstrings theories. In the absence of D-branes, they are theories of closed oriented strings with $\mathcal{N}=2$ supersymmetry in ten dimensions which differ in their chirality (Type IIB is chiral whereas Type IIA is not). These theories will be the main focus of the thesis. However, it is important to mention the existence of other three superstring theories that have consistent worldsheet constructions. 

\begin{itemize}
    \item \textbf{Heterotic Theories}: Theories of closed oriented strings with $\mathcal{N}=1$ supersymmetry. They are built from the union of the right sector of closed superstrings and the left sector of closed bosonic strings. Consequently, in the light-cone gauge the right sector is composed of $8$ bosons $X^i_R(\tau-\sigma)$ and $8$ fermionic superpartners, while the left sector contains $24$ bosons $\{X^i_L(\tau+\sigma),X^I_L(\tau+\sigma)\}$, with $i=2,\dots, 9$ and $I=1,\dots,16$. Heterotic theories live in $10$ dimensions, which means that the additional $16$ left-moving bosons cannot be linked to physical spacetime dimensions. Instead, they are understood as the parameters of a compactified 16-dimensional torus with size $R=\sqrt{\alpha'}$. 
    
    Modular invariance and anomaly cancellation greatly restrict the possible arrangements of the  massless states arising from the additional bosonic left sector. The momenta associated with the oscillators $X^I_L$ must be vectors of a 16-dimensional even self-dual lattice and there are only two options that satisfy those conditions: the lattice of the group $E_8\times E_8$ and the lattice of $SO(32)$. The massless states coming from these degrees of freedom constitute gauge bosons with respect to one of the two aforementioned gauge symmetry groups.  
    
    Therefore, there are two distinct heterotic theories, namely heterotic $E_8\times E_8$ and heterotic $SO(32)$. Their massless spectrum differs from type II theories in that the Ramond-Ramond fields are substituted by their respective non-abelian spacetime gauge field and their superpartners, which together fill vector multiplets of $\mathcal{N}=1$ 10 dimensional supersymmetry.

    \item \textbf{Type I theory}: Chiral theory of closed and open unoriented strings. It is built through the introduction of a orientifold quotient that truncates the Hilbert space of Type IIB  merging left- and right-moving degrees of freedom. Cancellation of the RR tadpole is possible due to the modification of the set of worldsheet diagrams that contribute to the one-loop amplitude. More specifically, the Chan-Paton indices have $N=32$ possible values. The resulting massless spectrum contains the standard graviton supermultiplet of $\mathcal{N}=1$ 10-dimensional supergravity arising from the closed string sector and a $SO(32)$ $\mathcal{N}=1$ vector supermultiplet in 10 dimensions.
\end{itemize}

\section{Non perturbative states and dualities}
\label{basic-sec: non perturbative and dualities}

So far we have focused on one-dimensional strings, their consistency properties, their perturbative expansion and the massless field spectrum they produce as low energy effective theories. However, despite its name, String Theory is not only a theory of strings. As was hinted at when considering open strings with Dirichlet boundary conditions,  there are a variety of non-perturbative objects of higher dimensions called branes. Their dynamics and behaviour under different compactifications provide a path to understanding String Theory beyond the low energy region, a challenge that has yet many open questions. Through the use of branes, dualities and compactifications a net of relations has been built connecting the five different superstrings theories as limits of an underlying 11-dimensional theory, named M-theory.

\subsection{Branes}
\label{basic-subsec: branes}
\subsubsection*{D-brane generalities}

From the supergravity point of view at weak coupling, Dp-branes are solitonic solutions that describe localized p+1 dimensional hyperplanes (p spatial dimensions plus time) on spacetime $\mathcal{M}_{10}$. They require the existence of open strings attached to them that describe their excitations. Thus, the brane is a dynamical topological defect whose degrees of freedom are encoded in the open strings with which it interacts. The field description of those degrees of freedom was obtained in \ref{basic-table: massless opens superstring spectrum DD+NN}. The scalars $\phi^m$ parametrize the fluctuations of the geometry of the brane and the gauge vectors $A_i$ describe the gauge fields that live confined in their interior.\footnote{Through these fields D-branes provide an elegant way of introducing non-abelian gauge symmetries in String Theory.} These bosonic fields are accompanied by their fermionic partners $\lambda_\alpha$, constituting a $U(1)$ vector supermultiplet in $\mathcal{N}=1$ supersymmetry in $p+1$ dimensions. Such supermultiplet can be understood as a dimensional reduction of $\mathcal{N}=1$ vector multiplet in $10$ dimensions. Furthermore, it is possible to check that supersymmetry extends beyond the massless states to the complete open string spectrum.

Recalling that the vacuum of Type II theories does not contain open strings, configurations of open strings together with branes should be understood as non-perturbative excited states that break Poincaré invariance.  Type II theories have $\mathcal{N}=2$ supersymmetry in $10$ dimensions, while D-branes only preserve half of the supersymmetry and are therefore 1/2 BPS objects. As a consequence, many of their properties are protected under continuous deformations, and they are thus preserved after introducing quantum and $\alpha'$ corrections.

As dynamical objects, branes are expected to have tension, i.e. mass to volume ratio. BPS objects are stable, which means that there has to be an additional force that compensates the gravitational force exerted by the brane's tension. This is achieved through the coupling to the gauge fields of the closed string sector. Hence, BPS branes are charged under these fields in such a way that charge repulsion compensates the gravitational pull. A geometrical analysis shows that a RR field $C_p$ can couple electrically to a (7-p)-brane\footnote{Note that this is an (8-p)-dimensional manifold.} and magnetically to a (p-1)-brane. Given the RR field content of type II theories (tables \ref{basic-table: type IIB spectrum} and \ref{basic-table: type IIA spectrum}), we conclude that Type IIA contains stable even Dp-branes and Type IIB contains stable odd Dp-branes.

\begin{tcolorbox}[breakable, enhanced, colback=uam!10!white, colframe=uam!85!black, title=Generalized Maxwell Theory]
\begin{small}
Given a p-form gauge field $A_p$ in a d-dimensional space, we can describe its dynamics in terms of its exterior derivative, the $p+1$ field strength form $F_{p+1}=dA_{p}$. In the absence of local charges, $F_{p+1}$ is exact, and hence $dF_{p+1}=0$. This standard relation, known as Bianchi identity, constitutes one of the two generalized Maxwell equations. The second is not trivial and arises from the introduction of a generalized Maxwell action $\int F\wedge \star F \Rightarrow d\star F=0$, where $\star$ is the Hodge dual operator. In the presence of sources, both relations are modified to
\begin{equation}
     d\star F_{p+1}=e \delta^{d-p}\,,\qquad dF_{p+1}=\mu \delta^{p+2}\,,
     \label{basic-eq: generalized Maxwell theory}
\end{equation}
where $\delta^p$ represents the Poincaré dual p-form to the (d-p)-cycle of the full spacetime in which  the source is localized. From here, it is straightforward to identify such cycle with a D-brane charged under the gauge field. The cycle associated to $\delta^{d-p}$ is a p-dimensional manifold magnetically charged under the gauge field $A_p$ while $\delta^{p+2}$ describes a $(d-p-2)$-manifold electrically charged under the same gauge field. Alternatively, instead of considering dual branes, one can identify dual fields associated with different branes carrying the same type of charge. Thus, given a field $C_{p}$ with flux strength $F_{p+1}$ and local electric charge source generated by a $(d-p-2)$-brane, one can consider a new field $C_{d-p-2}$ such that $F_{d-p-1}\equiv dC_{d-p-2}=\star F_{p+1}$. 

The value of the magnetic (electric) charge can be measured using Gauss Law through the integration of the (dual) flux in a codimension one sphere of the space transverse to the localized source. That is 
\begin{gather}
    Q_m=\int_{S^{p+1}} F_{p+1}=\int_{B^{p+2}} d F_{p+1}=\int_{B^{p+2}} \mu \delta^{p+2}=\mu\,,\\
    Q_e=\int_{S^{d-p-1}}\star F_{p+1}=\int_{B^{d-p}} d\star  F_{p+1}=\int_{B^{d-p}} \mu \delta^{d-p}=e\,,
\end{gather}
with $B^d$ the interior of the sphere $S^{d-1}$.

It is possible to verify that in a manifold with non-trivial homology p-cycles $\Sigma_p$ and (p-2)-branes charged under a (p-1)-form gauge field with minimal charge $Q_e$, having  uniquely defined quantum amplitudes requires the following relation for the flux integral
\begin{equation}
    Q_e\int_{\Sigma_p}F_p\in 2\pi\mathbb{Z}\,,
    \label{basic-eq: flux quantization}
\end{equation}
which means that fluxes must be elements of the integer cohomology group of the manifold $H^q(\mathcal{M},\mathbb{Z})$. Applying this results to a manifold with a localized magnetic source that acts as a non-trivial cycle, the above relation gives the generalized Dirac quantization condition 
\begin{equation}
    Q_eQ_m\in 2\pi \mathbb{Z}\,.
\end{equation}
\end{small}
\end{tcolorbox}

It is worth revisiting the discussion about the RR-tadpole anomaly and how it challenged the presence of open strings in type IIB. Now that Dp-branes are present, open strings no longer generate topological RR-tadpoles anomalies. In fact, open strings attached to a Dp-brane contribute to the tapdole corresponding to the RR-form $C_{p+1}$. Assuming $p<9$ (so there are non-compact transverse dimensions),  this form is dynamical and the equation of motion can be solved without inconsistencies. In other words, the presence of non-compact transverse directions allows the flux-lines to escape to infinity so no constraint is required.

\subsubsection*{Action}

After the intuitive review of the basic properties of the branes and their relation with the different elements of the theory, we are prepared to provide an effective action that describes the dynamics of branes in presence of the massless fields of the closed sector. Such action is divided into two components: the Dirac-Born-Infeld action, which depends on the NSNS sector of the closed string, and the Chern-Simons action, which describes the coupling to the RR fields of the closed sector. 

\begin{itemize}
    \item Dirac-Born-Infeld (DBI) action. It is a generalization of the Nambu-Goto action to higher dimensional objects that accounts for the presence of non-trivial backgrounds of graviton, the 2-form B and the dilaton. It can be derived from the study of the cylinder diagram describing the emission of closed strings between two parallel D-branes.
    \begin{equation}
 S_{\rm DBI}  =  - \mu_p \int_{W_{p+1}} d^{p+1}\xi e^{-\phi} \sqrt{-\det \left(P[G-B] - \frac{\ell_s^2}{2\pi}  F \right) }\, , 
 \label{basic-eq: DBI action}
\end{equation}
    where $P[G-B]$ is the pull-back of the spacetime tensor $G-B$ into the worldvolume geometry of the brane, $F$ is the field strength of the worldvolume gauge field $A$ and $\mu_p = \frac{2\pi}{\ell_s^{p+1}}$ parametrizes the brane tension. In a trivial background, it reduces to the integral of the worldvolume's volume $W_{p+1}$. 
    
    The pull-back introduces the dependence on the scalar fields of the open string sector $\phi^i$, that act as embedding functions of the worldvolume into the full spacetime. The combination of $B+\ell_s^2/2\pi F$ is required by gauge invariance. From their contribution to the worldsheet action, it can be seen that they are coupled under gauge transformations. Namely
    \begin{equation}
        B \rightarrow B+d\Lambda\,,\qquad A\rightarrow A-\frac{\ell_s^2}{2\pi}\Lambda\,.
    \end{equation}
    The above relation describes the fact that the open string, charged under $B$, deposits its charge on the brane, where it becomes a charge of the world volume gauge field $A$.
    
    The dependence with the dilaton of the DBI action shows that the tension of the D-brane scales with the string coupling like $1/g_s$. Therefore, it diverges at weak coupling and the brane becomes a rigid object, all in agreement with the non-perturbative nature of these states.

    \item Chern-Simons (CS) action. It describes the coupling to the RR fields. It is a purely topological term that does not depend on the metric and is given by
    \begin{equation}
         S_{CS}  =    \mu_p \int P\left[ {\sum_{q} C_q} \wedge e^{-B}\right] \wedge e^{- \frac{\ell_s^2}{2\pi}  F} \wedge \hat{A}(\mathcal{R})\, , 
         \label{basic-eq: CS action}
    \end{equation}
   where  $P[C_q]$ is the pullback of the RR form $C_q$ to the worldvolume and $\hat{A}$ is a polynomial of the curvature 2-form $\mathcal{R}$ whose two first terms are $\hat{A}\approx 1-1/(24\cdot8\pi^2) \rm{tr} \mathcal{R}^2$.
\end{itemize}

\subsubsection*{Dp-brane solutions}

As it has become patent throughout this chapter, in String Theory there is no spacetime action for the complete theory, rather only 10d supergravity massless fields effective actions. Classical solutions of such effective actions provide an approximation to the description of non-perturbative states. This approximation  becomes much more reliable for BPS states, since many of their properties are protected by supersymmetry beyond the regime of validity of the effective theory. Given that closed strings can interact with D-branes, it is to be expected that the presence of a Dp-brane generates a non-trivial background for the metric and RR fields.  The backreaction of $N$ Dp-branes with $p<6$ comes from solving a Poisson equation for type II 10d background fields. Denote by $x^\mu$, $\mu=0,\dots,p$ the dimensions along the worldvolume of the Dp-brane an $x^m$ $m=p+1,\dots, 9$ the dimensions transverse to it, the supergravity solution is given by \cite{Horowitz:1991cd} 
\begin{align}
    ds^2&=Z_p(r)^{-1/2}\eta_{\mu\nu}dx^\mu dx^\nu+Z_p(r)^{1/2}dx^mdx^m\,, \\
e^\phi&=g_sZ_p(r)^{\frac{3-p}{4}}\,, \\
H_{8-p}&=-g_s^{-1}(Z_p(r)^{-1}-1){\rm dvol}_{{ S}^{8-p}}\,,
\label{basic-eq: brane backreaction}
\end{align}
where $H_{8-p}$ is the flux around a $(8-p)$ sphere surrounding the object in the transverse (9-p)-dimensional space and it is thus related to $C_p$ through dualities (see discussion on Generalized Maxwell theory around \eqref{basic-eq: generalized Maxwell theory}). This will become much more apparent when the democratic formulation is introduced in section \ref{cy-sec: flux compactifications}. The function $Z_p$ is 
\begin{equation}
    Z_p(r)=1+g_sN\gamma\left(\frac{l_s}{r}\right)^{7-p},\qquad r^2=\sum_m (z^m)^2\,,
\end{equation}
with $\gamma$ some numerical factor that depends on $p$.

The above solution of the Poisson equation is not valid for low-codimension objects. In particular, for D8-branes \cite{Bergshoeff:2001pv}, which are charged under the Romans mass in massive Type IIA, we need to take $r=|x^9|$ and 
\begin{equation}
    Z_8=1-\frac{(N-8)g_sr}{2\pi\ell_s}\,.
\end{equation}
Therefore, the RR flux background reads
\begin{equation}
    G_0=-g_s^{-1}\partial_9 Z_8=\begin{cases}
    -\frac{N-8}{2\pi \ell_s}, \quad y^9>0\,,\\
    \frac{N-8}{2\pi \ell_s}, \quad y^9<0\,.
    \end{cases}
    \label{basic-eq: D8 solution}
\end{equation}

\subsubsection*{Chan-Paton indices and multiple branes}

The notion of branes recontextualizes the role of Chan-Paton indices. These degrees of freedom of the endpoints of open strings are labels that identify the brane to which that endpoint is attached. Thus, they become very useful for describing systems of multiple branes and studying the different gauge symmetries they give rise to.

The mass of open strings gets a contribution that depends on the transverse distance between the branes their endpoints are fixed to. Consequently, the symmetries will notably change depending on the spacetime distribution of the system of branes. If all $n$ branes are coincident (stacking over the same plane) the open string spectrum will contain $n^2$ massless sectors that generate a $U(N)$ symmetry. If all the branes are separated, the open string massless spectrum will be reduced to $n$ sectors coming from open strings with Chan-Paton indices of the form $aa$ and giving rise  $U(1)^n$ gauge bosons. The remaining $n^2-n$ sectors of the form $ab$ describe massive particles charged under $U(1)_a\times U(1)_b$.

\subsubsection*{Summary: Brane bestiary}

The discussion above explained how stable branes must be charged under background gauge fields from the closed bosonic sector, focusing on the Ramond-Ramond fields. In fact, for any (p+1)-form gauge potential there exists an extended object spanning across $p+1$ dimensions (including time) that is charged under it. The term Dp-brane is reserved for those that couple to the RR fields, but these are not the only ones. There is another gauge field that can act as background: the 2-form field $B$ from the NSNS gauge sector. Following the same reasoning, we can find a 5-brane magnetically charged under $B$, commonly known as the NS5-brane, and a 1-brane electrically charged under the same field. The latter is a 2-dimensional object coupled only to the NSNS closed string sector background: it is the fundamental string, denoted as $F1$.

Based on their field content, we can deduce the brane spectrum of the superstring theories:
\begin{itemize}
    \item Type II: NS5, F1 and Dp-branes with $p$ even in type IIA and odd in Type IIB.
    \item Heterotic: NS5 and F1 branes.
    \item Type I: Dp branes with $p=1,5$.
\end{itemize}

The fact that the fundamental strings can be described in the same terms as other branes which do not accept a perturbative description hints at the possibility that its prevalence is an artifact of the regime $g_s\ll 1$. As the value of the string coupling increases, all branes start to participate on equal footing and the theory becomes extremely complex. Therefore, it is logical to consider moving to different regions in parameter space, searching for limits in which the full theory simplifies again to a new effective theory characterized by another fundamental object.

\subsection{M-theory, F-theory and dualities}
\label{basic-subsec: dualities}

In general terms, a duality describes a quantum equivalence between two theories, so that a bijective map can be built relating the degrees of freedom and actions of both of them. This means that two dual theories are redundant and simply provide two different ways of presenting a single theory. Despite that, the existence of dual theories is far from trivial and the complex emergent phenomena in one of the theories can have a simple fundamental description in its dual. Thus, the relations that the duality establishes can prove to be a extremely powerful tool to improve the understanding of both facets of the unique underlying theory. That is the case when studying the different parametric regions of String Theory. In the previous section, we discovered the T-duality between type IIA and type IIB when they are compactified over $S^1$. In this section, we will explore other essential dualities and the insights they offer.

\subsubsection*{S-duality in Type IIB and F-theory}

It is possible to check that the type IIB action \eqref{basic-eq: type IIB action} once written in the Einstein frame\footnote{The frame used in \eqref{basic-eq: type IIB action} is called the string frame. The Einstein frame is defined so that in the low-energy effective action the Ricci scalar is not multiplied by the asymptotic value of the dilaton.} has an $SL(2,\mathbb{R})$ symmetry under the following transformation of the 2-forms and the complex string coupling $T^0\equiv C_0+i/g_s$
\begin{equation}
    T^0\rightarrow \frac{aT^0+b}{cT^0+d}\,,\qquad \left(\begin{array}{c}
          B_2 \\
          C_2
    \end{array}\right)\rightarrow \left(\begin{array}{cc}
        a & b \\
        c & d
    \end{array}\right)\left(\begin{array}{c}
          B_2 \\
          C_2
    \end{array}\right)\,,
\end{equation}
with $ad-bc=1$. The continuous symmetry is broken when considering the full theory due to charge quantization. However,  a discrete $SL(2,\mathbb{Z})$ symmetry prevails as a symmetry of the full type IIB theory. This is precisely the modular group (see \eqref{basic-eq: modular group} for a description of how the same group arises in a different context), generated by the transformations $T^0\rightarrow T^0+1$ and $T^0\rightarrow -1/T^0$. The latter, known as S-duality, is of particular interest. For the simple case $C_0=0$, it amounts to inverting the coupling constant $g_s\rightarrow g_s^{-1}$ and swapping the two 2-form fields. Thus, S-duality provides a map between weak and strong couplings. The fundamental object at weak coupling, the F1 string coupled to $B_2$, is mapped to the D1-brane coupled to $C_2$, which acts as the fundamental object of the strongly coupled theory. Similarly NS5 and D5 branes are also interchanged. Hence, it is possible to identify hybrid states in the spectrum that are charged under both fields. This leads to the consideration of (p,q)-strings and (p,q)-branes, which have a charge of $p$ under $B_2$ and $q$ under $C_2$.

The $SL(2,\mathbb{Z})$ creates a clear link between the transformations of the axio-dilaton $T^0$ in Type IIB and the complex structure of a torus. Such a relation can be established formally by building an elliptic fibration over the 10-dimensional spacetime $\mathcal{M}_{10}$. Deforming the fibre would correspond to going to different coupling limits. If the fibration is not trivial, the theory that results, called F-theory, generalizes the behaviour of type IIB. At this point, treating the axio-dilaton as the complex structure of a elliptic fibre might seem only a mathematical analogy. However, as we will see in chapter \ref{ch: Fintro} this identification is further supported by duality with M-theory and provides many interesting physical results. 

\subsubsection{Type IIA and M-theory}

We have observed that the strongly coupled limit of type IIB is dual to its own weak coupling limit. One may now wonder  if that is also the case with type IIA. To answer this question, we study how the brane tension behaves when $g_s\gg 1$. From the DBI action \eqref{basic-eq: DBI action} we see that the lightest state in that limit is the D0-brane, with mass $m=1/(g_s\sqrt{\alpha'})$. Consequently, the spectrum is dominated by bound states of $D0$ which form an infinite tower with equally spaced masses very reminiscent of a Kaluza-Klein tower that arises in a decompactification limit. This behaviour suggests the possibility that the strongly coupled type IIA theory can be described as the decompactification limit of 11-dimensional theory on $\mathcal{M}_{10}\times S^1$. To recover the mass spectrum of the D0, one just needs to consider the KK tower for the $S^1$ radius $R_{11}=g_s\sqrt{\alpha'}$. Thus, the best candidate to the low energy description of the strong coupling limit of type IIA is the unique 11-dimensional supergravity theory, which consists of a 11d metric $G_{MN}$, a 3-form $C_3$ and a 11d gravitino $\psi_{M,\alpha}$ and when compactified on a circle returns the 10d supergravity content of type IIA.  Both theories have two parameters which can be mapped from one to the other. Type IIA has the string coupling $g_s$ and the string tension $\alpha$. 11-dimensional supergravity compactified on a circle has the 11-dimensional gravitational coupling $\kappa_{11}$ and the radius of the circle $R_{11}$. We have already introduced the link between the radius and the type IIA constants. The gravitational coupling can be derived from dimensional reduction
\begin{equation}
    \kappa_{11}^2=2\pi R_{11}\kappa_{10}^2=\frac{1}{2}(2\pi)^8g_s^3(\alpha')^{9/2}\,.
\end{equation}

In the same way that non-chiral 10-dimensional supergravity is the low energy limit of type IIA string theory, there must exist a new 11-dimensional theory that analogously mirrors the full string theory description. This new 11-dimensional theory is called M-theory and it does not admit a perturbative expansion, which greatly complicates the formulation of its precise structure in the general region $g_s\sim 1$. In the low energy limit, however, it must be described by the well-understood unique 11-dimensional supergravity theory. The bosonic part of effective action is given by
\begin{equation}
    S_{11d}=\frac{1}{2\kappa_{11}^2}\int d^{11}x\sqrt{-G}R-\frac{1}{4}\star G_4\wedge G_4-\frac{1}{6}C_3 \wedge F_4\wedge F_4\,,
\end{equation}
with $F_4=dC_3$.

Since there is a single gauge field in the effective theory, there can only be two BPS objects: a 2-brane and a 5-brane, named M2 and M5 respectively. Studying how the tension of the different BPS objects in Type IIA can be written using the natural quantities of 11-dimensional supergravity, a map can be established with objects living in M-theory compactified on a circle. The results are summarized in table \ref{basic-table: M theory type IIa map}. Note the D8-brane has no well-defined lift to 11 dimensions, which casts some doubts on the validity of massive type IIA theory as a UV complete theory.
\begin{table}[htbp]
\begin{tabular}{c|cccccc}
Type II                    & D0       & F1                    & D2 & D4                   & NS5 & D6                   \\ \hline
M-theory on $S^1$ & KK modes & M2 on $ S^1$ & M2 & M5 on $S^1$ & M5  & KK magnetic monopole
\end{tabular}
\caption{Duality between BPS states in type IIA and M-theory compactified on a circle.}
\label{basic-table: M theory type IIa map}
\end{table}

\subsubsection*{Full picture}

Many more dualities have been established between the different theories, other notable cases being T-duality relating both heterotic theories or S-duality linking Type I with the heterotic $SO(32)$. The current understanding of these relations, summarized in figure \ref{basic-fig: duality web}, provides an intricate web of dualities and compactification limits connecting all string theories  and improving our understanding of each individual component. The five different superstrings theories are thus interpreted as distinct perturbative limits of the underlying, mostly unknown, 11-dimensional M-theory.

\begin{figure}[H]
    \centering
     \includegraphics[width=0.7\textwidth]{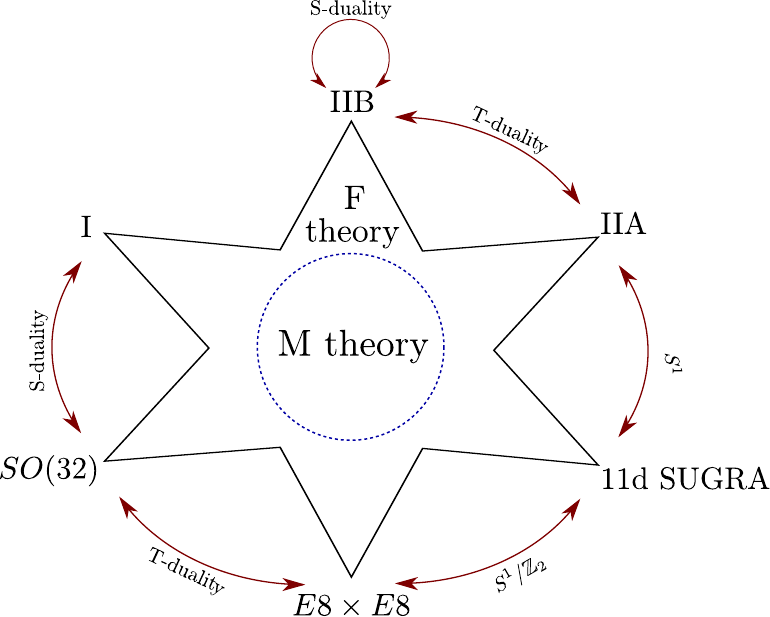}
    \caption{Representation of the web of dualities of the  string theories seen as different perturbative limits of  M-theory.}
    \label{basic-fig: duality web}
\end{figure}

\section{Swampland Program}
\label{basic-sec: swampland program}

In the previous sections we have reviewed the main features of String Theory as a 10-dimensional theory and discussed the many restrictions that arise due to consistency requirements. The final result showed that all allowed theories are connected into a single network based on a uniquely defined 11-dimensional theory. Nevertheless, this uniqueness is not preserved when constructing effective theories at lower dimensions. As we discussed briefly in section  \ref{basic-subsec: compactifications}, in order to provide a useful description of the Universe at the energy scales we are able to reach, String Theory must be compactified on a 6-dimensional manifold. Even though not every manifold is eligible for compactifications (in the next chapter we will see that a Calabi-Yau-like is usually required), the number of different possibilities seems to be colossal. Early estimations  already suggested that the number of inequivalent models could reach $10^{1500}$ \cite{Lerche:1986cx}. Such vast amount of possibilities was received with different degrees of acceptance. On the one hand, it could justify, together with the anthropic principle, the observed value of the cosmological constant. On the other hand,  the predictive power of the theory seemed greatly diminished since with so many vacua, any self-consistent effective theory appeared to be valid. The last point was challenged in \cite{Vafa:2005ui}, which started a paradigm shift from constructing specific effective models to searching for constraints provided by string theory and more generally quantum gravity arguments. This new perspective is known as the Swampland Program.

\subsection{Landscape vs Swampland}

The rich vacuum structure of low energy effective theories that arise as compactifications of String Theory is called the String Landscape. In recent years it has been observed that this set, although vast, is much smaller than it could seem. Many low energy theories that look consistent from different criteria, like anomaly cancellation, turned out to not be compatible with a coupling to quantum gravity. Following the Landscape metaphor, this set is named the Swampland. More specifically, it is defined as
\begin{tcolorbox}[colback=white]
\textbf{Swampland}: Set of apparently consistent effective theories that cannot be embedded  into a quantum gravity theory in the ultraviolet limit.
\end{tcolorbox}
Once a UV theory has been constructed, it is always possible to provide a low energy effective theory by integrating out the degrees of freedom beyond a certain scale $\Lambda_{\rm eff}$, resulting in a renormalizable sector and a tower of heavy non-renormalizable operators. The possibility of inverting the process and reconstruct the UV regime from the low energy theory is not guaranteed. Coupling a self-consistent effective field theory to quantum gravity will give a new energy scale $\Lambda_{Swamp}$ above which the theory must be modified in order to reach a consistent quantum gravity theory at high energies. Therefore, the EFT under consideration will be on the Swampland unless the cut-off of the theory is lower than the scale of the quantum gravity corrections, that is $ \Lambda_{\rm eff}\ll\Lambda_{Swamp}$. This second cut-off grows with the Planck Mass and therefore becomes more constraining as the theory goes to higher energies. Such behaviour is illustrated in figure \ref{basic-fig: landscape vs swampland}. The most extreme and interesting case is when $\Lambda_{Swamp}$ lies below any non-trivial energy scale of the effective theory, in which case the full theory would be on the Swampland. Therefore, through this new perspective, the Swampland program would have the potential to use quantum gravity criteria to restrict the set of allowed effective descriptions of our Universe.

\begin{figure}[htbp]
    \centering
    \includegraphics[width=0.8\textwidth]{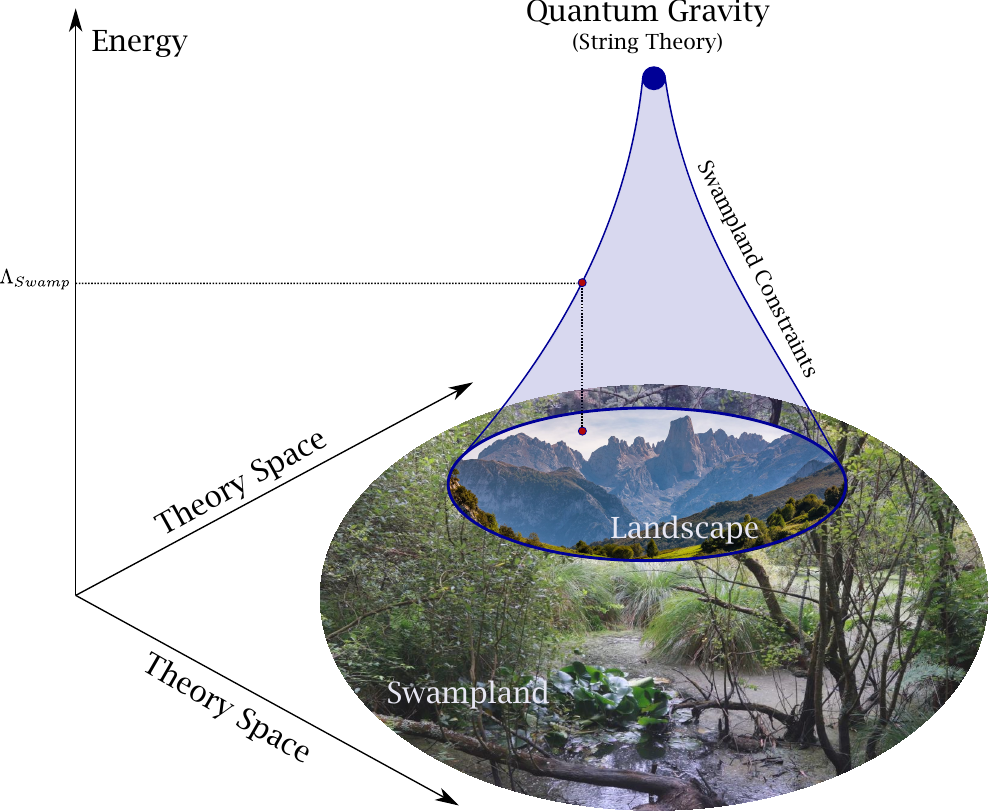}
    \caption{Schematic representation of the relation between the Landscape and the Swampland of EFT's with the UV theory of Quantum gravity. Adapted from \cite{Palti:2019pca, vanBeest:2021lhn,QuirantPellin:2022vyp}.}
    \label{basic-fig: landscape vs swampland}
\end{figure}

The borders between the Swampland and the Landscape are formulated in terms of conjectures that establish the characteristics that the effective theory needs to satisfy to be consistent with Quantum Gravity. They are derived from very different methods and address a varied arrange of topics, but they are connected through a net of logical implications which reinforces the global picture and hints at the existence of underlying quantum gravity principles that the Swampland Program could unravel.

Swampland conjectures are formulated from the point of view of the effective theory at low energies. Thus they do not assume a particular ultraviolet origin and in that sense they are more general than String Theory itself. However, together with Black Hole physics, String Theory, as a consistent quantum gravity framework, is the main tool to determine and gather evidence for the conjectures. 

It is worth noting that not all sources of data coming from String Theory have the same level of trust and, consequently, not all conjectures are on equal footing. In \cite{Palti:2019pca}, a distinction is drawn between string examples depending on their rigour: string-derived and string-inspired models. String-derived vacua are well understood through a full worldsheet description and provide solid evidence to Swampland conjectures. They are unfortunately a  relatively small set that generally requires supersymmetry and very simple geometries. On the other side, string-inspired examples are based on a large number of assumptions that have not been completely verified and thus should be considered quantum field theory constructions motivated (but not explicitly obtained) from String Theory. Many of the conjectures lie in a middle point of this classification, being verified by the most rigorous examples but failing once less trusted models are included. Determining the splitting point is thus partially subjective and two points of view regarding a given string-inspired example often coexist: it can be considered to be a counterexample of the conjecture or the conjecture might be informing that the example is not trustable. For this reason, a large amount of data from different sources is required in order to have a clear understanding of the conjecture and its implications.

\subsection{The Swampland Conjectures}
\label{basic-subsec: swampland conjectures}

Now we will overview some of the most important Swampland conjectures, focusing on those of greater relevance for this thesis. For a more detailed explanation of the subject, we refer the reader to the reviews \cite{Brennan:2017rbf,Palti:2019pca,vanBeest:2021lhn,Grana:2021zvf,Agmon:2022thq}.

\subsubsection*{No Global Symmetries}

A global symmetry is a transformation described by a local unitary operator that acts non-trivially in the space of physical states and that commutes with the Hamiltonian. 

\begin{tcolorbox}[colback=white]
    \textbf{No Global Symmetries Conjecture}: a theory with a finite number of states and consistently coupled to quantum gravity cannot have global symmetries. \cite{Banks:1988yz,Banks:2010zn}
\end{tcolorbox}

Therefore, any global symmetry must be broken or gauged at high energies. It is motivated by black hole dynamics, since the evaporation of black holes charged under this global symmetry (which cannot be radiated) would produce arbitrarily long lived remnants for any value of the charge resulting in a theory with infinite number of states. Its statement has been proved in AdS spacetimes using the AdS/CFT correspondence \cite{Harlow:2018jwu,Harlow:2018tng}.

This conjecture has been generalized to include any topological global charge using the notion of cobordisms. Two compact $d$-dimensional manifolds are said to be cobordant if their union is the boundary of another compact $d+1$-dimensional manifold. It is an equivalence relation and the subsequent quotient set $\Omega_d$ has a group structure under the disjoint union operation.

\begin{tcolorbox}[colback=white]
    \textbf{Cobordism conjecture}: The cobordism group of a D-dimensional quantum gravity compactified in $d$ dimensions must be trivial, that is $\Omega_d^{QG}=0$. \cite{McNamara:2019rup}
\end{tcolorbox}

This means that in a consistent quantum gravity theory all compactifications are related through interpolating manifolds (domain walls from the EFT perspective). If that were not the case, it would imply the existence of a topological global charge that generates a  global symmetry.

\subsubsection*{Weak Gravity Conjecture}

The No Global Symmetries conjecture is the most tested and best understood conjecture. However, it lacks predictive power at low energy levels due to its broad nature. The Weak Gravity Conjecture (WGC) aims to refine the relation between symmetries and gravity by providing bounds to the mass spectrum of charged states which can be tested directly on the effective field theories. It was first established by \cite{ArkaniHamed:2006dz} and has two different formulations: the electric and the magnetic versions. In addition to the general Swampland reviews, we recommend the specialized review \cite{Harlow:2022ich} for an in depth analysis. 

\begin{tcolorbox}[colback=white]
    \textbf{Weak Gravity Conjecture (Electric)}: Given a gauge theory weakly coupled to gravity, there exists an electrically charged state whose charge to mass ratio is greater than that of an extremal black hole. 
\end{tcolorbox}
For the case of a $U(1)$ symmetry with gauge coupling $g_{YM}$, a  black hole of mass $M$ in 4 dimensions must satisfy an extremality bound to avoid naked singularities
\begin{equation}
    M\geq \sqrt{2}g_{YM}q M_p\,,
\end{equation}
with $q$ the quantized charge of the black hole. An extremal black hole saturates the aforementioned bound. Therefore, the WGC imposes the existence of a state such that 
\begin{equation}
    m\leq \sqrt{2}g_{YM}qM_p\,.
\end{equation}
It can be generalized for an arbitrary gauge p-form in $d$ dimensions demanding the existence of a $p-1$ dimensional object (generally a brane) with tension $T_p$ and quantized charge $q_p$ verifying
\begin{equation}
    \frac{p(d-p-2)}{d-2}T^2\leq q_p^2 M_p^{d-2}\,,
\end{equation}
with $M_p$ the d-dimensional Planck mass.

The magnetic version is nothing more than the previous formulation applied to the magnetic dual gauge field. It provides an upper bound to the effective theory cut-off in terms of the gauge coupling.
\begin{tcolorbox}[colback=white]
    \textbf{Weak Gravity Conjecture (Magnetic)}: The cut-off $\Lambda_{\rm eff}$ of an effective theory with a p-form gauge field with gauge coupling $g_{YM}$ is bounded by
    \begin{equation}
        \Lambda_{eff}\leq (g_{YM}^2 M_p^{d-2})^{\frac{1}{2p}}\,. 
        \label{basic-eq: WGC magnetic}
    \end{equation}
\end{tcolorbox}

It has two main motivations. First, it obstructs the restoration of a global symmetry when taking the gauge coupling $g_{YM}$ to zero, since it would generate an infinite tower of light charged particles that the effective theory would not describe.  Consequently, its cut-off would also go to zero, as seen in \eqref{basic-eq: WGC magnetic}. The second argument is based on the requirement that extremal black holes can decay, which explicitly demands the existence of a particle with charge greater than the mass, such that its emission by the black hole does not violate the extremality bound and the Cosmic Censorship Conjecture.

When the WGC conjecture is saturated, a link between the mass and gauge coupling is established and thus it provides a relation between Poincaré and internal symmetries. By Coleman-Mandula theorem  \cite{Coleman:1967ad}, this is not possible unless supersymmetry is introduced. Therefore, if supersymmetry is absent, quantum corrections would be expected to prevent the physical states from saturating the bound. Such expectation was summarized in the Sharpenend Weak Gravity conjecture, first formulated by \cite{Ooguri:2016pdq}.

\begin{tcolorbox}[colback=white]
    \textbf{Sharpened Weak Gravity Conjecture}: The Weak Gravity Conjecture is only saturated by BPS states in a supersymmetric theory.
\end{tcolorbox}

The sharpened version of the WGC has significant implications in AdS spaces. If we have a non-supersymmetric d-dimensional AdS space supported by fluxes and consider the top form hodge dual to the flux quanta $F_d=dC_{d-1}$. Demanding that this gauge field $dC_{d-1}$ satisfies the Sharpenend WGC implies the existence of a charged $(d-2)$-brane with $T<QM_{p}^2$, and thus the tension is not strong enough to compensate the self-repulsion of the charge. Therefore, the system is unstable and the brane expands out to the boundary of AdS space, acting as a charged domain wall that transitions between two vacua with different values of the flux quanta. The preceding reasoning leads to the following conjecture \cite{Ooguri:2016pdq,Freivogel:2016qwc}.

\begin{tcolorbox}[colback=white]
 \textbf{Non-Supersymmetric AdS Instability Conjecture}:   Any non-supersymmetric AdS geometry supported by fluxes is unstable in a consistent quantum theory of gravity with low energy description in terms of the Einstein gravity coupled to a finite number of matter fields.
\end{tcolorbox}
Type IIA compactifications provide an excellent framework to study the Sharpened WGC and the AdS instability conjecture. We will explore them and test how they hold up in chapters \ref{ch: bionic} and \ref{ch: membranes}.

\subsubsection*{Distance Conjecture}

As it was briefly discussed in \ref{basic-subsec: compactifications} and will be extensively detailed in the following chapters, compactifications generate massless scalar fields, called moduli, whose vacuum expectation values control the parameters of the theory. The set of possible values of these fields forms the moduli space. In the 10-dimensional theory they can be understood as geometrical quantities, such as the compactification radius or the complex structure of a torus. They constitute a manifold equipped with a Riemannian metric and therefore with a notion of distance. From the low energy perspective, each point of the moduli space represents a different effective field theory with distinct parameters. Some EFTs defined in certain regions of moduli space can be more pathological than others. For example, a compactification over $S^1$ of radius $R$ develops a light tower of states that becomes massless in the decompactification limit $R\rightarrow \infty$, breaking the validity of the effective field theory. A similar scenario occurs when moving through moduli space to the limit in which the gauge coupling goes to zero, since the magnetic WGC demands that the cut-off of the theory vanishes as well.

From the above discussion, it becomes clear that moduli space can be a very useful construction to systematically study the behaviour of EFT as their parameters are modified. With this goal in mind, the Distance Conjecture was introduced in \cite{Ooguri:2006in} and can be divided into two parts.

\begin{tcolorbox}[breakable, enhanced, colback=white]
    \textbf{Swampland Distance Conjecture}: Given a theory coupled to gravity with  moduli space $\mathcal{T}$ with dimension greater than zero and metric function $d:\mathcal{T}\times \mathcal{T}\rightarrow \mathbb{R}$, then
    \begin{itemize}
        \item For any point $\varphi_0\in\mathcal{T}$ and positive number $c$ there exists another point $\varphi\in \mathcal{T}$ such that $d(\varphi,\varphi_0)>c$. Consequently, $\mathcal{T}$ cannot be compact and it admits at least one boundary point $\varphi_b\in \partial \mathcal{T}$ which is at infinite distance from any other point in $\mathcal{T}$.
        \item When approaching an infinite distance point $\varphi_b\in \partial\mathcal{T}$ there is an infinite tower of states that becomes exponentially light with the geodesic distance. That is, for a fixed $\varphi_0\in \mathcal{T}$ and $\varphi\rightarrow \varphi_b$,
        \begin{equation}
            M(\varphi)\sim M(\varphi_0)e^{-\lambda d(\varphi,\varphi_0)}\,,
        \end{equation}
        with $\lambda$ an unspecified real positive parameter, expected to be $\mathcal{O}(1)$.
    \end{itemize}
\end{tcolorbox}

Thus, the Distance Conjecture generalizes the observation made for decompactification limits in Kaluza-Klein compactifications. The infinite tower of states becomes exponentially light, signaling an inevitable breakdown of any effective field theory, as it is impossible to have an EFT description with an infinite number of degrees of freedom that is weakly coupled to Einstein gravity. The consequence is that the quantum gravity cut-off $\Lambda_{Swamp}$ decreases exponentially as well when approaching infinite distance points in moduli space.

The variations induced in the kinetic terms of a dynamical field can always be described by a properly defined metric, which allows to extend the study of infinite distance points beyond the moduli space to other field configurations. Using this approach, the Distance conjecture was generalized in \cite{Lust:2019zwm} to any non-compact Einstein Space. The result has a particularly interesting extension when applied to AdS spaces and the variations of their cosmological constant. Different values of the cosmological constant $\Lambda$ would correspond to different configurations in the field content. The generalization of the Distance Conjecture implies \cite{Lust:2019zwm}

\begin{tcolorbox}[colback=white]
   \textbf{AdS Distance Conjecture} (ADC): Any AdS vacuum has an infinite tower of states that becomes light in the flat limit $\Lambda\rightarrow 0$ satisfying (in Planck units)
    \begin{equation}
        m\sim |\Lambda|^\alpha\,,
    \end{equation}
    where $\alpha$ is a positive order-one number.
\end{tcolorbox}
A strong version  based on string-derived examples was proposed in the same paper.
\begin{tcolorbox}[colback=white]
\textbf{Strong AdS Distance Conjecture} : For supersymmetric AdS vacua $\alpha=1/2$ whereas for non-SUSY vacua $\alpha\geq 1/2$ .
\end{tcolorbox}

The strong version is heavily related with the notion of scale separation. The latter is a property of certain models with extra compact spacetime dimensions that requires the typical length scale of the non-compact dimensions (the anti-de Sitter in the current case) to be parametrically larger than the Kaluza-Klein scale of the extra dimensions. Since we inhabit four dimensions, scale separation must be a crucial aspect of our Universe. If the Strong AdS Distance Conjecture were to be true, taking $\alpha=1/2$ and assuming that the tower of states is the Kaluza-Klein tower, so $m\sim 1/R_{KK}$, it follows that $R_{KK}\sim R_{AdS}$. Therefore there could not be scale separation between the AdS scale and the compactification scale in supersymmetric AdS vacua. The result is formulated more broadly in the following conjecture.

\begin{tcolorbox}[colback=white]
    \textbf{AdS/KK scale separation conjecture} (ASSC): There is no family of AdS vacua in which parametric separation between the AdS and the lightest Kaluza-Klein scales can be achieved.
\end{tcolorbox}

The study of scale separation has been a recurring subject during the last decades, e.g. \cite{Duff:1986hr, Douglas:2006es, Tsimpis:2012tu, Gautason:2015tig, Font:2019uva, Montero:2022ghl}, due to its phenomenological importance. Many examples have been found in String Theory supporting the strong version of the AdS Distance Conjecture and the absence of scale separation. However, IIA compactified on a CY orientifold does not satisfy it, providing a promising avenue to construct phenomenologically interesting vacua. Even so, it is not exempted of caveats, since this kind of models belong to the string-inspired family of vacua that lacks a complete 10-dimensional description. To elucidate the problem, several research paths are currently active: alternate versions of the strong distance conjecture have been proposed \cite{Buratti:2020kda} and search for the conformal duals of scale separated AdS vacua is being conducted \cite{Conlon:2021cjk, Apers:2022tfm,Apers:2022zjx}. One of the main issues of Type IIA AdS vacua is the necessity to perform a smearing of the localized sources in order to obtain the solutions to the equations of motion. It is not known whether scale separation will be preserved once the backreaction of the sources is fully taken into account.  We will explore these questions in more detail in the following chapters.

\subsubsection*{de Sitter Conjecture}

Despite the vast extension of the String Landscape, constructing trustable de Sitter vacua has proven to be a very challenging task and there has not yet been found a fully string-derived de Sitter vacuum in a controllable regime. Given the observed accelerating expansion of our Universe, this poses a problem of paramount importance. 

Several no-go theorems have been established that rule out the possibility of building de Sitter vacua under certain assumptions, but no definitive answer has been found. Based on these difficulties (we refer to \cite{Danielsson:2018ztv, Andriot:2019wrs} for a detailed description of the open problems) as well as on the relation with other Swampland conjectures , the de Sitter conjecture was proposed in \cite{Obied:2018sgi}. It was later refined in \cite{Ooguri:2018wrx,Garg:2018reu}.

\begin{tcolorbox}[colback=white]
   \textbf{Refined de Sitter conjecture}: The scalar potential of a theory coupled to gravity must either satisfy
    \begin{equation}
        M_p\frac{|\nabla V|}{V}\geq c\,, \qquad \min(\nabla_i\nabla_j)\leq \frac{-c'V}{M_p^2}\,,
    \end{equation}
    with $c,c'$ two positive $\mathcal{O}(1)$ constants. 
\end{tcolorbox}
The refined version rules out de Sitter minimal but not critical points.  It has been tested in asymptoptic regions of moduli space \cite{Junghans:2018gdb,Banlaki:2018ayh,Grimm:2019ixq,Andriot:2022way} but due to its conflict with experimental observations it is one of the most controversial conjectures. In addition there are potential counterexamples, such as the KKLT construction \cite{Kachru:2003aw} and the Large Volume Scenario \cite{Balasubramanian:2005zx, Conlon:2005ki}. These are however string-inspired effective field theories which lack a 10-dimensional understanding and their validity is yet an open question. Recent developments include, the Transplanckian Cersorship conjecture, which was proposed in \cite{Bedroya:2019tba} and only forbids the existence of dS vacua in asymptotic regions of moduli space. In addition, the quintesence proposal provides a potential way to evade the conjecture by constructing an accelerated expanding universe with dynamical dark energy \cite{Calderon-Infante:2022nxb}.  

In the following chapters we will revisit this conjecture to test the possibility of finding de Sitter vacua in type IIA compactifications. We will see that the conjecture holds in the presence of NSNS and RR fluxes and also for the specific geometric fluxes we consider. However, the possibility of violating the conjecture is left open for other Ansatzs with (non)-geometric fluxes.

\ifSubfilesClassLoaded{%
\bibliography{biblio}%
}{}

\end{document}


\part[\textcolor{Teja}{Type IIA Compactifications}]{\scshape \textcolor{Teja}{\huge Type IIA Compactifications}}
\label{part: type IIA}


\ifSubfilesClassLoaded{%
\tableofcontents
}{}

\setcounter{chapter}{2}
\chapter{Calabi-Yau Compactifications in Type IIA}
\label{ch: calabi-yau}

In this chapter we provide an overview of the core concepts and results concerning flux compactifications with the final goal of generating a semi-realistic 4-dimensional vacua description.  We  focus the analysis on massive Type IIA theory, but many of the methods and conclusions can be applied to other compactifications. For a deeper take on the subject, we recommend the original references  \cite{Grimm:2004ua,Grimm:2004uq,Grimm:2005fa} as well as the insightful reviews \cite{Ibanez:2012zz, Grana:2005jc, Blumenhagen:2006ci, Denef:2007pq, Koerber:2010bx,  Tomasiello:2022dwe}.

We start section \ref{cy-sec: geometry} by introducing the fundamental requirements that the 4-dimensional effective description of a String Theory compactification will need to satisfy.  Poincaré invariance and minimal preservation of supersymmetry can be framed in the language of structures and holonomy groups  and will demand that the external space is Minkowski while the compact manifold is a Calabi-Yau (in the absence of flux backgrounds). We  study the properties of Calabi-Yau manifolds and how, from the 4-dimensional perspective, their geometrical parameters generate a space of massless scalar fields (moduli) endowed with a Kähler structure. Through these steps we mostly follow \cite{Grana:2005jc, Koerber:2010bx, Tomasiello:2022dwe}. We  close the first section with the realization that an  orientifold projection is required to obtain phenomenologically intersecting vacua and an analysis of the effects of this projection on the field content of the theory, summarizing the results from \cite{Grimm:2005fa} and adapting them to our conventions.

In section \ref{cy-sec: flux compactifications} we  combine the previous geometrical results with the addition of background fluxes in the context of a Type IIA orientifold and study the 10-dimensional equations of motion for the flux field strengths as well as the supersymmetric conditions. In doing so, we  find out that fluxes cannot  be arbitrarily turned on in compact spaces, as they give a positive contribution to the energy-momentum tensor that needs to be compensated by negative tension sources (orientifold planes) through a set of relations known as Bianchi identities. Even when these are satisfied, the backreaction of the fluxes and the local sources over the compact geometry forces the external space to become $AdS_4$ and break the Calabi-Yau structure, greatly increasing the complexity of the vacua analysis \cite{Lust:2004ig}. However, relatively simple supersymmetric $AdS_4$ solutions can be found in the limit of weak coupling and large compact volume, where the smearing of the localized sources can be implemented and an approximate Calabi-Yau structure can be recovered \cite{Acharya:2006ne}. We end the second section  providing a concise overview of the perturbative expansion that goes beyond the smearing approximation for these types of solutions, as explored in \cite{Marchesano:2020qvg,Junghans:2020acz}.

Finally, in section \ref{cy-sec: 4d eff action} we study how all the components come together to provide a 4-dimensional effective description of massive Type IIA supergravity in the smearing approximation. We  discuss how the flux background generates a potential for the moduli and how that potential can be treated as a bilinear \cite{Bielleman:2015ina, Herraez:2018vae}. Lastly, we  consider generic flux configurations of RR and NSNS fields and systematically describe the different branches of AdS vacua they generate, as detailed in \cite{Marchesano:2019hfb}. 

%

\section{Calabi-Yau manifolds and where to find them}
\label{cy-sec: geometry}

In section \ref{basic-subsec: compactifications}, we introduced the conceptual framework  employed in String Theory to hide the additional six spatial dimensions and provide an effective theory in a 4-dimensional spacetime that could potentially describe the world we observe.  The idea consists in splitting the 10-dimensional spacetime into two factorized sectors, isolating the extra dimensions in a compact manifold with sufficiently  small size to justify the lack of experimental detection: 
\begin{equation}
    \mathcal{M}_{10}=\mathcal{M}_4\times X_6\,.
    \label{cy-eq: spacetime factorization}
\end{equation}
We demand the extended manifold $\mathcal{M}_4$ to have maximal spacetime symmetry, so it can be either a Minkowski, de Sitter or anti-de Sitter space. The choice among these three options will depend on the properties and objects present on the compact manifold $X_6$, whose analysis will comprise the first part of this thesis. The 10-dimensional matrix factorizes accordingly
\begin{equation}
    ds^2_{10}=e^{2A(y)}g_{\mu\nu}^{(4)}dx^\mu dx^\nu+g_{mn}^{(6)}dy^m dy^n\,,\qquad \mu,\nu=0,\dots, 3\,, \quad m,n=1,\dots,6\,,
    \label{cy-eq: spacetime factorization metric}
\end{equation}
where we note that the two sectors are not completely decoupled. The requirement of maximal spacetime symmetry leaves open the possibility of a non-trivial warp factor $A(y)$ between the compact and extended dimensions.\footnote{This factor is trivial in Calabi-Yau compactifications but will become important when we consider the flux backreaction in terms of $SU(3)\times SU(3)$ structures.} Finally the Lorentz group of $\mathcal{M}_{10}$ is also decomposed. When $\mathcal{M}_4$ is just the Minkowski space, we simply have the following splitting
\begin{equation}
    SO(1,9)\rightarrow SO(1,3)\times SO(6)\,.
\end{equation}
When $\mathcal{M}_4$ is not Minkowski the analysis is more involved. Focusing on the surviving symmetry of the extended dimensions, one needs to consider the stabilizer group of the 4-dimensional projection $p$ of a 10-dimensional point $P$. For the three maximally symmetric spacetimes ($\mathbb{M}_4$, $AdS_4$, $dS_4$) it turns out to be the same \cite{Tomasiello:2022dwe}
\begin{equation}
    \rm{Stab}(p)=SO(1,3)\,.
\end{equation}

The next step is to consider how the different fields in type II theories must behave under the factorization of spacetime. Starting with the fermionic fields, the internal space of the gravitinos at a given point $P$ $(\psi_{M,\alpha}^{a}$ with $a=1,2)$ splits as 
\begin{equation}
    \psi^{a}_{M\alpha}(P)=\sum_{IJ} c_{IJ}\psi^{(4)I a}_{\tilde{\alpha}}\otimes \psi^{(6)Ja}_{M\hat{\alpha}}\,,
\end{equation}
where $\{\psi^{(4)Ia}_{\tilde{\alpha}}\}$ and $\{\psi^{(6)Ja}_{M\hat{\alpha}}\}$ are basis for the spinors and vector-spinors on $\mathcal{M}_4$ and $X_6$ respectively and $c_{IJ}$ are the coefficients of the gravitino expanded in that basis. The gravitinos should be invariant under the symmetries of $\mathcal{M}_4$, and hence under $SO(1,3)$, but no spinor satisfies that condition. Therefore we must demand that $\psi^{a}_M$ vanishes everywhere. A similar argument also holds for the dilatino, which means that maximal supersymmetry requires that the vacuum expectation value of all fermionic fields vanishes, leading to a purely bosonic background. The constraints for the bosons are less severe but still significant. Preserving the maximal symmetry of the 4-dimensional factorized spacetime requires the fluxes either to have no entries associated with the extended space $\mathcal{M}_4$ or to cover the four directions of this space. Therefore the NSNS flux $H$ will always be internal and the RR fluxes $F_n$ will satisfy the following decomposition
\begin{equation}
    F_n^{(10)}=\tilde{F}_n+\rm{Vol}_4\wedge \hat{F}_{n-4}\,,
    \label{cy-eq: p-form spliting}
\end{equation}
with $\tilde{F}_n$ and $n$-form and  $\hat{F}_{n-4}$ and $(n-4)$-form both living in the internal space $X_6$. 

\subsection{Supersymmetry and Calabi-Yau manifolds}

It is generally imposed that the low energy 4-dimensional theories resulting from the compactification preserve some residual supersymmetry from the 10-dimensional theories. There are many motivations for such requirement. First, they generate stable solutions with no tachyons and give rise to simplifications that allow for a simpler and more systematic study. Thus, they provide a good framework in which to gather examples and develop intuition. Secondly, supersymmetry is a restrictive enough condition to make the study of such solutions manageable, but broad enough to allow for phenomenological interesting results. More specifically, compactifications with unbroken $N=1$ supersymmetry in four dimensions are excellent tools for developing particle physics models. Larger unbroken algebras are less likely to describe realistic models since they do not allow the presence of chiral fermions. Agreement with observations would require an additional process of supersymmetry breaking for the remaining $\mathcal{N}=1$ algebra between the $TeV$ and the compactification scales.

Let us assume until stated otherwise that no fluxes are present so all form fields are set to zero. The 10-dimensional supersymmetry transformations are
\begin{equation}
    \begin{aligned}
    \delta \psi^a_M=D_{M
    } \epsilon^a\,,\qquad \delta \lambda^a=\partial_M \phi \gamma^M\epsilon^a\,,
    \end{aligned}
    \label{cy-eq: susy equations no fluxes}
\end{equation}
where $\psi_M^a$, $\lambda^a$ (with $a=1,2$ throughout all this discussion) are the gravitinos and the dilatinos respectively, $\phi$ is the dilaton and $D_M$ is the covariant spinor derivative. Supersymmetric solutions are defined as those with vanishing gravitino and dilatino variation. Thus we require
\begin{equation}
    D_M \epsilon^a=0\,,\qquad \partial_M \phi \gamma^M\epsilon^a=0\,.
\end{equation}
Therefore, supersymmetry demands the existence of globally defined spinors $\epsilon^1,\epsilon^2$ that satisfy the above relations. The number of independent spinors verifying such conditions determines the number of supercharges and hence the amount of 4-dimensional supersymmetry.

Given the factorization of spacetime \eqref{cy-eq: spacetime factorization}, the most general form for the spinors $\epsilon^a$ is given by a combination of products of 4-dimensional spinors $\zeta^a_{\pm}$ and 6-dimensional spinors $\eta^a_{\pm}$ of different chirality \cite{Tomasiello:2022dwe}
\begin{equation}
    \begin{aligned}
        \epsilon^1=&\sum_J \zeta^{1}_{+,J}\otimes \eta^{1}_{+,J}+\zeta^{1}_{-,J}\otimes \eta^{1}_{-,J}\,,\\
        \epsilon^2=&\sum_J \zeta^{2}_{+,J}\otimes \eta^{2}_{-,J}+\zeta^{2}_{-,J}\otimes \eta^{2}_{+,J}\,,
    \end{aligned}
    \label{cy-eq: susy spinor expansion}
\end{equation}
where in $\eta^{a}_{J}$ the index $a$ identifies the associated  SUSY deformation parameter and $J$ labels the set of 6-dimensional spinors that enter the expansion (each of these spinors is accompanied by a 4-dimensional one that shares its label). The chirality combinations are chosen to keep the non-chiral nature of type IIA theory. The relation between the spinors $\eta^{1}$ and $\eta^{2}$ is in principle undetermined, but at least one non-vanishing  spinor needs to be globally defined on the compact manifold. According to complex geometry results discussed in appendix \ref{ch: ap complex geometry}, this means that the compact manifold is endowed with a $SU(3)$ structure containing a globally defined decomposable and non-degenerate almost complex structure 3-form $\Omega$ and a compatible pre-symplectic 2-form $J$.

Substituting \eqref{cy-eq: susy spinor expansion} back into the supersymmetry equations \eqref{cy-eq: susy equations no fluxes}, we can consider the extended and  compactified dimensions independently. We start with the exterior dimensions, where $\eta^{a}_{\pm,J}$ behave like scalars. Requiring preservation of maximal spacetime symmetry in $\mathcal{M}_4$ implies that the 4-dimensional spinors $\zeta^{a}_{\pm,J}$ must satisfy \cite{Tomasiello:2022dwe}
\begin{subequations}
    \begin{align}
        D_\mu^{(4)}\zeta^{a}_{\pm,J}=\frac{\mu}{2}\gamma_\mu \zeta^{a}_{\mp,J} \,,\\
        \partial_m (\zeta^{a}_{-,J}+\zeta^{a}_{+,J})=0\,,
    \end{align}
\end{subequations}
where $\mu$ is a real coefficient relating the 4-dimensional curvature and metric $R_{\mu\nu}^{(4)}=-3\mu^2 g_{\mu\nu}^{(4)}$ that arises due to the spinor connection of the covariant derivative. Addressing now the compact dimensions and using the previous results, we find
\begin{subequations}
    \begin{align}
        D_i^{(4)}\eta^{a}_{+,J}=&0 \,, \label{cy-eq: holonomy equation}\\
        \mu e^{-A}\eta^{1}_{+,J}-\partial_mA\gamma^m\eta^{1}_{-,J}=&0\,,\\
        \mu e^{-A}\eta^{2}_{+,J}+ \partial_m A\gamma^m \eta^{2}_{-,J}=&0\,,
    \end{align}
    \label{cy-eq: spinor relations SUSY}
\end{subequations}
with $A$ the warp factor introduced in \eqref{cy-eq: spacetime factorization metric}. Relation \eqref{cy-eq: holonomy equation} requires the existence of a globally defined non-vanishing covariantly constant spinor on the compact manifold, which reduces the holonomy group to $SU(3)$. As discussed in appendix \ref{ch: ap complex geometry}, a manifold with $SU(3)$ structure and $SU(3)$ holonomy is known as Calabi-Yau.\footnote{Alternatively, one can construct the pre-symplectic form and complex structure form from the covariantly constant spinor using \eqref{ap.geo-eq: J and Omega from spinors} and observe that demanding \eqref{cy-eq: holonomy equation} requires that all torsion classes vanish.} Thus, we conclude that preservation of supersymmetry in the absence of background fluxes restricts the compact manifold to a Calabi-Yau.

Regarding the two remaining equations in \eqref{cy-eq: spinor relations SUSY}, one can check that $\eta^{a}_{+,J}$ and $\gamma^m\eta^{a}_{-,J}$ are linearly independent, so both coefficients $\mu$ and $dA$ have to vanish separately
\begin{equation}
    \mu=0\,,\qquad  dA=0\,, 
\end{equation}
and so $\mathcal{M}_4$ is Minkwoski and the warp factor is a constant that can be reabsorbed in the 4-dimensional metric. We conclude that the factorization \eqref{cy-eq: spacetime factorization} in the absence of fluxes simplifies to
\begin{equation}
    ds_{10}^2=ds_4^2+ds^2_{\rm CY}\,.
\end{equation}

One might naively think that given a manifold with a single covariantly constant internal spinor $\eta^1_+$ (and its complex conjugate $\eta^1_-$), the residual supersymmetry in the 4-dimensional theory would be $\mathcal{N}=1$. However, the conditions required for the external space $\mathcal{M}$ together with the internal gravitino supersymmetric equations imply the existence of two independently 4-dimensional spinors $\zeta^1$ and $\zeta^2$ \cite{Grana:2005jc}. Substituting back in \eqref{cy-eq: susy spinor expansion} we recover eight associated real supercharges and thus $\mathcal{N}=2$ in $d=4$.

The existence of one  covariantly constant internal spinor $\eta=\eta^1=\eta^2$ is the minimal requirement to preserve some supersymmetry, but it is possible to be more restrictive and consider compact manifolds that have several covariantly constant spinors $\eta^a$. If there are two independent spinors $\eta^1,\eta^2$, both the structure group and the cohomology group are reduced to $SU(2)$ and the amount of supersymmetry increases to $\mathcal{N}=4$ in the 4-dimensional picture. Any other independent internal spinor would make the holonomy trivial (the only compact manifold satisfying that requirement is the six-torus $T^6$) and provide a compactification with $\mathcal{N}=8$.

Since we are interested in building models with phenomenologically interesting properties we will consider compactifications that provide minimal supersymmetry and thus taken over general Calabi-Yau manifolds. As it was discussed, these constructions have $\mathcal{N}=2$, which means that further refinements  (such as adding fluxes and applying orientifold quotients) will be needed to reduce the supersymmetry to $\mathcal{N}=1$ and allow for realistic models.

\subsection{Calabi-Yau structure}

Having stated the importance of Calabi-Yau manifolds in supersymmetry-preserving compactifications, we will now proceed to study their main properties and underlying structure. As detailed in appendix \ref{ch: ap complex geometry}, a Calabi-Yau manifold is a particular class of Kähler manifolds. Therefore they have a complex structure that allows to define a set of holomorphic coordinates and  possess a pre-symplectic closed 2-form known as Kähler form. 

The existence of this type of manifold was established in the celebrated Calabi-Yau theorem \cite{yau1978ricci}. It states that, on a compact Kähler manifold, given any closed $(1,1)$-form $R$ representing the first Chern class there is exactly one Kähler metric in each Kähler class whose Ricci form is $R$. It can be applied to the subset of Kähler manifolds that have $SU(3)$ structure and hence a globally defined non-vanishing 3-form $\Omega$. The presence of a form $\Omega$ with such properties means that the canonical bundle is trivial and therefore the first Chern class vanishes. Consequently, the Kähler metric can be deformed to a Ricci flat metric preserving the original Kähler class, resulting in a manifold with $SU(3)$-holonomy: a Calabi-Yau manifold.

The transformation of the hermitian metric modifies the complex structure and thus the compatibility condition with the symplectic form (Kähler form). We use it to fix the  normalization condition for $J$ and $\Omega$ on a $SU(3)$-structure (see \eqref{ap.geo-eq: normalization J Omega})
\begin{subequations}
    \begin{gather}
        J\wedge \Omega=0\,,\\
        d\rm{vol}_6=-\frac{1}{6}J^3=-\frac{i}{8}\Omega\wedge \bar{\Omega}\,,
    \end{gather}
    \label{cy-eq: normalization J Omega}
\end{subequations}
which are sufficient to guarantee the compatibility condition required by a Hermitian metric.

\subsubsection{Calabi-Yau manifold characterization}

Since Calabi-Yau manifolds are complex, we can consider the Dolbeault operators $\partial,\bar{\partial}$ to split the space of closed forms into cohomology groups $H^{p,q}$. The complex dimensions of these spaces are known as Hodge numbers and are denoted by $h^{p,q}$. They offer a useful method to classify the different Calabi-Yau three-folds that can arise in string compactifications. On a general Kähler manifold of complex dimensions $d$, Hodge numbers have two symmetries: complex conjugation ($h^{p,q}=h^{q,p}$) and Poincaré duality ($h^{p,q}=h^{d-p,d-q}$). In Calabi-Yau manifolds there is an additional symmetry $h^{p,0}=h^{d-p,0}$, while in connected manifolds $h^{0,0}=1$ and the existence of a holomorphic everywhere non-vanishing d-form additionally imposes $h^{d,0}=1$. Furthermore, we restrict the scope of this analysis to simply connected manifolds.\footnote{Note that the torus, the simplest Calabi-Yau manifold, is not simply connected and can therefore have $h^{1,0}\neq 0$ and harmonic one-forms. This additional freedom can be seen in terms of the holonomy group. When the holonomy group is strictly $SU(3)$ the manifold is simply connected, but if it is a subgroup thereof that is no longer the case. $K3\times T^2$ has $SU(2)$ holonomy and $T^6$ holonomy is trivial. In such cases, non-trivial closed one-forms are present. In practice, when discussing Calabi-Yau manifolds strict $SU(3)$ holonomy is assumed. Toroidal compactifications are constructed by modding out discrete freely acting isometry groups to reduce the supersymmetry from $\mathcal{N}=4$ to $\mathcal{N}=1$, restoring a full $SU(3)$ holonomy. } These have a trivial fundamental group; therefore, by the Hurewicz theorem, also $h^{1,0}=0$. Finally, combining the previous result with Hirzebruch–Riemann–Roch theorem and the fact that $c_1=0$, we obtain $h^{p,0}=0$ for $p\neq d$. Thus, the dimension of the  cohomology classes of a Calabi-Yau three-fold only has two free parameters ($h^{1,1}$ and $h^{1,2}$)  and it can be summarized in the following Hodge diamond 
\begin{equation}
\begin{matrix}
 & & & 1 & & & & \\
 & &0 & & 0 & & & \\
 &0 & & h^{1,1} & & 0 & & \\
1 & \phantom{h^{2,1}} &h^{2,1} & &h^{2,1} & \phantom{h^{2,1}} &1 \\
&0 & & h^{1,1} & & 0 & \\
 & & 0 & & 0 & & \\
 & & & 1 & & & \\
\end{matrix}
\label{cy-eq: hodge diamond cy}
\end{equation}

\subsubsection*{Deformations of the metric}

Hodge numbers do not provide a one-to-one classification of the possible Calabi-Yau manifolds, since they do not completely identify the topology of the manifold. Even for a fixed topology, there is a continuous infinite family of manifolds with different metrics related by smooth deformations of a set of parameters, known as moduli, that characterize the size and shape of the compactification space. Their properties can be derived from the study of metric deformations. Yau's theorem guarantees that using the relation \eqref{ap.geo-eq: metric from kahler form}, we can write such deformation in terms of the Kähler form and the complex structure tensor
\begin{equation}
    \delta g_{mn}=\delta J_{mp}I_n{}^p+J_{mp}\delta I_n{}^{p} \quad \Rightarrow \quad \begin{cases}
        \delta g_{i\bar{j}}=-i\delta J_{i\bar{j}}\,\\
        \delta g_{\bar{i}\bar{j}}g^{\bar{j}k}\Omega_{klq}=-i(\delta \Omega)_{\bar{i}lq}\,
    \end{cases}
    \label{cy-eq: metric deformations}
\end{equation}
where $m,n,p$ are arbitrary indices and $i,j,k,q$ are associated to the holomorphic coordinates. We observe that the deformations of the hermitian metric can be directly mapped to deformations of the Kähler and holomorphic forms. In particular, scaling transformations that preserve the complex structure (the changes only affect the non-vanishing entries of a Hermitian metric, i.e. those with one holomorphic and one antiholomorphic index) are associated to deformations of the Kähler form. Meanwhile, modifications that break the Hermitian metric are parametrized by deformations of the holomorphic three-form, as expected. 

The deformations must preserve the Calabi-Yau nature of the manifold, which in particular means that \eqref{cy-eq: normalization J Omega} must hold.\footnote{An alternative way to proceed is to demand Ricci flatness. This leads to a set of equations for the metric deformations known as Lichnerowicz equation, whose analysis concludes that the forms $\delta J$ and $\delta \Omega$ must be harmonic. } First,  the Kähler form is a closed $(1,1)$-form and $J^3$ must generate the volume form. Therefore, $J$ cannot be exact and, by the Hodge theorem, it can be decomposed in a basis of $h^{1,1}$ harmonic forms $\{\omega_A\}$. Such decomposition must hold after performing a continuous deformation. Hence we have 
\begin{equation}
    J=\sum t^A\omega_A\,,
\end{equation}
where $t^a$ are the Kähler moduli, continuous real parameters whose changes characterize the deformations of the symplectic form. 

\begin{tcolorbox}[enhanced, breakable, colback=uam!10!white, colframe=uam!85!black, title=Kähler Cone and Mori Cone]
\small
The full picture of the moduli space of symplectic structures cannot be determined from infinitesimal deformations alone, since these do not capture the requirement of positive definiteness of the associated hermitian metric $g_{mn}=J_{mp}I_{n}{}^p$. More specifically, the volume of every cycle of the compact manifold must be non-negative. For the case of even cycles, this quantity is calibrated by $\Re(e^{i\theta}e^{-iJ})$ (more details on appendix \ref{ap.geo-subsec: polyforms}). Considering a 2-cycle $\Pi_2$ and choosing $\theta=\pi/2$
\begin{equation}
    {\rm Vol}(\Pi_2)=\int_{\Pi_2} J\,,
    \label{cy-eq: calibration of 2-cycles}
\end{equation}
is the smallest volume in the homology class $[\Pi_2]$, so it is sufficient to demand it to be positive in order to guarantee the positive definiteness of all members of the class. 

Cycles calibrated by $J$ are holomorphic, i.e. they are the zero loci of systems of holomorphic functions. The subspace of $H_2(Y,\mathbb{Z})$ generated by the classes with a holomorphic representative is known as the \textit{Mori Cone}. Then, the subspace of $J\in H^{1,1}(Y)$ that contains the allowed Kähler forms is given by those satisfying  \eqref{cy-eq: calibration of 2-cycles} for any $\Pi_2$ of the Mori cone. Such set is called the \textit{Kähler cone}. It is indeed a cone, since for any $J$ in the Kähler cone and $r>0$, $rJ$ also belongs to that subspace.
\end{tcolorbox}

Similarly, it can be proven that the complex structure deformations are parametrized by $h^{2,1}$ complex parameters $\mathcal{U}^K$ which are in one-to-one correspondence with harmonic $(2,1)$-forms $\chi_K$
\begin{equation}
    \chi_K=-\frac{1}{4}\Omega_{ijk}g^{k\bar{l}}\frac{\partial g_{\bar{l}\bar{q}}}{\partial \mathcal{U}^K} dz^i\wedge dz^j \wedge dz^{\bar{q}}\,,\qquad K\in\{1,\dots,h^{2,1}\}\,.
    \label{cy-eq: harmonic 2,1 complex structure}
\end{equation}
Thus, there are $h^{2,1}$ harmonic $(2,1)$-forms describing deformations of the pure antiholomorphic entries $g_{\bar{i}\bar{j}}$ and $h^{1,2}(=h^{2,1})$ harmonic $(1,2)$-forms describing deformations of the pure holomorphic sector $g_{ij}$.

\subsubsection*{Deformation of scalars and antisymmetric tensor fields}

As it was discussed in chapter \ref{ch: basics}, low energy effective actions for string theories contain massless scalars and p-form fields. It is then possible to consider deformations of such backgrounds in addition to the metric. 

Starting with a 10-dimensional free scalar on $\mathcal{M}_4\times X_6$, we can perform a Fourier decomposition that splits the internal and external profiles generalizing the procedure described in \ref{basic-subsec: compactifications}
\begin{equation}
    \phi(x^0,\dots, x^9)=\sum_{\vec{k}} \phi^{\vec{k}}_{6d}(y^m)\phi_{4d}^{\vec{k}}(x^\mu)\,,
\end{equation}
where $\phi^{\vec{k}}_{6d}$ are chosen to be the eigenfunctions of the Laplacian $\Delta_{X_6}$ (they form a basis of functions in $X_6$) and $\vec{k}$ labels the eigenspace with eigenvalue $\lambda^{\vec{k}}$. A massless free scalar in ten dimensions satisfy $\square_{10d}\phi=(\square_{4d}
+\Delta_{X_6})\phi=0$. Therefore, the 4-dimensional scalar field $\phi_{4d}^{\vec{k}}(x^\mu)$ verifies
\begin{equation}
    \square_{4d}\phi_{4d}^{\vec{k}}(x^\mu)-\lambda^{\vec{k}}\phi^{\vec{k}}_{4d}(x^\mu)=0\,,
\end{equation}
and so we conclude that the eigenvalue of the compact sector plays the role of a squared mass for the corresponding 4-dimensional field. Restricting to the low energy sector, we just need to consider solutions of the zero mode equation in $X_6$: harmonic functions. According to the Hodge diamond \eqref{cy-eq: hodge diamond cy}, the space of harmonic functions in a Calabi-Yau manifold (and actually in any compact manifold) is one-dimensional and corresponds to the space of constant functions.

A similar approach can be taken to describe deformations of the background of p-forms $C_p$. As we will see in more detail later, the equation of motion derived from the effective action is $\Delta_{10} C_p=(\Delta_{4d}+\Delta_{6d})C_p=0$. The $C_p$ can be expressed as a sum of terms factorized in the 4-dimensional and 6-dimensional space $C_p(x^M)=C_{p-q}(x^\mu)c_{p}(y^m)$ and therefore massless 4-dimensional fields arise from 10-dimensional p-forms whose compact part consists of harmonic q-forms with $q\leq p$ 
\begin{equation}
    \Delta_{6d}c_q(y^m)=0\,.
\end{equation}
Harmonic forms act as representatives of the cohomology classes. Thus, the number of 4-dimensional massless $(p-q)$-forms derived from the compactification of a 10-dimensional p-form is given by the Betti number $b_q(X_6)=\sum_{r=0}^q h^{r,q-r}$. We conclude that the space of massless deformations of the background p-form fields can be grouped in terms of the order of the 4-dimensional $(p-q)$-form they give rise to, with the multiplicity of zero modes in each of such families generated by the harmonic basis of $H^q(Y)$.

Let us consider first the NSNS 2-form $B$ present in both Type IIA and Type IIB theories. After compactification, it generates a 4-dimensional 2-form\footnote{This two-form must be closed in order to fulfill the requirement of maximal symmetry in $\mathcal{M}_4$ discussed around \eqref{cy-eq: p-form spliting}.} $B_{\mu\nu}$ ($b_0=1$), no 4-dimensional vectors $B_{\mu n}$ ($b_1=0$) and $h^{1,1}$ 4-dimensional scalars $B_{mn}$ ($b_2=h^{1,1}$). Thus we end up with the decomposition
\begin{equation}
    B_2=B(x^\mu)+\hat{B}(y^m)=B(x^\mu)+b^A(x^\mu)\omega_A\,,
    \label{cy-eq: N=2 B expansion}
\end{equation}
where in the last step we expanded the internal 2-form $\hat{B}^\mu$ in a basis of harmonic $(1,1)$-forms. The internal part can hence be linked to the Kähler deformation of the metric (also belonging to the cohomology group $H^{1,1}(X_6)$) through the definition of the complexified Kähler form
\begin{equation}
    J_c\equiv \hat{B}+iJ\,.
\end{equation}

Finally, we can consider the RR potentials $C_1$ and $C_3$ present in Type IIA. Since there are no harmonic 1-forms in $X_6$, $C_1$ must live in $\mathcal{M}_4$ and satisfy $dC_1=0$ according to $\eqref{cy-eq: p-form spliting}$, so it is not a dynamical field. The RR 3-form is much more interesting. It decomposes as 
\begin{equation}
    C_3=c_3(x^\mu)+A^A(x)\omega_A+\hat{C}_3\,,
    \label{cy-eq: N=2 C3 expansion}
\end{equation}
with $c_3$ a 4-dimensional 3-form, $A^A$ $h^{1,1}$ 4-dimensional 1-forms and $\hat{C}_3$ a 3-form of the internal space $X_6$ ($b_3=2+2h^{2,1}$ 4-dimensional scalars).

All this set of deformations studied from the compactified 4-dimensional point of view can be grouped to constitute the bosonic components of various supersymmetric multiplets of $\mathcal{N}=2$ in four dimensions. We refer the reader to \cite{Grimm:2005fa,Grana:2005jc} for a detailed discussion.

\subsection{Structure of Moduli Space}
\label{cy-subsec: structure moduli space}

In the previous paragraphs we have explained how the moduli space of deformations of the compact geometry has a product structure splitting the transformations of the Kähler and holomorphic form 
\begin{equation}
  M=  M^{2,1}_{cs}\times M^{1,1}_K\,,
 \label{cy-eq: moduli space factorization}
\end{equation}
with the dimension of each component given by the associated Hodge number. Now we would like to explore the geometric properties of the moduli space itself. The first step is to note that this space is endowed with a natural metric \cite{Hubsch:1992nu, tian1987smoothness, Candelas:1990pi}
\begin{equation}
    ds^2=\frac{1}{V}\int g^{i\bar{j}}g^{k\bar{l}}[\delta g_{ik}\delta g_{\bar{j}\bar{l}}+(\partial_{i\bar{l}}\partial g_{k\bar{j}}-\delta B_{i\bar{l}}\delta B_{k\bar{j}})]\sqrt{g}d^6y\,,
    \label{cy-eq: metric of deformations}
\end{equation}
where $V$ is the volume of the Calabi-Yau manifold and we have included the possibility to have deformations associated with the NSNS 2-form $B$ in addition to the hermitian metric. The first and second terms between brackets correspond to the distance on the complex structure moduli space and the Kähler space respectively.

Let us first discuss the complex structure sector. From \eqref{cy-eq: metric deformations} and \eqref{cy-eq: harmonic 2,1 complex structure} the metric can be written like
\begin{equation}
    ds^2_{cs}=2G_{K\bar{L}}\delta \mathcal{U}^K \delta \bar{\mathcal{U}}^{\bar{L}}\,\qquad G_{K\bar{L}}=-\frac{\int_{X_6}\chi_{K}\wedge\bar{\chi}_{\bar{L}}}{\int_{X_6} \Omega \wedge \bar{\Omega}}\,.
\end{equation}
Then, it is possible to prove using Kodaira's formula \cite{tian1987smoothness,Candelas:1990pi,Grimm:2005fa} that 
\begin{equation}
    \partial_K \Omega=\chi_{K}-\Omega\partial_K K_{\rm cs}\,,
\end{equation}
where $\partial_K=\partial/\partial \mathcal{U}^K$ and $K_{\rm cs}$ only depends on the complex structure moduli $\mathcal{U}^K$. Gathering all the information above, we see that the metric complex structure sector of the moduli space can be written as
\begin{equation}
    G_{K\bar{L}}=\partial_K\partial_{\bar{L}} K_{\rm cs}\,, \qquad K_{\rm cs}=-\log\left(-i\int_{X_6} \Omega\wedge \bar{\Omega}\right)\,.
    \label{cy-eq: CY kahler potential CS}
\end{equation}
Therefore we conclude that the function $K_{\rm cs}$ plays the role of a Kähler potential and thus the complex structure moduli space is a Kähler manifold on its own right.\footnote{A Kähler manifold requires both a Kähler form and a complex structure. The latter is directly inherited from the original Calabi-Yau. The compatibility condition between the two is guaranteed through the construction of the hermitian metric $G_{K\bar{L}}$ in terms of the Kähler potential.}

It is worth noting that $\Omega$ is only defined up to rescalings  by a holomorphic function $e^{-h(\mathcal{U})}$, which also modifies the associated Kähler potential.
\begin{equation}
    \Omega\rightarrow \Omega e^{-h(\mathcal{U})}\,,\qquad K_{\rm cs}\rightarrow K_{\rm cs} +h+\bar{h}\,.
    \label{cy-eq: Omega scale invariance}
\end{equation}

An analogous result can be derived for the Kähler sector of \eqref{cy-eq: metric deformations}. Making use of the complexified Kähler form $J_c$ and expanding it in a harmonic basis of (1,1)-forms $\omega_A$
\begin{equation}
    J_c=T^A\omega_A\,,
\end{equation}
we obtain
\begin{equation}
    ds^2_K\propto  G_{AB}\delta T^A \delta T^B \,,\qquad G_{AB}=\frac{1}{4V}\int\omega_A\wedge\star \omega_B\,. 
\end{equation}
Finally, using \eqref{cy-eq: normalization J Omega}, we know $J^3\propto d\rm{vol}_6$ and hence we can define a scalar function
\begin{equation}
    K_K=-\log\left(-\frac{4}{3}\int_{X_6} J^3\right)\,.
    \label{cy-eq: CY kahler potential Kahler}
\end{equation}
This function verifies $G_{AB}=\partial_{T^A}\partial_{\bar{T}^B}K_K$ and consequently performs the role of Kähler potential for the Kähler moduli. We conclude that both sectors of the moduli space in \eqref{cy-eq: moduli space factorization} are described by Kähler manifolds.

\subsection{Orbifolds and toroidal compactifications}
\label{cy-subsec: orbifolds}

By far, the most simple and well understood superstring compactification is the compactification on a flat torus. However, in compactifications from 10 to 4 dimensions, this example cannot be used for anything more than a proof of concept, since it leads to unappealing phenomenology. As discussed before, such compactifications give rise to $\mathcal{N}=8$ supersymmetry in 4 dimensions in addition to unrealistic gauge groups and matter representations. Thus, working with more general Calabi-Yau manifolds is required, which is far from a simple task, given that the explicit expression of the metric of almost all generic Calabi-Yau manifolds remains to be discovered. 

In this context, orbifold quotients of toroidal manifolds provide a good compromise between computational simplicity and phenomenologically realistic predictions. As such,  during the last decades they have become an essential tool for model building, developing intuition and testing general results \cite{Candelas:1985en,Dixon:1985jw}.  

An orbifold is obtained by dividing a smooth manifold $X$ by the non-free action of a discrete group $\Gamma$, non-free meaning that there exists some fixed point in the manifold which is left invariant under the action of the group.  For our purposes, we will assume $\Gamma$ to be abelian. Due to the aforementioned condition, orbifolds are not manifolds: it is not possible to build a smooth local map to an open set of $\mathbb{R}^n$ in the vicinity of the fixed point. Hence, while the torus is completely flat, toroidal orbifolds $X/\Gamma$ develop non-zero curvature in the fixed points under the action of the discrete group, where conical singularities arise.

As it was first introduced when discussing modular invariance in \ref{basic-susbsec: interaction and expansion}, a torus can be characterized by a lattice $\Lambda$. The torus is the space created by identifying points in $\mathbb{R}^n$ that differ by a vector in $\Lambda$
\begin{equation}
    T^n=\frac{\mathbb{R}^n}{\Lambda}\,.
\end{equation}
The length of the vectors connecting adjacent points of the lattice is associated with the Kähler form of the torus while its complex structure parametrizes the angles between them. To generate a toroidal orbifold  the group $\Gamma$ must be an automorphism of the torus lattice $\Lambda$. Furthermore, requiring that the resulting space has $SU(3)$ holonomy restricts $\Gamma$ to a subgroup of $SU(3)$. Consequently, the only available quotients are when $\Gamma=\mathbb{Z}_N$ with $N=3,4,6,7,8,12$ and $\Gamma=\mathbb{Z}_N\times \mathbb{Z}_M$ with $N,M=2,3,4,6$.

The underlying idea explaining how the quotient modifies the holonomy of the torus is simple. The identifications through the group action generate new loops around the singular points that are closed in the quotient space but not in the original. The holonomy transformations around such loops can be non-trivial, consequently modifying the holonomy group of the quotient.

Given the intuition from General Relativity, one could think that a theory defined on an orbifold would be singular due to the pathological properties of the fixed points. Surprisingly, the extended nature of strings solves the problem introducing a new set of states (known as twisted states) that, together with the ones projected from the original torus (untwisted), generate a well-behaved theory.

The twisted sector originates from the same cause that enlarged the holonomy group: there are new closed string states in the quotient space coming from open string states of the torus. Mathematically, those strings satisfy
\begin{equation}
    X^\mu(\sigma+2\pi,t)=gX^\mu(\sigma,t)\,,
    \label{cy-eq: twisted orbifold action}
\end{equation}
for some $g\in \Gamma$. Therefore they are localized around the fixed points and encode the information of the quotient. Their presence is crucial to guarantee the properties that make String Theory so promising (unitarity, finiteness, anomaly cancellation...), as they restore modular invariance in the worldsheet \cite{Dixon:1985jw}. To see this, one can consider the one-loop partition function  and a closed string invariant under the action of the group $\Gamma$. Its spatial component is parametrized by $\sigma$ and satisfies \eqref{cy-eq: twisted orbifold action} for $g=1$. Modular invariance interchanges the role of $\sigma$ and $t$. It can thus cause $\Gamma$ to act non-trivially on the resulting open string, which in turn requires the presence of such twisted states to have a modular invariant map. 

 Having discussed the well-behaved properties of the singular points, it is worth noting that toroidal orbifolds, like standard toroidal compactifications, are described by free 2-dimensional worldsheet theories, since in both cases strings propagate in a flat metric. Therefore, the quantization of the string is exact in the $\alpha'$ expansion.

An alternative way to handle orbifolds is to perform deformations or resolutions, which replace the conifold singularity by an $S^3$ or an $S^2$ manifold respectively, generating a smooth Calabi-Yau as a result \cite{Vafa:1994rv}. The fact that String Theory is well-behaved in the orbifold limit (when the volume of these spheres collapses to zero) hints at the fact that orbifolds are perfectly viable spaces corresponding to particular limit regions of moduli space. Orbifold spaces can be  seen as limit points on the boundary of the Kähler cone of the resolved manifold. For a detailed analysis of orbifold resolutions, we refer the reader to the seminal papers \cite{Aspinwall:1994ev, Lust:2006zh} as well as the comprehensive review \cite{Reffert:2007im}.



\subsection{Orientifolds, forms and fluxes }
\label{cy-subsec: orientifolds}

As we discussed previously, generic Calabi-Yau compactifications reduce the 4-dimensional supersymmetry to $\mathcal{N}=2$, whereas we would need $\mathcal{N}=1$ for phenomenological model building. The last step to arrive to such kind of constructions is to take an orientifold projection \cite{Angelantonj:1996uy, Berkooz:1996dw,Aldazabal:1998mr,Cvetic:2001nr}. This is achieved by modding out the Calabi-Yau $X_6$ by the orientifold action $\mathcal{O}=\Omega_p\mathcal{R}$ where $\Omega_p$ is the worldsheet parity and $\mathcal{R}$ is a $\mathbb{Z}_2$ symmetry of $X_6$.  Therefore $\Omega_p$ maps the left moving to the right moving sector and vice versa, resulting in a quotient space that admits the existence of unoriented strings. The action of $\mathcal{R}$ is chosen to preserve $\mathcal{N}=1$ supersymmetry. In Type IIA (which will be our main focus) it needs to be an antiholomorphic isometric involution of $X_6$ \cite{Acharya:2002ag}, which means $\mathcal{R}^2=1$ and
\begin{equation}
    \mathcal{R}J=-J\,.
\end{equation}
Due to the Calabi-Yau normalization condition \eqref{cy-eq: normalization J Omega}, it also implies $ \mathcal{R}\Omega=e^{2i\theta}\bar{\Omega}$ with $\theta$ a constant phase that we choose to be $\pi/2$ and so 
\begin{equation}
    \mathcal{R}\Omega=-\bar{\Omega} \,.
    \label{cy-eq: involution of Omega}
\end{equation}
Finally, to keep $\mathcal{N}=1$ supersymmetry an additional ingredient\footnote{In Type IIB orientifolds the isometric involution is holomorphic and this factor is optional, distinguishing between two different constructions. When its present the involution satisfies $\mathcal{R}\Omega=\Omega$ and the theory contains $O5$ and $O9$ planes. When it is absent $\mathcal{R}\Omega=-\Omega$ and the orientifold content consists on $O3$ planes and $O7$ planes \cite{Acharya:2002ag}.} must be added to account for the fermionic states in the particular case of Type IIA: a factor $(-1)^{F_L}$, with $F_L$ the spacetime fermion number in the left moving sector. We end up with
\begin{equation}
    \mathcal{O}=(-1)^{F_L}\Omega_p\mathcal{R}\,.
\end{equation}
 The sets of fixed points under the action of $\mathcal{R}$ are known as orientifold planes and are denoted by $Op$ with $p+1$ the spacetime dimension of the object. Orientifold planes are sources of the Ramond-Ramond fields with opposite charge to D-branes, satisfying $Q_{Op}=-2^{p-4}Q_{Dp}$. Their presence is thus an important tool to satisfy the tadpole constraints, as we will see in the following sections. Contrary to D-branes, whose coordinates are fields associated to open strings which could change modifying the profile of the extended object in the process, O-planes are completely fixed due to the involution action and are therefore non-dynamical. They can be seen as spacetime defects which carry a mass and charge density. 
 
Type IIA orientifolds generically contain $O6$ planes. To see this, first note that the involution $\mathcal{R}$ acts trivially in $\mathcal{M}_4$ and therefore orientifold planes are always spacetime filling. Regarding the internal space, the fact that the involution is antiholomorphic restricts the fixed set to three-cycles $\Pi_3$ satisfying
\begin{equation}
    J|_{\Pi_3}=0\,,\qquad \Re (\Omega)|_{\Pi_3}=0\,. 
\end{equation}
This means that fixed loci under the orientifold projection are spacetime filling O6 planes wrapping special Lagrangian 3-cycles \cite{Hitchin:1999fh} in the internal space. 

\subsubsection*{Basis of forms}

As we have recurrently observed through this section, the cohomology groups play a crucial role in characterizing the compactified space and the massless scalars that arise in the 4-dimensional theory.  Its structure, summarized in \eqref{cy-eq: hodge diamond cy}, is by $H^{1,1}(X_6)$, $H^{3}(X_6)$ and its duals.\footnote{The decomposition of 3-forms in the Dolbeault cohomology is not particularly useful as the holomorphic form $\Omega$ and its deformations mix different sectors.} The action of the involution $\mathcal{R}$ further splits such structure into even and odd forms $H^p(X_6)=H^p_+\oplus H_-^p$. In table \ref{cy-table: harmonic basis} we introduce a basis for each sector in terms of their harmonic representatives (note that Hodge duality imposes $h_+^{1,1}=h_-^{2,2}$ and $h_-^{1,1}=h_+^{2,2}$). 

\begin{table}[H]
\def\arraystretch{1.5}
\begin{center}
\begin{tabular}{|c|| c| c| c| c| c| c|} 
\hline
Cohomology group & $H^{1,1}_+$ & $H^{1,1}_-$ & $H^{2,2}_+$ & $H^{2,2}_-$ & $H^{3}_+$ & $H^{3}_-$ \\
\hline\hline
 Dimension & $h^{1,1}_+$ & $h^{1,1}_-$ & $h^{1,1}_-$ & $h^{1,1}_+$ & $h^{2,1}+ 1$ & $h^{2,1} + 1$ \\
 Basis & $\varpi_\alpha$ & $\omega_a$ & $\tilde\omega^a$ & $\tilde\varpi^\alpha$ & $\alpha_\mu$  & $\beta^\mu$  \\
 \hline
\end{tabular}
\end{center}
\caption{Representation of various harmonic forms in Type IIA orientifolds and their counting.  }
\label{cy-table: harmonic basis}
\end{table}

The basis is chosen to satisfy the following relations
\begin{align}
    \frac{1}{\ell_s^{6}}  \int_{Y} \om_a \wedge \tilde{\om}^b= \delta_a{}^b , \qquad   \frac{1}{\ell_s^{6}}  \int_{Y} \varpi_\alpha \wedge \tilde{\varpi}^\beta = {\delta}_\alpha{}^\beta , \qquad  \frac{1}{\ell_s^{6}} \int_{Y} \alpha_\mu \wedge \beta^\nu=\delta_\mu{}^\nu\,. 
    \label{cy-eq: basis relations}
\end{align}
It is also useful to define the intersection numbers of divisors dual to the harmonic $H^{1,1}$ basis of our orientifold \footnote{Due to our choice of volume form \ref{cy-eq: normalization J Omega} the triple intersection numbers must be defined with an additional minus sign compared to the more standard definition in the literature so that, whenever $\{[\ell_s^{-2}\omega_a]\}_a$ is dual to a basis of Nef divisors, ${\cal K}_{abc} \geq 0$. The same observation applies to the curvature correction term $K_a^{(2)}$ defined in \eqref{bio-Kcurv}.}
\begin{equation}
    \mathcal{K}_{abc}\equiv -\frac{1}{\ell_s^6}\int_{X_6} \omega_a\wedge\omega_b\wedge \omega_c\,,\qquad \hat{\mathcal{K}}_{a\alpha\beta}\equiv -\frac{1}{\ell_s^6}\int_{X_6}\omega_a\wedge \varpi_\alpha\wedge \varpi_\beta\,.
    \label{cy-eq: intersection numbers}
\end{equation}
\subsubsection*{Field decomposition}

Now that orientifold projection has been established together with its action on a basis of relevant harmonic forms for the Calabi-Yau geometry, let us consider how its presence affects the field content of Type IIA. First of all, in order for the 10-dimensional fields to remain invariant under $\mathcal{O}$, some of them will have to transform non-trivially under the involution $\mathcal{R}$ \cite{Brunner:2003zm}
\begin{equation}
    \mathcal{R}\phi=\phi\,,\quad \mathcal{R}g=g\,,\quad \mathcal{R}B=-B\,,\quad \mathcal{R}C_1=-C_1\,,\quad \mathcal{R}C_3=-C_3\,.
\end{equation}

Tracing back to \eqref{cy-eq: N=2 B expansion}, \eqref{cy-eq: N=2 C3 expansion} and the discussion therein we deduce the following:
\begin{itemize}
    \item The dilaton is simply a function of the 4-dimensional coordinates, as there are no non-trivial harmonic functions in a compact space. Similarly, the 1-form $C_1$ only lives $\mathcal{M}_4$, since there are no harmonic 1-forms. Given that  the involution only acts on the compact manifold, the orientifold projects out $C_1$
    \begin{equation}
        \phi(x^M)=\phi(x^\mu)\,,\qquad C_1=0\,.
    \end{equation}
    \item The four-dimensional 2-form sector of $B$ is projected out for the same reason as $C_1$. Therefore we have
    \begin{equation}
        B_2=b^a\omega_a \,, \qquad a\in\{1,\dots,h^{1,1}_-\}\,.
        \label{cy-eq: B orientifold decomposition}
    \end{equation}
    \item The 3-form $C_3$ still admits a rich expansion
    \begin{equation}
        C_3=c_3(x)+A^{\alpha}(x^\mu)\omega_\alpha +\hat{C}_3\,,\qquad \hat{C}_3\equiv \xi^\mu(x)\alpha_\mu\,,
        \label{cy-eq: C orientifold decomposition}
    \end{equation}
    with $\xi^\mu$ real 4-dimensional scalars associated to the internal 3-form $\hat{C}_3$, $A^\alpha$ 4-dimensional closed 1-forms and $c_3$ a 4-dimensional 3-form. Note that the latter carries no physical degrees of freedom and can be interpreted as a flux parameter.  
\end{itemize}

The remaining field is the graviton, which factorizes in the external and internal parts. The internal sector can be described in terms of the Kähler and holomorphic forms. Let us consider both of them in detail. The Kähler form is a closed $(1,1)$-form odd under the involution and can be decomposed as
\begin{equation}
    J=t^a\omega_a\,, \qquad a\in\{1,\dots, h_-^{1,1}\}\,,
\end{equation}
which using \eqref{cy-eq: normalization J Omega} and the basis \eqref{cy-table: harmonic basis} means
\begin{equation}
    -J\wedge J=\cK_{abc}t^at^b \tilde\omega^a \,.
\end{equation}
The Kähler moduli $t^a$ can be combined with the scalars $b^a$ in \eqref{cy-eq: B orientifold decomposition} to build $h^{1,1}_-$ $\mathcal{N}=1$ chiral multiplets. Consequently, we define the complexified Kähler form and the complexified Kähler moduli
\begin{equation}
    J_c\equiv B+iJ=T^a\omega_a\,,\qquad T^a=(b^a+it^a) \,.
\end{equation}

Meanwhile, the holomorphic 3-form is neither odd nor even, the involution maps it to its antiholomorphic complex conjugate, and so we have a decomposition of the form
\begin{equation}
    \Omega=\mathcal{Z}^\mu \alpha_\mu-\mathcal{F}_\mu\beta^\mu\,, \qquad \mu\in\{0,1,\dots,h^{2,1}\}\,.
    \label{cy-eq: Omega expansion Z and F}
\end{equation}
It can be shown \cite{Grimm:2005fa} that the parameters $\mathcal{Z}^\mu$ and $\mathcal{F}_\nu$ are related, which is a manifestation of the fact that the space of complex structure deformations is a special Kähler manifold. Indeed, for any Calabi-Yau orientifold\footnote{This result is valid for any Calabi-Yau, the orientifold projection is not required.} there exists a function $\mathcal{F}$ called the prepotential satisfying $\mathcal{F}_\mu=\partial \mathcal{F}/\partial \mathcal{Z}^\mu $.
Due  to the scale invariance of $\Omega$ \eqref{cy-eq: Omega scale invariance}, one of the parameters (conventionally chosen to be $Z^0$) is unphysical. We can thus identify the $h^{2,1}$ remaining $Z^K$ with the $h^{2,1}$ complex structure moduli $\mathcal{U}^K$  defining $\mathcal{U}^K=Z^K/Z^0$. There is an additional constraint to take into account, as \eqref{cy-eq: involution of Omega} further imposes
\begin{equation}
    \Re(\mathcal{Z}^\mu)=0\,,\qquad \Im(\mathcal{F}_\mu)=0\,.
    \label{cy-eq: truncation holomorphic form orientifold}
\end{equation}
The first set of equations introduces $h^{2,1}+1$ real conditions for $h^{2,1}$ complex scalars $\mathcal{U}^K$, one of which is redundant due again to the scale invariance of $\Omega$. We end up with $h^{2,1}$ real scalars $\Im(\mathcal{U_K})$ as moduli of the complex structure of the Calabi-Yau orientifold. 

Having this counting in mind, it is more convenient to keep the scale freedom \eqref{cy-eq: Omega scale invariance} unfixed  and introduce a compensator field
\begin{equation}
    \mathcal{C}=e^{-\phi}e^{\frac{1}{2}(K_{\rm cs}-K_K)}\,,
    \label{cy-eq: compensator field}
\end{equation}
with $\phi$ the 10-dimensional dilaton. Under a scaling compatible with the orientifold projection $\Omega\rightarrow \Omega e^{-\Re(h)} $, the compensator transforms as $\mathcal{C}\rightarrow\mathcal{C} e^{\Re(h)}$. This approach has the advantage of coupling the analysis of the dilaton modulus to the complex structure ones, which is convenient since they have to be paired together and with the internal 3-form $\hat{C}_3$ in \eqref{cy-eq: C orientifold decomposition} in order to recover a multiplet of $\mathcal{N}=1$ supersymmetry. We thus define a complexified holomorphic form
\begin{equation}
    \Omega_c\equiv \hat{C}_3+i\im(\mathcal{C}\Omega)\,,
\end{equation}
which encodes all the relevant information regarding the holomorphic form, the dilaton and the internal 3-form. Contrary to the original $\Omega$, $\Omega_c$ is even under the orientifold involution and can thus take the following expansion, that introduces the complexified complex structure moduli $U^\mu$
\begin{equation}
    \Omega_c= U^\mu \alpha_\mu\,,\qquad  U^\mu\equiv \xi^\mu+i u^\mu\,,\qquad \mu\in\{0,1,\dots, h^{2,1}\}\,.
\end{equation}

The full   massless spectrum of the bosonic sector of the $\mathcal{N}=1$ orientifold theory is summarized in table \ref{cy-table: N=1 field content type IIA}.

\begin{table}[htbp]
\def\arraystretch{1.5}
\center
\begin{tabular}{|c|c|c|}
\hline
Multiplet & Bosonic Field Content & Multiplicity \\ \hline
Gravity   & $g_{\mu\nu}$          & 1            \\
Vector    & $A^\alpha$            & $h_+^{1,1}$  \\
Chiral    & $b^a,t^a$             & $h_-^{1,1}$  \\
Chiral    & $\xi^\mu, u^\mu$      & $h^{2,1}+1$  \\ \hline
\end{tabular}
\caption{Bosonic content of the 4-dimensional $\mathcal{N}=1$ supergravity resulting from the compactification of Type IIA on a Calabi-Yau orientifold.}
\label{cy-table: N=1 field content type IIA}
\end{table}

The scalars $b^a$ and $\xi^\mu$ are called axions while their SUSY partners $t^a$ and $u^\mu$ are known as saxions. All these scalars are massless fields. It will be our aim in the following sections to to confer them mass  by adding background fluxes. The set of allowed values for the scalars $(b^a,t^a)$ and $(\xi^\mu, u^\mu)$ constitute two independent spaces at the classical level (moduli spaces), known as Kähler and complex structure moduli respectively.

\subsubsection*{Moduli Space structure}

In \ref{cy-subsec: structure moduli space} we discussed how both the Kähler and complex structure moduli spaces have a Kähler structure. Such structure is preserved after taking the orientifold projection and including the contribution from the $C_3$ field. Now we will briefly describe the new form of the Kähler potentials - see \cite{Grimm:2005fa} for a detailed derivation. Starting with the Kähler sector, we can expand \eqref{cy-eq: CY kahler potential Kahler} using \eqref{cy-eq: intersection numbers}. The truncation of the metric of the moduli space caused by the orientifold is trivial and thus we obtain
\begin{equation}
K_K \,  = \, -{\rm log} \left(\frac{i}{6} \CK_{abc} (T^a - \bar{T}^a)(T^b - \bar{T}^b)(T^c - \bar{T}^c) \right) \, = \,  -{\rm log} \left(\frac{4}{3} \cK\right) \, ,
\label{cy-eq: KK}
\end{equation}
where we have defined 
\begin{equation}
    \cK = \cK_{abc} t^at^bt^c = 6 {\rm Vol}_{X_6} \,,
\end{equation}
and for future convenience it is also useful to introduce
\begin{eqnarray}
    \cK_{ab}\equiv \cK_{abc}t^c\,,\qquad \cK_{a}=\cK_{abc}t^b t^c\,.
\end{eqnarray}
Similarly, we can give a compact expression for the complex structure Kähler potential using \eqref{cy-eq: CY kahler potential CS}, \eqref{cy-eq: Omega expansion Z and F} and the discussion surrounding the latter. The geometry of this sector is considerably more complicated, since the orientifold projection truncates half of the degrees of freedom (see expression \eqref{cy-eq: truncation holomorphic form orientifold}). The resulting Kähler potential  is 
\begin{equation}
 K_Q = -2 \log \left( - \frac{1}{4} \IM({\cal C} {\cal Z}^\mu) \RE({\cal C} {\cal F}_\mu) \right) = -  \log(e^{-4 \phi_4})\,,
 \label{cy-eq: KQ}
\end{equation}
where $\mathcal{C}$ is the compensator defined in \eqref{cy-eq: compensator field} and  $\phi_4$ is the 4-dimensional dilaton $e^{\phi_4}\equiv\frac{e^\phi}{\sqrt{{\rm Vol}_6}}$.

\section{Massive Type IIA Flux Compactifications}
\label{cy-sec: flux compactifications}

To preserve supersymmetry, the compact dimensions must have a Calabi-Yau structure when no background fluxes are present. As we have seen, the consequent 4-dimensional theory is plagued with massless scalars, the moduli, that include the dilaton and the parameter space that characterizes the geometry of the compact dimension. In the 4-dimensional effective theory, these massless scalars couple to matter and create long-range interactions. Since different kinds of matter can couple differently to each of the moduli fields, a 4d observer should be able to measure “fifth forces” causing various distinct accelerations to different objects. Such scenario would lead to violations in the principle of equivalence which for now have not been detected experimentally \cite{Grana:2005jc}. Therefore, the simplest explanation compatible with current observations is that the aforementioned interactions cannot be long range, which requires a mechanism that provides mass to the moduli. The most promising procedure to achieve that goal is the inclusion of non-trivial flux backgrounds, which induces potentials for the moduli.  The fields are stabilized at the minima of such potential.

\subsection{Democratic formulation}
\label{cy-subsec: democratic formulation}

In Type IIA superstring low energy limit, we have two kinds of fluxes: the Neveu-Schwarz flux $H_3=dB_2$ and the Ramond-Ramond fluxes $F_p=dC_{p-1}$ with $p=2,4$. In addition, massive type IIA also has a scalar parameter, the Romans mass, that can be interpreted as a contribution coming from a 0-form background field strength $F_0$. From the discussion of section \ref{basic-subsec: branes}, we know that  these fluxes are sourced by $D_p$ and $\rm{NS}$ branes. The map is not injective, however, as there are $D_p$ branes with $p=0,2,4,6,8$. Ignoring for now the $D8$ associated to the Romans mass, the others come in pairs $(D_{p-2},D_{8-p})$ of electrically and magnetically charged objects under a given field $C_{p-1}$. A convenient method to systematically work with D-branes and $F_p$ fluxes is the democratic formulation \cite{Bergshoeff:2001pv}, which doubles the degrees of freedom of the system providing RR-forms that are electrically charged under every brane. Therefore, we end up with fields $C_p$ for $p=1,3,5,7,9$. Since $C_9$ is not dynamical  ($dC_9$ is dual to the Romans mass and thus constant) and a $p$-form potential $C_p$ has the same degrees of freedom as a $(8-p)$-form potential, the democratic formulation doubles the size of the RR-sector. To recover the original theory, additional duality relations keeping track of the original electric-magnetic dualities need to be included. Writing \eqref{basic-eq: massive type IIA action} in the democratic formulation, we obtain that the 4-dimensional effective action for the bosonic sector is
\begin{equation}
    S=\frac{1}{2\kappa_{10}^2}\int d^{10}x\sqrt{-g}e^{-2\Phi}[R+4(\partial \phi)^2]-\frac{1}{4\kappa_{10}^2} e^{-2\phi}\int \star H\wedge  H-\frac{1}{8\kappa_{10}^2}\int \star {\bf G}\wedge {\bf G}+S_{\textrm{loc}}\,,
    \label{cy-eq: type IIA action}
\end{equation}
where $S_{\rm{loc}}$ corresponds to the contribution from DBI and Chern-Simons action from local sources (branes) discussed in section \ref{basic-subsec: branes} and ${\bf G}$ is a  polyform that groups the p-forms of the Ramond-Ramond sector
\begin{equation}
    {\bf G}=d_H{\bf C}+e^{B}\wedge \bar{\bf G}\,,
    \label{cy-eq: G definition}
\end{equation}
with $d_H=d-H\wedge$ the twisted external derivative, ${\bf C}=C_1+C_3+C_5+C_7+C_9$ and $\bar{\bf G}$ a formal sum of closed even p-forms in $X_6$ that represents the background fluxes. Finally, we can account for a background flux $\bar{H}$ on the NSNS sector generalizing the definition of the flux field strength $H=dB+\bar{H}$.

One must remember that \eqref{cy-eq: type IIA action} is only a pseudo-action, as it carries unphysical degrees of freedom. To recover the real dynamics, the duality conditions must be imposed externally
\begin{equation}
    {\bf G}=\star_{10} \lambda( {\bf G})\,,
\end{equation}
where $\lambda$ is the operator which reverses the indices of the form it is applied to. In this new framework, the condition \eqref{cy-eq: p-form spliting} derived from maximal spacetime 4-dimensional symmetry  amounts to
\begin{equation}
    {\bf G}=\textrm{dvol}_4\wedge  {\bf \tilde{G}}+{\bf \hat{G}}\,,
    \label{cy-eq: RR form splitting}
\end{equation}
with ${\bf\tilde{G}}$ and ${\bf\hat{G}}$ polyforms in the compact space. The duality condition in this case imposes the following relation in the compact manifold $X_6$
\begin{equation}
    {\bf\tilde{G}}=\star_6 \lambda({\bf \hat{G}})\,,
    \label{cy-eq: int-ext duality}
\end{equation}
and so the internal contribution ${\bf \hat{G}}$ contains all the relevant information regarding the RR fluxes.

\subsubsection*{Equations of motion}

The equations of motion can be derived from the action \eqref{cy-eq: type IIA action}. The full set of equations for the RR and NSNS fields is given by
 
 \begin{equation}
 \begin{aligned}
    0=&d_{H} \star_{10} {\bf G} - \sum_\alpha  \d({\Pi_\alpha}) \wedge e^{-\mathcal{F}_\alpha}\, ,\\
    0=&d(e^{-2\Phi}\star_{10}H)+\frac{1}{2}\sum_{p=2,4,6,8,10} \star_{10}{G}_p\wedge G_{p-2}-2\kappa_{10}^2\frac{\delta S_{\rm loc}}{\delta B}\,,
 \end{aligned}    
 \end{equation}
with $\delta(\Pi_\alpha)$ the bump-delta current that lives in the Poincaré dual class of a cycle $\Pi_\alpha$ hosting a D-brane source with quantized worldvolume flux $F_\alpha$ and  $\mathcal{F}=P[B]-\ell_s^2/2\pi \cdot F_\alpha$ a combination of the pullback of the 2-form $B$ and the worldvolume flux. 

Taking into account the decomposition \eqref{cy-eq: RR form splitting}, each of the equations for the RR fields can be split into two: one for the internal component $\hat{G}_p$ and another for the external component ${\rm dvol}_4\wedge \tilde{G}_p$. Using the duality \eqref{cy-eq: int-ext duality}, all the expressions can be written in terms of the internal sector $\hat{G}_p$. In this thesis, we will focus on local sources generated by the orientifold planes O6 and spacetime-filling D6-branes. Under these conditions, the equations of motion for RR fields $\hat{G}_p$  and the NSNS field $H$ become
  \begin{subequations}
 \begin{align}
    0=& d(\star_{10}\hat{G}_2)+H\wedge\star_{10}\hat{G}_4\,,\\
    0=&d(\star_{10}\hat{G}_4)+H\wedge\star_{10}\hat{G}_6\,,\\
    0=&d(\star_{10}\hat{G}_6)\,,\\
    0=&d(e^{-2\Phi}\star_{10}H)+\star_{10}\hat{G}_2\wedge \hat{G}_{0}+\star_{10}\hat{G}_4\wedge \hat{G}_{2}+\star_{10}\hat{G}_6\wedge \hat{G}_{4}\,. \label{bio-problem}
 \end{align}
 \label{cy-eq: NSNS RR eoms}
 \end{subequations}
 Meanwhile, the equations of motion for the dual fields $\tilde{G}_p$ act as a set of additional constraints for $\hat{G}_p$, known as Bianchi identities. They are analogous to the relation between the equation of motion for the electric field $F$ and its dual in Maxwell theory. We will soon consider them in detail, but, for completeness, let us first state the equations of motion for the dilaton and the metric (modified Einstein equation). They are respectively given by \cite{Lust:2008zd}
 \begin{align}
     0=& \nabla^2\phi -(d\phi)^2+\frac{1}{4}R-\frac{1}{8}H\cdot H-\frac{1}{4}\frac{\kappa_{10}^2e^{2\phi}}{\sqrt{-g}}\frac{\delta S_{\rm loc}}{\delta \phi}\,, \label{cy-eq: dilaton eom}\\
     0=& R_{MN}+2\nabla_M\nabla_N\phi-\frac{1}{2}\iota_M H\cdot \iota_N H-\frac{1}{4}e^{2\phi}\iota_M F\cdot \iota_N F\nonumber\\
     &-\kappa_{10}^2e^{2\phi}\left(-\frac{2}{\sqrt{-g}}\frac{\delta S_{\rm loc}}{\delta g^{MN}}+\frac{g_{MN}}{2\sqrt{-g}}\frac{\delta S_{\rm loc}}{\delta \phi}\right)\,. \label{cy-eq: Einstein eom}
 \end{align}

\subsubsection*{No-go de Sitter}

By taking a careful combination of the equations of motion of the dilaton and the metric, an equation can be derived for the cosmological constant $\Lambda$, which in the presence of non-trivial flux conf
igurations  can only be satisfied if $\Lambda$ is strictly negative, as long as orientifold planes (sources with negative tension) are not introduced \cite{Maldacena:2000mw}. Thus, it can be inferred that in flux compactifications there are no de Sitter or Minkowski vacua when there are no localized sources and higher curvature corrections to the equations of motion.

\subsubsection*{Bianchi identities}

Fluxes cannot be added to a manifold arbitrarily. There are geometrical constraints (Bianchi identities)  that need to be satisfied in relation with Stoke's theorem and the connection form of the principal bundle of the specific manifold. As we observed in the discussion of the equations of motion, they are a generalization of Gauss Law to higher dimensional forms in general manifolds. In their simplest form, the Bianchi identities demand the flux field strengths to be closed. Such constraints can be relaxed by including local sources, but in the case of compact manifolds, where the flux fields cannot escape to infinity, they are still far from trivial. The flux field strengths need to wrap cycles that are closed after accounting for the deformations in the internal geometry caused by the fluxes themselves.\footnote{Our first contact with such phenomena is the effect of the $B$-field over the metric, that requires the definition of the twisted exterior derivative $d_H=d-H\wedge$.} Integrating the Bianchi identities over the compact dimensions, the differential conditions become a set of diophantine equations (due to flux quantization) called \textit{Tadpole Constraints}. They formalize the requirement that the global sum of charges must add up to zero in the compact space. 

For the case at hand, the Bianchi identities can be derived by varying the action \eqref{cy-eq: type IIA action} including the local contributions. They read
\begin{equation}
        \ell_s^2d(e^{-B}\wedge{\bf G})=-\sum_\alpha \lambda(\delta(\Pi_\alpha))\wedge e^{\frac{\ell_s^2}{2\pi}F_\alpha}\,, \qquad dH=0\,,
    \label{cy-eq: BI def}
\end{equation}
with $\lambda$ the reverse index operator.

We will consider local sources given by D6-branes and O6-planes and no worldvolume flux. Taking into account that an O6-planes has minus four times the charge a D6, the Bianchi identities \eqref{cy-eq: BI def} become
\begin{equation}
\begin{gathered}
    d\hat{G}_0 = 0\, , \qquad d \hat{G}_2 = \hat{G}_0 H - 4 \d_{\rm O6} +   N_\a \d_{\rm D6}^\a \, ,  \qquad d \hat{G}_4 = \hat{G}_2 \wedge H\, , \\ 
    d\hat{G}_6 = 0\,\qquad  dH=0\,,
\label{cy-eq: BI expanded}
\end{gathered}
\end{equation}
where $\d_{\rm O6/D6}^\a\equiv \ell_s^{-2}\delta(\Pi_{\alpha})$ and $N_\alpha$ is the number of D-branes hosted by a 3-cycle $\Pi_\alpha$ in the internal space.  The Bianchi identity for $\hat{G}_2$ in particular implies that  the Poincaré dual of the cycle associated to the O6 lies in the same real cohomology class as $H$ when working with configurations without D6-branes or with D6-branes on top of the O6-plane. In terms of integer homology classes, we can then write
\begin{equation}
    P.D.[\ell_s^2 H]=h[\Pi_{O6}]=h[\Pi_{D6}]\,\quad {\rm with }\, \, h\in\mathbb{Q}\,.
\end{equation}

\subsection{SUSY equations}

The equations of motion for the RR and NSNS fluxes \eqref{cy-eq: NSNS RR eoms} are generally very challenging to solve. The problem becomes more tractable if we focus on supersymmetric solutions.  The vanishing condition for the supersymmetric variation of the fermions \eqref{cy-eq: susy equations no fluxes} gets modifications in the presence of fluxes. Accounting for those changes and imposing the constraints derived from the Bianchi identities is enough to fully describe supersymmetric vacua. The generalized complex structure $SU(3)\times SU(3)$ is a powerful tool to accomplish this goal in a broad framework.

Let us recall the expansion \eqref{cy-eq: susy spinor expansion} for the current case of  Calabi-Yau orientifold compactification. Since we have $\mathcal{N}=1$ 4-dimensional supersymmetry, there is only one preserved spinor in 4-dimensions, i.e, $\zeta^a_J=\zeta$. Meanwhile, in the compact 6-dimensional space with non-trivial holonomy we can have up to two pairs of independent spinors of two different chiralities ($\{\eta^a_{\pm}\}=\{\eta^1_{\pm},\eta^2_{\pm}\}$), which are parallel in the generic case of $SU(3)$-holonomy. Nevertheless, it is useful to leave the relation between these two spinors $\eta^1_+,\eta^2_{+}$ open to describe for more general scenarios. Thus the expansion \eqref{cy-eq: susy spinor expansion} becomes
\begin{equation}
    \begin{aligned}
        \epsilon^1&=\zeta_+\otimes \eta^1_++\zeta_-\otimes\eta^1_-\,,\\
        \epsilon^2&=\zeta_+\otimes \eta^2_-+\zeta_-\otimes \eta^2_+\,.
    \end{aligned}
\end{equation}

Appendix \ref{ch: ap complex geometry} shows that each of these spinors defines an $SU(3)$ structure. Furthermore, we can construct the pure forms
\begin{equation}
    \Phi_+\equiv \eta_+^1\otimes (\eta^2_+)^\dagger\,,\qquad \Phi_-=\eta^1_+\otimes(\eta^2_-)^\dagger\,,
 \end{equation}
which in turn define a $SU(3)\times SU(3)$ structure. 

The supersymmetry equations can be written in terms of these forms as follows \cite{koerber2007ten, Koerber:2007jb, Marchesano:2020qvg}
\begin{subequations}
\begin{align}
    d_H\Phi_+&=-2\mu e^{-A}\Re\Phi_-\,, \\
    d_H(e^A\im\Phi_-)&=-3\mu\im\Phi_+ +e^{4A}\star_6\lambda(\bf \hat{G})\,,
\end{align}
    \label{cy-eq: SUSY equations pure forms}
\end{subequations}
where we recall that $A$ is the wrap factor in \eqref{cy-eq: spacetime factorization metric} and $\mu$ is a real coefficient relating the 4-dimensional curvature and metric $R_{\mu\nu}^{(4)}=-3\mu^2 g_{\mu\nu}^{(4)}$. Note $\mu$ is a function of the 4-dimensional cosmological constant $\Lambda$: $\mu\equiv \sqrt{\Lambda/3}$.
 
To find a supersymmetric solution, one will only need to solve equations \eqref{cy-eq: SUSY equations pure forms} while satisfying the Bianchi identities \eqref{cy-eq: BI expanded}. It can be argued \cite{Lust:2004ig, Koerber:2007hd} that, in a purely bosonic supersymmetric background, satisfying this set of relations implies the dilaton and Einstein equations. The power of the $SU(3)\times SU(3)$ structure becomes apparent when considering different compactifications with varying ingredients. In that case, the formulas \eqref{cy-eq: SUSY equations pure forms} remain invariant and one only needs to change the specific expression for the pure forms $\Phi_{\pm}$. 

\subsection{AdS SUSY Vacua and the Smearing approximation}
\label{cy-subsec: smearing approx}

As we discussed in several instances, most vacua arising from 10-dimensional supergravity compactifications in the presence of fluxes have negative cosmological constant and therefore generate a macroscopic spacetime $\mathcal{M}_4={\rm AdS_4}$. Even though this type of solutions are not realistic in light of current cosmological observations, they provide a useful intermediate step to reach a fitting description of our Universe. Thanks to the AdS/CFT duality, these vacua are much better understood than any other construction.  They therefore can allow us to develop intuition on the general properties of the string Landscape. For these reasons, we will now focus on massive type IIA orientifold compactifications to a 4-dimensional anti-de Sitter space.

\subsubsection*{No sources}

Even when considering supersymmetric vacua, working with localized sources such as orientifold planes is extremely difficult and full solutions are generally not known. As a first take on the subject, let us  assume they are not present. In this case, solutions of the SUSY equations can be accommodated in the $SU(3)$ structure \cite{Lust:2004ig}. Therefore we can take $\psi=0$ in \eqref{ap.geo-eq: pure forms SU3xSU3} and so
\begin{equation}
    \Phi_+=e^{3A-\phi}e^{i\theta}e^{-iJ}\,,\qquad \Phi_-=e^{3A-\phi}\Omega\,.
\end{equation}
Plugging these forms in \eqref{cy-eq: SUSY equations pure forms} leads to
\begin{equation}
    dJ=-2\mu e^{-A}\sin\theta\re\Omega\,,\qquad d\Omega=\frac{4}{3}i\mu e^{-A}J\wedge J+i\im(W_2)\wedge J+dA\wedge\Omega\,,
\end{equation}
with $d\theta=0$ and $3dA=d\phi$. The fluxes satisfy the following relations \cite{Koerber:2010bx,Marchesano:2020qvg}
\begin{subequations}
    \begin{align}
        H&=2\mu e^{-A}\cos\theta\, \re\Omega\,,\\
        \hat{G}_0&=5\mu e^{-\phi-A}\cos\theta\,,\\
        \hat{G}_2&= \frac{1}{3}\mu e^{-\phi-A}\sin\theta\, J-J\cdot d(e^{-\phi\Im\Omega})\,,\\
        \hat{G}_4&=\frac{3}{2}\mu e^{-\phi-A}\cos \theta\, J\wedge J\,,\\
        \hat{G}_6&=3\mu e^{-\phi-A}\sin \theta {\rm dvol}_{X_6}\,,
    \end{align}
    \label{cy-eq: SU3 solutions theta neq 0}
\end{subequations}
where the product  $J\cdot$ is the internal product defined in the appendix \ref{ch: ap conventions}. It follows from the Bianchi identity for $\hat{G}_0$ ($d\hat{G}_0=0$) that $A$ and $\phi$ are independently constant and we can pick $A=0$ without loss of generality.\footnote{For a thoughtful insight on how one can have flux backgrounds in the absence of localized sources we refer the reader to \cite{Marolf:2000cb}.}

From the lens of the torsion analysis of $SU(3)$ introduced in \ref{ap.geo-table: torsion classes}, there are two classes that do not vanish: $\Im W_1\neq0$ and $\im W_2\neq 0$. Therefore, we conclude that including fluxes deforms the compact manifold's geometry and breaks the Calabi-Yau structure.

In the following chapters of this thesis we will narrow the scope of our vacua analysis to the case $\theta=0$ or $\pi$,\footnote{These two choices are related by a change on the Romans mass.} since it is the one that generates the best understood phenomenologically interesting vacua, like \cite{DeWolfe:2005uu}. For a study of the generic case of $\theta\neq 0$  we refer the reader to \cite{Saracco:2012wc, Tomasiello:2022dwe}. When $\theta=0$, we can write
\begin{equation}
    H=\frac{2}{5}e^{\phi}\hat{G}_0\re\Omega\,, \qquad \hat{G}_2=-e^{-\phi}W_2\,,\qquad \hat{G}_4=\frac{3}{10}\hat{G}_0J\wedge J\,,\qquad \hat{G}_6=0\,,
\end{equation}
\begin{equation}
    dJ=0\,,\qquad d\im\Omega= i W_2\wedge J\,,\qquad d\Re\Omega=0\,,
\end{equation}
and $\hat{G}_0=5e^{-\phi}\mu$.

\subsubsection{Smeared sources}

Adding sources disrupts the previous results. According to relation \eqref{basic-eq: brane backreaction}, the presence of a localized source causes a backreaction on the spacetime metric which generates a non-trivial profile for the dilaton and the warp factor.  Hence, in principle, the $SU(3)$ structure is no longer a good description for such kind of vacua.

In particular, using \eqref{cy-eq: SU3 solutions theta neq 0} when O6 planes are considered, the Bianchi identity for the flux $\hat{G}_2$ becomes \cite{Acharya:2006ne}
\begin{equation}
    e^{-\phi}\left[\frac{1}{4}|W_2|^2+e^{-2A}\mu^2\left(10\cos^2\theta-\frac{2}{3}\sin^2\theta\right)\right]\re\Omega=-\delta_{O6}\,.
\end{equation}
The smearing approximation, originally introduced in \cite{Acharya:2006ne}, proposes a clever way to deal with the above expression in the long-wavelength regime. This is achieved by replacing the localized source with a homogeneous distribution of the charge over the internal manifold described by a 3-form in the same cohomology class as the original cocycle.  Consequently, in the case at hand we make the following replacement
\begin{equation}
    \delta_{O6}\rightarrow - \hat{G}_0 H\,, 
\end{equation}
so $\hat{G}_2$ is closed according to \eqref{cy-eq: BI expanded}. Under such an assumption, we can find solutions satisfying all the equations and Bianchi identities with $W_2=\theta=0$. This solution will have $SU(3)$ structure with vanishing torsion classes  ($dJ=d\Omega=0$). Consequently, the smeared solution corresponds once again to a Calabi-Yau manifold. The flux vacua would then be given by

\begin{equation}
    H=\frac{2}{5}e^{\phi}\hat{G}_0\re\Omega\,, \qquad \hat{G}_2=0\,,\qquad \hat{G}_4=\frac{3}{10}\hat{G}_0J\wedge J\,,\qquad \hat{G}_6=0\,.
    \label{cy-eq: SU3 solutions}
\end{equation}

\subsection{AdS SUSY vacua beyond smearing}
\label{cy-subsec: beyond smearing}

The smearing approximation provides a useful method of solving the equations in a controlled setup that is meaningful in the limit of small cosmological constant, weak string coupling and large internal volume \cite{Saracco:2012wc, Marchesano:2020qvg, Junghans:2020acz}. However, by definition, it fails to describe the localized nature of the sources involved in the problem. To do so, one needs to depart from the $SU(3)$ structure and employ the language of $SU(3)\times SU(3)$ structures, repeating a similar analysis as the one performed in the previous section with the general expressions \eqref{ap.geo-eq: pure forms SU3xSU3}. For the case $\theta=0$ one obtains two constraints not involving the fluxes \cite{Saracco:2012wc, Marchesano:2020qvg,Tomasiello:2022dwe} 
\begin{subequations}
    \begin{gather}
    \re v=-\frac{e^A}{2\mu}d\log (\cos\psi e^{3A-\phi})\,,\\
    d(e^{3A-\phi}\cos\psi J_\psi)=0\,,
    \end{gather}
\end{subequations}
together with  the relations for the fluxes
\begin{subequations}
    \begin{align}
         \hat{H}&=2\mu e^{-A}\Re(iv\wedge\omega_\psi)\,,\\
         F_0&=-J_\psi\cdot d(\cos\psi e^{-\phi}\im v)+5\mu\cos\psi e^{-A-\phi}\,,\\
         F_2&=-J_\psi\cdot d\im (i\cos\psi e^{-\phi }v\wedge\omega_\psi)-2\mu\frac{\sin^2\psi}{\cos\psi}e^{-A--\phi}\im \omega_\psi\,,\\
         F_4&=J_\psi^2\left[\frac{1}{2}F_0-\mu\cos\psi e^{-A-\phi}\right]+J_\psi\wedge d\im(\cos\psi e^{-\phi}v)\,,\\
         F_6&=0\,,
    \end{align}
    \label{cy-eq: SU3xSU3 solutions exact}
\end{subequations}
where $\hat{H}$ and ${\bf F}$ are related to the physical fluxes through a B-transformation
\begin{equation}
    H=\hat{H}+d(\tan\psi\im \omega)\,,\qquad {\bf G}=e^{\tan\psi{\rm Im} \omega}\wedge{\bf F}\,.
\end{equation}

In \cite{Marchesano:2020qvg, Junghans:2020acz} they perform an expansion of \eqref{cy-eq: SU3xSU3 solutions exact} in the limit of small string coupling $g_s$. Note that this is equivalent to taking the limit of small cosmological constant or large internal volume since $\mu/\ell_s\sim g_s\sim {\rm Vol_Y}^{-1/2}$.  In the limit when $g_s\rightarrow 0$, the smeared solution is recovered and thus the system is described by a Calabi-Yau manifold with a $SU(3)$ structure. Following the reasoning of appendix \ref{ap.geo-sec: generalized complex structure}, the angle $\psi$ interpolates between the $SU(3)$ and the $SU(3)\times SU(3)$ structures. We will therefore require it to be at least of order $g_s$. Consistency arguments provide the scaling for the other quantities that define the structure.  Assuming as before that P.D.$[\ell_s^{-2}H] = h [\Pi_{\rm O6}] = h [\Pi_{\rm D6}]$, the leading term of the expansion is given by the smearing approximation, where the closed forms $J_{\rm CY}$ and $\Omega_{\rm CY}$ are defined. The first order corrections can be found through  a detailed analysis of the SUSY equations and the Bianchi identities. Since the content of sources we aim to describe in this thesis consists of O6 planes and D6 branes, the most relevant Bianchi identity is the one for the flux $\hat{G}_2$, which can be written in terms of the B-transformed field $F_2$. Using the Hodge decomposition on the latter 
\begin{equation}
    F_2=d^\dagger_{\rm CY}K+F_2^h+dC_1\,,
\end{equation}
with $F^h_2$ a Calabi-Yau harmonic, $d^\dagger_{\rm CY}$ the adjoint external derivative constructed with the Calabi-Yau metric and $K$ a 3-form current that satisfies
\begin{equation}
\ell_s^2 \Delta_{\rm CY} K =  \frac{2}{5} m^2 g_s \re \Omega_{\rm CY} + (N-4)\d(\Pi_{\rm O6})\, ,
\label{cy-eq: defK}
\end{equation}
where $\Delta_{\rm CY} = d^\dag_{\rm CY} d + d d^\dag_{\rm CY}$ is constructed from the CY metric and $m=\ell_s G_0$ is the Romans mass. The solution is of the form
\begin{equation}
 K = \varphi \re \Omega_{\rm CY}  + \re k \, ,
\label{cy-eq: formK}
\end{equation}
with $k$ a (2,1) primitive current and $\varphi$ is a real function that satisfies $\int_{X_6} \varphi = 0$ and
\begin{equation}
\Delta_{\rm CY}  \varphi = \frac{mh}{4}\left(\frac{{\cal V}_{\Pi_{\rm O6}}}{{\cal V}_{\rm CY}} - \delta^{(3)}_{\Pi_{\rm O6}}\right)  \ \implies \ \varphi \sim \cO(g_s^{1/3})\, ,
\end{equation}
where $\delta^{(3)}_{\Pi_{\rm O6}} \equiv \star_{\rm CY} (\im \Omega_{\rm CY} \wedge \d(\Pi_{\rm O6}))$. In terms of these quantities, we can describe the metric background and the varying dilaton profile as
\begin{subequations}	
	\label{cy-eq: solutionsu3}
\begin{align}
J & = J_{\rm CY} + \cO(g_s^2) \, , \qquad   \qquad  \Omega  = \Omega_{\rm CY} + g_s k +  \cO(g_s^2)\, , \\
e^{-A}  & = 1 + g_s \varphi + \cO(g_s^2) \, , \qquad e^{\phi}   = g_s \left(1 - 3  g_s \varphi\right) + \cO(g_s^3)\, ,
\end{align}
\end{subequations}   
where we have taken $g_s$ as the natural expansion parameter. Notice that $\varphi \sim -\frac{mh}{4r}$ near $\Pi_{\rm O6}$, and so as expected the 10d string coupling blows up and the warp factor becomes negative near that location. The function $\varphi$ indicates the region $\tilde{Y}_6 \equiv \{p \in X_6 | \, g_s |\varphi(p)| \ll 1\}$ in which the perturbative expansion on $g_s$ is reliable; beyond that point one may use the techniques of \cite{DeLuca:2021mcj} to solve  the 10d supersymmetry equations. The background fluxes are similarly expanded as
\begin{subequations}
	\label{cy-eq: solutionflux}
\begin{align}
 H & =   \frac{2}{5} \frac{m}{\ell_s} g_s \left(\re \Omega_{\rm CY} + g_s K \right) - \oh   d\re \left(\bar{v} \cdot \Omega_{\rm CY} \right) + \cO(g_s^{3}) \label{cy-eq: H3sol} \, , \\
 \label{cy-eq: G2sol}
 G_2 & =     d^{\dag}_{\rm CY} K  + \cO(g_s)  = - J_{\rm CY} \cdot d(4 \varphi \im \Omega_{\rm CY} - \star_{\rm CY} K) + \cO(g_s) \, , \\
G_4 & =  \frac{m}{\ell_s} J_{\rm CY} \wedge J_{\rm CY} \left(\frac{3}{10}  - \frac{4}{5} g_s \varphi \right)+   J_{\rm CY} \wedge g_s^{-1} d \im v + \cO(g_s^2) \, , \\
G_6 & = 0\, .
\end{align}
\end{subequations}   
Here $v$ is a (1,0)-form whose presence indicates that we are in a genuine $SU(3)\times SU(3)$ structure, as opposed to an $SU(3)$ structure. It is determined by
\begin{equation}
v  = g_s \p_{\rm CY} f_\star + \cO(g_s^3)\, , \qquad \text{with} \qquad \ell_s \Delta_{\rm CY} f_\star  = - g_s 8 m \varphi \, .
\end{equation} 
It is easy to see that \eqref{cy-eq: solutionsu3} and \eqref{cy-eq: solutionflux} reduce to the smeared solution in the limit $g_s \to 0$.  Moreover, as shown in \cite{Marchesano:2020qvg}, this background satisfies the supersymmetry equations and the Bianchi identities up to order $\cO(g_s^2)$.  As a cross-check of this result, we discuss in Appendix \ref{bio-ap:10deom} how the 10d equations of motion are satisfied  to the same level of accuracy.

\section{4d effective action and vacua}
\label{cy-sec: 4d eff action}

In the previous section we focused on the construction of consistent vacua for Type IIA 10-dimensional effective theory. We saw that solving the equations of motion is a very complex problem involving difficult differential geometry computations. An alternative approach to improve the understanding of the space of vacua described by our theory is to take the massive type IIA \eqref{cy-eq: type IIA action} action and directly apply dimensional reduction neglecting the warping and the source localization effects. As in the 10-dimensional case, we also omit worldsheet and D-brane instanton effects.

\subsection{Effective action and flux potential}
\label{cy-subsec: effective 4d potential}
\subsubsection*{Massless fields action}

As we explained in \ref{cy-subsec: orientifolds},  p-forms in the internal manifold $X_6$  behave as scalar fields after the compactification. The 4-dimensional bosonic field content of Type IIA $\mathcal{N}=1$ supergravity compactified on a Calabi-Yau orientifold in the absence of background sources is summarized in table \ref{cy-table: N=1 field content type IIA}. It is composed of the metric, $g_{\mu\nu}$, a set of vector fields, $A^\alpha$, and two sets of pairs of axionic and saxionic fields  $(b^a,t^a)$ and $(\xi^\mu, u^\mu)$. The elements of each pair are merged together in the complexified Kähler $J_{c}$ and holomorphic forms $\Omega_c$ respectively.
\begin{align}
    J_{c}&=B+i e^{i\phi} J=(b^a+it^a)\omega_a=T^a \omega_a\,, \label{cy-eq: complexified J expansion}\\
    \Omega_c&=C_3+i\im(\mathcal{C}\Omega)=(\xi^\mu+iu^\mu)\alpha_\mu = U^\mu \alpha_\mu\,, \label{cy-eq: complexified omega expansion}
\end{align}
where we have expressed the Kähler form $J$ in the Einstein frame and expanded in the basis of forms introduced in table \ref{cy-table: harmonic basis}.

One can then take the 10-dimensional type IIA action \eqref{cy-eq: type IIA action}, introduce the Kaluza-Klein reduction detailed in \ref{cy-subsec: orientifolds} and integrate taking into account the duality relations between the flux fields. The result is the 4-dimensional effective action of type IIA, which written in the Einstein frame takes the form

\begin{align}
    S_{\rm IIA}^{4d}=&\int -\frac{1}{2}R\star_4 1-K_{a\bar{b}}dT^a\wedge\star_4 d\bar{T}^b-K_{\mu\bar{\nu}}dU^\mu\wedge \star_4 d\bar{U}^\nu\nonumber\\ 
    &-\frac{1}{2}\im f_{\alpha\beta}F^\alpha\wedge F^\beta-\frac{1}{2}\re f_{\alpha\beta}F^\alpha\wedge \star F^\beta\,,
    \label{cy-eq: 4d type IIA effective action}
\end{align}
where  $K_{a\bar{b}}=\partial_{T^a}\partial_{\bar{T}^b} K_K$ and $K_{\mu\bar{\nu}}=\partial_{U^\mu}\partial_{U^\nu} K_{Q}$ are the metrics of the Kähler and complex structure sector,  $F^\alpha=d A^\alpha$ is the field strength of the vector field coming from the compactification of the RR 3-form and $f_{\alpha \beta}$ are the gauge kinetic functions associated to those field strengths
\begin{equation}
    2f_{\alpha\beta}=i\hat{\kappa}_{a\alpha\beta}T^a\,.
    \label{cy-eq: fg}
\end{equation}

\subsubsection*{Flux background: NSNS and RR fluxes}

The effective action \eqref{cy-eq: 4d type IIA effective action} was derived under the assumption that the background flux configuration was trivial, i.e., $\bar{\bf G}=0$ in \eqref{cy-eq: G definition}. When the background fluxes are turned on, they generate a scalar potential in the 4-dimensional effective theory. To see this, first rewrite RR fields ${\bf G}$ and NSNS field $H$ as 
\begin{equation}
    {\bf G}=e^B\wedge (d{\bf A}+\bar{\bf G})\,, \qquad H=dB+\bar{H}\,,
\end{equation}
with ${\bf A}={\bf C}\wedge e^{-B}$ and $\bar{\bf G}$ and $\bar{H}$ closed forms (or a sum thereof) that characterizes the static background. Imposing Page quantization condition \cite{Marolf:2000cb} one obtains
\begin{equation}
    \frac{1}{\ell_s^{2p-1}}\int_{\pi_{2p}} d\hat{A}_{2p-1}+\bar{G}_{2p}\in \mathbb Z\,,\qquad \frac{1}{\ell_s^2}\int_{\pi_3} dB+\bar{H}\in \mathbb{Z}\,,
\end{equation}
where $\pi_{2p}$ ($p=1,2,3$) and $\pi_3$ are internal cycles of $X_6$. In the absence of localized sources (smearing approximation), the gauge potentials ${\bf A}$ are globally well-defined and integrate to zero over a cycle. Therefore, from the 4-dimensional perspective, the flux background is characterized in terms of integer quanta. The cycles chosen for integration have to respect the orientifold projection and take into account how the field strength transforms. $\bar{G}_4$ is even under  the orientifold involution while $\bar{H}$ and $\bar{G}_2$ are odd. Thus, we can integrate over the corresponding de Rham duals of the basis of harmonic forms introduced in table \ref{cy-table: harmonic basis}. As an example, let us consider the NSNS flux. Being odd under the involution,  it will be characterized in the 4-dimensional space by a set of $h^{2,1}+1$ flux quanta $h^\mu$ coming from the integrals over cycles $\pi_3^\mu$ de Rham dual\footnote{A p-cycle $\pi_p$ and p-form $F_p$ are said to be de Rham duals if $\int_{\pi_p} F_p=1$. In the case at hand, this implies through \eqref{cy-eq: basis relations} that $-\alpha_\mu$ is the Poincaré dual to the cycle $\pi_3^\mu$.  We choose the opposite sign for convenience later on, so $\pi_3^\mu$ is the de Rham dual of $-\beta^\mu$.  Since that map is not affected by the orientifold the expansion of $H$ in that basis is perfectly valid. } to the basis $\beta^\mu$ of $H^{3}_-$ and so
\begin{equation}
    h_\mu=\frac{1}{\ell_s^2}\int_{\pi_3^\mu} H=\frac{1}{\ell_s^5}\int_{X_6} H \wedge \alpha_\mu \quad \rightarrow \quad \ell_s H=-h_\mu\beta^\mu\,.
\end{equation}
The flux quanta for the RR fields are similarly given by
\begin{equation}
\begin{aligned}
m \, =& \,  \ell_s G_0\, ,&  \quad  m^a\, =&\, \frac{1}{\ell_s^5} \int_{X_6} \bar{G}_2 \wedge \tilde \omega^a\, ,\\ 
 e_a\, =&\, - \frac{1}{\ell_s^5} \int_{X_6} \bar{G}_4 \wedge \omega_a \, ,& \quad e_0 \, =&\, - \frac{1}{\ell_s^5} \int_{X_6} \bar{G}_6 \, .
\end{aligned}
\label{cy-eq: RRfluxes}
\end{equation}

\subsubsection*{Flux background: Adding geometric and non-geometric fluxes}

This set of flux quanta can be enough to achieve full moduli stabilization \cite{Villadoro:2005cu,DeWolfe:2005uu,Camara:2005dc}. Even so, as pointed out in \cite{Shelton:2005cf}, one may consider a larger set of NS fluxes, related to each other by T-duality. Taking them into account results in a richer scalar potential, as analyzed in \cite{Aldazabal:2006up,Shelton:2006fd,Micu:2007rd,Ihl:2007ah,Wecht:2007wu,Robbins:2007yv}. We will give a short review of how they arise. For a detailed review, we refer the reader to \cite{Kachru:2002sk,Wecht:2007wu, Plauschinn:2018wbo}. 

In section \ref{basic-subsec: compactifications}, we introduced the notion of T-duality between Type IIA and Type IIB compactifications over circles with radius related by the transformation $R\leftrightarrow \alpha'/R$. In the presence of non-trivial background fluxes, such transformation mixes the field $B$ with the metric and the different RR p-forms among themselves, as illustrated in table \ref{cy-table: t-duality transform}. The specific map is given by the Buscher rules \cite{Buscher:1987sk}. We can apply T-duality to our geometry $\mathcal{M}_4\times X_6$ and study how the internal fluxes are affected. Since the RR forms are mapped to each other, we do not expect significant changes in that sector. The interesting aspect lies in the deformation of the metric through the action of the NSNS field. 

\begin{table}[h!]
\centering
\begin{tabular}{ccc}
Type IIA                & $\longleftrightarrow$ & Type IIB          \\ \hline
$G_{9M},B_{9M}$         &                   & $B_{9M}, G_{9M}$  \\
$C_9,C_M$               &                   & $a, C_{9M}$       \\
$C_{9MN},C_{MNL}$ &                   & $C_{MN},C_{9MNL}$
\end{tabular}
\caption{Schematic map of the T-duality transformation using along the coordinate $x^9$ using the notation of tables \ref{basic-table: type IIB spectrum} and \ref{basic-table: type IIA spectrum}.}
\label{cy-table: t-duality transform}
\end{table}

Let us consider as an example Type IIB compactified on a six-torus $T^6$ with non-vanishing NSNS flux $H_{abc}$. We choose the metric to be $ds^2=\sum_i (dy^i)^2$, with  the coordinates $y^i$ satisfying the toroidal identification $y^i\sim y^i+1$. Performing a T-duality along the direction $y^a$ and applying the Buscher rules, one recovers a new space with metric $ds^2=(dy^a-f^a_{bc}y^bdy^c)+\sum_{i\neq a} (dy^i)^2$. The coefficients $f^a_{bc}$ are integrally quantized as a function of the original $H_{abc}$. The resulting space corresponds to a twisted torus in which the coordinates now satisfy the periodic identification rule $(y^a,y^b,y^c)\sim (y^a+1,y^b,y^c)\sim (y^a,y^b+1,y^c) \sim (y^a+f^a_{bc}y^b,y^b,y^c+1)$ for all possible combinations of $b,c$.  Since the metric is globally well-defined in the final space, these new fluxes $f^a_{bc}$ are called geometric (or metric) fluxes. The T-duality guarantees that the resulting Type IIA compactification is well-defined and so it seems reasonable to consider compactifications in which they are present. 

Generically, geometric sources in 10 dimensions correspond to deformations of the curvature tensor. In such scenario, the harmonic forms are no longer the ones that are appropriate for performing field decompositions, since the Laplace equation also gets corrected. Laplacian harmonic forms are then replaced by globally well-defined forms that no longer need to be closed or co-closed under the original external derivative operator.

One can go on and perform another T-duality along a different direction. In the most general case, it results in a new space with a locally well-defined metric that fails to be invariant under periodic translations of the coordinates that have not been subjected to the transformation. The map describing the metric change is a $\beta$-transformation that arises in the framework of generalized complex geometry and is parametrized by a new set of flux quanta $Q^{ab}_c$, known as non-geometric fluxes.

It is possible to apply a third T-duality transformation, which yields a dual theory that does not even admit a local description of the internal space in terms of Riemannian geometry, and so the generated fluxes $R^{abc}$ are also non-geometric. We end up with the following picture
\begin{equation}
    H_{abc}\xrightarrow{T_a} f^a_{bc} \xrightarrow{T_b} Q^{ab}_c \xrightarrow{T_c}R^{abc}\,.
\end{equation}

When these fluxes are introduced, the Calabi-Yau structure is broken, even in the smearing approximation of the local sources. The best framework to systematically study geometric and non-geometric fluxes is, once again, generalized complex geometry, $SU(3)\times SU(3)$ structures and $SO(d,d)$ transformations. In appendix \ref{ap.geo-sec: generalized complex structure} we briefly comment how a B-transformation accounts for a non-trivial background flux $H$. Similarly, $A$-transformations (see \eqref{ap.geo-eq: b and beta generators}) give rise to geometric fluxes while $\beta$-transformations generate non-geometric fluxes.

We can consider Type IIA compactifications in which all these fluxes are present. They are captured in the description of the internal manifold through the definition of a twisted differential operator\footnote{In generalized complex geometry this comes from the study of the Courant bracket (extension of the Lie bracket) and its behaviour under $A,B$ and $\beta$-transformations.} \cite{Shelton:2006fd}
\begin{equation}
    {\cal D} = d - H \wedge  +\  f \triangleleft  +\ Q \triangleright  +\  R\, \bullet \, ,
    \label{cy-eq: twistedD}
\end{equation}
where $H$ is the NS three-form flux, $f$ encodes the geometric fluxes, $Q$ that of globally-non-geometric fluxes and $R$ is the locally-non-geometric fluxes. The action of various fluxes appearing in ${\cal D}$ is such that for an arbitrary $p$-form $A_p$, the pieces $H\wedge A_p$, $f \triangleleft A_p$, $Q \triangleright A_p$ and $R \bullet A_p$ denote a $(p+3)$, $(p+1)$, $(p-1)$ and $(p-3)$-form respectively. We describe their action on the basis of harmonic forms in Appendix \ref{ap.sys-sec: conv}.

\subsubsection{Superpotential}

The different flux backgrounds that we have discussed up until this point contribute to the moduli dynamics through a superpotential originally introduced in \cite{Gukov:1999ya, Taylor:1999ii}. It can be divided into the RR and the NSNS sector. The former is given by
\begin{equation}
     W_{\rm RR}=  -\frac{1}{\ell_s^6} \int_{X_6} e^{J_c} \wedge \bar{\bf G}\,, 
     \label{cy-eq: general RR superpotential}
\end{equation}
which after integrating using the expansion of the complexified Kähler form \eqref{cy-eq: complexified J expansion} and the definition of the flux quanta \eqref{cy-eq: RRfluxes} becomes
\begin{equation}
    \ell_s W_{\rm RR} =e_0 +  e_aT^a + \frac{1}{2}\, {\cal K}_{abc}  m^a T^b T^c  + \frac{m}{6}\, {\cal K}_{abc}\, T^a T^bT^c \, . 
    \label{cy-eq: general RR superpotential expanded}
\end{equation}
The superpotential regarding the NSNS sector can be adapted to include the geometric and non-geometric fluxes as follows \cite{Aldazabal:2006up,Shelton:2006fd}
\begin{equation}
    \label{cy-eq: general NS superpotential}
     W_{\rm NS} =   \frac{1}{\ell_s^6}  \int_{X_6} \Omega_c \wedge {\cal D}\left( e^{-J_c} \right)\,.
\end{equation}
Expanding the complexified Kähler and holomorphic forms \eqref{cy-eq: complexified J expansion}, \eqref{cy-eq: complexified omega expansion} and using the conventions from appendix \ref{ap.sys-sec: conv}, we get 
\begin{equation}
    \ell_s W_{\rm NS} =  U^\mu \Bigl[ h_\mu + f_{a\mu} T^a + \frac{1}{2} {\cal K}_{abc} \, T^b \, T^c \, Q^a{}_\mu + \frac{1}{6}\, {\cal K}_{abc} T^a T^b T^c \, R_\mu \Bigr] \,
    \label{cy-eq: general NS superpotential expanded}
\end{equation}

\subsubsection*{Scalar potential}

The background distribution for the RR and NSNS-fluxes induces a scalar potential for the geometric moduli (saxions) $t^a,u^\mu$ and the closed string sector moduli (axions) $b^a,\xi^\mu$. Under the assumption that background fluxes do not affect the K\"ahler potential pieces \eqref{cy-eq: KK} and \eqref{cy-eq: KQ},\footnote{The validity of this assumption should not be taken for granted and will depend on the particular class of vacua. The results in \cite{Junghans:2020acz,Buratti:2020kda,Marchesano:2020qvg} suggest that it is valid in the presence of only $p$-form fluxes $F_{\rm RR}$, $H$. However,  \cite{Font:2019uva} gives an example of compactification with metric fluxes in which the naive KK scale is heavily corrected by fluxes, and so should be the K\"ahler potential.} one can easily compute the F-term flux potential for closed string moduli via the standard supergravity expression
\begin{equation}
\label{cy-eq: VFgen}
\kappa_4^2\, V_F =  e^K \left(K^{{\cal A}\ov{\cal B}}\, D_{\cal A} W \, \ov{D}_{\ov{\cal B}^\prime} \ov{W} - 3 \, |W|^2\right),
\end{equation}
where $W=W_K+W_{\rm NS}$ and the index ${\cal A} = \{a, \mu\}$ runs over all  moduli.

 In addition, as pointed out in \cite{Ihl:2007ah,Robbins:2007yv},  geometric and non-geometric fluxes will generate a D-term contribution to the scalar potential when the even cohomology group $H_+^{1,1}$ is not trivial. This can be computed as
\begin{equation}
\label{cy-eq: VDgen}
V_D = \frac{1}{2}  \left( {\rm Re} f\right)^{-1\: \alpha\beta}\, D_\alpha \, D_\beta\, .
\end{equation}
In the expression above, $D_\alpha$ is the $D$-term for the $U(1)$ gauge group corresponding to a 1-form potential $A^\alpha$ that arises from decomposition \eqref{cy-eq: C orientifold decomposition} and is given by
\begin{equation}
D_\alpha = i \partial_{{\cal A}} K \, \delta_\alpha \varphi^{\cal A} + \zeta_\alpha \,,
\end{equation}
where $\delta_\alpha \varphi^{\cal A}$ is the variation of the scalar field $\varphi^{\cal A}$ (includes both axions and saxions) under a gauge transformation, and $\zeta_\alpha$ is the corresponding Fayet-Iliopoulos term.
In order to find the explicit expression of the D-term potential we perform a gauge transformation on the gauge bosons in \eqref{cy-eq: C orientifold decomposition}. We consider as well the dual field in the democratic formulation, $C_5$, which also admits a decomposition in the basis of table \ref{cy-table: harmonic basis} of the form $ C_5 =  C_{2\, \mu} \beta^\mu  + A_\alpha \tilde\varpi^\alpha $. Thus, we take
\begin{equation}
A^\alpha\, , A_\alpha \quad \longrightarrow  \quad A^\alpha + d \lambda^\alpha\, ,\, A_\alpha  \to A_\alpha + d \lambda_\alpha\, .
\end{equation}
The transformation of the RR $p$-form potential ${\bf C }  \equiv  C_1 + C_3 + C_5 + \dots $ can then be written in terms of the twisted differential ${\cal D}$ defined in  \eqref{cy-eq: twistedD}
\begin{eqnarray}
    {\bf C}   & \longrightarrow &  {\bf C} + {\cal D}\left(\lambda^\alpha \, \varpi_\alpha + \lambda_\alpha \, \tilde\varpi^{\alpha} \right) \label{cy-eq: C3change}\\
& = & \left(\xi^\mu + \lambda^\alpha \, \hat f_{\alpha}{}^\mu + \lambda_\alpha\, \hat{Q}^{\alpha \mu}\right) \, \alpha_\mu  + \dots \nonumber
\end{eqnarray}
where we have used the flux actions given in \ref{ap.sys-sec: conv}, with $\hat f_{\alpha}{}^\mu$, $\hat{Q}^{\alpha \mu}$ integers. This result shows that the scalar fields $\xi^\mu$ are not invariant under the gauge transformation, leading to the following shift in the ${\cal N} = 1$ coordinates $U^\mu$
\begin{equation}
\delta U^\mu =  \lambda^\alpha\, \hat f_{\alpha}{}^\mu +  \lambda_\alpha \, \hat{Q}^{\alpha \mu}\, ,
\label{cy-eq: transU}
\end{equation}
where we have again unified the NS fluxes under the index $\mu$. Note that due to the Bianchi identities \eqref{sys-eq: bianchids2} only the combinations of fields  $U^\mu$ invariant under \eqref{cy-eq: transU} appear in the superpotential and, as a result, the Fayet-Iliopoulos terms vanish.  Interpreting \eqref{cy-eq: transU} as gaugings of the U(1) gauge fields and their magnetic duals one obtains the D-terms
\begin{equation}
D_\alpha = \frac{1}{2} \partial_\mu K \, \left(\hat f_\alpha{}^\mu +\hat{\cal K}_{a\alpha\beta} b^a \hat{Q}^{\beta \mu}  \right) \, , \qquad   D^\alpha =\frac{1}{2}  \partial_\mu K \, \hat{Q}^{\alpha \mu} \, .
\end{equation}
Taking into account the kinetic couplings \eqref{cy-eq: fg} we end up with the following D-term scalar potential 
\begin{equation}
\label{cy-eq: DtermGen}
V_D =-\frac{1}{4}  \partial_\mu K \partial_\nu K \biggl({\rm Im}\, \hat{\cal K}^{-1\: \alpha\beta} \left(\hat f_\alpha{}^\mu +\hat{\cal K}_{a\alpha\gamma} b^a \hat{Q}^{\gamma \mu}  \right) \left(\hat f_\beta{}^\nu +\hat{\cal K}_{c\beta\delta} b^c \hat{Q}^{\delta \nu}  \right)
 +  {\rm Im}\, \hat{\cal K}_{\alpha\beta}  \hat{Q}^{\alpha \mu} \, \hat{Q}^{\beta \nu} \biggr) \, ,
\end{equation}
where $\hat{\cal K}_{\alpha\beta} = \hat{\cal K}_{a\alpha\beta} \, T^a$. Alternatively, one may obtain the same potential by following the tensor multiplet analysis of \cite{Grimm:2004uq,Louis:2004xi}.\footnote{This result is different from the type IIA D-term potential of \cite{Robbins:2007yv}, and recovers the expected discrete gauge symmetries related to $b$-field shifts. The same strategy can be applied to type IIB setups with non-geometric fluxes, recovering the full scalar  obtained by DFT dimensional reduction in   \cite{Blumenhagen:2015lta}.}

The total potential for the moduli is simply given by the sum of the two terms
\begin{equation}
    V=V_F+V_D\,.
\end{equation}
Vacua will be found by demanding that the derivatives of these potentials with respect to the moduli vanish, while the eigenvalues of the Hessian will provide their masses. Of special interest are vacua in which $\mathcal{N}=1$ supersymmetry remains unbroken. These correspond to points of moduli space where the F-terms (covariant derivatives of the superpotential) vanish.  The  expressions involved in these type of computations are generally very complicated. Fortunately, there are techniques that allow to write it as a sum of two bilinear contributions (one for the D-term and another for the F-term) in which the dependence on axions and saxions splits. We will review them in the next section in order to use it later on in chapters \ref{ch: systematics} and \ref{ch: Ftheory}.

\subsection{Bilinear formalism}
\label{cy-subsec: bilinear formalism}

Up until this point, we have developed the mathematical framework of string compactifications. We have identified how constraints motivated by 4-dimensional observations restrict the geometry of the internal manifold, discovered how massless scalar fields emerge as a result of the compactification of that manifold and the valuable role that flux background play in providing a potential that endows them with mass at the cost of deforming the compactification space. We have all the pieces required to determine how moduli stabilization depends on the choice of flux quanta and describe the landscape of flux compactifications. However, we now face a more practical problem: dealing with the scalar potential described in the previous section is a daunting task. To help in this subject, we will use the bilinear formalism introduced in \cite{Bielleman:2015ina, Carta:2016ynn, Herraez:2018vae} and extensively applied in \cite{Escobar:2018rna, Escobar:2018tiu, Marchesano:2019hfb}. It is a powerful tool that dramatically simplifies the systematic search of extrema in the scalar potential. 

The cornerstone of the formalism is the use of 4-forms to decompose and describe all flux contributions to the scalar potential, which generates a map between 4-dimensional Minkowski 4-forms and the flux quanta. The idea can be traced back to the paper Bousso and Polchinski \cite{Bousso:2000xa}, which tried to understand the small value of the 4-dimensional cosmological constant $\Lambda$ in terms of the contributions of non-propagating 3-forms $C_3^A$ coming from the RR and NSNS closed string sectors, building on the previous work of Brown and Teitelboim \cite{Brown:1987dd,Brown:1988kg}. The associated 4-form field strengths $F_4^A$ contribute to the vacuum energy through a potential of the form
\begin{equation}
V_{BP}=\sum_{A,B}Z_{AB}F_4^AF_4^B+\Lambda_0\,,
\end{equation}
where $Z_{AB}$ is a positive definite metric depending on all moduli and $\Lambda_0$ is the bare cosmological constant with a typically negative value of natural order close to the Planck mass. This way, if the number of 4-forms is sufficiently large, one can construct an arbitrarily small cosmological constant $V_{BP}$ without running into naturalness problems regarding the bare cosmological constant $\Lambda_0$.

This approach was recovered and expanded by \cite{Bielleman:2015ina} in the context of massive type IIA flux compactifications. There and in the following works, the authors showed that, up to boundary terms, the relevant contributions to the scalar potential appearing in the 4-dimensional effective action have the structure
\begin{eqnarray}
    -Z_{AB}F_4^A\wedge\star_4 F_4^B+2 F_4^A\rho_A-Z^{AB}\rho_A\rho_B\subset 16S_{4d}\,,
\end{eqnarray}
where the indices $A,B$ run over all the space of fluxes in the compactification, the functions $\rho_A$ are polynomials of the axions with coefficients given by the flux quanta and the topological data of the internal manifold  and $Z_{AB}$ is a matrix whose entries depend on the saxions as well as the internal manifold topology.

From the duality relation between the fluxes in the democratic formulation, one can prove that on-shell $\star_4 F_4^A=Z^{AB}\rho_B$ \cite{Herraez:2018vae}. Then the scalar potential becomes
\begin{equation}
    V=\frac{1}{8\kappa_4^2}Z^{AB}\rho_A\rho_B\,.
    \label{cy-eq: bilinear potential}
\end{equation}
We conclude that, without loss of generality, the scalar potential can be expressed in terms of a bilinear formula that factorizes the role of axions and flux quanta from the saxions, greatly simplifying the process of finding the vacua of the scalar fields.

\subsection{AdS Vacua}
\label{cy-subsec: AdS vacua}

The bilinear formalism was applied in the recent work \cite{Marchesano:2019hfb} to systematically study the  vacua of massive Type IIA orientifold compactifications in the presence of background RR and H fluxes as well as D6 branes, extending the results found in \cite{DeWolfe:2005uu} to far more general setups. They find that the scalar potential can be expressed as 
\begin{equation}
    V=\frac{1}{\kappa_4^2}\vec{\rho}^{\, t}\mathbf{Z}\vec{\rho}\,,
\end{equation}
where the flux and axion polynomials are given by 
\begin{subequations}
\begin{align}
    \ell_s\rho_0=&\, e_0+e_ab^a+\frac{1}{2}\mathcal{K}_{abc}m^ab^bb^c+\frac{m}{6}\mathcal{K}_{abc}b^ab^bb^c+h_\mu\xi^\mu\,,\\
    \ell_s\rho_a=&\, e_a+\mathcal{K}_{abc}m^bb^c+\frac{m}{2}\mathcal{K}_{abc}b^bb^c\,,\\
    \ell_s\tilde{\rho}^a=&\, m^a+mb^a\,,\\
    \ell_s\tilde{\rho}=&\, m\,,\\
    \ell_s\hat{\rho}_\mu=&\, h_\mu\,,
\end{align}
\end{subequations}
and the saxion dependent matrix takes the form
\begin{equation}
\mathbf{Z}=e^K\left(
    \begin{array}{ccccc}
        4 &  &   &   &   \\
        & K^{ab} &   &   &   \\
         &  & \frac{4}{9}\mathcal{K}^2K_{ab}  &   &   \\
          &  &   &  \frac{1}{9}\mathcal{K}^2 &\frac{2}{3}\mathcal{K}u^\mu   \\
           &  &   & \frac{2}{3}\mathcal{K}u^\nu  & K^{\mu\nu}  
\end{array}   \right)\,, 
\label{cy-eq: Z matrix ads landscpace}
\end{equation}
with $K_{ab}=\frac{1}{4}\partial_{t^a}\partial_{t^b}K_K$ and $K_{\mu\nu}=\frac{1}{4}\partial_{u^\mu}\partial_{u^\nu}K_Q$.

The solutions to the vacuum equations associated to this potential are characterized by the following relations between the axion polynomials and the saxions
\begin{equation}
    \rho_0=0\,,\qquad\hat{\rho}_\mu=\tilde{\rho}\mathcal{K}(\tilde{A}\partial_{u^\mu}K+\hat{\epsilon}^p_\mu)\,,\qquad \tilde{\rho}^a=\tilde{B}\tilde{\rho}t^a\,,\qquad \rho_a=\tilde{C}\tilde{\rho}\mathcal{K}_a \, ,
    \label{cy-eq: rho solutions}
\end{equation}
where each triplet of coefficients $(\tilde{A},\tilde{B},\tilde{C})$ characterizes a different family of vacua spawned by the Romans mass $\tilde{\rho}$. The parameter $\hat{\epsilon}^p_\mu$ is a function of the coefficients given by
\begin{equation}
    \hat{\epsilon}^p_0=\left(\frac{1}{8}-\frac{3\mathcal{E}}{2}\right)\partial_{u^0} K\,,\qquad \hat{\epsilon}_i=\left(\frac{\mathcal{E}}{2}-\frac{1}{24}\right)\partial_{u^i} K\,.
\end{equation}

There are only a small set of values that coefficients $(\tilde{A},\tilde{B},\tilde{C})$ can take to generate consistent vacua. They are summarized in table \ref{cy-table: summary ads vacua}. All the solutions were obtained assuming the smearing approximation in order to use the Calabi-Yau structure to describe the geometry of the internal space. The first branch was the one considered in \cite{DeWolfe:2005uu} and is the only one in which $\mathcal{N}=1$ supersymmetry remains unbroken.  The 10-dimensional theory described in \ref{cy-subsec: beyond smearing} corresponds to the uplift of this branch to a 10-dimensional theory beyond the Calabi-Yau limit - see \eqref{cy-eq: SU3 solutions}. The fact that an uplift for a 4-dimensional theory displaying scale separation like the one in \cite{DeWolfe:2005uu} exists lies in tension with the strong version of the AdS Distance conjecture discussed in section \ref{basic-subsec: swampland conjectures}.   

It could be that not all the branches found in table \ref{cy-table: summary ads vacua} admit a well-defined uplift, which would mean that they are not true String Theory compactifications but a spurious result of the information loss derived from the approximations taken to build the 4-dimensional effective theory. In chapters \ref{ch: bionic} and \ref{ch: membranes} we will explore the uplift of the second branch and study how it fits with the Swampland Conjectures. Notably, none of these branches describe a de Sitter space, in agreement with the no-go results presented in sections \ref{basic-subsec: swampland conjectures} and \ref{cy-subsec: democratic formulation}. Given the phenomenological motivation to achieve de Sitter, it is therefore interesting to see if more general choices of fluxes (geometric and non-geometric) can change that. We will address this topic in the next chapter.

Finally, the mass spectrum for the different branches is computed. The results show that the non-supersymmetric branches are perturbatively stable, which again is in tension with a Swampland prediction, i.e. the AdS instability conjecture. The problem could be circumvented by considering non-perturbative decays. We will explore this  path in chapters \ref{ch: bionic} and \ref{ch: membranes}.

\begin{table}[htbp]
\centering
\def\arraystretch{1.5}
\begin{tabular}{|c|c|c|c|c|c|c|}
\hline
{\bf Branch}                       & {\bf $\tilde{A}$ }            & {\bf $\tilde{B}$   }             & {\bf $\tilde{C}$      }                & {\bf $\mathcal{E}$} & {\bf $\kappa_4^2\Lambda$  }                          & {\bf SUSY } \\ \hline
{\bf A1-S1} & $\frac{1}{15}$  & 0                  & $\frac{3}{10}$           & $\frac{1}{12}$   & $-\frac{4e^K}{75}\mathcal{K}^2\tilde{\rho}^2$   & Yes  \\ \hline
{\bf A1-S1} & $\frac{1}{15}$  & 0                  & $-\frac{3}{10}$          & $\frac{1}{12}$   & $-\frac{4e^K}{75}\mathcal{K}^2\tilde{\rho}^2$   & No   \\ \hline
{\bf A1-S2} & $\frac{7}{120}$ & 0                  & $\pm\frac{\sqrt{6}}{10}$ & $\frac{1}{30}$  & $-\frac{8e^K}{225}\mathcal{K}^2\tilde{\rho}^2$  & No   \\ \hline
{\bf A1-S2} & $\frac{1}{24}$  & 0                  & 0                        & 0   & 0                                              & No   \\ \hline
{\bf A2-S1} & $\frac{1}{12}$  & $\pm \frac{1}{2}$  & $-\frac{1}{4}$           & $\frac{1}{12}$   & $-\frac{e^K}{18}\mathcal{K}^2\tilde{\rho}^2$    & No   \\ \hline
{\bf A2-S2} & $\frac{5}{84}$  & $\pm \frac{1}{14}$ & $-\frac{1}{4}$           & $\frac{1}{28}$   & $-\frac{11e^K}{294}\mathcal{K}^2\tilde{\rho}^2$ & No   \\ \hline
\end{tabular}
\caption{Different branches of solutions with the corresponding vacuum energy. Extracted from \cite{Marchesano:2019hfb}.}
\label{cy-table: summary ads vacua}
\end{table}



\ifSubfilesClassLoaded{%
\bibliography{biblio}%
}{}

\end{document}

\graphicspath{{Images/Systematics}}

\ifSubfilesClassLoaded{%
\tableofcontents
}{}

\setcounter{chapter}{3}
\chapter{Systematics of Type IIA moduli stabilization}
\label{ch: systematics}

One of the major challenges in the field of String Theory is to determine the structure of four-dimensional meta-stable vacua, a.k.a. the string Landscape. In this context,  type IIA flux compactifications with RR and NSNS fluxes have played a key role  in motivating and in testing many Swampland conjectures that restrict the domain of the Landscape. To some extent this is because, in appropriate regimes, type IIA moduli stabilization can be purely addressed at the classical level \cite{Derendinger:2004jn,Villadoro:2005cu,DeWolfe:2005uu,Camara:2005dc}, opening the door for a direct 10d microscopic description of such vacua, briefly discussed in sections \ref{cy-subsec: smearing approx} and  \ref{cy-subsec: beyond smearing}.

Despite all these key features, it is fair to say that the general structure of geometric type IIA flux compactifications is less understood than their type IIB counterpart \cite{Grana:2005jc,Douglas:2006es,Becker:2007zj,Denef:2007pq,Ibanez:2012zz}. Part of the problem is all the different kinds of fluxes that are present in the type IIA setup, which, on the other hand, is the peculiarity that permits to stabilize all moduli classically. Traditionally, each kind of flux is treated differently, and as soon as geometric fluxes are introduced the classification of vacua becomes quite involved. 

 The purpose of this chapter is to improve this picture by providing a unifying treatment of moduli stabilization in (massive) type IIA orientifold flux vacua with geometric and non-geometric contributions. Our main tool will be the bilinear form of the scalar potential $V = Z^{AB} \rho_A \rho_B$, introduced in section \ref{cy-subsec: bilinear formalism}. While this bilinear structure was originally found for the case of Calabi--Yau compactifications with $p$-form fluxes, building on \cite{Gao:2017gxk} we show that it can be extended to include the presence of geometric and non-geometric fluxes, even when these fluxes generate both an F-term and a D-term potential. 
 
 With this form of the flux potential, one may perform a systematic search for vacua, as already described for the Calabi--Yau case. We do so now for orientifold compactifications with $p$-form and geometric fluxes, which are one of the main sources of classical AdS$_4$ and dS$_4$ backgrounds in String Theory, and have already provided crucial information regarding Swampland criteria. On the one hand, the microscopic 10d description of AdS$_4$ geometric flux vacua has been discussed in several instances  \cite{House:2005yc,Grana:2006kf,Aldazabal:2007sn,Koerber:2008rx,Caviezel:2008ik,Koerber:2010rn}. On the other hand, they have provided several no-go results on de Sitter solutions \cite{Hertzberg:2007wc,Haque:2008jz,Caviezel:2008tf,Flauger:2008ad,Danielsson:2009ff,Danielsson:2010bc,Danielsson:2011au}, as well as examples of unstable de Sitter extrema that have served to refine the original de Sitter conjecture \cite{Andriot:2018wzk}.  Therefore, it is expected that a global, more exhaustive description of this class of vacua  and a systematic understanding of their properties  leads to further tests, and perhaps even refinements, of the de Sitter and AdS distance conjectures.
  
To perform our search for vacua we consider a certain pattern of on-shell F-terms, that is then translated into an Ansatz. Even if this F-term pattern is motivated from general stability criteria for de Sitter vacua \cite{GomezReino:2006dk,GomezReino:2006wv,GomezReino:2007qi,Covi:2008ea,Covi:2008zu}, one can show that de Sitter extrema are incompatible with such F-terms, obtaining a new kind of no-go result. Compactifications to AdS$_4$ are on the other hand allowed, and using our Ansatz we find both a supersymmetric  and a non-supersymmetric branch of vacua, intersecting at one point. In some cases we can check explicitly the perturbative stability of the non-SUSY AdS$_4$ branch, finding that the vacua are stable for a large region of the parameter space of our Ansatz, and even free of tachyons for a large subregion. We finally comment on the 10d description of this set of vacua.

This chapter is organized as follows. In section \ref{sys-sec: IIAorientifold} we consider the classical F-term and D-term potential of type IIA  compactifications with all kind of fluxes and express both potentials in a bilinear form. In section \ref{sys-sec: fluxpot} we propose an F-term pattern to avoid tachyons in de Sitter vacua, and build a general Ansatz from it. We also describe the flux invariants present in this class of compactifications. In section \ref{sys-sec: geovacua} we apply our results to configurations with $p$-form and geometric fluxes, in order to classify their different extrema. We find two different branches, that contain several previous results in the literature. In section \ref{sys-sec: stabalidity} we discuss which of these extrema are perturbatively stable, as well as their 10d description. We draw our conclusions in section \ref{sys-sec: conclu}. 

Some technical details have been relegated to the Appendices. Appendix \ref{ap.sys-sec: conv} contains several aspects regarding NS fluxes and flux-axion polynomials. Appendix \ref{ap.sys: curvature} develops the computations motivating our F-term Ansatz. Appendix \ref{ap.sys: Hessian} contains the computation of the Hessian for geometric flux extrema.


\section{The Type IIA general flux potential in the bilinear formalism}
\label{sys-sec: IIAorientifold}

In this section we consider the 4-dimensional $\mathcal{N}=1$ supergravity theory coming from the compactification on an orientifold of $\mathcal{M}_4\times X_6$, with $X_6$ a Calabi-Yau, and turning on the geometric and non-geometric fluxes introduced in section \ref{cy-subsec: effective 4d potential} over that setup. As discussed there, the potential of the effective theory is given by the sum of two contributions, dubbed F-term \eqref{cy-eq: VFgen} and D-term \eqref{cy-eq: VDgen}. Our goal will be to express them in the bilinear formalism \eqref{cy-eq: bilinear potential} splitting the dependence of saxion and axions.

\subsection{The F-term flux potential}

 As in \cite{Bielleman:2015ina,Herraez:2018vae}, one can indeed show that this F-term potential displays a bilinear structure of the form
\begin{equation}\label{sys-VF}
\kappa_4^2\, V_F  = {\rho}_\cA \, Z^{\cA\cB} \, {\rho}_\cB\, , 
\end{equation}  
where the matrix entries $Z^{\cA\cB}$ only depend on the saxions $\{t^a, n^\mu \}$, while the ${\rho}_\cA$ only depend on the flux quanta and the axions $\{b^a, \xi^\mu\}$. One can easily rewrite the results in \cite{Gao:2017gxk} to fit the above expression, obtaining the following result.

The set of axion polynomials with flux-quanta coefficients are
\begin{equation}
    \rho_\cA=\{\rho_0,\rho_a,\tilde{\rho}^a,\tilde{\rho},\rho_\mu,\rho_{a\mu},\tilde{\rho}^a_\mu ,\tilde{\rho}_\mu \}\, ,
    \label{sys-rhos}
\end{equation}
and are defined as
\bes
\label{sys-RRrhos}
\begin{align}
  \ell_s  \rho_0&=e_0+e_ab^a+\frac{1}{2}\mathcal{K}_{abc}m^ab^bb^c+\frac{m}{6}\mathcal{K}_{abc}b^ab^bb^c+\ell_s \rho_\mu\xi^\mu\, , \label{sys-eq: rho0}\\
 \ell_s   \rho_a&=e_a+\mathcal{K}_{abc}m^bb^c+\frac{m}{2}\mathcal{K}_{abc}b^bb^c+\ell_s\rho_{a\mu}\xi^\mu \, ,  \label{sys-eq: rho_a}\\
  \ell_s  \tilde{\rho}^a&=m^a+m b^a + \ell_s\tilde{\rho}^a_\mu\xi^\mu \, ,  \label{sys-eq: rho^a}\\
 \ell_s   \tilde{\rho}&=m+\ell_s\tilde{\rho}_\mu\xi^\mu \, ,   \label{sys-eq: rhom}
\end{align}   
\ees 
and
\bes
\label{sys-NSrhos}
\begin{align}    
\ell_s    \rho_\mu&=h_\mu+f_{a\mu}b^a+\frac{1}{2}\mathcal{K}_{abc}b^bb^cQ_\mu^a+\frac{1}{6}\mathcal{K}_{abc}b^ab^bb^cR_\mu \, , \\
 \ell_s   \rho_{a\mu}&=f_{a\mu}+\mathcal{K}_{abc}b^bQ^c_\mu+\frac{1}{2}\mathcal{K}_{abc}b^bb^cR_\mu \, ,  \label{sys-eq: rho_ak} \\
 \ell_s   \tilde{\rho}^a_\mu &=Q^a_\mu+b^aR_\mu \, , \\
 \ell_s   \tilde{\rho}_\mu &=R_\mu \, .
\end{align}
\ees
The polynomials \eqref{sys-NSrhos} are mostly new with respect to the Calabi--Yau case with $p$-form fluxes, as they highly depend on the presence of geometric and non-geometric fluxes. As in \cite{Herraez:2018vae}, both \eqref{sys-RRrhos} and \eqref{sys-NSrhos} have the interpretation of invariants under the discrete shift symmetries of the combined superpotential $W = W_{\rm RR} + W_{\rm NS}$. This invariance is more evident by writing $\ell_s \rho_\cA  = {\cal R}_\cA{}^\cB q_\cB$, where $q_\cA = \left\{e_0, \, e_b, \,  m^b, \, m, \, h_\mu, \, f_{b\mu}, \, Q^b{}_\mu, \, R_\mu \right\}$ encodes the flux quanta of the compactification and
\be
\label{sys-eq:invRmat}
{\cal R} = \begin{bmatrix}
    {\cal R}_0  \quad & \qquad {\cal R}_0 \, \, \, \xi^\mu \,  \\
  0  \quad  & \qquad {\cal R}_0 \,\, \, \delta_\nu^\mu
\end{bmatrix}\, , \quad 
 {\cal R}_0 = \begin{bmatrix}
 1 & \quad b^b & \quad \frac{1}{2} \, {\cal K}_{abc} \, b^a \, b^c & \quad \frac{1}{6}\, {\cal K}_{abc} \, b^a \, b^b \, b^c \\
 0 & \quad \delta_a^b & \quad {\cal K}_{abc} \, b^c & \quad \frac{1}{2} \, {\cal K}_{abc} \, b^b \, b^c \\
 0 & \quad 0 & \quad \delta_b^a & \quad b^a\\
 0 & \quad 0 & 0 & \quad 1 \\
\end{bmatrix}\, ,
\ee
is an axion-dependent upper triangular  matrix, see Appendix \ref{ap.sys-sec: conv} for details. Including curvature corrections will modify ${\cal R}_0$, such that discrete shift symmetries become manifest, and shifting an axion by a unit period can be compensated by an integer shift of $q_\cA$ \cite{Escobar:2018rna}.

As for the bilinear form $Z$, one finds the following expression
\be
\label{sys-eq: Z-matrix}
Z^{{\cal A}{\cal B}} =  e^K \, \begin{bmatrix}
   {\bf B}  \quad & \, \, {\cal O} \\
   {\cal O}^{\, t} \quad  & \, \, {\bf C}
\end{bmatrix}\, ,
\ee
where
\begin{equation}
    {\bf B} =\left(\begin{array}{cccc}
         4 & 0 & 0 & 0 \\
         0 & g^{ab} & 0 & 0\\
         0 & 0 & \frac{4\mathcal{K}^2}{9}g_{ab} & 0 \\
         0 & 0 & 0 & \frac{\mathcal{K}^2}{9}
    \end{array}
    \right)\, , \quad 
        \mathcal{O} =\left(\begin{array}{cccc}
         0 & 0 & 0 & -\frac{2\mathcal{K}}{3}u^\nu \\
         0 & 0 &  \frac{2\mathcal{K}}{3}u^\nu \delta^a_b & 0\\
         0 & -\frac{2\mathcal{K}}{3}u^\nu\delta^b_a & 0 & 0 \\
         \frac{2\mathcal{K}}{3}u^\nu & 0 & 0 & 0
    \end{array}
    \right)\, ,
    \label{sys-eq:GOmatrix}
\end{equation}
\begin{equation}
    {\bf C} =\left(\begin{array}{cccc}
         c^{\mu\nu} & 0 & -\tilde{c}^{\mu\nu}\frac{\mathcal{K}_b}{2} & 0 \\
         0 & \tilde{c}^{\mu\nu}t^at^b+ g^{ab}u^\mu u^\nu  &  0 & -\tilde{c}^{\mu\nu}t^a\frac{\mathcal{K}}{6}\\
         -\tilde{c}^{\mu\nu}\frac{\mathcal{K}_a}{2} & 0 & \frac{1}{4}\tilde{c}^{\mu\nu}\mathcal{K}_a\mathcal{K}_b+\frac{4\mathcal{K}^2}{9}g_{ab}u^\mu u^\nu  & 0 \\
         0 & -\tilde{c}^{\mu\nu}t^b\frac{\mathcal{K}}{6} & 0 & \frac{\mathcal{K}^2}{36}c^{\mu\nu}
    \end{array}
    \right)\, .
\end{equation}
Here $K = K_K + K_Q$, $g_{ab} = \frac{1}{4} \partial_{t^a} \partial_{t^b} K_K\equiv \frac{1}{4}\p_a\p_b K_K$, and $c_{\mu\nu} = \frac{1}{4} \partial_{u^\mu}\partial_{u^\nu} K_Q\equiv \frac{1}{4}\p_\mu\p_\nu K_Q$, while upper indices denote their inverses. Also $u^\mu = \IM U^\mu =(n^K, u_\Lambda)$ stands for the complex structure saxions, and we have defined  $\mk_{a}=\mk_{abc}t^bt^c$ and $\tilde{c}^{\mu\nu}=c^{\mu\nu}-4u^\mu u^\nu $. 

Compared to the Calabi--Yau case \eqref{cy-eq: Z matrix ads landscpace} (where only NSNS and RR fluxes were present) the matrices {\bf C} and ${\cal O}$ are more involved, again due to the presence of geometric and non-geometric fluxes. Interestingly, the off-diagonal matrix ${\cal O}$ has the same source as in the Calabi--Yau case, namely the contribution from the tension of the localized sources after taking into account  tadpole cancellation. Indeed, the contribution of background fluxes to the D6-brane tadpole is given by \cite{Aldazabal:2006up}
\be
\ell_s^2 {\cal D} \bar{\bf G} =  \left(m h_\mu  -  m^a f_{a \mu} +  e_aQ^a{}_\mu  -  e_0 R_\mu  \right)\, \beta^\mu \, ,
\label{sys-DFtadpole}
\ee
which generalizes the contribution found in \eqref{cy-eq: BI expanded} and can be easily expressed in terms of the $\rho_\cA$.
The corresponding absence of D6-branes needed to cancel such tadpole then translates into the following piece of the potential 
\be
\kappa_4^2 V_{\rm loc} = \frac{4}{3} e^K {\cal K}\, u^\mu  \left( \tilde\rho  \rho_\mu -  \tilde\rho^a  \rho_{a\mu}  +  \rho_a \tilde\rho^a{}_\mu - \rho_0 \, \tilde\rho_\mu \right)\, ,
\ee 
which is nothing but the said off-diagonal contribution.

Putting all this together, the final expression for the F-term potential reads
\begin{align}
   \kappa_4^2 V_F =\, &e^K\left[4\rho_0^2+g^{ab}\rho_a\rho_b+\frac{4\mathcal{K}^2}{9}g_{ab}\tilde{\rho}^a\tilde{\rho}^b+\frac{\mathcal{K}^2}{9}\tilde{\rho}^2+c^{\mu\nu}\rho_\mu\rho_\nu+\left(\tilde{c}^{\mu\nu}t^at^b+g^{ab}u^\mu u^\nu \right)\rho_{a\mu}\rho_{b\nu}\right.\nonumber\\
    &+\left(\tilde{c}^{\mu\nu}\frac{\mathcal{K}_a}{2}\frac{\mathcal{K}_b}{2}+\frac{4\mathcal{K}^2}{9}g_ {ab}u^\mu u^\nu \right)\tilde{\rho}^a_\mu\tilde{\rho}^b_\nu+\frac{\mathcal{K}^2}{36}c^{\mu\nu}\tilde{\rho}_\mu\tilde{\rho}_\nu-\frac{4\mathcal{K}}{3}u^\nu \rho_0\tilde{\rho}_\nu+\frac{4\mathcal{K}}{3}u^\nu \rho_a\tilde{\rho}^a_\nu\nonumber\\
    &\left.-\frac{4\mathcal{K}}{3}u^\nu \tilde{\rho}^a\rho_{a\nu}+\frac{4\mathcal{K}}{3}u^\nu \tilde{\rho}\rho_\nu-\tilde{c}^{\mu\nu}\mathcal{K}_a\rho_\mu\tilde{\rho}^a_\nu-\tilde{c}^{\mu\nu}t^a\frac{\mathcal{K}}{3}\rho_{a\mu}\tilde{\rho}_\nu\right]\, .
    \label{sys-eq:potential}
\end{align}
This expression expands on the results of \cite{Herraez:2018vae} and can be easily connected to other known formulations of (non-)geometric potentials in the type IIA literature, e.g. \cite{Villadoro:2005cu,Flauger:2008ad,Blumenhagen:2013hva,Shukla:2019akv}.

\subsection{The D-term flux potential}

Geometric and non-geometric fluxes can couple to the $U(1)$ symmetries that arise from the even cohomology group $H_+^{1,1}$ in the closed string sector. The coupling generates the D-term contribution to the scalar potential introduced in \eqref{cy-eq: VDgen} and conveniently rewritten in \eqref{cy-eq: DtermGen}. It turns out that one can present this expression in a bilinear form similar to \eqref{sys-VF} by defining the following flux-axion polynomials
\be
\label{sys-eq:NSorbitsNew2}
\ell_s \hat{\rho}_\alpha{}^\mu = \hat f_\alpha{}^\mu + \hat{\cal K}_{a\alpha\beta}\, b^a \,  \hat{Q}^{\beta \mu}\,,\qquad
 \ell_s\tilde\rho^{\alpha \mu} = \hat{Q}^{\alpha \mu} \, ,
\ee
so that one has 
\bea\nonumber
 \kappa_4^2 V_D& = &\frac{1}{4}
\begin{bmatrix}
\hat{\rho}_\alpha{}^\mu & \, \,  \tilde\rho^{\alpha \mu} \\
\end{bmatrix} . \begin{bmatrix}
  \frac{3}{2\mk} g^{\alpha\beta}\, \partial_\mu K \partial_\nu K  & \quad 0 \\
0 &\frac{2\mk}{3}  g_{\alpha\beta} \partial_\mu K \partial_\nu K  \\
\end{bmatrix}
 . \begin{bmatrix}
\hat{\rho}_\beta{}^\nu \\
\tilde\rho^{\beta \nu} \\
\end{bmatrix}
\\ 
& = &\frac{1}{4}
\partial_\mu K \partial_\nu K  \left(  \frac{3}{2\mk}g^{\alpha\beta}  \hat{\rho}_\alpha{}^\mu \hat{\rho}_\beta{}^\nu +\frac{2\mk}{3} \, g_{\alpha\beta} \,\, \tilde\rho^{a \mu} \tilde\rho^{\beta \nu} \right)\, , 
\label{sys-eq:D-terms-new}
\eea
with $g_{\a\b} = -\frac{3}{2\mk}{\rm Im}\, \hat{\cal K}_{\alpha\beta}$ and $g^{\a\b}$ its inverse. It is then easy to see that the full flux potential $V = V_F + V_D$ can be written of the bilinear form \eqref{sys-VF}, by simply adding \eqref{sys-eq:NSorbitsNew2} to the polynomials \eqref{sys-rhos} and enlarging $Z$ accordingly. 


\section{Analysis of the potential}
\label{sys-sec: fluxpot}

While axion polynomials allow for a simple, compact expression for the flux potential, finding its vacua in full generality is still quite a formidable task. In this section we discuss some general features of this potential that, in particular, will lead to a simple Ansatz for the search of vacua. In the following section we will implement these observations for the case of compactifications with geometric fluxes. As the D-term piece of the potential will not play a significant role, in this section we will neglect its presence by considering compactifications such that $h_+^{1,1} = 0$.  Nevertheless, the whole discussion can be easily extended to a more general case.

\subsection{Stability and F-terms}\label{sys-ss:fterms}

Given the F-term potential \eqref{sys-eq:potential}, one may directly compute its first derivatives to find its extrema and, subsequently, its second derivatives to check their perturbative stability. However, as (meta)stability may be rather delicate to check for non-supersymmetric vacua, it is always desirable to have criteria that simplify the stability analysis. 

A simple criterium to analyze vacua metastability for F-term potentials in 4-dimensional supergravity was developed in \cite{GomezReino:2006dk,GomezReino:2006wv,GomezReino:2007qi,Covi:2008ea,Covi:2008zu}, with particular interest on de Sitter vacua. As argued in there, the sGoldstino direction in field space is the one more likely to become tachyonic in generic de Sitter vacua. Therefore, a crucial necessary condition for metastability is that such a mass is positive. Interestingly, the stability analysis along the sGoldstino direction can essentially be formulated in terms of the K\"ahler potential, which allows analysing large classes of string compactifications simultaneously. 

Following the general discussion in \cite{GomezReino:2006dk,GomezReino:2006wv,GomezReino:2007qi,Covi:2008ea,Covi:2008zu}  the sGoldstino masses can be estimated by
\be
m^2 = (3m_{3/2}^2 + \kappa_4^2 V)\, \hat{\sig}-  \frac{2}{3} \kappa_4^2 V\, ,
\label{sys-sgoldmass}
\ee
where $m_{3/2} = e^{K/2} |W|$ is the gravitino mass, and
\be
\hat{\sig} = \frac{2}{3}-R_{A\bar B C \bar D} f^{A} f^{\bar B} f^{C} f^{\bar D}\, ,
\label{sys-sigma}
\ee
is a function of the normalized F-terms $f_A = \frac{G_A}{(G^AG_A)^{1/2}}$ with $G_A = D_{A} W$, and the Riemann curvature tensor $R_{A\bar{B}C\bar{D}}$. Therefore, if $V$ is positive so must be $\hat{\sig}$, or else the extremum will be unstable. Reversing the logic, the larger $\hat{\sigma}$ is, the more favorable will be a class of extrema to host metastable  vacua. 

It is quite instructive to compute $\hat{\sigma}$ in our setup. Notice that because the Riemann curvature tensor only depends on the K\"ahler potential, the analysis can be done independently of which kind of fluxes are present. Moreover, because the moduli space metric factorizes, $R_{A\bar B C \bar D} \neq 0$ only if all indices correspond to either K\"ahler or complex structure directions. As a consequence, the normalized F-terms can be expressed as
\be
f_A = \left({\rm cos}\, \b\, g_a, {\rm sin}\, \b\, g_\mu\right)\, 
\ee
where $g_a = \frac{G_a}{(G^aG_a)^{1/2}}$, $g_\mu  = \frac{G_\mu}{(G^\mu G_\mu)^{1/2}}$ are the normalized F-terms in the K\"ahler and complex structure sectors, respectively, and ${\rm tan}\, \b = \frac{(G^\mu G_\mu)^{1/2}}{(G^aG_a)^{1/2}}$. Therefore we have that
\be
\hat{\sig} = \frac{2}{3}-  \left({\rm cos}\, \b\right)^4 R_{a\bar b c \bar d}\, g^{a} g^{\bar b} g^{c} g^{\bar d} -  \left({\rm sin}\, \b\right)^4 R_{\mu\bar{\nu} \sigma\bar{\rho}}\, g^{\mu} g^{\bar \nu} g^{\sigma} g^{\bar \rho} \, .
\label{sys-sigma2}
\ee
Following the discussion of Appendix \ref{ap.sys: curvature}, one finds that the terms $R_{a\bar b c \bar d}\, g^{a} g^{\bar b} g^{c} g^{\bar d}$ and $R_{\mu\bar{\nu} \sigma\bar{\rho}}\, g^{\mu} g^{\bar \nu} g^{\sigma} g^{\bar \rho}$ are respectively minimized  by
\be
{g}_a = \frac{\g_K}{\sqrt{3}} K_a\, , \quad  {g}_\mu = \frac{\g_Q}{2} K_\mu\, ,
\label{sys-partialmax}
\ee
where $\g_K, \g_Q \in \C$ are such that $|\g_K|^2 = |\g_Q|^2 = 1$.  In this case we have that
\be
\hat{\sigma} = \frac{2}{3} - \left({\rm cos}\, \b\right)^4 \frac{2}{3} - \left({\rm sin}\, \b\right)^4 \frac{1}{2}\, ,
\label{sys-sigma3}
\ee
and it is positive for any value of $\b$. The choice \eqref{sys-partialmax} corresponds to F-terms of the form
\be
G_A =\left\{G_a, G_\mu \right\}=\left\{\a_K K_a,\a_Q K_\mu \right\}\, ,
\label{sys-solsfmax}
\ee
with  $\a_K, \a_Q \in \C$, the maximum value of \eqref{sys-sigma3} being attained for $\a_K = \a_Q$ or equivalently $\tan\beta=2/\sqrt{3}$. Remarkably, the explicit branches of vacua obtained in \cite{Marchesano:2019hfb} and summarized at the en the previous chapter have this F-term pattern.\footnote{More precisely, {\bf S1} vacua branches in \cite{Marchesano:2019hfb} are of the form \eqref{sys-solsfmax}. The solutions found within the branches {\bf S2} correspond to cases where the complex structure metric factorizes in two, and so their F-terms are specified in terms of a third constant $\a$. Finally, F-terms for Minkowski vacua with D6-brane moduli also have a similar structure, except that \eqref{sys-solsfmax} should be written in terms of contravariant F-terms \cite{Escobar:2018tiu}.} In the following we will explore type IIA flux vacua whose  F-terms are of the form \eqref{sys-solsfmax}, assuming that they include a significant fraction of perturbatively stable vacua. It would be interesting to extend our analysis to other possible maxima of  $\hat\sigma$ not captured by \eqref{sys-partialmax}.

\subsubsection*{An F-term Ansatz}

As it turns out, \eqref{sys-solsfmax} can be easily combined with the bilinear formalism used in the previous section.
Indeed, as pointed out in \cite{Herraez:2018vae} and summarized in the previous chapter, F-terms can be easily expressed in terms of the axion polynomials $\rho_\cA$. The expressions in \cite{Herraez:2018vae} are  generalized to the more involved flux superpotential \eqref{cy-eq: general RR superpotential expanded} and \eqref{cy-eq: general NS superpotential expanded}, obtaining that
\begin{align}
G_a =&\left[\rho_a-\mk_{ab}\tr^b_\mu u^\mu-\frac{3}{2}\frac{\mk_a}{\mk}\left(t^b\rho_b+u^\mu\rho_\mu-\frac{1}{2}\mathcal{K}_b\tilde{\rho}^b_\mu u^\mu+\frac{1}{6}\mk\tr\right)\right]\nonumber\\ +&i\left[\mk_{ab}\tr^b+\rho_{a\mu}u^\mu+\frac{3}{2}\frac{\mk_a}{\mk}\left(\rho_0-t^bu^\mu\rho_{b\mu}-\frac{1}{2}\mk_b\tr^b-\frac{1}{6}\mk \tilde{\rho}_\mu u^\mu\right)\right]\, \label{sys-eq: F-Ta}\, ,\\
G_\mu =&\left[\rho_\mu-\frac{1}{2}\mk_a\tr^a_\mu+\frac{\p_\mu K}{2}\left(t^a\rho_a+u^\nu\rho_\nu-\frac{1}{2}\mathcal{K}_b\tilde{\rho}^b_\nu u^\nu-\frac{1}{6}\mk\tr\right)\right]\nonumber \\ +&i\left(t^a\rho_{a\mu}-\frac{1}{6}\mk \tilde{\rho}_\mu-\frac{\p_\mu K}{2}\left(\rho_0-t^au^\nu\rho_{a\nu}-\frac{1}{2}\mk_b\tr^b+\frac{1}{6}\mk \tilde{\rho}_\nu u^\nu\right)\right)\, .
\label{sys-eq: F-Umu}
\end{align}

Therefore, to realize \eqref{sys-solsfmax}, one needs to impose the following on-shell conditions
\bes
\label{sys-proprho}
\begin{align}
   \rho_a-\mk_{ab}\tr^b_\mu u^\mu & = \ell_s^{-1} {\mathcal P}\, \partial_a K \label{sys-eq: f-term prop rho_a}\, ,\\
       \mathcal{K}_{ab}\tilde{\rho}^b+\rho_{a\mu}u^\mu & =  \ell_s^{-1} {\mathcal Q}\, \partial_a K \label{sys-eq: f-term prop rho^a}\, ,\\
    \rho_\mu-\frac{1}{2}\mk_a\tr^a_\mu & =  \ell_s^{-1} \cM\, \partial_\mu K\, ,\\
   t^a\rho_{a\mu}-\frac{1}{6}\mk \tilde{\rho}_\mu  & =  \ell_s^{-1}\cN\, \partial_\mu K\, , \label{sys-eq: f-term prop rhoak}
\end{align}
\ees
where ${\mathcal P}$, ${\mathcal Q}$, $\cM$, $\cN$ are real functions of the moduli. In the next section we will impose these conditions for compactifications with geometric fluxes, obtaining a simple Ansatz for the search of type IIA flux vacua.

\subsection{Moduli and flux invariants}
\label{sys-ss:invariants}

If instead of the above Ansatz we were to apply the more standard strategy of \cite{Marchesano:2019hfb}, we would compute the first and second derivatives of the potential \eqref{sys-eq:potential}, to classify its different families of extrema and determine the perturbative stability of each of them. As pointed out in \cite{Herraez:2018vae} for the Calabi--Yau case, the derivatives of the axion polynomials \eqref{sys-RRrhos} and \eqref{sys-NSrhos} are themselves combinations of axion polynomials, see Appendix \ref{ap.sys-sec: conv} for the expressions in our more general setup. As a result, all the derivatives of the potential are functions of the saxions $\{t^a, u^\mu\}$ and the $\rho_\cA$, and in particular the extrema conditions $\p V|_{\rm vac} =0$ amount to algebraic equations involving both:
\be
\left(\p_{\a} V\right) (t^a, u^\mu, \rho_\cA)|_{\rm vac} = 0 \, ,
\label{sys-extrema}
\ee
where $\a$ runs over the whole set of moduli $\{b^a, \xi^\mu, t^a, u^\mu\}$. The fact that the extrema equations depend on the quantized fluxes $q_\cA$ only through the $\rho_\cA$ is not surprising, as these are the gauge invariant quantities of the problem  \cite{Bielleman:2015ina,Carta:2016ynn}. In addition, because in our approximation the axions $\{b^a, \xi^\mu\}$ do not appear in the K\"ahler potential and in the superpotential they appear polynomially, they do not appear explicitly in \eqref{sys-extrema}, but only through the $\rho_\cA$ as well. Therefore, finding the extrema of the F-term potential amounts to solve a number of algebraic equations on $\{t^a, u^\mu, \rho_\cA\}$. 

This simplifying picture may however give the impression that the more fluxes that are present, the less constrained the system of equations is. Indeed, \eqref{sys-extrema} always amounts to $2 (1+ h^{1,1}_- + h^{2,1})$ equations, while the number of unknowns is $1 + h^{1,1}_- + h^{2,1} + n_q$, with $n_q$ the number of different $\rho$'s, which depends on the fluxes that we turn on. For Calabi--Yau with $p$-form fluxes $n_q = 3 + 2h^{1,1}_- + h^{2,1}$, while by including geometric and non-geometric fluxes we can increase it up to $n_q = 2 (2 + h^{2,1}) (1 + h^{1,1}_-)$. From this counting, it would naively seem that the more fluxes we have, the easier it is to solve the extrema equations. This is however the opposite of what is expected for flux compactifications. 

The solution to this apparent paradox is to realize that the $\rho_\cA$ are not fully independent variables, but are constrained by certain relations that appear at linear and quadratic order in them. Such relations turn out to be crucial to properly describe the different branches of vacua. In the following we will describe them for different cases in our setup.

\subsubsection*{Calabi--Yau with $p$-form fluxes}

Let us consider the case where only the fluxes $G_{2n}$, $H$ are turned on, while $f = Q = R = 0$. The moduli stabilization analysis reduces to that in \cite{Marchesano:2019hfb}, and the extrema conditions reduce to $2 h^{1,1}_- + h^{2,1} + 2$ because only one linear combination $h_\mu\xi^\mu$ of complex structure axions appears in the F-term potential. In this case the vector of axion polynomials $\rho_\cA=(\rho_0,\rho_a,\tilde{\rho}^a,\tilde{\rho},\rho_\mu)$ has $3 + 2h^{1,1}_- + h^{2,1}$ entries, but several are independent of the axions. Indeed, at the linear level 
\be 
 \tilde{\rho}=\ell_s ^{-1} m\, ,\qquad \quad    \rho_\mu=  \ell_s^{-1} h_\mu\, ,
 \label{sys-invCYl}
\ee
are axion-independent, while at the quadratic level
\be
 \tilde{\rho}\rho_{a}-\frac{1}{2}\mathcal{K}_{abc}\tilde{\rho}^b\tilde{\rho}^c\, =\, \ell_s^{-2}\left(m e_a -  \frac{1}{2}\mathcal{K}_{abc} m^bm^c\right)\, ,
  \label{sys-invCYq}
\ee
is also independent of the axions. If we fix the flux quanta $q_\cA = (e_0,  e_b,   m^b,  m,  h_\mu)$, the value of \eqref{sys-invCYl} and \eqref{sys-invCYq} will be fixed, and $\rho_\cA$ will take values in a $(1 + h^{1,1}_-)$-dimensional orbit. This orbit corresponds to the number of axions that enter the F-term potential, and so taking these constraints into account allows to see \eqref{sys-extrema} as a determined system. 

Interestingly, the quadratic invariant \eqref{sys-invCYq} was already identified in \cite{DeWolfe:2005uu} as the quantity that determines the value of the K\"ahler saxions in supersymmetric vacua of this kind. In fact, this is also true for non-supersymmetric vacua \cite{Marchesano:2019hfb}. One has that
\be
m e_a -  \frac{1}{2}\mathcal{K}_{abc} m^bm^c = \tilde{A} \CK_a\, ,
\ee
with $\tilde{A} \in \mathbb{R}$ fixed for each branch of vacua. Moreover, for the branches satisfying \eqref{sys-solsfmax}, the complex structure saxions are fixed in terms of the fluxes as $h_\mu = \hat{A} \cK \p_\mu K$, with $\hat{A}$ constant. Therefore the fluxes fix both the saxions and the allowed orbit for the $\rho_{\cA}$. Finding the latter in terms of \eqref{sys-extrema} is equivalent to finding the values of $b^a$ and $h_\mu\xi^\mu$.

\subsubsection*{Adding geometric fluxes}

Let us now turn to compactifications with fluxes $G_{2n}$, $H$, $f$, while keeping $Q = R = 0$. The number of axions $\xi^\mu$ that enter the scalar potential now corresponds to the dimension of the vector space spanned by $\langle h_\mu, f_{a\mu}\rangle$, for all possible values of $a$. If we see $f_{a\mu}$ as a $h^{1,1}_-  \times (h^{2,1}+1)$ matrix of rank $r_f$, the number of relevant entries on $\rho_\cA=(\rho_0,\rho_a,\tilde{\rho}^a,\tilde{\rho},\rho_\mu, \rho_{a\mu})$ is $2 + (2+ r_f)h^{1,1}_- + (1+ r_f)(1 +h^{2,1})- r_f^2$. At the linear level the invariants are
\be 
 \tilde{\rho}=\ell_s ^{-1} m\, ,\qquad \quad    \rho_{a\mu}=  \ell_s^{-1} f_{a\mu}\, ,
 \label{sys-invmetl}
\ee
while at the quadratic level we have
\be
 \tilde{\rho}\rho_\mu-\tilde{\rho}^a\rho_{a\mu} = \ell_s^{-2} \left(mh_\mu - m^a f_{a\mu}\right)  \, , \qquad  c^{a} \left(\tilde{\rho}\rho_{a}-\frac{1}{2}\mathcal{K}_{\bar{abc}}\tilde{\rho}^b\tilde{\rho}^c\right)\, .
 \label{sys-invmetq}
\ee
Here the $c^a \in \mathbb{Z}$ are such that $c^a \rho_{a\mu} = 0$ $\forall \mu$, so there are $h^{1,1}_- - r_f$ of this last class of invariants. Taking all these invariants into account we find that $\rho_\cA$ takes values in a $(1 + h^{1,1}_- + r_f)$-dimensional orbit,\footnote{If $d^a f_{a\mu} = h_\mu$ for some $d^a \in \R$, then the $\rho_\cA$ draw a  $(h^{1,1}_- + r_f)$-dimensional orbit, and one less axion is stabilized. As a result one can define an additional flux invariant.  See next section for an example.} signalling the number of stabilized axions. In other words, with the inclusion of metric fluxes the orbit of allowed $\rho_\cA$ increases its dimension, which implies that more moduli, in particular more axions $\xi^\mu$ are fixed by the potential. As in the CY case, the saxions are expected to be determined in terms of these invariants.

\subsubsection*{Adding non-geometric fluxes}

The same kind of pattern occurs when non-geometric fluxes are included. If one sets $R = 0$, the invariants at the linear level are $\tilde{\rho}$ and $\tilde{\rho}_\mu^a$, as well the combinations $c^a d^\mu \rho_{a\mu}$ with $c^a, d^\mu \in \mathbb{Z}$ such that $c^a d^\mu \cK_{abc}Q^c_\mu = 0$, $\forall b$. At the quadratic level, the first invariant in \eqref{sys-invmetq} is replaced by 
\be
\tilde{\rho}\rho_\mu-\tilde{\rho}^a\rho_{a\mu} + \rho_a \tilde{\rho}^a_\mu \, ,
   \label{sys-invngq}
\ee
where we have taken into account the Bianchi identity $f_{a[\mu}\, Q^a{}_{\nu]} = 0$.
Additionally, the second invariant  in \eqref{sys-invmetq} may also survive if there are choices of $c^a \in \mathbb{Z}$ such that $c^a \rho_{a\mu} \xi^\mu = 0$ $\forall \xi^\mu$.
Finally, when all kind of fluxes are nonvanishing, the only invariant at the linear level is $R_\mu$, and some particular choices of $\tilde{\rho}^a_\mu$ and $\rho_{a\mu}$. At the quadratic level we have the generalization of \eqref{sys-invngq}
\be
\tilde{\rho}\rho_\mu-\tilde{\rho}^a\rho_{a\mu} + \rho_a \tilde{\rho}^a_\mu -\rho_0\tilde{\rho}_\mu\, ,
   \label{sys-invngr}
\ee
where we have imposed the Bianchi identity $\rho_{[\mu} \, \tilde{\rho}_{\nu]} - \rho_{a[\mu}\, \tilde{\rho}^a{}_{\nu]} = h_{[\mu} \, R_{\nu]} - f_{a[\mu}\, Q^a{}_{\nu]} = 0$, see Appendix \ref{ap.sys-sec: conv}. Notice that this invariant and its simpler versions are nothing but  the D6-brane tadpole \eqref{sys-DFtadpole} induced by fluxes. We also have the new invariants
\be
\tilde{\rho}_{[\mu}^a \tilde{\rho}_{\nu]}\, ,\qquad \qquad   \rho_{a(\mu}\tilde{\rho}_{\nu)}-\mathcal{K}_{abc}\tilde{\rho}^b_\mu\tilde{\rho}^c_\nu\, ,
\label{sys-finalNGinv}
\ee
where as above $(\ )$ and $[\ ]$ stand for symmetrisation and anti-symmetrisation of indices, respectively. Finally, if the second invariant in \eqref{sys-finalNGinv} vanishes, or in other words if we have $f_{a(\mu}Q_{\nu)}=\mathcal{K}_{abc}Q^b_\mu Q^c_\nu$, then
\be
\rho_{a(\mu}\tilde{\rho}_{\nu)}^a - 3 \rho_{(\mu} \tilde{\rho}_{\nu)}\, ,
\label{sys-ffinalNGinv}
\ee
is also an invariant.\footnote{Remarkably, both \eqref{sys-ffinalNGinv} and the second invariant in \eqref{sys-finalNGinv} vanish if the ``missing" Bianchi identities $f_{a(\mu}Q_{\nu)}=\mathcal{K}_{abc}Q^b_\mu Q^c_\nu$ and $f_{a(\mu}Q_{\nu)}^a = 3 h_{(\mu} R_{\nu)}$ proposed in \cite{Gao:2018ayp} turn out to hold generally.}


\section{Geometric flux vacua}
\label{sys-sec: geovacua}

In this section we would like to apply our previous results to the search of vacua in type IIA flux compactifications. For concreteness, we focus on those configurations with $p$-form and geometric fluxes only, leaving the systematic search of non-geometric flux vacua for the future. Then, as we will see,  the Ansatz formulated in the last section, which amounts to impose on-shell F-terms of the form \eqref{sys-solsfmax},  
forbids de Sitter solutions. In contrast, we find six branches of AdS extrema corresponding to our Ansatz, coming in mixed pairs of  supersymmetric and non supersymmetric vacua. Out of the six, two of them, which could be considered as the most generic branches of the initial Ansatz and are associated to nearly-Kähler geometries,  will be the focus of this chapter. The perturbative stability of the non-supersymmetric branch of this first pair will be analyzed in the next section.  The other branches correspond to half-flat compact geometries that were not considered in the original paper \cite{Marchesano:2020uqz}. We will briefly mention their properties but leave their detailed analysis for a later work \cite{systematicsfollowup}. 

\subsection{The geometric flux potential}

Let us first of all summarize our previous results and restrict them to the case of $p$-form and geometric fluxes. The scalar potential reads $V = V_F + V_D$, with
\begin{align}
   \kappa_4^2 V_F =\, &e^K\left[4\rho_0^2+g^{ab}\rho_a\rho_b+\frac{4\mathcal{K}^2}{9}g_{ab}\tilde{\rho}^a\tilde{\rho}^b+\frac{\mathcal{K}^2}{9}\tilde{\rho}^2\right.\nonumber\\
    + &\left.c^{\mu\nu}\rho_\mu\rho_\nu+\left(\tilde{c}^{\mu\nu}t^at^b+g^{ab}u^\mu u^\nu \right)\rho_{a\mu}\rho_{b\nu}-\frac{4\mathcal{K}}{3}u^\nu \tilde{\rho}^a\rho_{a\nu}+\frac{4\mathcal{K}}{3}u^\nu \tilde{\rho}\rho_\nu\right]\, ,
    \label{sys-eq:potentialgeom}\\
\kappa_4^2 V_D = \, &\frac{3}{8\mathcal{K}} \partial_\mu K \partial_\nu K \,  g^{\alpha\beta} \, \hat{\rho}_\alpha{}^\mu \hat{\rho}_\beta{}^\nu\, .\label{sys-eq: D-potentialgeom}
\end{align}
The definitions for $g^{ab}$,  ${c}^{\mu\nu}$, $\tilde{c}^{\mu\nu}$ and $g^{\alpha\beta}$ are just as in section \ref{sys-sec: IIAorientifold}, while the $\rho_\cA$ simplify to
\bes
\label{sys-RRrhosgeom}
\begin{align}
  \ell_s  \rho_0&=e_0+e_ab^a+\frac{1}{2}\mathcal{K}_{abc}m^ab^bb^c+\frac{m}{6}\mathcal{K}_{abc}b^ab^bb^c+\ell_s\rho_\mu\xi^\mu\, , \label{sys-eq:rho0g}\\
 \ell_s   \rho_a&=e_a+\mathcal{K}_{abc}m^bb^c+\frac{m}{2}\mathcal{K}_{abc}b^bb^c+\ell_s\rho_{a\mu}\xi^\mu \, ,  \label{sys-eq:rho_ag}\\
  \ell_s  \tilde{\rho}^a&=m^a+m b^a \, ,  \label{sys-eq:rho^ag}\\
 \ell_s   \tilde{\rho}&=m \, ,   \label{sys-eq:rhomg}\\
 \ell_s    \rho_\mu&=h_\mu+f_{a\mu}b^a \, , \\
 \ell_s   \rho_{a\mu}&=f_{a\mu} \, ,  \label{sys-eq: ho_ak} \\
 \ell_s   \hat{\rho}_{\a}^{\mu}&=\hat{f}_{\a}^{\mu} \, . 
\end{align}   
\ees 

Using these explicit expressions one may compute the first order derivatives of the scalar potential with respect to the axions $\{\xi^\mu, b^a\}$ and saxions $\{u^\mu, t^a\}$ of the compactification. As expected the extrema conditions are of the form \eqref{sys-extrema}. In Planck units this amounts to:

\vspace*{.5cm}

\textbf{Axionic directions}

\bes
\label{sys-paxions}
\begin{equation}
\label{sys-paxioncpx}
  e^{-K}\frac{\partial V}{\partial \xi^\mu}=8\rho_0\rho_\mu +2g^{ab}\rho_a\rho_{b\mu} \, ,
\end{equation}
\begin{equation}
\label{sys-paxionk}
 e^{-K}\frac{\partial V}{\partial b^a}= \ 8\rho_0\rho_a+\frac{8}{9}\mathcal{K}^2g_{ac}\tilde{\rho}\tilde{\rho}^c  +2\mathcal{K}_{abd}g^{bc}\rho_c\tilde{\rho}^d+2c^{\mu\nu}\rho_{a\mu}\rho_\nu\, ,
\end{equation}
\ees

\vspace*{.5cm}

\textbf{Saxionic directions}

\bes
\label{sys-psaxions}
\begin{eqnarray}
\label{sys-psaxioncpx}
 e^{-K}  \frac{\partial V}{\partial u^\mu} & =  & e^{-K} V_F\partial_\mu K+\frac{4}{3}\mathcal{K}\tilde{\rho} \rho_\mu +\partial_\mu c^{\kappa\sigma}\rho_\kappa\rho_\sigma - \frac{4}{3}\mathcal{K}\tilde{\rho}^a\rho_{a\mu} +2g^{ab}\rho_{a\mu}\rho_{b\nu}u^\nu\\ \nonumber
    & & + t^at^b(\partial_\mu c^{\kappa\sigma}\rho_{a\kappa}\rho_{b\sigma}-8\rho_{a\mu}\rho_{b\nu}u^\nu )
    + \frac{3}{4\mathcal{K}}e^{-K}\partial_\mu \partial_\sigma K \partial_\nu K \,  g^{\alpha\beta} \, \hat{\rho}_\alpha{}^\sigma \hat{\rho}_\beta{}^\nu\, ,
\end{eqnarray}
\begin{eqnarray}
\label{sys-psaxionk}\nonumber
e^{-K} \frac{\partial V}{\partial t^a} &= & e^{-K} V_F\partial_{a}K+\partial_{a}\left(\frac{4}{9}\mathcal{K}^2\tilde{\rho}^b\tilde{\rho}^c
 g_{bc}\right)+\partial_{a}g^{cd}\rho_c\rho_d+\mathcal{K}_a\tilde{\rho}\left(\frac{2}{3}\mathcal{K}\tilde{\rho}+4u^\mu{\rho}_\mu \right)\\
 & & - 4 \mathcal{K}_a \tilde{\rho}^b\rho_{b\nu}u^\nu +2\tilde{c}^{\mu\nu}t^c\rho_{a\mu}\rho_{c\nu}+\partial_a g^{bc}\rho_{b\mu}u^\mu \rho_{c\nu}u^\nu\nonumber\\  & & +\frac{3}{8\mathcal{K}}e^{-K} \partial_\mu K \partial_\nu K \,  \p_a g^{\alpha\beta} \, \hat{\rho}_\alpha{}^\mu \hat{\rho}_\beta{}^\nu\,-\frac{9\mathcal{K}_a}{8\mathcal{K}^2}e^{-K} \partial_\mu K \partial_\nu K \,   g^{\alpha\beta} \, \hat{\rho}_\alpha{}^\mu\hat{\rho}_\beta^\nu\, .
\end{eqnarray}
\ees

\subsection{de Sitter no-go results revisited}
\label{sys-subsec: no-go's}

From \eqref{sys-psaxions} one can obtain the following off-shell relation
\begin{align}
\nonumber
    & u^\mu \partial_{u^\mu} V +x \, t^a \partial_{t^a} V=- (4+3x)V_F - (2+x) V_D + 4e^K\left[x\left(\frac{1}{2}g^{bc}\rho_b\rho_c+\frac{4\mathcal{K}^2}{9}g_{bc}\tilde{\rho}^b \tilde{\rho}^c +\frac{\mathcal{K}^2}{6}\tilde{\rho}^2\right) \right. \\
    & +\left. \frac{1}{2}c^{\mu\nu}\rho_\mu\rho_\nu  +\left(\frac{1}{3}+x\right)\mathcal{K}u^\nu \left(\tilde{\rho}\rho_\nu -\tilde{\rho}^b\rho_{b\nu}\right)  +\frac{1}{2}(1+x)(\tilde{c}^{\mu\nu}t^bt^c+g^{bc}u^\mu u^\nu )\rho_{b\mu}\rho_{c\nu}\right]\, ,
   \label{sys-eq: arbitrary combination of partial derivatives}
\end{align}
with $x \in \R$ an arbitrary parameter. Different choices of $x$ will lead to different equalities by which one may try to constrain the presence of extrema with positive energy, in the spirit of  \cite{Hertzberg:2007wc,Flauger:2008ad}. In practice it is useful to rewrite this relation as
\begin{equation}
    u^\mu \partial_{u^\mu} V +x t^a \partial_{t^a} V = - 3V + \Xi_x\, ,
\end{equation}
where, for instance, the choice $x=1/3$ leads to 
\begin{equation}
    \label{sys-eq: cosmo const relation}
   \Xi_{1/3} = \frac{2}{3} V_D  +  4e^K\left[-2\rho_0^2-\frac{1}{3}g^{bc}\rho_b\rho_c-\frac{2}{27}\tilde{\rho}^b\tilde{\rho}^c\mathcal{K}^2g_ {bc}+\frac{1}{6}(t^at^b\tilde{c}^{\mu\nu}+g^{ab}u^\mu u^\nu )\rho_{a\mu}\rho_{b\nu}\right] ,
\end{equation}
while the choice $x=1$ gives
\begin{equation}
 \Xi_1 = 4e^K\left[\frac{\mathcal{K}^2}{18}\tilde{\rho}^2-4\rho_0^2-\frac{1}{2}g^{ab}\rho_a\rho_b-\frac{1}{2}c^{\mu\nu}\rho_\mu\rho_\nu\right]\, .
    \label{sys-eq: cosmo cons relation 2}
\end{equation}
Extrema of positive energy require $\p V =0$ and $V >0$, and so necessarily both \eqref{sys-eq: cosmo const relation} and \eqref{sys-eq: cosmo cons relation 2} should be positive. It is easy to see that this requires that both the Romans' parameter $\tilde{\rho}$ and geometric fluxes (either $\rho_{a\mu}$ or $\hat{\rho}_\alpha^\mu$)  are present, in agreement with previous results in the literature \cite{Haque:2008jz,Caviezel:2008tf,Flauger:2008ad,Danielsson:2009ff,Danielsson:2010bc,Danielsson:2011au}. In that case, it is unlikely that the potential satisfies an off-shell inequality of the form proposed in \cite{Obied:2018sgi}, at least at the classical level. 

In our formulation one can make more precise which kind of fluxes are necessary to attain de Sitter extrema. For this, let us express the last term of \eqref{sys-eq: cosmo const relation} as
\be
(t^at^bc^{\mu\nu}+g^{ab}u^\mu u^\nu -4t^at^bu^\mu u^\nu )\rho_{a\mu}\rho_{a\nu} = \left[t^at^bc_{\rm P}^{\mu\nu}+u^\mu u^\nu g_{\rm P}^{ab} -\frac{5}{3} t^at^bu^\mu u^\nu \right]\rho_{a\mu}\rho_{a\nu}\, ,
\label{sys-ccgeomterm}
\ee
where $g_{\rm P}^{ab}$, $c_{\rm P}^{\mu\nu}$ are the primitive components of the K\"ahler and complex structure metric, respectively. That is 
\be
\label{sys-eq: primitive metric}
g_{\rm P}^{ab} = \frac{2}{3}\left(t^at^b - \mathcal{K}\mathcal{K}^{ab}\right)\, , \qquad \qquad c_{\rm P}^{\mu\nu} = \frac{1}{3}u^\mu u^\nu - 4G_Q G_Q^{\mu\nu}\, ,
\ee
where $G_Q = e^{-K_Q}$ and $G_Q^{\mu\nu}$ is the inverse of $\p_\mu \p_\nu G_Q$. These metric components have the property that they project out the K\"ahler potential derivatives along the overall volume and dilaton directions, namely $g_{\rm P}^{ab} \p_b K = c_{\rm P}^{\mu\nu} \p_\nu K =0$. So in order for the bracket in \eqref{sys-eq: cosmo const relation} to be positive, the geometric fluxes $\rho_{a\mu}$ not only must be non-vanishing, but they must also be such that
\be
t^a \rho_{a\mu}\, t^b \rho_{a\nu}\, c_{\rm P}^{\mu\nu}+  \rho_{a\mu} u^\mu \, \rho_{a\nu} u^\nu\, g_{\rm P}^{ab} \neq 0\, .
\label{sys-primcond}
\ee
In other words, either the vector $\rho_{a\mu} u^\mu$ is not proportional to $\p_a K$ or the vector $t^a \rho_{a\mu}$ is not proportional to $\p_\nu K$. The condition is likely to be satisfied at some point in field space, but in order to allow for a de Sitter extremum it must be satisfied on-shell as well. 

Remarkably, we find that the F-term Ansatz of section \ref{sys-ss:fterms} forbids de Sitter extrema. Indeed, if we impose that the on-shell relations \eqref{sys-proprho} are satisfied with the non-geometric fluxes turned off (cf. \eqref{sys-proprhog} below) we obtain that, on-shell
\be
t^a \rho_{a\mu}\, t^b \rho_{a\nu}\, c_{\rm P}^{\mu\nu}+  \rho_{a\mu} u^\mu \, \rho_{a\nu} u^\nu\, g_{\rm P}^{ab} = \frac{4}{9} \cK^2 g^{\rm P}_{ab} \tilde{\rho}^a \tilde{\rho}^b \, ,
\label{sys-primvalue}
\ee
with $g^{\rm P}_{ab}$ the inverse of $g_{\rm P}^{ab}$ in the primitive sector. Even if this term is positive, it can never be bigger than the other negative contributions within the bracket in \eqref{sys-eq: cosmo const relation}. In fact, after plugging \eqref{sys-primvalue} in \eqref{sys-eq: cosmo const relation} there is a partial cancellation between the third and fourth term of the bracket, that then becomes semidefinite negative:
\be
4e^K\left[-2\rho_0^2-\frac{1}{3}g^{ab}\rho_a\rho_b-\frac{2}{27}\tilde{\rho}^a\tilde{\rho}^b\mathcal{K}^2g_ {ab}^{\rm NP}-\frac{5}{18}t^at^bu^\mu u^\nu \rho_{a\mu}\rho_{b\nu}\right]\, ,
\ee
with $g_ {ab}^{\rm NP} = g_{ab} -  g^{\rm P}_{ab} = \frac{3}{4}\frac{\cK_a \cK_b}{\cK^2} $ the non-primitive component of the K\"ahler moduli metric.

Even if the bracket in \eqref{sys-eq: cosmo const relation} is definite negative,  there is still the contribution from the piece $\frac{2}{3}V_D$, which is positive semidefinite. However, one can see that with the Ansatz \eqref{sys-solsfmax} this contribution vanishes. Indeed, using the Bianchi identity $f_{a\mu} \hat{f}_\alpha{}^\mu = 0$ and \eqref{sys-eq: f-term prop rhoak geom}, or alternatively $h_\mu \hat{f}_\alpha{}^\mu = 0$ and \eqref{sys-eq: f-term prop rhomu geom}. one can see that the D-term $D_\alpha =\frac{1}{2} \partial_\mu K \, \hat f_\alpha{}^\mu$ vanishes, and so does $V_D$.

To sum up, for type IIA geometric flux configurations, in any region of field space in which the F-terms are of the form \eqref{sys-solsfmax} we have that the F-term potential satisfies
\be
u^\mu \partial_{u^\mu} V+\frac{1}{3}t^a \partial_{t^a} V \leq -3V\, , 
\label{sys-eq: no-go geom inequality}
\ee
and so de Sitter extrema are excluded. In other words:
\begin{center}
{\em In type IIA geometric flux compactifications, classical  de Sitter extrema \\  are incompatible with F-terms of the form \eqref{sys-solsfmax}.}
\end{center}
In section \ref{sys-sec: 10d} we will interpret this result from a geometrical viewpoint. 
It would be interesting to extend this discussion to non-geometric flux compactifications, along the lines of \cite{deCarlos:2009fq,Shukla:2019dqd}, to see if this result applies there as well.

\subsection{Imposing the Ansatz}\label{sys-imposing}

Besides the cosmological constant sign, let us see other constraints that the on-shell condition \eqref{sys-solsfmax} leads to. By switching off all non-geometric fluxes, \eqref{sys-proprho} simplifies to
\bes
\label{sys-proprhog}
\begin{align}
   \rho_a & = \ell_s^{-1} {\mathcal P}\, \partial_a K \label{sys-eq: f-term prop rho_a geom}\, ,\\
       \mathcal{K}_{ab}\tilde{\rho}^b+\rho_{a\mu}u^\mu & =  \ell_s^{-1} {\mathcal Q}\, \partial_a K \label{sys-eq: f-term prop rho^a geom}\, ,\\
    \rho_\mu & =  \ell_s^{-1}\cM\, \partial_\mu K\, ,  \label{sys-eq: f-term prop rhomu geom} \\
   t^a\rho_{a\mu } & =  \ell_s^{-1} \cN\, \partial_\mu K\, , \label{sys-eq: f-term prop rhoak geom}
\end{align}
\ees
where again ${\mathcal P}$, ${\mathcal Q}$, $\cM$, $\cN$ are real functions of the moduli. Such functions and other aspects of this Ansatz are constrained by the extrema conditions \eqref{sys-paxions} and \eqref{sys-psaxions} with which they must be compatible. Indeed, plugging \eqref{sys-proprhog} into \eqref{sys-paxions} and \eqref{sys-psaxions} one obtains  
\bes
\label{sys-paxionsA}
\begin{equation}
\label{sys-paxioncpxA}
8 \left(\rho_0\cM -  {\mathcal P}\cN\right) \partial_\mu K = 0 \, ,
\end{equation}
\begin{equation}
\label{sys-paxionkA}
\left[ 8  {\mathcal P} (\rho_0 -  {\mathcal Q}) - \frac{1}{3} \tilde{\rho}  \cK \left(-2 {\mathcal Q} + 8 \cN \right)   \right]  \partial_a K   + \left[ \frac{4}{3} \CK\tilde{\rho} + 8  {\mathcal P} - 8 \cM \right] \rho_{a\mu} u^{\mu} = 0 \, ,
\end{equation}
\ees
\bes
\label{eq: psaxionsA}
\begin{equation}
\label{eq: psaxioncpxA}
\left(4\rho_0^2+12\mathcal{P}^3+3\mathcal{Q}^2+8\mathcal{M}^2+8\mathcal{N}^2+\frac{\cK^2}{9}\tilde{\rho}^2-20\mathcal{QN}-4\mathcal{M}\cK\tilde{\rho}\right) \partial_\mu K = 0 \, ,
\end{equation}
\begin{equation}
\label{eq: psaxionkA}
\left[4\rho_0^2+4\mathcal{P}^2-\mathcal{Q}^2+24\mathcal{QN}+16\mathcal{M}^2-\frac{\cK^2}{9}\tilde{\rho}^2\right]\partial_a K   + \left[ \frac{8}{3} \mathcal{Q} -56  {\mathcal N}  \right] \rho_{a\mu} u^{\mu} = 0 \, .
\end{equation}
\ees
which must be satisfied on-shell. For generic choices of flux quanta we do not expect both pairs of brackets in \eqref{sys-paxionkA} and \eqref{eq: psaxionkA} to vanish independently and thus
\be
\rho_{a\mu} u^{\mu} \propto \p_a K\, , \qquad {\rm and} \qquad \tilde{\rho}^a \propto t^a\, ,
\label{sys-eq: ansatz refinement relation}
\ee
simplifying the Ansatz. In the main body of the chapter we will restrict to this choice, which provides the two generic branches of solutions mentioned at the beginning of the section. One may wonder, however,  if there are non-trivial solutions when both brackets in \eqref{sys-paxionsA} vanish independently. The answer is affirmative and in fact requiring supersymmetry implies such property. Nevertheless, SUSY vacua is still compatible with \eqref{sys-eq: ansatz refinement relation} even if it is not the most general scenario. The cases where \eqref{sys-eq: ansatz refinement relation} is not satisfied give rise to four new branches (one supersymmetric and one non-supersymmetric) that have some phenomenologically interesting properties. These more exotic branches have been explored in the context of toroidal orientifold compactifications in \cite{Cribiori:2021djm}, where solutions displaying scale separation were found. We will briefly comment on this topic at the end of this section.

Going back to the case in which  \eqref{sys-eq: ansatz refinement relation} holds,  we are led to the following on-shell relations
\begingroup
\allowdisplaybreaks
\bes
\label{sys-Ansatz}
\begin{align}
   \ell_s \rho_0&= A \CK\, , \label{sys-eq: ans rho0}\\
    \ell_s\rho_a&= B \cK \partial_a K  \, ,  \label{sys-eq: ans rho_a}\\
    \ell_s\tilde{\rho}^a&= C t^a  \, ,  \label{sys-eq: ans rho^a}\\
    \ell_s\tilde{\rho}&= D \, ,  \label{sys-eq: ans rhotilde}\\
    \ell_s\rho_\mu&=E\cK \partial_\mu K  \, ,   \label{sys-eq: nsns}\\
    \ell_s\rho_{a\mu} t^a &= \frac{F}{4}  \cK\partial_\mu K  \, ,  \label{sys-eq: geoma}\\
     \ell_s\rho_{a\mu} u^\mu &= \frac{F}{3}  \CK \partial_a K   \, , \label{sys-eq: geomu}
\end{align}
\ees
\endgroup
where $A, B, C, D, E, F$ are functions of the saxions. We have extracted a factor of $\CK$ in some of them so that the expression for the on-shell equations simplifies. These coefficients generalize the triplet $(\tilde{A},\tilde{B},\tilde{C})$ in \eqref{cy-eq: rho solutions} to a far richer structure that among other cases includes $\tilde{\rho}=0$. The map is given by 
\begin{equation}
    \tilde{A}=\frac{E}{\tilde{\rho}}\,,\qquad \tilde{B}=\frac{C}{\tilde{\rho}}\,,\qquad \tilde{C}=-\frac{3B}{\tilde{\rho}}\,.
\end{equation}
In terms of \eqref{sys-Ansatz} we have that the vanishing of \eqref{sys-paxions} amounts to
\bes
\label{sys-paxionsAA}
\begin{gather}
\label{sys-paxioncpxAA}
 4 AE -  BF =  0 \, ,\\
\label{sys-paxionkAA}
 3AB  - \frac{1}{12} CD  + B C  -  EF = 0 \, ,
\end{gather}
\ees
assuming that at each vacuum $\p_\mu K \neq 0 \neq \p_a K$. Similarly, the vanishing of \eqref{sys-psaxions} implies
\bes
\label{sys-psaxionsAA}
\begin{gather}
\label{sys-psaxioncpxAA}
 4A^2 + 12B^2 +\frac{1}{3} C^2 + \frac{1}{9}D^2 + 8 E^2 -\frac{5}{6} F^2 +CF -4DE = 0\, ,\\
\label{sys-psaxionkAA}
4A^2 + 4 B^2 - \frac{1}{9} C^2 - \frac{1}{9}D^2 + 16E^2 -\frac{5}{9}F^2  = 0\, ,
\end{gather}
\ees
where we have used the identities in \cite[Appendix A]{Marchesano:2019hfb}.

Expressing the extrema equations in terms of the Ansatz \eqref{sys-Ansatz} has the advantage that we recover a system of algebraic equations. Nevertheless, eqs.\eqref{sys-paxionsAA} and \eqref{sys-psaxionsAA} may give the wrong impression that we have an underdetermined system, with four equations and six unknowns $A, B, C, D, E, F$. Notice, however, that these unknowns are not all independent, and that relations among them arise when the flux quanta are fixed. Indeed, let us first consider the case without geometric fluxes, which sets $F=0$. In this case, AdS vacua require that the Roman's parameter $m$ is non-vanishing so we may assume that $D \neq 0$. Because the LHS of \eqref{sys-paxionsAA} and \eqref{sys-psaxionsAA}  are  homogeneous polynomials of degree two, we may divide each of them by $D^2$ to obtain four equations on four variables: $A_D = A/D$, $B_D = B/D$, $C_D = C/D$, $E_D = E/D$. The solutions correspond to $A_D= 0$ and several rational values for $B_D, C_D, E_D$, which reproduce the different {\bf S1} branches found in \cite{Marchesano:2019hfb}.\footnote{To compare to \cite{Marchesano:2019hfb} one needs to use the  dictionary: $B_D = - C_{\rm MQ}/3$, $C_D = B_{\rm MQ}$, $E_D = A_{\rm MQ}$.} Finally, the variable $D = m$ is fixed when the flux quanta are specified. 

The analysis is slightly more involved in the presence of geometric fluxes. Now we may assume that $F \neq 0$, since otherwise we are back to the previous case. Our Ansatz implies that the first flux invariant in \eqref{sys-invmetq} is a linear combination of the vectors $(f_a)_\mu = f_{a\mu}$, as
\be
m\hat{h}_\mu \equiv m h_\mu - m^a f_{a\mu} = \left(DE - \frac{CF}{4}\right)  \cK \p_\mu K  =   \left( \frac{4DE}{F}  - C \right)  t^a  f_{a\mu} \, ,
\label{sys-hath}
\ee
where $\cK$, $ \p_\mu K$, $ t^a$ correspond to the value of the K\"ahler saxions in the corresponding extremum, etc.  One can write the above relation as
\be
m\hat{h}_\mu = d^a f_{a\mu}\, ,
\label{sys-condhhat}
\ee
where the constants $d^a$ are fixed once that we specify the fluxes $m$, $h_\mu$, $m^a$, $f_{a\mu}$. As a consequence, the number of stabilized complex structure axions $\xi^\mu$ is $r_f = {\rm rank}\, f_{a\mu}$, while the rest may participate in St\"uckelberg mechanisms triggered by the presence of D6-branes \cite{Camara:2005dc}.\footnote{Microscopically, \eqref{sys-condhhat} means that $h_\mu$ is in the image of the matrix of geometric fluxes $f_{a\mu}$, and as such it is cohomologically trivial. Macroscopically, it means that the number of independent complex structure axions entering the scalar potential are $ {\rm dim} \langle h_\mu, f_{1\mu}, f_{2\mu}, \dots \rangle= {\rm rank} f_{a\mu} \equiv r_f $, and not $r_f+1$.} Strictly speaking, $d^a$ is only fixed up to an element in the kernel of $f_{a\mu}$, but this is irrelevant for our purposes. Indeed, notice that due to our Ansatz
\bea\nonumber
m\hat{e}_a \equiv  me_a -  \frac{1}{2}\mathcal{K}_{abc} m^bm^c  &= & \left(BD + \frac{C^2}{6} \right)  \cK \p_a K   - m f_{a\mu} \xi^\mu \\ 
 & = & \left[\left(\frac{3BD}{F} + \frac{C^2}{2F} \right)  u^\mu   -  D \xi^\mu  \right] f_{a\mu}  \, ,
 \label{sys-hatea}
\eea
where again $\cK$, $u^\mu$, $\xi^\mu$ stand for the vevs at each extremum. This implies several things. First, the second set of invariants in \eqref{sys-invmetq} vanishes identically. Second, the combination $md^a \hat{e}_a$ is fully specified by the flux quanta, without any ambiguity.  Finally in terms of 
\begin{equation} 
m^2 \hat{e}_0 \equiv m^2 e_0 - m m^a e_a  + \frac{1}{3} \mathcal{K}_{abc}m^am^bm^c  \, ,
\label{sys-hate0}
\end{equation} 
we can define the following cubic flux invariant
\be
m^2 \hat{e}_0 -  m d^a  \hat{e}_a  =  \cK  \left[AD^2  + 3BCD + \frac{C^3}{3} + \left( \frac{4DE}{F}  - C\right) \left(3BD + \frac{C^2}{2} \right)  \right] \, .
\label{sys-extrainv}
\ee
The existence of this additional invariant is expected from the discussion of section \ref{sys-ss:invariants}. As we now show, $\cK$ is fixed at each extremum by the choice of the flux quanta and the Ansatz' variables. Therefore \eqref{sys-extrainv} and $D=m$ provide two extra constraints on these variables, which together with \eqref{sys-paxionsAA} and \eqref{sys-psaxionsAA}  yield a determined system of algebraic equations. 

To show how  $\cK$ is specified, let us first see how the saxionic moduli are determined. First \eqref{sys-hath} determines $(4DE-CF) \cK  \partial_\mu K $ in terms of the flux quanta, which is equivalent to determine $(4DE-CF)^{-1}  u^\mu/\cK$. Plugging this value into \eqref{sys-eq: geomu} one fixes $(4DE/F-C)^{-1}  \partial_a K$ in terms of the fluxes, which is equivalent to fix $(4DE/F-C)  t^a$. Therefore at each extremum we have that
\be
\left(\frac{4DE}{F}-C\right)^3  \cK  \, ,
\label{sys-vevK}
\ee
is specified by the flux quanta. Notice that this is compatible with \eqref{sys-hath}, and we can actually use this result to fix the definition of $d^a$, by equating \eqref{sys-vevK} with $\cK_{abc}d^ad^bd^c$.

\subsection{Branches of vacua}\label{sys-branchvacu}

Let us analyze the different solutions to the algebraic equations \eqref{sys-paxionsAA} and \eqref{sys-psaxionsAA}. 
Following the strategy of the previous subsection, we assume that $F \neq 0$ and define $A_F = A/F$, $B_F = B/F$, $C_F = C/F$, $D_F = D/F$, $E_F = E/F$. Then, from \eqref{sys-paxioncpxAA} we obtain
\begin{equation}
    B_F=4A_FE_F\, ,
    \label{sys-eq: B_F}
\end{equation}
which substituted into \eqref{sys-paxionkAA} gives the following relation
\begin{equation}
  C_FD_F= 12 E_F(12A_F^2+4A_FC_F-1)\, .
    \label{sys-eq: C_FD_F}
\end{equation}
Then, multiplying \eqref{sys-psaxionkAA} by $C_F^2$ and using \eqref{sys-eq: C_FD_F} we obtain
\be
144 E_F^2 \Delta_F = C_F^2\left[36 A_F^2 - C_F^2 - 5 \right]  \, ,
  \label{sys-eq: E_F Delta_F}
\ee
where
\be
\Delta_F =(12A_F^2+4A_FC_F-1)^2 - 4 A_F^2 C_F^2 - C_F^2\, .
   \label{sys-eq: DeltaF def}
\ee
We have two possibilities, depending on whether $\Delta_F = 0$ or not. Let us consider both:

\begin{itemize}

\item $\Delta_F = 0$

In this case, from \eqref{sys-eq: E_F Delta_F} and \eqref{sys-eq: DeltaF def}, we find four different real solutions for $(A_F, C_F)$:
\bes
\label{sys-delta=0sols}
   \begin{align}
   \label{sys-SUSYsol}
            A_F=-\frac{3}{8}\, , \ \ \ & \ \ \ C_F=\frac{1}{4}\, ,\\
            A_F=\frac{3}{8}\, , \ \ \ & \ \ \ C_F=-\frac{1}{4}\, ,
            \label{sys-nonSUSYdelta=0}\\
            A_F=\pm \frac{1}{2\sqrt{3}}\, , \ \ \ & \ \  \ C_F= 0 \, .
            \label{sys-nonSUSYdelta=0C=0}
        \end{align}
\ees
Given the solution \eqref{sys-SUSYsol}, one can solve for $D_F$ in \eqref{sys-eq: C_FD_F} and check that \eqref{sys-psaxioncpxAA} and \eqref{sys-psaxionkAA} are automatically satisfied. We then find that:
\be
\eqref{sys-SUSYsol} \ \raw \ B_F = -\frac{3}{2} E_F\, , \qquad D_F = 15E_F\, ,
\label{sys-SUSYbranch}
\ee
with $E_F$ unfixed. Thus, at this level $(E,F)$ are free parameters of the solution. As we will see below, this case corresponds to the supersymmetric branch of solutions. The remaining solutions can be seen as limiting cases of the following possibility:

\item $\Delta_F \neq 0$

Under this assumption we can solve for $E_F$ in \eqref{sys-eq: E_F Delta_F}:
     \begin{equation}
            E_F^2=\frac{C_F^2}{144\Delta_F} \left[36 A_F^2 - C_F^2 - 5 \right] 
            \label{sys-eq: E_F delta neq0}
        \end{equation}
Then we see that \eqref{sys-psaxioncpxAA} and \eqref{sys-psaxionkAA} amount to solve the following relation:
      \begin{align}
       &\frac{8A_F^{2}C_F^{4}}{3}+4A_FC_F^{4}-\frac{7C_F^{4}}{6}+64A_F^{3}C_F^{3}+48A_F^{2}\,C_F^{3}-\frac{16A_FC_F^{3}}{3}-4C_F^{3} +576A_F^{4}C_F^{2}
       \nonumber\\ \nonumber 
           & +144A_F^{3}C_F^{2}-\frac{296A_F^{2}C_F^{2}}{3}-4A_FC_F^{2}+\frac{7C_F^{2}}{3}+2304A_F^{5}C_F -592A_F^{3}C_F+24A_F^{2}C_F\\
          & +\frac{100A_FC_F}{3}-2C_F+3456A_F^{6}-1176A_F^{4}+124A_F^{2}-\frac{25}{6} = 0\, ,
         \label{sys-eq: megaeqF}
        \end{align}
which selects a one-dimensional family of solutions in the $(A_F,C_F)$-plane. We only consider those such that  \eqref{sys-eq: E_F delta neq0} is non-negative, see figure \ref{sys-fig: generalsol}.
One can check that all values in \eqref{sys-delta=0sols} are also solutions of \eqref{sys-eq: megaeqF}. Even if for them $\Delta_F =0$, we have that 
\be
D_F^2 = \left(1+ \frac{C_F^2(4A_F^2+1)}{\Delta_F}\right) \left[ 36A_F^2 -C_F^2-5\right]\, ,
 \label{sys-eq: D_F delta neq0}
\ee
as well as \eqref{sys-eq: E_F delta neq0}, attain regular limiting values that solve the equations of motion. Because \eqref{sys-eq: megaeqF} constrains one parameter in terms of the other, we have two free parameters, say $(C,F)$, unfixed by the equations \eqref{sys-paxionsAA} and \eqref{sys-psaxionsAA}. 
\end{itemize}

\begin{center}    
\begin{figure}[H]
    \centering
    \includegraphics[width=0.6\textwidth]{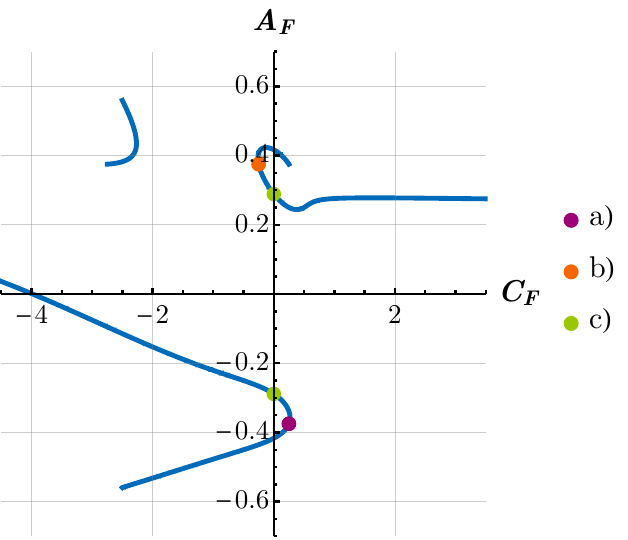}
    \caption{Set of points that verify \eqref{sys-eq: megaeqF} (blue curve) and have $E_F^2\geq 0$. The coloured dots correspond to the particular solutions \eqref{sys-delta=0sols}. Both curves tend asymptotically to $A_F=1/4$ for $C_F\rightarrow\pm \infty$. }
    \label{sys-fig: generalsol}
\end{figure}
\end{center}

\subsection{Full Picture}\label{sys-sec: summary}

Let us summarize our results so far. Given the on-shell F-terms \eqref{sys-solsfmax} and the restriction \eqref{sys-eq: ansatz refinement relation}, we find two branches of vacua, summarized in table \ref{sys-vacuresul}. Naively, each branch seems to contain two continuous parameters. However, after choosing a specific set of flux quanta, two extra constraints will be imposed on these solutions, due to the fact that $D=m$ and eq.\eqref{sys-extrainv}.
 Then, as we scan over different choices of flux quanta, we will obtain a discretum of values for the parameters of the Ansatz, within the above continuous solutions. In other words, the two branches become a discrete set of points once that flux quantization is imposed.

\begin{table}[H]
\def\arraystretch{1.5}
\begin{center}
\scalebox{1}{%
    \begin{tabular}{| c ||     c | c | c | c |}
    \hline
  Branch & $A_F$  & $B_F$  & $C_F$  & $D_F$  \\
  \hline \hline
  \textbf{SUSY}  & $-\frac{3}{8}$ &  $-\frac{3}{2}E_F$  & $\frac{1}{4}$  & $15E_F$     \\ \hline
  \textbf{non-SUSY}   & eq.\eqref{sys-eq: megaeqF}  & $4A_FE_F$  & eq.\eqref{sys-eq: megaeqF}  & $ \sqrt{\frac{\Delta_F}{C_F^2}  + (4A_F^2+1)} \, 12E_F $       \\
      \hline
    \end{tabular}}      
\end{center}
\caption{Branches of solutions in terms of the quotients $A_F = A/F$, etc. of the parameters of the Ansatz \eqref{sys-Ansatz}. In the SUSY branch $E_F$ is not constrained by the equations of motion, while in the non-SUSY extrema it is given by \eqref{sys-eq: E_F delta neq0}. Moreover $\Delta_F$ is given by \eqref{sys-eq: DeltaF def}, being always zero in the SUSY branch.  \label{sys-vacuresul}}
\end{table}

As we show below, the branch where  $A_F = -3/8$, $C_F =1/4$ and $E_F$ is not constrained by the vacuum equations corresponds to supersymmetric vacua, while the other branch contains non-supersymmetric ones. Remarkably, both branches intersect at one point. The non-supersymmetric branch splits into three when imposing the physical condition $E_F^2 \geq 0$, as can be appreciated from figure \ref{sys-fig: generalsol}. Each point of these blue curves contains two solutions, corresponding to the two values $E_F = \pm \frac{C_F}{12} \sqrt{\Delta_F^{-1}(36A_F^2 - C_F^2 - 5)}$.

\bigskip

\noindent
\textbf{F-terms}
\medskip

\noindent
One can recast the F-terms for each of these extrema as
\bes
\begin{align}
    G_{a}&=\left[\left(-\frac{1}{2}B_F-2E_F+\frac{1}{12}D_F\right)+i\left(-\frac{1}{12}C_F-\frac{1}{2}A_F-\frac{1}{6}\right)\right]F \, \mathcal{K}^2\partial_a K\, ,\\
   G_{\mu}&=\left[\left(-\frac{3}{2}B_F-\frac{1}{12}D_F-E_F\right)+i\left(-\frac{1}{4}-\frac{1}{2}A_F+\frac{1}{4}C_F\right)\right]F\, \mathcal{K}^2\partial_\mu K\, ,
\end{align}
\ees
and one can see that requiring that they vanish is equivalent to impose \eqref{sys-SUSYsol} and \eqref{sys-SUSYbranch}.  Therefore, the branch \eqref{sys-SUSYsol} corresponds to supersymmetric vacua, while general solutions to \eqref{sys-eq: megaeqF} represent non-supersymmetric extrema of the potential.

\bigskip
\noindent
\textbf{Vacuum energy and KK scale}
\medskip

\noindent
Using \eqref{sys-eq: cosmo const relation} and imposing the extremization of the potential, one can see that the vacuum energy has the following expression in the above branches of solutions:
\begin{equation}
   4\pi \kappa_4^4  V|_{\rm vac}= - \frac{4}{3}e^K\mathcal{K}^2F^2 \left(2A_F^2+64A_F^2E_F^2+\frac{1}{18}C_F^2+\frac{5}{18}\right)\, .
\end{equation}
In the supersymmetric branch this expression further simplifies to
\begin{equation}
    4\pi \kappa_4^4  V|_{\rm vac}^{\rm SUSY}=-  e^K\mathcal{K}^2F^2\left(12E_F^2+\frac{3}{4}\right)\, .
\end{equation}
So essentially we recover that the AdS$_4$ scale in Planck units is of order
\be
\frac{\Lambda_{\rm AdS}^2}{M_{\rm P}^2}  \sim e^{4-d} V_{X_6}   F^2  \sim \frac{t^3}{u^4}  F^2 \chi\, ,
\ee
where in the last step we have defined  $\chi \equiv 2A_F^2+64A_F^2E_F^2+\frac{1}{18}C_F^2+\frac{5}{18}$. This is to be compared with the KK scale
\be
\frac{M_{\rm KK}^2}{M_{\rm P}^2} \sim   e^{2D} V_{X_6}^{-1/3} \sim  t^{-1} u^{-2} \, ,
\ee
obtaining the quotient
\be
\frac{\Lambda_{\rm AdS}^2}{M_{\rm KK}^2} \sim  e^{2D} V_{X_6}^{4/3} F^2 \sim \frac{t^{4}}{u^2} F^2 \chi \, .
\label{sys-quot}
\ee
Scale separation will occur when this quotient is small, which seems hard to achieve parametrically, unlike in \cite{DeWolfe:2005uu,Marchesano:2019hfb}. Indeed, unless some fine tuning occurs, at large $t$, $u$ one expects that $e^K|W|^2 \sim e^K|W_{\rm RR}|^2 + e^K |W_{\rm NS}|^2$, which in supersymmetric vacua dominates the vacuum energy. If both terms are comparable, then in type IIA setups with bounded geometric fluxes and Romans mass  $u \sim t^2$, and there is no separation due to the naive modulus dependence in \eqref{sys-quot}. If one term dominates over the other the consequences are even worse, at least for supersymmetric vacua.\footnote{With specific relations between flux quanta parametric scale separation at the 4-dimensional level is possible \cite{Font:2019uva}. Remarkably, it was there found that this naive 4-dimensional scale separation did not occur at the 10d level. \label{sys-f:ssep}} Because $\chi$ is at least an order one number, the most promising possibility for achieving scale separation is that $F$ scales down with $t$. While this scaling is compatible with \eqref{sys-hath}, we have not been able to find examples where this possibility is realized.\footnote{In particular, in the  SUSY branch of toroidal compactifications we have not found any flux configuration with naive scale separation beyond the case of \cite{Font:2019uva} mentioned in footnote \ref{sys-f:ssep}.} Even if $F$ does not scale with the moduli, it would seem that generically $F \lesssim \cO(0.1)$  is a necessary condition to achieve a vacuum at  minimal scale separation. This is perhaps to be expected because in the limit $F \raw 0$ we recover the analysis of \cite{Marchesano:2019hfb}, where parametric scale separation occurs, at least from the present 4-dimensional perspective. 

In fact, the case $F=0$ also displays vacua at parametric large volume and small string coupling. While in our setup we have not been able to find families of vacua with such behaviour, one can see that small values of $F$ also favour vacua in the large volume-weak coupling regime, where the K\"ahler potential used in our analysis can be trusted. Indeed, notice that the LHS of \eqref{sys-hath} corresponds to the contribution to the fluxes to the tadpoles and so it is a bounded integer number. As such, large values of the K\"ahler moduli will be linked to small values of $(4D_FE_F - C_F) F$. Using the scaling $u \sim t^2$, a similar conclusion can be drawn for weak coupling.

\begin{tcolorbox}[breakable, enhanced,  colback=uam!10!white, colframe=uam!85!black]
\small
    As mentioned before, there is another set of branches of solutions that arises in the case where both brackets of \eqref{sys-paxionkA} and \eqref{eq: psaxionkA} vanish independently and has the potential to display scale separation. It has been explored by \cite{Cribiori:2021djm} in the context of toroidal orientifold compactifications with metric fluxes but without Roman mass or H flux. In that scenario, an uplift to M-theory was also provided. Such results can be understood and generalized in our language by setting $\mathcal{M}=\mathcal{P}=\tilde{\rho}=0$ in \eqref{sys-proprhog}, which provides a refined Ansatz to explore this new branch in a non-trivial but manageable way. 
    
    We now briefly address how to find and extend these solutions. The key point lies in demanding the brackets \eqref{sys-paxionkA} and \eqref{eq: psaxionkA} to vanish, which poses severe constraints in the parameters of the original Ansatz \eqref{sys-proprho}.
The two possible options are summarised in the table below, where we have defined for clarity a new quantity
\begin{equation}
    \mathcal{S}\equiv 3+4\frac{\mathcal{P}^2}{\mathcal{N}^2}\, .
    \label{eq: param S def}
\end{equation}

\begin{center}
\begin{tabular}{|c||c|c|c|c|}
\hline
\diagbox[height=1.2cm,width=4.2cm]{Branch}{Parameters} & $\ell_s\rho_0$ & $\mathcal{Q}$ & $\ell_s\tilde{\rho}\mathcal{K}$ & $\mathcal{M}$ \\ \hhline{|=||=|=|=|=|}
 SUSY&
     $-\frac{3}{2}\mathcal{N}$    &    $\mathcal{N}$            & $-10\mathcal{P}$                          & $-\frac{2}{3}\mathcal{P}$\\   \hline  non-SUSY & $-\frac{\mathcal{N}}{2}\left(1-\frac{12}{\mathcal{S}}\right) $  & $\mathcal{N}$               &          $-6\mathcal{P}\left(1-\frac{4}{\mathcal{S}}\right)$                  &  $\frac{4\mathcal{P}}{\mathcal{S}}$             \\\hline
\end{tabular}
\end{center}
\vspace{0.5 cm}

By construction, the parameters of the table satisfy the equations of motion of the Kähler sector. Demanding that this restricted Ansatz also solves the equations of the complex structure sector  further constrains the non-SUSY branch by imposing the vanishing of the Romans mass. As a result, we arrive at the four branches of solutions displayed in the table below.

\vspace{0.5cm}

\begin{center}
\begin{tabular}{|c||c|c|c|c|c|c|}
\hline
\diagbox[height=1.2cm,width=4.2cm]{Branch}{Parameters} & $\mathcal{P}$ & $\mathcal{S}$ &$\ell_s \rho_0$ & $\mathcal{Q}$ & $\ell_s \tilde{\rho}\, = m$ & $\mathcal{M}$ \\ \hhline{|=||=|=|=|=|=|=|}
 SUSY&
    Free & \eqref{eq: param S def} &$-\frac{3}{2}\mathcal{N}$    &    $\mathcal{N}$            & $-10\frac{\mathcal{P}}{\cK}$                          & $-\frac{2}{3}\mathcal{P}$\\   \hline \multirow{3}{*}{non-SUSY}& $0$ & $3$ & \multirow{3}{*}{$-\frac{\mathcal{N}}{2}\left(1-\frac{12}{\mathcal{S}}\right)$}   &\multirow{3}{*}{$\mathcal{N}$}     &    \multirow{3}{*}{$0$  }   &  \multirow{3}{*}{$\frac{4\mathcal{P}}{\mathcal S}$}             \\\cline{2-3}  & $+\frac{\mathcal{N}}{2}$ & 4 &    &      &              & \\\cline{2-3} & $-\frac{\mathcal{N}}{2}$ &  4 &   &              &     &   \\\hline
\end{tabular}
\end{center}
\vspace{0.5cm}

By evaluating the above solutions in the F-terms equations \eqref{sys-eq: F-Ta} and \eqref{sys-eq: F-Umu} one can check that first row of the previous table corresponds to a supersymmetric branch of vacua. Actually, these results generalize the SUSY branch found in table \ref{sys-vacuresul} beyond the case \eqref{sys-proprho}. In addition, we also find three new non-SUSY families of solutions. From the 10d perspective, these four branches describe half-flat manifolds, in contrast with the nearly-Kähler geometry arising when \eqref{sys-proprho} is satisfied, which we discuss in section \ref{sys-sec: 10d}. These branches and their properties will be explored in more detail in a future work \cite{systematicsfollowup}.

\end{tcolorbox}

\subsection{Relation to previous results}

In order to verify the validity of our formalism and the results we have obtained, we proceed to recover some of the existing results in the literature. As argued in the next section, from the viewpoint of SU(3)-structure manifolds our vacua correspond to nearly-K\"ahler compactifications. We will therefore focus on examples that fit within that class, and mainly on two papers whose results we will link with ours. To simplify the comparisons we will omit the factors $\ell_s$ in this section.

\bigskip
\noindent
\textbf{Comparison to Camara et al. \cite{Camara:2005dc}}
\medskip

\noindent
    This reference studies RR, NS and metric fluxes on a $T^6/(\Omega(-1)^{F_L}I_3)$ Type IIA orientifold. We are particularly interested in section 4.4, where $\CN=1$ AdS vacua in the presence of metric fluxes are analyzed. One can easily use our SUSY branch (see table \ref{sys-vacuresul}), the definitions of the flux polynomials \eqref{sys-RRrhosgeom} and our Ansatz \eqref{sys-Ansatz} to reproduce their relations between flux quanta and moduli fixing. We briefly discuss the most relevant ones.
    
    In \cite{Camara:2005dc} they study the particular toroidal geometry in which all three complexified Kähler moduli are identified. This choice greatly simplifies the potential and the flux polynomials.  To reproduce the superpotential in \cite[eq.(3.15)]{Camara:2005dc} we consider the case $T^a=T$,  $\forall a$, so that there is only one K\"ahler modulus and the Kähler index $a$ can be removed. The flux quanta $\{e_0, e_a, m^a,m,h_\mu,\rho_{a\mu}\}$ are such that  $e_a=3c_1$, $m^a=c_2$ and 
\begin{equation}
    \rho_{a\mu}=
    \begin{cases}
    3a&\hspace{1cm} \mu=0\, ,\\
    b_\mu &\hspace{1cm} \mu\neq 0\, , 
    \end{cases}
    \qquad a, b_\mu \in \mathbb{Z}\, .
\end{equation}
Imposing the constraint $D=m$ on the SUSY Ansatz we have
\begin{align}
    A&=-\frac{3}{8}F\, ,   &    B&=-\frac{m}{10}\, , &    C&=\frac{1}{4}F\, ,    &   D&=m=15E\, .
\end{align}

        The first step is to use the invariant combinations of fluxes and axion polynomials together with the Ansatz to fix the value of the saxions. Notice that because we only have one K\"ahler modulus, $\rho_{a\mu}$ has necessarily rank one, and so \eqref{sys-hath} fixes $t$ as function of the fluxes and the parameter $F$:
\begin{equation}
    \left(\frac{4ED}{F}-C\right)\rho_{a\mu}t^a=mh_\mu-\rho_{a\mu}m^a\longrightarrow \begin{cases} \left(\frac{4m^2}{15F}-\frac{1}{4}F\right)3a t=mh_0-3ac_2 \textrm{\hspace{1cm}if $\mu=0$}\, ,\\
    \left(\frac{4m^2}{15F}-\frac{1}{4}F\right)b_\mu t=mh_\mu-b_\mu c_2 \textrm{\hspace{1cm}if $\mu\neq0$\, .}
    \end{cases}
\end{equation}
This relation provides a constraint for the fluxes in order for this family of solutions to be realized (cf. \cite[eq.(4.32)]{Camara:2005dc}).
The complex structure saxions are instead determined in terms of  $\rho_{a\mu}$:
\begin{equation}
    \rho_{a\mu} t^a=\frac{F}{4}\mathcal{K}\partial_\mu K\longrightarrow \begin{cases}3a t=-\frac{F\mathcal{K}}{4 u^0}\, ,\\
    b_\mu t=-\frac{F\mathcal{K}}{4 u^\mu}\, ,
    \end{cases}
    \label{sys-eq: camara u fix}
\end{equation}
which reproduces the relation  \cite[eq.(4.31)]{Camara:2005dc}.

To obtain the remaining relations of \cite[section 4.4]{Camara:2005dc}, we take into account that $\mathcal{K}=6t^3$ and take advantage of the particularly simple dependence of our Anstaz when considered on an isotropic torus. Using that $F=4C$ we can go back to \eqref{sys-eq: camara u fix} to eliminate the $F$ dependence of the complex structure moduli. 
\begin{equation}
     \rho_{a\{\mu=0\}} t^a=F\mathcal{K}\partial_{\mu=0} K=-\frac{6t^3F}{4u^0}=-C\frac{6t^3}{u^0}=-\frac{6t^2}{u^0}\tilde{\rho}^a \longrightarrow 3atu^0=-6t^2(c_2+vm)\, ,
     \label{sys-eq: camara u fix 2}
\end{equation}
which, up to redefinition of the parameters, is just relation  \cite[eq.(4.34)]{Camara:2005dc}. Similarly, we have
\begin{equation}
    \rho_{\mu=0}=E\mathcal{K}\partial_{\mu=0}K\longrightarrow h_0+3av=-\frac{m}{15}\frac{6t^3}{u^0}\, .
    \label{sys-eq: camara t fix}
\end{equation}
Replacing $u_0$ using \eqref{sys-eq: camara u fix 2} in the above expression leads to
\begin{equation}
    t^2=\frac{5(h_0+3av)(c_2+mv)}{am}\, ,
    \label{sys-eq: camara u-t fix}
\end{equation}
which is equivalent to \cite[eq.(4.41)]{Camara:2005dc} and provides an alternative way to fix the K\"ahler moduli $t$. 

To fix the complex structure axions $\xi^\mu$ we note that
\begin{equation}
    \rho_a=B\mathcal{K}\partial_a K =-\frac{3}{2}E\mathcal{K}\partial_a K= \frac{3u^0}{2}\rho_{\mu=0}\partial_a K \longrightarrow \rho_at^a=-\frac{9}{2}(h_0+3av)u^0\, .
\end{equation}
Expanding $\rho_a$ and replacing $t$ using \eqref{sys-eq: camara u fix 2} we arrive at
\begin{equation}
    3c_1+6c_2v+3mv^2+3a\xi^0+\sum_\mu b_\mu \xi^\mu=\frac{9}{a}(c_2+mv)(h_0+3av)\, ,
    \label{sys-eq: complex axion combination camara}
\end{equation}
and hence we derive an analogous relation to  \cite[eq.(4.33)]{Camara:2005dc}. We observe that it  only fixes one linear combination of complex structure saxions. This was to be expected, since by construction the geometric fluxes are of rank one. Finally, we can fix the  K\"ahler axion $b$ using the flux polynomial $\rho_0$
\begin{equation}
    \rho_0=A\mathcal{K}=-\frac{3C}{2}\mathcal{K}=-\frac{3}{2t}\tilde{\rho}^a\mathcal{K}\longrightarrow\rho_0=-9(c_2+mv)t^2\, ,
\end{equation}
which after replacing the complex axions using \eqref{sys-eq: complex axion combination camara} and substituting $t$ using \eqref{sys-eq: camara t fix} and \eqref{sys-eq: camara u-t fix} leads to the same equation for the K\"ahler axion as the one shown in \cite[eq.(4.40)]{Camara:2005dc}.

\bigskip
\noindent
\textbf{Comparison to Dibitetto et al. \cite{Dibitetto:2011gm}}
\medskip

\noindent
In this reference the vacuum structure of isotropic $\mathbb{Z}_2\times\mathbb{Z}_2$ compactifications is analyzed, combining algebraic geometry and supergravity techniques. We are particularly interested in the results shown in \cite[section 4]{Dibitetto:2011gm}, where they consider a setup similar to \cite[section 4.4]{Camara:2005dc}, but go beyond supersymmetric vacua.\footnote{It is worth noting that in order to solve the vacuum equations, \cite{Dibitetto:2011gm} follows a complementary approach to the standard one. Typically, one starts from the assumption that the flux quanta have been fixed and then computes the values of the axions and saxions that minimize the potential. Ref.\cite{Dibitetto:2011gm} instead fixes a point in field space, and reduces the problem to find the set of consistent flux backgrounds compatible with this point being an extremum of the scalar potential. Both descriptions should be compatible.} More concretely, in this section they study type IIA orientifold compactifications on a $\mathbb{T}^6/(\mathbb{Z}_2\times \mathbb{Z}_2)$ isotropic orbifold in the presence of metric fluxes. Hence, they have an $STU$ model with the axiodilaton $S$, the overall K\"ahler modulus $T$ and the overall complex structure modulus $U$.

They obtain sixteen critical points with one free parameter and an additional solution with two free parameters. This last case is not covered by our Ansatz, since the associated geometric fluxes do not satisfy \eqref{sys-eq: geoma} and \eqref{sys-eq: geomu}. Therefore it should correspond to a non-supersymmetric vacuum with F-terms different from \eqref{sys-solsfmax}. The remaining sixteen critical points are grouped into four families and summarized in \cite[table 3]{Dibitetto:2011gm}. Taking into account their moduli fixing choices, we can relate their results for the flux quanta with the parameters of our Ansatz as follows:

\begin{itemize}
    \item When $s_2=1$, solution $1$ from  \cite[table 3]{Dibitetto:2011gm} corresponds to a particular point of the SUSY branch in our table \ref{sys-vacuresul}, with $E_F=\pm \frac{1}{4\sqrt{15}}$ (sign given by $s_1$). 
    \item When $s_2=-1$, solution $1$ of  \cite[table 3]{Dibitetto:2011gm} corresponds to the limit solution \eqref{sys-nonSUSYdelta=0} of the non-SUSY branch (point (b) in figure \ref{sys-fig: generalsol}). We confirm the result of \cite{Dibitetto:2011gm} regarding stability: similarly to the SUSY case, this is a saddle point with tachyonic mass $m^2=-8/9|m^2_{BF}|$ (for a detailed analysis on stability check section \ref{sys-persta} and Appendix \ref{ap.sys: Hessian}).
    
    \item Solution $2$ from \cite[table 3]{Dibitetto:2011gm} corresponds to a limit point $C_F=0$ of the non-SUSY branch with $\Delta_F\neq0$ and $A_F=\pm5/12$. Such solution was not detailed in our analysis of section \ref{sys-branchvacu} since, despite being a limit point, it still verifies \eqref{sys-eq: E_F delta neq0}, \eqref{sys-eq: megaeqF} and \eqref{sys-eq: D_F delta neq0}. In \cite[table 4 ]{Dibitetto:2011gm} it is stated that this solution is perturbatively unstable, in agreement with our results below (see figure \ref{sys-fig: excludsol}).
    
    \item Solution $3$ from \cite[table 3]{Dibitetto:2011gm} is a particular case of the non-SUSY branch, corresponding to $A_F=s_1/4$ and $C_F=s_1/2$ (with $s_1=\pm1$). This specific point falls in the stable region of figure \ref{sys-fig: excludsol}. The analysis of section \ref{sys-persta} reveals that the mass spectrum has two massless modes, confirming the results of \cite{Dibitetto:2011gm}.

    \item Solution $4$ of \cite[table 3]{Dibitetto:2011gm} is not covered by our Ansatz since, similarly to the two-dimensional solution, our parameter $F$ is not well-defined under this combination of geometric fluxes. We then expect F-terms not of the form \eqref{sys-solsfmax}.
\end{itemize}

Hence, the results of \cite{Dibitetto:2011gm} provide concrete examples of solutions for both the supersymmetric and non-supersymmetric branches of table \ref{sys-vacuresul}.

\bigskip
\noindent
\textbf{Examples of de Sitter extrema}
\medskip

\noindent
In \cite{Caviezel:2008tf}, the authors study the cosmological properties of type IIA compactifications on orientifolds of manifolds with geometric fluxes. They apply the no-go result of \cite{Flauger:2008ad} to rule out de Sitter vacua in all the scenarios they consider except for the manifold $SU(2)\times SU(2)$, where they find a de Sitter extremum, albeit with tachyons. One can check that the fluxes considered in section 4.2 of \cite{Caviezel:2008tf} do not satisfy condition \eqref{sys-condhhat}. Therefore, this example lies outside of our Ansatz and so relation \eqref{sys-eq: no-go geom inequality} does not hold. 

More generally, geometric examples of de Sitter extrema are built from compactifications on SU(3)-structure manifolds which are not nearly-K\"ahler. As we will see in section \ref{sys-sec: 10d}, our Ansatz \eqref{sys-Ansatz} implies that the internal manifold is nearly-K\"ahler, in the approximation of smeared sources. Therefore, our analysis does not capture the attempts to find extrema in manifolds with torsion class ${\cal W}_2 \neq 0$, see e.g. \cite{Koerber:2008rx,Caviezel:2008ik,Caviezel:2008tf,Danielsson:2009ff,Danielsson:2010bc}. 


\section{Stability and 10d description}
\label{sys-sec: stabalidity}
Given the above set of 4d AdS extrema some questions arise naturally. First of all, one should check which of these points are \textit{actual} vacua, meaning stable in the perturbative sense. In other words, we should verify that they do not contain tachyons violating the BF bound \cite{Breitenlohner:1982bm}. As it will be discussed below, for an arbitrary geometric flux matrix $f_{a\mu}$ it is not possible to perform this analysis without the explicit knowledge of the moduli space metric. Nevertheless, the problem can be easily addressed if we restrict to the case in which $f_{a\mu}$ is a rank-one matrix, which will be the case studied in section  \ref{sys-persta}. On the other hand, one may wonder if these 4-dimensional solutions have a 10d interpretation. We will see that our Ansatz can be described as an approximate SU(3)-structure background, which we will match with known 10d solutions in the literature.

\subsection{Perturbative stability}\label{sys-persta}

 Following the approach in \cite{Marchesano:2019hfb} we will compute the physical eigenvalues of the Hessian by decomposing the K\"ahler metrics (both for the complex structure and K\"ahler fields) into their primitive and non-primitive pieces. This decomposition together with the Ansatz \eqref{sys-Ansatz} reduces the Hessian to a matrix whose components are just numbers and whose eigenvalues are proportional to the physical masses of the moduli. The explicit computations and details are given in Appendix \ref{ap.sys: Hessian}, whose main results we will summarize in here. To simplify this analysis we will initially ignore the contribution of the D-term potential, that is, we will set $\hat{\rho}_\alpha{}^\mu=0$. We will briefly discuss its effect at the end of this section.

As mentioned above, we will consider the case in which $f_{a\mu} = \ell_s \rho_{a\mu}$ has rank one, since the case with a higher rank cannot be solved in general. Let us see briefly why. One can show that the Ansatz \eqref{sys-Ansatz} implies:
\begin{align}
\label{sys-rhoproblem}
    f_{a\mu}&=-\frac{F K}{12}\p_a K\p_\mu K+\tilde f_{a\mu},    &   &\text{with}    &    & t^a \tilde f_{a\mu}=0= u^\mu \tilde f_{a\mu}\, ,
\end{align}
and so $\tilde f_{a\mu}$ must be spanned by $ t_a^\bot\otimes u_\mu^\bot$, where the  $\left\{t_a^\bot\right\}$ form a basis of the subspace orthogonal to $ t^a$, and similarly for  $ u_\mu^\bot$. The contribution of the first term of \eqref{sys-rhoproblem} to the Hessian can be studied in general. The contribution of the second term depends, among other things, on  how both the $ t_a^{\bot}$ and $ u_\mu^{\bot}$ are stabilized, which can only be studied if the explicit form of the internal metric is known. Therefore, in the following we will set  $\tilde f_{a\mu}=0$. Notice that, for this case, our Ansatz implies that just one linear combination of axions is stabilized, since from \eqref{sys-Ansatz} it follows that $\rho_\mu \propto \rho_{a\mu}, \forall a$.

\vspace*{.5cm}
\noindent
\textbf{SUSY Branch}
\\
As expected, the SUSY case is perturbatively stable. The results can be summarized as: 

\begin{table}[H]
\def\arraystretch{1.5}
\begin{center}
\scalebox{1}{%
    \begin{tabular}{| c ||     c | c | c | c |}
    \hline
  Branch & Tachyons (at least)  & Physical eigenvalues  & Massless modes (at least)  \\
  \hline \hline
  \textbf{SUSY}  & $h^{2,1}$ &  $m^2_{tach}=\frac{8}{9} m_{BF}^2$  & $h^{2,1}$  \\
  \hline
    \end{tabular}}      
\end{center}
\caption{Massless and tachyonic modes  for the supersymmetric minimum\label{sys-susytach}.}
\end{table}
Let us explain the content of the table and especially the meaning of ``at least''. All the details of this analysis are discussed in appendix \ref{ap.sys: Hessian}

\begin{itemize}
    \item Since the potential only depends on a linear combination of complex structure axions and the dilaton, the other $h^{2,1}$ axions of this sector are seen as flat directions. Their saxionic partners, which pair up with them into complex fields,  are tachyonic directions with mass $\frac{8}{9}m_{BF}^2$. Both modes are always present for any value of $E_F$ so we refer to them with the ``at least" tag. This is expected form general arguments, see e.g. \cite{Conlon:2006tq}. 
    
    \item For $E_F\lesssim 0.1$ there appear new tachyons with masses above the BF bound, in principle different from $\frac{8}{9}m_{BF}^2$. The masses of these modes change continuously with $E_F$, and so they become massless before becoming tachyonic.
    \item Finally, there are also modes which have a positive mass for any $E_F$.
\end{itemize} 

\vspace*{.5cm}
\noindent
\textbf{Non-SUSY branch}
\\
This case presents a casuistry that makes it difficult to summarize in just one table. As discussed in section \ref{sys-branchvacu}, the non-SUSY vacuum candidates are described  by  the physical solutions of eq.\eqref{sys-eq: megaeqF}, represented in figure \ref{sys-fig: generalsol}. On top of this curve one can represent the regions that are excluded  at the perturbative level: 
\label{sys-ss:stability}
\begin{center}    
\begin{figure}[H]
    \centering
    \includegraphics[scale=0.8]{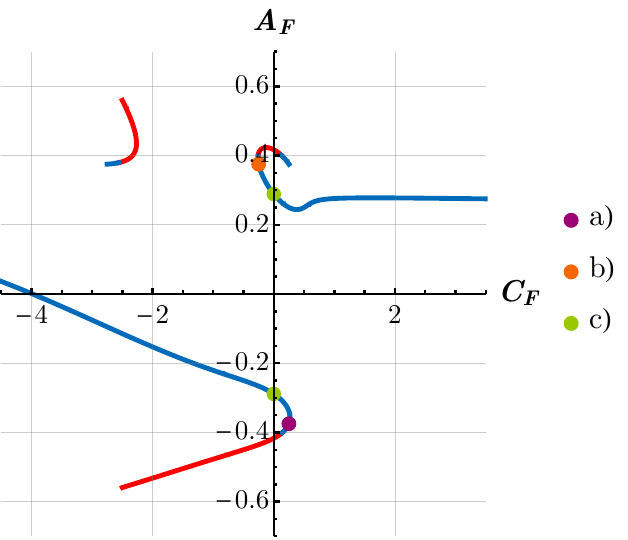}
    \caption{Set of points that verify \eqref{sys-eq: megaeqF} with $E_F^2\geq 0$ and:  have no tachyons violating the BF bound and therefore are perturbatively  stable (blue curve);  have tachyons violating the BF bound and therefore are perturbatively unstable (red curve). The colored dots correspond to the particular solutions \eqref{sys-delta=0sols}.}
    \label{sys-fig: excludsol}
\end{figure}
\end{center}
Some comments are in order regarding the behaviour of the modes:
\begin{itemize}
    \item In the regions with $|A_F|\gtrsim 0.4$ there is always a tachyon whose mass violates the BF bound. This corresponds to the red pieces of the curves in figure \ref{sys-fig: excludsol}.
    \item On the blue region of the curves, tachyons appear only in the vicinity of the red region, while away from it all the masses are positive. For instance, in the curve stretching to the right  there are no tachyons for $C_F\gtrsim 1.5$.
\end{itemize}

The explicit computation of the modes and their masses is studied in appendix \ref{ap.sys: Hessian}.

\vspace*{.5cm}
\noindent
\textbf{D-term contribution}

As announced in the introduction, let us finish this section by commenting on the effect of the D-terms on stability. The first thing one has to notice is that, although $V_D=0$ once we impose the Ansatz $\eqref{sys-Ansatz}$, the Hessian $H_D$ associated to the D-terms is generically different from zero -see \eqref{sys-dhessian}-. Indeed one can show that the matrix $H_D$ is a  positive semidefinite matrix. Therefore, splitting the contribution of  $V_F$ and  $V_D$ to the Hessian into $H=H_F+H_D$ and using the inequalities collected in \cite{bhatia2013matrix}, one can prove that the resulting eigenvalues of the full Hessian $H$ will always be equal or greater than the corresponding $H_F$ eigenvalues. Physically, what this means is that the D-terms push the system towards a more stable regime. In terms of the figure \eqref{sys-fig: excludsol} and taking into account the directions affected by $H_D$ -see again \eqref{sys-dhessian}-, one would expect that, besides having no new unstable points (red region), some of them do actually turn into stable ones (blue points) once the D-terms come into play.

\subsection{10d interpretation}
\label{sys-sec: 10d}

For those geometric vacua that fall in the large-volume regime, one may try to infer a microscopic description in terms of a 10d background AdS$_4 \times X_6$. In this section we will do so by following the general philosophy described in section \ref{cy-subsec: smearing approx} and appendix \ref{ap.geo-subsec: torsion classes}, by interpreting our 4-dimensional solution in terms of an internal manifold $X_6$ with SU(3)-structure. We hasten to stress that this does not mean that the internal metric of $X_6$ corresponds to a SU(3)-structure. As in the 10d uplift of the 4-dimensional supersymmetric vacua \cite{DeWolfe:2005uu}, recently analyzed in \cite{Junghans:2020acz,Marchesano:2020qvg} and described in the previous chapter, it could be that the actual 10d background displays a more general $SU(3)\times SU(3)$-structure that is approximated by an $SU(3)$-structure in some limit.  Based on the lessons learnt from the (approximate) Calabi--Yau case, one should be able to describe the 4-dimensional vacua from a 10d SU(3)-structure perspective if the localized sources are smeared, so that the Bianchi identities amount to the tadpole conditions derived from \eqref{sys-DFtadpole}, already taken into account by our analysis. Even in the smearing approximation keeping the $SU(3)$ description is not trivial, since we know from section \ref{cy-subsec: effective 4d potential} that the precise framework in which to work with geometric and non-geometric fluxes is also the generalized structure $SU(3)\times SU(3)$. Nevertheless, given the simple configuration of fluxes selected by our Ansatz, staying in the $SU(3)$ structure regime is possible, as we will now see.

Following the notation and results introduced in chapter \ref{ch: calabi-yau} and the reasoning of \cite[Section 5.2]{Marchesano:2019hfb}, one may translate our Ansatz into 10d backgrounds in terms of the gauge invariant combination of fluxes
\be
{\bf G} \,=\, d_H {\bf C} + e^{B} \wedge {\bf \hat{G}} \, ,
\label{sys-bfG}
\ee
where $d_H = d - H \wedge$. From here one reads
\begin{align}
\label{sys-solutionsu3}
\ell_s G_6= -6 A\, d{\rm vol}_{X_6}\, , \quad \ell_s G_4 = -3 B\, J \wedge J \, , \quad \ell_s G_2 = C\, J \, , \quad \ell_s H = 6 E \,  g_s \re (\Omega)\, ,
\end{align}
and $ \ell_s G_0 = D$. A vanishing D-term $D_\alpha = \frac{1}{2} \partial_\mu K \hat{f}_\alpha{}^\mu$ implies no contribution from $\hat{f}_\alpha{}^\mu$ to the torsion classes, as in the setup in \cite{Camara:2011jg}. Conversely, \eqref{sys-eq: geoma} and \eqref{sys-eq: geomu} imply that
\begin{equation}
    dJ =  \frac{3}{2}  \frac{F g_s}{\ell_s} \re ( \Omega)\, , \qquad d \im ( \Omega) = - \frac{F g_s}{\ell_s} J \wedge J\ ,
    \label{sys-geomsu3}
\end{equation}
which translate into the following  $SU(3)$ torsion classes defined in \eqref{ap.geo-eq: torsion classes} as
\be
W_1 = -i \frac{ g_s  F}{\ell_s} \, , \qquad W_2 = W_3 = W_4 = W_5 = 0\, .
\label{sys-torsionsu3}
\ee
Therefore, in terms of an internal SU(3)-structure manifold, our vacua correspond to nearly-K\"ahler compactifications.  

With this dictionary, it is easy to interpret our SUSY branch of solutions in terms of the general SU(3)-structure solutions for ${\cal N} = 1$ AdS$_4$ type IIA  vacua \cite{Behrndt:2004km,Lust:2004ig}. We can compare with the parametrization of \cite[eq.(4.24)]{Koerber:2010bx}, and see that the relations \eqref{sys-SUSYsol} and \eqref{sys-SUSYbranch} fit perfectly upon identifying
\be
\ell_s |W_0|  e^{- A - i\hat{\theta}} =  3g_s\left(E + i \frac{F}{4}\right) \, ,
\label{sys-SUSYdict}
\ee
where $|W_0|$ is the AdS$_4$ scale from the 10d frame, and $\hat{\theta}$ a phase describing the solution. 

One can in fact use this dictionary to identify some solutions in the non-supersymmetric branch with 10d solutions in the literature, like e.g. those in \cite{Lust:2008zd}. Indeed, let us in particular consider \cite[section 11.4]{Lust:2008zd}, where $\mathcal{N}=0$ AdS$_4$ compactifications are constructed by extending integrability theorems for 10d supersymmetric type II backgrounds. We first observe that the second Bianchi identity in \cite[eq.(11.29)]{Lust:2008zd} describes our first vacuum equation \eqref{sys-paxioncpxAA}. Similarly \cite[eqs.(11.31),(11.35),(11.36)]{Lust:2008zd} are directly related to \eqref{sys-psaxioncpxAA}, \eqref{sys-psaxionkAA} and \eqref{sys-paxionkAA} respectively.

Using these relations three classes of solutions are found in \cite[section 11.4]{Lust:2008zd}:
\begin{enumerate}
    \item The first solution \cite[(11.38)]{Lust:2008zd} is a particular case of the non-SUSY branch, corresponding to $A_F=\pm1/4$ and  $C_F=\pm1/2$, with $A_FC_F>0$.
    \item The second solution \cite[(11.39)]{Lust:2008zd} corresponds  the limit solution of the non-SUSY branch with $C_F=0$ and $\Delta_F\neq0$.
    \item The third solution \cite[(11.40)]{Lust:2008zd} describes a point in the SUSY branch characterized by $E_F=\pm \frac{1}{4\sqrt{15}}$. 
\end{enumerate}

To sum up, the results of \cite{Lust:2008zd} provide concrete 10d realisation of solutions for both the supersymmetric and non-supersymmetric branches of table \ref{sys-vacuresul}.

Finally, this 10d picture allows us to understand our no-go result of section \ref{sys-subsec: no-go's} from a different perspective. Indeed, given the torsion classes \eqref{sys-torsionsu3} the Ricci tensor of the internal manifold $X_6$ reads \cite{BEDULLI20071125,Ali:2006gd}
\be
{\cal R}_{mn} = \frac{5}{4} g_{mn} |W_1|^2\, ,
\ee
and so it corresponds to a manifold of positive scalar curvature, instead of the negative curvature necessary to circumvent the obstruction to de Sitter solutions \cite{Silverstein:2007ac}.


\section{Summary}
\label{sys-sec: conclu}

In this chapter we have taken a systematic approach towards moduli stabilization in 4d type IIA orientifold flux compactifications. The first step has been to rewrite the scalar potential, including both the F-term and D-term contributions, in a bilinear form, such that the dependence on the axions and the saxions of the compactification is factorized. This bilinear form highlights the presence of discrete gauge symmetries on the compactification, which correspond to simultaneous discrete shifts of the axions and the background fluxes. This structure has been already highlighted for the F-term piece of the potential in Calabi-Yau compactifications with $p$-form fluxes \cite{Bielleman:2015ina,Carta:2016ynn,Herraez:2018vae}, and in here we have seen how it can be extended to include general geometric and non-geometric fluxes as well. 

Besides a superpotential, these new fluxes  generate a D-term potential, which displays the same bilinear structure. The D-term potential arises from flux-induced  St\"uckelberg gaugings of the U(1)'s of the compactification by some axions that do not appear in the superpotential, and that generate conventional discrete gauge symmetries arising from $B \wedge F$ couplings. Such discrete symmetries are unrelated to the ones in the F-term potential. However, the D-term potential itself depends on the B-field axions $b^a$, because they appear in the gauge kinetic function $f_{\alpha\beta}$, and these axions do appear as well in the F-term potential, participating in its discrete symmetries. It would be interesting to understand the general structure of discrete shift symmetries that one can have in flux compactifications with both F-term and D-term potentials. In addition, it would be interesting to complete the analysis by including the presence of D6-branes with moduli and curvature corrections, along the lines of \cite{Carta:2016ynn,Herraez:2018vae,Escobar:2018tiu,Escobar:2018rna}.

As in \cite{Bielleman:2015ina,Carta:2016ynn,Herraez:2018vae}, it is the presence of discrete shift symmetries that is behind the factorization of the scalar potential into the form \eqref{sys-VF}, where $ Z^{\cA\cB}$ only depends on the saxionic fields, and ${\rho}_\cA$ are gauge invariant combinations of flux quanta and axions. With the explicit form of the ${\rho}_\cA$ one may construct combinations that are axion independent, and therefore invariant under the discrete shifts of the compactification. In any class of compactifications, some of the fluxes are invariant by themselves, while others need to be combined quadratically to yield a flux invariant. We have analyzed the flux invariants that appear in type IIA Calabi--Yau, geometric and non-geometric flux compactifications,\footnote{The bilinear formulation may be extended to non-geometric type IIB orientifolds with O3/O7 planes along the lines of \cite{Shukla:2019wfo} which could subsequently help in performing a systematic type IIB vacua analysis.} their interest being that they determine the vev of the saxions at the vacua of the potential. Therefore, in practice, the value of these flux invariants will control whether the vacua are located or not in regions in which the effective field theory is under control. 

Another important aspect when analysing flux vacua is to guarantee their stability, at least at the perturbative level. Guided by the results of \cite{GomezReino:2006dk,GomezReino:2006wv,GomezReino:2007qi,Covi:2008ea,Covi:2008zu}, we have analyzed the sGoldstino mass estimate in our setup, imposing that it must be positive as a necessary stability criterium to which de Sitter extrema are particularly sensitive. Our analysis has led us to the simple Ansatz \eqref{sys-solsfmax} for the F-terms on-shell, which can be easily translated to relations between the $\rho_\cA$ and the value of the saxions at each extremum, cf. \eqref{sys-proprho}.

The next step of our approach has been to find potential extrema based on this Ansatz, a systematic procedure that we have implemented for the case of geometric flux compactifications. This class of configurations is particularly interesting because they contain de Sitter extrema and are therefore simple counterexamples of the initial de Sitter conjecture \cite{Obied:2018sgi}, although so far seem to satisfy its refined version \cite{Garg:2018reu,Ooguri:2018wrx}. In this respect, we have reproduced previous de Sitter no-go results in the literature \cite{Hertzberg:2007wc,Flauger:2008ad} with our bilinear expression for the potential, but with two interesting novelties. First, when imposing that the F-terms are of the form \eqref{sys-solsfmax} either on-shell or off-shell, we recover an inequality of the form \eqref{sys-eq: no-go geom inequality} that forbids de Sitter extrema. We find quite amusing that this result is recovered after imposing an Ansatz inspired by de Sitter metastability. Second, our analysis includes a flux-induced D-term potential, and so the possibility of D-term uplifting, typically considered in the moduli stabilization literature, does not seem to work in the present setting. We see our result as an interesting product of integrating several de Sitter criteria, and it would be interesting to combine it with yet other no-go results in the literature, like for instance those in \cite{Andriot:2018ept,Andriot:2019wrs,Grimm:2019ixq}. 

As is well known, type IIA  orientifold compactifications with geometric fluxes provide a non-trivial set of AdS$_4$ vacua, which we have analyzed from our perspective. We have seen that, by imposing the on-shell Ansatz \eqref{sys-solsfmax}, the equations of motion translate into four algebraic equations. By solving them, we have found two different sets of branches of vacua. We focused on the most generic set, consisting on one supersymmetric branch and another non-supersymmetric one. We have shown how they include many of the vacua found in the geometric flux compactification literature. This link with previous results can be made both with references that perform a 4-dimensional analysis and those that solve the equations of motion at the 10d level, which is particularly interesting for the rather scarce non-supersymmetric solutions. Regarding 10d configurations, we have seen that our Ansatz corresponds to a nearly-K\"ahler geometry in the limit of smeared sources. This implies, in particular, that geometric flux compactifications that can be deformed to a non-trivial torsion class ${\cal W}_2$, correspond to F-terms that deviate from \eqref{sys-solsfmax}. It would be interesting to work out the phenomenological consequences of this fact.

In any event, we hope to have demonstrated that with our systematic approach one may be able to obtain an overall picture of classical type IIA flux vacua. Our strategy not only serves to find and characterize different metastable vacua, but also to easily extract the relevant physics out of them, like the F-terms, vacuum energy and light spectrum of scalars. A global picture of this sort is essential to determine what the set of string theory flux vacua is and it is not, and the lessons that one can learn from it. Hopefully, our results will provide a non-trivial step towards this final picture.




\ifSubfilesClassLoaded{%
\bibliography{biblio}%
}{}

\end{document}
\end{document}

\graphicspath{{Images/Bionic}}

\ifSubfilesClassLoaded{%
\tableofcontents
}{}

\setcounter{chapter}{4}
\chapter{Non-perturbative instabilities in Non-SUSY AdS Vacua}
\label{ch: bionic}

Now that we have an improved understanding of the behaviour of the 4-dimensional AdS effective theories coming from Type IIA supergravity compactifications, we will take a step back and return to the analysis of 10-dimensional theories with the goal of contrasting their properties with the predictions provided by the Swampland conjectures.

Out of the different aspects of the Swampland Program \cite{Vafa:2005ui,Brennan:2017rbf,Palti:2019pca,vanBeest:2021lhn,Grana:2021zvf} one of the most far-reaching is the interplay between quantum gravity and supersymmetry breaking. In the specific context of non-supersymmetric vacua, several proposals for Swampland criteria put severe constraints on their stability,  as we briefly addressed in section \ref{basic-subsec: swampland conjectures}. In particular, the AdS Instability Conjecture \cite{Ooguri:2016pdq,Freivogel:2016qwc} proposes that all $\cN=0$ AdS$_d$ vacua are at best metastable, with  bubble nucleation always mediating some non-perturbative decay. The motivation for this proposal partially arises from a refinement of the Weak Gravity Conjecture (WGC) stating that the WGC inequality is only saturated in supersymmetric theories \cite{Ooguri:2016pdq}. Applied to $(d-2)$-branes, this implies a specific decay mechanism for $\cN=0$ AdS$_d$ vacua supported by $d$-form background fluxes, in which a probe superextremal $(d-2)$-brane nucleates and expands towards the AdS$_d$ boundary, as in \cite{Maldacena:1998uz}.

These proposals have been tested in different contexts, and in particular for type II setups in which the AdS solution is supported by fluxes \cite{Gaiotto:2009mv,Antonelli:2019nar,Apruzzi:2019ecr,Bena:2020xxb,Suh:2020rma,Guarino:2020jwv,Guarino:2020flh,Basile:2021vxh,Apruzzi:2021nle,Bomans:2021ara}. Remarkably, compactifications of the form AdS$_4 \times X_6$, where $X_6$ admits a Calabi--Yau metric \cite{Villadoro:2005cu,DeWolfe:2005uu,Camara:2005dc}, known in the literature as DGKT-like vacua, remain elusive of the conjecture, because so far  the decays  observed for perturbatively stable $\cN=0$ vacua are marginal \cite{Narayan:2010em}, and the corresponding membranes saturate the WGC inequality. A better understanding of these constructions seems thus crucial to the Swampland Program: Their non-supersymmetric version challenges the AdS Instability Conjecture, and more precisely the WGC for membranes, while the supersymmetric settings challenge the strong version of the AdS Distance Conjecture \cite{Lust:2019zwm}. As pointed out in \cite{Buratti:2020kda}, the tension with the AdS Distance Conjecture could be solved by taking into account the discrete symmetries related to 4-dimensional membranes, so the spectrum and  properties of 4-dimensional membranes seem to be at the core of both issues.  Finally, the constructions in \cite{DeWolfe:2005uu,Camara:2005dc} are particularly interesting phenomenologically, since besides supersymmetry breaking they incorporate key features like scale separation and chiral gauge theories supported on D6-branes wrapping intersecting three-cycles of $X_6$.

Nevertheless, we recall that the constructions in \cite{Villadoro:2005cu,DeWolfe:2005uu} have an important caveat: they do not solve the 10d equations of motion and Bianchi identities, unless localized sources like D6-branes and O6-planes are smeared over the internal dimensions \cite{Acharya:2006ne}. In section \ref{cy-subsec: beyond smearing} we reviewed how the problem can be addressed  by formulating the geometrical configuration as a perturbative expansion, of which the leading term is the smeared-source Calabi--Yau approximation, and where the expansion parameter is essentially the AdS$_4$ cosmological constant \cite{Saracco:2012wc}. The first-order correction to the smeared background was found in \cite{Junghans:2020acz,Marchesano:2020qvg}, displaying localized sources and a natural expansion parameter $R^{-4/3} \sim g_s^{4/3}$, where $R$ is the AdS$_4$ radius in string units and $g_s$ is the average 10d string coupling.\footnote{Another caveat surrounding these constructions is that they combine O6-planes and a non-vanishing Romans mass, which makes difficult to understand them microscopically. However, T-dual versions of the solutions in \cite{Junghans:2020acz,Marchesano:2020qvg} have been constructed in \cite{Cribiori:2021djm} with similar properties, vanishing Romans mass and an 11d description.}  

The aim of this chapter is to revisit the stability of the AdS$_4$ vacua in \cite{DeWolfe:2005uu,Camara:2005dc,Narayan:2010em}, with the vantage point of the more precise 10d description, summarized in section \ref{cy-subsec: beyond smearing}. We consider $\cN=1$ and $\cN=0$ vacua which, in the smearing approximation,  are related by an overall sign flip of the internal four-form flux $G_4$. These were considered in \cite{DeWolfe:2005uu,Narayan:2010em} for $X_6 = T^6/ (\Z_3 \times \Z_3)$, and generalized to arbitrary Calabi--Yau geometries in \cite{Marchesano:2019hfb}, corresponding to the first two rows of table \ref{cy-table: summary ads vacua} (\textbf{A1-S1} branch). In section \ref{cy-subsec: beyond smearing} we presented a rather explicit 10d-solution for the SUSY branch in terms of an $SU(3)\times SU(3)$-structure deformation of the Calabi--Yau metric. For their non-supersymmetric cousins we use the approach in \cite{Junghans:2020acz} to provide a solution at the same level of approximation. 
In this setup, we consider 4-dimensional membranes that come from wrapping D$(2p)$-branes on $(2p-2)$-cycles of $X_6$. These membranes couple to  fluxes that support the AdS$_4$ background, more precisely to the dynamical fluxes of the 4-dimensional theory \cite{Lanza:2019xxg,Lanza:2020qmt}. Therefore, even if there could be other non-perturbative decay channels, the $\cN=0$ sharpening of the WGC suggests that at least one of these membranes or a bound state of them should be superextremal, and thus a candidate to yield an expanding bubble. Note that these AdS$_4$ backgrounds have  not been constructed as near-horizon limits of a backreacted black brane solutions, so it is a priori not clear which membrane is the most obvious candidate to fulfil the conjecture. 

It was argued in \cite{Aharony:2008wz,Narayan:2010em} that D4-branes wrapping either holomorphic or anti-holomorphic cycles of $X_6$ saturate a BPS bound for the $\cN=1$ and $\cN=0$ vacua mentioned above, while D2-branes and D6-branes wrapping four-cycles never do. By looking at each of their couplings to the fluxes supporting the AdS$_4$ background and their tension we recover the same result. Remarkably, we not only do so for the smeared-source Calabi--Yau  approximation considered in \cite{Aharony:2008wz,Narayan:2010em}, but also when the first-order corrections to this background are taken into account. It follows that, at this level of approximation, such (anti-)D4-branes give rise to extremal objects that can at most mediate marginal decays. This extends to bound states of D6, D4 and D2-branes, in the sense that they do not yield any superextremal 4-dimensional membrane. 

We then turn to consider D8-branes wrapping $X_6$. Due to a Freed--Witten anomaly generated by the $H$-flux, D6-branes must be attached to the D8 worldvolume. From the 4-dimensional perspective, these are membranes that not only change the Romans mass flux $F_0$ when crossing them, but also the number of space-time filling D6-branes, so that the tadpole condition is still satisfied. It turns out that the presence of attached D6-branes acts as a force on the D8-branes, and exactly cancels the effect of their charge and tension in supersymmetric vacua, as it should happen for a BPS object. This provides a rationale  for the precise relation between $F_0$, $R$, $g_s$ found in \cite{DeWolfe:2005uu}. In $\cN=0$ vacua the energetics of D8-branes is more interesting, because curvature corrections induce D4-brane charge and tension on their worldvolume. The induced tension is in general negative, implying that the D8-brane is dragged towards the boundary of $\cN=0$ AdS$_4$.  As we argue, this corresponds to a superextremal 4-dimensional membrane that mediates a decay to another non-supersymmetric vacuum with larger $|F_0|$ and fewer D6-branes, in agreement with the sharpened Weak Gravity Conjecture.

This picture is however incomplete, since it relies on the smeared description. First-order corrections to the Calabi--Yau background modify the D8-brane action by terms comparable to an induced D4-brane tension. In fact, beyond the smearing approximation the D8/D6 system should be treated as a BIon-like solution, whose tension differs from the sum of D8 and D6-brane tensions. We compute this difference for $X_6 = T^6/(\Z_2 \times \Z_2)$, and find that this new correction is comparable to curvature-induced effects. Nevertheless, for simple D-brane configurations we find that it is also negative, and so the D8-branes are still dragged towards the $\cN=0$ AdS$_4$ boundary. If the same is true in more general setups, then the combined effect of curvature and BIon-like corrections provide a non-perturbative instability for $\cN=0$ AdS$_4$ vacua with space-time filling D6-branes, in line with the AdS Instability Conjecture.

The chapter addresses the subject in increasing level of complexity as follows. In section \ref{bio-s:memb} we discuss the energetics of membranes in AdS$_4$ backgrounds with four-form fluxes, which we then use as a criterion for membrane extremality. In section \ref{bio-s:dgkt} we review the $\cN=1$ AdS$_4$ Calabi--Yau orientifold vacua with fluxes in the smearing approximation, and classify BPS membranes that come from wrapped D-branes. Section \ref{bio-s:nonsusy} does the same for non-supersymmetric AdS$_4$, finding superextremal membranes thanks to curvature corrections, and section \ref{bio-s:insta} argues that they mediate actual decays in the 4-dimensional theory. Section \ref{bio-s:nonsmeared} describes the 10d background with localized sources for $\cN=1$ and $\cN=0$ AdS$_4$ vacua, and shows that D4-branes saturate a BPS bound in both cases. Section \ref{bio-s:bion} describes D8/D6-brane systems as BIons, and shows that they are BPS in $\cN=1$ but feel a net force in $\cN=0$ vacua. We finally present our conclusions in section \ref{bio-s:conclu}.

Several technical details have been relegated to the appendices.  Appendix \ref{bio-ap:10deom} shows that the  backgrounds of section \ref{bio-s:nonsmeared} satisfy the 10d equations of motion. Appendix \ref{bio-ap:dbi} shows how the BIon profile of section \ref{bio-s:bion} linearizes the  DBI action. Appendix \ref{bio-ap:IIBion} relates this profile to 4-dimensional strings in type IIB warped Calabi--Yau compactifications and to SU(4) instantons in Calabi--Yau four-folds. 


\section{Membranes in \texorpdfstring{AdS$_4$}{AdS4}}
\label{bio-s:memb}

In a 4-dimensional Minkowski background with $\cN=1$ supersymmetry, simple examples of static BPS membranes are  3d hyperplanes of $\pr^{1,3}$ including the time-like direction. Analogous objects in anti-de Sitter can be described by considering the Poincar\'e patch of AdS$_4$, whose metric reads
\be
ds^2_4 =e^{\frac{2z}{R}} (-dt^2 + d\vec{x}^2) + dz^2\, ,
\label{bio-PPatch}
\ee
with $R$ the AdS length scale, $\vec{x} = (x^1, x^2)$, and all coordinates range over $\pr$. In such coordinates, the AdS$_4$ boundary is located at $z = \infty$. Similarly to the Minkowski case, one may consider a membrane that spans the coordinate $t$ and a surface within $(x^1, x^2, z)$. Particularly simple is the case where the surface is the plane $z=z_0$, with $z_0 \in \pr$ fixed. While this object may look like the BPS membranes of Minkowski, the tension of such a membrane decreases exponentially as we take $z_0 \to -\infty$. Therefore, if we place such an object in AdS$_4$ and take the probe approximation, it will inevitably be driven away from the boundary and it cannot be BPS. 

This can be avoided if on top of the AdS$_4$ metric we consider a four-form flux background $F_4$, to whose three-form potential $C_3$ the membrane couples as $-\int C_3$. Indeed, if we have
\be
\langle F_4 \rangle = -\frac{3Q}{R} {\rm vol}_4
\qquad \Longrightarrow \qquad \langle C_3 \rangle = Q\, e^{\frac{3z}{R}} dt \wedge dx^1 \wedge dx^2 \, ,
\label{bio-3form}
\ee
and $Q$ coincides with the tension of the membrane $T$, then the variation of the tension when moving in the $z$ coordinate is compensated by the potential energy $-\int \langle C_3 \rangle$ gained because of its charge. Moving along this coordinate is then a flat direction and the membrane may be BPS. If $Q > T$ one may still find BPS membrane configurations, but they cannot be parallel to the boundary. We instead have that force cancellation occurs for embeddings of the form
\be
\left\{t, x^1, x^2 = \pm \frac{R}{\sqrt{\frac{Q^2}{T^2}-1}} \, e^{-\frac{z}{R}} + c\right\}\, , \qquad c \in \pr\, .
 \label{bio-QnotT}
\ee

Four-form flux backgrounds are ubiquitous in AdS$_4$ backgrounds obtained from string theory, and in particular in those with 4-dimensional $\cN=1$ supersymmetry or $\cN=0$ spontaneously broken. The membrane profiles $z=z_0$ and \eqref{bio-QnotT} were found in \cite{Koerber:2007jb} in the context of $\cN=1$ AdS$_4$ backgrounds obtained from type II string theory, but from the above discussion it follows that they can also be present in backgrounds with supersymmetry spontaneously broken by fluxes.  One can in fact see that  the set of 4-dimensional fluxes arising from the compactification is directly related to the spectrum of BPS branes, as well as to the internal data specifying the supersymmetry generators. 

In the following we will be chiefly concerned with those membranes whose profile is given by $z=z_0$. As argued in \cite{Koerber:2007jb}, for $Q=T$ and at $z \to \infty$ they capture the BPS bound of a spherical membrane in global coordinates at asymptotically large radius. It is precisely the domain walls that correspond to spherical membranes near the AdS boundary that determine if the non-perturbative decay of one vacuum to another with lower energy is favourable or not. Thus, by considering the energetics of membranes in the Poincar\'e patch with $z = z_0 \to \infty$ we may detect if there could be some domain wall triggering such a decay. If all membranes satisfy $T > Q$ such a decay should not occur, if $Q=T$ it should be marginal, and  if $T<Q$ the AdS background may develop a non-perturbative instability. According to the conjectures in \cite{Ooguri:2016pdq,Freivogel:2016qwc}, any $\cN=0$ AdS background of this sort should have at least one non-perturbative instability towards a new vacuum, and therefore a membrane with $T<Q$. In the following sections we will consider the membranes that appear from wrapping D-branes on internal cycles in backgrounds of the form AdS$_4 \times X_6$, where $X_6$ admits a Calabi--Yau metric, and compute $T$ and $Q$ for them. In particular we will consider the $\cN=1$ vacua of \cite{DeWolfe:2005uu} and some of the non-supersymmetric vacua found in \cite{Camara:2005dc,Narayan:2010em,Marchesano:2019hfb}, which are stable at the perturbative level. We will not only consider the Calabi--Yau approximation of these references, but also the solutions with localized sources found in \cite{Junghans:2020acz,Marchesano:2020qvg}. As we will see, for non-supersymmetric vacua the answer is not the same once this more precise picture is taken into account.


\section{Supersymmetric \texorpdfstring{AdS$_4$}{AdS4} orientifold vacua}
\label{bio-s:dgkt}

Examples of membranes satisfying $Q=T$ are typically found in supersymmetric AdS$_4$ backgrounds, where the equality follows from saturating a BPS bound. In this section we analyze for which membranes this condition is met for the supersymmetric type IIA flux compactifications of \cite{DeWolfe:2005uu}, for an arbitrary Calabi--Yau geometry $X_6$, in the approximation of smeared sources \cite{Acharya:2006ne}. With the simple criterion $Q=T$ one can reproduce the results of \cite{Aharony:2008wz} for membranes arising from D2, D4 and D6-branes wrapping internal cycles of $X_6 = T^6/(\Z_3 \times \Z_3)$, and extend them to any Calabi--Yau manifold. Furthermore, one may detect an additional set of BPS membranes, namely those coming from D8-branes wrapping $X_6$, to which space-time filling D6-branes are attached. This last feature makes such membranes quite special, particularly when one considers them for non-supersymmetric AdS$_4$ backgrounds and beyond the smearing approximation.

\subsection{10d background in the smearing approximation}
\label{bio-ss:smeared}

We consider type IIA string theory compactified in an orientifold of $\mathcal{M}_4 \times X_6$, where $X_6$ is a compact Calabi--Yau three-fold, following the definitions and conventions detailed in sections \ref{cy-subsec: orientifolds} and \ref{cy-subsec: democratic formulation}. Let us recapitulate the core elements.

In the absence of localized sources, each $p$-form within ${\bf \bar{G}}$ is quantized, so one can define the internal RR flux quanta in terms of the following integer numbers
\begin{equation}
m \, = \,  \ell_s G_0\, ,  \quad  m^a\, =\, \frac{1}{\ell_s^5} \int_{X_6} \bar{G}_2 \wedge \tilde \omega^a\, , \quad  e_a\, =\, - \frac{1}{\ell_s^5} \int_{X_6} \bar{G}_4 \wedge \omega_a \, , \quad e_0 \, =\, - \frac{1}{\ell_s^5} \int_{X_6} \bar{G}_6 \, ,
\label{bio-RRfluxes}
\end{equation}
with $\omega_a$, $\tilde \omega^a$ defined in table \ref{cy-table: harmonic basis} and
\be
J_{\rm CY} = t^a \omega_a\, , \qquad - J_{\rm CY} \wedge J_{\rm CY} = {\cal K}_a \tilde{\omega}^a\, . 
\ee
Here ${\cal K}_a \equiv {\cal K}_{abc} t^bt^c$, with ${\cal K}_{abc} = - \ell_s^{-6} \int_{X_6} \omega_a \wedge \omega_b \wedge \omega_c$ the Calabi--Yau triple intersection numbers and $-\frac{1}{6} J_{\rm CY}^3 = - \frac{i}{8} \Omega_{\rm CY} \wedge \bar{\Omega}_{\rm CY}$ its volume form.

The fixed locus $\Pi_{\rm O6}$ of the orientifold involution ${\cal R}$ is one or several 3-cycles of $X_6$ in which O6-planes are located. Further localized sources may include D6-branes wrapping three-cycles and coisotropic D8-branes \cite{Font:2006na}. Together with the contribution of background fluxes they must cancel the O6-plane RR charge. Thus, in the presence of D6-branes and O6-planes the Bianchi identities for the RR fluxes \eqref{cy-eq: BI def} read
\be
d\hat{G}_0 = 0\, , \qquad d \hat{G}_2 = \hat{G}_0 H - 4 \d_{\rm O6} +   N_\a \d_{\rm D6}^\a \, ,  \qquad d \hat{G}_4 = \hat{G}_2 \wedge H\, , \qquad d\hat{G}_6 = 0\, ,
\label{bio-BIG}
\ee
where  we have defined $\d_{\rm D6/O6}\equiv \ell_s^{-2}  \d(\Pi_{\rm D6/O6})$. This in particular implies that
\be
{\rm P.D.} \left[4\Pi_{\rm O6}- N_\a \Pi_{\rm D6}^\a\right] = m [\ell_s^{-2} H] \, ,
\label{bio-tadpole}
\ee
constraining the quanta of Romans parameter and NS flux. Let us in particular choose P.D.$[\ell_s^{-2}H] = h [\Pi_{\rm O6}] = h [\Pi_{\rm D6}^\a]$, $\forall \a$. We then find the constraint
\be
mh +N = 4\, ,
\label{bio-tadpole2}
\ee
with $N$ the number of D6-branes wrapping $\Pi_{\rm O6}$. Supersymmetry in addition implies that $mh$ and $N$ are non-negative, yielding a finite number of solutions.\footnote{In several instances (e.g., toroidal orbifolds) $[\Pi_{\rm O6}]$ may be an integer multiple $k$ of a three-cycle class. In those cases $h, N$ need not be integers, but instead $kh, kN \in \mathbb{Z}$, allowing for a richer set of solutions to \eqref{bio-tadpole2}.} 

The constraint on sign$(mh)$ can be seen by means of a 4-dimensional analysis of the potential generated by background fluxes, following \cite{DeWolfe:2005uu,Camara:2005dc}. Such a potential was obtained in \cite{Grimm:2004ua} by combining the superpotential generated by the RR and NS flux quanta and the classical K\"ahler potential of Calabi--Yau orientifolds without fluxes, as we reviewed in the previous chapter. The most important result for the current study is the existence of a discretum of $\cN=1$ AdS$_4$ vacua, associated to an internal Calabi--Yau manifold $X_6$ such that the internal fluxes satisfy
\be
\ell_s [ H ]  = \frac{2}{5} m g_s [\re \Omega_{\rm CY} ] \, , \qquad  \hat{G}_2  =  0\, ,  \qquad  \ell_s \hat{G}_4  = -\hat{e}_a  \tilde{\omega}^a  = -\frac{3}{10} m \, \cK_a \tilde{\omega}^a  \, , \qquad  \hat{G}_6 =  0\, , 
\label{bio-intfluxsm}
\ee
where we have defined
\be
\hat{e}_a = e_a - \oh \frac{\cK_{abc} m^bm^c}{m}\, .
\label{bio-hate}
\ee
This is equation \eqref{cy-eq: SU3 solutions} from the 10-dimensional perspective and the first row of table \ref{cy-table: summary ads vacua} from the 4-dimensional point of view.

Care should however be taken when interpreting such relations from the viewpoint of the actual 10d supergravity solution, since the presence of fluxes and localized sources will deform the internal geometry away from the Calabi--Yau metric, and a $G_2$ and $G_4$ of the above form will never satisfy the Bianchi identities \eqref{bio-BIG}, as we discussed in section \ref{cy-subsec: smearing approx}.\footnote{Additionally, in the presence of D6-brane moduli the integral of $\bar{G}_2$ will depend on them. This can  be dealt with by translating such a dependence into a superpotential involving both open and closed string moduli \cite{Marchesano:2014iea,Carta:2016ynn}.} The standard way to deal with both issues is to see \eqref{bio-intfluxsm} as a formal solution in which all localized sources have been smeared \cite{Acharya:2006ne}. This so-called smeared solution is then the leading term in a perturbative series that should converge to the actual background \cite{Saracco:2012wc}, with expansion parameter $g_s^{4/3}$, and where sources are localized \cite{Junghans:2020acz,Marchesano:2020qvg}.  Instead of \eqref{bio-intfluxsm}, the relations that this background must satisfy are
\be
[ H ]  = \frac{2}{5} \hat{G}_0 g_s  [\re \Omega_{\rm CY} ] \, , \quad \int_{X_6} \hat{G}_2 \wedge \tilde{\omega}^a =  0\, ,  \quad \frac{1}{\ell_s^6} \int_{X_6} \hat{G}_4  \wedge \omega_a  =  - \frac{3}{10} \hat{G}_0 {\cal K}_a \, , \quad  \hat{G}_6  =  0\, , 
\label{bio-intflux}
\ee
where $g_s$ is the average value of $e^{\phi}$, with $\phi$ a varying 10d dilaton. This value determines the AdS$_4$ length scale in the 10d string  frame $R$, from the following additional relation
\be
\frac{\ell_s}{R} = \frac{1}{5} |m| g_s  \, .
\label{bio-Rads}
\ee
There is in addition a non-trivial warp factor, and the Calabi--Yau metric on $X_6$ is deformed to an $SU(3) \times SU(3)$-structure metric, as described in chapter \ref{cy-subsec: beyond smearing}. We will discuss this more accurate background in section \ref{bio-ss:mpqt}. For now we focus on the smearing approximation. 

It follows from such a description that the Calabi--Yau volume 
\be
{\cal V}_{\rm CY}  = -\frac{1}{6\ell_s^6} \int_{X_6} J_{\rm CY}^3 =  \frac{1}{6} {\cal K}_{abc}t^at^bt^c \equiv  \frac{1}{6} {\cal K}\, ,
\ee
depends on $m$ and $\hat{e}_a$, grwowing larger when we increase their absolute value. One can then for instance see that $1/R$ grows as we increase $h$ or $m$, and decreases as we increase $\hat{e}_a$. A more precise result can be obtained from the 4-dimensional analysis, which yields the following 4-dimensional Einstein frame vacuum energy 
\be
\Lambda = - \frac{16\pi}{75\kappa_4^4} e^K {\cal K}^2 m^2 \, ,
\label{bio-lambda}
\ee
where $K$ is the K\"ahler potential, given by \eqref{cy-eq: KK}. One can then see that $\Lambda$ scales like $|m|^{5/2}$, as in the explicit toroidal solutions in \cite{DeWolfe:2005uu,Camara:2005dc}. Recall however that the allowed values for $m$ are bounded by the tadpole condition \eqref{bio-tadpole2}.

Finally, one can include the effect of curvature corrections to the 4-dimensional analysis, following \cite{Palti:2008mg,Escobar:2018rna}. We will only include those corrections dubbed $K^{(1)}_{ab}$ and $K_a^{(2)}$ in \cite{Palti:2008mg,Escobar:2018rna}, given by
\be
K^{(1)}_{ab} = \frac{1}{2} {\cal K}_{aab} \, , \qquad K_a^{(2)} = -\frac{1}{24\ell_s^6} \int_{X_6} c_2(X_6) \wedge \omega_a\, ,
\label{bio-Kcurv}
\ee
which respectively correspond to $\cO(\alpha')$ and $\cO(\alpha'^2)$ corrections, since higher orders will be beyond the level of accuracy of our analysis. If $\{[\ell_s^{-2}\omega_a]\}_a$ is dual to a basis of Nef divisors, then $K_a^{(2)} \geq 0$ \cite{Miyaoka1987}. The effect of such corrections is to redefine the background flux quanta as follows
\be
e_0 \to e_0 - m^a K_a^{(2)} \, , \qquad \quad e_a \to e_a  - K_{ab}^{(1)} m^b + m K_a^{(2)} \, ,
\label{bio-curvflux}
\ee 
so in particular they modify the flux combinations \eqref{bio-hate} that determine the K\"ahler moduli vevs. This modification makes more involved the scaling of $\Lambda$ with $m$, but since in the regime of validity we have that ${\cal K}_a \gg K_a^{(2)}$, it turns out that $\Lambda \sim |m|^{5/2}$ is still a good approximation. 

\subsection{4d BPS membranes}
\label{bio-ss:4dmem}

Given a type II flux compactification to $\CN=1$ AdS$_4$, one may study the spectrum of BPS D-branes via $\kappa$-symmetry or pure spinor techniques, as in \cite{Aharony:2008wz,Koerber:2007jb}, and in particular determine those D-branes that give rise to BPS membranes from the 4-dimensional perspective. In the following we will take the more pedestrian viewpoint of section \ref{bio-s:memb} to identify such BPS membranes. This criterion will also be useful when considering non-supersymmetric AdS$_4$ vacua. 

An analysis of 4-dimensional BPS membranes parallel to the AdS$_4$ boundary in the Poincar\'e patch was carried out in \cite{Aharony:2008wz}, for the particular case $X_6 = \mathbb{T}^6/(\Z_3 \times \Z_3)$ of \cite{DeWolfe:2005uu}, in the smearing approximation. It was found that D4-branes wrapping holomorphic cycles are BPS, while D2 and D6 branes cannot be so. Let us see how to recover such results and extend them to general Calabi--Yau geometries using the picture of section \ref{bio-s:memb}. For this we recall that in the type IIA democratic formulation the RR background fluxes take the form
\be
{\bf G} = {\rm vol}_4 \wedge \tilde{\bf G} + \hat{\bf G}\, ,
\label{bio-demoflux}
\ee
where ${\rm vol}_4$ is the AdS$_4$ volume form and $\tilde{G}$ and $\hat{\bf G}$ only have internal indices, satisfying the relation $\tilde{\bf G} = - \lambda ( *_6 \hat{\bf G})$. Therefore from \eqref{bio-intfluxsm} and \eqref{bio-Rads} we find the following fluxes that translate into a 4-dimensional four-form background 
\be
G_6 =  - \frac{3\eta}{Rg_s} {\rm vol}_4 \wedge J_{\rm CY}\, , \qquad G_{10} =  - \frac{5\eta}{6 R g_s} {\rm vol}_4 \wedge J^3_{\rm CY}\, ,
\label{bio-610fluxes}
\ee
with $\eta =  {\rm sign }\, m$. Contrarily, no component of ${\rm vol}_4$ appears in $G_4$ or $G_8$. We hence deduce the following  couplings for 4-dimensional membranes arising from D(2$p$)-branes wrapping (2$p-$2)-cycles of $X_6$:
\be
Q_{\rm D2} = 0 \, , \qquad Q_{\rm D4} =  e^{K/2} \frac{\eta}{\ell_s^2} \int_\Sigma J_{\rm CY} \, , \qquad Q_{\rm D6} = 0\, , \qquad Q_{\rm D8} =  -\frac{5}{3} e^{K/2} \eta\, q_{\rm D8} {\cal V}_{\rm CY}\, ,
\label{bio-QDGKT}
\ee
expressed in 4-dimensional Planck units. Here $\Sigma$ is the two-cycle wrapped by the D4-brane, and $q_{\rm D8} = \pm 1$ specifies the orientation with which the D8-brane wraps $X_6$. This implies that for $\eta=1$ a BPS D4-brane must wrap a holomorphic two-cycle with vanishing worldvolume flux $\cF = B + \frac{\ell_s^2}{2\pi} F$ to be BPS, so that $e^{K/2} \ell_s^{-2}\int_\Sigma J_{\rm CY} = e^{K/2} {\rm area}(\Sigma)/\ell_s^2 \equiv T_{\rm D4}$, while for $\eta = -1$ the two-cycle must be anti-holomorphic. This choice of orientation for $\Sigma$ can be understood from looking at how the four-form varies when crossing the D4-brane from $z = \infty$ to $z = -\infty$. In both cases, due to \eqref{cy-eq: BI def} and the choice of orientation for $\Sigma$ one decreases the absolute value of the four-form flux quanta $\hat{e}_a$, and therefore the vacuum energy. This is consistent with our expectations, as it permits to have a BPS domain-wall solution mediating a marginal decay from a vacuum with higher energy (at $z = \infty$) to one with lower energy (at $z = -\infty$). Considering this set of BPS membranes allows us to scan over the set of vacua with different four-form flux quanta.  Differently, D6-branes wrapping four-cycles of $X_6$ and D2-branes can never yield 4-dimensional BPS membranes. This indeed reproduces and generalizes the results found in \cite{Aharony:2008wz}, adapted to our conventions. 

It however remains to understand the meaning of $Q_{\rm D8}$, which naively does not seem to allow for BPS membranes that come from wrapping (anti-)D8-branes on $X_6$. On general grounds one would expect that such BPS membranes exist as well, in order to scan over the different values of $m$. In particular, one would expect that for $\eta=1$ D8-branes $(q_{\rm D8} =1)$ wrapping $X_6$ are BPS, while for $\eta=-1$ the same occurs for anti-D8-branes $(q_{\rm D8} = -1)$. Indeed, when crossing the corresponding domain wall from $z = \infty$ to $z = -\infty$ the value of $|m|$ increases and the vacuum energy decreases in both setups, paralleling the case for D4-branes. However, the factor of $5/3$ and a sign prevent achieving the necessary BPSness condition $Q_{\rm D8}=T_{\rm D8} \equiv e^{K/2} {\cal V}_{\rm CY}$.

The resolution to this puzzle comes from realising that D8-branes wrapping $X_6$ cannot be seen as isolated objects. Instead, D6-branes must be attached to them, to cure the Freed--Witten anomaly generated on the (anti-)D8-brane by the NS flux background $H$\footnote{From the definitions introduced in the DBI action of a D-brane \eqref{basic-eq: DBI action}, we know that the quantity $\mathcal{F}$, that combines the B-field along the D8 with the brane worldvolume flux,  must satisfy
\begin{equation}
d\mathcal{F}_{D8}=H|_{D8}\,.
\end{equation}
Since the $D8$ covers the full internal volume we can choose any 3-cycle  in the dual class of $[H]$ and integrate the above expression. The left hand side will vanish, but the right hand will not, as we are demanding the presence of a non-trivial $H$ background motivated by moduli stabilization requirements. This conflict, known as the Freed-Witten anomaly, can be solved by adding external sources.}. In the above setup the D6-branes will be wrapping a three-cycle of $X_3$ on the Poincar\'e dual class to $\eta [\ell_s^{-2} H] = |h| [\Pi_{\rm O6}]$, and extend along the 4-dimensional region of AdS$_4$ $(t,x^1,x^2) \times [z_0 , \infty)$ that is bounded by the 4-dimensional membrane. More generally, we need an excess of space-time filling D6-branes wrapping $\Pi_{\rm O6}$ on the interval $[z_0 , \infty)$ to the right of the (anti-)D8-brane, as compared to the ones in the left-interval $(-\infty, z_0]$ to cancel the said Freed--Witten anomaly: 
\be
N_{\rm right} - N_{\rm left} = |h| \, ,
\label{bio-excessD6}
\ee
see figure \ref{bio-fig:D8D6}. Since $m$ jumps by $\eta$ when crossing the membrane from right to left (as a consequence of \eqref{basic-eq: D8 solution}), $mh$ jumps by $|h|$, and so \eqref{bio-excessD6} guarantees that the tadpole condition \eqref{bio-tadpole2} is satisfied at both sides. 

\begin{figure}[htbp]
    \centering
    \includegraphics[width=10cm]{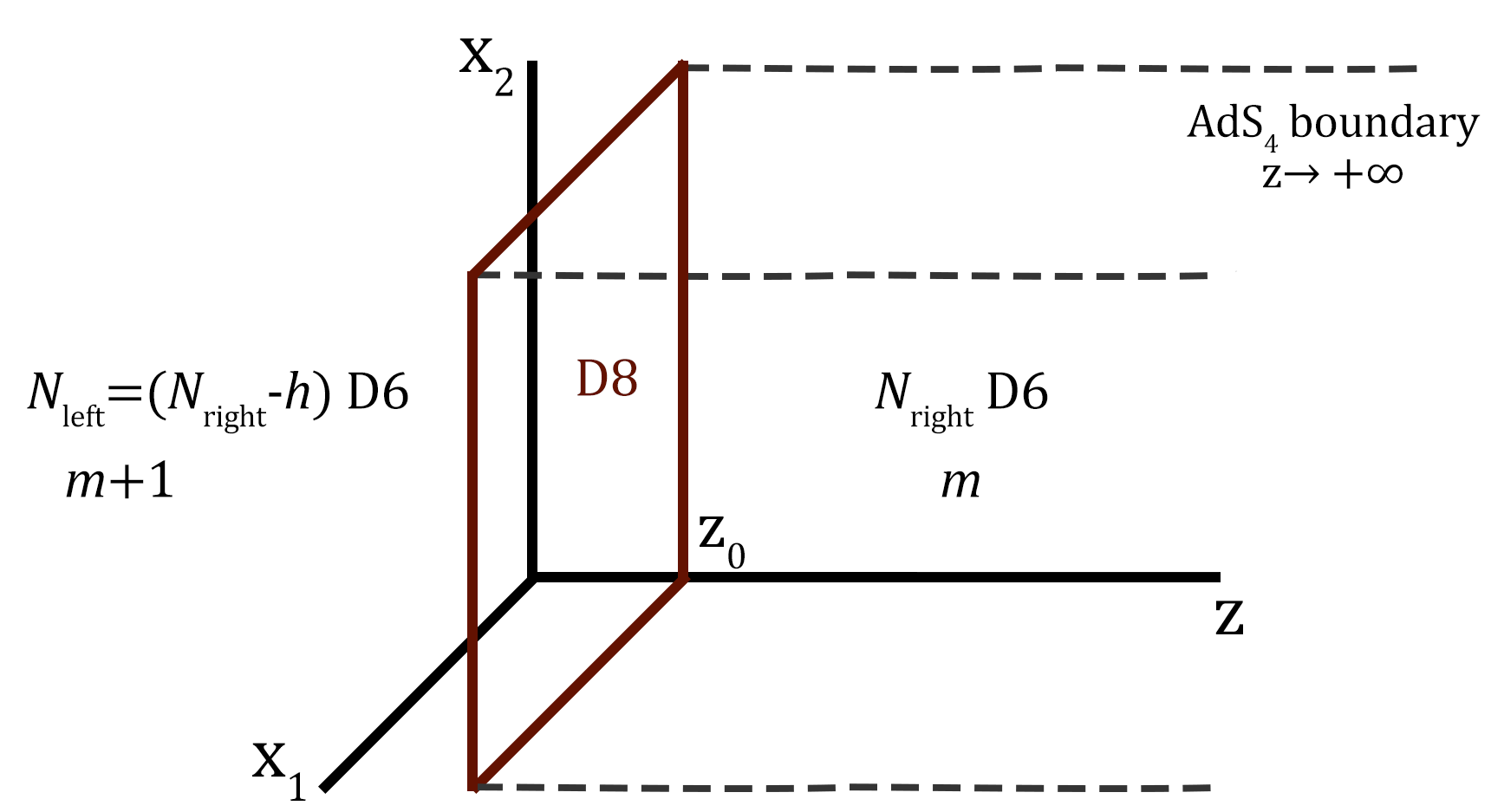}
    \caption{To cure the Freed--Witten anomaly induced by the $H$-flux on the D8-brane worldvolume, an excess of $|h|$ space-time filling D6-branes must be attached from its position to the AdS$_4$ boundary. We take $m, h >0$ in the figure. Note the jump of the value of Romans mass can be derived from the standard D-brane supergravity solution \cite{Bergshoeff:2001pv} described in \eqref{basic-eq: D8 solution}.}
    \label{bio-fig:D8D6}
\end{figure}

Since the number of space-time filling D6-branes is different at both sides of the D8-brane, their presence will induce an energy dependence in terms of the D8-brane position. Indeed, if we decrease $z_0$ and move the D8-brane away from the AdS$_4$ boundary the region of AdS$_4$ filled by $N_{\rm right}$ D6-branes will grow, and so will the total energy of the system. As a result, the D6-brane jump induced by the Freed--Witten anomaly pulls the D8-branes towards the boundary of AdS$_4$. It turns out that this effects precisely cancels the effect of the tension $T_{\rm D8}$ and coupling $Q_{\rm D8}$ of the D8-brane, which both drag the 4-dimensional membrane away from the AdS boundary. 

One can derive such a cancellation via a microscopic calculation of the DBI+CS action for the D8/D6 system, dimensionally reduced to 4d. Of course, from the viewpoint of the 4-dimensional membrane the tension of space-time filling D6-branes extended along $(-\infty, z_0]$ and $[z_0, \infty)$ is infinite. Nevertheless, one may compute how the energy of the system varies  as we modify the D8-brane position $z_0$. Indeed, the DBI contribution to the action is given by the sum of the following two terms:
\begin{align}
 \label{bio-DBID8}
    S_{\rm DBI}^{\rm D8} =& -\frac{1}{g_s} {\cal V}_{\rm CY}e^{\frac{3z_0}{R}} \frac{2\pi}{\ell_s^3} \int  dt dx^1 dx^2\, , \\
    S_{\rm DBI}^{\rm D6} = &
    -\frac{1}{g_s}  {\cal V}_{\Pi_{\rm O6}} \frac{2\pi}{\ell_s^4} \left(N_{\rm left} \int_{-\infty}^{z_0} dz'e^{\frac{3z'}{R}} +N_{\rm right} \int_{z_0}^\infty dz'e^{\frac{3z'}{R}}\right) \int  dt dx^1 dx^2\, ,
    \label{bio-DBID6}
    \end{align}
with
\be
{\cal V}_{\Pi_{\rm O6}} = \frac{1}{\ell_s^3} \int_{\Pi_{\rm O6}} \im\, \Omega_{\rm CY} = \frac{1}{h \ell_s^5}  \int_{X_6}  \im \Omega_{\rm CY} \wedge H  = \frac{8}{5} \frac{m}{h} g_s {\cal V}_{\rm CY} = \frac{8\ell_s}{|h|R}{\cal V}_{\rm CY} \, ,
\ee
where we have used that in our conventions O6-planes and BPS D6-branes are calibrated by $\im\, \Omega_{\rm CY}$, and then the relations \eqref{bio-intflux} and \eqref{bio-Rads}. Let us now consider an infinitesimal variation $z_0 \to z_0 + \ell_s \epsilon$.  The variation of these actions is
\begin{align}
\label{bio-DBID8var}
\delta_\epsilon  S_{\rm DBI}^{\rm D8} & = -\frac{3}{Rg_s } {\cal V}_{\rm CY}\, e^{\frac{3z_0}{R}}   \frac{2\pi}{\ell_s^2} \int   dt dx^1 dx^2\, ,\\
\delta_\epsilon S_{\rm DBI}^{\rm D6} & = -\frac{8}{Rg_s} \frac{N_{\rm left} - N_{\rm right}}{|h|} {\cal V}_{\rm CY}\, e^{\frac{3z_0}{R}}   \frac{2\pi}{\ell_s^2} \int   dt dx^1 dx^2 =  \frac{8}{Rg_s} {\cal V}_{\rm CY}\, e^{\frac{3z_0}{R}}   \frac{2\pi}{\ell_s^2} \int   dt dx^1 dx^2\, .
\label{bio-DBID6var}
\end{align}
That is, the dragging effect of the D6-branes ending on the D8-brane overcomes the effect of its tension, acting like an additional coupling $Q^{\rm eff}_{\rm D6} =  \frac{8}{3} e^{K/2} {\cal V}_{\rm CY}$.  This precisely compensates the coupling of the 4-dimensional membrane made up from a D8-brane in the case $\eta =1$ and from an anti-D8-brane in the case $\eta =-1$, as claimed. Microscopically, this cancellation is seen from the variation of the (anti-)D8-brane Chern-Simons action. By evaluating the coupling to the RR potential $C_9$ that corresponds to \eqref{bio-610fluxes} and integrating over $X_6$ one obtains:
\be
S_{\rm CS}^{D8} = q_{\rm D8} \frac{2\pi}{\ell_s^9} \int C_9  =  -q_{\rm D8}\,  \eta \frac{5}{3} g_s^{-1}  {\cal V}_{\rm CY}  e^{\frac{3z_0}{R}} \, \frac{2\pi}{\ell_s^3} \int  dt dx^1 dx^2 \, .
\label{bio-CSD8}
\ee
It is then easy to see that for $q_{\rm D8} = \eta$ the variation $\delta_\epsilon  S_{\rm CS}^{\rm D8}$ precisely cancels \eqref{bio-DBID8var}+\eqref{bio-DBID6var}. Therefore, the effect of the D6-branes can be understood as generating an effective coupling $Q_{\rm D8/D6}^{\rm eff} = Q_{\rm D8} + Q^{\rm eff}_{\rm D6} =  \eta q_{\rm D8}  e^{K/2} {\cal V}_{\rm CY}$. Indeed, notice that if one chose $q_{\rm D8} = - \eta$ then the Freed--Witten anomaly would be opposite and the D6-branes would be extending along $z \in (-\infty, z_0]$. This would result into $Q_{\rm D8/D6}^{\rm eff} = - e^{K/2} {\cal V}_{\rm CY}$, destabilizing the system towards $z_0 \to -\infty$. 

\subsubsection*{Considering bound states}

In general, the Chern-Simons action of a D8-brane reads
\be
S_{\rm CS}^{\rm D8} =  \frac{2\pi}{\ell_s^9} \int P\left[ {\bf C} \wedge e^{-B}\right] \wedge e^{- \frac{\ell_s^2}{2\pi}  F} \wedge \sqrt{\hat{A}({\cal R})}\, ,
\ee
where ${\bf C} = C_1 + C_3 + C_5 + C_7 + C_9$ and $\hat{A}({\cal R}) = 1 + \frac{1}{24} \frac{\Tr R^2}{8\pi^2} + \dots $ is the A-roof genus. These couplings encode that in the presence of a worldvolume flux and/or curvature, we actually have a bound state of a D8 with lower-dimensional D-branes. If the bound state is BPS, then its tension will be a sum of D8 and D4-brane tensions. Taking also into account the effect of the D6-branes ending on it we have that
\be
T_{\rm D8}^{\rm total} = T_{\rm D8} + \left(K^F_a - K_a^{(2)}\right)  T_{\rm D4}^a\, ,
\label{bio-TD8tot}
\ee
where we have defined
\begin{align}
    T_{\rm D4}^a &= e^{K/2} t^a\,,\\
    K^F_a &= \frac{1}{2\ell_s^6} \int_{X_6} \cF \wedge \cF \wedge \omega_a\,,
    \label{bio-eq: KF def}
\end{align}
and $K_a^{(2)}$ was introduced in \eqref{bio-Kcurv}. Similarly, the  Chern-Simons action of this bound state will give, upon dimensional reduction 
\be
q_{\rm D8} Q_{\rm D8}^{\rm total} = Q_{\rm D8/D6}^{\rm eff} + \left(K^F_a - K_a^{(2)}\right)  Q_{\rm D4}^a = \eta T_{\rm D8}^{\rm total}\, ,
\label{bio-QD8tot}
\ee
where $ Q_{\rm D4}^a = \eta e^{K/2}   t^a$. Hence, again for $\eta=1$ a D8-brane will satisfy the BPS condition $Q=T$, while for $\eta =-1$ this will occur for an anti-D8-brane. One important aspect of these corrections is that the induced D4-brane tension in \eqref{bio-TD8tot} is in general negative. Indeed, the curvature term $K_a^{(2)} t^a = -\frac{1}{24\ell_s^6} \int_{X_6} c_2(X_6) \wedge J_{\rm CY}$ is positive in the interior of the K\"ahler cone for Calabi--Yau geometries, inducing a negative D4-brane tension. In the present case this is compensated by an induced negative D4-brane charge in \eqref{bio-QD8tot}. However, such a compensation will no longer occur for the non-supersymmetric AdS$_4$ flux backgrounds that we now turn to discuss.


\section{Non-supersymmetric \texorpdfstring{AdS$_4$}{AdS4} vacua}
\label{bio-s:nonsusy}

The type IIA flux potential obtained in  \cite{Grimm:2004ua} has, besides the supersymmetric vacua found in \cite{DeWolfe:2005uu}, further non-supersymmetric families of vacua. This can already be seen by the toroidal analysis of \cite{DeWolfe:2005uu,Camara:2005dc}, and its generalization to any Calabi-Yau was reviewed in section \ref{cy-subsec: AdS vacua}, corresponding to the second row of table \ref{cy-table: summary ads vacua}. A subset of such vacua was analyzed in \cite{Narayan:2010em} in terms of perturbative and non-perturbative stability, for the particular case of $X_6 = \mathbb{T}^6/(\Z_3 \times \Z_3)$. It was found that one particular family of vacua, dubbed type 2 in \cite{Narayan:2010em}, was stable both at the perturbative and non-perturbative level.\footnote{As pointed out in \cite{Marchesano:2019hfb} the remaining non-supersymmetric families (type 3 - type 8) found in \cite{Narayan:2010em} are not actual extrema of the flux potential, and only seem so when the potential is linearized as in \cite{Narayan:2010em}.} In the following we will extend this analysis to general Calabi--Yau geometries in the smearing approximation, and to new membranes like those arising from the D8/D6 configuration considered above. 

The non-supersymmetric vacua dubbed type 2 in \cite{Narayan:2010em} are in one-to-one correspondence with supersymmetric vacua, by a simple sign flip of the internal four-form flux $\hat{G}_4 \to - \hat{G}_4$. Because $\hat{G}_4$ enters quadratically in the 10d supergravity Lagrangian, the energy of such a vacuum is similar to its  supersymmetric counterpart and, as argued in \cite{Narayan:2010em}, one expects it to share many of its nice properties. It was indeed found in \cite{Marchesano:2019hfb} that such non-supersymmetric vacua, dubbed {\bf A1-S1} therein (see table \ref{cy-table: summary ads vacua}), exist for any Calabi--Yau geometry, and that they are  stable at the perturbative level. Instead of the (smeared) supersymmetric relations \eqref{bio-intfluxsm} we now have
\be
\ell_s [ H ]  = \frac{2}{5} m g_s [\re \Omega_{\rm CY} ] \, , \qquad  G_2  =  0\, ,  \qquad  \ell_s G_4  = \hat{e}_a  \tilde{\omega}^a  = \frac{3}{10} m \, \cK_a \tilde{\omega}^a  \, , \qquad  G_6 =  0\, , 
\label{bio-intfluxsmnosusy}
\ee
and most features are analogous to the supersymmetric case. In particular, the AdS$_4$ radius and vacuum energy are also given by \eqref{bio-Rads} and \eqref{bio-lambda}, respectively. 

Because the energy dependence with the flux quanta is the same, one should be looking for similar non-perturbative transitions that jump to a vacuum of lower energy: Those that decrease $|\hat{e}_a|$ and those that increase $|m|$ or $|h|$. The objects that will implement such jumps will again be 4-dimensional membranes that come from (anti-)D4-branes and (anti-)D8-branes. Because of the sign flip in $G_4$, the role of the D4-branes will be exchanged with that of anti-D4-branes with respect to the supersymmetric case. 

Indeed, the relations \eqref{bio-intfluxsmnosusy} imply that \eqref{bio-610fluxes} is replaced by
\be
G_6 =   \frac{3\eta}{Rg_s} {\rm vol}_4 \wedge J_{\rm CY}\, , \qquad G_{10} =  - \frac{5\eta}{6 R g_s} {\rm vol}_4 \wedge J^3_{\rm CY}\, ,
\label{bio-610fluxesnosusy}
\ee
with no further external fluxes. As a result we find the following 4-dimensional membrane couplings:
\be
Q_{\rm D2} = 0 \, , \qquad Q_{\rm D4}^{\rm ns} =  -  e^{K/2} \frac{\eta}{\ell_s^2} \int_\Sigma J_{\rm CY} \, , \qquad Q_{\rm D6} = 0\, , \qquad Q_{\rm D8} =  -\frac{5}{3}  \eta\, q_{\rm D8} e^{K/2} {\cal V}_{\rm CY}\, .
\label{bio-QDGKTnosusy}
\ee
By analogy with the supersymmetric case, we now find that the equality $Q =T$ is realized by D4-branes wrapping anti-holomorphic two-cycles, for $\eta =1$, and holomorphic two-cycles for $\eta=-1$. This essentially amounts to exchanging the roles of D4-brane and anti-D4-brane, as advanced. If we chose the object with opposite charge (e.g. a D4-brane wrapping a holomorphic two-cycle for $\eta=1$) then we would have that $Q = -T$ and the effects of the tension and the coupling to the flux background would add up, driving the membrane away from the boundary. In general, it is not possible to find a D4-brane such that $Q >T$, just like it is not possible to find it in supersymmetric vacua. This reproduces the result of  \cite{Narayan:2010em} that D4-brane decays are, at best, marginal. Regarding D8-branes, the naive story is essentially the same as for ${\cal N} =1$ vacua. Since $Q_{\rm D8}$ remains the same, $Q^{\rm eff}_{\rm D8/D6}$ will compensate $T_{\rm D8}$ for $\eta=q_{\rm D8}$. 

Now, the interesting case occurs when we consider bound states of D8 and D4-branes, by introducing the effect of worldvolume fluxes and/or curvature corrections. In a D8-brane configuration similar to the one in the supersymmetric case the tension is the same:
\be
T_{\rm D8}^{\rm total} = T_{\rm D8} + \left(K^F_a - K_a^{(2)}\right)  T_{\rm D4}^a\, .
\label{bio-TD8totnosusy}
\ee
In the large volume approximation $T_{\rm D8} \gg T_{\rm D4}^a$, and so just like in the supersymmetric case we need to consider a D8-brane whenever $\eta =1$, or else $T>Q$. The coupling of the corresponding 4-dimensional membranes is now different from \eqref{bio-QD8tot}, and reads
\be
q_{\rm D8} Q_{\rm D8}^{\rm total} = Q_{\rm D8/D6}^{\rm eff} + \left(K^F_a - K_a^{(2)}\right)  Q_{\rm D4}^{{\rm ns}, a} = \eta \left[ T_{\rm D8} -  \left(K^F_a - K_a^{(2)}\right)  T_{\rm D4}^a\right]  \, .
\label{bio-QD8totnosusy}
\ee
As a result we find that
\be
Q_{\rm D8}^{\rm total} - T_{\rm D8}^{\rm total} = 2 \left(K_a^{(2)} -K^F_a\right)  T_{\rm D4}^a\, ,
\label{bio-QTsmnosusy}
\ee
where we have imposed $\eta=q_{\rm D8}$. On the one hand, by assumption the D8-brane worldvolume flux induces pure D4-brane charge, which means that $K^F_aT_{\rm D4}^a >0$.\footnote{If we consider  diluted fluxes that induce pure anti-D4-brane charge, their contributions would cancel in \eqref{bio-QTsmnosusy}.} On the other hand,  generically $K_a^{(2)} T_{\rm D4}^a > 0$, since for a Calabi--Yau \cite{Miyaoka1987}
\be
-\int_{X_6} c_2(X_6) \wedge J_{\rm CY} \geq 0\, ,
\label{bio-condcurv}
\ee
with the equality occurring only at the boundary of the K\"ahler cone. This means that the curvature corrections are inducing negative D4-brane charge and tension on the D8-brane. The effects of such negative tension and charge add up in the present non-supersymmetric background, and drag the D8-brane towards the AdS$_4$ boundary.\footnote{Notice that this mechanism is analogous to the one in  \cite{Maldacena:1998uz}, in which a D5-branes wraps the $K3$ in AdS$_3 \times S^3 \times K3$.} So if the worldvolume fluxes are absent or give a smaller contribution, we will have that $Q_{\rm D8}^{\rm total} > T_{\rm D8}^{\rm total}$ and the energy of the configuration will be minimized at $z_0 \to \infty$. As such, these D8/D4 bound states are clear candidates to realize the AdS instability conjecture of \cite{Ooguri:2016pdq,Freivogel:2016qwc}. In the next section we will argue that this is indeed the case.

While a remarkable result, one must realize that it does not apply to all non-supersymmetric vacua of this sort. It only applies to those flux vacua which contain space-time filling D6-branes, that is those with $N>0$ in \eqref{bio-tadpole2}. If $N=0$ we cannot consider a transition like the above in which $m$ increases its absolute value. In other words, then the D8-brane configuration described above cannot exist.\footnote{Or it could at the expense of introducing anti-D6-branes, which would introduce a whole new set of instabilities.} These are precisely the kind of vacua considered in \cite{Narayan:2010em} which, even with these new considerations, would a priori remain marginally stable. Moreover, if \eqref{bio-condcurv} vanished at some boundary of the K\"ahler cone, there would be a priori no instability triggered by D8/D4-brane bound states, which would be marginal.  In fact, this last statement is not true, but only a result of the smearing approximation. As we will see, when describing the same setup but in terms of a background that admits localized sources, corrections to the D8-brane tension will appear, modifying the above computation. 


\section{\texorpdfstring{AdS$_4$}{AdS4} instability from the 4d perspective}
\label{bio-s:insta}

The results of the previous section suggest that non-supersymmetric AdS$_4 \times X_6$ vacua with a flux background of the form \ref{bio-intfluxsmnosusy} develop non-perturbative instabilities if they contain space-time filling D6-branes. From the 4-dimensional perspective such an instability would be mediated by a membrane that arises from wrapping a D8-brane on $X_6$, since it becomes a membrane with $Q > T$ upon dimensional reduction. However, the link between the inequality $Q > T$ and a non-perturbative gravitational instability typically follows an analysis similar to \cite{Maldacena:1998uz}, implicitly relying on the thin-wall approximation. As pointed out in \cite{Narayan:2010em}, D8-branes are not in the thin-wall approximation unless the value of $|m|$ is very large, which is not generically true. Therefore in this section we would like to provide an alternative argument of why these vacua are unstable. 

For this we will make use of the symmetry between supersymmetric and non-supersymmetric vacua mentioned in section \ref{bio-s:nonsusy}. That is, for the same value of the fluxes $m$, $h$ and $|\hat{e}_a|$ the saxion vevs are stabilized at precisely the same value in both supersymmetric and non-supersymmetric vacua, and the vacuum energy \eqref{bio-lambda} is also the same. For simplicity let us consider a pair of supersymmetric and non-supersymmetric vacua in which $e_0 = m^a=0$ and
\be
m^{\rm susy} = m^{\cancel{\rm susy}} > 0\, , \qquad  h^{\rm susy} = h^{\cancel{\rm susy}} > 0\, , \qquad \hat{e}_a^{\rm susy} = - \hat{e}_a^{\cancel{\rm susy}} > 0\, .
\ee
In both backgrounds, a D8-brane without worldvolume fluxes will induce the following shift of flux quanta as we cross it from $z = \infty$ to $z = -\infty$ as
\begin{align}
m^{\rm susy} \to m^{\rm susy} + 1\, , \qquad & |\hat{e}_a^{\rm susy}| \to |\hat{e}_a^{\rm susy}  + K_a^{(2)}| \, , \\
m^{\cancel{\rm susy}} \to m^{\cancel{\rm susy}} + 1\, , \qquad & |\hat{e}_a^{\cancel{\rm susy}}| \to  |\hat{e}_a^{\cancel{\rm susy}} + K_a^{(2)}| = |\hat{e}_a^{{\rm susy}}  - K_a^{(2)}|\, .
\end{align}
Because the absolute value of the four-form flux quanta $\hat{e}_a$ are different after the jump for the supersymmetric and the non-supersymmetric case, so are the vevs of the K\"ahler moduli and the vacuum energy. To fix this, let us add to the supersymmetric setup a D4-brane wrapping a holomorphic two-cycle in the Poincar\'e dual class to $2K_a^{(2)}[\tilde{\omega}^a]$. The resulting 4-dimensional membrane can create a marginal bound state with the one coming from the D8-brane, implementing the combined jump
\begin{align}
m^{\rm susy} \to m^{\rm susy} + 1\, , \qquad  |\hat{e}_a^{\rm susy}| \to |\hat{e}_a^{\rm susy}  - K_a^{(2)}| \, .
\end{align}
Now both supersymmetric and non-supersymmetric jumps are identical, in the sense that the variation of the scalar fields from the initial to the final vacuum is the same, and so are the initial and final vacuum energies. As a result, the energy stored in the field variation of both solutions should be identical. What is different is the tension of the membranes. We have that
\be
T_{\rm susy} = T_{\rm D8} +  K_a^{(2)}  T_{\rm D4}^a > T_{\rm D8} -  K_a^{(2)}  T_{\rm D4}^a = T_{\cancel{\rm susy}}\, ,
\ee
assuming as before that \eqref{bio-condcurv} is met. Therefore, because the supersymmetric decay is marginal, the non-supersymmetric one should be favoured energetically, rendering the non-supersymmetric vacuum unstable. 


\section{Beyond the smearing approximation}
\label{bio-s:nonsmeared}

The Calabi--Yau flux backgrounds of section \ref{bio-s:dgkt} and \ref{bio-s:nonsusy} can be thought of as an approximation to the actual 10d solutions to the equations of motion and Bianchi identities, in which O6-planes and D6-branes are treated as localized sources. More precisely, the smeared Calabi--Yau solution can be recovered from the actual solution in the limit of small cosmological constant, weak string coupling and large internal volume, as discussed in section \ref{cy-subsec: beyond smearing}. Any of these quantities can be used to define an expansion parameter, so that the actual 10d solution can be described as a perturbative series, of which the smeared solution is the leading term. While a solution for the whole series (i.e. the actual 10d background) has not been found yet, the next-to-leading term of the expansion was found in \cite{Marchesano:2020qvg} for the case of the supersymmetric vacua  and presented in \eqref{cy-eq: solutionflux}. We will first comment on this case and then construct a similar background with localized sources for the non-supersymmetric vacua of section \ref{bio-s:nonsusy} at the same level of accuracy. As we will see, these more precise backgrounds do not affect significantly the energetics of 4-dimensional
membranes made up from D4-branes. However, as it will be discussed in the next section, they
yield non-trivial effects for membranes that correspond to D8/D6 systems.

\subsection{Supersymmetric \texorpdfstring{AdS$_4$}{AdS4}}
\label{bio-ss:mpqt}

Given the background described in section \ref{cy-subsec: beyond smearing}, one may reconsider the computation of the tension and coupling made in the smearing approximation. Let us for instance consider a D4-brane wrapping a two-cycle $\Sigma$. Instead of the expression for $G_6$ in \eqref{bio-610fluxes} we obtain
\bea\nonumber
G_6 & = & -  {\rm vol}_4 \wedge \left[ J_{\rm CY}  \frac{m}{5\ell_s } \left(3 - 8 g_s \varphi \right) - \oh \star_{\rm CY} d \left(J_{\rm CY} \wedge d^c f_\star \right)  \right] e^{4A}+ \cO(g_s^2) \\ \nonumber
& =& -  {\rm vol}_4 \wedge \left[ J_{\rm CY}  \frac{m}{5\ell_s } \left(3 - 20 g_s \varphi \right) - \oh \left( \Delta_{\rm CY} - dd^\dag_{\rm CY}\right) \left(f_\star  J_{\rm CY}  \right)  \right] + \cO(g_s^2) \\
& =&  -  {\rm vol}_4 \wedge \left[ J_{\rm CY} \frac{3\eta}{Rg_s} + \oh dd^\dag_{\rm CY} \left(f_\star  J_{\rm CY}  \right)  \right] + \cO(g_s^2)\, ,
\label{bio-6fluxesloc}
\eea
where $d^c \equiv i(\bar{\p}_{\rm CY} - \p_{\rm CY})$ and we have used that $J_{\rm CY} \wedge d^c f  = \star_{\rm CY} d(J_{\rm CY} f)$. Since the only difference with respect to the smearing approximation is an exact contribution, the membrane coupling $Q_{\rm D4}$ remains unchanged, and it is still given by $Q_{\rm D4} =  \eta e^{K/2} \int_\Sigma J_{\rm CY}$. As before, D4-branes  wrapping holomorphic ($\eta=1$) and anti-holomorphic ($\eta=-1$) two-cycles will be BPS, and will feel no force in the above AdS$_4$ background, as expected from supersymmetry.

\subsection{Non-supersymmetric \texorpdfstring{AdS$_4$}{AdS4}}
\label{bio-ss:nonsmearednonsusy}

Just like for supersymmetric vacua, one would expect a 10d description of the non-supersymmetric vacua of section \ref{bio-s:nonsusy} compatible with localized sources. Again, the idea would be that the smeared background is the leading term of an expansion in powers of $g_s$. In the following we will construct a 10d background with localized sources which can be understood as a first-order correction to the smeared Calabi--Yau solution \eqref{bio-intfluxsmnosusy} in the said expansion. 

The main feature of the non-supersymmetric background \eqref{bio-intfluxsmnosusy} is that it flips the sign of the RR four-form flux $G_4$, while it leaves the remaining fluxes, metric, dilaton and vacuum energy invariant. This means that the Bianchi identities \eqref{bio-BIG} do not change at leading order, and in particular the leading term for two-form flux $G_2$ should have the same form \eqref{cy-eq: G2sol} as in the supersymmetric case. Moreover, the localized solution is likely to be described in terms of the quantities $\varphi$ and $k$ that arise from the Bianchi identity of $G_2$, at least at the level of approximation that we are seeking. Because of this, it is sensible to consider a 10d metric and dilaton background similar to the supersymmetric case, namely \eqref{cy-eq: solutionsu3}. 

Regarding the background flux $G_4$, there should be a sign flip on its leading term, but it is clear that this cannot be promoted to an overall sign flip, because the co-exact piece of $G_4$, that contributes to the Bianchi identity, must be as in the supersymmetric case. Since the harmonic and co-exact pieces of the fluxes are fixed by the smearing approximation and the Bianchi identities, the question is then how to adjust their exact pieces to satisfy the equations of motion. Using the approach of \cite{Junghans:2020acz}, we find that the appropriate background reads
\begin{subequations}
	\label{bio-solutionfluxnnosusy}
\begin{align}
 H & =   \frac{2}{5} \frac{m}{\ell_s} g_s \left(\re \Omega_{\rm CY} - 2g_s K \right) + \frac{1}{10}   d\re \left(\bar{v} \cdot \Omega_{\rm CY} \right) + \cO(g_s^{3})  \, , \\
 G_2 & =     d^{\dag}_{\rm CY} K  + \cO(g_s)   \, , \\
G_4 & =  -\frac{m}{\ell_s} J_{\rm CY} \wedge J_{\rm CY} \left(\frac{3}{10}  + \frac{4}{5} g_s \varphi \right) -\frac{1}{5}   J_{\rm CY} \wedge g_s^{-1} d \im v + \cO(g_s^2) \, , \\
G_6 & = 0\, ,
\end{align}
\end{subequations} 
with the same definition for the (1,0)-from $v$. In Appendix \ref{bio-ap:10deom} we show that this background satisfies the 10d equations of motion up to order $\cO(g_s^2)$, just like the supersymmetric case.

With this solution in hand, one may proceed as in the supersymmetric case and recompute the 4-dimensional membrane couplings and tensions. If the result is different from the one in the smearing approximation the difference could be interpreted as a $g_s$ correction. To begin, let us again consider a D4-brane wrapping a two-cycle $\Sigma$. The coupling of such a brane can be read from the six-form RR flux with legs along AdS$_4$
\bea\nonumber
G_6 & = & {\rm vol}_4 \wedge \left[ J_{\rm CY}  \frac{m}{5\ell_s } \left(3 + 8 g_s \varphi \right) + \frac{1}{10} \star_{\rm CY} d \left(J_{\rm CY} \wedge d^c f_\star \right)  \right] e^{4A}+ \cO(g_s^2) \\ \nonumber
& =& {\rm vol}_4 \wedge \left[ J_{\rm CY}  \frac{m}{5\ell_s } \left(3 - 4 g_s \varphi \right) + \frac{1}{10} \left( \Delta_{\rm CY} - dd^\dag_{\rm CY}\right) \left(f_\star  J_{\rm CY}  \right)  \right] + \cO(g_s^2) \\
& =&  {\rm vol}_4 \wedge \left[ J_{\rm CY} \frac{3\eta}{Rg_s} - \frac{1}{10} dd^\dag_{\rm CY} \left(f_\star  J_{\rm CY}  \right)  \right] + \cO(g_s^2)\, .
\label{bio-6fluxeslocnosusy}
\eea
Remarkably, we again find that the first non-trivial correction to the smearing approximation is an exact form, and so it vanishes when integrating over $\Sigma$. As a result, the 4-dimensional membrane couplings $Q_{\rm D4}^{\rm ns} =  - \eta e^{K/2} \ell_s^{-2}\int_\Sigma J$ remain uncorrected at this level of the expansion, and there is a force cancellation for D4-branes wrapping anti-holomorphic ($\eta=1$) and holomorphic ($\eta=-1$) two-cycles, just like in our discussion of section  \ref{bio-s:nonsusy}. Presumably, by looking at higher-order corrections one may find one that violates the equality $Q_{\rm D4}^{\rm ns} = T_{\rm D4}^{\rm ns}$ in one way or the other, which would be a non-trivial test of the conjecture in \cite{Ooguri:2016pdq}. Such a computation is however beyond the scope of the present work. Instead, we will focus on membranes whose coupling and tension departure from the smeared result already at this level of approximation, namely those membranes arising from D8/D6 systems. To see how this happens, one must first take into account that beyond the smearing approximation such systems are described by BIonic configurations, as we now discuss. 


\section{BIonic membranes}
\label{bio-s:bion}

A D$p$-brane ending on a D$(p+2)$-brane to cure a Freed--Witten anomaly constitutes a localized source for gauge theory on the latter. When going beyond the smearing approximation one should take this into account, and describe the combined system as a BIon-like solution \cite{Gibbons:1997xz}. In this section we do so for the D8/D6-brane system, and compute the  tension and flux coupling of the associated 4-dimensional membrane for both the supersymmetric and non-supersymmetric backgrounds of the last section. As we will see, the BIonic nature of the membrane will modify their coupling and tension of the membrane with respect to the smeared result.

\subsection{Supersymmetric \texorpdfstring{AdS$_4$}{AdS4}}
\label{bio-ss:bionsusy}

Let us consider a D8-brane wrapping $X_6$ with orientation $q_{\rm D8} = \pm 1$ and extended along the plane $z = z_0$ in the Poincar\'e patch of AdS$_4$. As pointed out above, due to the non-trivial $H$-flux background we must have an excess of $h$ D6-branes wrapping $\Pi_{\rm O6}$ and extended to the right of the D8-brane, namely along $(t,x^1,x^2) \times [z_0 , \infty) \subset$ AdS$_4$. This setup implies a Bianchi identity for the D8-brane worldvolume flux $\cF = B + \frac{\ell_s^2}{2\pi} F$ of the form
\be
d{\cal F} = H - \frac{h}{\ell_s} \delta(\Pi_{\rm O6})\, .
\label{bio-dFD8}
\ee
Because by construction the rhs is trivial in cohomology, this equation always has a solution. Moreover, if we are in the smearing approximation, we have that the rhs of \eqref{bio-dFD8} vanishes, and so $\cF$ must be closed. The energy-minimizing configurations then correspond to solving the standard F-term and D-term-like equations for $\cF$ \cite{Koerber:2007jb}, which in our setup means that $\cF$ is a harmonic (1,1)-form of $X_6$ such that $3\cF \wedge J_{\rm CY}^2  = \cF^3$. When we see such a D8-brane as a membrane in four dimensions, this harmonic worldvolume flux is the one responsible for the contribution $K_a^F Q^a_{\rm D4}$ to their flux coupling and tension. 

If we describe our system beyond the smearing approximation, the D8-brane worldvolume flux can no longer be closed. Instead, it must satisfy a Bianchi identity that is almost identical to the one of the RR two-form flux. Even when the harmonic piece of $\cF$ vanishes, we find that
\be
\cF =  \frac{G_2}{G_0} = \frac{\ell_s}{m} d^{\dag}_{\rm CY} K  + \cO(g_s) \, .
\label{bio-cfsol}
\ee
assuming that the D6-branes are equally distributed on top of the O6-planes before and after the jump, see \cite{Casas:2022mnz} for more general setups.  BPS configurations with D$p$-branes ending on D$(p+2)$-branes, inducing a non-closed worldvolume flux on the latter are usually described by BIon-like solutions \cite{Gibbons:1997xz}, in which the D$(p+2)$-brane develops a spike along the direction in which the D$p$-branes are extended. A relatively simple configuration of this sort is given by the D5/D3 system in type IIB flux compactifications, that was analyzed in \cite{Evslin:2007ti} from the viewpoint of calibrations. In this setup a D5-brane wraps a special Lagrangian three-cycle $\Lambda$ of a warped Calabi--Yau compactification, and extends along the plane $x^3 = x^3_0$ of $\pr^{1,3}$. If $\int_{\Lambda} H = - N$, then $N$ space-time filling D3-branes must end on the D5-brane, stretched along $(t,x^1,x^2) \times [x^3_0 , \infty) \subset \pr^{1,3}$ and located at a point $p \in \Lambda$. This induces an internal worldvolume flux on the D5-brane, solving the equation $d\cF = N\left(\delta(p) - \frac{d{\rm vol}_\Lambda}{{\rm Vol}(\Lambda)}\right)$. To render the configuration BPS it is necessary to give a non-trivial profile to the D5-brane position field $X^3$, such that $dX^3 = \star_{\Lambda} \cF$. The resulting profile features a spike $X^3 \sim \frac{N}{r}$ around the point $p$, which represents the $N$ D3-branes ending on the D5. The D5-brane BIon configuration accounts for the whole energy of the D5/D3 system. 

Our D8/D6 setup can be seen as a six-dimensional analogue of the D5/D3 system. The presence of the worldvolume flux \eqref{bio-dFD8} can be made compatible with a BPS configuration if we add a non-trivial profile for the D8-brane transverse field $Z$. The relation with the worldvolume flux is now given by
\be
\star_{\rm CY} dZ = q_{\rm D8} \im \Omega_{\rm CY} \wedge \cF + \cO(g_s) \, .
\label{bio-BIonrel}
\ee
This expression can be motivated in a number of ways. In Appendix \ref{bio-ap:dbi} we show that, upon imposing it, the DBI action is linearized at the level of approximation that we are working, as required by a BPS configuration. In Appendix \ref{bio-ap:IIBion} we describe a similar configuration in type IIB flux compactifications, that can then be mapped to the BPS Abelian SU(4) instantons of Calabi--Yau four-folds \cite{Donaldson:1996kp}. Finally, notice that \eqref{bio-BIonrel} implies that
\be
\Delta_{\rm CY} Z = \ell_s q_{\rm D8} h \left( \delta^{(3)}_{\Pi_{\rm O6}} -\frac{{\cal V}_{\Pi_{\rm O6}}}{{\cal V}_{\rm CY}} \right)\, ,
\ee
and so whenever $q_{\rm D8} h = |h|$ we recover a spike profile of the form $Z \sim \frac{|h|\ell_s}{r}$ near $\Pi_{\rm O6}$, as expected. In fact we can draw the more precise identification
\be
Z = z_0 - \frac{4\ell_s \varphi}{|m|}\, ,
\ee
where we have imposed the BPS relation $q_{\rm D8} = \eta \equiv {\rm sign}\, m$. Notice that this identifies the spike profile of the BIon solution towards the AdS$_4$ boundary with the strong coupling region near the O6-plane location, where our perturbative  expansion on $g_s$ is no longer trustable, see fig. \ref{bio-fig:D8D6bion}.

\begin{figure}[htbp]
    \centering
    \includegraphics[width=10cm]{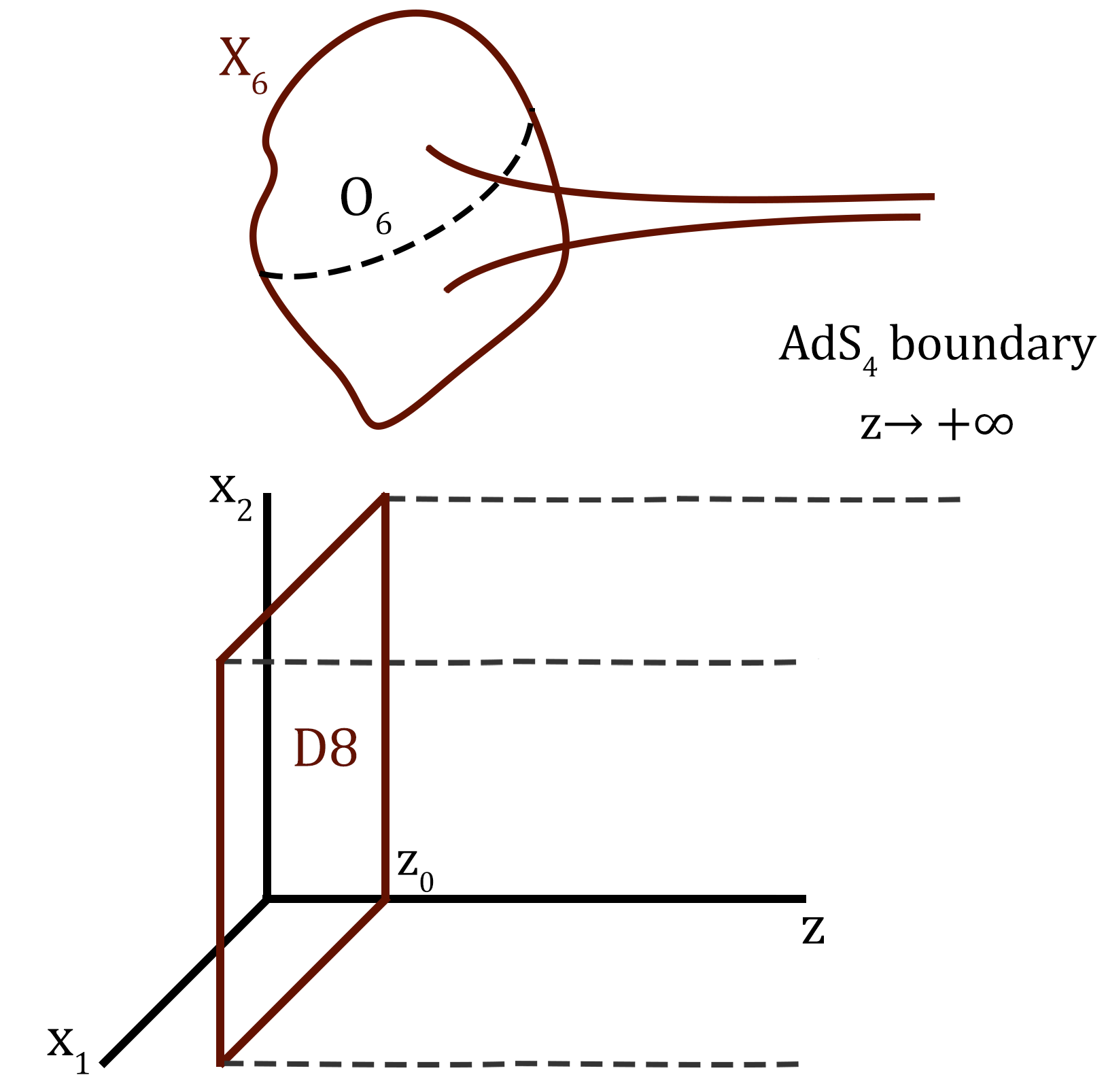}
    \caption{Beyond the smearing approximation, the D8/D6 system of figure \ref{bio-fig:D8D6} becomes a BIon-like solution for the D8-brane, with a BIon profile that peaks at the O6-plane location.}
    \label{bio-fig:D8D6bion}
\end{figure}

The relation \eqref{bio-BIonrel} implies that the DBI action of the BIon can be computed in terms of calibrations. Indeed, ignoring  curvature corrections, the calibration for a D8-brane wrapping $X_6$ and with worldvolume fluxes is given by 
\be
- \im \Phi_+ = - g_s^{-1} q_{\rm D8} \im e^{-iJ_{\rm CY}} + \cO(g_s) = g_s^{-1} q_{\rm D8}\left(- \frac{1}{6} J_{\rm CY}^3 + J_{\rm CY}\right) + \cO(g_s)\, ,
\label{bio-imPhi+}
\ee
while that for D6-branes wrapping a three-cycle of $X_6$ is 
\be
\im \Phi_- =  g_s^{-1} \left( \im v + \Im \Omega - \oh \psi \im \omega \wedge \im \Omega  \right) + \cO(g_s)  = g_s^{-1} \Im \Omega_{\rm CY} + \cO(g_s^0)\, ,
\label{bio-imPhi-}
\ee
at leading order in our expansion.  Here $\psi$ and $\omega_0$ are a complex function and 2-form which describe the $SU(3)\times SU(3)$ structure, and such that $\Omega = \frac{i}{\psi} v \wedge \omega + \cO(g_s^2)$, see \cite{Marchesano:2020qvg} for details. Applying the general formulas of \cite{Evslin:2007ti}, we find that the BIon DBI action reads 
\bea
\label{bio-psp}
dS_\text{DBI}^{\rm D8} &= & dt \wedge dx^1 \wedge dx^2 \wedge e^{\frac{3Z}{R}} q_{\rm D8} \left(  \im \Phi_+ - dZ \wedge e^A \im \Phi_-  \right) \wedge e^{-\cF} \\
& \simeq & dt \wedge dx^1 \wedge dx^2 \wedge g_s^{-1} e^{\frac{3z_0}{R}}  \left( \frac{1}{6}J_{\rm CY}^3  - \oh J_{\rm CY} \wedge \cF^2 + q_{\rm D8}dZ \wedge \im \Omega_{\rm CY} \wedge \cF \right)  \\
& = &   - dt \wedge dx^1 \wedge dx^2 \wedge g_s^{-1} e^{\frac{3z_0}{R}}  \left(- \frac{1}{6}J_{\rm CY}^3  + \oh J_{\rm CY} \wedge \cF^2 + \star_{\rm CY} dZ \wedge dZ \right)\, .
\label{bio-finaldbi}
\eea
The last line coincides with our result of Appendix \ref{bio-ap:dbi}, and with what is expected for a BIon solution. Indeed, the first two terms of \eqref{bio-finaldbi} correspond to the DBI action of the magnetised D8-brane, while the third one corresponds to the D6-branes that stretch towards the AdS$_4$ boundary. Nevertheless, notice that the middle term $\oh J_{\rm CY} \wedge \cF^2$ gives an extra contribution to the DBI action compared to the smearing approximation of section \ref{bio-ss:4dmem}. Indeed, when $\cF$ is a harmonic form this term accounts for the contribution $K_a^FT_{\rm D4}^a$ in \eqref{bio-TD8tot}. When going away from the smearing approximation $\cF$ will also have a co-exact piece, given by \eqref{bio-cfsol}, that will contribute to the DBI even if $\cF^{\rm harm} =0$. Because it induces a non-trivial D4-brane charge, one may interpret this extra contribution to the D8-brane tension as a curvature correction induced by the non-trivial BIon profile, as opposed to D6-branes sharply ending on the D8-brane, although it would be interesting to derive this expectation from first principles. As we will see, this additional contribution to the tension does not play much of a role in the present supersymmetric setup, but it is crucial for the dynamics of Bionic membranes in non-supersymmetric backgrounds. 

Eq.\eqref{bio-psp} suggests how to generalize \eqref{bio-BIonrel} to a relation describing the BIon profile to all orders in $g_s$. The natural choice is 
\be
  \star_6 dZ =  -  q_{\rm D8} e^{\phi-2A} \left. \im \Phi_- \wedge  e^{-\cF}\right|_5 \, ,
  \label{bio-stardZ}
\ee
where the Hodge star is performed with the exact, non-Calabi--Yau metric of $X_6$, and $|_5$ means that we are only keeping the five-form component of the polyform on the rhs. With this choice the BIon DBI action would read
\be
dS_\text{DBI}^{\rm D8} =
 dt \wedge dx^1 \wedge dx^2 \wedge  e^{\frac{3Z}{R}} q_{\rm D8} \left(  \im \Phi_+  \wedge e^{-\cF} - e^{3A-\phi} \star_{6} dZ \wedge dZ \right)\, ,
\label{bio-finaldbiex}
\ee
as expected on general grounds. In addition, \eqref{bio-psp} encodes the force cancellation observed for the D8/D6 system in the smearing approximation, which can now be derived for the single object which is the BIonic D8-brane, and in the exact background. For this, notice that the Chern-Simons part of the D8-brane action reads
\be  
dS_\text{CS}^{\rm D8}  = - dt \wedge   dx^1 \wedge dx^2 \wedge \frac{R}{3} e^{\frac{3Z}{R}} e^{4A} q_{\rm D8}  \star_6 \lambda \hat{G}  \wedge e^{-\cF}  \, ,
\ee
where $\hat{G}$ is defined as in \eqref{bio-demoflux}. Putting both contributions together and using the bulk supersymmetry equation
\be
 d_H  \left( e^A \im \Phi_- \right) +  \frac{3}{R} \im \Phi_+ =  e^{4A} \star_6 \lambda \hat{G}\, ,
 \label{bio-susy1}
\ee
and \eqref{bio-dFD8} one finds that 
\begin{align}
\label{bio-DBICS}
  dS_\text{DBI}^{\rm D8}  & + dS_\text{CS}^{\rm D8} = - dt \wedge dx^1 \wedge dx^2 \wedge \frac{R}{3} q_{\rm D8}  \left[ de^{\frac{3Z}{R}} \wedge  e^A \im \Phi_- +  e^{\frac{3Z}{R}}  d_H   \left( e^A \im \Phi_- \right)  \right]  \wedge e^{-\cF} \\ \nonumber
& =   - dt \wedge dx^1 \wedge dx^2 \wedge \frac{R}{3} q_{\rm D8}  \left[ d \left(e^{\frac{3Z}{R}}   e^A \im \Phi_- \wedge e^{-\cF} \right) + \frac{h}{\ell_s} e^{\frac{3Z}{R}}  \delta(\Pi_{\rm O6})  \wedge e^A \im \Phi_- \wedge  e^{-\cF} \right]  \, .
\end{align}
The first term of the second line is a total derivative that will vanish when integrating over $X_6$, while the second term is an infinite contribution to the action, that accounts for the DBI action of the $|h|$ D6-branes extending along $[z_0, \infty)$. Indeed, it is easy to see that the leading piece of this term is of the form $ |h| g_s^{-1} e^{\frac{3Z}{R}} \delta(\Pi_{\rm O6})  \wedge \im \Omega_{\rm CY} = |h|g_s^{-1} {\cal V}_{\Pi_{\rm O6}} e^{\frac{3Z_\infty}{R}}$, with $Z_\infty \equiv Z|_{\Pi_{\rm O6}} = \infty$. The relevant point is that $Z_{\infty}$ is independent of $z_0$, and therefore this second term is independent of the D8-brane transverse position. Therefore, the total energy of the BIonic 4-dimensional membrane will be independent of $z_0$, even if contains some infinite contributions. This matches the results obtained in the smearing approximation, and is equivalent to the BPS equilibrium relation $Q_{\rm D8}^{\rm BIon} = T_{\rm D8}^{\rm BIon}$. 

The above computation is quite general, and essentially follows from some  general observations made in \cite{Koerber:2007jb} applied to the present setup. It is nevertheless instructive to see how \eqref{bio-susy1}, which is a key relation to achieve force cancellation for our BIonic D8-brane, is satisfied for the background \eqref{cy-eq: solutionsu3} and \eqref{cy-eq: solutionflux}, in preparation for the non-supersymmetric case. We have that
\begin{align}
&d_H  \left( e^A \im \Phi_- \right)  = \frac{1}{2}d d^c f_\star + \star_{\rm CY} G_2 - \frac{2}{3} G_0 \left(\frac{2}{5}- g_s\varphi \right)J_{\rm CY}^3  +  \cO(g_s^{5/3})\, , \\
 &\frac{3}{R} \im \Phi_+  = \frac{3}{5} q_{\rm D8}  |G_0| \left( - J_{\rm CY} + \frac{1}{6} J_{\rm CY}^3\right) + \cO(g_s^{2})\, , \\ 
 & e^{4A} *_6 \lambda \hat{G}  = -\frac{1}{2} dd^\dag_{\rm CY} \left(f_\star  J_{\rm CY}  \right) -\frac{3}{5} G_0 J_{\rm CY} - \star_{\rm CY} G_2 - \frac{1}{6} G_0 \left(1 - 4g_s \varphi\right) J_{\rm CY}^3 + \cO(g_s^{5/3})\, , 
\end{align}
and so one only has to impose $\eta=q_{\rm D8} $ and use that $d d^c f = - dd^\dag_{\rm CY} \left(f  J_{\rm CY}  \right)$  to show the equality. 

\subsection{Non-supersymmetric \texorpdfstring{AdS$_4$}{AdS4}}
\label{bio-ss:bionnosusy}

Let us now consider the D8-brane BIon in the non-supersymmetric AdS$_4$ background of section \ref{bio-ss:nonsmearednonsusy}. Notice that the metric and dilaton background are similar to  the supersymmetric case, and that the $H$-flux only changes by an exact piece at subleading order, so that \eqref{bio-cfsol} remains intact. Because of this, the DBI action of the BIon should be identical to the supersymmetric case, at least to the level of approximation that we are working, and so should be the BIon profile \eqref{bio-BIonrel}. One may thus run a very similar argument to \eqref{bio-DBICS} to see whether the D8-brane is in equilibrium or not with the background. If not, the same computation will determine whether it is dragged towards the boundary or away from it. 

The key relation to look at is again the bulk supersymmetry equation \eqref{bio-susy1}. If  satisfied, the BIonic membrane will be at equilibrium for any choice of transverse position $z_0$. In the smearing approximation we have already seen that there is no equilibrium whenever there is a non-trivial D4-brane charge induced by curvature or worldvolume fluxes, c.f. \eqref{bio-QTsmnosusy}, so we do not expect \eqref{bio-susy1} to be satisfied. Evaluating the background \eqref{cy-eq: solutionsu3} and \eqref{bio-solutionfluxnnosusy} one finds that
\begin{align}
&d_H  \left( e^A \im \Phi_- \right)  = \frac{1}{2}d d^c f_\star + \star_{\rm CY} G_2 - \frac{2}{15} G_0 \left(2 +  g_s\varphi \right)J_{\rm CY}^3  +  \cO(g_s^{4/3})\, , \\
 &\frac{3}{R} \im \Phi_+  = \frac{3}{5} q_{\rm D8}  |G_0| \left( - J_{\rm CY} + \frac{1}{6} J_{\rm CY}^3\right) + \cO(g_s^{2})\, , \\ 
 & e^{4A} *_6 \lambda \hat{G}  = \frac{1}{10} dd^\dag_{\rm CY} \left(f_\star  J_{\rm CY}  \right) + \frac{3}{5} G_0 J_{\rm CY} - \star_{\rm CY} G_2 - \frac{1}{6} G_0 \left(1 - 4g_s \varphi\right) J_{\rm CY}^3 + \cO(g_s^{5/3})\, , 
\end{align}
which results in\footnote{In the language of \cite{Lust:2008zd,Held:2010az}, this corresponds to a background where gauge BPSness is not satisfied, and as a result some space-time filling D-branes may develop tachyons. One can however check that D6-branes wrapping special Lagrangians of $X_6$, and in particular those on top of the orientifold, do not develop any instability. It would be interesting to see if D8-branes wrapping coisotropic five-cycles \cite{Font:2006na} could develop them.}
\be
 d_H  \left( e^A \im \Phi_- \right) +  \frac{3}{R} \im \Phi_+ -  e^{4A} *_6 \lambda \hat{G} = - \frac{3}{5} dd^\dag \left(f_\star  J_{\rm CY}  \right) -\frac{6}{5} G_0 J_{\rm CY}  - \frac{4}{5}  G_0 g_s \varphi J_{\rm CY}^3  +\cO(g_s^{4/3})\, .
\ee
Plugged into the DBI and CS actions, and again ignoring curvature terms, this translates into
\begin{align}
\nonumber
  dS_\text{DBI}^{\rm D8}   + dS_\text{CS}^{\rm D8} &= - dt \wedge dx^1 \wedge dx^2 \wedge \frac{R}{3} q_{\rm D8}e^{\frac{3Z}{R}}  \left[\frac{3}{10} \left( dd^\dag \left(f_\star  J_{\rm CY}  \right) + 2 G_0 J_{\rm CY}\right) \wedge \cF^2  + \frac{4}{5}  G_0 g_s \varphi J_{\rm CY}^3 \right] + \dots\\
  & = - dt \wedge dx^1 \wedge dx^2 \wedge \frac{R}{3} e^{\frac{3z_0}{R}}   \left[\frac{3}{5} |G_0| J_{\rm CY} \wedge \cF^2 + \frac{4}{5}  |G_0| g_s \varphi J_{\rm CY}^3 \right] + \dots
  \label{bio-DBICSns}
\end{align}
where we have neglected terms that do not depend on $z_0$, and in the second line we have only kept terms up to order $\cO(g_s^{4/3})$. Out of the two remaining terms, one of them will vanish when integrating over $X_6$, since  $\int_{X_6} \varphi = 0$. The other one finally gives
\be
 Q_{\rm D8}^{\rm BIon, ns} - T_{\rm D8}^{\rm BIon, ns} =  - e^{K/2} \frac{1}{\ell_s^{6}}\int_{\rm X_6} J_{\rm CY} \wedge \cF^2 + \cO(g_s^2) \, .
\label{bio-QTbionnosusy}
\ee
This result is perhaps not very surprising, because it reproduces the result \eqref{bio-QTsmnosusy} of the smearing approximation when curvature corrections are omitted and $\cF$ is a harmonic form. However remember that in the present setup $\cF$ is always non-vanishing, even when the harmonic piece of $\cF$ is set to zero. Therefore, 
\be
2\Delta_{\rm D8}^{\rm Bion} \equiv - e^{K/2} \frac{1}{\ell_s^{6}}\int_{\rm X_6} J_{\rm CY} \wedge \cF^2
\label{bio-QTbionnosusyexp}
\ee
constitutes a correction to the previous result \eqref{bio-QTsmnosusy}. Since a vanishing harmonic piece for $\cF$ is always a choice, there will always be some BIonic membrane whose charge-to-tension ratio will be fixed by the curvature term $2K_a^{(2)} T_{\rm D4}^a$ plus  \eqref{bio-QTbionnosusyexp}. 

One may thus wonder what is the magnitude of $\Delta_{\rm D8}^{\rm Bion}$ compared to $2K_a^{(2)} T_{\rm D4}^a$, as well as its sign. For this notice that \eqref{bio-cfsol} is suppressed as $\cO(g_s^{2/3})$ compared to a harmonic two-form. Therefore $\Delta_{\rm D8}^{\rm Bion}$ gets an relative suppression of $\cO(g_s^{4/3}) \sim {\cal V}_{\rm CY}^{-2/3}$, just like both terms in \eqref{bio-QTsmnosusy}. In other words, $\Delta_{\rm D8}^{\rm Bion}$ and  $2K_a^{(2)} T_{\rm D4}^a$ scale similarly with the string coupling. As for the sign, it will be the result of two competing quantities, since
\be
2\Delta_{\rm D8}^{\rm Bion} =  e^{K/2} \frac{1}{\ell_s^{6}} \int_{X_6}  \star_{\rm CY} \cF_2 \wedge \cF_2 -   \star_{\rm CY} \cF_1 \wedge \cF_1\, ,
\ee
where $\cF_1 \equiv  \cF^{(1,1)}$ and $\cF_2 \equiv \cF^{(2,0)+(0,2)}$. If we assume \eqref{bio-cfsol} we obtain
\begin{align}
\label{bio-cF1}
\cF_1 & = \frac{i}{2G_0} J_{\rm CY} \cdot \bar{\partial} k = G_0^{-1}J_{\rm CY} \cdot d\left(\star_{\rm CY} K - 2 \varphi \im \Omega_{\rm CY} \right) \, , \\
\cF_2 &  = - G_0^{-1} J_{\rm CY} \cdot d\left( 2\varphi \im \Omega_{\rm CY} \right) \, .
\label{bio-cF2}
\end{align}
Intuitively, a $(1,1)$ component of $\cF$ induces D4-brane charge on the BIon worldvolume, and drags it away from the boundary, while a $(2,0)+(0,2)$ component induces anti-D4-brane charge and therefore the opposite effect. So if the integrated norm of $\cF_2$ wins over that of $\cF_1$ the BIonic membrane suffers an additional force that draws it towards the boundary of AdS$_4$, providing a source of instability for the non-supersymmetric vacuum.


\section{Summary}
\label{bio-s:conclu}

In this chapter we have revisited the non-perturbative stability of type IIA $\cN=0$ AdS$_4 \times X_6$ orientifold vacua, where $X_6$ has a Calabi--Yau metric in the smeared-source approximation. For our analysis we have used the results of  \cite{Junghans:2020acz,Marchesano:2020qvg},  which give a description of these backgrounds beyond the Calabi--Yau approximation. Such a description is quite accurate in the large-volume, weak-coupling regime, at least at regions of $X_6$ away from the O6-plane location. However, as already pointed out, we are still working with an approximate solution which will have further corrections at higher orders in the expansion. At such a higher level of accuracy, and specially in non-supersymmetric settings, there will be additional corrections that one should take into account, and which are beyond the scope of the present analysis. 

Given our results, there are several open questions to be addressed. First, we have unveiled a potential decay channel for $\cN=0$ AdS$_4$ vacua with space-time filling D6-branes, triggered by nucleating D8-branes that take the system to a new $\cN=0$ vacuum with larger $|F_0|$ and fewer D6-branes. There are two quantities that determine if this decay channel exists, namely the curvature correction term $K_a^{(2)} T_{\rm D4}^a$ to the D8-brane action and the BIon correction $\Delta_{\rm D8}^{\rm Bion}$ defined in \eqref{bio-QTbionnosusyexp}. The sharpened WGC for membranes \cite{Ooguri:2016pdq} predicts that $K_a^{(2)} T_{\rm D4}^a+\Delta_{\rm D8}^{\rm Bion} > 0$, securing the decay channel. In the next chapter we will test this relation in several toroidal orbifold examples. In particular it will be interesting to see if the two terms always add up to yield a positive quantity, the key question being how $\Delta_{\rm D8}^{\rm Bion}$ behaves in general. Because $\cF$ is a non-closed but nevertheless quantized two-form, it could be that $\Delta_{\rm D8}^{\rm Bion}$ is determined by the topological data of the problem.

More generally, the instabilities that we have discussed only apply to vacua with space-time filling D6-branes. For instance, the explicit vacua described in \cite{DeWolfe:2005uu,Narayan:2010em} were based on toroidal orbifolds, but the $H$-flux and $F_0$ quanta were chosen such that no D6-branes were present. For these vacua and others alike, our results find no superextremal membrane that could mediate the decay, since D4-branes saturate a BPS bound in the same sense that they do in the smeared-source approximation analysis. It would be interesting to see if pushing our analysis to the next term in the expansion one could find that $Q_{\rm D4} \neq T_{\rm D4}$ in $\cN=0$ backgrounds, or if some other kind of corrections sourced by supersymmetry-breaking effects creates an imbalance. If not, one may consider more exotic classes of processes where four-form flux is discharged, like decays involve a mixture of bubbles of nothing and D4-brane charge (see e.g. \cite{Bomans:2021ara}) to fully test the sharpened WGC for membranes. 

In any event, we believe that the decay processes that we have studied are interesting per se, and deserve further study. Notice for instance that after bubble nucleation the AdS$_4$ flux dual to the Romans mass is not discharged, as in \cite{Maldacena:1998uz}, but on the contrary it increases. And the same happens with the 4-dimensional four-form flux dual to $G_4$. From the 4-dimensional viewpoint there is nothing wrong with this fact, as we jump to a new $\cN=0$ vacuum with lower vacuum energy. Indeed, we have argued in \ref{bio-s:insta} that these decays are favourable from the 4-dimensional viewpoint, even when we are away from the thin-wall approximation. It would however be interesting to carry a more detailed 4-dimensional analysis of this process, as well as to build the explicit 4-dimensional solution. Moreover, it would be important to analyze the superextremality of the membranes from a standard 4-dimensional viewpoint, like the analysis of the WGC for membranes carried out in \cite{Lanza:2020qmt}.

From the microscopic viewpoint, it would be interesting to see if our computations can be generalized to other string theory settings. Obvious candidates are the class of type IIA orientifold compactifications studied in \cite{Villadoro:2005cu,Banks:2006hg,Cribiori:2021djm}, which share many similar properties with the ones considered in this chapter. But one may also consider other compactifications which share key ingredients like scale separation and non-Abelian chiral gauge theories, and see if similar results are obtained. After all, our results hint that $\cN=0$ 4-dimensional EFTs with non-trivial gauge sectors are more susceptible to decay to vacua where such gauge sectors are absent. If true in general, this would have deep implications for string theory model building, and probably result into a new branch of implications of the Swampland Program.




\ifSubfilesClassLoaded{%
\bibliography{biblio}%
}{}

\end{document}

\graphicspath{{Images/Membranes_in_WGC}}

\ifSubfilesClassLoaded{%
\tableofcontents
}{}

\setcounter{chapter}{5}
\chapter{Membranes in \texorpdfstring{$AdS_4$}{AdS4} orientifold vacua and their Weak Gravity Conjecture}
\label{ch: membranes}

As we have discussed in depth throughout the previous chapters, in order to properly describe the string Landscape one not only needs to provide the set of string vacua, but also specify some key properties like their stability. In this sense, the AdS Instability Conjecture \cite{Ooguri:2016pdq,Freivogel:2016qwc}, that proposes that all $\cN=0$ AdS$_d$ vacua are at best metastable, is a very powerful statement. DGKT type vacua represent an important contender against this conjecture, as they provide a 4-dimensional perturbatively stable spectrum featuring scale separation and $\mathcal{N}=1,0$ supersymmetry (see \ref{cy-subsec: AdS vacua}). Both results could be made compatible by proving the existence of non-perturbative stabilities in the non-supersymmetric vacua, like a bubble nucleation process. Given that these are examples of AdS$_4$ vacua supported by 4-dimensional fluxes, the proposal of \cite{Ooguri:2016pdq} gives clear candidates to mediate non-perturbative decays, namely 4-dimensional membranes coupled to such fluxes, with a charge $Q$ and tension $T$ such that $Q >T$. In the chapter \ref{ch: bionic} we explored that possibility for $SU(3)\times SU(3)$ compactifications obtained from a perturbative expansion around the Calabi-Yau geometry that accounts for the backreaction of the localized sources. 

The most obvious candidate for the decay, i.e. D4-branes wrapping (anti)holomorphic two-cycles of $X_6$, was found to have $Q=T$ after including the first term of the expansion related to one-loop corrections. Thus, this kind of process corresponds to a marginal decay and not an actual instability. However, we found a better candidate: a potential decay channel mediated by a BIon made from a D8-brane wrapping the internal manifold $X_6$ and space-time filling D6-branes attached to it. As we concluded, at leading order they satisfy the BPS equality $Q=T$, but at the level of one-loop corrections and for $\cN =0$ vacua this is no longer true, there being two sources of correction to this equality. The first source is the correction to the D8-brane worldvolume action due to the curvature of $X_6$, that induces a negative D4-brane charge and tension specified by the second Chern class of $X_6$. The first source is the correction to the D8-brane worldvolume action due to the curvature of $X_6$, that induces a negative D4-brane charge and tension specified by the second Chern class of $X_6$. For the $\cN =0$ vacua of interest this correction is such that $\Delta^{\rm curv}_{\rm D8} (Q-T) > 0$, favouring the nucleation of the membrane towards the AdS$_4$ boundary. The second correction, given by \eqref{bio-QTbionnosusy} is harder to compute, as it involves the worldvolume flux induced by the BIon-like  backreaction of localized objects, namely the D6-branes ending on the D8-brane. 

In this chapter we undertake a deeper study of  $\Delta^{\rm BIon}_{\rm D8}$, considering orientifolds of the form $X_6 = (T^2 \times T^2 \times T^2)/\Gamma$ with different orbifold groups $\Gamma$ and D6-brane configurations. Remarkably, we find that for certain D6-brane configurations $\Delta^{\rm BIon}_{\rm D8}  < 0$, even in the simple geometry $X_6 = T^6/(\IZ_2 \times \IZ_2)$. The key ingredient to achieve this negative sign seems to be the presence of localized sources that do not intersect, and, in particular, non-intersecting O6-planes. 

The chapter is organized as follows. In section \ref{mem-s:nonsusy} we review the AdS$_4$ compactifications of interest and the computation of 4-dimensional membrane charges and tensions in them, reintroducing the definitions and notations seen throughout the previous chapter in a slightly modified way that makes more explicit the contribution from the localized sources and the importance of the homology groups to which they belong. In section \ref{mem-s:Delta} we summarize how to compute the BIonic excess charge $\Delta^{\rm BIon}_{\rm D8}$ in toroidal orientifolds, based on the explicit computations of section \ref{mem-s:examples}. Given this expression for $\Delta^{\rm BIon}_{\rm D8}$ we provide a simple example in which $\Delta^{\rm curv}_{\rm D8} +\Delta^{\rm BIon}_{\rm D8} < 0$. Due to flux quantization conditions, such an example must be engineered in a blown-up $T^6/(\IZ_2 \times \IZ_2)$ geometry, discussed in appendix \ref{mem-ap:Z2xZ2}, and whose second Chern class is computed in Appendix \ref{mem-ap:Z2xZ2curv}. We finally review in section \ref{mem-sec: current status} some of the most recent developments on the subject, that were obtained following this work, and draw our conclusions in section \ref{mem-s:conclu}. 


\section{\texorpdfstring{$AdS_4$}{AdS4} orientifold vacua}
\label{mem-s:nonsusy}

We will consider the same type IIA String Theory compactified on a Calabi--Yau three-fold $X_6$ described in the previous chapters. The fixed locus $\Pi_{\rm O6}$ of ${\cal R}$ is made of one or several smooth 3-cycles of $X_6$, hosting O6-planes. The presence of O6-planes reduces the background supersymmetry to 4d $\CN=1$, and induces an RR tadpole that can be cancelled by a combination of D6-branes wrapping special Lagrangian three-cycles \cite{Blumenhagen:2005mu,Blumenhagen:2006ci,Marchesano:2007de,Ibanez:2012zz}, D8-branes wrapping coisotropic cycles with fluxes \cite{Font:2006na}, and background fluxes including the Romans mass. For simplicity, in the following we will consider that the D-brane content consists of D6-branes placed on top of the O6-planes or in another representative of the same homology class. The remaining RR tadpole is then cancelled by the presence of background fluxes, yielding either a 4d $\CN=1$ or $\CN=0$ vacuum.

As in the previous chapter, we consider the RR flux polyform \eqref{cy-eq: G definition} and its associated Bianchi identities \eqref{cy-eq: BI def}
\begin{equation}\label{mem-IIABI}
\ell_s^{2} \,  d (e^{-B} \wedge {\bf G} ) = - \sum_\a \lambda \left[\delta (\Pi_\alpha)\right] \wedge e^{\frac{\ell_s^2}{2\pi} F_\alpha} \, ,  \qquad d H = 0 \, ,
\end{equation} 
where $\Pi_\alpha$ hosts a D-brane source with a quantized worldvolume flux $F_\alpha$, and $\delta(\Pi_\alpha)$ is the bump $\delta$-function form with support on $\Pi_\alpha$ and indices transverse to it, such that $\ell_s^{p-9} \d(\Pi_\a)$ lies in the Poincar\'e dual class to $[\Pi_\a]$. O6-planes contribute to the Binachi identities as D6-branes but with minus four times their charge and $F_\alpha \equiv 0$. Finally, $\lambda$ is the operator that reverses the order of the indices of a $p$-form.

In the presence of D6-branes and O6-planes the Bianchi identities for the RR fluxes are given by \eqref{bio-BIG} and thus we recover the tadpole relation \eqref{bio-tadpole}
\be
{\rm P.D.} \left[4\Pi_{\rm O6}- N_\a \Pi_{\rm D6}^\a\right] = m [\ell_s^{-2} H] \, ,
\label{mem-tadpole}
\ee
where  $N_\a$ is the number D6-branes wrapping a three-cycle in the  homology class $[\Pi^{\rm D6}_\a]$. 

We keep our focus in the first two branches of  4-dimensional  vacua described in table \ref{cy-table: summary ads vacua}, which correspond to the following conditions on the internal background fluxes:
\be
[ H ]  = \frac{2}{5} G_0 g_s  [\re \Omega_{\rm CY} ] \, , \quad \int_{X_6} G_2 \wedge \tilde{\omega}^a =  0\, ,  \quad \frac{1}{\ell_s^6} \int_{X_6} G_4  \wedge \omega_a  =  \epsilon \frac{3}{10} G_0 {\cal K}_a \, , \quad  G_6  =  0\, , 
\label{mem-intflux}
\ee
where $\omega_a$, $\tilde \omega^a$ are the elements of the harmonic basis \ref{cy-table: harmonic basis} and, as before, $\CK_a = - \int_{X_6} J_{\rm CY} \wedge J_{\rm CY} \wedge \omega_a=\cK_{abc}t^at^b$. Here  $\epsilon = -1$ describes supersymmetric backgrounds, while $\epsilon = 1$ generates the non-supersymmetric vacua. 

Let us recover once again the formalism introduced in section \ref{cy-subsec: beyond smearing} to write an approximate solution to the 10d massive type IIA  equations of motion for both the supersymmetric and non-supersymmetric case.  Indeed, recovering the discussion around \eqref{cy-eq: defK}, let us express the RR two-form flux in terms of a three-form current $K$ as $G_2 = d^\dag_{\rm CY} K$, so that its Bianchi identity reads
\begin{equation}
    \Delta_{\rm CY} K = G_0H +  \delta_{\rm O6+D6}    = \frac{2}{5} m^2 g_s \ell_s^{-2} \re \Omega_{\rm CY}+  \delta_{\rm O6+D6}  +  \cO(g_s^2)\, ,
    \label{mem-eq: K equation}
\end{equation}
where we have defined $\Delta_{\rm CY} = d^\dag_{\rm CY} d + d d^\dag_{\rm CY}$ and $\delta_{\rm O6+D6} = - 4\delta_{\rm O6} + N_\a \delta_{\rm D6}^\a$, and we have used  the leading term in the expansion of $H$, see below. This equation has a solution if \eqref{mem-tadpole} is satisfied, and it is particularly simple at leading order in $g_s$ if the D6-branes wrap special Lagrangian three-cycles $\Pi^{\rm D6}_\a$ that are mutually BPS with $\Pi_{\rm O6}$. At this level $H$ is a harmonic three-form, which means that we can decompose the leading term of the rhs of \eqref{mem-eq: K equation} as
\begin{equation}
  \ell_s^{-2} \sum_{\a,\eta} q_{\a, \eta} \left(  H_\a - \delta(\Pi_{\a,\eta}) \right) \, .
    \label{mem-Krhsum}
\end{equation}
Here $\Pi_{\a, \eta}$ is a three-cycle hosting a localized source, either D6-brane or O6-plane, and $q_{\a, \eta} \in \IZ$ minus its charge in D6-brane units. The index $\eta$ labels different three-cycles that correspond to the same homology class: $[\Pi_{\a,\eta}] = [\Pi_{\a}]$, $\forall \eta$.  Finally, $H_\a$ is the harmonic representative of the Poincar\'e dual class to $\ell_s^3 [\Pi_{\a}]$. Then, using that $\Pi_{\a,\eta}$ are special Lagrangian three-cycles calibrated by $\im \Omega_{\rm CY}$, one can show that the Laplace equations
\begin{equation}
    \ell_s^2  \Delta_{\rm CY} K_{\a,\eta} = H_\a - \delta(\Pi_{\a,\eta})\, .
    \label{mem-eq: Kalpha equation}
 \end{equation}
 have a solution of the form \cite{Hitchin:2010qz,Marchesano:2020qvg}
 \be
 K_{\a,\eta} = \varphi_{\a,\eta} \re \Omega_{\rm CY}  + \re k_{\a,\eta} \, ,
\label{mem-formKalpha}
\ee
and by linearity of the equation \eqref{mem-eq: K equation} one can express $K$ as
\be
 K =  \sum_{\a,\eta} q_{\a, \eta} K_{\a,\eta} = \varphi \re \Omega_{\rm CY}  + \re k \, ,
\label{mem-formK}
\ee
and so the quantities $\varphi$ and $k$ that determine the background \eqref{cy-eq: solutionsu3} are given by $ \varphi = \sum_{\a,\eta} q_{\a, \eta} \varphi_{\a,\eta}$ and $ k =  \sum_{\a,\eta} q_{\a, \eta} k_{\a,\eta}$, respectively. In particular we have that
\be
\Delta_{\rm CY}  \varphi_{\a,\eta} = \left(\frac{{\cal V}_{\Pi_\a}}{{\cal V}_{\rm CY}} - \delta^{(3)}_{\a,\eta}\right)  \ \implies \ \varphi \sim \cO(g_s^{1/3})\, ,
\ee
where $\delta^{(3)}_{\a,\eta} \equiv \star_{\rm CY} (\im \Omega_{\rm CY} \wedge \delta(\Pi_{\a,\eta}))$, ${\cal V}_{\rm CY} = -\frac{1}{6}\ell_s^{-6} \int_{X_6} J_{\rm CY}^3$ is the Calabi--Yau volume and ${\cal V}_{\Pi^{\rm O6}_\a} = \ell_s^{-3} \int_{\Pi_\a} \im \Omega_{\rm CY}$. As a result $\varphi \sim - \frac{q_{\a, \eta}}{r}$ in the vicinity of a $\Pi^{\rm O6}_{\a,\eta}$. If the localized charge is negative it describes a small region where the 10d string coupling blows up, the warp factor becomes negative and, as expected, the supergravity approximation cannot be trusted. 

Let us consider a simplified setup in which all localized sources wrap three-cycles determined by the O6-plane locus. We describe the O6-plane locus  as a union of several smooth three-cycles
\begin{equation}
    \Pi_{\rm O6} = \bigcup_{\a,\eta} \Pi_{\a,\eta}\, , \quad \text{with} \quad    [\Pi_{\rm O6}] = \sum_\a p_\a [\Pi_{\a}]\, ,
    \label{mem-splitO6}
\end{equation}
where  the index $\alpha$ runs over different homology classes and $\eta$ over the $p_\a$ different representatives of the same homology class: $[\Pi^{\rm O6}_{\a,\eta}] = [\Pi^{\rm O6}_{\a,\eta'}] \equiv [\Pi^{\rm O6}_{\a}]$. Then we consider D6-branes that wrap three-cycles on the same homology classes, that is we take $[\Pi^{\rm D6}_{\a}] = [\Pi^{\rm O6}_{\a}]$. One may further assume that all D6-branes lie on top of O6-planes, so $\Pi^{\rm D6}_{\a,\eta} = \Pi^{\rm O6}_{\a,\eta}$. An advantage of this further simplification is that on top of the O6-planes one can always have a vanishing worldvolume flux for the D6-brane, which is a necessary condition for a vacuum. If we displace such a D6-brane away from the O6-plane location the presence of the $H$-flux will generically induce a $B$-field in its worldvolume, that will generate a dynamical tadpole.\footnote{In general there will be a discretum of other representatives within $ [\Pi^{\rm O6}_{\a}]$ besides the O6-plane locus where the D6-brane worldvolume flux can vanish, similarly to the open string landscape in \cite{Gomis:2005wc}. Our discussion below can be easily extended to include those D6-brane locations as well.} Then, in an analogous fashion to \cite{Mininno:2020sdb}, the WGC could be violated due to the lack of equilibrium. Our choice avoids such a possibility. 

To sum up, we consider a setup in which the three-cycles $\Pi_{\a, \eta}$ in \eqref{mem-Krhsum} correspond to those in \eqref{mem-splitO6}. As a result 
\begin{equation}
   \ell_s^{2} \delta_{\rm O6+D6} = - \sum_{\a,\eta}  q_{\a,\eta} \delta(\Pi^{\rm O6}_{\a,\eta})\, ,
   \label{mem-O6D6source}
\end{equation}
where $q_{\a,\eta} = 4 - N_{\a, \eta}$ is minus the localized charge on each three-cycle. We also choose P.D.$[\ell_s^{-2}H] = h [\Pi_{\rm O6}]$ and $N_\a \equiv \sum_{\eta} N_{\a,\eta}  = N p_\a$, which leads to the simple tadpole constraint
\be
mh = 4 - N  \, .
\label{mem-tadpole2}
\ee
Here notice that $h$ and $N$ need not be integers, because a consistent configuration only requires that $h[\Pi_{\rm O6}]$ and $N [\Pi_{\rm O6}]$ are integer homology classes. So if $[\Pi_{\rm O6}] = M [\hat{\Pi}_{\rm O6}]$, with $M \in \IZ$ and $[\hat{\Pi}_{\rm O6}] \in H_{3}(X_6, \IZ)$, we only need to require that $hM, NM \in \IZ$, as will happen in the toroidal orientifold geometries that we will analyze in the following sections. Additionally, the 4-dimensional analysis on vacua conditions requires that $mh$ and $N, N_{\a,\eta}$ are non-negative, so that there is a finite number of solutions to the tadpole equation. 

The approximate flux background is also described in terms of $\varphi$ and $k$. From our knowledge of section \ref{cy-sec: flux compactifications} and chapter \ref{ch: bionic} we have that
\begin{subequations}
	\label{mem-solutionflux}
\begin{align}
 H & =   \frac{2}{5} G_0 g_s \left(\re \Omega_{\rm CY} + R g_s K \right) -\frac{S}{2}  d\re \left(\bar{v} \cdot \Omega_{\rm CY} \right) + \cO(g_s^{3}) \label{mem-H3sol} \, , \\
 \label{mem-G2sol}
 G_2 & =     d^{\dag}_{\rm CY} K  + \cO(g_s)  = - J_{\rm CY} \cdot d(4 \varphi \im \Omega_{\rm CY} - \star_{\rm CY} K) + \cO(g_s) \, , \\
G_4 & =  -\epsilon G_0 J_{\rm CY} \wedge J_{\rm CY} \left(\frac{3}{10}  + \epsilon \frac{4}{5} g_s \varphi \right)+  S J_{\rm CY} \wedge g_s^{-1} d \im v + \cO(g_s^2) \, , \\
G_6 & = 0\, ,
\end{align}
\end{subequations}   
where in the supersymmetric case
\be
\label{mem-susypar}
\epsilon = -1 \, , \qquad R = 1\, , \qquad S = 1\, ,
\ee
and in the non-supersymmetric case
\be
\label{mem-nonsusypar}
\epsilon = 1 \, , \qquad R = -2 \, , \qquad S =  - \frac{1}{5}\, .
\ee
Finally, $v$ is a (1,0)-form determined by
\be
v  = g_s \p_{\rm CY} f_\star + \cO(g_s^3)\, , \qquad \text{with} \qquad \Delta_{\rm CY} f_\star  = - g_s 8 G_0 \varphi \, .
\ee 

\subsubsection*{4d membranes}

In this background, one may consider branes that correspond to membranes in 4d. There are three different kinds of such membranes that are BPS objects in $\cN=1$ vacua. D8-branes wrapping the whole internal manifold $X_6$, NS5-branes wrapping special Lagrangian three-cycles of $X_6$ and D4-branes wrapping (anti)holomorphic two-cycles of $X_6$.

Let us summarize the results for D4-brane wrapping an (anti)holomorphic two-cycle $\Sigma$ of $X_6$ described in chapter \ref{ch: bionic}. Crossing such a membrane in 4d induces a change in the quanta of the internal four-form flux, scanning over the infinite family of flux vacua found in \cite{DeWolfe:2005uu}. To see if such a membrane induces a non-perturbative instability  one can dimensionally reduce the DBI+CS action of the D4-brane in the probe approximation, as done in \cite{Aharony:2008wz,Narayan:2010em}. This can be interpreted as computing the 4-dimensional membrane charge $Q$ and tension $T$, and if $Q >T$ one expects an instability similar to the one of \cite{Maldacena:1998uz}. This computation was performed in \cite{Aharony:2008wz,Narayan:2010em} for D4-branes in both cases $\epsilon = \pm 1$, in the smearing approximation, corresponding to only consider the leading terms of the background expansion \eqref{cy-eq: solutionsu3} and \eqref{mem-solutionflux}, which yield a Calabi--Yau metric, and more precisely to set $\varphi = k =0$ in those expressions. In the previous chapter the computation was extended to the corrected backgrounds in \cite{Marchesano:2021ycx}, which can be interpreted as a one-loop correction to the DBI+CS expressions of \cite{Aharony:2008wz,Narayan:2010em}, and more precisely to the effect of a crosscap diagram between such   D4-branes and the O6-planes. At this level of accuracy it was shown in \eqref{bio-QDGKT}, \eqref{bio-QDGKTnosusy} and the discussion around those expressions that in 4-dimensional Planck units 
\be
T_{\rm D4} =  e^{K/2} \frac{1}{\ell_s^2} \left| \int_\Sigma J_{\rm CY} \right|\, , \quad Q_{D4} =   e^{K/2} \frac{\epsilon \eta}{\ell_s^2} \int_\Sigma J_{\rm CY} -  \frac{a5}{3 G_0} dd^\dag_{\rm CY} \left(f_\star  J_{\rm CY}  \right) = e^{K/2} \frac{\epsilon \eta}{\ell_s^2} \int_\Sigma J_{\rm CY}\, ,
\ee
where $\eta = {\rm sign} \, G_0$, $\eps,a$ are as in \eqref{mem-susypar} and \eqref{mem-nonsusypar} and $K$ is the 4-dimensional K\"ahler potential. By appropriately choosing the orientation of $\Sigma$, or equivalently by considering D4-branes or anti-D4-branes on holomorphic cycles, one can get $ Q_{\rm D4}=T_{\rm D4}$, which correspond to marginal domain walls, but not $ Q_{\rm D4}>T_{\rm D4} $. Thus, in order to check the refinement of the Weak Gravity Conjecture made in \cite{Ooguri:2016pdq} one should compute further terms in the background expansion given above.

As we detailed in the previous chapter, in models with background D6-branes, that is with $N > 0$ in \eqref{mem-tadpole2}, there is second kind of 4-dimensional membranes obtained from D-branes that are BPS in $\CN=1$ vacua. These are D8-branes wrapped on the whole of $X_6$, whose description is more involved than those of D4-branes. First, they can host harmonic (1,1) primitive worldvolume fluxes $\cF_h$, which together with the curvature corrections modify the DBI+CS action and induce D4-brane charge and tension. Taking these two effects into account one obtains a total tension of the form \eqref{bio-TD8tot}
\be
T_{\rm D8}^{\rm total} = T_{\rm D8} + \left(K^F_a - K_a^{(2)}\right)  T_{\rm D4}^a\, ,
\label{mem-TD8tot}
\ee
with $T_{\rm D8} = e^{K/2} {\cal V}_{\rm CY}$ and   $T_{\rm D4}^a = e^{K/2} t^a$, where $J_{\rm CY} = t^a \omega_a$ defines the K\"ahler moduli. Also from \eqref{bio-Kcurv} and \eqref{bio-eq: KF def} we know
\be
 K_a^{(2)} = - \frac{1}{24\ell_s^{6}} \int_{X_6} c_2(X_6) \wedge \omega_a \qquad \text{and} \qquad K^F_a = \frac{1}{2\ell_s^6} \int_{X_6} \cF_h \wedge \cF_h \wedge \omega_a\, .
\label{mem-Kcorr}
\ee
In this case, motivated by the insight obtained in the previous chapter, we make explicit the distinction between the different contributions to the worldvolume flux of the D8. It is important to notice that in our conventions both $ K_a^{(2)} T_{\rm D4}^a$ and $K^F_a T_{\rm D4}^a$ are non-negative quantities. In addition, one can always set $K^F_a=0$ via setting $\cF_h =0$. 

A second important feature of these D8-branes is that they have D6-branes ending on them, to cure the Freed-Witten anomaly induced by the $H$-flux \cite{Maldacena:2001xj}. In 4-dimensional terms, a membrane of this sort induces a jump in the flux quantum $m$ when crossing its worldvolume, so there should be a corresponding jump in $N$ in order to satisfy \eqref{mem-tadpole2} at both sides of the membrane.  For a single D8-brane we have the following transition\footnote{As we will see, such D8-branes oftentimes go in pairs. However, their  jump \eqref{mem-jump} should be considered separately.} 
\begin{equation}
    m \to m + 1 \,\implies \, q_{\a, \eta} \to  q_{\a, \eta} + \hat q_{\a, \eta} \quad \text{with} \quad \sum_{\eta} \hat q_{\a, \eta} =  h p_\a\, ,
    \label{mem-jump}
\end{equation}
where  $\hat q_{\a, \eta} \geq 0$, and the upper bound $q_{\a, \eta} \leq 4$ should always be respected.  
At the level of accuracy with which we are describing the 10d background, this feature manifests itself as a BIon-like profile developed by the D8-brane, as shown in chapter \ref{ch: bionic}. This profile is slightly more involved than the simplest examples \cite{Gibbons:1997xz,Evslin:2007ti}, but it contains similar features. We have a non-closed piece of the D8-brane worldvolume flux that reads\footnote{For the simplest configuration in which D6-branes are equally distributed on top of the O6-plane components before and after the jump, that is $\hat{q}_{\a,\eta} = h$, $\forall \a, \eta$, we have that $ \mathcal{F}_{\rm BIon} =  G_0^{-1} d^{\dag}_{\rm CY} K+ \cO(g_s)$, as assumed in the previous chapter.}
\begin{equation}
    \mathcal{F}_{\rm BIon} =  \sum_{\a, \eta}  \hat{q}_{\a, \eta} \cF_{\a, \eta}  + \cO(g_s) \, , \qquad  \cF_{\a, \eta}  = d^\dag K_{\a, \eta}\, ,
    \label{mem-eq: worldvolume flux}
\end{equation}
We also have  a non-trivial profile for the D8-brane transverse coordinate
\be
Z = z_0 - \ell_s \sum_{\a, \eta} \hat q_{\a, \eta} \varphi_{\a, \eta}\, , \qquad z_0/\ell_s \in \mathbb{R}\, . 
\ee
This BIon-like profile also contributes to the D8-brane DBI+CS action, and therefore modifies the 4-dimensional membrane charge and tension. In terms of the latter, we have an extra term in \eqref{mem-TD8tot} 
\be
 T_{\rm D8}^{\rm BIon} =  e^{K/2} \frac{1}{2\ell_s^6} \int_{ X_6} J_{\rm CY} \wedge \cF_{\rm BIon}^2 + \cO(g_s^2) \, ,
\label{mem-Tbion}
\ee
which resembles the term $K_a^F T^a_{\rm D4}$, except that it involves a different component of the worldvolume flux. In the supersymmetric background and for a BPS D8-brane, the three corrections to $T_{\rm D8}$ also appear in the 4-dimensional membrane charge, yielding as expected that $T_{\rm D8}^{\rm total}  = Q_{\rm D8}^{\rm total}$. For the non-supersymmetric background with $\epsilon = 1$ the same D8-brane develops these corrections but with opposite charge. That is
\be
Q_{\rm D8}^{\rm total} = T_{\rm D8} - \epsilon \left(K^F_a - K_a^{(2)} \right)  T_{\rm D4}^a - \epsilon  T_{\rm D8}^{\rm BIon} \, .
\label{mem-QD8tot}
\ee
 As a result, the excess charge for such membranes  reads
\be
 Q_{\rm D8}^{\rm total} - T_{\rm D8}^{\rm total} = \left(1+\epsilon\right) \left[K_a^{(2)} T_{\rm D4}^a - K^F_a T_{\rm D4}^a - T_{\rm D8}^{\rm BIon} \right]  \, .
\label{mem-QTtotalnosusy}
\ee
If the term in brackets is positive for some 4-dimensional  membrane the refined WGC of \cite{Ooguri:2016pdq} is verified, signalling a non-perturbative instability of the non-supersymmetric vacuum. As mentioned before, the first term inside the bracket is always non-negative, and in fact it is positive away from the boundary of the K\"ahler cone. The second one is non-positive, but it can always be chosen to vanish by appropriate choice of worldvolume fluxes. It is thus the third one that remains to analyze, which will be the subject of the next section. For concreteness we define the quantity
\be
\Delta_{\rm D8}^{\rm Bion} \equiv - e^{K/2} \frac{1}{2\ell_s^6} \int_{ X_6}  J_{\rm CY} \wedge \cF^2_{\rm BIon}\, , 
\label{mem-QTbion}
\ee
that we dub as the BIonic excess charge of the membrane. A priori this quantity is comparable to the effect of curvature corrections, and it is in fact larger for Calabi--Yau geometries near a toroidal orbifold limit. In the next sections we will analyze $\Delta_{\rm D8}^{\rm Bion}$ precisely for those geometries. Remarkably, we find a very simple expression, that suggests generalization to arbitrary Calabi--Yau geometries of the form $\Delta_{\rm D8}^{\rm Bion} = D_a T_{\rm D4}^a$, where $D_a$ depend on discrete data.


\section{Toroidal orientifolds}
\label{mem-s:Delta}

In this section we specify the above setup to toroidal Abelian orbifolds of the form $T^6/\IZ_N$ or $T^6/(\IZ_N \times \IZ_M)$, where the covering space is a factorizable six-torus $T^6 = (T^2)_1 \times (T^2)_2 \times (T^2)_3$ and the orbifold action respects the factorization. As we show in the next section, for these geometries one can compute the quantity \eqref{mem-QTbion} explicitly, obtaining a simple general expression. In the following we will summarize this expression and discuss its consequences for the stability of AdS$_4$ vacua with different D6-brane configurations.

\subsection{The BIonic excess charge}
\label{mem-ss:BIec}

In toroidal Abelian orbifolds of the form $(T^2)_1 \times (T^2)_2 \times (T^2)_3/\Gamma$, with $\Gamma = \IZ_N$ or $\Gamma= \IZ_N \times \IZ_M$, the O6-plane content in the covering space $T^6$ is characterized by a set of factorizable three-cycles, which in homology read
\be
[\Pi_{\rm O6}] = \sum_{\alpha,\eta}  [\Pi^{\rm O6}_{\a,\eta}] = \sum_\alpha  p_\a [\Pi^{\rm O6}_\a] = \sum_\alpha p_\a  \left[(n_\a^1, m_\a^1) \times (n_\a^2, m_\a^2)\times (n_\a^3, m_\a^3)\right] \, .
\label{mem-O6sum}
\ee
Here $\a$ runs over different homology classes in the covering space, specified by the wrapping numbers $(n_\a^i, m_\a^i) \in \IZ^2$ of each factorizable three-cycle on $(T^2)_i$. The index $\eta$ runs over different representatives in the same homology class, giving rise to a multiplicity $p_\a$. If we place the existing D6-branes on top of the O6-planes, more precisely $N_{\a,\eta}$ of them on top of $\Pi^{\rm O6}_{\a,\eta}$, the background RR two-form flux is of the form $G_2 = d^\dag_{\rm CY} K$, where
\be
K =  \sum_{\a,\eta} q_{\a, \eta} K_{\a,\eta}\, , \qquad  \ell_s^2  \Delta_{\rm CY} K_{\a,\eta} = H_\a - \delta(\Pi^{\rm O6}_{\a,\eta})\, ,
\label{mem-Ktorus}
\ee
with $ q_{\a, \eta} = 4 - N_{\a,\eta}$ and
\be
H_\a = \ell_s^3 \left(m_\a^1 dx^1 - n_\a^1dy^1\right) \wedge \left(m_\a^2 dx^2 - n_\a^2dy^2\right) \wedge \left(m_\a^3 dx^3 - n_\a^3dy^3\right)\, ,
\ee
where $(x^i,y^i)$ are the period-one coordinates of $(T^2)_i$. From here one can extract the quantities $\varphi$ and $k$ that appear in \eqref{mem-formK}, and describe the full background \eqref{cy-eq: solutionsu3} and \eqref{mem-solutionflux}. 

Additionally, given a D8-brane-mediated flux jump of the form \eqref{mem-jump}, the BIon-like solution that describes the D8/D6-brane system features a coexact worldvolume flux of the form  \eqref{mem-eq: worldvolume flux}.  As a consequence we have that \eqref{mem-QTbion} is of the form
\begin{equation}
    \Delta_{\rm D8}^{\rm Bion} =  \oh \sum_{\alpha,\beta, \eta, \zeta} \hat{q}_{\a,\eta} \hat{q}_{\b,\zeta}\, {\Delta}_{\alpha,\eta;\beta,\zeta}\, , \qquad {\Delta}_{\alpha,\eta;\beta,\zeta} = - e^{K/2} \frac{1}{\ell_s^{6}} \int_{ X_6}  J_{\rm CY} \wedge \cF_{\a,\eta} \wedge \cF_{\b,\zeta} .
    \label{mem-Deltasum}
\end{equation}

From our explicit computations in the next section we moreover obtain the following results:

\begin{itemize}
    \item The integral in \eqref{mem-Deltasum} is non-zero only when the intersection number $I_{\alpha\beta} = [\Pi^{\rm O6}_\alpha] \cdot [\Pi^{\rm O6}_\beta] = 0$ and $[\Pi^{\rm O6}_\alpha] \neq [\Pi^{\rm O6}_\beta]$, which in particular implies that $\Delta_{\alpha,\eta;\alpha,\zeta} = 0$.  In practice, this means that non-vanishing contributions  to \eqref{mem-Deltasum} come from $\cN =2$ sectors of the compactification, that is from pairs of D6-branes wrapping three-cycles related by an $SU(2)$ rotation. In our setup, this translates into wrapping numbers $(n_\a^i, m_\a^i)$, $(n_\b^i, m_\b^i)$,  that are similar in one two-torus $(T^2)_i$ and different in the other two. We denote these pairs of three-cycles as $\cN=2$ pairs, see figure \ref{mem-fig: separation and intersection} for examples.

\begin{figure}[htb]
\centering
\begin{subfigure}[b]{.85\textwidth}
  \centering
     \includegraphics[scale=0.35]{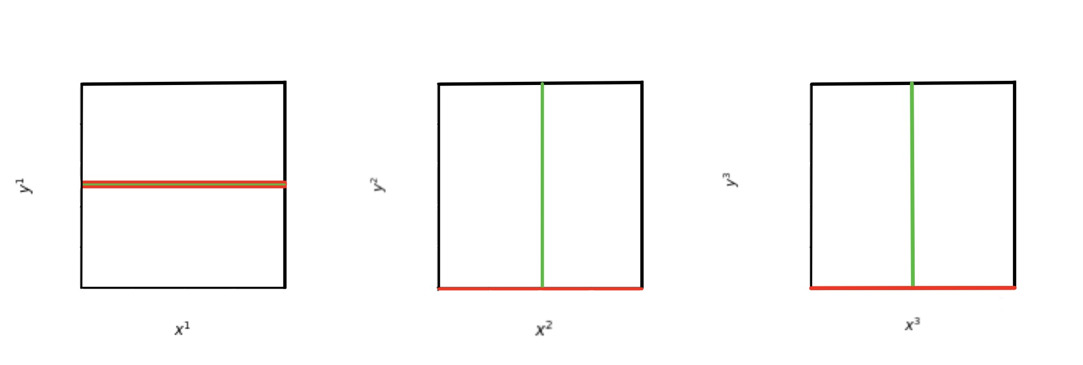}
    \caption{Diagram corresponding to an $\cN=2$ pair with one intersection over a one-cycle.}
 \end{subfigure}\\
\begin{subfigure}[b]{.85\textwidth}
    \centering
    \includegraphics[scale=0.348]{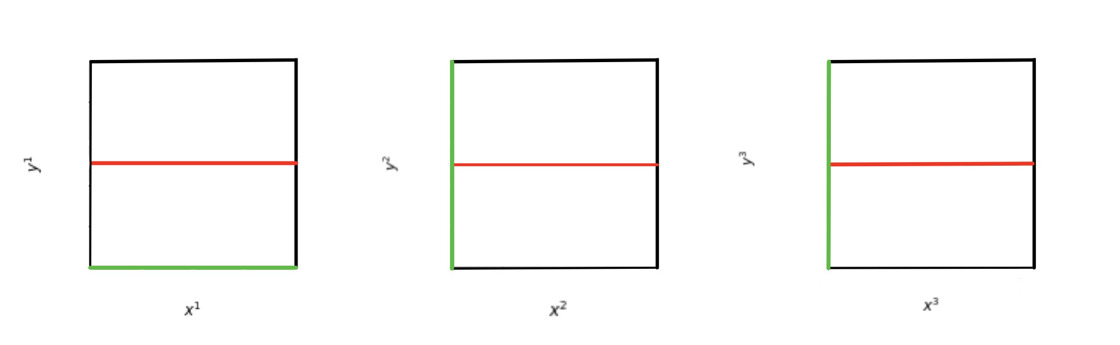}
    \caption{Diagram corresponding to an $\cN=2$ pair with  no intersection.}
    \end{subfigure}
    \caption{Configuration of 3-cycles projected over $T^2\times T^2\times T^2$ that contribute to \eqref{mem-Deltasum} in the $T^6/\mathbb{Z}_2\times\mathbb{Z}_2$ orbifold.}
    \label{mem-fig: separation and intersection}
\end{figure}
 \item Given a $\cN=2$ pair $(\a,\eta;\b,\zeta)$, the integral in \eqref{mem-Deltasum} depends separately on the indices $\a,\b$ that describe the homology classes $ [\Pi^{\rm O6}_\alpha]$ and $[\Pi^{\rm O6}_\beta]$, and the indices $\eta,\zeta$ that specify the representatives.  The dependence in $\a,\b$ corresponds to the number of regions of minimal separation between $\Pi^{\rm O6}_\alpha$ and $\Pi^{\rm O6}_\beta$, which we dub $\cN=2$ subsectors. For instance, if $\Pi^{\rm O6}_\alpha$ and $\Pi^{\rm O6}_\beta$ intersect over one-cycles, the number of $\cN=2$ subsectors is the number of intersections. To measure this number we define
 \begin{equation}
    \#(\Pi_\a \cap \Pi_\b)_i = |n_\a^j m_\b^j - n_\b^jm_\a^j| \times |n_\a^k m_\b^k - n_\b^km_\a^k|\, ,
    \label{mem-interdef}
\end{equation}
where $i \neq j \neq k$. When $\Pi^{\rm O6}_\alpha$ and $\Pi^{\rm O6}_\beta$ have parallel one-cycles in $(T^2)_i$ but they do not coincide, \eqref{mem-interdef} does not count intersections, but instead regions of minimal separation between the two three-cycles. In both cases, \eqref{mem-interdef} amounts to the number of `intersections' in the two two-tori where $\Pi^{\rm O6}_\alpha$ and $\Pi^{\rm O6}_\beta$ are not parallel, it is non-vanishing for a single choice of $i$, and because each $\CN=2$ subsector contributes equally to the integral in \eqref{mem-Deltasum}, $\Delta_{\a,\eta;\b,\zeta}$ is proportional to this number. 

\item The dependence on the indices $\eta,\zeta$ arises because $\Delta_{\a,\eta;\b,\zeta}$ is different if   $\Pi^{\rm O6}_{\alpha,\eta}$ and $\Pi^{\rm O6}_{\beta,\zeta}$ intersect or not. In general, the contribution of each $\CN=2$ subsector to the integral in \eqref{mem-Deltasum} is proportional to $t^i$, which is the area of the $(T^2)_i$ selected by \eqref{mem-interdef}, or in other words the two-torus where $\Pi^{\rm O6}_{\alpha,\eta}$ and $\Pi^{\rm O6}_{\beta,\zeta}$ are parallel. The coefficient of the contribution depends on whether  these two three-cycles intersect or not. If they intersect over a one-cycle on $(T^2)_i$, each 
$\CN=2$ subsector contributes to the integral $-\ell_s^{-6}\int_{T^6} J_{\rm CY} \wedge \cF_{\a,\eta} \wedge \cF_{\b,\zeta}$  over the covering space  as
\begin{equation}
   \frac{t^i}{12} \,. 
   \label{mem-intercont}
\end{equation}
If instead $\Pi^{\rm O6}_{\alpha,\eta}$ and $\Pi^{\rm O6}_{\beta,\zeta}$ do not overlap, but they are only parallel in $(T^2)_i$ we obtain\footnote{In the toroidal orientifold geometries that we consider in the next section, an $\cN=2$ pair of O6-planes that do not intersect are separated at mid-distance in their common transverse space in $(T^2)_i$. When we consider D6-branes wrapped in the same homology classes $[\Pi^{\rm O6}_{\alpha}]$ and $[\Pi^{\rm O6}_{\beta}]$ but not on top of orientifold planes in $(T^2)_i$, their BPS locations form a discretum analogous to the ones in \cite{Gomis:2005wc,Marchesano:2006ns}, because the presence of $H$-flux implies that only at certain discrete locations the D6-brane worldvolume flux $\cF = B|_{\Pi_{\rm D6}} +\frac{\ell_s^2}{2\pi}F$ can vanish. In this case, the separation between three-cycles is of the form $\frac{\ell_s^2 t_i}{L} \frac{k}{2P}$, where $L$ the length of the one-cycle wrapped in $(T^2)_i$, $P \in \mathbb{N}$ is determined by the quanta of $H$-flux, and $0\leq k \leq 2P$ is an integer. Given this separation, the contribution of this $\CN=2$ D6-brane pair to the integral $-\ell_s^{-6}\int_{T^6} J_{\rm CY} \wedge \cF_{\a,\eta} \wedge \cF_{\b,\zeta}$  is given by 
\begin{equation}
\nonumber
    \oh\left(\frac{1}{6} - \frac{k}{2P}\left(1-\frac{k}{2P}\right)\right) t^i\, ,
\end{equation}
which reduces to \eqref{mem-intercont} for $k=0$ and to \eqref{mem-interpar} for $k=P$.\label{mem-OSL}}
\begin{equation}
   -\frac{t^i}{24} \,. 
   \label{mem-interpar}
\end{equation}
Integrating over $X_6$, we divide both results by the orbifold group $\Gamma$ order,  dubbed $N_\Gamma$.

\end{itemize}

Adding all these results together, we end up with the following expression for the BIonic contribution to the 4-dimensional membrane excess charge:
\begin{equation}
     \Delta_{\rm D8}^{\rm Bion} = \frac{1}{24N_\Gamma} \sum_{(\a,\eta;\b,\zeta)\in \cN=2} \hat{q}_{\a,\eta} \hat{q}_{\b,\zeta}\, \varepsilon_{\eta\zeta} \, \#(\Pi_\a \cap \Pi_\b)_i T^i_{\rm D4}\, .
     \label{mem-finalDelta}
\end{equation}
Here $T^i_{\rm D4} = e^{K/2} t^i$ corresponds to the 4-dimensional membrane tension of a  D4-brane wrapped around $(T^2)_i$, while $\varepsilon_{\eta\zeta}=2$ for intersecting $\cN=2$ pairs and $\varepsilon_{\eta\zeta}=-1$ for those at mid-distance. Note that in the above expression the factor of $2$ associated to the exchange of $\mathcal{F}_{\alpha,\eta}$ and $\mathcal{F}_{\beta,\zeta}$ in \eqref{mem-Deltasum} has already been accounted for, so that we sum over each $\cN=2$ pair only once. 

Finally, we find that in general a D8-brane with a worldvolume flux is not invariant under the orientifold action, and therefore we need to consider two of them. This reflects the fact that in Calabi-Yau orientifolds oftentimes the quantum of Romans mass must be even. In fact, if we insist of working with a toroidal orbifold geometry the quantization conditions for $m$ and other background fluxes become even more restrictive, as we now turn to discuss.

\subsection{Flux quantization and blow-up modes}
\label{mem-ss:fluxquant}

In the absence of localized sources the Bianchi identities \eqref{mem-IIABI} are quite trivial, in the sense that $e^{-B} \wedge {\bf C}$ is globally well-defined. Then the quantization condition for NS and RR fluxes read
\begin{equation}
\frac{1}{\ell_s^{p}} \int_{\Pi_{p+1}} \bar{G}_{p+1}   \in  \IZ\, , \qquad  \frac{1}{\ell_s^{2}}\int_{\Pi_3}  H \in \IZ\, .
\label{mem-Pqu}
\end{equation} 
When we include localized sources like D-branes, we need to substitute these conditions by Page charge quantization \cite{Marolf:2000cb}. Nevertheless, we can still make use of the quanta defined in \eqref{mem-Pqu}, which are in fact the flux quanta used to describe the compactification in the smearing approximation. 

Additionally, the presence of O-planes can affect the quantization of those fluxes that are not sourced by any localized object. Indeed, as pointed out in \cite{Frey:2002hf}, in type IIB orientifold compactifications that only contain O3-planes with negative charge and tension (dubbed O3$^-$) the quanta of NS and RR background three-form fluxes must be even integers. This observation was applied to toroidal orbifold geometries in \cite{Blumenhagen:2003vr,Cascales:2003zp}, where it was found that three-form flux quanta in the  covering space should be multiples of $2M$ if no flux along collapsed three-cycles was to be involved, with $M \in \IZ$ depending on the particular orbifold. 

Clearly, these type IIB orientifold constraints must have a counterpart in our type IIA setup. Let us for instance take the type IIB setup of \cite{Frey:2002hf}, with 64 O3$^-$ on a $T^6$. An NS flux of the form $H = h\, dy^1 \wedge dy^2 \wedge dy^3$ is consistent if $h \in 2\IZ$. By performing three T-dualities along $\{x^1, x^2, x^3\}$ one recovers type IIA on $T^6$ with 8 O6$^-$ that extend along such coordinates. Assuming a factorized metric, this T-duality does not affect the $H$-flux that we have considered, and so one concludes that a type IIA  $H$-flux integrated over a three-cycle that intersects an even number of O6$^-$ must be quantized in terms of even integers. The same reasoning can be applied by T-dualising the type IIB RR three-form flux along any three-cycle of $T^6$. By doing so, we recover that $G_0$, $\bar{G}_2$, $\bar{G}_4$, $\bar{G}_6$ should also correspond to even integer quanta in the said type IIA background. In general, we expect a similar statement to apply in a smooth Calabi-Yau geometry $X_6$, whenever a $p$-cycle intersects an even number of O6$^-$. 

The orbifold geometries $X_6 = T^6/\Gamma$ that we consider in the next section do contain O6$^-$, but their homology classes are more involved than that of $T^6$. The difference mostly resides in the orbifold twisted sector, which corresponds to a set of cycles that are collapsed in the orbifold limit of a smooth Calabi--Yau. Since they are collapsed, the approximation of diluted fluxes that leads to the solution \eqref{cy-eq: solutionsu3} and \eqref{mem-solutionflux} is justified as long as the background fluxes do not have components on the twisted sector. Here is where the logic of \cite{Blumenhagen:2003vr,Cascales:2003zp} applies, and as a result the flux quanta computed in the covering space $T^6$ must be multiples of $2M$, for some $M \in \IZ$. In the following we will discuss how these quantization conditions look like in the case of the $\IZ_2 \times \IZ_2$ orientifolds mirror dual to the ones considered in  \cite{Blumenhagen:2003vr,Cascales:2003zp}. 

\subsubsection*{The $\IZ_2 \times \IZ_2$ orbifold}

Let us consider a $\IZ_2 \times \IZ_2$ orbifold over the factorizable six-torus $T^6= (T^2)_1 \times (T^2)_2 \times (T^2)_3$. The complex coordinate describing each two-torus is given by
\begin{equation}
    z^i=2\pi R_i(x^i+ i u_i y^i)\, ,
    \label{mem-coordZ2Z2}
\end{equation}
with $x^i$ and $y^i$ real coordinates of unit periodicity, $u_i \in \mathbb{R}$ describing the complex structure and $t^i = 4\pi^2 \ell_s^{-2} R_i^2 u_i$ the K\"ahler moduli of each $T^2$. The generators of the orbifold group act as
\be
\theta\, :\, (z^1,z^2,z^3) \mapsto (-z^1,-z^2,z^3)\, , \qquad \omega\, :\, (z^1,z^2,z^3) \mapsto (z^1,-z^2,-z^3)\, ,
\label{mem-eq: z2z2 action}
\ee
leaving fixed the coordinate values $x^i, y^i = \{0, 1/2\}$. Such coordinates correspond to the orbifold twisted sector, which can be interpreted as a set of collapsed cycles. The nature of these cycles depends on the choice of discrete torsion \cite{Font:1988mk,Vafa:1994rv}, which specifies how $\omega$ acts on the fixed point set of $\theta$, and so on. With one choice of discrete torsion the twisted sector corresponds to 48 collapsed two-cycles and 48 collapsed four-cycles, and the orbifold cohomology amounts to $(h^{1,1}, h^{2,1})_{\rm orb} = (51,3)$, while for the second choice it correspond to 96 collapsed three-cycles and  $(h^{1,1}, h^{2,1})_{\rm orb} = (3,51)$. These two choices are related to each other by mirror symmetry. 

We can now apply the orientifold quotient $\Omega_p (-1)^{F_L}{\cal R}$, with
\be
{\cal R} \, :\, (z^1,z^2,z^3) \mapsto (\bar{z}^1,\bar{z}^2,\bar{z}^3)\, .
\ee
This generates four different kinds of O6-planes:
\begin{subequations}	
	\label{mem-O6Z2Z2}
\begin{align}
[\Pi^{\rm O6}_{\cal R}] & = \left[(1, 0) \times (1, 0) \times (1, 0)\right]\, , \\
[\Pi^{\rm O6}_{{\cal R}\theta}] & = \left[(0, 1) \times (0, -1)\times (1, 0)\right]\, , \\
[\Pi^{\rm O6}_{{\cal R}\omega}] & = \left[(1, 0) \times (0, 1) \times (0,-1)\right]\, , \\
[\Pi^{\rm O6}_{{\cal R}\theta\omega}] & = \left[(0, -1) \times (1,0) \times (0,1)\right]\, ,
\end{align}
\end{subequations}   
each labelled by the orientifold group element that leaves them fixed. The multiplicity of each O6-plane class is $p_\a =8$, and they go over the different orbifold fixed points, so the index $\eta$ is better represented by the vector $\vec\eta = (\eta_1, \eta_2,\eta_3)$ with $\eta_i = 0, 1/2$. While the fixed loci are the same, the O6-plane nature is different for both choices of discrete torsion. For $(h^{1,1}, h^{2,1})_{\rm orb} = (51,3)$ all of them are O6$^-$, while for $(h^{1,1}, h^{2,1})_{\rm orb} = (3,51)$ one of the four classes in \eqref{mem-O6Z2Z2} has to correspond to O6$^+$-planes \cite{Angelantonj:1999ms}. Thus, in this second case, by placing D6-branes on top of the O6-planes one will never be able to construct a model absent of NS tadpoles, even in the presence of fluxes.\footnote{One could do so by introducing D6-branes at angles \cite{Marchesano:2004xz,Blumenhagen:2005tn}, but these more involved configurations will not be considered here.} For this reason in the following we will focus on the case where $(h^{1,1}, h^{2,1})_{\rm orb} = (51,3)$. 

Let us now see what is the appropriate flux quantization in the  $\IZ_2 \times \IZ_2$ orientifold with $(h^{1,1}, h^{2,1})_{\rm orb} = (51,3)$. In the absence of orientifold projection one can use the results of \cite{Blumenhagen:2003vr}, that show that the integral lattice of three-cycles is of the form $2[\Pi_\a]$, where $[\Pi_\a] = \left[(n_\a^1, m_\a^1) \times (n_\a^2, m_\a^2)\times (n_\a^3, m_\a^3)\right]$ is an integer three-cycle in the covering space $T^6$. If we now apply our  criterion for flux quantization in the presence of O6$^-$-planes we obtain that the $H$-flux must be quantized in units of 4 from the viewpoint of $T^6$. That is, $[\ell_s^{-2} H] = \sum_\a 4 h_\a  {\rm P.D.} [\Pi_\a]$, with $h_\a \in \IZ$. In particular, if as before we consider a flux of the form  $[\ell_s^{-2} H] =  h  {\rm P.D.} [\Pi_{\rm O6}]$, we find that $h \in \IZ/2$.  

This quantization in units of four is quite reminiscent of a similar condition for D6-branes. Indeed, for this choice of discrete torsion the minimal amount of covering-space three-cycles needed to build a consistent boundary state is two \cite{Douglas:1998xa,Gomis:2000ej}. Then, when introducing the orientifold projection and placing the D6-branes on top of an O6-plane one finds that its gauge group is $USp(2N)$, which means that each D6-brane in the orientifolded theory corresponds to four D6-branes in the covering space \cite{Cvetic:2001nr}. In other words, the charges $q_{\a,\eta}$ that appear in \eqref{mem-O6D6source} are quantized in units of 4. 

Let us finally turn to the quantization of internal RR fluxes. In this case one can directly use the results of \cite{Cascales:2003zp} on a type IIB mirror symmetric orientifold, because both the RR fluxes and the D-branes that generate them have a simple behaviour under T-duality. It was found in  \cite{Cascales:2003zp} that covering-space RR three-form fluxes must be quantized in units of 8 if one does not want to turn them on along twisted three-cycles. In our type IIA setup, this means that the quanta of Romans mass $m$ and that of four-form flux must also be quantized in units of 8 if one wants to maintain the orbifold geometry $T^6/(\IZ_2 \times \IZ_2)$. From the type IIA perspective the quantization in units of 8 of the Romans mass may seem surprising, but one can understand it in terms of the D-brane object that generates $G_0 =\ell_s^{-1}m$, namely a D8-brane wrapped on the internal space. Such a D8-brane will have induced D4-brane charge in the twisted sector, due to the curvature corrections and the non-trivial B-field at the orbifold point. The results of \cite{Blumenhagen:2003vr,Cascales:2003zp}  imply that, in order to construct a D8-brane boundary state with no induced twisted charges, one needs four of them in the covering space to form the regular representation of the orbifold group. The orientifold then doubles this number to eight D8-branes. In terms of fluxes, if one wants to have a non-vanishing Romans mass without inducing any four-form flux on the orientifold twisted sector one must impose that $m$ is a multiple of 8. 

Notice that these flux quantization conditions are quite constraining when imposing the tadpole equation \eqref{mem-tadpole2}, as they only allow for the solution
\be
m = 8\, , \qquad h = \oh\, , \qquad N=0\, ,
\label{mem-orbisolBI}
\ee
which contains no D6-branes at all. Thus, a domain-wall transition of the form \eqref{mem-jump} is not allowed starting from this orientifold vacuum, because the quantum of Romans  mass cannot be any larger, and this applies to both supersymmetric and non-supersymmetric vacua. 

Nevertheless, one can apply the same philosophy of \cite{DeWolfe:2005uu} and consider orientifold vacua in which the K\"ahler moduli of the twisted sector have been blown up due to the presence of a four-form flux along them, see Appendix \ref{mem-ap:Z2xZ2}. In this case we no longer need to impose that $m$ is a multiple of 8, but only impose the orientifold constraint that sets it as an even integer. Therefore we have a richer set of solutions to the tadpole constraint \eqref{mem-tadpole2}, like the family 
\be
m = 2k\, , \qquad h = \oh\, , \qquad N=4-k\, ,\qquad k = 1,2,3,4\, ,
\label{mem-blowsolBI}
\ee
or
\be
m = 2k\, , \qquad h = 1 \, , \qquad N=4-2k\, ,\qquad k = 1,2\, .
\label{mem-blowsolBI2}
\ee
Moreover, if as in \cite{DeWolfe:2005uu} we make a choice of four-form flux such that the blow-up K\"ahler moduli are much smaller than the toroidal ones, then the result \eqref{mem-finalDelta} should be a good approximation for the BIonic D8-brane excess-charge in $\cN=0$ vacua. Indeed, when twisted K\"ahler moduli are blown up both $J_{\rm CY}$ and ${\cal F}$ will be modified and so will be $\Delta_{\rm D8}^{\rm Bion}$, but one expects an effect that is of the order of the size of the blown-up two-cycles. Therefore, if we blow up the twisted two-cycles but their size remains much smaller than the toroidal K\"ahler moduli, we expect \eqref{mem-finalDelta} to give us a good approximation of the BIonic D8-brane excess charge. 

Given the value of $\Delta_{\rm D8}^{\rm Bion}$, one should finally compare it with $\Delta_{\rm D8}^{\rm curv} \equiv K_a^{(2)} T^a_{\rm D4}$, which one can again compute in the orbifold limit. For this computation the relevant intersection number is $c_2(X_6) . R_i$, where $R_i$ is the sliding divisor defined in Appendix \ref{mem-ap:Z2xZ2}. Using the results of \cite{Denef:2005mm} one finds that $c_2(X_6) . R_i = 24$ and therefore
\be
\Delta_{\rm D8}^{\rm curv} = \oh  \left(T^1_{\rm D4} + T^2_{\rm D4} + T^3_{\rm D4}\right)\, ,
\label{mem-D8curvz2xz2}
\ee
see Appendix \ref{mem-ap:Z2xZ2curv} for details. Recall that in order to satisfy the refined WGC for 4-dimensional membranes, it should be that $\Delta_{\rm D8}^{\rm curv} + \Delta_{\rm D8}^{\rm Bion} > 0$ for any BIon configuration.

\subsection{BIon configurations and the WGC}

One can check that \eqref{mem-finalDelta} reproduces the result  obtained in \cite{Marchesano:2021ycx} for the orientifold $T^6/(\Z_2 \times \IZ_2)$ and a transition \eqref{mem-jump}  in which $\hat{q}_{\a,\eta} = h$, $\forall \a, \eta$.  Indeed, there are six different pairs of different homology classes. For each combination there are 64 $\cN=2$ pairs, 32 of which intersect and 32 which do not, and each of them with a single $\cN=2$ subsector. The parallel one-cycles correspond to the basis $\{[(1,0)^i], [(0,1)^i]\}$, $i=1,2,3$ of $H_1(T^6, \IZ)$, and so each $T^i_{\rm D4}$ is selected twice in the sum \eqref{mem-finalDelta}. Applying all these data we obtain
\begin{equation}
     \Delta_{\rm D8}^{\rm Bion}(T^6/(\IZ_2\times\IZ_2)) = \frac{h^2}{24 \cdot 4} 32  (2-1) 2 \left(T^1_{\rm D4} + T^2_{\rm D4} + T^3_{\rm D4}\right)\, .
     \label{mem-DeltaZ2xZ2}
\end{equation}
 However, such a transition is never realized as a jump between AdS$_4$ vacua. Indeed, we have seen that $q_{\a,\eta}$ must be a multiple of 4, so if we have the same number of D6-branes on top of each orientifold it means that the negative charge and tension of each O6$^-$-plane is cancelled, and necessarily $mh=0$ in \eqref{mem-tadpole2}. In other words, we are in a 4d Minkowski vacuum. The second option for this equal distribution of D6-branes is to have none at all, which takes us back to an AdS$_4$ vacuum in which $mh=4$, like the one in \eqref{mem-orbisolBI}. A transition between these two $\Z_2 \times \IZ_2$ orientifold vacua is not mediated by 4-dimensional membrane arising from a BIonic D8-brane, but instead from a bound state of D8-brane, D4-brane and NS5-brane. The 4-dimensional vacuum of larger energy is $\cN=1$ Minkowski, and the membrane bound state is BPS and satisfies a no-force condition regardless of whether we jump to a $\cN=1$ or $\cN=0$ AdS$_4$ vacuum, as expected from the general results of \cite{Herraez:2020tih,Lanza:2020qmt}. 

Transitions mediated by a BIonic D8-brane for instance arise when increasing the value of $k$ in the family of vacua \eqref{mem-blowsolBI} and \eqref{mem-blowsolBI2} which, as explained, take us away from the orbifold limit. If we are in a non-supersymmetric vacuum of the sort discussed in section \ref{mem-s:nonsusy}, the BIon excess charge should be computed to a good approximation by \eqref{mem-finalDelta}, which will depend on how the D6-branes are arranged before and after the jump. In general we will have $8(4-2kh)$ D6-branes distributed in groups of 4 on the three-cycles $\Pi^{\rm O6}_{\a,\eta}$ within each homology class in \eqref{mem-O6Z2Z2}. 

For simplicity, we may consider the case where for each value of $\a$ all $8(4-2kh)$ D6-branes are on a single three-cycle, that is in a given choice of $\eta$. For instance, one may consider the case that such D6-branes are on top of the four O6-planes that go through the origin, which corresponds to selecting $\vec{\eta} = (0,0,0)$ for each value of $\a$, as represented in figure \ref{mem-fig: z2z2 intersecciones}. Then one can apply \eqref{mem-finalDelta} to compute the BIon excess charge of a single D8-brane, without taking into account its orientifold image. In this case we have that 
\be
\hat{q}_{\a, (0,0,0)} = 8h\, , \ \hat{q}_{\a, \vec{\eta} \neq (0,0,0)} = 0\quad  \forall \a\, , \qquad \varepsilon_{(0,0,0),(0,0,0)} = 2 \, , \qquad  \#(\Pi_\a \cap \Pi_\b) = 1 \, ,
\label{mem-eq: intersecting D6branes example}
\ee
and that each two-torus is selected twice by the pairwise intersection. Therefore
\begin{equation}
     \Delta_{\rm D8}^{\rm Bion} = \frac{8h^2}{3} \left(T^1_{\rm D4} + T^2_{\rm D4} + T^3_{\rm D4}\right)\, ,
     \label{mem-DeltaZ2xZ2int}
\end{equation}

\begin{figure}[h]
\centering
\begin{subfigure}[b]{.415\textwidth}
  \centering
    \hspace{-0.32cm} \includegraphics[width=\textwidth]{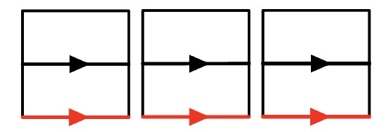}
    \caption*{$\Pi^{\rm O6}_{\cal R}  $}
 \end{subfigure}\hfill
\begin{subfigure}[b]{.415\textwidth}
  \centering
 \includegraphics[width=\textwidth]{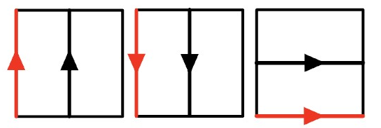}    \hspace{-0.36cm}
    \caption*{$\Pi^{\rm O6}_{{\cal R}\theta} $}
\end{subfigure}\\
\begin{subfigure}[b]{.4\textwidth}
    \centering
    \includegraphics[width=\textwidth]{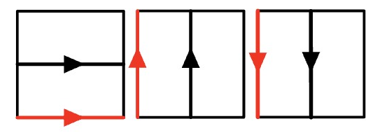}
    \caption*{$\Pi^{\rm O6}_{{\cal R}\omega}  $}
\end{subfigure}\hfill
\begin{subfigure}[b]{.4\textwidth}
    \centering
    \includegraphics[width=\textwidth]{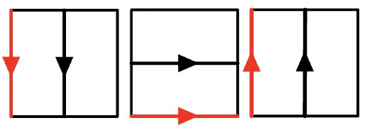}
    \caption*{$\Pi^{\rm O6}_{{\cal R}\theta\omega} $}
    \end{subfigure}
\caption{D6-brane configuration leading to \eqref{mem-eq: intersecting D6branes example}. In red are the O6-planes with D6-branes on top of them.}
\label{mem-fig: z2z2 intersecciones}
\end{figure}

\begin{figure}[h]
\centering
\begin{subfigure}[h]{.4\textwidth}
  \centering
    \includegraphics[width=\textwidth]{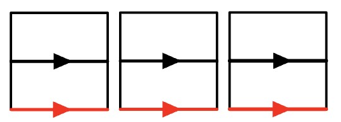}
    \caption*{$\Pi^{\rm O6}_{\cal R}$}
 \end{subfigure}\hfill
\begin{subfigure}[h]{.4\textwidth}
  \centering
 \includegraphics[width=\textwidth]{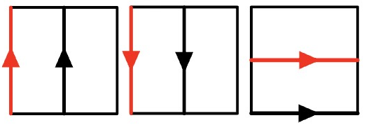}   
    \caption*{$\Pi^{\rm O6}_{{\cal R}\theta}$}
\end{subfigure}\\
\begin{subfigure}[h]{.4\textwidth}
    \centering
    \includegraphics[width=\textwidth]{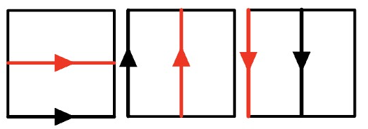}
    \caption*{$\Pi^{\rm O6}_{{\cal R}\omega}  $}
\end{subfigure}\hfill
\begin{subfigure}[h]{.4\textwidth}
    \centering
    \includegraphics[width=\textwidth]{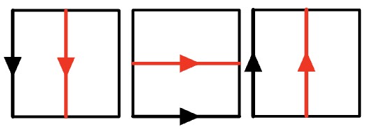}
    \caption*{$\Pi^{\rm O6}_{{\cal R}\theta\omega} $}
    \end{subfigure}
    \caption{D6-brane configuration that leads to \eqref{mem-eq: non intersecting D6branes example}. In red are the O6-planes with D6-branes on top of them.}
    \label{mem-fig: z2z2 no intersecciones}
\end{figure}
\noindent
signalling an instability of the vacuum. One can also consider a configuration in which the D6-branes do not intersect among each other, like for instance in figure \ref{mem-fig: z2z2 no intersecciones}. Then
\be
\hat{q}_{{\cal R}, (0,0,0)} =  \hat{q}_{{\cal R}\theta, (0,0,\oh)} =   \hat{q}_{{\cal R}\omega, (\oh,\oh,0)} =  \hat{q}_{{\cal R}\theta\omega, (\oh,\oh,\oh)} = 8h\, 
\label{mem-eq: non intersecting D6branes example}
\ee
with all other $\hat{q}_{\a,\vec\eta}$ vanishing. Because there is no pair of BIon sources that intersect, $\varepsilon_{\vec\eta, \vec\zeta}=-1$ and the contributions to \eqref{mem-finalDelta} are all negative, and more precisely we recover
\begin{equation}
     \Delta_{\rm D8}^{\rm Bion} = - \frac{4h^2}{3} \left(T^1_{\rm D4} + T^2_{\rm D4} + T^3_{\rm D4}\right)\, .
      \label{mem-DeltaZ2xZ2noint}
\end{equation}
Taking into account the curvature correction effect \eqref{mem-D8curvz2xz2}, one concludes that, for $h=1$, $\Delta_{\rm D8}^{\rm curv} + \Delta_{\rm D8}^{\rm Bion} < 0 $.  In this case there is an excess tension for the corresponding 4-dimensional membrane, which then does not satisfy the inequality of the Weak Gravity Conjecture. As far as D8/D6-systems are concerned, such a 4-dimensional non-supersymmetric vacuum seems non-perturbatively stable.

\subsubsection*{Caveats}

The result $\Delta_{\rm D8}^{\rm curv} + \Delta_{\rm D8}^{\rm Bion} < 0$ is surprising from the viewpoint of the WGC for 4-dimensional membranes. Indeed,  the set of $\cN=0$ vacua corresponding to \eqref{mem-eq: intersecting D6branes example} and \eqref{mem-eq: non intersecting D6branes example} have several independent decay channels. One consists of decreasing the four-form flux quanta via nucleation of D4-branes on two-cycles. A second one is to increase $k$ in \eqref{mem-blowsolBI} or \eqref{mem-blowsolBI2}, mediated by BIonic D8-branes. A third one would be to leave $m$ fixed and increase the $H$-flux quantum $h$ whenever the tadpole conditions permits, mediated by an NS5-brane wrapping a special Lagrangian three-cycle. Out of these three possibilities, only the first one is available when $k$ takes it maximal value in \eqref{mem-blowsolBI} or \eqref{mem-blowsolBI2}. In that case from the intuition developed in  \cite{Ooguri:2016pdq} one would expect that at least some D4-brane nucleation is favoured, leading to a non-perturbative instability. If that is the case, all vacua of this sort, including those with space-time filling D6-branes, are likely to be unstable via D4-brane nucleation, and so the AdS Instability Conjecture would be verified for this setup. As mentioned before, at this level of approximation $Q_{\rm D4} = T_{\rm D4}$, and it remains as an open problem to see whether or not  $Q_{\rm D4} > T_{\rm D4}$ after further corrections are taken into account.

Whenever we have several possible decay channels involving independent 4-dimensional membrane charges, we would expect that several 4-dimensional membranes satisfy the refined WGC $Q >T$, or more precisely a Convex Hull Condition \cite{Cheung:2014vva} adapted to 4-dimensional membranes. For the vacua of the sort \eqref{mem-eq: intersecting D6branes example} and \eqref{mem-eq: non intersecting D6branes example} this includes at least one 4-dimensional membrane with D8-brane charge. However for $h=1$ in \eqref{mem-blowsolBI2} we find that depending on the D6-brane positions we have either $Q_{\rm D8}^{\rm total} > T_{\rm D8}^{\rm total}$ or  $Q_{\rm D8}^{\rm total} < T_{\rm D8}^{\rm total}$. This contradicts  our WGC-based expectations, because in both cases the transition is very similar energetically. Indeed, the vacuum energy at tree level reads 
\be
V|_{\rm vac} = - \frac{16\pi}{75} e^K \cK^2 m^2 \simeq - \frac{243\pi}{50} \sqrt{\frac{3}{5}} \frac{\kappa^{3/2} h^4 |m|^{5/2}}{|\hat{e}_1\hat{e}_2 \hat{e}_3|^{3/2}}\, ,
\label{mem-vacuumen}
\ee
where $\hat{e}_i$ are defined as in \eqref{mem-Kahler} and correspond to the flux combinations that fix the untwisted K\"ahler moduli (that have triple intersection number  ${\cal K}_{123} = \kappa = 2$), and in the second equality we have neglected the contribution coming from blown-up two-cycles. A jump of the form $k \to k+1$ in \eqref{mem-blowsolBI} or \eqref{mem-blowsolBI2} not only translates into a change in $m$ but also in $\hat{e}_i$, which are negative numbers for $\cN=0$ vacua with $m>0$, see Appendix  \ref{mem-ap:Z2xZ2}. Given that $\Delta_{\rm D8}^{\rm Bion} = D_i T_{\rm D4}^i$, it seems reasonable to assume that the full flux jump is given by
\be
m \to m + 2\, , \qquad \hat{e}_i \to \hat{e}_i +  2 K_i^{(2)} + 2 D_i \simeq  \hat{e}_i  + 1 + 2 D_i\, ,
\label{mem-cavejump}
\ee
where for simplicity we have set $m^a=0$ in \eqref{mem-Kahler}, and again neglected fluxes along twisted cycles. Or results above imply that $D_i = 8h^2/3$ for \eqref{mem-eq: intersecting D6branes example} and $D_i = -4h^2/3$ for \eqref{mem-eq: non intersecting D6branes example}, so in all cases $|\hat{e}_i|$ decreases except when $\Delta_{\rm D8}^{\rm curv} + \Delta_{\rm D8}^{\rm Bion} < 0$. While this effect increases the vacuum energy in such a case, for large values of $|\hat{e}_i|$ it is a subleading effect with respect to the increase in $|m|$. So we always decrease the vacuum energy when we perform the jump $k \to k+1$, and so there is a priori no reason why in one vacuum D8-brane nucleation is favoured and not in the other. 

In light of these considerations, let us discuss some possible loopholes in our derivation of \eqref{mem-DeltaZ2xZ2noint}, or in its interpretation as a violation of the WGC for 4-dimensional membranes:

\begin{enumerate}

\item As mentioned above, the results \eqref{mem-DeltaZ2xZ2int} and \eqref{mem-DeltaZ2xZ2noint} are approximations, because they are computed in terms of an integral in the orbifold covering space $T^6$. However, in order to have a transition that increases $k$ in \eqref{mem-blowsolBI} it is necessary to consider Calabi--Yau geometries in which the twisted cycles have been blown up. This will modify the integral that leads to the general result \eqref{mem-finalDelta}, but one expects the correction to be suppressed as the quotient $t^{\rm tw}/t^{\rm untw}$, between the typical size of a blown-up two-cycle $t^{\rm tw}$ and that of an untwisted two-cycle $t^{\rm untw}$. As follows from the analysis of Appendix \ref{mem-ap:Z2xZ2}, this quotient can be arbitrarily small, and so it is consistent to neglect the corresponding correction to $\Delta_{\rm D8}^{\rm Bion}$. Similarly, as we blow up the twisted cycles, the excess charge \eqref{mem-QTtotalnosusy} will receive a different contribution from the term $K_a^{(2)} T_{\rm D4}^a$, as it follows from eq.\eqref{mem-c2J}. Again, this correction should be suppressed as $t^{\rm tw}/t^{\rm untw}$ compared to \eqref{mem-finalDelta}, and can be neglected in the same way that they were neglected in \eqref{mem-vacuumen}. In particular, it is highly unlikely that any of these corrections will flip the sign of  $\Delta_{\rm D8}^{\rm curv} + \Delta_{\rm D8}^{\rm Bion}$ computed in the orbifold limit.

\item The 10d supergravity solution \eqref{cy-eq: solutionsu3} and \eqref{mem-solutionflux} is a perturbative expansion that fails near the O6-planes, and this could affect significantly the D8-brane BIon solution. Corrections to the integrals in \eqref{mem-Deltasum} could come from such regions, which we treat via the regularization scheme used  in the next section. This however seems unlikely in the examples at hand, because the regions in which the BIon solution blows up and needs to be regularized are those in which the D6-brane charge cancels the O6-plane negative charge or even flips it, and the 10d background is at weak coupling and well-behaved.\footnote{Notice that in addition the D6-brane configuration \eqref{mem-eq: non intersecting D6branes example}, which is the problematic one for the WGC, displays no intersecting sources, and so it is more reliable with respect to the computation of $\Delta_{\rm D8}^{\rm Bion}$.} 

\item Assuming that the sign in \eqref{mem-DeltaZ2xZ2noint} is correct, there could be  another D8-brane that mediates a decay and has $\Delta_{\rm D8}^{\rm Bion} > 0$. For instance one could consider a  BIon profile different from \eqref{bio-BIonrel}, with lower tension. It would however be problematic if such a BIon solution existed as it would mean that, in a supersymmetric setup one would find a D8-brane with the same charges and lower tension than a BPS object.

\item The expression for the vacuum energy \eqref{mem-vacuumen} is a tree-level result, and it is subject to one-loop corrections. In particular there will be corrections coming from open string states stretching between different D6-branes. The masses of these objects are the main difference between the two configurations \eqref{mem-eq: intersecting D6branes example} and \eqref{mem-eq: non intersecting D6branes example}. In the first case they include light modes that will appear in the effective theory, while in the second case they are all massive modes above the compactification scale that need to be integrated out. The resulting threshold corrections will therefore be different and this could imply a change in the vacuum energy such that the decay is no longer energetically favoured in the second case. While this is an exciting possibility, it could also be that such threshold corrections to the vacuum energy are captured by the different values of $D_i$ in \eqref{mem-cavejump}. In that case for large values of $\hat{e}_i$ the effect on the vacuum energy would be significantly suppressed and nothing would change. 

\item The 4-dimensional membranes made up from D8-branes belong to the set of EFT membranes defined in \cite{Lanza:2020qmt} (see also \cite{Lanza:2019xxg}), and so their domain wall solutions can be described in 4-dimensional EFT terms. However such solutions are a priori not captured by the thin wall approximation. It could then be that because of the significant variation of the scalar fields, the criterion $Q_{\rm D8}^{\rm total} > T_{\rm D8}^{\rm total}$ is not the appropriate one to detect a non-perturbative instability. Nevertheless, if the expression $\Delta_{\rm D8}^{\rm Bion} = D_i T_{\rm D4}^i$ does translate into the flux jump \eqref{mem-cavejump} when crossing the 4-dimensional membrane, one could apply the reasoning of \cite[Section 5]{Marchesano:2021ycx} and conclude that when  $Q_{\rm D8}^{\rm total} < T_{\rm D8}^{\rm total}$ there is no membrane nucleation. 

\item Finally, it could be that a more complicated bound state 4-dimensional membrane charges mediates the decay. Adding harmonic worldvolume fluxes to the D8-brane would not help, at least in the diluted flux approximation, as these switch on a positive $K_a^F$ in \eqref{mem-QTtotalnosusy} and render $Q_{\rm D8}^{\rm total} - T_{\rm D8}^{\rm total}$ even more negative, so one should perhaps look at more exotic flux configurations away from the diluted flux limit. This point was addressed in \cite{Marchesano:2022rpr} and we will discuss it in more detail in section \ref{mem-sec: current status}.  A different option is to involve NS5-branes.  It follows from \eqref{mem-vacuumen} that in order to decrease the energy we need to increase the $H$-flux quantum $h$, which is not always an option. Indeed, if we increase $k=1 \to 2$ in \eqref{mem-blowsolBI2} there is no room to also increase $h$ without violating the tadpole condition. 
\end{enumerate}

\section{Examples}
\label{mem-s:examples}

In this section we present several examples of toroidal orbifolds, that illustrate how the different elements of formula describing the BIonic excess charge work together to provide the final result. We mainly focus on the $\IZ_2 \times \IZ_2$, $\mathbb{Z}_4$ and $\mathbb{Z}_3\times \mathbb{Z}_3$ orbifold groups, for which we perform the computations explicitly. We also consider, more schematically, the $\mathbb{Z}_6$ and $\mathbb{Z}_2\times \mathbb{Z}_4$ orbifolds.

In order to compute the integral $\int_{X} \mathcal{F}\wedge \mathcal{F}\wedge J$ we need to find an explicit expression for  the world-volume flux. As a first step we identify the different O6-planes and perform a Fourier expansion of the bump $\delta$-forms that describe them. The motivation for this being that the world-volume flux is determined by a set of 3-form currents $K_{\a,\eta}$ as in  \eqref{mem-eq: worldvolume flux}, and such 3-form currents are defined through the Laplace equation \eqref{mem-Ktorus}. Therefore, to find concrete expressions for $K_{\a,\eta}$ we need to build currents whose Laplacian returns  bump $\delta$-forms. Expanding in Fourier modes will prove to be an extremely useful tool to make this construction while controlling at the same time the connection with the smeared limit of our solution. Once these aspects are known, it is immediate to compute $\mathcal{F}_{\alpha,\eta}$ and evaluate the BIonic corrections using \eqref{mem-Deltasum}.

\subsection{\texorpdfstring{$T^6/\mathbb{Z}_2\times\mathbb{Z}_2$}{T6/Z2xZ2}}

We start by revisiting in greater detail the orbifold discussed in the previous section, that is a $\mathbb{Z}_2\times\mathbb{Z}_2$ orbifold with periodic coordinates given by \eqref{mem-coordZ2Z2} and orbifold action acting as \eqref{mem-eq: z2z2 action}. The metric and the Kähler form are
\begin{align}
    g =&\, 4\pi^2 \ell_s^2\, {\rm diag} \left(\hat{R}_1^2, \hat{R}_2^2, \hat{R}_3^2, u_1^2\hat{R}_1^2, u_2^2\hat{R}_2^2, u_3^2\hat{R}_3^2 \right)  \, ,
    \\
    J=& \, \ell_s^2( t^1 dx^1\wedge dy^1 +t^2 dx^2\wedge dy^2 +t^3 dx^3\wedge dy^3). \label{mem-eq: kahler form z2z2}
\end{align}
where we have defined the dimensionless radii $\hat{R}_i=R_i/\ell_s$ and the K\"ahler moduli $t^i=4\pi^2 \hat{R}_{i}^2 u_i$.

It is worth noting that the choice of complex structure  \eqref{mem-coordZ2Z2} is not the only one compatible with the $\mathbb{Z}_2\times \mathbb{Z}_2$ symmetry. For each of the two-tori we are free to choose the complex structure as $\tau^i=iu_i$ or $\tau^i=1/2+iu_i$. From this point onward we will focus on the case where all the tori follow the former choice, as in \eqref{mem-coordZ2Z2}. Results are similar for the other possible choices. 

The orientifold planes are given by the fixed points of the orientifold involution $\sigma(z)=\bar{z}$, up to orbifold action identifications. Consequently, we have the four different kinds of orientifold planes, summarized in table \ref{mem-table: z2z2 orientifolds} and already introduced in \eqref{mem-O6Z2Z2}. They are schematically represented as the arrow segments (both red and black) in figures \ref{mem-fig: z2z2 intersecciones} and \ref{mem-fig: z2z2 no intersecciones}.

\renewcommand{\arraystretch}{0.9}
\begin{table}[H]
$$
\begin{array}{|l|l|c|}
\hline \Pi_{\alpha} & \text { Fixed point equation } & \text { O6-plane position }  \\
\hline \Pi_{0} & \sigma\left(z^{a}\right)=z^{a} & y^{1} \in\left\{0,\frac{1}{2}\right\} \quad y^{2} \in\left\{0,\frac{1}{2}\right\} \quad y^{3} \in\left\{0,\frac{1}{2}\right\} \\
\Pi_{1} & \sigma\left(z^{a}\right)=\theta\left(z^{a}\right) & x^{1} \in\left\{0,\frac{1}{2}\right\} \quad x^{2} \in\left\{0,\frac{1}{2}\right\} \quad y^{3} \in\left\{0,\frac{1}{2}\right\} \\
\Pi_{\mathcal{R}\omega} & \sigma\left(z^{a}\right)=\omega\left(z^{a}\right) & y^{1} \in\left\{0,\frac{1}{2}\right\} \quad x^{2} \in\left\{0,\frac{1}{2}\right\} \quad x^{3} \in\left\{0,\frac{1}{2}\right\} \\
\Pi_{\mathcal{R}\theta\omega} & \sigma\left(z^{a}\right)= \theta\omega\left(z^{a}\right) & x^{1} \in\left\{0,\frac{1}{2}\right\} \quad y^{2} \in\left\{0,\frac{1}{2}\right\} \quad x^{3} \in\left\{0,\frac{1}{2}\right\} \\
\hline
\end{array}
$$
\caption{O6-planes in $T^6/\mathbb{Z}_2\times \mathbb{Z}_2$.}
\label{mem-table: z2z2 orientifolds}
\end{table}
The above content of O6-planes can be expressed in terms of invariant bulk three-cycles. This is quite simple for the current case, but it will become more nuanced in the following examples. Let $\pi_{2i-1}$ and $\pi_{2i}$ constitute a basis of fundamental one-cycles on the torus $(T^2)_i$ $(i=1,2,3)$, i.e. one-cycles winded once around the directions used for the periodic identifications that parametrized the torus in \eqref{mem-coordZ2Z2}. Then we define the following set of toroidal three-cycles:
\be
\pi_{IJK}=\pi_I \otimes \pi_J \otimes \pi_K\, .
\ee
with $I=1,2$, $J=3,4$ and $K=5,6$. From \cite{Blumenhagen:2003vr} we know that the smallest integer toroidal cycles are
\begin{equation}
\begin{array}{ll}
\rho_{1} \equiv 2\pi_{135}, & \rho_{2} \equiv 2\pi_{136}\,,\\
\rho_{3} \equiv 2\pi_{145}, & \rho_{4} \equiv 2\pi_{146}\,,\\
\rho_{5} \equiv 2\pi_{235}, & \rho_{6} \equiv 2\pi_{236}\,,\\
\rho_{7} \equiv 2\pi_{245}, & \rho_{8} \equiv 2\pi_{246}\,.\\
\end{array} \label{mem-eq: bulk cycles z2}
\end{equation}
Then, the orientifold plane content can be expressed in terms of these invariant cycles as %
\begin{equation}
    \Pi_{\rm O6}=4\rho_1-4\rho_7-4\rho_4-4\rho_6\,.
\end{equation}
The next step will be to construct the $\delta$-like bump functions living in the factorized orbifold structure. Taking the O6-plane positions from  Table \ref{mem-table: z2z2 orientifolds} a delta bump function can be expressed as a product of  conventional Fourier expansions for each $T^2_{i}$ with support on the fixed loci $\Pi_{\alpha}$.

\begin{subequations}
\begin{align}
    \delta(\Pi_\mathcal{R})=&\ell_s^3\sum_{\vec{\eta}}\left[\sum_{n_1\in\mathbb{Z}} e^{2\pi i n_1 (y^1-\eta_1)} dy^1 \right]\wedge \left[\sum_{n_2\in\mathbb{Z}} e^{2\pi i n_2 (y^2-\eta_2)} dy^2 \right]\wedge \left[\sum_{n_3\in\mathbb{Z}} e^{2\pi i n_3 (y^3-\eta_3)} dy^3 \right]\, ,\\
    \delta(\Pi_{1})=&\ell_s^3\sum_{\vec{\eta}}\left[\sum_{n_1\in\mathbb{Z}} e^{2\pi i n_1 (x^1-\eta_1)} dx^1 \right]\wedge \left[\sum_{n_2\in\mathbb{Z}} e^{2\pi i n_2 (x^2-\eta_2)} dx^2 \right]\wedge \left[\sum_{n_3\in\mathbb{Z}} e^{2\pi i n_3 (y^3-\eta_3)} dy^3 \right]\, ,\\
    \delta(\Pi_{\mathcal{R}\omega})=&\ell_s^3\sum_{\vec{\eta}}\left[\sum_{n_1\in\mathbb{Z}} e^{2\pi i n_1 (y^1-\eta_1)} dy^1 \right]\wedge \left[\sum_{n_2\in\mathbb{Z}} e^{2\pi i n_2 (x^2-\eta_2)} dx^2 \right]\wedge \left[\sum_{n_3\in\mathbb{Z}} e^{2\pi i n_3 (x^3-\eta_3)} dx^3 \right]\, ,\\
      \delta(\Pi_{\mathcal{R}\theta\omega})=&\ell_s^3\sum_{\vec{\eta}}\left[\sum_{n_1\in\mathbb{Z}} e^{2\pi i n_1 (x^1-\eta_1)} dx^1 \right]\wedge \left[\sum_{n_2\in\mathbb{Z}} e^{2\pi i n_2 (y^2-\eta_2)} dy^2 \right]\wedge \left[\sum_{n_3\in\mathbb{Z}} e^{2\pi i n_3 (x^3-\eta_3)} dx^3 \right]\, ,
\end{align}
\label{mem-eq: deltas z2z2}
\end{subequations}
where $\vec{\eta}=(\eta_1,\eta_2,\eta_3)$ has entries that are 0 or $\frac{1}{2}$. With all this information, we can then build the three-forms $K_{\a,\eta}$ satisfying \eqref{mem-Ktorus}:
\begin{subequations}
\begin{align}
    K_{\mathcal{R},\eta}=-\ell_s^3\sum_{0\neq \vec{n}\in \mathbb{Z}^3}\frac{e^{2\pi i \vec{n}[(y^1,y^2,y^3)-\vec{\eta}]}}{|\vec{n}|^2} dy^1\wedge dy^2\wedge dy^3\, ,\\
    K_{\mathcal{R}\theta,\eta}=\ell_s^3\sum_{0\neq \vec{n}\in \mathbb{Z}^3}\frac{e^{2\pi i \vec{n}[(x^1,x^2,y^3)-\vec{\eta}]}}{|\vec{n}|^2} dx^1\wedge dx^2\wedge dy^3\, ,\\
    K_{\mathcal{R}\omega,\eta}=\ell_s^3\sum_{0\neq \vec{n}\in \mathbb{Z}^3}\frac{e^{2\pi i \vec{n}[(y^1,x^2,x^3)-\vec{\eta}]}}{|\vec{n}|^2} dy^1\wedge dx^2\wedge dx^3\, ,\\
    K_{\mathcal{R}\theta\omega,\eta}=\ell_s^3\sum_{0\neq \vec{n}\in \mathbb{Z}^3}\frac{e^{2\pi i \vec{n}[(x^1,y^2,x^3)-\vec{\eta}]}}{|\vec{n}|^2} dx^1\wedge dy^2\wedge dx^3\, ,
\end{align}
\end{subequations}
with the indices $\alpha,\eta$  associated to the orientifold planes $\Pi_{\alpha,\eta}$ and $|\vec{n}|^2 = n_1^2/\hat{R}_1^2+n_2^2/\hat{R}_2^2+n_3^2/\hat{R}_3^2$. The relative signs between the different $K_{\alpha}$ are chosen so that $\Im\Omega$ calibrates all the orientifold planes. 

At this stage, we can present the relation in cohomology between the flux $H$ and the orientifold planes derived from \eqref{mem-Krhsum}, so that by using the equations of motion \eqref{mem-intflux} we can fix the complex structure moduli $u_i$. This implies
\begin{equation}
     [\ell_s^{-2}H]=8h \left([\beta^0]-[\beta^1]-[\beta^2]-[\beta^3]\right)\, ,
\end{equation}
where the $\beta^i$ are elements of the following basis of bulk 3-forms:
\bea\nonumber
\a_0 = dx^1 \wedge dx^2 \wedge dx^3\, , & \quad & \b^0 = dy^1 \wedge dy^2 \wedge dy^3 \, ,\\ \nonumber
\a_1 = dx^1 \wedge dy^2 \wedge dy^3\, , & \quad & \b^1 = dy^1 \wedge dx^2 \wedge dx^3 \, ,\\ \nonumber
\a_2 = dy^1 \wedge dx^2 \wedge dy^3\, , & \quad & \b^2 = dx^1 \wedge dy^2 \wedge dx^3 \, ,\\ \nonumber
\a_3 = dy^1 \wedge dy^2 \wedge dx^3\, , & \quad & \b^3 = dx^1 \wedge dx^2 \wedge dy^3 \, .
\eea
Defining $\rho=8\pi^3 \hat{R}_1\hat{R}_2\hat{R}_3$ and considering our choice of complex structure, the holomorphic (3,0)-form $\Omega$ is given by
\bea
\re \Om_{\rm CY} & = &  \ell_s^3 \rho \left(u_1u_2u_3 \b^0 - u_1 \b^1 - u_2 \b^2 - u_3 \b^3 \right) \, ,\\
\im \Om_{\rm CY} & = &  \ell_s^3 \rho \left( \a_0 - u_2u_3 \a_1 - u_1u_3 \a_2 - u_1u_2 \a_3 \right) \, .
\eea
Then, a solution to the first equation in \eqref{mem-intflux} can be accomplished if all the complex structure moduli are fixed to $u_i=1$, and   $\mu=\ell_s^{-1}4h/\rho$.

In light of all this, keeping the complex structure unfixed, we can construct $\mathcal{F}_{\alpha,\eta}=\ell_s d^{\dagger}K_{\alpha,\eta}$. We arrive at:
\begin{subequations}
\begin{align}
    \mathcal{F}_{\mathcal{R},\eta}=\frac{i\ell_s^2}{2\pi}\sum_{0\neq \vec{n}\in \mathbb{Z}^3}\frac{e^{2\pi i \vec{n}[(y^1,y^2,y^3)-\vec{\eta}]}}{|\vec{n}|^2} \left(\frac{n_1}{\hat{R}_1^2}dy^2\wedge dy^3-\frac{n_2}{\hat{R}_2^2}dy^1\wedge dy^3+\frac{n_ 3}{\hat{R}_3^2}dy^1\wedge dy^2\right)\, ,\label{mem-eq: F0 z2z2} \\
    \mathcal{F}_{\mathcal{R}\theta,\eta}=-\frac{i\ell_s^2}{2\pi}\sum_{0\neq \vec{n}\in \mathbb{Z}^3}\frac{e^{2\pi i \vec{n}[(x^1,x^2,y^3)-\vec{\eta}]}}{|\vec{n}|^2} \left(\frac{n_1}{\hat{R}_1^2}dx^2\wedge dy^3-\frac{n_2}{\hat{R}_2^2}dx^1\wedge dy^3+\frac{n_ 3}{\hat{R}_3^2}dx^1\wedge dx^2\right)\, ,\label{mem-eq: F1 z2z2}\\
     \mathcal{F}_{\mathcal{R}\omega,\eta}=-\frac{i\ell_s^2}{2\pi}\sum_{0\neq \vec{n}\in \mathbb{Z}^3}\frac{e^{2\pi i \vec{n}[(y^1,x^2,x^3)-\vec{\eta}]}}{|\vec{n}|^2} \left(\frac{n_1}{\hat{R}_1^2}dx^2\wedge dx^3-\frac{n_2}{\hat{R}_2^2}dy^1\wedge dx^3+\frac{n_ 3}{\hat{R}_3^2}dy^1\wedge dx^2\right)\, ,\\
      \mathcal{F}_{\mathcal{R}\theta\omega,\eta}=-\frac{i\ell_s^2}{2\pi}\sum_{0\neq \vec{n}\in \mathbb{Z}^3}\frac{e^{2\pi i \vec{n}[(x^1,y^2,x^3)-\vec{\eta}]}}{|\vec{n}|^2} \left(\frac{n_1}{\hat{R}_1^2}dy^2\wedge dx^3-\frac{n_2}{\hat{R}_2^2}dx^1\wedge dx^3+\frac{n_ 3}{\hat{R}_3^2}dx^1\wedge dy^2\right)\, .
\end{align}
\label{mem-eq: F z2z2}
\end{subequations}

Finally, we would like to compute $\int \mathcal{F}_{\alpha,\eta}\wedge\mathcal{F}_{\beta,\xi}\wedge J_{CY}$. To perform this integral we regularize it by interchanging the order between summation and integration. The physical interpretation of this procedure corresponds to smearing the O6-plane over a region of radius $\sim \ell_s$, which is the region of $X_6$ where the supergravity approximation cannot be trusted. 
In practice this corresponds to a truncation of the summation over the Fourier modes labelled by $\vec{n}$. In a finite sum we are able to swap summation and integration freely. We then take the limit when the cut-off of the sum diverges, returning to our original system with a localized source. 

At this point we can check some of the statements made in the last section. First of all, we verify that $\Delta_{\alpha,\eta;\alpha,\zeta}=0$. We focus on the simplest case and consider the contribution from two components of  $\Pi^{\rm O6}_\mathcal{R}$. In particular we choose $\alpha=0$ and $\eta=\zeta=(0,0,0)$ and compute
\begin{equation}
    \Delta_{\mathcal{R},\vec{0};\mathcal{R},\vec{0}} = - e^{K/2} \frac{1}{\ell_s^{6}} \int_{X_6}  J_{\rm CY} \wedge \cF_{\mathcal{R},\vec{0}} \wedge \cF_{\mathcal{R},\vec{0}} =0\, .
\end{equation}
Using \eqref{mem-eq: kahler form z2z2} and \eqref{mem-eq: F0 z2z2}  we immediately see that the contribution vanishes, since there is always a wedge product of repeated one-forms. Note that this is independent on the value of $\vec{\eta}$ in \eqref{mem-eq: F0 z2z2}. Therefore we conclude that $\Delta_{\mathcal{R},\eta;\mathcal{R},\zeta}=0$ for any $\eta$ and $\zeta$. Similar cancellations occur for all contributions of this nature involving other cohomology classes. 

We now focus on the remaining possible contributions, which belong to the $\mathcal{N}=2$ sectors of the compactification and are characterized by D6-branes that have similar wrapping numbers in one of the two-tori and different in the other two. For concreteness we consider two examples: one in which the D6-branes intersect over a one-cycle, and one in which there is no intersection. Starting with the former we build the configuration from \eqref{mem-eq: intersecting D6branes example} and evaluate the contribution from the pair of D6-branes associated to  $\Pi^{\rm O6}_\mathcal{R}$ and $\Pi^{\rm O6}_{\mathcal{R}\theta}$. As depicted in figure \ref{mem-fig: z2z2 intersecciones}, the branes intersect over $(T^2)_3$. The associated BIon contribution is 
\begin{align}
     \Delta_{\mathcal{R},\vec{0};\mathcal{R}\theta,\vec{0}}=& - e^{K/2} \frac{1}{\ell_s^{6}} \int_{X_6}  J_{\rm CY} \wedge \cF_{\mathcal{R},\vec{0}} \wedge \cF_{\mathcal{R}\theta,\vec{0}}\nonumber\\
     =&- e^{K/2}\frac{t^3}{4\pi^2\ell_s^6}\int_{X_6} \sum_{0\neq \vec{n},\vec{m}\in\mathbb{Z}^3} \frac{e^{2\pi i\vec{n}(y^1,y^2,y^3)}e^{2\pi i\vec{m}(x^1,x^2,y^3)}}{|\vec{n}|^2|\vec{m}|^2}\frac{n_3m_3}{\hat{R}_3^2}\, \Phi_6\nonumber\\
     =&-e^{K/2}\frac{t^3}{4\pi^2 N_\Gamma}\sum_{0\neq \vec{n},\vec{m}\in\mathbb{Z}^3} \delta_{n_1}\delta_{n_2}\delta_{m_1}\delta_{m_2}\delta_{n_3+m_3}\frac{1}{|\vec{n}|^2|\vec{m}|^2}\frac{n_3m_3}{\hat{R}_3^4}\nonumber\\
     =&\, e^{K/2}\frac{t^3}{4\pi^2 N_\Gamma}\sum_{n_3\neq 0}\frac{n_3^2}{n_3^4}
     = e^{K/2}\frac{t^3}{4\pi^2 N_\Gamma}2\frac{\pi^2}{6} = \frac{T^3_{\rm D4}}{12N_\Gamma}\, ,
     \label{mem-eq: delta z2 intersection}
\end{align}
where we have defined $\Phi_6=\ell_s^6 dx^1\wedge dx^2\wedge dx^3\wedge dy^1\wedge dy^2\wedge dy^3$. To go from the second to the third line we have used the regularization procedure stated above. It is easy to repeat the same computation for any $ \Delta_{\mathcal{R},\vec\eta;\mathcal{R}\theta,\vec\zeta}$ such that $\eta_3=\zeta_3$ (in order to preserve the intersection along $(T^2)_3$). The new exponential factors arising from \eqref{mem-eq: F z2z2} vanish once the Kronecker deltas are considered. Similarly, the same result is obtained for intersections involving other cohomology classes. Hence, we verify that  an $\mathcal{N}=2$ sector in which  D6-branes intersect over a one-cycle in $(T^2)_i$ contribute as $\frac{T^i}{12N_\Gamma}$ to \eqref{mem-Deltasum}.

Finally we test the case in which the D6-branes do not overlap but run parallel over the one two-torus. To do so, we build the configuration described in \eqref{mem-eq: non intersecting D6branes example} (see figure \ref{mem-fig: z2z2 no intersecciones}) and evaluate the contribution from the D6-brane associated to the $\Pi^{\rm O6}_\mathcal{R}$ and $\Pi^{\rm O6}_{\mathcal{R}\theta}$ as before. We obtain
\begin{align}
     \Delta_{\mathcal{R},\vec{0};\mathcal{R}\theta,(0,0,1/2)}=& - e^{K/2} \frac{1}{\ell_s^{6}} \int_{X_6}  J_{\rm CY} \wedge \cF_{\mathcal{R},\vec{0}} \wedge \cF_{\mathcal{R}\theta,(0,0,1/2)}\nonumber\\
     =&- e^{K/2}\frac{t^3}{4\pi^2 \ell_s^6}\int_{X_6} \sum_{0\neq \vec{n},\vec{m}\in\mathbb{Z}^3} \frac{e^{2\pi i\vec{n}(y^1,y^2,y^3)}e^{2\pi i\vec{m}(x^1,x^2,y^3)}e^{i\pi m_3}}{|\vec{n}|^2|\vec{m}|^2}\frac{n_3m_3}{\hat{R}_3^2} \Phi_6\nonumber\\
     =&-e^{K/2}\frac{t^3}{4\pi^2 N_\Gamma}\sum_{0\neq \vec{n},\vec{m}\in\mathbb{Z}^3} \delta_{n_1}\delta_{n_2}\delta_{m_1}\delta_{m_2}\delta_{n_3+m_3}\frac{(-1)^{m_3}}{|\vec{n}|^2|\vec{m}|^2}\frac{n_3m_3}{\hat{R}_3^4}\nonumber\\
     =&\, e^{K/2}\frac{t^3}{4\pi^2 N_\Gamma}\sum_{n_3\neq 0}\frac{(-1)^{n_3}}{n_3^2}
     = e^{K/2}\frac{t^3}{4\pi^2 N_\Gamma}2\frac{-\pi^2}{12} 
     = -\frac{T^3_{\rm D4}}{24N_\Gamma}\, ,
\end{align}
and so we recover \eqref{mem-interpar}.

It is worth noting that even though \eqref{mem-intercont} and \eqref{mem-interpar} are correct for all the examples we consider, they do not describe the most general scenario we can think of, see footnote \ref{mem-OSL}. For a generic $\mathcal{N}=2$ configuration in which the D6-branes run parallel along the $(T^2)_a$ over one-cycles of length $L$ and separated by a distance $\frac{\ell_s^2t^a}{L} \eta$, one can generalize the computations above to see that the contribution to \eqref{mem-Deltasum} is given in terms of the dilogarithmic function as
\begin{equation}
     \frac{T^a_{\rm D4}}{2\pi^2N_\Gamma}\Re\left[\textrm{Li}_2(e^{2\pi i \eta})\right] = \frac{T^a_{\rm D4}}{N_\Gamma} \oh \left(\frac{1}{6} - \eta(1-\eta) \right)\, .
\end{equation}

\subsection{\texorpdfstring{$T^6/\mathbb{Z}_4$}{T6/Z4}}

Let us now consider the $\mathbb{Z}_4$ orbifold over a factorizable six-torus $T^6=(T^2)_1\times (T^2)_2 \times (T^2)_3$, as discussed in \cite{blumenhagen2000supersymmetric,blumenhagen2003supersymmetric}, see also \cite{Ihl:2006pp}. The two-dimensional lattice that defines each 2-torus is generated by the basis of the complex plane $e_{i1}=2\pi R_i$ and $e_{i2}=2\pi R_i\tau_i$, where  $R_i$ are the radii of $(T^2)_i$ and $\tau_i=a_i+iu_i$ is its complex structure. The complex coordinate for each 2-torus is
\begin{equation}
    z^i=2\pi R_i(x^i+ \tau_i y^i)\,, \qquad x^i,y^i\in \mathbb{R}\,.
    \label{mem-coordinates}
\end{equation}
The action of the $\mathbb{Z}_4$ group over $T^6$ is generated by an element $\theta$ that acts as follows
\begin{equation}
    \theta(z^i)=e^{2\pi i v_i} z^i\,,
\end{equation}
with $v_i=(1/4,1/4,1/2)$. The action of this group severely constrains the complex structure. In fact, the complex structure of the first two $T^2$'s is fixed. For the third torus, in which the $\mathbb{Z}_4$ action has an orbit of order $2$, the constraints are less severe. There are two options available, commonly denoted by  AAA and AAB \cite{blumenhagen2003supersymmetric,Forste:2001gb}, and both of them have $u_3$ free. The AAA case is characterized by the choice $a_3=0$, whereas the $AAB$ has $a_3=1/2$. Therefore, in the $\mathbb{Z}_4$ orbifold  there is always one unconstrained complex structure modulus.

For concreteness let us consider the choice AAA. All the steps of the analysis can be replicated in the AAB scenario to arrive to the same results. In the present case, we have $\tau_1=\tau_2=i$ and $\tau_3=i u_3$. The basis of the lattice that generates the torus is orthogonal and gives the following identifications 
\begin{subequations}
\begin{align}
    z^1 &\sim z^1+2\pi R_1 \sim z^1+2\pi i R_1\, ,\\
    z^2 &\sim z^2+2\pi R_2 \sim z^2+2\pi i R_2\, ,\\
    z^3 &\sim z^3+2\pi R_3 \sim z^3+2\pi i u_3 R_3\, .
\end{align}
\label{mem-eq: periodic identification Z4}
\end{subequations}
Up to the constraints on the complex structure, the covering space metric and the Kähler form are the same as in the $\mathbb{Z}_2\times \mathbb{Z}_2$ case.
\begin{align}
    g =&\, 4\pi^2 \ell_s^2\, {\rm diag} \left(\hat{R}_1^2, \hat{R}_2^2, \hat{R}_3^2, \hat{R}_1^2, \hat{R}_2^2, u_3^2\hat{R}_3^2 \right)\,   ,  \\
    J=&\, \ell_s^2( t^1 dx^1\wedge dy^1 +t^2 dx^2\wedge dy^2 +t^3 dx^3\wedge dy^3)\, ,
    \label{mem-eq: J z4}
\end{align}
where again we  defined the dimensionless radii $\hat{R}_i=R_i/\ell_s$ and the K\"ahler moduli $t^i=4\pi^2 \hat{R}_{i}^2 u_i$.

The orientifold planes are given by the fixed points of the orientifold involution $\sigma(z)=\bar{z}$, up to orbifold action identifications. Consequently,  we find the orientifold planes summarized in table \ref{mem-table: z4 orientifolds} and represented in figure \ref{mem-figure: Z4-planes}.
\renewcommand{\arraystretch}{0.9}
\begin{table}[H]
$$
\begin{array}{|l|l|c|}
\hline \Pi_{\alpha} & \text { Fixed point equation } & \text { O6-plane position } \\
\hline \Pi_{0} & \sigma\left(z^{a}\right)=z^{a} & y^{1} \in\left\{0,\frac{1}{2}\right\} \quad y^{2} \in\left\{0,\frac{1}{2}\right\} \quad y^{3} \in\left\{0,\frac{1}{2}\right\} \\
\Pi_{1} & \sigma\left(z^{a}\right)=\theta\left(z^{a}\right) & y^{1}- x^{1} =0 \quad y^{2}- x^{2} =0 \quad x^{3} \in\{0,\oh\} \\
\Pi_{2} & \sigma\left(z^{a}\right)=\theta^{2}\left(z^{a}\right) & x^1\in\{0,\oh\} \quad x^{2}\in\{0,\oh\} \quad y^{3}\in\{0,\oh\} \\
\Pi_{3} & \sigma\left(z^{a}\right)= \theta^{3}\left(z^{a}\right) & y^{1} +x^1=1 \quad y^{2}+ x^{2} =1 \quad x^{3}\in\{0,\oh\} \\
\hline
\end{array}
$$
\caption{O6-planes in $T^6/\mathbb{Z}_4$.}
\label{mem-table: z4 orientifolds}
\end{table}
\begin{figure}[h!]
  \includegraphics[width=\textwidth]{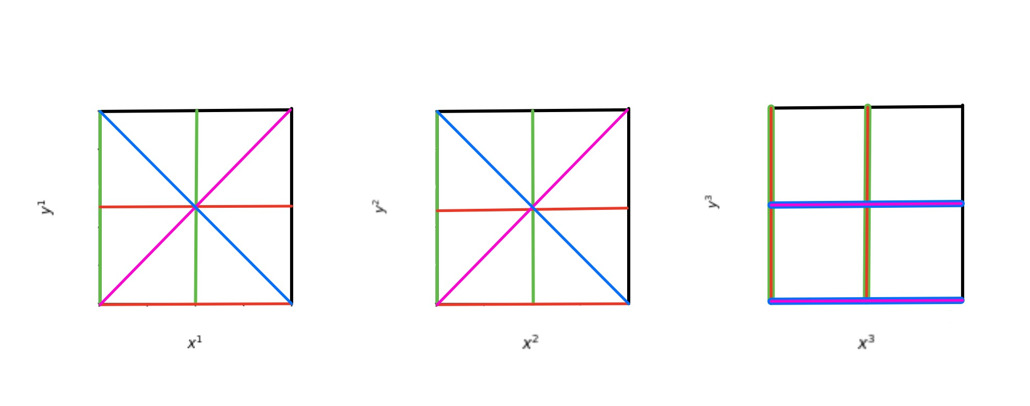}
    \caption{Orientifold planes projected over $T^2\times T^2\times T^2$ in the $\mathbb{Z}_4$ orbifold. Planes $\Pi_\mathcal{R}$, $\Pi_{1}$, $\Pi_{2}$, $\Pi_{3}$ are represented by the colours red, pink, green and blue respectively.}
    \label{mem-figure: Z4-planes}
\end{figure}
The above content of O6-planes can be expressed in terms of invariant bulk three-cycles following the same reasoning as in the $\IZ_2\times\IZ_2$ case. Let $\pi_{2i-1}$ and $\pi_{2i}$ constitute a basis of fundamental one-cycles on the torus $T^2_i$ $(i=1,2,3)$, i.e. cycles winded once along the periodic directions given by the identifications that defined our tori in \eqref{mem-eq: periodic identification Z4}. We  used them to build the following three-cycles
\be
\pi_{IJK}=\pi_I \otimes \pi_J \otimes \pi_K\, ,
\ee
with $I=1,2$, $J=3,4$ and $K=5,6$. For the $\IZ_4$ orientifold the minimal invariant bulk three-cycles are given by \cite{blumenhagen2003supersymmetric}
\begin{equation}
\begin{array}{ll}
\rho_{1} \equiv 2\left(\pi_{135}-\pi_{245}\right), & \bar{\rho}_{1} \equiv 2\left(\pi_{136}-\pi_{246}\right)\, , \\
\rho_{2} \equiv 2\left(\pi_{145}+\pi_{235}\right), & \bar{\rho}_{2} \equiv 2\left(\pi_{146}+\pi_{236}\right).
\end{array} \label{mem-eq: bulk cycles z4}
\end{equation}
The factor of 2 in \eqref{mem-eq: bulk cycles z4} is due to the fact that $\theta^2$ acts trivially over $\pi_{ijk}$. Hence, the O6-planes content can be expressed as
\be
\Pi_{\rm O6}=4\rho_1 - 2\bar{\rho}_2\, .
\ee

As we have seen, due to the factorized structure of the orbifold, the orientifold three-cycles are also factorized as products of one-cycles in the covering space, each one defined in each of the two-tori. A $\delta$-function supported on these one-dimensional objects can be expressed using the conventional Fourier expansion for the $\delta$-function distribution:
\begin{equation}
    \delta(w)=\frac{1}{S}\sum_{n\in\mathbb{Z}}e^{2\pi i n w/S}\, ,
    \label{mem-eq: fourier delta def}
\end{equation}
where $w$ denotes the direction transverse to the cycle normalized to unit norm and $S$ is the periodicity of the configuration along such a transverse direction. Therefore, in order to build the bump $\d$-functions for factorizable three-cycles, we need to find the transverse periodicity $S$ of the respective one-cycles, which we define as  the distance that separates two consecutive intersection points between the loci of the cycle (given by the linear equations of table \ref{mem-table: z4 orientifolds}) projected over the two-torus we are considering  and the transverse direction to the cycle in that same two-torus. As a general rule, if we have a minimal-length one-cycle of length $L$ on a two-torus of area $A$, the dimensionless transverse period $S$ that appears in \eqref{mem-eq: fourier delta def} will be $S=A /\ell_s L$.

We did not have to worry about this factor in the $\mathbb{Z}_2\times \mathbb{Z}_2$ example, since all the cycles had periodicity one in the normalized coordinates. That will no longer be the case in general for the rest of our examples. We illustrate this reasoning by building the $\delta$-like bump functions with support on to the loci $\Pi_i$ introduced in table \ref{mem-table: z4 orientifolds}. The factor $S$ will be crucial to properly define the $\delta$-bump function describing the orientifold planes that do not decompose as a single product of fundamental one-cycles, such as $\Pi_1$.
\begin{subequations}
\begin{align}
    \delta(\Pi_{0})=&\ell_s^3\sum_{\vec{\eta}}\left[\sum_{n_1\in\mathbb{Z}} e^{2\pi i n_1 (y^1-\eta_1)} dy^1 \right]\wedge \left[\sum_{n_2\in\mathbb{Z}} e^{2\pi i n_2 (y^2-\eta_2)} dy^2 \right]\wedge \left[\sum_{n_3\in\mathbb{Z}} e^{2\pi i n_3 (y^2-\eta_2)} dy^3 \right]\, ,\\
    \delta(\Pi_{1})=&\ell_s^3\left[\sqrt{2}\sum_{n_1\in\mathbb{Z}} e^{2\sqrt{2}\pi i n_1 \hat{y}^1} d\hat{y}^1 \right]\wedge \left[\sqrt{2}\sum_{n_2\in\mathbb{Z}} e^{2\sqrt{2}\pi i n_2 \hat{y}^2} d\hat{y}^2 \right]\wedge \left[\sum_{\eta_3}\sum_{n_3\in\mathbb{Z}} e^{2\pi i n_3 (x^3-\eta_3)} dx^3 \right]\, ,\\
    \delta(\Pi_{2})=&\ell_s^3\sum_{\vec{\eta}}\left[\sum_{n_1\in\mathbb{Z}} e^{2\pi i n_1 (x^1-\eta_1)} dx^1 \right]\wedge \left[\sum_{n_2\in\mathbb{Z}} e^{2\pi i n_2 (x^2-\eta_2)} dx^2 \right]\wedge \left[\sum_{n_3\in\mathbb{Z}} e^{2\pi i n_3 (y^3-\eta_3)} dy^3 \right]\, ,\\
    \delta(\Pi_{3})=&\ell_s^3\left[\sqrt{2}\sum_{n_1\in\mathbb{Z}} e^{2\sqrt{2}\pi i n_1 \tilde{y}^1} d\tilde{y}^1 \right]\wedge \left[\sqrt{2}\sum_{n_2\in\mathbb{Z}} e^{2\sqrt{2}\pi i n_2 \tilde{y}^2} d\tilde{y}^2 \right]\wedge \left[\sum_{\eta_3}\sum_{n_3\in\mathbb{Z}} e^{2\pi i n_3 (x^3-\eta_3)} dx^3 \right]\, ,
\end{align}
\end{subequations}
where we have defined $\hat{y}^i=\frac{1}{\sqrt{2}}(x^i-y^i)$, $\tilde{y}^i=\frac{1}{\sqrt{2}}(x^i+y^i)$ and $\vec{\eta}$ has entries that are either $0$ or $1$. With all this information it is straightforward to build the three-form $K$ satisfying \eqref{mem-eq: K equation} through the introduction of the following set of three-form currents defined in \eqref{mem-Ktorus}:

\begin{subequations}
\begin{align}
    &K_{0,\eta}=-\ell_s^3\sum_{0\neq \vec{n}\in \mathbb{Z}^3}\frac{e^{2\pi i \vec{n}[(y^1,y^2,y^3)-\vec{\eta}]}}{|\vec{n}|^2} dy^1\wedge dy^2\wedge dy^3\, ,\\
    &K_{1,\eta_3}=-2\ell_s^3\sum_{0\neq \vec{n}\in \mathbb{Z}^3}\frac{e^{2\pi i \vec{n}[(\sqrt{2}\hat{y}^1,\sqrt{2}\hat{y}^2,x^3)-(0,0,\eta_3)]}}{|\vec{n}|^2} d\hat{y}^1\wedge d\hat{y}^2\wedge dx^3\,,\\
    &K_{2,\eta}=\ell_s^3\sum_{0\neq \vec{n}\in \mathbb{Z}^3}\frac{e^{2\pi i \vec{n}[(x^1,x^2,y^3)-\vec{\eta}]}}{|\vec{n}|^2} dx^1\wedge dx^2\wedge dy^3\, ,\\
   &K_{3,\eta_3}=2\ell_s^3\sum_{0\neq \vec{n}\in \mathbb{Z}^3}\frac{e^{2\pi i \vec{n}[(\sqrt{2}\tilde{y}^1,\sqrt{2}\tilde{y}^2,x^3)-(0,0,\eta_3)]}}{|\vec{n}|^2} d\tilde{y}^1\wedge d\tilde{y}^2\wedge dx^3\, ,
\end{align}
\end{subequations}
 where $K_{\alpha,\eta}$ is the function associated to $\Pi_{\alpha ,\eta}$ and  $|\vec{n}|^2=n_1^2/S_1^2 \hat{R}_1^2+n_2^2/S_2^2\hat{R}_2^2+n_3^2/S_3^2\hat{R}_3^2$, with $S_i$ the transverse period of the one-cycle obtained from projecting the three-cycle $\Pi$ over $(T^2)_i$. Note that $|\vec{n}|$ changes for each function $K_{\alpha}$, since each one is describing a different three-cycle. Also, as before, the relative signs between the different $K_{\alpha}$ are chosen so that $\Im\Omega$ calibrates all the orientifold planes.

At this point, we introduce the cohomology relation $[\ell_s^{-2}H]=h\rm P.D[\Pi_{\rm O6}]$, which implies
\begin{equation}
\begin{split}
     \ell_s^{-2} [H]=&h\left( 8[\beta^0]  +4[d\hat{y}^{1}\wedge d\hat{y}^{2}\wedge dx^3] - 4[d\tilde{y}^{1}\wedge d\tilde{y}^{2}\wedge dx^3]  - 8[\beta^3]\right) \\
      = &\ 8h\left([\beta^0]-\frac{1}{2}[\beta^1]-\frac{1}{2}[\beta^2]-[\beta^3]\right)\, .        
\end{split}
\end{equation}
Now we can impose the equation of motion using \eqref{mem-intflux}. Defining $\rho=8\pi^3 \hat{R}_1\hat{R}_2\hat{R}_3$ and taking into account our choice of complex structure, the holomorphic form $\Omega$ is
\bea
\re \Om_{\rm CY} & = &  \ell_s^3 \rho \left(u_3 \b^0 -  \b^1 - \b^2 - u_3 \b^3 \right) \, ,\\
\im \Om_{\rm CY} & = &  \ell_s^3 \rho \left( \a_0 -  u_3\a_1 - u_3 \a_2 -  \a_3 \right) \, .
\eea
In order to satisfy \eqref{mem-intflux}  the remaining complex structure modulus must be fixed to $u_3=2$, while $\mu$ is given by $\mu=\ell_s^{-1}4h/\rho u_3$. 

Along the lines of the $\IZ_2\times\IZ_2$ case, let us turn to the appropriate flux quantization condition in the $\IZ_4$ orientifold. Taking the results from \cite{blumenhagen2003supersymmetric}, the minimal integral lattice of three-cycles is defined as in \eqref{mem-eq: bulk cycles z4}. Applying the flux quantization criterion for the $H$ flux once we consider the presence of O6-planes we find that $[\ell_s^{-2}H] =2h{\rm P.D} [2\rho_1 - \bar{\rho}_2] = h{\rm P.D} [\Pi_{\rm O6}]$ with $h\in \IZ$.

This quantization condition is more constraining than in the $\IZ_2 \times \IZ_2$ orbifold, allowing solutions to the tadpole involving only a single jump in the quantum of Roman mass 
\begin{equation}
    m = 2k\, , \qquad h = 1\, , \qquad N=4-2k\,,\qquad k = 1,2. \label{mem-eq: tadpole z4}
\end{equation}

Next, we compute the different components of $\mathcal{F}$ in \eqref{mem-eq: worldvolume flux} as $\mathcal{F}_{\alpha,\eta}=\ell_s d^\dagger K_{\alpha,\eta}$:
\begin{subequations}
\begin{align}
   \mathcal{F}_{0,\eta}=&\frac{i\ell_s^2}{2\pi}\sum_{0\neq \vec{n}\in \mathbb{Z}^3}\frac{e^{2\pi i \vec{n}[(y^1,y^2,y^3)-\vec{\eta}]}}{|\vec{n}|^2} \left(\frac{n_1}{\hat{R}_1^2}dy^2\wedge dy^3-\frac{n_2}{\hat{R}_2^2}dy^1\wedge dy^3+\frac{n_ 3}{\hat{R}_3^2}dy^1\wedge dy^2\right)\, , \label{mem-eq: F0 Z4}\\
   \mathcal{F}_{1,\eta}=&\frac{i\ell_s^2}{2\pi}2\sum_{0\neq \vec{n}\in \mathbb{Z}^3}\frac{e^{2\pi i \vec{n}[(\sqrt{2}\hat{y}^1,\sqrt{2}\hat{y}^2,x^3)-(0,0,\eta_3)]}}{|\vec{n}|^2}\nonumber\\
   &\times\left(\frac{\sqrt{2}n_1}{\hat{R}_1^2}d\hat{y}^2\wedge dx^3-\frac{\sqrt{2}n_2}{\hat{R}_2^2}d\hat{y}^1\wedge dx^3+\frac{n_ 3}{\hat{R}_3^2}d\hat{y}^1\wedge d\hat{y}^2\right)\, , \\
    \mathcal{F}_{2,\eta}=&-\frac{i\ell_s^2}{2\pi}\sum_{0\neq \vec{n}\in \mathbb{Z}^3}\frac{e^{2\pi i \vec{n}[(x^1,x^2,y^3)-\vec{\eta}]}}{|\vec{n}|^2} \left(\frac{n_1}{\hat{R}_1^2}dx^2\wedge dy^3-\frac{n_2}{\hat{R}_2^2}dx^1\wedge dy^3+\frac{n_ 3}{\hat{R}_3^2}dx^1\wedge dx^2\right)\, , \\
    \mathcal{F}_{3,\eta}=&-\frac{i\ell_s^2}{2\pi}2\sum_{0\neq \vec{n}\in \mathbb{Z}^3}\frac{e^{2\pi i \vec{n}[(\sqrt{2}\tilde{y}^1,\sqrt{2}\tilde{y}^2,x^3)-(0,0,\eta_3)]}}{|\vec{n}|^2} \nonumber\\
    &\times\left(\frac{\sqrt{2}n_1}{\hat{R}_1^2}d\tilde{y}^2\wedge dx^3-\frac{\sqrt{2}n_2}{\hat{R}_2^2}d\tilde{y}^1\wedge dx^3+\frac{n_ 3}{\hat{R}_3^2}d\tilde{y}^1\wedge d\tilde{y}^2\right)\, .\label{mem-eq: F3 Z4}
\end{align}
\end{subequations}
Now we would like to compute $\int J_\cy\wedge \mathcal{F}_{\alpha,\eta}\wedge\mathcal{F}_{\beta,\zeta}$. To perform this integral we regularize it as before, interchanging the order between summation and integration.  Similarly to the $\mathbb{Z}_2\times \mathbb{Z}_2$ orbifold, this allows us to obtain Kronecker deltas from the following relations:
\begin{align}
    \int_{T^2}  e^{2\pi i ny^1}e^{2\pi^i my^1} dx^1 dy^1=&\, \delta_{n+m}\, ,\\
    \int_{T^2} e^{2\sqrt{2}\pi i n\tilde{y}^1}e^{2\sqrt{2}\pi^i m\tilde{y}^1} dx^1 dy^1=&\, \delta_{n+m}\, ,\\
    \int_{T^2}  e^{2\sqrt{2}\pi i n\hat{y}^1}e^{2\sqrt{2}\pi^i m\hat{y}^1} dx^1 dy^1=&\, \delta_{n+m}\, ,\\
    \int_{T^2} e^{2\pi i ny^1}e^{2\sqrt{2}\pi^i m\tilde{y}^1} dx^1 dy^1=&\, \delta_n\delta_m\, ,\\
    \int_{T^2} e^{2\pi i ny^1}e^{2\sqrt{2}\pi^i m\hat{y}^1} dx^1 dy^1=&\, \delta_n\delta_m\, ,\\
    \int_{T^2} e^{2\sqrt{2}\pi i n\tilde{y}^1}e^{2\sqrt{2}\pi^i m\hat{y}^1} dx^1 dy^1=&\, \delta_n\delta_m\, .
\end{align}
With all this information we can finally evaluate the different terms that contribute to \eqref{mem-Deltasum}. Many of them will be exactly as in the $\mathbb{Z}_2\times \mathbb{Z}_2$ orbifold, but there are also some new kinds of  contributions. First of all, we can consider pairs of three-cycles with non-vanishing intersection number. Let us for instance choose  $\Delta_{0,\vec{0};3,\vec{0}}$. From figure \ref{mem-figure: Z4-planes} we see that the three-cycles intersect at a single point. Using \eqref{mem-eq: J z4}, \eqref{mem-eq: F0 Z4} and \eqref{mem-eq: F3 Z4} we obtain

\begin{align}
\Delta_{0,\vec{0};3,0}=& -  \frac{e^{K/2}}{\ell_s^{6}} \int_{X_6}  J_{\rm CY} \wedge \cF_{0,\vec{0}} \wedge \cF_{3,0}\nonumber\\
=&- \frac{e^{K/2}}{4\pi^2\ell_s^{6}}\int_{X_6}\sum_{0\neq \vec{n},\vec{m}\in \mathbb{Z}^3}\frac{2e^{2\pi i [\vec{n}(y^1,y^2,y^3)+ \vec{m}(\sqrt{2}\tilde{y}^1,\sqrt{2}\tilde{y}^2,x^3)]}}{|\vec{n}|^2|\vec{m}|^2}\nonumber\\
&\times\left(\frac{n_1m_1t^1}{\hat{R}_1^4}+\frac{n_2m_2 t^2}{\hat{R}_2^4}+\frac{n_3m_3t^3}{2\hat{R}_3^4}\right)\Phi_6\nonumber\\
=&-e^{K/2}\frac{1}{4\pi^2N_\Gamma} \sum_{0\neq \vec{n},\vec{m}\in \mathbb{Z}^3}\frac{1}{|\vec{n}|^2|\vec{m}|^2}\nonumber\\
&\times\left(\frac{2n_1m_1t^1}{\hat{R}_1^4}+\frac{2n_2m_2 t^2}{\hat{R}_2^4}+\frac{n_3m_3t^3}{\hat{R}_3^4}\right)\delta_{n_1}\delta_{n_2}\delta_{n_3}\delta_{m_1}\delta_{m_2}\delta_{m_3}\nonumber\\
=&\, 0   \, ,
\end{align}
where we defined $\Phi_6=\ell_s^6 dx^1\wedge dx^2\wedge dx^3\wedge dy^1\wedge dy^2\wedge dy^3$. Therefore, we observe once more that the only non-trivial contributions come from the $\mathcal{N}=2$ sector. For the case of $\mathbb{Z}_4$ orbifold, the aforementioned sector is richer and more diverse than the $\mathbb{Z}_2\times\mathbb{Z}_2$ orbifold. In addition to pairs of branes of the form \eqref{mem-eq: delta z2 intersection} we must also consider contributions involving cycles that do not run along the fundamental periodic directions. Let us focus on $\Delta_{1,0;3,0}$. In figure \ref{mem-figure: Z4-planes} we can observe the involved three-cycles intersect over a one-cycle on $(T^2)_3$. We find that
\begin{align}
\Delta_{1,0;3,0}=& - e^{K/2} \frac{1}{\ell_s^{6}} \int_{X_6}  J_{\rm CY} \wedge \cF_{1,0} \wedge \cF_{3,0}\nonumber\\
=&-e^{K/2} \frac{1}{4\pi^2\ell_s^{6}}\int_{X_6}\sum_{0\neq \vec{n},\vec{m}\in \mathbb{Z}^3}\frac{e^{2\pi i \vec{n}(\sqrt{2}\hat{y}^1,\sqrt{2}\hat{y}^2,x^3)}}{|\vec{n}|^2}\frac{e^{2\pi i \vec{m}(\sqrt{2}\tilde{y}^1,\sqrt{2}\tilde{y}^2,x^3)}}{|\vec{m}|^2}\frac{4n_3m_3t^3}{\hat{R}_3^4}\Phi_6\nonumber\\
=&-e^{K/2}\frac{1}{4\pi^2N_\Gamma} \sum_{0\neq \vec{n},\vec{m}\in \mathbb{Z}^3}\frac{1}{|\vec{n}|^2|\vec{m}|^2}\frac{4n_3m_3t_3}{\hat{R}_3^4}\delta_{n_1}\delta_{n_2}\delta_{m_1}\delta_{m_2}\delta_{n_3+m_3}\nonumber\\
=&-e^{K/2}\frac{1}{4\pi^2N_\Gamma} \sum_{0\neq n_3}\frac{\hat{R}_3^4}{4n_3^4}\frac{-4n_3^2t_3}{\hat{R}_3^4}=\frac{T^3_{\rm D4}}{12N_\Gamma}\, .
\end{align}
 The result again agrees with \eqref{mem-intercont}. Similarly, we can consider cycles that do not intersect, but run parallel along one of the two-torus. We take, for instance, $\mathcal{F}_{1,0}$ and $\mathcal{F}_{3,1/2}$, obtaining
\begin{align}
     \Delta_{1,0;3,1/2}=& - e^{K/2} \frac{1}{\ell_s^{6}} \int_{X_6}  J_{\rm CY} \wedge \cF_{1,0} \wedge \cF_{3,1/2}\nonumber\\
     =&-e^{K/2} \frac{1}{4\pi^2\ell_s^{6}}\int_{X_6}\sum_{0\neq \vec{n},\vec{m}\in \mathbb{Z}^3}\frac{e^{2\pi i \vec{n}(\sqrt{2}\hat{y}^1,\sqrt{2}\hat{y}^2,x^3)}}{|\vec{n}|^2}\frac{e^{2\pi i \vec{m}(\sqrt{2}\tilde{y}^1,\sqrt{2}\tilde{y}^2,x^3)}e^{i\pi m_3}}{|\vec{m}|^2}\frac{4n_3m_3t^3}{\hat{R}_3^4}\Phi_6\nonumber\\
    =&-e^{K/2}\frac{1}{4\pi^2N_\Gamma} \sum_{0\neq \vec{n},\vec{m}\in \mathbb{Z}^3}\frac{(-1)^{m_3}}{|\vec{n}|^2|\vec{m}|^2}\frac{4n_3m_3t_3}{\hat{R}_3^4}\delta_{n_1}\delta_{n_2}\delta_{m_1}\delta_{m_2}\delta_{n_3+m_3}\nonumber\\
    =&e^{K/2}\frac{1}{4\pi^2N_\Gamma} \sum_{0\neq n_3}\frac{(-1)^{n_3}t^3}{n_3^2}=-\frac{T^3_{\rm D4}}{24N_\Gamma}\, .
\end{align}
Putting all the contributions together we conclude that
\begin{equation}
    \Delta_{\rm D8}^{\rm BIon} = \frac{1}{24N_\Gamma}\, \left(\sum_{\alpha,\beta}\hat{q}_{0,\alpha}\,\hat{q}_{2,\beta}\,\varepsilon_{\alpha\beta}  +\sum_{\sigma,\rho}4\hat{q}_{1,\sigma}\,\hat{q}_{3,\rho}\,\varepsilon_{\sigma\rho}\right)T^3_{\rm{D4}} \, ,
\end{equation}
with $\varepsilon_{\alpha\beta}$ defined as in \eqref{mem-finalDelta}. Taking as an example the family of solutions defined in \eqref{mem-eq: tadpole z4} we can provide again a configuration of D6-branes with negative $\Delta_{\rm D8}^{\rm Bion}$. For instance, let us consider a configuration such that for each value of $\alpha$ all the corresponding $p_{\alpha}(4-2k)$ D6-branes are wrapping a single three-cycle. In particular, one can take
\begin{equation}
    \hat{q}_{0,(0,0,0)}=\hat{q}_{2,(0,0,\frac{1}{2})}= 8, \hspace{3em} \hat{q}_{1,(0,0,0)}=\hat{q}_{3,(0,0,\frac{1}{2})}=2.
\end{equation}
With such a configuration we obtain
\begin{equation}
    \Delta_{\rm D8}^{\rm Bion}= -\frac{5}{6} T^{3}_{\rm D4}\, .
\end{equation}
Therefore, this result signals again a BIonic excess tension for the 4-dimensional membrane, which could imply a possible failure of the WGC inequality. Indeed, a naive computation\footnote{For all the $\mathbb{Z}_4$ orbifolds studied in \cite[Appendix B]{Reffert:2006du} one obtains the relations $c_2(X_6).R_i =0$, $R_i \simeq 4D_{i\a} +\dots$ $i=1,2$ and $c_2(X_6).R_3 =0$, $R_3 \simeq 2D_{3\a} +\dots$, where the dots represent exceptional divisors. From here one can deduce that $\Delta_{\rm D8}^{\rm curv} = \oh  T^{3}_{\rm D4}$, following the same reasoning as in the $\Z_2 \times Z_2$ orbifold.} gives 
$\Delta_{\rm D8}^{\rm curv} = \oh  T^{3}_{\rm D4}$ in the orbifold limit, which implies that  $\Delta_{\rm D8}^{\rm curv} + \Delta_{\rm D8}^{\rm Bion} < 0$. Hence, this vacuum also seems to be in tension with the WGC for 4-dimensional membranes. 

Repeating the analysis for the choice AAB provides the same results.

\subsection{\texorpdfstring{$T^6/\mathbb{Z}_3\times \mathbb{Z}_3$}{T6/Z3xZ3}}

We consider now the case where the internal space is an orientifold of the orbifold $T^6/\mathbb{Z}_3^2$ described in \cite{Strominger:1985ku,DeWolfe:2005uu,Junghans:2020acz}. In order to be consistent with our choice of orientifold involution, we will slightly change the notation of the aforementioned references.

We will again work in the covering space, which is a factorizable six torus, $T^6=(T^2)_1\times (T^2)_2 \times (T^2)_3$   with complex coordinates $z^i$ given by \eqref{mem-coordinates}. 
The $\mathbb{Z}_3\times\mathbb{Z}_3$ orbifold action reads 
\begin{equation}
    \theta: z^i \rightarrow \alpha^2 z^i\,,\qquad  \omega:z^i\rightarrow \alpha^{2i}z^i+\left(\frac{1}{2}i+\frac{\sqrt{3}}{2}\right)\, ,
\end{equation}
with $\alpha=e^{i\pi/3}$. 

The above symmetries, together with the orientifold involution, are more constraining that those introduced in the $\mathbb{Z}_2\times \mathbb{Z}_2$
or $\mathbb{Z}_4$ orbifolds and they fully fix the complex structure to
\begin{equation}
    \tau_1=\tau_2=\tau_3=\frac{\sqrt{3}}{2}+\frac{1}{2}i\, .
\end{equation}
Hence, the factor $(T^2)_i$ can be described as a quotient of $\mathbb{C}$ by a lattice generated by $e_{i1}=2\pi R_i i$ and $e_{i2}=2\pi R_i(\sqrt{3}/2+i/2)$.  They provide the following periodic identifications:
\begin{equation}
    z^i\sim z^i+2\pi R_i i \sim z^i+2\pi R_i \left(\frac{1}{2}i+\frac{\sqrt{3}}{2}\right)\,.
    \label{mem-eq: periodic identification Z3z3}
\end{equation}

It is worth noting that only the  generator $\theta$ of the first $\mathbb{Z}_3$ preserves the lattice generated by these vectors. The trick of this orbifold is that we are not taking the quotient simultaneously. $Q$ is not a symmetry of $T^6$ by itself, but it emerges as a symmetry of the quotient $T^6/\mathbb{Z}_3^{\theta}$. This construction was described in detail in \cite{Strominger:1985ku}.  Using the periodic coordinates, the metric and the Kähler form are 
\begin{align}
    g =&\, 4\pi^2\ell_s^2\left(\begin{array}{cccccc}
        \hat{R}_1^2 & 0 & 0 & \frac{\hat{R}_1^2}{2} & 0 & 0 \\
        0 & \hat{R}_2^2 & 0 & 0 &  \frac{\hat{R}_2^2}{2} & 0 \\
        0 & 0 & \hat{R}_3^2 & 0 & 0 &  \frac{\hat{R}_3^2}{2} \\
         \frac{\hat{R}_1^2}{2} & 0 & 0 & \hat{R}_1^2 & 0 & 0 \\
        0 &  \frac{\hat{R}_2^2}{2} & 0 & 0 & \hat{R}_2^2 & 0 \\
        0 & 0 &  \frac{\hat{R}_3^2}{2} & 0 & 0 & \hat{R}_3^2 
    \end{array}\right)\,,\\
    \nonumber\\
    J=&\, \ell_s^2(t^1 dx^1\wedge dy^1 +t^2 dx^2\wedge dy^2 + t^3 dx^3\wedge dy^3)\,,\label{mem-eq: kahler form z3z3}    
\end{align}
where  we  defined the dimensionless radii $\hat{R}_i=R_i/\ell_s$ and the K\"ahler moduli $t^i=4\pi^2\sqrt{3}/2 \hat{R}_{i}^2$.

The orientifold planes are given by the set of points that satisfy $\sigma(z)=z$ up to the action of the orbifold. This gives nine different loci, summarized in table \ref{mem-table: z3xz3 orientifolds} and represented schematically in figure \ref{mem-figure: tilted torus}. Using the expression $S=A /\ell_s L$, it is easy to see that the transverse period of all the one-cycles involved in the problem is $S=1/2$. 

\renewcommand{\arraystretch}{0.95}
\begin{table}[H]
$$
\begin{array}{|l|l|lll|}
\hline \Pi_{\alpha} & \text { Fixed point equation } & & \text { O6-plane position }  & \\
\hline \Pi_{0} & \sigma\left(z^{a}\right)=z^{a} & x^1+2y^{1} \in\left\{1, 2\right\} \quad &x^2+2y^{2} \in\left\{1, 2\right\} &x^3+2y^{3} \in\left\{1, 2\right\} \\
\Pi_{1} & \sigma\left(z^{a}\right)=\theta\left(z^{a}\right) & y^{1}+2 x^{1} \in\{1,2\} &y^{2}+2 x^{2} \in\{1,2\} &y^{3}+2 x^{3} \in\{1,2\} \\
\Pi_{2} & \sigma\left(z^{a}\right)=\theta^{2}\left(z^{a}\right) & y^{1}- x^{1}=0 &y^{2}- x^{2}=0 &y^{3}- x^{3}=0 \\
\Pi_{3} & \sigma\left(z^{a}\right)=\omega \theta^{2}\left(z^{a}\right) & x^1+2y^{1} \in\left\{1, 2\right\} &y^{2}+2 x^{2} \in\{1,2\} &y^{3}-x^{3}=0 \\
\Pi_{4} & \sigma\left(z^{a}\right)=\omega^{2} \theta\left(z^{a}\right) & x^1+2y^{1} \in\left\{1, 2\right\} &y^{2}- x^{2}=0 &y^{3}+2 x^{3} \in\{1,2\} \\
\Pi_{5} & \sigma\left(z^{a}\right)=\omega \theta\left(z^{a}\right) & y^{1}- x^{1}=0 &x^2+2 y^{2} \in\left\{1, 2\right\} &y^{3}+2 x^{3} \in\{1,2\} \\
\Pi_{6} & \sigma\left(z^{a}\right)=\omega^{2} \theta^{2}\left(z^{a}\right) & y^{1}+2 x^{1} \in\{1,2\} &x^2+2y^{2} \in\left\{1, 2\right\} &y^{3}- x^{3}=0 \\
\Pi_{7} & \sigma\left(z^{a}\right)=\omega\left(z^{a}\right) & y^{1}+2 x^{1} \in\{1,2\} &y^{2}- x^{2}=0 &x^3+2y^{3} \in\left\{1, 2\right\} \\
\Pi_{8} & \sigma\left(z^{a}\right)=\omega^{2}\left(z^{a}\right) & y^{1}- x^{1}=0 & y^{2}+2 x^{2} \in\{1,2\} & x^3+2 y^{3} \in\left\{1, 2\right\} \\
\hline
\end{array}
$$
\caption{O6-planes in $T^6/\mathbb{Z}_3\times \mathbb{Z}_3$.}
\label{mem-table: z3xz3 orientifolds}
\end{table}

\begin{figure}[h!]
  \includegraphics[width=\textwidth]{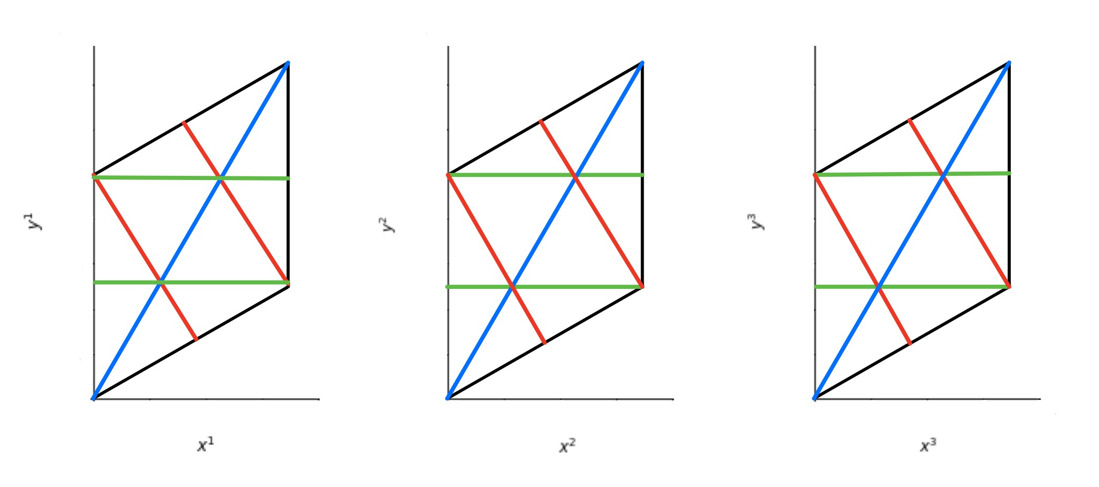}
    \caption{Fundamental domain of $T^2\times T^2\times T^2$ and the fixed loci for $T^6/\mathbb{Z}_3\times \mathbb{Z}_3$. Planes $\Pi_0$, $\Pi_{1}$, $\Pi_{2}$ are represented by the colours green, red and blue respectively. }
    \label{mem-figure: tilted torus}
\end{figure}
The above O6-plane content can be expressed in terms of bulk three-cycles $\rho_i$. Consider again the three-cycles inherited from the covering space $T^6$. Let us define the basis of fundamental one-cycles $\pi_{2i-1}$ and $\pi_{2i}$ of the tilted torus $(T^2)_i$, i.e. cycles winded once along the periodic directions given by the identifications that defined our tori in \eqref{mem-eq: periodic identification Z3z3}.

Then, summing over the orbits of two three-cycles, say $\pi_{135}$ and $\pi_{136}$, we obtain the following two invariant three-cycles $\rho_1$ and $\rho_2$, which are used to build the orientifold $\Pi_{\rm O6}$:
\begin{align}
   & \rho_1=3\,(\pi_{135} + \pi_{246} - \pi_{245}-\pi_{236} -\pi_{146}), \label{mem-eq: bulk cycles z3z3 1}\\
   &\rho_2=3\,(\pi_{235} + \pi_{145} - \pi_{245} + \pi_{136} - \pi_{236} - \pi_{146})\, ,\label{mem-eq: bulk cycles z3z3 2}\\
   &\Pi_{\rm O6} = 6\rho_1 - 3\rho_2 \, .\label{mem-eq: O6 z3z3}
\end{align}

With all this information we can repeat a similar reasoning as in the previous cases. Therefore, we build the following functions $K_{\alpha}$. Note that no $\eta$ index is needed to label the $\IZ_3\times\IZ_3$ orientifold planes.

{\allowdisplaybreaks\begin{subequations}
\begin{align}
    K_0=&-8\ell_s^3\sum_{0\neq\vec{n}\in\mathbb{Z}^3} \frac{e^{4\pi i\vec{n}(\bar{y}^1,\bar{y}^2,\bar{y}^3)}}{|\vec{n}|^2}\quad  d\bar{y}^1\wedge d\bar{y}^2\wedge d\bar{y}^3 \,,\\
    K_{1}=& 8\ell_s^3\sum_{0\neq\vec{n}\in\mathbb{Z}^3} \frac{e^{4\pi i\vec{n}(\tilde{y}^1,\tilde{y}^2,\tilde{y}^3)}}{|\vec{n}|^2} \quad d\tilde{y}^1\wedge d\tilde{y}^2\wedge d\tilde{y}^3\,,\\
    K_{2}=& 8\ell_s^3\sum_{0\neq\vec{n}\in\mathbb{Z}^3} \frac{e^{4\pi i\vec{n}(\hat{y}^1,\hat{y}^2,\hat{y}^3)}}{|\vec{n}|^2}\quad d\hat{y}^1\wedge d\hat{y}^2\wedge d\hat{y}^3\,,\\
    K_{3}=&-8\ell_s^3\sum_{0\neq\vec{n}\in\mathbb{Z}^3} \frac{e^{4\pi i\vec{n}(\bar{y}^1,\tilde{y}^2,\hat{y}^3)}}{|\vec{n}|^2}\quad  d\bar{y}^1\wedge d\tilde{y}^2\wedge d\hat{y}^3\,,\\
    K_{4}=& -8\ell_s^3 \sum_{0\neq\vec{n}\in\mathbb{Z}^3} \frac{e^{4\pi i\vec{n}(\bar{y}^1,\hat{y}^2,\tilde{y}^3)}}{|\vec{n}|^2} \quad  d\bar{y}^1\wedge d\hat{y}^2\wedge d\tilde{y}^3\,,\\
    K_{5}=& -8\ell_s^3 \sum_{0\neq\vec{n}\in\mathbb{Z}^3} \frac{e^{4\pi i\vec{n}(\hat{y}^1,\bar{y}^2,\tilde{y}^3)}}{|\vec{n}|^2}\quad  d\hat{y}^1\wedge d\bar{y}^2\wedge d\tilde{y}^3\,,\\
    K_{6}=&  -8\ell_s^3\sum_{0\neq\vec{n}\in\mathbb{Z}^3} \frac{e^{4\pi i\vec{n}(\tilde{y}^1,\bar{y}^2,\hat{y}^3)}}{|\vec{n}|^2}\quad d\tilde{y}^1\wedge d\bar{y}^2\wedge d\hat{y}^3\,,\\
    K_{7}=& -8\ell_s^3 \sum_{0\neq\vec{n}\in\mathbb{Z}^3} \frac{e^{4\pi i\vec{n}(\tilde{y}^1,\hat{y}^2,\bar{y}^3)}}{|\vec{n}|^2} \quad d\tilde{y}^1\wedge d\hat{y}^2\wedge d\bar{y}^3\,,\\
    K_{8}=&-8\ell_s^3 \sum_{0\neq\vec{n}\in\mathbb{Z}^3} \frac{e^{4\pi i\vec{n}(4\hat{y}^1,\tilde{y}^2,\bar{y}^3)}}{|\vec{n}|^2} \quad d\hat{y}^1\wedge d\tilde{y}^2\wedge d\bar{y}^3 \,,
\end{align}
\label{mem-eq: P functions Z3xZ3}
\end{subequations}}
where we have defined $\hat{y}^i=(-x^i+y^i)/2$, $\bar{y}^i=(x^i+2y^i)/2$ and $\tilde{y}^i=(2x^i+y^i)/2$.
Note again that the relative signs in the above expression have been chosen so that the volume of the orientifolds is calibrated by $\Im \Omega$.

With all this information we introduce the cohomology relation $[\ell_s^{-2}H]=h\rm P.D[\Pi_{\rm O6}]$, which implies
\begin{align}
    [\ell_s^{-2}H]=&8h\left([d\bar{y}^1\wedge d\bar{y}^2\wedge d\bar{y}^3]-[d\tilde{y}^1\wedge d\tilde{y}^2\wedge d\tilde{y}^3]-[d\hat{y}^1\wedge d\hat{y}^2\wedge d\hat{y}^3]\right.\nonumber\\ 
    &+[d\bar{y}^1\wedge d\tilde{y}^2\wedge d\hat{y}^3]
   +[d\bar{y}^1\wedge d\hat{y}^2\wedge d\tilde{y}^3]+[d\hat{y}^1\wedge d\bar{y}^2\wedge d\tilde{y}^3]\nonumber\\
   &\left.+[d\tilde{y}^1\wedge d\bar{y}^2\wedge d\hat{y}^3]+[d\tilde{y}^1\wedge d\hat{y}^2\wedge d\bar{y}^3]+[d\hat{y}^1\wedge d\tilde{y}^2\wedge d\bar{y}^3]\right)\nonumber\\
   =&9h(-2[\alpha_0]+[\alpha_1]+[\alpha_2]+[\alpha_3]+2[\beta_0]-[\beta_1]-[\beta_2]-[\beta_3]) \, .
\end{align}
Now, in a similar reasoning to the previous cases the flux quantization condition for the $\IZ_3\times\IZ_3$ orientifold will be given applying the quantization criterion for the $H$-flux. Taking the invariant bulk three-cycles \eqref{mem-eq: bulk cycles z3z3 1},\eqref{mem-eq: bulk cycles z3z3 2} along with the O6-planes content \eqref{mem-eq: O6 z3z3} we arrive to $[\ell_s^{-2}H]=2\tilde{h}\rm P.D [2\rho_1 - \rho_2]$ with $\tilde{h}\in \IZ$. Then, the possible values for $h$ are restricted to $h\in \frac{2}{3}\IZ$. This constraints the possible solutions for the tadpole equation. One family of solutions is of the form

\begin{equation}
    m = 2k\, , \qquad h = \frac{2}{3}\, , \qquad N=4-\frac{4k}{3}\, ,\qquad k = 1,2,3.
\end{equation}
We can now provide the different components of $\mathcal{F}$:
{\allowdisplaybreaks\begin{subequations}
\begin{align}
    &\mathcal{F}_{0}= \frac{h\ell_s^216i}{2\pi}\, \sum_{0\neq\vec{n}\in\mathbb{Z}^3} \frac{e^{4\pi i\vec{n}(\bar{y}^1,\bar{y}^2,\bar{y}^3)}}{|\vec{n}|^2}\left(\frac{n_1}{\hat{R}_1^2}  d\bar{y}^2\wedge d\bar{y}^3-\frac{n_2}{\hat{R}_2^2}  d\bar{y}^1\wedge d\bar{y}^3+\frac{n_3}{\hat{R}_3^2}  d\bar{y}^1\wedge d\bar{y}^2\right) \,,\\
   &\mathcal{F}_{1}= -\frac{h\ell_s^216i}{2\pi}\,\sum_{0\neq\vec{n}\in\mathbb{Z}^3} \frac{e^{4\pi i\vec{n}(\tilde{y}^1,\tilde{y}^2,\tilde{y}^3)}}{|\vec{n}|^2}\left(\frac{n_1}{\hat{R}_1^2}  d\tilde{y}^2\wedge d\tilde{y}^3-\frac{n_2}{\hat{R}_2^2}  d\tilde{y}^1\wedge d\tilde{y}^3+\frac{n_3}{\hat{R}_3^2}  d\tilde{y}^1\wedge d\tilde{y}^2\right) \,,\\
    &\mathcal{F}_{2}= -\frac{h\ell_s^216i}{2\pi}\,\sum_{0\neq\vec{n}\in\mathbb{Z}^3} \frac{e^{4\pi i\vec{n}(\hat{y}^1,\hat{y}^2,\hat{y}^3)}}{|\vec{n}|^2}\left(\frac{n_1}{\hat{R}_1^2}  d\hat{y}^2\wedge d\hat{y}^3-\frac{n_2}{\hat{R}_2^2}  d\hat{y}^1\wedge d\hat{y}^3+\frac{n_3}{\hat{R}_3^2}  d\hat{y}^1\wedge d\hat{y}^2\right) \,,\\
    &\mathcal{F}_{3}= \frac{h\ell_s^216i}{2\pi}\,\sum_{0\neq\vec{n}\in\mathbb{Z}^3} \frac{e^{4\pi i\vec{n}(\bar{y}^1,\tilde{y}^2,\hat{y}^3)}}{|\vec{n}|^2}\left(\frac{n_1}{\hat{R}_1^2}  d\tilde{y}^2\wedge d\hat{y}^3-\frac{n_2}{\hat{R}_2^2}  d\bar{y}^1\wedge d\hat{y}^3+\frac{n_3}{\hat{R}_3^2}  d\bar{y}^1\wedge d\tilde{y}^2\right) \,,\\
  &\mathcal{F}_{4}= \frac{h\ell_s^216i}{2\pi}\,\sum_{0\neq\vec{n}\in\mathbb{Z}^3} \frac{e^{4\pi i\vec{n}(\bar{y}^1,\hat{y}^2,\tilde{y}^3)}}{|\vec{n}|^2} \left(\frac{n_1}{\hat{R}_1^2}  d\hat{y}^2\wedge d\tilde{y}^3-\frac{n_2}{\hat{R}_2^2}  d\bar{y}^1\wedge d\tilde{y}^3+\frac{n_3}{\hat{R}_3^2}  d\bar{y}^1\wedge d\hat{y}^2\right) \,,\\
   &\mathcal{F}_{5}= \frac{h\ell_s^216i}{2\pi}\, \sum_{0\neq\vec{n}\in\mathbb{Z}^3} \frac{e^{4\pi i\vec{n}(\hat{y}^1,\bar{y}^2,\tilde{y}^3)}}{|\vec{n}|^2}\left(\frac{n_1}{\hat{R}_1^2}  d\bar{y}^2\wedge d\tilde{y}^3-\frac{n_2}{\hat{R}_2^2}  d\hat{y}^1\wedge d\tilde{y}^3+\frac{n_3}{\hat{R}_3^2}  d\hat{y}^1\wedge d\bar{y}^2\right) \,,\\
   &\mathcal{F}_{6}= \frac{h\ell_s^216i}{2\pi}\,\sum_{0\neq\vec{n}\in\mathbb{Z}^3} \frac{e^{4\pi i\vec{n}(\tilde{y}^1,\bar{y}^2,\hat{y}^3)}}{|\vec{n}|^2}\left(\frac{n_1}{\hat{R}_1^2}  d\bar{y}^2\wedge d\hat{y}^3-\frac{n_2}{\hat{R}_2^2}  d\tilde{y}^1\wedge d\hat{y}^3+\frac{n_3}{\hat{R}_3^2}  d\tilde{y}^1\wedge d\bar{y}^2\right) \,,\\
   &\mathcal{F}_{7}= \frac{h\ell_s^216i}{2\pi}\, \sum_{0\neq\vec{n}\in\mathbb{Z}^3} \frac{e^{4\pi i\vec{n}(\tilde{y}^1,\hat{y}^2,\bar{y}^3)}}{|\vec{n}|^2}\left(\frac{n_1}{\hat{R}_1^2}  d\hat{y}^2\wedge d\bar{y}^3-\frac{n_2}{\hat{R}_2^2}  d\tilde{y}^1\wedge d\bar{y}^3+\frac{n_3}{\hat{R}_3^2}  d\tilde{y}^1\wedge d\hat{y}^2\right) \,,\\
   &\mathcal{F}_{8}= \frac{h\ell_s^216i}{2\pi}\, \sum_{0\neq\vec{n}\in\mathbb{Z}^3} \frac{e^{4\pi i\vec{n}(\hat{y}^1,\tilde{y}^2,\bar{y}^3)}}{|\vec{n}|^2}\left(\frac{n_1}{\hat{R}_1^2}  d\tilde{y}^2\wedge d\bar{y}^3-\frac{n_2}{\hat{R}_2^2}  d\hat{y}^1\wedge d\bar{y}^3+\frac{n_3}{\hat{R}_3^2}  d\hat{y}^1\wedge d\tilde{y}^2\right) \,.
\end{align}
\label{mem-eq: F Z3Z3}
\end{subequations}}
The last step will be to compute $\int_{X_6} J_{CY}\wedge\mathcal{F}_{\alpha,\eta}\wedge\mathcal{F}_{\beta,\xi}$. To do so we will face six different families of integrals that we regularize by exchanging integration and summation following the same line of reasoning as in the previous cases. We also make use the following relations that allows us to obtain Kronecker deltas
\begin{eqnarray}
    \int_{T^2}  e^{4\pi i n\bar{y}_1}e^{4\pi^i m\bar{y}_1} dx^1 dy^1=&\, \delta_{n+m}\, , \qquad
    \int_{T^2} e^{4\pi i n\tilde{y}_1}e^{4\pi^i m\tilde{y}_1} dx^1 dy^1=&\, \delta_{n+m}\,,\nonumber\\
    \int_{T^2}  e^{4\pi i n\hat{y}_1}e^{4\pi^i m\hat{y}_1} dx^1 dy^1=&\, \delta_{n+m}\,,\qquad
    \int_{T^2}  e^{4\pi i n\bar{y}_1}e^{4\pi^i m\tilde{y}_1} dx^1 dy^1=&\, \delta_n\delta_m\,,\\
    \int_{T^2}  e^{4\pi i n\bar{y}_1}e^{4\pi^i m\hat{y}_1} dx^1 dy^1=&\, \delta_n\delta_m\,,\qquad
    \int_{T^2}  e^{4\pi i n\tilde{y}_1}e^{4\pi^i m\hat{y}_1} dx^1 dy^1=&\, \delta_n\delta_m\,. \nonumber
\end{eqnarray}
It is worth noting that the different terms contributing to \eqref{mem-Deltasum} always intersect along one-cycles in contrast to earlier results where parallel cycles appear as in \eqref{mem-eq: non intersecting D6branes example}. As we have shown previously and as maintained here, intersecting cycles only provide positive contributions, thus leaving $\Delta_{\rm D8}^{\rm BIon} \geq 0$ for any configuration of D6-branes on top of O6-planes. To illustrate this feature, we can consider pairs of branes that intersect over one-cycles on the third two-torus. Let us compute, for instance, $\Delta_{0,7}$. In figure \ref{mem-figure: tilted torus} and with the help of table \ref{mem-table: z3xz3 orientifolds} we can observe the preceding pair of branes.
\begin{align}
    \Delta_{0,7}=&-e^{-K/2}\frac{1}{\ell_s^6}\int_{X_6}J_{CY}\wedge\mathcal{F}_{0}\wedge\mathcal{F}_{7} \\\nonumber
    =&-e^{-K/2}\frac{144}{4\pi^2\ell_s^6}\int_{X_6}\sum_{0\neq\vec{n},\vec{m}\in \IZ^3}\frac{e^{4\pi i\vec{n}(\bar{y}^1,\bar{y}^2,\bar{y}^3)}}{|\vec{n}|^2}\frac{e^{4\pi i\vec{m}(\hat{y}^1,\tilde{y}^2,\bar{y}^3)}}{|\vec{m}|^2}\frac{t^3n_3m_3}{\hat{R}_3^4}\Phi_6\\\nonumber
    =&-e^{-K/2}\frac{36}{N_{\Gamma}\pi^2}\sum_{0\neq\vec{n},\vec{m}\in \IZ^3}\frac{1}{|\vec{n}|^2|\vec{m}|^2}\frac{t^3n_3m_3}{\hat{R}^4_3}\delta_{n1m1}\delta_{n2m2}\delta_{n3+m3} \\\nonumber
    =& \, e^{-K/2}t^3\frac{9}{4N_{\Gamma}}\sum_{0\neq n_3\in \IZ}\frac{1}{n_{4}^2}= \frac{3T^3_{\rm D4}}{4N_\Gamma}\, ,
\end{align}
where again we have defined $\Phi_6=\ell_s^6 dx^1\wedge dx^2\wedge dx^3\wedge dy^1\wedge dy^2\wedge dy^3$.

Iterating the previous procedure we can compute \eqref{mem-Deltasum} for the most general configuration of D6-branes. We to arrive to
\begin{equation}
     \begin{aligned}
     \Delta_{\rm D8}^{\rm BIon}=& \frac{9}{12N_\Gamma}\left[\left(\hat{q}_0\hat{q}_{4}+\hat{q}_{0}\hat{q}_{3}+\hat{q}_{3}\hat{q}_{4} +\hat{q}_{1}\hat{q}_{6} +\hat{q}_{1}\hat{q}_{7} + \hat{q}_{6}\hat{q}_{7} + \hat{q}_{2}\hat{q}_{5} +\hat{q}_{2}\hat{q}_{8} + \hat{q}_{5}\hat{q}_{8} \right) T^1_{\rm D4}\right.\\
     &\left(\hat{q}_{0}\hat{q}_{5} + \hat{q}_{0}\hat{q}_{6} + \hat{q}_{5}\hat{q}_{6} + \hat{q}_{1}\hat{q}_{3} + \hat{q}_{1}\hat{q}_{8} + \hat{q}_{3}\hat{q}_{8} + \hat{q}_{2}\hat{q}_{4} + \hat{q}_{2}\hat{q}_{7} + \hat{q}_{4}\hat{q}_{7}\right)T^2_{\rm D4}\\
     &\left.\left(\hat{q}_{0}\hat{q}_{7} + \hat{q}_{0}\hat{q}_{8} + \hat{q}_{7}\hat{q}_{8} + \hat{q}_{1}\hat{q}_{4} + \hat{q}_{1}\hat{q}_{5} + \hat{q}_{4}\hat{q}_{5} + \hat{q}_{3}\hat{q}_{4} + \hat{q}_{3}\hat{q}_{6} + \hat{q}_{4}\hat{q}_{6}\right)T^3_{\rm D4}\, \right]\, ,
     \end{aligned}
\end{equation}
where the factor $1/12$ comes from \eqref{mem-intercont} and the factor $9$ from \eqref{mem-interdef} (invariant throughout all contributions for the present case).

\subsection{Other orbifolds}

We can extend the same analysis to other orbifolds. We briefly summarize our results below.

\subsubsection*{\texorpdfstring{$T^6/\mathbb{Z}_6$}{T6/Z6}}

 We work with the orbifold described in \cite{blumenhagen2000supersymmetric, lust2007moduli} adapted to our conventions. We start by introducing in a lattice generated by  $e_{i1}=2\pi R_i(a_i+iu_i)$  and $e_{i2}=2\pi iR_i $, with $a_i=\sqrt{3}/2$, $u_i=1/2$ $\forall i$.  Hence, we have the same complex structure as in the $\mathbb{Z}_3\times \mathbb{Z}_3$ example 
\begin{equation}
    z^1=2\pi R_1(iy^1+\tau_1 x^1)\,,\qquad z^2=2\pi R_2(iy^2+\tau_2 x^2)\,,\qquad z^3=2\pi R_3(iy^3+\tau_2 x^3)\,,
\end{equation}
with $\tau_i=\sqrt{3}/2+1/2i$. The action of $\mathbb{Z}_6$ over $T^6$ is generated by an element $\theta$ that acts as 
\begin{equation}
    \theta(z^i)=e^{2\pi i v_i} z^i\,,
\end{equation}
where $v_i=(1/6,1/6,-1/3)$. The orientifold planes associated to this symmetry are summarized in table \ref{mem-table: z6 orientifolds}. Following the same steps as in the previous computations we arrive to
\begin{align}
    \Delta_{\rm D8}^{\rm BIon} =& \frac{2\sqrt{3}\pi^2 e^{-K/2}}{3N_\Gamma}\left(\hat{q}_{0}\hat{q}_{3} + \hat{q}_1\hat{q}_{4} + \hat{q}_2\hat{q}_{5}\right)\hat{R}_3^2 = \frac{\ell_s^6}{3}\left(\hat{q}_\mathcal{R}\hat{q}_{3} + \hat{q}_1\hat{q}_{4} + \hat{q}_2\hat{q}_{5}\right)T^3_{D4}\, .
\end{align}
\renewcommand{\arraystretch}{0.95}
\begin{table}[H]
$$
\begin{array}{|l|l|lll|}
\hline \Pi_{\a} & \text { Fixed point equation } & &\text { O6-plane position }  & \\
\hline \Pi_{0} & \sigma\left(z^{a}\right)=z^{a} & x^1+2y^{1} \in\left\{1, 2\right\}\quad  & x^2+2y^{2} \in\left\{1, 2\right\} & x^3+2y^{3} \in\left\{1, 2\right\} \\
\Pi_1 & \sigma\left(z^{a}\right)=\theta\left(z^{a}\right) & x^{1}+ y^{1} =1 & x^{2}+ y^{2} =1& y^{3}-x^{3} =0 \\
\Pi_{2} & \sigma\left(z^{a}\right)=\theta^{2}\left(z^{a}\right) & 2 x^{1}+ y^{1}\in\{1,2\} & 2 x^{2}+ y^{2}\in\{1,2\} & 2 x^{3}+ y^{3}\in\{1,2\} \\
\Pi_{3} & \sigma\left(z^{a}\right)=\theta^{3}\left(z^{a}\right) & x^1=0 & x^{2}=0 & x^3+2y^{3}\in\{1,2\} \\
\Pi_{4} & \sigma\left(z^{a}\right)=\theta^4\left(z^{a}\right) & y^{1}-x^1=0  & y^{2}-x^2=0 & y^{3}-x^3=0 \\
\Pi_{5} & \sigma\left(z^{a}\right)=\theta^5\left(z^{a}\right) & y^1=0 & y^2=0 & 2x^{3}+ y^{3} \in\{1,2\} \\
\hline
\end{array}
$$
\caption{O6-planes in $T^6/\mathbb{Z}_6$.}
\label{mem-table: z6 orientifolds}
\end{table}

\subsubsection*{$T^6/\mathbb{Z}_2\times \mathbb{Z}_4$}

Lastly, we consider the $\mathbb{Z}_2\times \mathbb{Z}_4$ orbifold described in \cite{lust2007moduli, forste2001supersymmetric}. We work in a lattice generated by  $e_{i1}=2\pi R_i$  and $e_{i2}=2\pi iR_i u_i $, with $u_i=(1,1,u_3)$. Consequently we have the same complex structure as in the $\mathbb{Z}_4$ example, with $z^i=2\pi R_i(x^i+i u_i y^i)$.  The action of the $\mathbb{Z}_2\times \mathbb{Z}_4$ group over our $T^6$ is generated by an order four element $\theta$ and an order two element $\omega$ that act as 
\begin{equation}
   \theta(z^i)=e^{2\pi i v_i} z^i\,, \quad \omega(z^i)=e^{2\pi i w_i}z^i\, ,
\end{equation}
where $v_i=(1/4,-1/4,0)$ and $w_i=(0,1/2,-1/2)$. With this action we find the orientifold planes summarized in table \ref{mem-table: z2xz4 orientifolds}. They lead to the following result

\begin{align}
     \Delta_{\rm D8}^{\rm BIon} =& \frac{1}{24N_\Gamma}\left[ \left(\sum_{\alpha,\beta}\hat{q}_{0,\alpha}\hat{q}_{4,\beta}\varepsilon_{\alpha\beta} + \sum_{\sigma,\rho}\hat{q}_{2,\sigma}\hat{q}_{6,\rho}\varepsilon_{\sigma\rho}  + 4\sum_{\omega,\gamma}\hat{q}_{3,\omega}\hat{q}_{7,\gamma} + 4\sum_{\epsilon,\delta}\hat{q}_{1,\epsilon}\hat{q}_{5,\delta}\right)T^1_{\rm D4}\right.\nonumber\\
     &+\left(\sum_{\alpha,\beta}\hat{q}_{0,\alpha}\hat{q}_{6,\beta}\varepsilon_{\alpha\beta} + \sum_{\sigma,\rho}\hat{q}_{2,\sigma}\hat{q}_{4,\rho}\varepsilon_{\sigma\rho}  + 4\sum_{\omega,\gamma}\hat{q}_{1,\omega}\hat{q}_{7,\gamma} + 4\sum_{\epsilon,\delta}\hat{q}_{3,\epsilon}\hat{q}_{5,\delta}\right)T^2_{\rm D4}\nonumber\\
     &+\left.\left(\sum_{k,m,\nu,\mu}\hat{q}_{k,\nu}\hat{q}_{m,\mu}\varepsilon_{k\nu,m\mu}+ 4\sum_{\sigma,\rho}\hat{q}_{1,\sigma}\hat{q}_{3,\rho}\varepsilon_{\sigma\rho} +4\sum_{\omega,\gamma}\hat{q}_{5,\omega}\hat{q}_{7,\gamma}\varepsilon_{\omega\gamma}\right)T^3_{\rm D4}\right]\, ,
\end{align}
where $(k,m)=[(0,1),(0,2),$  $(0,3),(1,2),(2,3),(4,5),(4,6), (4,7),(5,6),(6,7)]$.
\renewcommand{\arraystretch}{0.95}
\begin{table}[H]
$$
\begin{array}{|l|l|ll|}
\hline \Pi_{i} & \text { Fixed point equation } & & \text { O6-plane position }  \\
\hline \Pi_{0} & \sigma\left(z^{a}\right)=z^{a} & y^{1} \in\left\{0, \frac{1}{2}\right\} & y^{2} \in\left\{0, \frac{1}{2}\right\} \quad y^{3} \in\left\{0, \frac{1}{2}\right\} \\
\Pi_{1} & \sigma\left(z^{a}\right)=\theta\left(z^{a}\right) & x^{1}+y^1=1 & x^2-y^2=0 \quad y^3\in \{0,\frac{1}{2}\} \\
\Pi_{2} & \sigma\left(z^{a}\right)=\theta^{2}\left(z^{a}\right) & x^1\in\{0,\frac{1}{2}\} & x^2\in\{0,\frac{1}{2}\} \quad y^3\in\{0,\frac{1}{2}\} \\
\Pi_{3} & \sigma\left(z^{a}\right)=\theta^{3}\left(z^{a}\right) & y^1-x^1=0 & x^{2}+y^2=1\quad y^{3}\in\{0,\frac{1}{2}\} \\
\Pi_{4} & \sigma\left(z^{a}\right)=\omega\left(z^{a}\right) & y^1\in\{0,\frac{1}{2}\}  & x^2\in\{0,\frac{1}{2}\} \quad x^3\in\{0,\frac{1}{2}\} \\
\Pi_{5} & \sigma\left(z^{a}\right)=\omega\theta\left(z^{a}\right) & x^1+y^1=1 & x^2+y^2=1 \quad x^3\in\{0,\frac{1}{2}\} \\
\Pi_{6} & \sigma\left(z^{a}\right)=\omega\theta^2\left(z^{a}\right) & x^{1}\in\{0,\frac{1}{2}\}  & y^2\in\{0,\frac{1}{2}\} \quad x^3\in\{0,\frac{1}{2}\} \\
\Pi_{7} & \sigma\left(z^{a}\right)=\omega\theta^3\left(z^{a}\right) & x^1-y^1=0 & x^2-y^2=0 \quad x^3\in\{0,\frac{1}{2}\} \\
\hline
\end{array}
$$
\caption{O6-planes in $T^6/\mathbb{Z}_2\times \mathbb{Z}_4$.}
\label{mem-table: z2xz4 orientifolds}
\end{table}
\section{Current Status}
\label{mem-sec: current status}

The puzzling question regarding the non-perturbative stability of non supersymmetric AdS vacua was addressed and partially solved in \cite{Marchesano:2022rpr}. There, the authors successfully provided the 10d uplift of the last pair of $S1$ branches  detailed in the fifth row of table \ref{cy-table: summary ads vacua} beyond the smearing approximation, building on top of the results of \cite{Marchesano:2021ycx} and \cite{Casas:2022mnz} described in the present and previous chapters. For the sake of completeness we will briefly sketch their results in this section. 

The internal flux profiles that solve the 10d equations of motion \eqref{cy-eq: solutionflux} and \eqref{bio-solutionfluxnnosusy} can be merged with the expressions corresponding  to the last pair of branches as
\begin{align}
    H&=6\tilde{A}G_0g_s(\Re\Omega_{\rm CY}+Rg_s K)-\frac{S}{2}d \re(\bar{v}\cdot\Omega_{\rm CY})+\mathcal{O}(g_s^3)\,,\\
    G_2&=\tilde{B}G_0J_{\rm CY}-J_{\rm CY}\cdot d(4\varphi \im\Omega_{\rm CY}-\star_{\rm CY} K)+\mathcal{O}(g_s)\,,\\
    G_4&=G_0J_{\rm CY}\wedge J_{\rm CY}(\tilde{C}-12\tilde{A}g_s\varphi)+SJ_{\rm CY}\wedge g_s^{-1}d\im v+\mathcal{O}(g_s^2)\,,\\
    G_6&=0\,,
\end{align}
where $\tilde{A},\tilde{B},\tilde{C}$ are the parameters of table \ref{cy-table: summary ads vacua} and $R,S\in \mathbb{R}$ are the coefficients that distinguish between the uplifts of the different branches. Their values are summarized in table \ref{mem-table: uplift coefficients}. 

\begin{table}[htbp]
\centering
\def\arraystretch{1.5}
\begin{tabular}{|c|c|c|c|c|c|c|c|c|}
\hline
Branch                               & $\tilde{A}$            & $\tilde{B}$              & $\tilde{C}$             & $R$  & $S$            & $\mu$                       & SUSY & pert. stable \\ \hline
\textbf{A1-S1}$+$   & $\frac{1}{15}$ & $0$              & $\frac{3}{10}$  & $1$  & $1$            & $\frac{1}{5}G_0g_s$         & Yes  & Yes          \\ \hline
\textbf{A1-S1}$-$   & $\frac{1}{15}$ & $0$              & $-\frac{3}{10}$ & $-2$ & $-\frac{1}{5}$ & $\frac{1}{5}G_0g_s$         & No   & Yes          \\ \hline
\textbf{A2-S1}$\pm$ & $\frac{1}{12}$ & $\pm\frac{1}{2}$ & $-\frac{1}{4}$  & $-1$ & $0$            & $\frac{1}{\sqrt{24}}G_0g_s$ & No   & Yes          \\ \hline
\end{tabular}
\caption{Branches of \textbf{S1} solutions from table \ref{cy-table: summary ads vacua} beyond the smearing approximation. Extracted from \cite{Marchesano:2022rpr}.}
\label{mem-table: uplift coefficients}
\end{table}


In order to study the non-perturbative stability, in \cite{Marchesano:2022rpr} the authors consider more exotic brane configurations where the D8 and D6 branes are allowed to have a non-primitive worldvolume flux contribution along the internal directions (we assumed them to be zero). These are considered exotic states because in the large volume regime, they carry a large lower-dimensional D-brane charge.  They work with the following options
\begin{subequations}
 \begin{align}
    \textrm{(anti-)D6-brane on $\mathcal{S}$ with} & & \mathcal{F}^2&=J^2_{\rm CY}|_{\mathcal{S}}\,, \\
     \textrm{D8-brane on $X_6$ with} & & \mathcal{F}^2\wedge J_{\rm CY}&=cJ^3_{\rm CY}\,, c\leq 0\,, \quad {\rm and} \quad 3\mathcal{F}\wedge J_{\rm CY}^2=\mathcal{F}^3\,,\\
     \textrm{anti-D8-brane on $X_6$ with} & & \mathcal{F}^2&=3J^2_{\rm CY}\,.
\end{align}   
\end{subequations}
It is important to note that these states might not exist depending on the values of the complexified Kähler moduli, as there are several non trivial relations and quantization constraints that need to be satisfied. 

The charge of such kind of brane configurations wrapping a 2p-cycle is given by
\begin{eqnarray}
    Q=\frac{\eta e^{K/2}}{\ell_s^{2p}}\int_{2p}e^{-\mathcal{F}}\wedge \mathbf{Q}\,, \qquad  \mathbf{Q}=\sum_p \frac{q_p}{p!}J_{\rm CY}^p\,,
\end{eqnarray}
with the coefficient $\eta={\rm sign} (m)$ and the coefficients $q_p$ corresponding to charge to tension ratios $Q_{D(2p+2)}/T_{D(2p+2)}$, which depend on the values of $\tilde{B}$ and $\tilde{C}$ from the previous table.

As expected, the contribution of the internal worldvolume flux in the SUSY branch is the same for the charge and the tension and thus stability is preserved. For the \textbf{A1-S1}$-$ branch discussed in the last two chapters, the relevant states are D8-branes with worldvolume flux of the form
\begin{equation}
    \mathcal{F}=\pm \sqrt{3}J_{\rm CY}+\mathcal{F}_p\,,
\end{equation}
where we note that in our analysis only the term $\mathcal{F}_p$ was present (the BIonic contribution). The additional term can be understood as a bound state of a D8-brane and $N\sim 9T_{D8}/T_{D4}$ anti-D4-branes. Then, choosing $m>0$, they find out
\begin{equation}
    Q-T=2(1-||\tilde{\delta}||^4_0+\dots)T_{D8}\,, 
\end{equation}
where $||\tilde{\delta}||_0$ encodes the BIonic correction
\begin{equation}
    ||\tilde{\delta}||=\frac{1}{2}\sqrt{\mathcal{F}_{p,ab}\mathcal{F}_p^{ab}}\sim \mathcal{O}\left(\frac{M}{V_{CY}^{1/3}}\right)\,, \qquad ||\tilde{\delta}||^n_0\equiv \int_{X_6}||\tilde{\delta}||^n/\mathcal{V}_{\rm CY}\,.
\end{equation}
Therefore, the result of considering exotic branes is a new contribution to the charge of the D8 that scales proportionally to $T_{D8}$ instead of the $T_{D4}$ correction given by the BIonic construction. In the large volume limit, where our approximation is well defined, the term with $T_{D8}$ will always dominate. For the Non-SUSY branch \textbf{A1-S1}  the new contribution to the charge in the exotic brane is positive and thus gives an overall relation $Q>T$ which makes the configuration unstable and provides a non-perturbative decay channel as predicted by the WGC. 

A similar analysis is performed for the third branch (\textbf{A2-S1}). Despite the greater complexity of the process due to the explicit dependence of the worldvolume flux on the Kähler moduli, the authors obtain non-perturbative decay channels in this branch as well. The current state of non-perturbative stability of the AdS vacua introduced in \ref{cy-table: summary ads vacua} is condensed in table \ref{mem-table: current status}.

\begin{table}[htbp]
\centering
\def\arraystretch{1.5}
\begin{tabular}{|c|c|c|c|c|c|}
\hline
Branch     & SUSY & pert. stable & sWGC D4  & sWGC D8 & non-pert. stable      \\ \hline
\textbf{A1-S1}$+$   & Yes  & Yes          & Yes      & Yes     & Yes                   \\ \hline
\textbf{A1-S1}$-$   & No   & Yes          & Marginal & Yes     & Unclear if $N_{D6}=0$ \\ \hline
\textbf{A2-S1}$\pm$ & No   & Yes          & Yes      & Yes     & No                    \\ \hline
\end{tabular}
\caption{Stability of the uplifted versions of the branches of AdS of table \ref{cy-table: summary ads vacua} and their relation with the sharpened WGC.}
\label{mem-table: current status}
\end{table}

\section{Summary}
\label{mem-s:conclu}

In this chapter we have analyzed type IIA AdS$_4$ flux vacua with O6-planes and D6-branes. These vacua can be either $\cN=1$ and $\cN=0$, and the latter can be subject to non-perturbative instabilities via membrane nucleation, in line with the AdS Instability Conjecture \cite{Ooguri:2016pdq,Freivogel:2016qwc}. We have analyzed those instabilities that correspond to 4-dimensional membranes made up from D8-branes wrapping the compact manifold $X_6$, building on the results from the previous chapter. As pointed out therein, one should be able to determine whether $Q > T$ or not for this class of membranes with our current, approximate description of a family of $\cN=0$ vacua that are closely related to supersymmetric ones. Now we have expanded on this observation by analysing such D8-brane charge and tension in several orientifold backgrounds with different space-time filling D6-brane configurations. We have considered D6-branes that lie on top of O6-planes, which always solve the vacua conditions. 

As pointed out in chapter \ref{ch: bionic} in settings where the worldvolume flux has a trivial non-primitive term $Q_{\rm D8} = T_{\rm D8}$ at leading order, and then there are three corrections that can tip the scales to one side or the other, represented in \eqref{mem-QTtotalnosusy}. Out of these three corrections two of them are unavoidable, namely the curvature correction $\Delta_{\rm D8}^{\rm curv} = K_a^{(2)} T_{\rm D4}^a$ and the BIon correction $\Delta_{\rm D8}^{\rm Bion} = - T_{\rm D8}^{\rm BIon}$. It turns out that $K_a^{(2)} T_{\rm D4}^a$ always favours $Q_{\rm D8}^{\rm total} > T_{\rm D8}^{\rm total}$, while $\Delta_{\rm D8}^{\rm Bion}$ can have both signs and it is sensitive to the D6-brane configuration. Therefore requiring that $Q_{\rm D8}^{\rm total} > T_{\rm D8}^{\rm total}$ in $\CN=0$ vacua, as the refined WGC for membranes does, translates into the non-trivial constraint $\Delta_{\rm D8}^{\rm curv} + \Delta_{\rm D8}^{\rm Bion} > 0$ for any D6-brane configuration. We have computed $\Delta_{\rm D8}^{\rm Bion}$ in toroidal orbifold geometries, finding that the simple expression \eqref{mem-finalDelta} that indeed shows that this correction can be either positive or negative. A negative value is favoured when we have pairs of D6-branes that do not intersect in the internal dimensions, so that open strings stretched between them lead to a spectrum with masses above the compactification scale. By choosing the D6-brane positions one can build configurations where $\Delta_{\rm D8}^{\rm Bion} < 0$. In this way, we have been able to engineer vacua where $\Delta_{\rm D8}^{\rm curv} + \Delta_{\rm D8}^{\rm Bion} < 0$,  therefore naively violating the WGC inequality for 4-dimensional membranes. They are however not necessarily in tension with the AdS Instability Conjecture, since there could be other channels, in particular D4-brane nucleation, that could mediate a non-perturbative decay to an $\cN=0$ vacuum of lower energy. 

We have pointed out some caveats that could reconcile our results with our expectations from the WGC for 4-dimensional membranes. From these, the most promising ones are considering more exotic bound states involving D8-branes, or one-loop threshold corrections to the vacuum energy, which just like $\Delta_{\rm D8}^{\rm Bion}$ depend on the D6-brane positions, and could decrease the vacuum energy such that the controversial decay channels are no longer energetically favoured. 

In particular, the results developed in this chapter were used as a stepping stone in \cite{Marchesano:2022rpr}. There, they considered bound states of D8-branes with non-diluted worldvolume fluxes that have non-primitive $(1,1)$ components. Through the addition of this new ingredient, a new term to the D8 charge is found. For the non-supersymmetric vacuum branch considered in this chapter, the new contribution,  parametrically larger than the BIon contribution, favors the instability and thus provides a new decay channel in support of the predictions stated by the AdS instability conjecture and the WGC.

In any case, taken at face value, our results suggest that $\cN=0$ AdS$_4$ vacua with a gauge sector without zero/light modes charged under it are more stable than those that contain charged light modes. Showing whether or not this is true is an interesting challenge, as well as to unveil the would-be implications for our understanding of the string Landscape.


\ifSubfilesClassLoaded{%
\bibliography{biblio}%
}{}

\end{document}


\part[\textcolor{Teja}{F-theory Compactifications}]{\scshape \textcolor{Teja}{\huge F-theory Compactifications}}
\label{part: F and B}


\ifSubfilesClassLoaded{%
\tableofcontents
}{}

\setcounter{chapter}{6}
\chapter{Type IIB and F-theory overview}
\label{ch: Fintro}

In this chapter and the following ones, we  move on from Type IIA  and consider moduli stabilization in different String Theory setups, namely Type IIB and F-theory. The tools, techniques and general intuition developed up to this point will prove to be very useful when considering these new corners of the space of theories described in \ref{basic-fig: duality web}. 

We  start by providing a general overview of Type IIB flux compactifications, emphasizing the common and distinct aspects with respect to Type IIA, as well as describing the powerful relation between both: Mirror Symmetry. Then we explain how Type IIB can be understood as a particular limit of a 12-dimensional theory, the so-called F-theory \cite{Vafa:1996xn}, and provide a schematic review of the main properties of the latter that will become relevant when considering moduli stabilization. For an in-depth analysis of the topic, we refer the reader to the reviews \cite{Denef:2008wq, Weigand:2010wm, Weigand:2018rez,Wiesner:2021pgd}.

\section{Type IIB Compactifications}
\label{fb-sec: typeIIB compactafications}

Type IIB theory was introduced alongside Type IIA in \ref{basic-subsec: Type II theories}. There, we saw that both are theories of closed superstrings whose massless content includes a graviton, a dilaton, an NSNS 2-form field, and several RR p-forms. They also allow for the presence of D-branes that, in turn, provide the structure required to define open strings. Despite these similarities, they have notably distinct properties and behaviours derived from their different field content and chiral nature.

\subsection{Field content and Moduli Space}

Most of the discussion of chapter \ref{ch: calabi-yau} regarding the geometrical properties of compactifications was formulated in terms of 4-dimensional phenomenological requirements. Therefore, it applies to Type IIB as well.  In particular, we will keep working with Calabi-Yau 3-fold compactifications. Adapting to Type IIB amounts to changing the field content and the orientifold projection.

\subsubsection*{Orientifold Projection}

 In this part of the thesis we will focus on the O3/O7 orientifold and so we have the following action
\begin{equation}
        \mathcal{O}=\Omega_p\mathcal{R}\,,
\end{equation}
where $\Omega_p$ is the worldsheet parity operator and $\mathcal{R}$ is a holomorphic involution that satisfies
\begin{equation}
    \mathcal{R}J=J\,,\qquad \mathcal{R}\Omega= -\Omega\,.
\end{equation}
Based on the above definition of the orientifold, we can split the basis of the cohomology groups of the Calabi-Yau into even and odd parts, similarly to the decomposition performed for Type IIA in table \ref{cy-table: harmonic basis}. This time we end up with table \ref{fb-table: harmonic basis type IIB}. We note that the $H^3$ sector splits asymmetrically, since now the action of the orientifold over $\Omega$ does not mix the Dobeault cohomology groups $h^{3,0}$ and $h^{0,3}$.
\begin{table}[H]
\def\arraystretch{1.5}
\begin{center}
\begin{tabular}{|c|| c| c| c| c| c| c|} 
\hline
Cohomology group & $H^{1,1}_+$ & $H^{1,1}_-$ & $H^{2,2}_+$ & $H^{2,2}_-$ & $H^{3}_+$ & $H^{3}_-$ \\
\hline\hline
 Dimension & $h^{1,1}_+$ & $h^{1,1}_-$ & $h^{1,1}_-$ & $h^{1,1}_+$ & $2h^{2,1}_+$ & $2h^{2,1}_- + 2$ \\
 Basis & $\varpi_\alpha$ & $\omega_a$ & $\tilde\omega^a$ & $\tilde\varpi^\alpha$ & $(\alpha_{\hat{I}},\beta^{\hat{I}})$  & $(\alpha_I,\beta^I)$  \\
 \hline
\end{tabular}
\end{center}
\caption{Representation of various harmonic forms in Type IIB orientifolds and their counting.  }
\label{fb-table: harmonic basis type IIB}
\end{table}
\noindent where the real symplectic basis $(\alpha_I,\beta^I)$ again satisfies
\begin{equation}
    \int \alpha_I\wedge \beta^J=\delta_I^J\,.
\end{equation}
The massless field content of Type IIB was introduced in \ref{basic-table: type IIB spectrum}. They transform under the orientifold involution as follows
\begin{equation}
    \mathcal{R}\phi=\phi\,,\quad \mathcal{R}g=g\,,\quad \mathcal{R}B=-B\,,\quad \mathcal{R}C_0=C_0\,,\quad \mathcal{R}C_2=-C_2\,, \quad \mathcal{R}C_4=C_4\,.
\end{equation}
These fields will need to decompose under a metric factorization of the form \eqref{cy-eq: spacetime factorization}. Thus applying the same arguments of 4-dimensional Poincaré invariance and orientifold truncation, we end up with 
 \cite{Grimm:2005fa} 
\begin{gather}
    J=v^\alpha \varpi_ \alpha\,,\qquad \alpha\in \{1,\dots,h^{1,1}_+\}\,,\\ 
    \Omega=Z^I\alpha_I-\mathcal{F}_I\beta^I\,, \qquad I\in\{0,1,\dots, h^{2,1}_-\}\,.
    \label{fb-eq: holomorphic form expansion}
\end{gather}
A similar reasoning to the one described below \eqref{cy-eq: Omega expansion Z and F} can be employed, enabling to define the coordinates of the complex structure moduli space $z^i\equiv Z^i/Z^0$ with $i\in \{1,\dots,h^{2,1}_-\}$ and a prepotential $\mathcal{F}$ satisfying $\mathcal{F}_I=\partial\cF /\partial Z^I$. The other fields decompose as
\begin{equation}
    \begin{gathered}
        B=b^a\omega_a\,, \qquad  C_2=c^a\omega_a\,, \qquad a\in\{1,\dots h^{1,1}_-\}\,,\\
        C_4=D_2^\alpha\varpi_\alpha V^{\hat{I}}\wedge \alpha_{\hat{I}}+U_{\hat{I}}\wedge\beta^{\hat{I}}+\rho_\alpha \tilde{\varpi}^\alpha\,.
\end{gathered}
\end{equation}
And so we obtain three sets of 4-dimensional scalar fields $c^a,b^a,\rho_\alpha$, two sets of 4-dimensional 1-forms $V^{\hat{I}}$ and $U_{\hat{I}}$ and a set of 4-dimensional 2-forms $D_2^\alpha$. They can be grouped into different 4-dimensional $\mathcal{N}=1$ multiplets, as summarized in table \ref{fb-table: N=1 field content type IIB}

\begin{table}[htbp]
\def\arraystretch{1.5}
\center
\begin{tabular}{|c|c|c|}
\hline
Multiplet & Bosonic Field Content & Multiplicity \\ \hline
Gravity   & $g_{\mu\nu}$          & 1            \\
Vector    & $V^{\hat{I}}$            & $h_+^{2,1}$  \\
Chiral    & $b^a,c^a$             & $h_-^{1,1}$  \\
Chiral    & $z^i$      & $h^{2,1}_-$  \\ 
Chiral    & $(v^\alpha,\rho_\alpha)$      & $h^{1,1}_+$  \\ 
Chiral    & $(\phi,C_0)$      & $h^{2,1}_-$  \\ \hline
\end{tabular}
\caption{Bosonic content of the 4-dimensional $\mathcal{N}=1$ supergravity resulting from the compactification of Type IIB on a Calabi-Yau $O3/O7$ orientifold.}
\label{fb-table: N=1 field content type IIB}
\end{table}

\subsubsection*{Kähler structure}

The content of massless scalars (moduli) can be grouped into a Kähler and a complex structure sectors, just as in type IIA. This can be achieved by defining the following complexified quantities \cite{Grimm:2005fa}
\begin{equation}
\begin{gathered}
    \tau=C_0+ie^{-\phi}\,,\qquad G^a=c^a-\tau b^a\,,\\
    \mathcal{T}_\alpha=i(\rho_\alpha-\frac{1}{2}\mathcal{K}_{\alpha ab}c^ab^b)+\frac{1}{2}e^{-\phi}\cK_{\alpha}-\zeta_\alpha\,,
\end{gathered}
\end{equation}
where $\cK_{ABC}$ are the intersection numbers obtained from the triple intersection of elements of the basis $H^2$ in table \ref{fb-table: harmonic basis type IIB} and
\begin{equation}
    \cK_{\alpha}=\cK_{\alpha\beta\gamma}v^\beta v^\gamma\,,\qquad \zeta_\alpha=-\frac{i}{2(\tau-\bar{\tau})}\cK_{\alpha bc}G^b(G^c-\bar{G}^c)\,.
\end{equation}
The set $(\tau,z^a)$ constitutes the complex structure sector while $(G^a,\mathcal{T}_\alpha)$ are the components of the Kähler sector. As it was the case in Type IIA, both moduli space sectors factorized and each is endowed with a Kähler structure of its own. The full Kähler potential describing the structure of the moduli space of the 4-dimensional effective theory is  
\begin{equation}
    \kappa_4^2 K=-\log\left[i\int\Omega(z)\wedge \bar{\Omega}(\bar{z})\right]-\log[2e^{-4\phi_4}]\,,
\end{equation}
where $\phi_4$ is the 4-dimensional dilaton and is a function of $(\phi, G^a,\mathcal{T}_\alpha)$. In the large volume limit with $\mathcal{T}_\alpha \sim \mathcal{T}\gg 1$ this dependence can be written more explicitly as
\begin{equation}
    \kappa_4^2 K=-\log\left[i\int\Omega(z)\wedge \bar{\Omega}(\bar{z})\right]-\log(-i(\tau-\bar{\tau}))-3\log(-i(\mathcal{T}-\bar{\mathcal{T}}))\,,
\end{equation}
The last term of the Kähler potential ($K_K$) has an essential property: it satisfies the no-scale condition \cite{Grimm:2004uq, Ellis:1983sf}
\begin{equation}
    \partial_A K_K \partial_{\bar{B}} K_K K^{A\bar{B}}_K=3\,,
\end{equation}
for $A$ labelling $(G^a,\mathcal{T}_\alpha)$. It will become important when considering flux compactifications, since it greatly simplifies the leading term of the induced scalar potential, at the expense of leaving the Kähler moduli as flat directions.

\subsubsection*{Background fluxes}

We follow the standard practice in the literature and consider Type IIB compactified on a Calabi-Yau orientifold with $H$ and $F_3$ background fluxes \cite{Michelson:1996pn, Taylor:1999ii, Giddings:2001yu}. First, we consider the consistency conditions for the flux field strengths. We assume that there are no sources for the fields we have turned on and so they satisfy the Bianchi identities
\begin{equation}
    dH=0\,,\qquad dF_3=0\,.
\end{equation}
On the other hand, the Bianchi identity for the 5-form flux is
\begin{equation}
    d\title{F}_5=d\star \tilde{F}_5=H_3\wedge F_3 +2\kappa_{10}^2\mu_3\rho_3^{\rm local}\,,
\end{equation}
where $\rho_{3}^{\rm local}$ is the localized source contribution coming from D3-branes and O3-planes and $\mu_3$ was defined below \eqref{basic-eq: DBI action}. 

The effect of the non-trivial flux background can be described in terms of a superpotential\footnote{Contrary to what we observed in Type IIA with geometric flux compactifications, the current choice of fluxes does not induce D-term contributions for the 4-dimensional vector fields $V^{\hat{I}}$.} \cite{Gukov:1999ya}
\begin{equation}
    W\equiv\int G_3\wedge \Omega\,,\qquad G_3\equiv F_3-\tau H_3\,.
\end{equation}
Note that the 5-form flux $\tilde{F}_5$ defined in \eqref{basic-eq: Type IIB form flux definitions} is not affected by this background fluxes, since the terms $H\wedge C_2$ and $B\wedge F_3$ are projected out by the orientifold.

With the superpotential and Kähler potential, we can derive the scalar potential for the moduli of type IIB using the standard supergravity formula \eqref{cy-eq: VFgen}. It is then possible to show that the no-scale condition described above induces a cancellation between the terms associated with the Kähler sector, which yields the simple expression
\begin{equation}
    V=e^K\left(K^{I\bar{J}} D_I WD_{\bar {J}} W\right)\,.
\end{equation}
Since $K^{I\bar{J}}$ is the inverse Kähler metric, this scalar potential is positive definite. Vacua are located at moduli space points satisfying $D_I W=0$ and thus correspond to Minkwoski vacua. If in addition we demand $D_{\mathcal{T}_\alpha}W=0$ (so the covariant derivatives of the Kähler sector also vanish) we obtain supersymmetric vacua.

The above structure does not allow to fix the Kähler moduli. To solve this problem, one would need to include more refined terms to the supergravity approximation, like perturbative and non-perturbative $\alpha'$-corrections which would spoil the non-scale structure.

\subsection{Mirror Symmetry}

In section \ref{basic-sec: non perturbative and dualities}, we discussed how Type IIA and Type IIB theories are related through T-dualities in 9 dimensions. More specifically, both theories compactified over $S^1$ are dual upon inversion of the compactification radius $R\rightarrow \alpha'/R$ and  exchange of the winding and Kaluza-Klein modes. Now we are considering a far more elaborated compactification, involving a curved space of 6 dimensions instead of a 1-dimensional circle, but one can wonder whether a similar property is still present. The answer is affirmative and is given by Mirror Symmetry. In this section we will give a short review of the concept and refer the reader to reference \cite{Hori:2003ic} for an exhaustive take on the subject.

Mirror symmetry informs us that for each Calabi-Yau manifold $X_6$ there exists another Calabi-Yau manifold\footnote{There can be some pathological cases in which the mirror is not a Calabi-Yau. Manifolds with $h^{2,1}=0$ are such examples, since they would be mapped to manifolds with $h^{1,1}=0$, that cannot be Calabi-Yau.} $Y_6$, named mirror manifold, such that the compactification of type IIA string theory on $X_6$ is equivalent to the compactification of type IIB string theory on $Y_6$. They satisfy the following relation
\begin{equation}
    H^{p,q}(X_6)=H^{3-p,q}(Y_6)\,,
\end{equation}
which in particular implies $h^{1,1}(X_6)=h^{2,1}(Y_6)$ and so the Kähler and complex structure degrees of freedom are exchanged. The complex structure moduli space of $X_6$ is identified with the Kähler moduli space of $Y_6$  and vice versa. The same map applies to their respective prepotentials and to each independent supermultiplet. From the language of pure spinors and $SU(3)\times SU(3)$ structures it can be understood as the map
\begin{equation}
    \begin{aligned}
    \Phi_+ & & \leftrightarrow &  & \Phi_-\\
    e^{B+iJ} & & \leftrightarrow &  &  \Omega
\end{aligned}
\end{equation}
where in the second line we particularized to the case of $SU(3)$ manifolds (see appendix \ref{ch: ap complex geometry} for more details).

On a more practical level, we can use mirror symmetry to identify the complexified Kähler moduli $T^a=b^a+it^a$ with $a\in\{1,\dots,h^{1,1}(X_6)\}$ of type IIA compactified on $X_6$ with the periods $Z^I$ with $I=0,a$ that describe the complex structure moduli space in the mirror Type IIB on $Y_6$ (see \eqref{fb-eq: holomorphic form expansion}). As it happens with most useful dualities, quantities that are easy to compute in one limit are challenging to tackle on the other, providing a great insight on both fronts. For example, the Kähler sector of type IIA on $X_6$ receives $\alpha'$ corrections, whereas the  complex structure sector of type IIB on $Y_6$ is exactly computable from classical geometry in supergravity.

An even more powerful method to phrase and exploit this property of String Theory compactifications is the so-called homological mirror symmetry  \cite{Kontsevich:1994dn} which propagates the symmetry to maps between a special class of branes, known as A and B branes. These branes are defined in terms of calibrations. A-branes are calibrated by the holomorphic form $\Omega$ while B-branes are calibrated by $(-1)^{k/2}/k! J^{k}$ for $k=1,2,3$. Since mirror symmetry exchanges pure forms, it will also map A-branes to B-branes and vice versa. Using such property, the central charge of a Type IIB A-brane $L$ wrapped on a special Lagrangian cycle $\sigma\subset Y_6$, given by the periods of $Y_6$
\begin{equation}
    Z(L)=\int_\Sigma \Omega\,,
\end{equation}
can be related to the central charge of B-branes in type IIA. The latter can be described by a complex element $\mathcal{E}\in D^b(Y_6)$ in the bounded derived category of coherent sheaves  on $X_6$ \cite{Mayr:1996sh, Douglas:2000gi, Cota:2019cjx} (for information on sheaves we refer the reader to \cite{Hori:2003ic, huybrechts2005complex}) and its central charge is \cite{iritani2009integral}
\begin{equation}
    Z(\mathcal{E})=\int_{X_6} e^{J_c}\Gamma_{\mathbb{C}}(X_6)\lambda({\rm ch}(\mathcal{E}))\,,
    \label{fb-eq: central charges Klemm formula Type IIB}
\end{equation}
where $\Gamma_{\mathbb{C}}(X_6)$ is the $\Gamma$-class of $X_6$ and has the expansion
\begin{equation}
    \Gamma_{\mathbb{C}}(X_6)=1+\frac{1}{24}c_2(X_6)+\frac{\zeta(3)}{(2\pi i)^3}c_3(X_6)\,,
\end{equation}
with $c_i(X_6)$ the i-th Chern class of $X_6$, ${\rm ch}(\mathcal{E})$ the Chern character of $\mathcal{E}$ and $\lambda$ the operator that reverses the indices of a form  (so for $\beta\in H^{p,p}(X_6)$, $\lambda(\beta)=(-1)^p\beta$).

Thus, evaluating \eqref{fb-eq: central charges Klemm formula Type IIB} along D6-branes of type IIA on $X_6$, one can obtain the polynomial corrections to the prepotential of Type IIB compactified on $Y_6$.

\subsection{Moduli Stabilization ingredients}
\label{fb-subsec: Type IIB ingredients}

\subsubsection*{The prepotential}

Now that we have a solid grasp of the similarities, differences and relations between Type IIA and Type IIB compactifications, let us study in more detail the different quantities that play a role in moduli stabilization. Thus, let us consider a symplectic basis $\{A^I,B_I\}$, $I=0,\dots,h^{2,1}_-$ of $H_3(Y_6,\Z)$ dual to $(\beta_I, \alpha^I)$ of table \ref{fb-table: harmonic basis type IIB}. With such basis, the periods of the Calabi--Yau $(3,0)$-form $\Omega$ are encoded in the vector
\begin{equation}
\Pi^t\equiv(\F_I,Z^I)=\left(\int_{B_I}\Omega,\int_{A^I}\Omega\right)\ ,
\end{equation}
where $t$ stands for the transpose. The complex structure moduli fields are defined to be $z^i\equiv Z^i/Z^0$, $i=1,\dots,h^{2,1}$ and the $\F_I$ components are expressed as derivatives of the \emph{prepotential} $\F$. Setting the gauge $Z^0=1$, the period vector takes the following form:
\begin{equation}
\label{IIB-eq:period}
    \Pi = 
    \begin{pmatrix}
     2\mathcal{F} - z^i \partial_i \F \\
    \partial_i \F \\
     1 \\
    z^i 
    \end{pmatrix}.
\end{equation}

In the LCS regime, the prepotential reads
\begin{equation}
\F=-\frac{1}{6}\kappa_{ijk}z^iz^jz^k-\oh a_{ij}z^iz^j+\hat{a}_iz^i+\oh\kappa_0+\F_{\rm inst}\ .
\label{IIB-eq:full_prepotential}
\end{equation}
The instanton contribution $\F_{\rm inst}$ is subleading in the LCS regime and can be expressed as sum of polylogarithm $\text{Li}_p(q)\equiv \sum_{k>0}\frac{q^k}{k^p}$ ponderated by Gopakumar-Vafa invariants $n_{\vec d}$ labeled by $\vec d\in(\mathbb{Z}^+)^{h^{2,1}}$ \cite{Cicoli:2013cha},
\begin{equation}
\F_{\rm inst}=-\frac{i}{(2\pi)^3}\sum_{\vec d}n_{\vec d}\, \text{Li}_3[e^{-2\pi d_iz^i}]\ .
\end{equation}
The coefficients $\kappa_{ijk}$, $a_{ij}$ and $\hat{a}_i$ can be computed from topological data of the mirror manifold $X_6$ of the Calabi--Yau $Y_6$, while $\kappa_0$ depends on the Euler characteristic of $X_6$. More precisely, we have \cite{Mayr:2000as}
\begin{align}
\begin{split}
&\kappa_{ijk}\equiv\int_{X_6}\omega_i\wedge\omega_j\wedge\omega_k\ ,\qquad a_{ij}\equiv-\oh\int_{X_6}\omega_i\wedge i_*\text{c}_1(\text{P.D}[w_j])\ ,\\
&\hat{a}_i\equiv\frac{1}{24}\int_{X_6}\omega_i\wedge \text{c}_2(X_6)\ ,\qquad \kappa_0\equiv\frac{\zeta(3)\chi(X_6)}{(2\pi i)^3}\ = i \, \frac{\zeta (3)}{4\pi^3} (h^{1,1} - h^{2,1})\ ,
\end{split}
\label{IIB-eq: mirror quantities def}
\end{align}
where $\omega_i$, $i=1,\dots,h^{1,1}(X_6)$ form a basis of $H^2(X_6,\Z)$, $i_*$ denotes the pushforward of the embedding $i$ of the divisors into $X_6$, P.D stands for Poincaré Dual and $\text{c}_1$ and $\text{c}_2$ denote the first and second Chern classes respectively. It can further be shown \cite{Cicoli:2013cha} that $a_{ij}$ can be rewritten in terms of the triple intersection numbers as follows
\begin{equation}
    a_{ij} = -\frac{1}{2} \int_{X_6} \omega_i \wedge \omega_j \wedge \omega_j  \mod \mathbb{Z}\ .
\end{equation}
Finally, it is important to note that both $\hat{a}_i$ and $a_{ij}$ are defined only modulo $\mathbb{Z}$, since shifts on these parameters correspond to different choices for the symplectic basis of 3-cycles of $X_6$. This leads to significant restrictions on their values when considering the transformation properties of the period vector under monodromies  $z^i\to z^i+v^i$, $v^i\in\Z$ at LCS. More concretely, the coefficients of the prepotential must satisfy the following conditions \cite{Mayr:2000as}: 
\begin{align}
    a_{ij} + \frac{1}{2} \kappa_{ijj} \in \mathbb{Z}\quad\text{ and }\quad
    2\hat{a}_i + \frac{1}{6} \kappa_{iii} \in \mathbb{Z}\label{IIB-consistency}\ .
\end{align}
The first equation can also be generalized to take the form
\begin{equation}
\label{IIB-freedom}
a_{ij}v^j+\oh\kappa_{ijk}v^jv^k= 0 \!\!\!\mod \mathbb{Z}\ .
\end{equation}
Note that we can make use of the redundancy of $a_{ij}$ to shift its value like $a_{ij} \to a_{ij} + n_{ij}$, $n_{ij} \in \mathbb{Z}$ so that the LHS of \eqref{IIB-freedom} is actually 0.

\subsubsection*{Kähler potential}

The tree-level Kähler potential is given by 
\begin{equation}
K= K_{\text{k}} + K_{\text{dil}} + K_{\text{cs}} = -2\log(\V)-\log(-i(\tau-\bar\tau))-\log(-i\Pi^\dagger\cdot\Sigma\cdot\Pi)\ ,
\end{equation}
where $\V$ is the volume of $X_6$, $\tau$ is the axio-dilaton and we have defined the canonical symplectic $(2h^{2,1}+2)\times(2h^{2,1}+2)$ matrix
\begin{equation}
\Sigma\equiv\begin{pmatrix}
0 & \mathds{1}\\ -\mathds{1} & 0
\end{pmatrix}.
\end{equation}
The Kähler potential at the approximation of large complex structure can be shown to read
\begin{align}
    K_{\text{cs}} &= - \log \left( \frac{i}{6} \kappa_{ijk} (z^i-\bar{z}^i) (z^j-\bar{z}^j) (z^k-\bar{z}^k) - 2 \, \Im (\kappa_0) \right) \nonumber \\
    &= - \log \left( \frac{4}{3} \kappa_{ijk} t^i t^j t^k - 2 \, \Im (\kappa_0) \right)\ ,
\label{IIB-eq: complex Kahler potential}
\end{align}
where we have defined $z^i \equiv b^i + i t^i$ and, for later use, we also introduce $\tau \equiv b^0 + i t^0$. 

It will be important to develop some of the derivatives of the Kähler potential, for future reference. The most relevant ones are the following:
\begin{align}
    K_{\tau} &= - \frac{1}{\tau - \bar{\tau}} = \frac{i}{2 t^0}\ , \label{IIB-eq:Kt}\\
    K_{\tau \bar{\tau}} &= - \frac{1}{(\tau - \bar{\tau})^2} = \frac{1}{4 (t^0)^2}\ , \\
    K_i &= - \frac{i}{2} \mathring{\kappa}_{ijk} (z^j - \bar{z}^j) (z^k - \bar{z}^k) = 2 i \mathring{\kappa}_{ijk} t^j t^k\ , \label{IIB-eq:Ki}\\
    K_{i\bar{j}} &= i \mathring{\kappa}_{ijk} (z^k - \bar{z}^k) + \frac{1}{4} \mathring{\kappa}_{imn} \mathring{\kappa}_{jpq} (z^m - \bar{z}^m) (z^n - \bar{z}^n)  (z^p - \bar{z}^p) (z^q - \bar{z}^q) \nonumber \\
    &= - 2 \mathring{\kappa}_{ijk} t^k + 4 \mathring{\kappa}_{imn} \mathring{\kappa}_{jpq} t^m t^n t^p t^q\ ,\label{IIB-eq:Kij}
\end{align}
where we have defined $\mathring{\kappa}_{ijk} \equiv e^{K_{\rm cs}} \kappa_{ijk}$ and the indices $\tau$ and $i$ denote derivatives of the Kähler potential with respect to the axio-dilaton and the complex structure moduli $z^i$ respectively (barred indices naturally denote derivatives with respect to the complex conjugate fields).

Intuitively, the LCS regime establishes how the cubic term inside the previous logarithm compares with the constant contribution $\kappa_0$. Thus, we introduce the following \emph{LCS parameter} to measure how close to the LCS point a given solution is:
\begin{align}
\label{IIB-eq:def_xi}
    \xi \equiv \frac{- 2 \, \Im (\kappa_0)}{\frac{4}{3} \kappa_{ijk} t^i t^j t^k} = \frac{-2 e^{K_{\text{cs}}} \Im (\kappa_0)}{1 + 2 e^{K_{\text{cs}}} \Im (\kappa_0)}\ .
\end{align}
By definition, the Large Complex Structure limit is the regime where $t^i \to \infty, \forall i \in \lbrace 1, \ldots, h^{2,1} \rbrace$. This implies that the LCS point is located at $\xi = 0$.  In what follows, we will regard the limit $\xi \to 0$ as the LCS limit; however, one should note that this correspondence may not always be  applicable, since one may obtain $\xi \approx 0$ with some saxions remaining small, giving rise to non-negligible exponential corrections. In any case, we consider the condition $\xi \to 0$ to be sufficiently constraining as to become a good indicator of how close to the LCS point a vacuum can be located and, thus, how small exponential corrections can be.

On the other hand, it can be checked that in those geometries where $h^{2,1} > h^{1,1}$, we obtain negative eigenvalues in the field-space metric $K_{i\bar{j}}$ if $\xi > 1/2$, thus rendering those solutions unphysical; as for geometries with $h^{2,1} > h^{1,1}$, solutions with $\xi<-1$ will suffer from the same problem\footnote{Note that for the definition \eqref{IIB-eq:def_xi} to be useful, we require $\kappa_0$ to be non-zero, which implies $\chi(X_6) \neq 0$ or, equivalently, $h^{1,1} \neq h^{2,1}$. In what follows, we will assume the models under study to satisfy this property.} \cite{Blanco-Pillado:2020wjn}.

\subsubsection*{Flux superpotential}

With these definitions, we can express the usual Gukov-Vafa-Witten (GVW) superpotential $W$ \cite{Gukov:1999ya}, induced by fluxes threading the compact geometry. 
We first introduce the flux vector
\begin{equation}
N\equiv f-\tau h\ \ \ \ \text{with}\ \ \ \ 
f\equiv\begin{pmatrix}
\int_{B^I} F_3 \\ 
\int_{A_I} F_3
\end{pmatrix} 
\equiv
\begin{pmatrix}
f^B_0 \\ f^B_i \\ f_A^0 \\ f_A^i 
\end{pmatrix}
\ \ \ \ \text{and}\ \ \ \
h\equiv\begin{pmatrix}
\int_{B^I}H_3\\
\int_{A_I}H_3
\end{pmatrix}
\equiv
\begin{pmatrix}
h^B_0 \\ h^B_i \\ h_A^0 \\ h_A^i 
\end{pmatrix}.
\label{fb-eq: G3 flux expansion}
\end{equation}
These fluxes induce a D3-tadpole Ramond-Ramond charge in the compact space, which has to be cancelled by negatively charged objects, like orientifold planes. The full D3-charge $\Nf$ induced by these fluxes is shown to be
\begin{equation}
    \Nf = f^T \cdot \Sigma \cdot h = - \frac{N^{\dag}\cdot \Sigma \cdot N}{\tau - \bar{\tau}}\ .
\end{equation}

The GVW superpotential can then be easily expressed as\footnote{Note that we deliberately forget a factor $1/\sqrt{4\pi}$ since it will be irrelevant for the vacuum equations and everything we will compute.} \cite{Gukov:1999ya}
\begin{equation}
W\equiv\int (F_3-\tau H_3)\wedge \Omega=N^T\cdot\Sigma\cdot\Pi\ .
\end{equation}
From this equation we can obtain the full expression for the superpotential, which reads
\begin{align}
\begin{split}
W = &- \frac{1}{6} N_A^0 \kappa_{ijk} z^i z^j z^k + \frac{1}{2} \kappa_{ijk} N_A^i z^j z^k + \left( N_A^j a_{ij} + N_i^B - N_A^0 \hat{a}_i \right) z^i\\
&- \kappa_0 N_A^0 - N_A^i \hat{a}_i + N_0^B\ .
\label{IIB-eq:Wfull}
\end{split}
\end{align}

\subsubsection*{Vacuum equations}

At tree-level, type IIB Calabi--Yau compactifications with three-form fluxes yield 4d Minkowski vacua.  Since the 4d EFT features a no-scale structure in the Kähler sector ($K^{\rho\sigma}K_\rho K_\sigma=3$ where $\rho, \sigma$ run over Kähler moduli), the corresponding vacua equations are given by $D_A W\equiv\partial_AW+K_AW= 0, \ A\in\left\lbrace \tau,z^i \right\rbrace$. Let us write these equations explicitly:
\begin{align}
D_\tau W &= \left[ - h  - \frac{1}{\tau - \bar{\tau}} (f - \tau h) \right]^T \cdot \Sigma \cdot \Pi = - \frac{1}{\tau - \bar{\tau}} \bar{N}^T \cdot \Sigma \cdot \Pi = 0\ , \\[5pt]
D_i W &= N^T \cdot \Sigma \cdot D_i \Pi = 0\ ,
\end{align}
which translate into
\begin{align}
\label{IIB-eq:DtW_DiW_full}
& - \frac{1}{6} \bar{N}_A^0 \kappa_{ijk} z^i z^j z^k + \frac{1}{2} \kappa_{ijk} \bar{N}_A^i z^j z^k + \left( \bar{N}_A^j a_{ij} + \bar{N}_i^B - \bar{N}_A^0 \hat{a}_i \right) z^i - \kappa_0 \bar{N}_A^0 - \bar{N}_A^i \hat{a}_i + \bar{N}_0^B = 0\ , \nonumber\\[5pt]
& - \frac{1}{2} N_A^0 \kappa_{ijk} z^j z^k + \kappa_{ijk} N_A^j z^k + \left( N_A^j a_{ij} + N_i^B - N_A^0 c_i \right) + K_i W = 0\ .
\end{align}
Supersymmetric vacua are realized if, in addition, the covariant derivatives of the superpotential with respect to the Kähler moduli are zero. Since they are proportional to $W$, the superpotential should vanish to yield a supersymmetric vacuum. Namely, with $\sigma$ referring to the Kähler sector:
\begin{equation}
\text{Supersymmetric condition: }D_\sigma W=K_\sigma W=0 \Longleftrightarrow W=0\ .
\end{equation}

\section{Basics of F-theory}
\label{fb-sec: f-theory}

F-theory offers a fascinating insight into the Landscape of String Theory compactifications thanks to its deep relation with algebraic and arithmetic geometry, which enables the incorporation of non-perturbative analysis while allowing for sufficient control to perform computations.  One way to think about F-theory is as a supersymmetric compactification of the strong coupling limit of Type IIB orientifolds in the presence of 7-branes. These localized sources backreact on the geometry, inducing variations on the axio-dilaton profile at different points of the compactification space.  Due to the $SL(2,\mathbb{Z})$ non-perturbative invariance of the axio-dilaton discussed in section \ref{basic-subsec: dualities}, it is possible to describe such non-trivial profile in terms of an elliptic fibration of a torus over the 6-dimensional compact space. We will discuss how this notion arises, its relation with M-theory and its applications in string compactifications.

\subsection{From Type IIB to F-theory}

We recall that Type IIB is invariant under $SL(2,\mathbb{Z})$ transformation acting like  $T^0\equiv C_0+i/g_s$
\begin{equation}
    \tau \rightarrow \frac{aT^0+b}{cT^0+d}\,,\qquad \left(\begin{array}{c}
          B_2 \\
          C_2
    \end{array}\right)\rightarrow M\left(\begin{array}{c}
          B_2 \\
          C_2
    \end{array}\right)\,,\qquad M= \left(\begin{array}{cc}
        a & b \\
        c & d
    \end{array}\right) \in SL(2,\mathbb{Z})\,,
\end{equation}
with $\tau=C_0+ie^{-\phi}$. The action over $\tau$  is formally identical to the behaviour of a complex structure of an elliptic curve under a modular transformation (see discussion around \eqref{basic-eq: modular group}). To improve our understanding of this relation, it is useful to describe the torus as a hypersurface of a complex space of complex dimension two instead of a quotient of $\mathbb{C}$ by a lattice as we have been doing until now. Therefore, we will consider the torus as a projective subvariety of the weighted projective space $\mathbb{P}_{231}$\footnote{This space is defined by as the quotient of $\mathbb{C}^3\backslash \{0,0,0\}$ under the equivalence relation $(x,y,z)\sim(\lambda^2 x,\lambda^3 y,\lambda z)$ with $\lambda\in \mathbb{C}$.} defined through the vanishing loci of the Weierstrass polynomial
\begin{equation}
    P_W\equiv y^2-x^3-fxz^4-gz^6\,.
\end{equation}
The coefficients of the Weierstrass polynomial can be mapped to functions of the complex structure of the torus through the Eisenstein series $g_2$ and $g_3$
\begin{align}
    f(\tau)&\equiv -4^{1/3}g_2(\tau)=-4^{1/3} 60\sum_{(m,n)\in\mathbb{Z}^2\neq(0,0)} (m+n\tau)^{-4}\,,\\
    g(\tau)&\equiv -4 g_3(\tau)=-560\sum_{(m,n)\in\mathbb{Z}^2\neq(0,0)} (m+n\tau)^{-6}\,.
    \label{fb-eq: weirstrass coefficients}
\end{align}
Inversely, the functions $f$ and $g$ can be used to identify the complex structure via the Jacobi function
\begin{equation}
    j(\tau)=4\frac{24^3 f(\tau)^3}{\Delta}\,,\qquad \Delta=4f^3(\tau)+27g^2(\tau)\,.
    \label{fb-eq: discriminant D7}
\end{equation}
The important thing to note about the last expression is that it becomes singular when the discriminant $\Delta$ vanishes.

Going back to our goal of describing how the torus geometry encodes the variation of the axio-dilaton in type IIB compactifications, we introduce a factorization of the 10-dimensional space of the form
\begin{equation}
    \mathcal{M}_{1,9}=\mathbb{R}^{1,9-2n}\times B_n\,,
\end{equation}
with $B_n$ a compact manifold of complex dimension $n$. Supersymmetry requirements demand $B_n$ to be Kähler and $C_0$ to enter holomorphically in the axio-dilaton $\tau$, so $\bar{\partial}\tau=0$ \cite{Greene:1989ya}. Consequently, the holomorphic variation of $\tau$ defines a holomorphic line bundle over $B_n$, denoted by $\mathcal{L}$.  Furthermore, one can use Einstein equations to relate the curvature of the internal manifold with the changes of the dilaton, which is translated into a relation between the first Chern classes of the line bundle and the internal manifold \cite{Bianchi:2011qh}
\begin{equation}
    c_1(B_n)=c_1(\mathcal{L})\,.
    \label{fb-eq: chern class relation}
\end{equation}
The above relation thus establishes a correspondence between the variation of the axio-dilaton and the Ricci-curvature of $B_n$

At this point, we have all the ingredients needed to uniquely define an elliptic fibration over $B_n$: a line bundle $\mathcal{L}$ over $B_n$ and a choice of section of $\mathcal{L}^4$ and $\mathcal{L}^6$. More specifically, by taking the coefficients $f$ and $g$ introduced in \eqref{fb-eq: weirstrass coefficients} as sections of $\mathcal{L}$, one has $f\in \Gamma(B_n,\mathcal{L}^4)$ and  $g\in\Gamma(B_n,\mathcal{L}^6)$ and the associated fibre elliptic fibre $\mathbb{E}_\tau$ has the axio-dilaton $\tau$ as complex structure parameter.

The final picture is a torus fibration described by the following diagram
\begin{eqnarray}
    \pi: \hspace{1cm} & \mathbb{E}_\tau \rightarrow & X_{n+2} \nonumber\\
     & & \hspace{0.3cm} \downarrow\\
     & & \hspace{0.2cm} B_n \nonumber
\end{eqnarray}
The resulting elliptic fibration is a Kähler manifold whose first Chern class satisfies $c_1(X_{n+2})=c_1(B_n)-c_1(\mathcal{L})$. Then \eqref{fb-eq: chern class relation} means it has vanishing first Chern class and it is therefore a Calabi-Yau four-fold.   

In $X_{n+2}$, the presence of type IIB localized sources (D7/O7-planes) is entirely captured by the geometry of the fibration. The points of the base $B_n$ in which the determinant $\Delta$ in \eqref{fb-eq: discriminant D7} vanishes corresponds to the location of divisors $D_i$ being wrapped by D7-branes. Taking into account that the first Chern class $c_1(\mathcal{L})$ represents the zeros of a generic section of the line bundle, we have
\begin{equation}
    c_1(X_{n+2})=c_1(B_n)-\sum_{i}\frac{1}{12}n_i \delta[D_i]\,,
\end{equation}
where $\delta[D_i]$ is the Poincaré dual form of the divisor $D_i$ and $n_i$ is the order of vanishing of $\Delta$ on that divisor. The Calabi-Yau condition for $X_{n+2}$ then reads
\begin{eqnarray}
    \sum_i n_i \delta[D_i]=12 c_1(B_n)\,,
\end{eqnarray}
which is  the analogue of the type IIB charge cancellation condition of RR charges. Therefore we are able to encode all the information of the compactification in terms of the geometrical quantities relating the base and the elliptic fibre.

\subsection{M-theory and F-theory duality}

Alternatively,  F-theory can be introduced starting from M-theory and using its relation with Type IIA and T-duality. This approach has the advantage of providing a natural way to introduce the toroidal fibre. We briefly sketch the underlying reasoning.  In section \ref{basic-subsec: dualities}, we shortly explained how M-theory compactified on a circle $S_A^1$ with radius $R_A\rightarrow  0$ yields Type IIA string theory. One can now compactify the resulting theory on a different circle $S_B$ to obtain a 9-dimensional theory describing $M$-theory compactified on a 2-torus $S_A\times S_B$, or, in the limit of vanishing radius $R_A$, Type IIA compactified on a circle of radius $R_B$· Then, using T-duality and keeping the complex structure of the torus $\tau\sim R_B/R_A$, the latter is equivalent to Type IIB string theory compactified on a circle of radius $\alpha'/R_B$. All these relations lead us to state that M-theory compactified on a torus in the limit of vanishing volume $V=R_AR_B\rightarrow 0$ is dual to type IIB compactified on a circle in the decompactification limit $1/R_B\rightarrow \infty$ (so the dual theory grows a new dimension). The complex structure in the elliptic fibre of the M-theory side $\tau$ is not affected by the vanishing volume limit, and it is mapped to the dilaton of type IIB.

The above construction describes a duality between M-theory compactified on  $T^2$ with vanishing volume and 10-dimensional type IIB with constant dilaton. The  generalization of this duality to any elliptic fibration  yields F-theory. We conclude that F-theory on an elliptic fibre can be defined as the zero area limit of M-theory on that elliptic fibre, and corresponds to type IIB compactified on the base with a non-trivial $\tau$ profile determined by the geometry of the fibration.

\subsection{Flux Compactifications}
\label{fb-subsec: ftheory compactifications}

The relation with M-theory is very useful to understand the field content and moduli structure of F-theory compactifications. We recall from section \ref{basic-subsec: dualities} that the bosonic content of M-theory is simply the 11-dimensional metric and a 3-form field. We can then perform the usual game and expand the different forms under the desired factorization of spacetime over which we wish to compactify. For the M-theory case, that is $\mathcal{M}_3\times X_8$ with $X_8$ a Calabi-Yau 4-fold which is also an elliptic fibration. Thus, the internal part of the metric has an associated Kähler and complex structure form. 

Starting with the Kähler form $J$, one has the decomposition
\begin{equation}
    J=v^0[S_0]+v^a\alpha[D_a]\,,
\end{equation}
where $[D_0],[D_a]$ are the 2-forms Poincaré dual to the base $B_6$ and the vertical divisors  $D_\alpha=\pi^{*}(D^b_a)$, the latter being inherited from the divisors of the base $D^b_\alpha$. Therefore,  the Kähler moduli space has dimension $h^{1,1}(B_6)+1$. One can perform a similar decomposition of $C_3$ and join the resulting moduli together to build complexified Kähler moduli. The Kähler potential for the Kähler sector is
\begin{equation}
    K^M=-3\log \mathcal{V}\,,\qquad \mathcal{V}=\frac{1}{4!}\int_{X_4}J\wedge J\wedge J\wedge J\,.
    \label{fb-eq: kahler sector potential f-theory}
\end{equation}
There are some subtleties regarding how the moduli combine and their up-lift to F-theory.  For further details, we refer to \cite{Grimm:2010ks}. 

In addition to the Kähler sector, there are also complex structure moduli related to the holomorphic 4-form $\Omega$. As in the three-fold case, the moduli are given by the periods of this form and the associated Kähler potential is
\begin{equation}
    K_{\rm cs}=-\log \int_{X^8}\Omega \wedge \bar{\Omega}\,.
    \label{fb-eq: cs Kahler potential f-theory}
\end{equation}

As we mentioned before, the presence of D7-branes in the type IIB theory perspective is represented by the degeneration of the elliptic fiber. Generally, such degeneration leads to a singular manifold, greatly complicating the analysis. To solve the problem, the singularities can be blown up replacing the singular fiber with a $\mathbb{P}^1$ manifold, in a similar process as the one described in \ref{cy-subsec: orbifolds}. We denote the blown-up manifold by $\hat{X}_8$.

In the last part of this thesis we will focus our efforts on the stabilization of the complex structure sector, which can be achieved by turning on the background of the 4-form flux $G_4$ associated to $C_3$ of M-theory. This field encodes simultaneously the RR-fluxes of type IIB and the D7-brane worldvolume. To adequately describe its role in F-theory, the corresponding lift of $G_4$ will need to be studied. 

The background 4-form flux has to satisfy three main properties. First, it needs to verify the quantization condition \cite{Witten:1996md}
\begin{equation}
    G_4+\frac{1}{2}c_2(\hat{X}_8)\in H^4(\hat{X}_8,\mathbb{Z})\,,
\end{equation}
which means that in general the flux is half-integer quantized. Second, it must preserve some supersymmetry in the compactified space, leading to the following constraints \cite{Becker:1996gj,Gukov:1999ya, Haack:2001jz}
\begin{equation}
    G_4\in H^{2,2}(\hat{X}_8,\mathbb{R})\cap H^4(\hat{X}_8,\mathbb{Z}/2)\,,\qquad J\wedge G_4 =0\,.
\end{equation}
Finally, compatibility with the F-theory lifting requires 4-dimensional Poincaré invariance. This imposes several transversality relations
\begin{equation}
    [G_4]\cdot[S_0]\cdot \pi^*[D^b_a]=[G_4]\cdot \pi^*[D^b_b]\cdot \pi^*[D_b^b]=0\,.
\end{equation}
Therefore we will need a good understanding of the middle cohomology group $H^4(\hat{X}_8,\mathbb{C})$ and more specifically $H^{2,2}(\hat{X}_8,\mathbb{C})$. In this context, it is useful to distinguish between the horizontal middle cohomology (the one built from the holomorphic forms and its derivatives) and the vertical cohomology (originated from products of $(1,1)$-forms). For a  Calabi-Yau 4-fold, the horizontal piece has the orthogonal decomposition
\begin{equation}
    H^4_H(\hat{X}_8,\mathbb{C})=H^{4,0}\oplus H^{3,1}\oplus H^{2,2}_H\oplus H^{1,3}\oplus H^{0,4}\,.
\end{equation}
Contrary to the 3-fold case, in 4-folds these two groups, although relevant, do not factorize perfectly the cohomology group $H^{2,2}$. Instead we have
\begin{eqnarray}
    H^{2,2}(\hat{X}_8,\mathbb{C})=H^{2,2}_{\rm hor}(\hat{X}_8,\mathbb{C})\oplus H^{2,2}_{\rm vert}(\hat{X}_8,\mathbb{C})\oplus H^{2,2}_{\rm rem}(\hat{X}_8,\mathbb{C})\,,
\end{eqnarray}
with $H_{\rm rem}(\hat{X}_8,\mathbb{C})$ a sector that is neither a product of $(1,1)$-forms nor obtained from variations of $\Omega$.

The background $G_4$ flux generates two superpotentials. One involves the Kähler moduli and generates a D-term potential, while the second one, commonly known as Gukov-Vafa-Witten superpotential \cite{Gukov:1999ya}, refers to the complex structure moduli and gives rise to an F-term potential of the form \eqref{cy-eq: VFgen}
\begin{equation}
W_D=\int_{\hat{X}_8} J\wedge J\wedge G_4\, \qquad W_F=\int_{\hat{X}_8}\Omega\wedge G_4\,.
\end{equation}

In the following chapter we will focus on the latter and denote it simply by $W$. The associated scalar potential can be rewritten as
\begin{equation}
 V_F=\frac{1}{4\mathcal{V}^4}\left[G_4\wedge \star G_4-\int G_4\wedge G_4\right]  \,.
\end{equation}
Such potential is positive semidefinite and thus yields Minkowski vacua, which is  obtained when $D_i W=0$. This requirement amounts to demanding that the 4-form flux is self-dual
\begin{equation}
    G_4=\star G_4\,.
\end{equation}
 This is considerably difficult to check in general. An alternative is to directly study the superpotential and expand it in terms of the periods of the holomorphic form, which can be determined using homological mirror symmetry analogously to the Type IIB case described around \eqref{fb-eq: central charges Klemm formula Type IIB}.

 Thus, the idea is to take M-theory on $\hat{X}_8$ and compactify on an additional circle $S^1$ to obtain a 2-dimensional effective theory of type IIA. Mirror symmetry maps this theory to Type IIA\footnote{Note that the mirror symmetry map changes with the number of dimensions of the compactification space. Each complex dimension is related to a single T-duality that exchanges Type IIA and Type IIB. When working with Calabi-Yau 3-folds, this number is odd and the map takes Type IIA to Type IIB. Now it is even and thus mirror symmetry acts as a map between two different Type IIA compactifications.} compactified on a different Calabi-Yau 4-fold $\hat{Y}_8$. An A-brane wrapping a special Lagrangian cycle on $\hat{X}_8$ is related to a B-brane in $\hat{Y}_8$, i.e. an element of the bounded derived category of coherent sheaves\footnote{We will only work in the large volume approximation. In that case, B-branes can be thought of as standard D(2p)-branes wrapping (2p)-cycles.} $\mathcal{E}\in D^b(\hat{Y}_8)$. One can identify the periods of $\Omega$ with the central charges of the A-branes, which match the central charges of B-branes in the mirror. The latter can be computed by \cite{Cota:2017aal}
\begin{eqnarray}
    Z(\mathcal{{E}})=\int_{\hat{Y}_8} e^{J_c}\Gamma_{\mathbb{C}}(\hat{Y}_8)\lambda({\rm ch}\mathcal{E})\,,
\end{eqnarray}
where $\Gamma_{\mathbb{C}}(\hat{Y}_8)$ gets an additional term with respect to its type IIB counterpart
\begin{equation}
    \Gamma_{\mathbb{C}}(\hat{Y}_8)=1+\frac{1}{24}c_2(\hat{Y}_8)+\frac{\zeta(3)}{(2\pi i)^3}+\frac{1}{5760}(7c_2(\hat{Y}_8)^2-4c_4(\hat{Y}_8))\,.
\end{equation}

Finally, it is important to mention that there is an additional constraint that the fluxes need to satisfy in order to generate well-defined vacua: the Bianchi identity for the M-theory 3-form. In its integrated version (Tadpole Constraint) it amounts to \cite{Dasgupta:1999ss,Sethi:1996es}
\begin{equation}
    -\frac{1}{2}\int_{\hat{X}_8}G_4\wedge G_4+\frac{1}{24}\chi_4(\hat{X}_8)=N_{M_2}\geq 0\,,
    \label{fb-eq: mtheory tadpole}
\end{equation}
where $\chi_4(\hat{X}_8)$ is the Euler characteristic of $\hat{X}_8$ and $N_{M2}$ denotes the number of spacetime filling M2-branes. This number matches the number of D3-branes in the dual F-theory vacuum and the stability of Minkowski vacua requires it to be positive.

\subsection{Tadpole Conjecture}
\label{fb-subsec: tadpole conjecture}

Although the landscape is vastly large, we have repeatedly emphasized that not all solutions are of interest to us. For instance, generic compactifications are filled with moduli that come from complex structure and Kahler deformations. One often restricts one’s attention to effective four-dimensional theories with few or no massless scalar fields by compactifying on Calabi-Yau three-folds in the presence of fluxes that generate a potential for the moduli. Since the number of flux quanta grows with the number of complex structure deformations, naively one could expect that the landscape of this type of compactifications would be dominated by CY manifolds with the largest number of such moduli. However, the fluxes that stabilize the moduli also source electric brane charges, which must sum up to zero on a compact manifold. Hence, brane tadpole-cancellation conditions place upper bounds on the amount of these fluxes. This picture can dramatically change the perception of the landscape distribution and suggests that, actually, manifolds with low number of moduli should be the dominant ones.
Such point of view was formalized by the Tadpole Conjecture \cite{Bena:2020xrh}, which puts an upper bound in the number of moduli stabilized in a given manifold. Thus, there is a balance to be achieved between two competing approaches: the requirement of fluxes to obtain moduli stabilization and the constraints that these fluxes impose on the tadpole cancellation conditions. 

The core idea of the Tadpole Conjecture is most easily framed in F-theory. There, we know that the tadpole constraint for the contribution of the flux $G_4$ to the charge, $N_{\rm flux}$, is of the form
\begin{equation}
    N_{\rm flux}+N_{D3}=\frac{\chi(\hat{Y}_8)}{24}\,.
\end{equation}
The number of D3-branes must be positive and vacua equations impose self-duality of $G_4$. Both facts combined imply
\begin{equation}
    0\leq N_{\rm flux}\leq \frac{\chi(\hat{Y}_8)}{24}\,.
\end{equation}
Now, the Euler characteristic is given and $\chi(\hat{Y}_8)=6(8+h^{1,1}+h^{3,1}-h^{2,1})$ and in the large number of moduli limit one expects the term $h^{3,1}$ to dominate and so
\begin{eqnarray}
    \frac{\chi(\hat{Y}_8)}{24}\sim\frac{h^{3,1}}{4}\,.
    \label{fb-eq: euler characteristic scaling}
\end{eqnarray}
The tadpole conjecture states that for sufficiently large number of moduli, the flux contribution to the tadpole satisfies
\begin{equation}
    N_{\rm flux}>\alpha \, n_{\rm stab}\,,\qquad {\rm with \ } \alpha>\frac{1}{3}\,,
\end{equation}
where $n_{\rm stab}$ the number of stabilized moduli. If the conjecture is correct, \eqref{fb-eq: euler characteristic scaling} means that at large number of moduli there will always be remaining flat directions and, consequently, full moduli stabilization is not possible.

The conjecture has been tested positively in several examples \cite{Bena:2021wyr, Coudarchet:2023mmm, Braun:2023pzd} and has been proved in the strict large complex structure regime using asymptotic Hodge structure theory in \cite{Grana:2022dfw}. However, examples deep in the bulk of moduli space seem to be in tension with the conjecture \cite{Lust:2022mhk}. We will come across the tadpole conjecture in the next chapter.

\ifSubfilesClassLoaded{%
\bibliography{biblio}%
}{}

\end{document}


\ifSubfilesClassLoaded{%
\tableofcontents
}{}

\setcounter{chapter}{7}
\chapter{F-theory flux vacua at large complex structure}
\label{ch: Ftheory}

A powerful feature of F-theory compactifications is that they provide an overall picture of the set of string vacua, since they are directly connected to most string theory constructions via dualities, as we discussed in \ref{basic-subsec: dualities} and in the previous chapter. This trait is particularly significant in the context of compactifications to four dimensions, where they are in addition endowed with a notably simple and efficient mechanism to stabilize moduli. Indeed, complex structure moduli fixing in F-theory  through the presence of background four-form fluxes is a paradigmatic framework to remove unwanted neutral scalars from the low energy effective theory \cite{Grana:2005jc,Douglas:2006es,Blumenhagen:2006ci,Becker:2007zj,Marchesano:2007de,Denef:2008wq}. It is from this framework that we have developed our current understanding of the string Landscape. 

Given the vast size of F-theory flux Landscape, it is not obvious how to describe all the information encoded in complex structure moduli stabilization. One possible approach is to treat the set of flux vacua as an ensemble, and apply statistical methods to extract their physical properties \cite{Denef:2007pq}. An  alternative strategy is to assume that complex structure moduli are fully fixed at a very high scale, and so one can safely integrate out all of them to analyze the physics of K\"ahler moduli and localized degrees of freedom \cite{Ibanez:2012zz,Quevedo:2014xia,Baumann:2014nda}. The information of complex structure moduli stabilization is then encoded in a set of parameters that appear in the effective theory below the flux scale, and which are oftentimes  assumed to be tunable in terms of an appropriate choice  of Calabi--Yau geometry and flux quanta.

It has however been pointed out that there could be more to it than this generic picture of complex structure moduli stabilization. On the one hand, some works have questioned the idea that one can generically fix all complex structure moduli and at the same time satisfy the tadpole consistency conditions of the compactification \cite{Braun:2020jrx,Bena:2020xrh,Bena:2021wyr}(see section \ref{fb-subsec: tadpole conjecture}). On the other hand, it has been shown that at asymptotic limits in complex structure field space the flux potential simplifies and its form can be classified in terms of robust Calabi--Yau data \cite{Grimm:2019ixq}, leading to certain no-go results and general arguments in favour of the finiteness of flux vacua \cite{Grimm:2020cda}. 

Clearly, these recent results point towards a rich structure underlying F-theory flux potentials that is yet to be unveiled. To uncover this structure, it is important to gain analytic control over F-theory flux potentials and its corresponding set of vacua. Ideally, given a Calabi--Yau four-fold and a choice of four-form fluxes, one would like to understand directly from these data how many complex structure moduli are stabilized by the potential, at which point in field space they are fixed, and what is their mass spectrum. 

It is thus the purpose of this chapter to take a non-trivial step in this direction, by providing an explicit, analytic description of F-theory flux potentials and their vacua. We do so by focusing on regions of large complex structure of smooth Calabi--Yau four-folds. In this regime, we are able to provide an explicit expression for the F-theory F-term potential for any four-fold $Y_8$, up to exponentially-suppressed terms. At this level of approximation, the only data that are needed to specify the potential are the flux quanta and certain topological numbers of the mirror four-fold $X_8$. This simplicity allows us to perform a general analysis of the vacua conditions for an arbitrary number of complex structure fields, and eventually uncover different families in which such vacua are arranged.   

An important ingredient of our analysis is the fact that at moderate and large complex structure the 4d K\"ahler potential displays a number of axionic shift symmetries, only broken by the exponentially-suppressed terms that we neglect. Because of this, each complex structure field splits  into an axionic and a saxionic component. Microscopically, the periodicity of the axions corresponds to the monodromies around the large complex structure point that act non-trivially both on the periods of the holomorphic  $\Omega$ and flux $G_4$ four-forms. It turns out that in terms of these real variables the scalar potential takes a very simple form analogous to the one described in \ref{cy-subsec: bilinear formalism} for type IIA, namely $V =\half Z^{AB} \rho_A \rho_B$, with $\rho_A$  monodromy-invariant combinations of fluxes and axions, and $Z^{AB}$ only depending on the saxions. Since the potential is  positive semi-definite and only yields Minkowski vacua, the on-shell equations amount to $Z^{AB}\rho_B =0$ $\forall A$, and so they can be solved algebraically.

Using these on-shell equations, one is able to rewrite the flux contribution to the D3-brane tadpole $N_{\rm flux}$ as a sum of positive terms, and from there derive that certain flux quanta must vanish at large complex structure in order to find vacua in this regime. Depending on which quanta vanish we distinguish different families of flux vacua, which we then analyze. In the most generic family, which is present in any Calabi--Yau four-fold $Y_8$, the number of stabilized moduli depends on the choice of fluxes, an effect that we characterize with explicit formulas. Remarkably, even in the most favourable case full moduli stabilization is not that easy to observe: It is only manifest when the entries of  $Z^{AB}$ are computed to certain accuracy. In practice, one may compute them {\it i)} in the strict asymptotic limit \cite{Grimm:2019ixq}, {\it ii)} by approximating the periods of $\Omega$ with their leading behaviour (section \ref{Ft-sec:leading}) , and {\it iii)} by including all the polynomial corrections to such periods, neglecting only exponentially-suppressed terms (section \ref{Ft-sec:poly}). For this family of vacua only with this third description full complex structure moduli stabilization is manifest. Less accurate descriptions yield potentials that typically have at least one flat direction. As a consequence, most vacua cannot exist at parametrically large complex structure. In fact, we find that the saxion vevs are bounded from above by roughly $K^{(3)} N_{\rm flux}^{p + \oh}$, where $K^{(3)}$ represents the minor polynomial correction to the potential, $N_{\rm flux}$ is the flux contribution to the D3-brane tadpole, and $p \leq h^{3,1}(Y_8)$ is bounded by the number of complex structure moduli.    

In this generic scheme, the condition to achieve full moduli stabilization depends on those flux quanta that contribute to $N_{\rm flux}$. It is therefore possible that in some instances $N_{\rm flux}$ grows as we increase the number of moduli, as recently proposed by the Tadpole Conjecture in \cite{Bena:2020xrh}. Our framework allows us to propose a formula that tests this statement, and that can be checked in any compactification. Regardless of whether this happens or not, we find a second family of vacua that is in tension with the Tadpole Conjecture. This new family of vacua arises whenever a complex structure saxion appears at most linearly in $e^{-K}$ (with $K$ the K\"ahler potential) and the superpotential, a setup which we dub the {\em linear scenario}. Examples of this are Calabi--Yau four-folds $Y_8$ whose mirror $X_8$ is a fibration of a Calabi--Yau over a $\mathbb{P}^1$, and in particular type IIB orientifold compactifications. The new set of vacua appears at large values of the linear saxion, with $N_{\rm flux}$  a simple product of two flux quanta. The remaining non-vanishing flux quanta are such that they fix all complex structure moduli. Remarkably, in the particular case of  type IIB compactifications the polynomial corrections identified as $K^{(3)}$ are also needed to implement this full moduli stabilization and, in fact, this family of vacua are mirror dual of the Minkowski type IIA flux vacua originally found in \cite{Palti:2008mg}. The necessity of polynomial corrections is however not a universal feature in other F-theory realisations of the linear scenario, as we show with an explicit example. This indicates that it is this more exotic family of vacua, and maybe new ones yet to be discovered, that dominate the landscape of F-theory vacua at regions of parametrically large complex structure.

The analysis performed in this chapter is structure as as follows. In section \ref{Ft-s:potential} we compute the flux scalar potential for arbitrary four-folds, first using the leading terms of the periods of $\Omega$ and then including all polynomial terms. In section \ref{Ft-s:vacua} we analyze the resulting vacua equations, and in particular how a finite D3-brane tadpole affects the existence of vacua. From here we obtain the most generic family of flux vacua in the large complex structure regime, which nevertheless cannot exist at parametrically large complex structure. In section \ref{Ft-s:IIB} we apply our results to the special case of type IIB orientifold compactifications, matching them with the existing literature. In particular, we identify a family of flux vacua which is different from the generic one, in which the expression for $N_{\rm flux}$ is independent of the number of moduli. Section \ref{Ft-s:linear} upgrades this family of vacua to a genuine F-theory setup, which we dub linear scenario. In section \ref{Ft-s:examples} we illustrate our findings with explicit constructions of Calabi--Yau four-folds, whose mirror are smooth fibrations. We finally present our conclusions in section \ref{Ft-s:conclu}.


\section{The F-theory potential at large complex structure}
\label{Ft-s:potential}

In a region of sufficiently large complex structure, the moduli space geometry of F-theory on a Calabi--Yau (CY) four-fold simplifies, in the sense that each complex structure field splits into an axionic and a saxionic real components. This not only constrains the form of the 4d effective K\"ahler potential, but also of the superpotential induced by background four-form fluxes. In this section we will compute both, and from there provide an explicit bilinear expression for the F-term scalar potential, on which we will base our subsequent analysis. In section \ref{Ft-sec:leading} we will consider the leading form of the potential at large saxion values, from which one can infer most of the intuition regarding the ensemble of flux vacua, and in section \ref{Ft-sec:poly} we will include the polynomial corrections to these leading terms. As we will see in section \ref{Ft-s:vacua}, such corrections turn out to be crucial to fully understand moduli stabilization in F-theory. 

\subsection{The leading flux potential}
\label{Ft-sec:leading}

Let us consider F-theory compactified on a Calabi--Yau four-fold $Y_8$, which is a smooth elliptic fibration over a three-fold base $C_6$. In section \ref{fb-sec: f-theory} we learnt that the presence of an internal background four-form flux $G_4$ generates a scalar potential for both the complex structure and K\"ahler moduli of $Y_8$. The potential for K\"ahler moduli can be seen as a D-term potential while the potential for the complex structure moduli can be understood as an F-term potential, with Gukov-Vafa-Witten superpotential \cite{Gukov:1999ya}
\be
W = \int_{Y_8} G_4 \wedge \Omega\, ,
\label{Ft-GVW}
\ee
where $\Omega$ is the holomorphic (4,0)-form of $Y_8$, in terms of which we define its complex structure moduli. At large volume the K\"ahler potential splits into a Kähler sector contribution given by \eqref{fb-eq: kahler sector potential f-theory}, and a complex structure sector contribution \eqref{fb-eq: cs Kahler potential f-theory}, which, given its relevance for this chapter, we rerwrite for completeness,
\be
K_{\rm cs} = - \log \int_{Y_8} \Omega \wedge \bar{\Omega} \, .
\label{Ft-kahler}
\ee
Both potentials are positive semi-definite, and select global, 4d Minkowski minima at those points in moduli space where the Hodge self-duality condition is satisfied \cite{Haack:2001jz}
\be
G_4 = \star G_4\, .
\label{Ft-SDG4}
\ee
Those minima in which $G_4$ is a primitive (2,2)-form are, moreover, supersymmetric \cite{Becker:1997cp}. 

Our goal is to provide an explicit expression for the F-term scalar potential in terms of the complex structure moduli of the four-fold. To do so one must first determine a basis for the lattice $\Lambda_W$ of quantized fluxes that enters \eqref{Ft-GVW}, and then compute the corresponding periods of $\Omega$. It turns out that the first part of this problem is quite subtle. This lattice pairs up via \eqref{Ft-GVW} with the horizontal subspace of the middle cohomology of the four-fold  $H_H^4(Y_8) \subset H^4(Y_8)$, which is generated by $\Omega$ and its derivatives \cite{Strominger:1990pd,Greene:1993vm}. We have that $\dim H_H^4(Y_8) = 2 + 2 h^{3,1}(Y_8) + \dim H_H^{2,2}(Y_8)$, with the embedding $ H_H^{2,2}(Y_8) \subset  H^{2,2}(Y_8)$ being quite involved \cite{Braun:2014xka}. As a consequence, in a four-fold there is no clear link between the number of complex structure moduli, which is given by $h^{3,1}(Y_8)$, and the number of fluxes that enter the superpotential.\footnote{Recall that for type IIB  on a Calabi--Yau three-fold we have $b_3/2$ complex fields on the complex structure and axio-dilaton sectors, and a real lattice of background three-form fluxes of dimension $2b_3$. In sections \ref{Ft-s:linear} and \ref{Ft-s:examples} we will consider F-theory constructions that reproduce the same sort of relation.} 

Fortunately, one may implement the strategy of \cite{CaboBizet:2014ovf,Cota:2017aal} to overcome these difficulties and find concrete expressions for the F-term potential. The main idea, reviewed in section \ref{fb-subsec: ftheory compactifications},  is to use homological mirror symmetry and consider the mirror four-fold of $Y_8$, which we denote as $X_8$. Then one may compactify type IIA on $X_8$, and identify the periods of $\Omega$ in $Y_8$ with the central charges of topological B-branes on $X_8$, which generate the mirror of the lattice $\Lambda_W$. In the large volume regime, this lattice can be understood as D$(2p)$-branes wrapping holomorphic $2p$-cycles, with $p=0,1,2,3,4$. The subtleties alluded above translate into constructing a basis of holomorphic 4-cycles, a set that can be generated by intersecting pairs of divisors of $X_8$. This basis can be constructed explicitly when $X_8$ is a smooth fibration, see \cite{Cota:2017aal} and the discussion in sections \ref{Ft-s:linear} and  \ref{Ft-s:examples}. An element of the corresponding lattice will have a central charge of the form $\int_{X_8} e^{J_c} \wedge F_{RR}$, where $F_{RR}$ is a closed even polyform and $J_c = B + iJ$ is the complexified K\"ahler form of $X_8$. It follows that, under these assumptions, the F-theory superpotential \eqref{Ft-GVW} can be identified with a 2d analogue of the 4d type IIA RR flux superpotential \cite{Taylor:1999ii}. 

The leading order term for the central charge $\Pi_{2p}$ of a D$(2p)$-brane wrapping a holomorphic $2p$-cycle on $X_8$ in the large volume limit is 
\begin{subequations}
\label{Ft-eq:periods}
\begin{align}
        \Pi_0&=1\, , \\
        \Pi_2^i&=-T^i\, ,\\
        \Pi_{4\, \mu} &=\oh \eta_{\mu\nu} \zeta^\nu_{ij}T^iT^j\, ,\\
        \label{Ft-period4}
        \Pi_{6\, i}&=-\frac{1}{6}\mathcal{K}_{ijkl}T^jT^kT^l\, ,\\ 
        \Pi_8&=\frac{1}{24}\mathcal{K}_{ijkl}T^iT^jT^kT^l\, ,
\end{align}
\end{subequations}
where $T^i = b^i + i t^i$, $i = 1, \dots, h^{1,1}(X_8)$ stand for the complexified K\"ahler moduli of $X_8$, and $\CK_{ijkl}$ for its quadruple intersection numbers. The index $\mu$ in $\Pi_{4\, \mu}$ runs over a basis of four-cycles generating all the intersections of a basis of Nef divisor classes $[D_i]$ on $X_8$. As a result we can write the class of their intersection as $[\g_{ij}] =  [D_i . D_j] = \zeta^\mu_{ij} [\sigma_\mu]$ for some set of integral four-form classes $[\sigma_\mu]$ and some $\zeta^\mu_{ij} \in \mathbb{Z}$. Finally $\eta_{\mu\nu} = [\sigma_\mu] \cdot [\sigma_\nu]$ is the intersection matrix of this sector, which must satisfy
\be
\CK_{ijkl} = \zeta_{ij}^\mu \eta_{\mu\nu}\zeta^\nu_{kl}  = \zeta_{ij}^\mu\zeta_{\mu, kl}\, .
\label{Ft-interrel}
\ee
where in the second equality we have defined $\zeta_{\mu, kl} \equiv [\sigma_\mu] \cdot [D_k] \cdot [D_l]$.

Applying the mirror symmetry map, the $\{T^i\}$ become the complex structure moduli of $Y_8$, where now $i = 1, \dots, h^{3,1}(Y_8)$. The set of holomorphic $2p$-cycles classes of $X_8$ becomes a lattice of horizontal four-cycles in $Y_8$, such that $[\sigma_\mu] \mapsto [\sigma_\mu^Y]$. The central charges \eqref{Ft-eq:periods} become the leading terms for the periods of the four-form $\Omega$ in the large complex structure limit, where it admits an expansion of the form
\begin{equation}
    \Omega=\alpha \pi_0+\alpha_i\pi_2^i+\sigma_\mu^Y \pi^\mu_4 +\beta^i\pi_{6i}+\beta\pi_8\, .
    \label{Ft-Omega}
\end{equation}
Here $\{ \a, \a_i, \sigma_\mu^Y, \b^i, \b\}$ represent a set of harmonic four-forms which is also an integral basis for $H^4_H(Y_8)$. Their moduli-dependent coefficients are given by
\begin{equation}
\label{Ft-eq:coeff}
        \pi_0=1\, , \quad
        \pi_2^i=T^i\, ,\quad
        \pi_{4}^\mu =\oh  \zeta^\mu_{ij}T^iT^j\, ,\quad
        \pi_{6i}=\frac{1}{6}\mathcal{K}_{ijkl}T^jT^kT^l\, ,\quad
        \pi_8=\frac{1}{24}\mathcal{K}_{ijkl}T^iT^jT^kT^l\, .
\end{equation}
The classical intersection numbers for their Poincar\'e dual four-cycles are 
\begin{equation}
    \int_{Y_8}\alpha \wedge \beta=1,\hspace{1cm} \int_{Y_8}\alpha_i\wedge \beta^j=-\delta_{i}^j,\hspace{1cm}
\int_{Y_8}\sigma_\mu^Y\wedge \sigma_\nu^Y=\eta_{\mu\nu}\, .
\label{Ft-intersection}
\end{equation}
In fact the intersection matrix for $\{ \a, \a_i, \sigma_\mu^Y, \b^i, \b\}$ is more involved, as \eqref{Ft-intersection} receive corrections that destroy its block-anti-diagonal form and which, in the mirror four-fold $X_8$, arise due to curvature terms. We discuss such corrections in subsection \ref{Ft-sec:poly}, where we show that they can be absorbed in a redefinition of the $G_4$-flux quanta. Thus, for the purpose of providing an explicit expression for the F-term potential, one may still work with these naive intersection numbers.

To compute the flux superpotential we only need to expand the flux $G_4$ in the same basis of four-forms
\be
G_4 =  m \a  - m^i \a_i  +\hat{m}^{\mu} \sigma_\mu^Y - e_i \b^i  + e \b\, ,
\label{Ft-G4}
\ee
where $m, m^i, \hat{m}^\mu, e_i, e \in \Z$ represent the flux quanta. Using \eqref{Ft-intersection} we obtain that the superpotential takes the form
\be
W  =  e +  e_iT^i + \frac{1}{2}\, \hat{m}^{\mu} \zeta_{\mu,kl}   T^k T^l  + \frac{1}{6}\, {\cal K}_{ijkl}\, m^i T^j T^k T^l +  \frac{m}{24}\, {\cal K}_{ijkl}\, T^i T^j T^k T^l \, .
\label{Ft-supo}
\ee
One can obtain a more symmetric expression by considering a set of integers $m^{ij}$ that satisfy
\be
\hat{m}^{\mu} = \oh  \zeta_{ij}^\mu m^{ij}\, ,
\label{Ft-hatm}
\ee
so that the superpotential becomes
\be
W  =  e +  e_iT^i + \frac{1}{4}\, {\cal K}_{ijkl}  m^{ij} T^k T^l  + \frac{1}{6}\, {\cal K}_{ijkl}\, m^i T^j T^k T^l +  \frac{m}{24}\, {\cal K}_{ijkl}\, T^i T^j T^k T^l \, .
\label{Ft-supalt}
\ee
In general the choice of $m^{ij}$ is not unique, but it is easy to see that any choice will yield the same final expression. We will predominantly use the form of the superpotential \eqref{Ft-supo}, although in some instances it will be more convenient to use the auxiliary expression \eqref{Ft-supalt} that involves the redundant set of fluxes $m^{ij}$.

Notice that this superpotential is nothing but a linear combination of the central charges $\Pi_{2p}$ in \eqref{Ft-eq:periods} which, upon mirror symmetry becomes a linear combination of the periods of $\Omega$. Indeed, we have that
\be
W = \vec{q}^{\, t} \Sigma \vec{\Pi} = e \Pi_0 - e_i \Pi_2^i + \hat{m}^\mu \Pi_{4\, \mu} - m^i \Pi_{6\, i} + m \Pi_8\, , 
\label{Ft-supoprod}
\ee
which clearly reproduces \eqref{Ft-supo}. Here we have defined the vector of fluxes $\vec{q}^{\, t} = (m, m^i,  \hat{m}^\mu, e_i, e)$, the vector of periods $\vec{\Pi}^{\, t} = (\Pi_0, \Pi_2^i, \Pi_4^\mu, \Pi_{6\, i}, \Pi_8)$ and the pairing matrix
\begin{align}
    \Sigma = \left(\begin{matrix} 0&0&0&0&1 \\ 0&0&0&-\delta_j^i &0 \\ 0&0&\d_{\mu \nu} &0&0\\ 0&-\delta_i^j &0&0&0 \\ 1&0&0&0&0   \end{matrix}\right)\, .
    \label{Ft-pairingm}
\end{align}

We can also use \eqref{Ft-eq:periods}, \eqref{Ft-Omega} to compute the piece of the K\"ahler potential \eqref{Ft-kahler}. We have that
\be
    K_{\rm cs}=-\log\left[2\re (\pi_0\bar{\pi}^8)\int_{Y_8}\alpha\wedge {\beta}+2\re (\pi_2^i\bar{\pi}^6_j)\int_{Y_8}\alpha_i\wedge {\beta}^j
 +  \pi_4^\mu\bar{\pi}_4^\nu \int_{Y_8}\sigma_\mu^Y\wedge \sigma_\nu^Y \right]\, ,
  \label{Ft-Kcscomp}
\ee
from where we obtain
\begin{equation}
    K_{\rm cs}=-\log(\frac{2}{3}\mathcal{K}_{ijkl}t^it^jt^kt^l)\, .
    \label{Ft-Kcs}
\end{equation}
As expected, in this large complex structure limit the leading term of the K\"ahler potential only depends on $t^i \equiv \im T^i$, and so the field space metric displays abundant continuous shift symmetries. As we will see below, polynomial corrections to the periods \eqref{Ft-eq:periods} do modify \eqref{Ft-Kcs}, but they do not introduce a dependence on $b^i \equiv \re T^i$. This can be expected from considering type IIA compactified on the mirror manifold $X_8$, where the $b^i$ correspond to integrals of the B-field. In the large volume limit these fields can be considered as axions, since the only terms breaking the continuous shift symmetry are generated by world-sheet instanton effects and are therefore suppressed as $e^{2\pi i T^in_i  }$, $n_i \in \Z$. The same statement applies to our setup, where the periodic nature of the fields $b^i$ translates into the familiar set of monodromies $\cT_i$ around the large  complex structure point, which act non-trivially on the basis $\{ \a_0, \a_i, \sigma_\mu, \b^i, \b^0\}$, the periods $\Pi_{2p}$ and the flux quanta, but leave $\Omega$ and $G_4$ invariant. 

This large set of axionic variables allows us to derive a simple, analytic expression for the F-term scalar potential. The main observation is that one should express the scalar potential in terms of a set of axion polynomials $\rho_A$ linear on the flux quanta, which are invariant under the action of the monodromies $\cT_i$. Because at the two-derivative level the scalar potential is quadratic in the fluxes, one recovers an expression of the form
\be
V = \oh Z^{AB} \rho_A \rho_B \, ,
\label{Ft-bilinear}
\ee
where $\rho_A \equiv \rho_A(b)$ are independent of the saxions $t^i$. The matrix entries $Z^{AB}$ do not depend on the fluxes, and so they can only depend on the axions through periodic functions. However, such periodic functions necessarily enter the periods of $\Omega$ through terms of the form $e^{2\pi i T^in_i  }$, which are exponentially suppressed in the large complex structure regime. Therefore under our assumptions we have that $Z^{AB} \equiv Z^{AB}(t)$ only depends on the saxions of the compactification, providing a simple, factorized bilinear structure for the F-term scalar potential. This same strategy was applied for type IIA 4d flux compactifications in \cite{Bielleman:2015ina,Carta:2016ynn,Herraez:2018vae,Marchesano:2020uqz}, where a potential with the structure \eqref{Ft-bilinear} was obtained, in agreement with general EFT considerations \cite{Farakos:2017jme,Bandos:2018gjp,Lanza:2019xxg}. As shown in \cite{Escobar:2018tiu,Escobar:2018rna,Marchesano:2019hfb,Marchesano:2020uqz}, this bilinear structure allows one to characterize the set of vacua in a simple, systematic manner, and even to determine the behaviour of the system away from them \cite{Valenzuela:2016yny,Grimm:2019ixq,Grimm:2020ouv}.  In section \ref{Ft-s:vacua} we will use the form \eqref{Ft-bilinear} of the F-theory F-term potential to classify the set of flux vacua at large complex structure. Finally, as pointed out in \cite{Grimm:2019ixq}, the same bilinear expression \eqref{Ft-bilinear} holds near other points at infinite distance in complex structure field space, and so in principle our strategy could be extended to these regions as well. 

To find the bilinear expression \eqref{Ft-bilinear} one must use the well-known no-scale properties of F-theory compactifications to simplify the Cremmer et al. \cite{Cremmer:1982en} formula for the F-term potential. In particular, the fact that the K\"ahler moduli do not appear in the superpotential translates into the following simplified expression \cite{Haack:2001jz,Giddings:2001yu}
\begin{equation}
    V=e^{K}\sum_{i,j} K^{i\bar{j}}D_{i}W D_{\bar{j}}\overline{W}\, ,
    \label{Ft-Cremmer}
\end{equation}
where $i, j = 1, \dots , h^{3,1}(Y_8)$ run over the complex structure moduli of $Y_8$. Here $D_i = \p_i + (\p_i K)$ stands for the supergravity covariant derivative, while $K^{i\bar{j}}$ is the inverse of the K\"ahler metric $K_{i\bar{j}} \equiv \p_i\p_{\bar{j}} K$. Because the K\"ahler potential is independent of the complex structure axions, it is more convenient to express both in terms of tensors with real indices $g_{ij} \equiv \frac{1}{4} \p_{t^i}\p_{t^j} K =  K_{i\bar{j}}$. These read
\be
g_{ij} = 4\frac{\mathcal{K}_i\mathcal{K}_j}{\mathcal{K}^2}-3\frac{\mathcal{K}_{ij}}{\mathcal{K}}\, \qquad g^{ij} =\frac{4}{3}t^it^j-\frac{1}{3}\mathcal{K}\mathcal{K}^{ij} \, ,
  \label{Ft-metric}
\ee
with $\cK^{ij}$ the inverse of $\cK_{ij}$, and we have defined the contractions
\be
 \mathcal{K}\equiv\mathcal{K}_{ijkl}t^it^jt^kt^l\, , \quad \mathcal{K}_i\equiv \mathcal{K}_{ijkl}t^jt^kt^l\, , \quad \mathcal{K}_{ij}\equiv \mathcal{K}_{ijkl}t^kt^l\, , \quad \mathcal{K}_{ijk}\equiv \mathcal{K}_{ijkl}t^l\, .
\ee

The expression \eqref{Ft-Cremmer} is already positive semi-definite and  bilinear, but still not of the form \eqref{Ft-bilinear}. To make explicit the factorization between axions and saxions, one must define the flux-axion polynomials $\rho_A$, which capture the discrete symmetries of the superpotential, and whose geometric interpretation and general definition is given in appendix \ref{Ft-ap:georho}. In our setup they read
\begin{subequations}
\label{Ft-rhos}
\begin{align}
    \rho=&\ e+e_ib^i+\frac{1}{2}\hat{m}^\mu\zeta_{\mu,kl} b^k b^l +\frac{1}{6}\mathcal{K}_{ijkl}m^ib^jb^kb^l+\frac{1}{24}m \mathcal{K}_{ijkl}b^ib^jb^kb^l\, , \\
    \rho_i=&\ e_i+\hat{m}^\mu\zeta_{\mu,il} b^l+\frac{1}{2}\mathcal{K}_{ijkl}m^jb^kb^l+\frac{1}{6}m\mathcal{K}_{ijkl}b^jb^kb^l\\
    \hat{\rho}^\mu=&\ \hat{m}^\mu+\zeta_{ij}^\mu b^im^j+\frac{1}{2}\zeta_{ij}^\mu b^ib^j\ , \\
    \tilde{\rho}^i=&\ m^i+mb^i\, ,\\
    \tilde{\rho}=&\ m\, .
    \end{align}
    \end{subequations}
As pointed out in \cite{Herraez:2018vae}, these polynomials are related to each other via derivatives, leading to a convenient way to express for the superpotential and F-terms. For the case at hand we have
\begin{subequations}
    \label{Ft-eq: superpotential and deriv}
\begin{align}
\label{Ft-eq:suporho}
    W=&\rho+i\rho_it^i-\frac{1}{2}\zeta_\mu\hat{\rho}^\mu-\frac{i}{6}\mathcal{K}_i\tilde{\rho}^i+\frac{\mathcal{K}}{24}\tilde{\rho}\, ,\\
    \partial_{i}W=&\rho_i+i\zeta_{\mu i}  \hat{\rho}^\mu-\frac{1}{2}\mathcal{K}_{ij}\tilde{\rho}^j-\frac{i}{6}\mathcal{K}_i\tilde{\rho}\, ,
\end{align}
\end{subequations}
together with $\partial_j K = 2 i \cK_j/\cK$, and where we have defined the contractions $\zeta_\mu \equiv \zeta_{\mu,ij} t^it^j$ and $\zeta_{\mu i} \equiv \zeta_{\mu,ij} t^j$. Plugging these expressions into \eqref{Ft-Cremmer} and using the properties of the metrics \eqref{Ft-metric} one finds the following expression for the F-theory flux potential
\begin{align}
    V= e^{K} \left[4\left(\rho-\frac{\mathcal{K}}{24}\tilde{\rho}\right)^2+g^{ij}\left(\rho_i+\frac{\mathcal{K}}{6}g_{ik}\tilde{\rho}^k\right)\left(\rho_j+\frac{\mathcal{K}}{6}g_{jl}\tilde{\rho}^l\right)+
    g^{ij}_{P}\zeta_{\mu i}\zeta_{\nu j} \hat{\rho}^\mu  \hat{\rho}^\nu
    \right]\, ,
    \label{Ft-scalarpot}
\end{align}
where $g_P^{ij}$ is the primitive component of the inverse metric, i.e. $g_P^{ij}=\frac{1}{3}(t^it^j-\mathcal{K}\mathcal{K}^{ij})$. This expression for the potential is one of the main results of this section. It reproduces the bilinear, factorized structure in \eqref{Ft-bilinear} as a sum of three positive semi-definite terms, that correspond to a block-diagonal structure for the saxion-dependent matrix $Z$. Indeed, if we arrange the flux-axion polynomials in a vector of the form
\be
\vec{\rho}^{\, t} = \left(\tilde{\rho}, \tilde{\rho}^i,   \hat{\rho}^{\mu}, \rho_i,   \rho   \right) \, ,
\label{Ft-vecrho}
\ee
then the said matrix reads
\begin{align}
\label{Ft-ZAB}
Z^{AB} =
\frac{e^K\cK}{3}
\begin{pmatrix}
\frac{\cK}{24} & & & & -1 \\
&  \frac{\cK}{6} g_{ij} & &  \d^i_j &  \\
& & \frac{6}{\cK}  g^{ij}_{P}\zeta_{\mu i}\zeta_{\nu j} & &  \\
&  \d^i_j & &  \frac{6}{\cK} g^{ij} &  \\
-1 & & & &  \frac{24}{\cK} \\
\end{pmatrix} \, ,
\end{align}
which can be easily taken to a block-diagonal form. Notice that each block is singular, and that their ranks add up to $2h^{3,1}(Y_8)$. Therefore, generically the vacua equations $Z^{AB}\rho_B =0$ amount to impose $2h^{3,1}(Y_8)$ conditions on the same amount of unknowns, namely the complex structure real fields. Finally, note that we can rewrite this expression as 
\begin{align}
\label{Ft-ZABdiag}
2{\cal V}_3^2 Z =  {\rm diag} \left(\frac{\cK}{24},  \frac{\cK}{6} g_{ij},  g_{\mu\nu},  \frac{6}{\cK}g^{ij},  \frac{24}{\cK} \right)
- 
\chi_0\, ,
\end{align}
where $g_{\mu\nu} \equiv \eta_{\mu\nu} -2 (\cK^{ij}-\cK^{-1} t^it^j) \zeta_{\mu i}\zeta_{\nu j}$ and 
\begin{align}\label{Ft-eq:chi0}
    \chi_0 = \left(\begin{matrix} 0&0&0&0&1 \\ 0&0&0&-\delta_j^i &0 \\ 0&0&\eta_{\mu\nu} &0&0\\ 0&-\delta_i^j &0&0&0 \\ 1&0&0&0&0   \end{matrix}\right)\, ,
\end{align}
encodes the intersection numbers  \eqref{Ft-intersection}.
 As it follows from the results of appendix \ref{Ft-ap:georho}, splitting $Z$ in these two terms corresponds to the well-known expression for the scalar potential
\be
V = \frac{1}{4{\cal V}_3^2} \left[ \int_{Y_8} G_4 \wedge \star G_4 -   \int_{Y_8} G_4 \wedge G_4\right]\, ,
\ee
at this level of approximation. As we will see below, the polynomial corrections to the scalar potential will respect the factorization between axions and saxions, and therefore the bilinear structure \eqref{Ft-bilinear}. On the one hand, the corrections to the intersection numbers \eqref{Ft-intersection} will modify $\vec{\rho}$ but not $Z$. On the other hand, the corrections to the K\"ahler potential  \eqref{Ft-Kcs}  will leave $\vec{\rho}$ invariant but destroy the block-diagonal structure of $Z$.

It is instructive to compare the above results with previous analysis in the literature. For instance, one would recover the F-theory flux potential analyzed in \cite{Marsh:2015zoa} by setting $m^i = \hat{m}^\mu = e_i = e =0$ and keeping only $m$ as a non-vanishing quantum of flux. The scalar potential would still look the same, but the axion dependence in \eqref{Ft-rhos} would become very simple. As we will see in section \ref{Ft-s:vacua}, vacua with $m \neq 0$ are not allowed at sufficiently large complex structure, in agreement with the result of \cite{Marsh:2015zoa}. Including the remaining flux quanta does a priori allow us to find non-trivial extrema of the potential, as we will also study in the next section.

One may also compare \eqref{Ft-scalarpot} with the asymptotic potentials analyzed in \cite{Grimm:2019ixq} restricted to the particular case of the  large complex structure limit. In the language of \cite{Grimm:2019ixq}, the approximation that leads to the expression \eqref{Ft-scalarpot} lies in between those that result in the asymptotic form of the potential and its strictly asymptotic form. To achieve the latter, one must take the expression \eqref{Ft-ZABdiag} and replace each of the entries in $ {\rm diag} \left(\frac{\cK}{24},  \frac{\cK}{6} g_{ij},  g_{\mu\nu},  \frac{6}{\cK}g^{ij},  \frac{24}{\cK} \right)$ by its leading term on the complex structure saxions $t^a$, which amounts to replace the Hodge star operator by its strictly asymptotic approximation $C_{\rm sl(2)}$. The plain asymptotic form of the potential (that is, replacing $\star$ by $C_{\rm nil}$) is achieved by adding further polynomial corrections to \eqref{Ft-scalarpot}, which we now turn to analyze. As we will see, full moduli stabilization is only achieved when these corrections are taken into account. Moreover, their presence leads to important restrictions on the space of flux vacua, which remain undetected if only the strictly asymptotic form of the potential is used.

\subsection{Polynomial corrections}
\label{Ft-sec:poly}

The leading form of the potential \eqref{Ft-scalarpot} receives several corrections of different nature, which can be classified in terms of corrections to the superpotential and K\"ahler potential. In the following we will address those that depend on the complex structure sector and are polynomial corrections to $W$ and $e^{-K}$. 
These can be treated like perturbative corrections to the leading potential, as opposed to exponentially-suppressed corrections. 
Taking these polynomial corrections into account permits to extend our analysis to regions where the complex structure saxions are only moderately large, so that the exponential corrections of the form $e^{2\pi i T^in_i }$ can still be neglected. The reader not interested in the details of the following derivation may only focus on the results \eqref{Ft-supofinal} and \eqref{Ft-Kcscorr}, that summarize the polynomial corrections for the superpotential and K\"ahler potential, and proceed to the next section.

To compute the said corrections let us again consider type IIA compactified in the mirror four-fold $X_8$. Here the polynomial corrections that arise in the K\"ahler sector are due to curvature corrections, while the exponential corrections that we will neglect arise from world-sheet instanton effects. The polynomial corrections are encoded in the asymptotic expression for the D$(2p)$-brane central charges, as computed in \cite{Gerhardus:2016iot} and reviewed in appendix \ref{Ft-sap:corrper}. They correct the leading terms in  \eqref{Ft-eq:periods} as
\begin{subequations}
\label{Ft-eq:corrper}
\begin{align}
    \Pi_0^{\rm corr}&=1\, ,\\
     \Pi_{2}^{i\, {\rm corr}}&=-T^i\, ,\\
     \Pi_{4\, ij}^{\rm corr} &=\frac{1}{2}\cK_{ijkl}T^kT^l+\frac{1}{2}\left(\CK_{iijk}+\CK_{ijjk} \right) T^k + \frac{1}{12}\left(2\CK_{iiij} + 3\CK_{iijj} + 2\CK_{ijjj} \right) +  K_{ij}^{(2)} \, ,\\
    \Pi_{6\, i}^{\rm corr} &=-\frac{1}{6}\cK_{ijkl}T^jT^kT^l-\frac{1}{4}\cK_{iijk}T^jT^k-\frac{1}{6}\cK_{iiij}T^j - K^{(2)}_{ij}T^j  + \half K^{(2)}_{ii} +i K_i^{(3)} \, ,\\
    \Pi_8^{\rm corr}&=\frac{1}{24}\cK_{ijkl}T^iT^jT^kT^l+ \half K_{ij}^{(2)} T^iT^i - iK_i^{(3)} T^i + K^{(0)}\, , 
\end{align}
\end{subequations}
where we have defined
\be
K^{(2)}_{ij} = \frac{1}{24}\int_{X_8} c_2(X_8) \wedge D_i \wedge D_j\, , \qquad K_i^{(3)} = -\frac{\zeta(3)}{8\pi^3}\int_{X_8} c_3(X_8)\wedge D_i \, ,
\label{Ft-K23}
\ee
and
\be
K^{(0)} = \frac{1}{5760}\int_{X_8} 7c_2(X_8)^2-4c_4(X_8)\, .
\label{Ft-K0}
\ee
Notice that here we are working with the redundant set of four-cycles $\g_{ij}= D_i .  D_j$. 

From these expressions it is easy to compute how the corrected version of the F-theory superpotential \eqref{Ft-supo} looks like. Indeed, mirror symmetry translates \eqref{Ft-eq:corrper} into the corrected periods of $\Omega$ in $Y_8$, and so one simply needs to multiply them by the $G_4$ flux quanta, as in \eqref{Ft-supoprod}. In this case it is more convenient to work with the auxiliary flux quanta $m^{ij}$ defined in \eqref{Ft-hatm}, and therefore to extend the  flux vector to $\vec{q}^{\, \prime\, t} = (m, m^i,  m^{ij}, e_i, e)$. One then finds that
\be
W^{\rm corr} = \vec{q}^{\, \prime\, t} \Sigma \vec{\Pi}^{\rm corr} = e \Pi_0 - e_i \Pi_2^i + \half m^{ij} \Pi_{4\, ij}^{\rm corr} - m^i \Pi_{6\, i}^{\rm corr} + m \Pi_8^{\rm corr}\, ,
\label{Ft-supoprodcorr}
\ee
where $\Sigma$ is the obvious extension of \eqref{Ft-pairingm} to the auxiliary flux basis. This is a rather involved expression, but it becomes more manageable if one distinguishes between two classes of corrections that appear in the periods of $\Omega$. The first one corresponds to corrections to the intersection numbers \eqref{Ft-intersection}, and the second one to the K\"ahler potential \eqref{Ft-Kcs}. As we will see, each of these corrections has a different effect on the F-term scalar potential, which becomes more transparent when it is written in the bilinear form \eqref{Ft-bilinear}.

To compute the corrections to the intersection numbers \eqref{Ft-intersection}, one may again consider type IIA compactified on the mirror manifold $X_8$. There, two D$(2p)$-branes wrapping holomorphic cycles on $X_8$ of complementary dimension have a natural topological intersection number, that can be thought of as the mirror dual to \eqref{Ft-intersection}. Then, on a D-brane wrapping a $2p$-cycle with $p \geq 2$, a non-trivial curvature may induce lower-dimensional D-brane charges. This affects the index that counts the open strings stretching between the two D-branes, and which in the absence of induced charges amounts to the intersection number between cycles. The curvature-corrected open string index between two B-branes $\mathcal{E}$ and $\mathcal{F}$ reads
\begin{align}\label{Ft-intersectionmatrix}
    \chi(\mathcal{E}, \mathcal{F}) = \int_{X_8} \text{Td}(X_8)\lambda\left(\text{ch}\, \mathcal{E}\right) \left(\text{ch}\, \mathcal{F}\right)\,,
\end{align}
where $\text{ch}\, \mathcal{E}$ is the Chern character of $\mathcal{E}$, and the Todd class for a Calabi--Yau four-fold is
\begin{align}\label{Ft-eq: Todd}
    \text{Td}(X_8) = 1 + \frac{c_2}{12} + \frac{3c_2^2 - c_4}{720} \,. 
\end{align}
Finally, for an element $\beta \in H^{2k}(Y,\mathbb{Z})$ we define $\lambda(\beta) = (-1)^k \beta$ (operator that reverts the order of the indices of a p-form). It is the topological index \eqref{Ft-intersectionmatrix} that is well-behaved under the mirror map, and gives the actual intersection numbers of the four-forms that appear in \eqref{Ft-Omega}, instead of \eqref{Ft-intersection}. Nevertheless, it turns out that, upon applying the proper redefinitions, one can still use the intersection matrix \eqref{Ft-intersection}.

Indeed, the open string index for holomorphic $2p$-cycles on $X_8$ is computed in appendix \ref{Ft-sap:corrper}, with the result
\begin{align}\label{Ft-eq:chicorr}
    \chi = \Lambda^T \chi_0 \Lambda \,,
\end{align}
where $\chi_0$ is defined as in \eqref{Ft-eq:chi0} and
\begin{align}\label{Ft-eq:Lambda}
    \Lambda = \left(\begin{matrix} 1&0&0&0&0 \\ 0&\delta_i^j & 0&0&0 \\ \frac{1}{24} c_2^\mu &-\frac{1}{2}\zeta^\mu_{ii} & \zeta^\mu_{kl} & 0&0 \\ 0 & \frac{1}{6}\cK_{jjji} + K^{(2)}_{ji} & -\frac{1}{2}\left(\cK_{jkkl} +\cK_{jkll}\right) &\delta_j^i&0 \\ 
   K^{(0)} & -\frac{1}{24} \cK_{iiii} - \half K^{(2)}_{ii} & \lambda_{kl} &0&1 \end{matrix}\right)\,,
\end{align}
contains the corrections induced by the curvature. Here we have defined $c_2(X_8) = c_2^\mu \sigma_\mu$ and $ \lambda_{kl}= \frac{1}{12}\left(2\cK_{kkkl} + 3\cK_{kkll} +2 \cK_{klll}\right) + K^{(2)}_{kl}$. Notice that $\Lambda$ is independent of $K_i^{(3)}$.

With these expressions at hand, it is easy to see that the superpotential \eqref{Ft-supoprodcorr} can be rewritten as
\be
W^{\rm corr} = \left(\Lambda \sigma \vec{q}^{\, \prime}\right)^{\, t}\cdot  \chi_0 \cdot \left(\Lambda \vec{\pi}^{\rm corr}\right) \, ,
\label{Ft-suporot}
\ee
where $\sigma$ is a diagonal matrix with entries $1$ and $-1$ chosen so we keep the relative signs as in the expansion \eqref{Ft-G4} and recover an analogue expression to the standard superpotential \eqref{Ft-supo}. $\Lambda \vec{\pi}^{\rm corr}$ is given by
 
\be
\Lambda \vec{\pi}^{\rm corr} = \left(\begin{matrix}1&0&0&0&0 \\ 0&\d^j_i&0&0&0 \\0&0&\d_\mu^\nu&0&0 \\ -iK_i^{(3)}&0&0&\d_j^i&0 \\0&-iK_i^{(3)} &0&0&1 \end{matrix}\right) \left(\begin{matrix} \pi_0 \\ \pi^i_2 \\ \pi^{\mu}_4 \\ \pi_{6i} \\ \pi_8 \end{matrix} \right) \, .
\ee
The components of $\vec{\pi}^{\rm corr}$ can be interpreted as the corrected moduli-dependent coefficients of $\Omega$ in the expansion \eqref{Ft-Omega}. Here we will not need the precise expression of such components, because the quantities of interest only depend on $\Lambda \vec{\pi}^{\rm corr}$. The expression \eqref{Ft-suporot} implies that, when taking into account the polynomial corrections in our F-theory setup, one can still use the classical intersection numbers \eqref{Ft-intersection} if one makes the replacements
\be
\vec{q} \to \sigma^{-1} \Lambda \sigma \vec{q}^{\, \prime} \, , \qquad \vec{\pi} \to \Lambda \vec{\pi}^{\rm corr}\, ,
\ee
in all the computations of the previous subsection. That is, in \eqref{Ft-Omega} we perform the replacements
\be
\pi_{6i} \to \pi_{6i} -i K_i^{(3)} \, , \qquad \pi_8 \to \pi_8 -i  K_i^{(3)} T^i\, ,
\label{Ft-picorr2}
\ee
and in \eqref{Ft-G4} we replace the flux quanta by
\begin{subequations}\label{Ft-eq:fluxshift}
\begin{align}
    \bar{m}^\mu &=  \hat{m}^{\mu} + \frac{1}{2} \zeta^\mu_{ii} m^i +\frac{m}{12} c_2^\mu \,,\\
    \bar{e}_j &= e_j +\frac{m^i}{6} \cK_{jjji} + m^i K_{ij}^{(2)}+\frac{1}{4} \left(\cK_{jkkl}+\cK_{jkll}\right) m^{kl}\,,\\
    \bar{e} &= e +\frac{1}{2} m^{jk} \lambda_{jk} +m^i\left(\frac{1}{24} \cK_{iiii} + \oh K_{ii}^{(2)}\right) + m K^{(0)}\,. 
\end{align}
\end{subequations}
To sum up, the corrected expression for the GVW superpotential takes the form
\be
W^{\rm corr}  =  \bar{e} +  \bar{e}_iT^i + \frac{1}{2}\,  \bar{m}^{\mu} \zeta_{\mu,kl}  T^k T^l  + \frac{1}{6}\, {\cal K}_{ijkl}\, m^i T^j T^k T^l +  \frac{m}{24}\, {\cal K}_{ijkl}\, T^i T^j T^k T^l -i  K_i^{(3)} \left(m^i + m T^i\right)  \, . 
\label{Ft-supofinal}
\ee

This strategy to rewrite the superpotential not only gives a more manageable expression, but also yields the corrected K\"ahler potential as a byproduct. Indeed, it follows that the corrections to \eqref{Ft-Kcs} can be computed from the expression \eqref{Ft-Kcscomp}, by performing the replacements \eqref{Ft-picorr2}. One then finds that
\be
K_{\rm cs}^{\rm corr}=-\log\left(\frac{2}{3}\mathcal{K}_{ijkl}t^it^jt^kt^l + 4 K_i^{(3)} t^i \right)\, .
\label{Ft-Kcscorr}
\ee
Notice that this expression still respects the continuous shift symmetry of the axionic fields $b^i$ and it only depends on the type IIA $\alpha'^3$-corrections that correspond to the third Chern class of $X_8$. It is also a natural generalization of the $\alpha'^{3}$-correction to the K\"ahler potential in type IIA compactifications in Calabi--Yau three-folds, see e.g. \cite{Palti:2008mg,Escobar:2018rna}. In appendix \ref{Ft-sap:corrka} we rederive the same expression using a different method, as a cross-check of our results.

From these expressions one can derive the corrections to the F-term scalar potential \eqref{Ft-scalarpot}. For this, it is useful to write the superpotential and its derivatives in terms of shifted axion polynomials. We have that 
\begin{subequations}
    \label{Ft-eq:supo and der corr}
\begin{align}
    W^{\rm corr}&=\, \bar{\rho}+i\bar{\rho}_it^i-\frac{1}{2}\zeta_\mu \bar{\rho}^{\mu}-i \left(\frac{1}{6}\mathcal{K}_i + K_i^{(3)}\right)\tilde{\rho}^i + \left(\frac{\mathcal{K}}{24} +  K_i^{(3)}t^i \right)\tilde{\rho}  \, ,\\
    \partial_{i}W^{\rm corr}&=\bar{\rho}_i+i\zeta_{\mu i}  \bar{\rho}^\mu-\frac{1}{2}\mathcal{K}_{ij}\tilde{\rho}^j -i\left(\frac{\mathcal{K}_i}{6}+K_i^{(3)}\right)\tilde{\rho}\, \, ,
\end{align}
\end{subequations}
where
\begin{subequations}
\label{Ft-corrhos}
\begin{align}
     \bar{\rho}=&\ \bar{e}+\bar{e}_ib^i+\frac{1}{2}\bar{m}^\mu\zeta_{\mu,kl} b^k b^l +\frac{1}{6}\mathcal{K}_{ijkl}m^ib^jb^kb^l+\frac{1}{24}m \mathcal{K}_{ijkl}b^ib^jb^kb^l\, , \\
    \bar{\rho}_i=&\ \bar{e}_i+\bar{m}^\mu\zeta_{\mu,il} b^l+\frac{1}{2}\mathcal{K}_{ijkl}m^jb^kb^l+\frac{1}{6}m\mathcal{K}_{ijkl}b^jb^kb^l\\
    \bar{\rho}^\mu=&\ \bar{m}^\mu+\zeta_{ij}^\mu b^im^j+\frac{1}{2}m\zeta_{ij}^\mu b^ib^j\ , \\
        \tilde{\rho}^i=&\ m^i+mb^i\, ,\\
    \tilde{\rho}=&\ m\, .
    \end{align}
    \end{subequations}
Notice that if we take $K^{(3)}_i \to 0$ the corrected scalar potential reduces to \eqref{Ft-scalarpot}, except for the flux redefinition \eqref{Ft-eq:fluxshift} that only replaces the components of \eqref{Ft-vecrho} by \eqref{Ft-corrhos}. As we show in appendix \ref{Ft-sap:corrFterm}, the effect of a non-vanishing $K^{(3)}_i$ is to modify the matrix \eqref{Ft-ZAB}, inducing  new non-vanishing entries that destroy its block-diagonal structure. Due to its complicated form, it is easier to characterize the corrections to the vacua equations in terms of the vanishing conditions for the corrected F-terms, as we do in appendix \ref{Ft-sap:corrvac}.

\subsubsection*{Monodromies}

The above expressions allow us to connect the definition of $\vec{\rho}$ with the monodromies that act on the periods of $\Omega$. For this it is useful to describe the superpotential in terms of the vector of auxiliary fluxes  $\vec{q}^{\, \prime\, t} = (m, m^i,  m^{ij}, e_i, e_0)$, as in \eqref{Ft-supoprodcorr}. Then one can rewrite this expression as 
\be
W^{\rm corr} = \left(\hat{R} \vec{q}^{\, \prime}\right)^{\, t}  \hat{R}^{t\, -1} \Sigma \vec{\Pi}^{\rm corr}\, ,
\label{Ft-Wmono}
\ee
where
\be
\hat{R} \equiv \hat{\Lambda}^{-1} R(b) \hat{\Lambda} = e^{b^i \hat{P}_i} \, ,
\ee
with $\hat{\Lambda}$ the extension of $\Lambda$ to a square matrix, as defined in \eqref{Ft-hatLambda}, and 
\begin{align}
\label{Ft-eq: axion matrix}
    R(b) &= \,  \begin{pmatrix}
    1&0&0&0&0 \\ b^i&\delta_k^i & 0&0&0 \\   b^ib^j & b^i\delta_k^j + b^j\delta_k^i  & \delta^i_k \delta^j_l & 0&0 \\ \frac{1}{6}\mathcal{K}_{ijkl}b^jb^kb^l & \frac{1}{2}\mathcal{K}_{ijkl}b^jb^l & \frac{1}{2}\mathcal{K}_{ijkl} b^j &\delta_j^i&0 \\ 
   \frac{1}{24} \mathcal{K}_{ijkl}b^ib^jb^kb^l & \frac{1}{6}\mathcal{K}_{ijkl}b^ib^jb^l & \frac{1}{4}\mathcal{K}_{ijkl}b^ib^j &b^i&1
    \end{pmatrix}\, , \\
  \hat{P}_n  & = \,   \hat{\Lambda}^{-1} P_n \hat{\Lambda} = 
  \hat{\Lambda}^{-1} 
   \begin{pmatrix}
    0&0&0&0&0 \\ \delta^i_n &0 & 0&0&0 \\   0 & \delta^i_n \delta^j_k +   \delta^i_k \delta^j_n  & 0 & 0&0 \\ 0 & 0 & \half\mathcal{K}_{inkl}  & 0 &0 \\ 
  0 &0 & 0 & \delta^i_n &0
    \end{pmatrix}
   \hat{\Lambda}\, .
\end{align}
Here $R(b)$ is the axion-dependent rotation matrix which transforms the flux vector into the vector of flux-axion polynomials as  $R\vec{q}^{\, \prime} = \vec{\rho}^{\, \prime}$, where $\vec{\rho}^{\, \prime\, t} = \left(\tilde{\rho}, \tilde{\rho}^i,   {\rho}^{ij}, \rho_i,   \rho   \right)$ is the extension of \eqref{Ft-vecrho} to include the polynomials $  \rho^{ij}= m^{ij}+m^ib^j+m^jb^i+mb^ib^j$. The matrices $P_i$ are the generators of such a rotation. 

One can check that $\hat{R}^{t\, -1} \Sigma \vec{\Pi}^{\rm corr}$ does not depend on the axions $b^i$, and so that \eqref{Ft-Wmono} expresses the superpotential as a product of an axion-dependent and a saxion-dependent vector. From \eqref{Ft-eq:corrper} one obtains that the monodromy action on the periods
\begin{align}
    \vec{\Pi}^{\rm corr}(T^j+1) = {\cal T}_j \cdot \vec{\Pi}^{\rm corr}(T^j)\,,
\end{align}
is given by
\begin{align}
    {\cal T}_j = \left(\begin{matrix} 1 & 0 &0&0&0\\ -\delta_{j}^k & \delta_{i}^k &0&0&0\\ 0 & -\delta^i_j \delta^k_l  &\delta_{i}^k\delta_{j}^l &0&0\\ 0&0& -\half\cK_{ijkl} & \delta_{i}^k &0\\ 0 &0& \frac{\cK_{iijk}+ \cK_{ijkk}}{2}+\frac{\cK_{jjki}}{4}&-\delta_{i}^j &1 \end{matrix}\right)\,. 
\end{align}
This action is fully encoded in the rotation matrix $\hat{R}$, and more precisely in its generators $\hat{P}_i$. In particular we have that
\be
{\cal T}_i = \Sigma\, e^{\hat{P}_i^t}\, \Sigma = e^{-\hat{P}_i} \, .
\label{Ft-monoexp}
\ee


\section{Tadpoles and vacua}
\label{Ft-s:vacua}

With an explicit form for the F-term scalar potential in the large complex structure regime one may characterize the set of vacua in that region. We will pay particular attention to the fact that the flux contribution to the tadpole $N_{\rm flux}$ is bounded from above, something that forbids the presence of certain flux vacua at arbitrarily large complex structure. As we will see, this tadpole constraint leads to different moduli stabilization scenarios, classified by which flux components are turned on. In this section we will analyze the most generic of these scenarios, in which one can clearly see that the corrections $K_i^{(3)}$ to the K\"ahler potential are crucial to stabilize all moduli. As a direct consequence, one finds an upper bound for the vev of the complex structure saxions, that depends both on $K_i^{(3)}$ and $N_{\rm flux}$. One can also consider a quite different setup in which such a bound is absent, whose general discussion we leave for section \ref{Ft-s:linear}.

\subsection{General flux vacua}

Armed with the explicit form of the potential at large complex structure, one may now analyze its set of vacua. Let us first consider the leading flux potential \eqref{Ft-scalarpot}. Since it is a sum of three positive semi-definite terms and its dependence on the K\"ahler moduli only enters through the overall factor $e^{K} \cK \propto {\cal V}_3^{-2}$, its minima correspond to Minkowski vacua where these three terms vanish. In other words, we must impose the following set of on-shell conditions
\begin{subequations}
\label{Ft-eq:Mink}
\begin{empheq}[box=\widefbox]{align}
    \rho&=\frac{1}{24}\mathcal{K}\tilde{\rho} \label{Ft-eq:Mink 0}\\
    \rho_i&=-\frac{1}{6}\mathcal{K}g_{ij}\tilde{\rho}^j \label{Ft-eq:Mink i} \\
    0&= \left(  \cK \zeta_{\mu i}- \mathcal{K}_i\zeta_{\mu}\right) \hat{\rho}^\mu 
    \label{Ft-eq:Mink mu}
\end{empheq}
\end{subequations}
where the general solution for \eqref{Ft-eq:Mink mu} reads
\be
\hat{\rho}^\mu = A \zeta^\mu + C^\mu\, , \qquad \zeta_{\mu i} C^\mu =0\quad \forall i\, , 
\label{Ft-splitmu}
\ee
with $A$, $C^\mu$ moduli-dependent quantities. For those vacua that preserve supersymmetry, we need to impose that $W=0$ on-shell. From \eqref{Ft-eq:suporho} we see that this implies two additional conditions:
\be
t^i\rho_i = 0\, , \qquad  \zeta_\mu\hat{\rho}^\mu = \frac{\CK}{6}\tilde{\rho} \, .
\label{Ft-eq:SUSY}
\ee

From our discussion in the previous section it follows that, in order to implement the polynomial corrections that correspond to $K^{(0)}$ and $K_{ij}^{(2)}$, we only need to perform the replacement 
\be
(\rho, \rho_i, \hat{\rho}^\mu) \to (\bar\rho, \bar\rho_i, \bar{\rho}^\mu)
\label{Ft-reprho}
\ee 
in \eqref{Ft-eq:Mink} and \eqref{Ft-eq:SUSY}, with the new quantities given by \eqref{Ft-corrhos}. Therefore the above equations essentially hold whenever it is a good approximation to neglect the correction due to  $K_i^{(3)}$ in the K\"ahler potential \eqref{Ft-Kcscorr}. The vacua equations that follow from including  such a  correction to the K\"ahler potential are discussed in appendix \ref{Ft-sap:corrvac}. In here we simply collect the result, approximated to linear order in $\eps_i = 6K^{(3)}_i/\CK$:
\begin{subequations}
    \label{Ft-eq:Minkcorr}
\begin{align}
 \bar\rho- \frac{1}{24}\mathcal{K}\tilde{\rho}&  = - \frac{1}{48}\eps_it^i \left[ \cK\tilde{\rho} + 18 \zeta_\mu \bar\rho^{\mu} \right] \, , \\
 \label{Ft-eq:Minkcorrrhoi}
  \bar{\rho}_i+\frac{1}{6}\mathcal{K}g_{ij}\tilde{\rho}^j & =  \frac{1}{3}  \CK_i  \left( \eps_j -  \epsilon_k t^k \frac{ \cK_j}{\cK} \right)\tilde{\rho}^j   - \frac{1}{6}\epsilon_i \cK_j \tilde \rho^j   \, , \\
  \left(\zeta_{\mu i}- \frac{\mathcal{K}_i}{\CK} \zeta_{\mu}\right) \bar{\rho}^\mu& = \frac{1}{8} \left(\eps_i - \epsilon_k t^k \frac{\cK_i }{\cK} \right)   \left( \cK \tilde \rho +  2 \zeta_\mu \bar\rho^{\mu}\right)  \, .
 \end{align}
\end{subequations}
Finally, those vacua that are supersymmetric will satisfy the additional conditions
\bea
t^i \bar{\rho}_i = \frac{1}{4} \left( \cK \eps_i\tilde{\rho}^i  -  \epsilon_k t^k \cK_j\tilde{\rho}^j \right) \, ,
\qquad 
\zeta_\mu \bar{\rho}^{\mu} =\frac{\cK}{6}\left(1 + \eps_it^i \right) \tilde{\rho} \, ,
\label{Ft-eq:SUSYcorr}
\eea
up to quadratic terms in $\eps_i$. 

\subsection{The tadpole constraint}
\label{Ft-sec:tadpole}

We recall from \eqref{fb-eq: mtheory tadpole}, that in any consistent F-theory compactification on a four-fold $Y_8$ one must satisfy the D3-brane tadpole condition
\be
N_{\rm flux} = \oh \int_{Y_8} G_4 \wedge G_4 = \frac{\chi(Y_8)}{24} - N_{\rm D3}\, ,
\label{Ft-tadpole}
\ee
where $\chi(Y_8)$ is the Euler characteristic of $Y_8$, and $N_{\rm D3}$ is the number of space-time filling D3-branes. The number $\chi(Y_8)$ can take a range of values depending on the four-fold, but since stability of Minkowski vacua requires $N_{\rm D3}>0$, \eqref{Ft-tadpole} sets an upper bound for $N_{\rm flux}$. Thus, as we found in section \ref{fb-subsec: tadpole conjecture}, the allowed range of the flux contribution to the tadpole is $0\leq N_{\rm flux} \leq \chi(Y_8)/24$ for any Minkowski flux vacuum. To understand what this implies in our setup, one may easily compute the value of $N_{\rm flux}$ in terms of the expressions of section \ref{Ft-s:potential}. Starting from \eqref{Ft-G4} one finds 
\be
N_{\rm flux} \equiv \bar{e} m - \bar{e}_i m^i + \frac{1}{2} \eta_{\mu\nu} \bar{m}^{\mu} \bar{m}^{\nu}\, ,
\label{Ft-Nflux}
\ee
where the barred flux quanta are defined in \eqref{Ft-eq:fluxshift} and their presence arises from the corrections to the naive intersection numbers \eqref{Ft-intersection}. 

The interesting observation is that this expression for $N_{\rm flux}$ equals a bilinear of flux-axion polynomials, namely
\be
N_{\rm flux} = \bar{\rho} \tilde{\rho} - \bar{\rho}_i \tilde{\rho}^i + \frac{1}{2} \eta_{\mu\nu} \bar{\rho}^{\mu} \bar{\rho}^{\nu}\, .
\label{Ft-Nfluxrho}
\ee
One can check this identity directly, or by realising that the flux contribution to the tadpole \eqref{Ft-Nflux} is one of the flux monodromy-invariants that constrain the orbit of values that $\vec{\rho}$ can take. In fact, since the entries of $\vec{\rho}$ are invariant under monodromies as well, their on-shell value can only depend on such flux invariants and, because of \eqref{Ft-eq:Mink}, the same holds for the saxion vevs. The invariants that arise in generic F-theory flux  compactifications are listed in appendix \ref{Ft-ap:invariants}.

This last expression for $N_{\rm flux}$ can be evaluated at each vacuum via the on-shell conditions derived above. For simplicity, let us assume that we are in a sufficiently large complex structure regime such that the K\"ahler potential correction term $K^{(3)}_it^i$ in \eqref{Ft-Kcscorr} can be neglected. Then one may use \eqref{Ft-eq:Mink} with the replacement \eqref{Ft-reprho} to obtain
\be
N_{\rm flux} \stackrel{\rm vac}{=} \frac{\CK}{24} \left(\tilde{\rho}^2 + 4 g_{ij} \tilde{\rho}^i\tilde{\rho}^j \right)  +  \frac{1}{2} g_{\mu\nu} \bar{\rho}^{\mu} \bar{\rho}^{\nu}\, ,
\label{Ft-Nfluxvac}
\ee
where $g_{\mu\nu}$ is defined as in \eqref{Ft-ZABdiag}, and we have used that for a vector of the form \eqref{Ft-splitmu} we have that  $\eta_{\mu\nu} \hat{\rho}^\nu =  g_{\mu\nu} \hat{\rho}^\nu$, see appendix \ref{Ft-ap:georho} for details.

Along any limit of large complex structure we have that $\CK \to \infty$, because otherwise $\CK_i \to 0$ for at least some $i$, which takes us away from the regime of validity of our analysis. Then the question is if along these limits all terms on the rhs of \eqref{Ft-Nfluxvac} remain bounded from above. If they did not, no vacua would be found at sufficiently large complex structure, for any value of $\chi(Y_8)$. Since all terms are positive definite, they need to be bounded separately. 

The first term on the rhs of \eqref{Ft-Nfluxvac} is clearly unbounded, so we must impose $\tilde{\rho} = m = 0$, which then implies $\tilde{\rho}^i = m^i$. For the second term, the question is whether $\CK g_{ij} m^i m^j = (4 \CK_i\CK_j/\CK -3 \CK_{ij})m^im^j$ remains bounded or not along the different large complex structure limits. Those choices of $m^i$ where it is not bounded should be set to zero in order to find a consistent vacuum. This depends crucially on the topology of $Y_8$ through the quadruple intersection numbers $\CK_{ijkl}$ of its mirror $X_8$. A full classification of all possibilities should follow from the techniques developed in \cite{Grimm:2019ixq} applied to the special case of large complex structure limits. Here, we take a simplified approach by asking whether $\CK g_{jj}$ remains bounded or not in the case that we blow up a single modulus $t^i\rightarrow \infty$. If it does not, one should set $m^j=0$ to find vacua in that regime.

We can distinguish four different cases: 
\begin{itemize}
    \item[$(i)$] The modulus $t^i$ appears with a quartic term in the K\"ahler potential, i.e. $\cK_{iiii}\ne 0$. In this case the component $\CK g_{ii}$ is not bounded since 
    \begin{subequations}\label{Ft-Kgii1}
    \begin{align}
        {\cK} g_{ii}  \sim (t^i)^2 \rightarrow \infty\,.
    \end{align}
    In addition, for those indices $j \neq i$ such that $\cK_{iiij}\ne 0$, the diagonal term $\cK g_{jj}$ scales as
    \begin{align}
         \cK g_{jj} = \frac{4\cK_j\cK_j}{\cK}-3\cK_{jj}  \sim (t^i)^2 \rightarrow \infty\,, 
    \end{align}
    \end{subequations}
     and it is therefore also unbounded.  
    \item[$(ii)$] The modulus $t^i$ appears only cubic in the K\"ahler potential, i.e. $\cK_{iiii}=0$ but $\cK_{iiik}\ne 0$ for some $k\ne i$. In this case the component $\cK g_{ii}$ is unbounded as 
    \begin{subequations}\label{Ft-Kgii2}
    \begin{align}
        {\cK} g_{ii}  \sim \cK_{iiik} t^i t^k \rightarrow \infty\,,
    \end{align}
    with no summation involved. 
    If in addition $\cK_{iijk}\ne 0$ for some $k$, also the component $\cK g_{jj}$ is unbounded, as it scales at least as 
    \begin{align}
        \cK g_{jj} \sim t^i \rightarrow \infty \,. 
    \end{align}
    \end{subequations}
    \item[$(iii)$] The K\"ahler potential depends quadratically on the modulus $t^i$ which corresponds to $\cK_{iiij}=0, \forall j$ but $\cK_{iikl}\ne 0$ for some $k,l\ne i$. In this case the metric component $\cK g_{ii}$ does not scale:
    \begin{subequations}\label{Ft-Kgii3}
    \begin{align}
        \cK g_{ii} \sim \cK_{iikl} t^k t^l\sim \text{const.}
    \end{align}
    But the components $\cK g_{jj}$ are still unbounded, since generically they scale as 
    \begin{align}
        \cK g_{jj}\sim (t^i)^2 \rightarrow \infty\,,
    \end{align}
    \end{subequations}
    as long as $\cK_{iijk}\ne 0$ for some $k$. 
    \item[$(iv)$] Finally, if the K\"ahler potential is only linear in $t^i$, i.e. $\cK_{iikl}=0, \forall k,l$, but $\cK_{ijkl}\ne0$ for $j,k,l\ne i$ the diagonal component $\cK g_{ii}$ vanishes asymptotically as
    \begin{subequations}\label{Ft-Kgii4}
    \begin{align}
        \cK g_{ii}  \sim \frac{\cK_{ijkl} t^j t^k t^l}{t^i} \rightarrow 0 \,.
    \end{align}
    The other components $\cK g_{jj}$ are nevertheless unbounded as, generically 
    \begin{align}
         \cK g_{jj}\sim t^i \rightarrow \infty\,.
    \end{align}
    \end{subequations}
\end{itemize}

Given this behaviour of the tensor $\cK g_{ij}$, one would expect to find very few vacua in which $m^i \neq 0$ for some $i$ in regions where $t^i \gtrsim \oh \sqrt{\chi(Y_8)}, \forall i$. Exceptions to this rule may for instance happen if the index $i$ appears only linearly in the quadruple intersection numbers $\cK_{ijkl}$, and if we consider the regime $t^i \gg t^j, \forall j \neq i$. In that case one may satisfy the tadpole constraint for $m^i$ arbitrary and $m^j =0, \forall j \neq i$. A clear setup where this happens is when we consider a factorized geometry like $Y_8 = Y_6 \times \mathbb{T}^2$, that can be interpreted as a type IIB flux compactification, and identify $T^i$ with the complex structure of $\mathbb{T}^2$. The type IIB setup will be analyzed in section  \ref{Ft-s:IIB}, while the more general linear setup will be discussed in section \ref{Ft-s:linear}. In the next subsection we will consider the more generic case in which we need to set $m = m^i =0, \forall i$ in order to find vacua in the region $t^i \gtrsim \oh \sqrt{\chi(Y_8)}$, keeping in mind that in some special cases this constraint could be stronger than necessary. For smaller saxion values these restricted flux quanta will also give rise of vacua, but there they will coexist with vacua with other flux patterns, see e.g. \cite{Denef:2005mm,Honma:2017uzn}.

\subsection{Moduli stabilization}
\label{Ft-sec:moduli}

Motivated by the above discussion, let us restrict our attention to flux vacua at large complex structure such that 
\begin{align}
   \vec{q}^{\, t} = (m, m^i, \hat m^\mu, \bar e_i, \bar e) = (0,0, \hat m^\mu, \bar e_i, \bar e) \, ,
    \label{Ft-truncflux}
\end{align}
which implies that $\tilde{\rho} = \tilde{\rho}^i = 0$ and that $\bar{\rho}^\mu = \hat{m}^\mu$. In this case the flux contribution to the D3-brane tadpole reads
\be
N_{\rm flux} =  \frac{1}{2} \eta_{\mu\nu} \hat{m}^{\mu} \hat{m}^{\nu}\, .
\label{Ft-Nfluxgen}
\ee

Plugged into \eqref{Ft-eq:Mink}, the restricted fluxes \eqref{Ft-truncflux} imply
\begin{align}
\label{Ft-eq: vacuum equations m=ma=0}
\begin{cases}
    \bar{\rho}=0\\
    \bar{\rho}_i=0\\
   \cK \zeta_{\mu i} \hat{m}^\mu  = \mathcal{K}_i\zeta_{\mu}\hat{m}^\mu
\end{cases}
\end{align}
where we recall that the last equation is equivalent to the decomposition \eqref{Ft-splitmu} for $\hat{m}^\mu$. This system has the simplifying property that the  equations for axions and saxions decouple. From the first two equations we obtain
\begin{subequations}
\label{Ft-barrhoeqs}
\begin{align}
    \bar\rho&=0 \implies \bar{e}+\bar{e}_ib^i+\frac{1}{2}\hat{m}^\mu\zeta_{\mu,kl} b^k b^l=0 \stackrel{\eqref{Ft-barrhoieq}}{\implies} \bar{e}=-\frac{1}{2}\bar{e}_ib^i\, ,
\label{Ft-barrhoeq}\\
   \bar\rho_i&=0 \implies \hat{m}^\mu\zeta_{\mu,ij} b^j =-\bar{e}_i\, .
   \label{Ft-barrhoieq}
    \end{align}
 \end{subequations}

To analyze the implication of these two equations let us define the matrix $M_{ij} \equiv \hat{m}^\mu\zeta_{\mu,ij}$, and let $r$ be its rank. From \eqref{Ft-barrhoieq} we obtain a system of $r$ equations with $h^{3,1}(Y_8)$ unknowns. This system will only have a solution if the vector $\bar{e}_i$ lies in the image of $M$, which will impose $h^{3,1}(Y_8) - r$ constraints on these fluxes. Only when these constraints are met we will be able to find a vacuum, and in this case only $r$ axions will be stabilized. In particular, notice that then only $r$ complex structure fields appear in the superpotential \eqref{Ft-supofinal}. This suggests that several saxionic directions will not be stabilized either, as one can see from the third equation in \eqref{Ft-eq: vacuum equations m=ma=0}. Indeed, in general we have that $\zeta_\mu \neq 0$, as this corresponds to the volume of a holomorphic four-cycle in the mirror four-fold $X_8$, but also that it only depends on $r$ saxionic directions, and so the remaining ones are unfixed by the vacuum equations. Moreover this third equation is such that contracted with $t^i$ becomes trivial and so, in fact, it only stabilizes $r-1$ saxions. Therefore at least one saxionic direction is left unconstrained, even in the case of maximal rank. 

Coming back to \eqref{Ft-barrhoeqs}, we see that only those axions $b^i$ that are fixed by \eqref{Ft-barrhoieq} will appear in \eqref{Ft-barrhoeq}, which translates into an additional constraint that must be imposed on the fluxes in order to achieve a vacuum.  This time, however, the constraint is removed when corrections to the K\"ahler potential are taken into account, similarly to the effect observed  in \cite{Palti:2008mg,Escobar:2018tiu,Escobar:2018rna} in the context of Minkowski type II flux compactifications on three-folds. Indeed, including the corrections to the K\"ahler potential couples the equations for axions and saxions, which in turn changes the counting of stabilized moduli. This can already be seen from the vacua equations corrected at linear order in the parameter $\eps_i = 6K^{(3)}_i/\CK$, see \eqref{Ft-eq:Minkcorr}, which adapted to the present case read
\begin{subequations}
    \label{Ft-eq:Minkcorrmodu}
\begin{align}
\label{Ft-corr1}
 \bar\rho&  = - \frac{3}{8}\eps_it^i  \zeta_\mu \hat m^{\mu}  \, , \\
 \label{Ft-corr2}
  \bar{\rho}_i & = 0   \, , \\
  \label{Ft-corr3}
   \left(\CK\zeta_{\mu i}-\mathcal{K}_i \zeta_{\mu}\right) \hat{m}^\mu& = \frac{1}{4} \left(\CK\eps_i - \epsilon_k t^k \cK_i  \right)   \zeta_\mu \hat m^{\mu}  \, .
 \end{align}
\end{subequations}

Notice that \eqref{Ft-corr2} is the same as before, and therefore gives $r$ equations on the axions. Similarly, \eqref{Ft-corr3} becomes trivial when contracted with $t^i$ and so, even if modified, still yields $r-1$ equations for the saxions. The main difference comes from \eqref{Ft-corr1}, which couples axions and saxions and using \eqref{Ft-corr2} becomes
\be\label{Ft-corr1+corr2}
 \bar{e}+\frac{1}{2}\bar{e}_ib^i = - \frac{3}{8}\eps_it^i  \zeta_\mu \hat m^{\mu} \, .
\ee
On the one hand, this equation no longer sets a constraint for the flux $\bar{e}$. On the other hand, plugging in the value for $b^i$ obtained from \eqref{Ft-barrhoieq} one obtains an additional equation for the saxions which, together with \eqref{Ft-corr3}, fixes the vev for $r$ of them. Using the results of appendix \ref{Ft-sap:corrvac}, one can check that this structure is in fact preserved at all orders in the correction parameter $\eps_i$, and so the counting holds at the level of polynomial terms in the scalar potential.  

To sum up, we obtain a system with only $r = \rank (M)$ complex structure fields fixed by the above vacua equations. Fixing the remaining ones would necessarily imply taking into account the exponentially-suppressed corrections that we are neglecting in our analysis. It is beyond the scope of our work to determine whether full moduli stabilization would then be achieved or not, although in any event such fields would be extremely light in this regime. 

In general we will consider those cases in which the rank of  $M_{ij} \equiv \hat{m}^\mu\zeta_{\mu,ij}$ equals $h^{3,1}(Y_8)$, which a priori can be achieved by choosing an appropriate flux $\hat{m}^\mu$. Since in this scheme $N_{\rm flux} = \oh \eta_{\mu\nu} \hat{m}^\mu \hat{m}^\nu$, one may wonder if such flux choices restrict the possible values of $N_{\rm flux}$. Let us for instance consider the case in which the choice of $\hat{m}^\mu$ is  such that $r = h^{3,1}(Y_8)$ implies
\begin{eqn}\label{Ft-originTC}
 \zeta_{\mu, ij}M^{ij}  = \frac{1}{2\gamma}  \eta_{\mu\nu} \hat{m}^\nu + \beta_\mu \, ,
\end{eqn}
where $M^{ik}M_{kj} = \delta^i_j$, $\gamma$ is a real  function of the fluxes with a lower bound $\alpha>0$ and $\hat{m}^\mu \beta_\mu \leq 0$. Then we have that $N_{\rm flux} \geq \alpha h^{3,1}(Y_8)$, which is the sort of behaviour proposed by the Tadpole Conjecture in \cite{Bena:2020xrh}. Whenever \eqref{Ft-originTC} holds, and depending on the precise value for $\alpha$, a large number of moduli could be in tension with satisfying the upper bound for $N_{\rm flux}$, as pointed out in \cite{Bena:2020xrh}. It would be thus interesting  to determine in which cases \eqref{Ft-originTC} occurs.

We can go a step further in our analysis and impose bounds on the saxion vevs by recalling the leading solution for $\hat{\rho}^\mu$, see \eqref{Ft-splitmu}. Since now $m^i=m=0$ we have
\begin{equation}
    \hat{m}^\mu=A\zeta^\mu+C^\mu +\cO(\epsilon_i)\, ,
\end{equation}
with $C^\mu\zeta_{\mu i}=0$. Therefore, the tadpole is given by
\begin{equation}
\label{Ft-eq: tadpole constraint anstaz}
    N_{\rm flux}=\frac{1}{2} g_{\mu\nu}\hat{m}^\mu \hat{m}^\nu=\frac{1}{2}A^2\cK +\frac{1}{2}C^\mu C^\nu g_{\mu\nu}+\cO(\epsilon_i)\geq \frac{1}{2} A^2\cK+\cO(\epsilon_i)\, .
\end{equation}
On the other hand, substituting in \eqref{Ft-corr1} we obtain
\begin{equation}
    A=-\frac{4\bar{\rho}}{9 K^{(3)}_i t^i}\, .
    \label{Ft-astiA}
\end{equation}
Looking now at the equation \eqref{Ft-corr2}
\begin{equation}
    \bar{e_i}=-\hat{m}^\mu \zeta_{\mu, il}b^l\equiv - M_{il} b^l\, ,
    \label{Ft-Meq}
\end{equation}
we can infer that $\bar{\rho}$ behaves like $\bar{\rho}\sim q/P(\hat{m}^\mu)$ for some  integer $q$ and some polynomial $P(\hat{m}^\mu)$ of degree $r = \rank M$ in the $\hat{m}^\mu$. For instance, when $M$ is invertible and so $r=h^{3,1}(X_8)$, the matrix $M_{il}$ has integer combinations of the $\hat m^\mu$ as coefficients, and thus its inverse
\begin{align}
    M^{-1} = \frac{1}{\text{det}M} \sum_{s=0}^{h^{3,1}-1}M^s \sum_{k_1,\dots k_{h^{3,1}-1}} \prod_{l=1}^{h^{3,1}} \frac{(-1)^{k_l+1}}{l^{k_l}k_l!}\left(\Tr{M^l}\right)^{k_l},\quad s+\sum_{l=1}^{h^{3,1}-1} l k_l=h^{3,1}-1\,,
\end{align}
depends inversely on det $M$, which is a degree $h^{3,1}$ polynomial on the fluxes $\hat{m}^\mu$. The remaining terms appearing in $M^{-1}$ are polynomials of the integers $\hat{m}^\mu$, up to combinatoric factors.  Because in this case
\begin{eqn}
\bar{\rho} = \bar{e} - \oh M^{ij} \bar{e}_i  \bar{e}_j \, ,
\end{eqn}
with $M^{ij}$ the inverse of $M_{ij}$, we can estimate that there exists an integer $p\leq h^{3,1}(X_8)$ satisfying $N_{\rm flux}^p \bar{\rho}\gtrsim d^{2p-1}$, with $d \equiv {\rm g.c.d} \{m^\mu\}$. When $M$ is not invertible, we instead have that $p \leq r = \rank M$. Finally, using \eqref{Ft-eq: tadpole constraint anstaz}, we conclude that
\begin{equation}
\label{Ft-eq: K3 saxion bound}
   \cK <  (N_{\rm flux})^{2p+1} d^{2-4p} (K_i^{(3)}t^i)^2 \, .
\end{equation}
For a given choice of fluxes, this relation sets an upper bound on the possible values of the complex structure saxions. The details of this constraint will  heavily depend  on the topology of the mirror four-fold, through its intersection numbers and the $\alpha'^3$-correction terms $K_i^{(3)}$. For instance, notice that for a saxionic direction $t^i$ along which $\cK$ grows linearly \eqref{Ft-eq: K3 saxion bound} does not really set a bound, in agreement with our results of section \ref{Ft-s:linear}. As a very rough estimate, \eqref{Ft-eq: K3 saxion bound}  sets an overall bound for the complex structure saxion vevs of the form
\be
t^i \lesssim N_{\rm flux}^{p+\frac{1}{2}} d^{1-2p} | K_i^{(3)}|\, .
\label{Ft-boundgen}
\ee
Remarkably, our reasoning applies also when some fields are not fixed at the polynomial level.

Finally note that, even when $M$ has maximal rank, this moduli stabilization scheme suggests that there is a saxionic field direction whose mass is suppressed by $\eps_it^i$ compared to the other ones, as it is only stabilized when the corrections to the K\"ahler potential are taken into account. To check whether the scalar mass spectrum is hierarchical or not  one should describe the potential in terms of canonically normalized fields, which we will not attempt to do in this work. Nevertheless, we already see that the key ingredient for such a potential hierarchy is the mixing between different blocks in the saxion-dependent matrix \eqref{Ft-ZAB}, which only appears due to $K_i^{(3)}$ corrections, and so by construction it is suppressed in the large complex structure regime.


\section{The type IIB limit}
\label{Ft-s:IIB}

A celebrated moduli stabilization setup corresponds to type IIB orientifold compactifications with background three-form fluxes. In this section we specify our results to this case, neglecting the presence of D7-brane moduli and worldvolume fluxes. As we will see, our findings  imply not only a simple form for the scalar potential at large complex structure and weak coupling, but also two different moduli stabilization schemes with an upper bound for the complex structure vevs. One of these schemes challenges the behaviour expected by the Tadpole Conjecture of \cite{Bena:2020xrh} (see section \ref{fb-subsec: tadpole conjecture}). Such a scheme will be generalized to genuine F-theory compactifications in section \ref{Ft-s:linear}.

\subsection{The flux potential}

Type IIB compactifications with background three-form fluxes can be understood as F-theory on $(C_6 \times \mathbb{T}^2)/\Z_2$, with $C_6$ a Calabi--Yau three-fold, provided that the presence of D7-branes can be neglected for the bulk dynamics. We can then apply the results of the previous two sections by splitting the index counting complex structure moduli as $i = \{0, a\}$, where $T^0$ represents the complex structure of $\mathbb{T}^2$ and $T^a$, $a = 1, \dots, h^{2,1}(C_6)$ the complex structure moduli of the three-fold. We also impose
\be
\CK_{0abc} = \kappa_{abc}\, ,
\label{Ft-interT2}
\ee
where $\kappa_{abc}$ are the triple intersection numbers of the mirror three-fold $B_6$. From \eqref{Ft-supalt} we obtain a leading-order superpotential of the form 
\begin{align}\nonumber
 W =&\ e +  e_0 T^0 + e_a T^a + \frac{1}{2} m^{ab} \kappa_{abc} T^c T^0 + \frac{1}{2} m^{0a} \kappa_{abc} T^bT^c \\
   & +\frac{m^0}{6} \kappa_{abc} T^aT^bT^c + \frac{1}{2} m^a  \kappa_{abc} T^b T^cT^0  + \frac{m}{6}  \kappa_{abc}T^aT^b T^cT^0 \, .
   \label{Ft-supoIIBF}
\end{align}
This expression does not fully correspond to the superpotential of type IIB flux compactifications, due to the redundancy associated to the quanta $m^{ij}$. A one-to-one correspondence between flux quanta is achieved when we consider an expression of the form \eqref{Ft-supo}, which involves specifying a basis of holomorphic four-cycles classes $\{[\sigma_\mu]\}$ in the mirror four-fold $X_8 = (B_6 \times \mathbb{T}^2)/\Z_2$. 

In this case the basis $\{[\sigma_\mu]\}$ can be constructed explicitly, as follows. We first consider the $B_6$ Mori cone generators $[\mathcal{C}'^a]$,  $a = 1, \dots, h^{1,1}(B_6)$, and the divisor classes $[D_a']$, that generate its K\"ahler cone and specify its triple intersection numbers as $\kappa_{abc} = [D_a']\cdot [D_b']\cdot [D_c']$. The K\"ahler cone of $X_8$ is generated by $[D_a] = [D_a' \times \mathbb{T}^2]$, and by the class of $B_6$, which we denote as $[D_0]$. Following section \ref{Ft-s:potential}, we consider the set of holomorphic four-cycles
\be
\gamma_{ij}  = D_i . D_j \, , \quad i = \{0, a\}\, ,
\label{Ft-gamma}
\ee
that correspond to the quanta $m^{ij}$ in \eqref{Ft-supoIIBF}. The elements of this set are not independent in homology, as opposed to the following ones
\be
H_a = D_a' \,, \qquad H_{\hat{a}} = \mathcal{C}^{\hat{a}} \times \mathbb{T}^2\, ,
\label{Ft-4formbasis}
\ee
which form the holomorphic four-form basis $\{[\sigma_\mu]\} = \{[H_a], [H_{\hat{a}}]\}$. In other words, the index $\mu$ in  \eqref{Ft-supo} splits as $\mu = \{a, \hat{a}\}$, with $a, \hat{a} = 1, \dots, h^{1,1}(B_6)$. The intersection matrix for \eqref{Ft-4formbasis} is
\be
\eta_{a \hat{a}} = [H_a] \cdot [H_{\hat{a}}] = \delta_{a \hat{a}}\, ,
\ee
with the remaining entries vanishing. The relation with the redundant set \eqref{Ft-gamma} is given by
\be
\zeta^a_{0b} = \zeta^a_{b0}  = \delta_{ab}\, , \qquad   \zeta_{a, bc}\equiv \zeta^{\hat{a}}_{bc} \eta_{\hat{a}a}= \kappa_{abc}\, ,
\label{Ft-Koszul}
\ee
with vanishing remaining entries. One can then easily check that  \eqref{Ft-interT2} is recovered from \eqref{Ft-interrel}. 

Having fixed $\{[\sigma_\mu]\}$, the superpotential for the mirror four-fold $Y_8 = (C_6 \times \mathbb{T}^2)/\Z_2$ reads
\begin{align}\nonumber
 W =&\ \bar{e} +  \bar{e}_0 T^0 + \bar{e}_a T^a +  \bar{m}_a T^a T^0 + \frac{1}{2} \hat{m}^{a} \kappa_{abc} T^bT^c  +\frac{m^0}{6} \kappa_{abc} T^aT^bT^c \\
   & + \frac{1}{2} m^a  \kappa_{abc} T^b T^cT^0  + \frac{m}{6}  \kappa_{abc}T^aT^b T^cT^0 -i  K_0^{(3)} \left(m^0 + m T^0\right) \, .
   \label{Ft-supoIIBF2}
\end{align}
where we have applied \eqref{Ft-hatm}, defined $m_a \equiv \delta_{a\hat{a}} \hat{m}^{\hat{a}}$ and already taken into account the polynomial corrections of  section \ref{Ft-sec:poly}. Notice that for the case of $Y_8 = (C_6 \times \mathbb{T}^2)/\Z_2$ we have that $K^{(0)} = K^{(2)}_{00} = K^{(2)}_{ab} = K^{(3)}_a = 0$. We similarly obtain the corrected K\"ahler potential 
\be
K_{\rm cs}^{\rm corr}=- \log (2t^0) -\log\left(\frac{4}{3}\kappa_{abc}t^at^bt^c+ 2 K_0^{(3)} \right)\, ,
\label{Ft-KcscorrIIB}
\ee
which, after including the dilaton into the complex structure sector, matches \eqref{IIB-eq: complex Kahler potential}, since from the definitions \eqref{IIB-eq: mirror quantities def} and \eqref{Ft-K23} $K_0=-\im(\kappa_0)$.  One may continue to connect these expressions with the more standard formulation of type IIB flux compactifications on Calabi--Yau orientifolds. We start with the superpotential  \cite{Gukov:1999ya}
\begin{align}
W_{\rm IIB}=\int_{C_6} \Omega_3 \wedge G_3\,,
\end{align}
where $G_3=F_3-\tau H_3$ is the complexified three-form flux, with $\tau = C_0 - ig_s^{-1}$ the axio-dilaton. The holomorphic three-form $\Omega_3$ of the Calabi--Yau $C_6$, can be expanded in the symplectic basis of harmonic three-forms on $C_6$ $(\alpha_I, \beta^I)$ as \eqref{fb-eq: holomorphic form expansion}. Meanwhile, the prepotential is given by \eqref{IIB-eq:full_prepotential} and in the large complex structure limit one can ignore the instanton corrections. By introducing the projective coordinates $z^a =Z^a/Z^0$ we can write the holomorphic three-form as 
\begin{align}
\Omega_3= \alpha_0 + z^a \alpha_a +\left(\frac{1}{2}\kappa_{abc}z^bz^c + a_{ab} z^b\hat{a}_a  \right) \beta^a- \left( \frac{1}{6} \kappa_{abc}z^a z^bz^c + \hat{a}_a z^a + \kappa_0 \right) \b^0\, .
\label{Ft-O3exp}
\end{align}
Similarly, we can expand the $G_3$ flux following \eqref{fb-eq: G3 flux expansion} and arrive to the superpotential \eqref{IIB-eq:Wfull}. One can see that this expression matches \eqref{Ft-supoIIBF2} upon performing the identifications summarized in table \ref{Ft-table: type IIB F-theory dictionary}. Additionally, from \eqref{Ft-O3exp} one also reproduces \eqref{Ft-KcscorrIIB}, as already shown in \cite{Palti:2008mg,Escobar:2018rna}.

\begin{table}[htbp]
\centering
\def\arraystretch{1.25}
\begin{tabular}{|c|c|ccc}
\cline{1-2} \cline{4-5}
\textbf{F-theory} & \textbf{Type IIB}                       & \multicolumn{1}{c|}{\textbf{}} & \multicolumn{1}{c|}{\textbf{F-theory}} & \multicolumn{1}{c|}{\textbf{Type IIB}} \\ \cline{1-2} \cline{4-5} 
$\bar{e}$         & $f_0^B-\hat{a}_if^i_A$                  & \multicolumn{1}{c|}{}          & \multicolumn{1}{c|}{$T^0$}             & \multicolumn{1}{c|}{$\tau$}            \\
$\bar{e}_0$       & $-h_0^B+\hat{a}_ih^i_A$                 & \multicolumn{1}{c|}{\phantom{aaa}}          & \multicolumn{1}{c|}{$T^a$}             & \multicolumn{1}{c|}{$t^i$}             \\
$\bar{e}_a$       & $f_i^B+\kappa_{ij}f^j_A-\kappa_i f^0_A$ & \multicolumn{1}{c|}{}          & \multicolumn{1}{c|}{$K_{0abc}$}    & \multicolumn{1}{c|}{$\kappa_{ijk}$}    \\
$\hat{m}^a$       & $f^i_A$                                 & \multicolumn{1}{c|}{}          & \multicolumn{1}{c|}{$K_{ab}^{(1)}$}   & \multicolumn{1}{c|}{$-a_{ij}$}          \\
$\bar{m}_a$       & $-h_i^B+a_{ij}h^j_A+\hat{a}_ih^0_A$     & \multicolumn{1}{c|}{}          & \multicolumn{1}{c|}{$K_a^{(2)}$}       & \multicolumn{1}{c|}{$\hat{a}_i$}       \\
$m^a$             & $h^i_A$                                 & \multicolumn{1}{c|}{}          & \multicolumn{1}{c|}{$K^{(3)}$}         & \multicolumn{1}{c|}{$i\kappa_0$}      \\
$m^0$             & $-f^0_A$                                & \multicolumn{1}{c|}{}          & \multicolumn{1}{c|}{$\epsilon$}        & \multicolumn{1}{c|}{$\xi$}             \\ \cline{4-5} 
$m$               & $h^0_A$                                 &                                &                                        &                                        \\ \cline{1-2}
\end{tabular}
\caption{Dictionary between type IIB notation (chapter \ref{ch: Fintro}) and F-theory. Note that in Type IIB we use the indices $i,j,k$ to label the entries of cohomology class $h^{2,1}(C_6)$, while in F-theory these label $h^{1,1}(X_8)$ and we use the indices $a,b,c$ to refer to the entries of $h^{1,1}$ of the 3-fold base $B_6$ of the elliptically fibered $X_8$. Mirror symmetry implies $h^{2,1}(C_6)=h^{1,1}(B_6)$.}
\label{Ft-table: type IIB F-theory dictionary}
\end{table}

Using the results of section \ref{Ft-s:potential} one may give a compact expression for the resulting F-term scalar potential. The flux-axion polynomials are\footnote{For alternative definitions of flux-axion invariants in the type IIB compactifications see \cite{Bielleman:2015ina,Shukla:2019wfo,Blanco-Pillado:2020wjn}.}
\begin{align}
\nonumber \bar{\rho} &= \bar{e} + \bar{e}_0 b^0+  \bar{e}_ab^a+ \bar{m}_ab^ab^0 +  \kappa_{abc} \left(\frac{1}{2} \hat{m}^{a} b^bb^c  +\frac{1}{2} m^ab^bb^cb^0 +\frac{1}{6}m^0 b^ab^bb^c +\frac{1}{6}m b^ab^bb^c b^0\right)\,,\\
\nonumber\bar{\rho}_0&= \bar{e}_0+\bar{m}_ab^a +\kappa_{abc} \left(\frac{1}{2}m^a b^bb^c +\frac{1}{6} m b^a b^bb^c\right)\,,\\
\nonumber\bar{\rho}_a&= \bar{e}_a + \bar{m}_ab^0 +  \kappa_{abc}\left( \hat{m}^{b}b^c  + m^b b^0 b^c +\frac{1}{2}m^0 b^bb^c + \frac{1}{2} m b^b b^c b^0\right)\,,\\
\bar{\rho}_a^\prime&= \bar{m}_a +  \kappa_{abc} \left(m^b b^c +\frac{1}{2} m b^b b^c\right) \,,\\
\nonumber\hat{\rho}^{a}&= \hat{m}^{a} + m^a b^0 +m^0 b^a + m b^0 b^a \,,\\
\nonumber\tilde \rho^a&= m^a + m b^a \,,\\ 
\nonumber \tilde \rho^0&=m^0 + mb^0 \,,\\ 
\nonumber\tilde \rho&=m\, ,
\end{align}
in terms of which the potential takes the form \eqref{Ft-bilinear}.  At leading order, the saxion-dependent matrix $Z$ reads
\begin{align}
\label{Ft-ZABIIB}
Z^{AB} =
\frac{4}{3}e^K t^0\kappa
\begin{pmatrix}
\frac{1}{6}t^0\kappa & & & & & & & -1 \\
&  \frac{1}{6}\frac{\kappa}{t^0} & & & & & 1  &  \\ 
& & \frac{2}{3} t^0 \kappa g_{ab}^\kappa  & & & \delta^a_b & & \\
& & &\frac{2}{3} \frac{\kappa}{t^0} g_{ab}^\kappa  & - \delta^a_b & & & \\
& & &  -  \delta^a_b & \frac{3}{2}\frac{t^0}{\kappa} g^{ab}_\kappa & & 
\\
& &  \delta^a_b   & & &  \frac{3}{2}\frac{1}{t^0\kappa}g^{ab}_\kappa & &  \\
&1  & & & & & 6\frac{t^0}{\kappa} &  \\
-1 & & & & & & & \frac{6}{t^0\kappa} \\
\end{pmatrix} \, ,
\end{align}
where $\vec{\rho}^{\, t} = \left(\tilde{\rho}, \tilde{\rho}^0, \tilde{\rho}^a,   \hat{\rho}^{a},  \bar{\rho}_a^\prime, \bar{\rho}_a, \bar{\rho}_0,   \bar\rho   \right)$ and we have defined
\be
\kappa \equiv \kappa_{abc}t^at^bt^c\, ,\qquad \kappa_a \equiv \kappa_{abc}t^bt^c\, , \qquad \kappa_{ab} \equiv \kappa_{abc}t^c\, ,
\ee
and
\be
g_{ab}^\kappa = \frac{3}{2\kappa} \left(\frac{3\kappa_a\kappa_b}{2\kappa} - \kappa_{ab}\right)\, , \qquad  g^{ab}_\kappa =  2t^at^b - \frac{2}{3} \kappa \kappa^{ab}\, .
\ee
Notice that in this case the matrix $Z$ has the structure
\be
Z = 
\begin{pmatrix}
A & B \\ B^t & B^t A^{-1} B \\
\end{pmatrix} \, , \qquad A = A^t\, ,
\label{Ft-blockZ}
\ee
with $A$, $B$ non-singular $2h^{3,1} \times 2h^{3,1}$ matrices. This form is  preserved by polynomial corrections.

\subsection{Tadpoles and moduli stabilization}
\label{Ft-sec:moduliIIB}

Let us analyze the conditions for Minkowski vacua and the implications of the tadpole constraint in the type IIB orientifold limit. If we consider a large complex structure regime such that the effect of the correction $K^{(3)}$ can be neglected, the vacua conditions read
\begin{subequations}
\label{Ft-eq:MinkIIB}
\begin{empheq}[box=\widefbox]{align}
    \bar\rho&=\frac{1}{6}t^0 \kappa\tilde{\rho} \label{Ft-eq:MinkIIB 0}\\
    \bar\rho_0&=-\frac{1}{6}\frac{\kappa}{t^0}\tilde{\rho}^0 \label{Ft-eq:MinkIIB 00}\\
    \bar{\rho}_a&=-\frac{2}{3}t^0 \kappa g_{ab}^\kappa\tilde{\rho}^b \label{Ft-eq:MinkIIB i} \\
    \bar{\rho}'_a &= \frac{2}{3} \frac{\kappa}{t^0} g_{ab}^\kappa \hat \rho^b
    \label{Ft-eq:MinkIIB mu}
\end{empheq}
\end{subequations}
All these equations are a straightforward application of the general result \eqref{Ft-eq:Mink} to the type IIB limit, except perhaps \eqref{Ft-eq:MinkIIB mu}. To see how it arises from  \eqref{Ft-eq:Mink mu} notice that
\begin{equation}
\label{Ft-vacua mu strategy}
    \begin{cases}
4\left(  \zeta_{\mu 0}- \frac{\mathcal{K}_0}{ \cK} \zeta_{\mu}\right) \bar{\rho}^\mu =  2  t^a \bar{\rho}_a^\prime - \frac{ \kappa_a \hat{\rho}^a}{t^0}  = 0\, ,\\
4\left(  \zeta_{\mu a}- \frac{\mathcal{K}_a}{ \cK} \zeta_{\mu}\right) \bar{\rho}^\mu = 4t^0 \bar{\rho}_a^\prime + 4\kappa_{ab}\hat{\rho}^b - \frac{3\kappa_a}{\kappa} \left(2t^0t^b  \bar{\rho}_b^\prime + \kappa_b \hat{\rho}^b\right) = 0\, ,
    \end{cases}
\end{equation}
where we used that $\zeta_{\hat{a}, 0b} = \delta_{\hat{a}b}$ and $\zeta_{a, bc} = \kappa_{abc}$. Together, these two conditions imply  \eqref{Ft-eq:MinkIIB mu}.

As in the general case, when turning on the correction  $K^{(3)}$ the above vacuum equations are corrected. In particular, for the type IIB limit we find that the equations \eqref{Ft-eq:Minkcorr} yield 
\begin{subequations}
 \label{Ft-eq:MinkcorrIIB}
\begin{align}
    \bar \rho -\frac{t^0\kappa }{6} \tilde \rho &= -\eps\, \frac{\kappa t^0 }{12} \left[\tilde \rho + \frac{9}{\kappa t^0}  \kappa_a \hat \rho^a \right]
    \,,\\
    \bar \rho_0 +\frac{1}{6}\frac{\kappa}{t^0}\tilde \rho^0 &= \eps\, \frac{1}{12} \left(\frac{\kappa}{t^0} \tilde\rho^0 - 9  \kappa_a \tilde\rho^a \right)\,,\\
    \bar \rho_a +\frac{2}{3}t^0 \kappa g_{ab}^\kappa \tilde \rho^a &= \eps\, \frac{3}{4}  \kappa_a \left({\tilde \rho^0} - \frac{t^0\kappa_b}{\kappa}\tilde \rho^b \right) \,,\\
    \bar{\rho}'_a - \frac{2}{3} \frac{\kappa}{t^0} g_{ab}^\kappa \hat \rho^b & = \eps \frac{3}{8}  \frac{\kappa_a}{\kappa}  \left(2\kappa \tilde{\rho} + \frac{\kappa_b}{t^0} \hat \rho^b + 2 t^b \bar{\rho}'_b \right)\,,
\end{align}
\end{subequations}
where we have defined  $\epsilon \equiv \frac{3 K^{(3)}}{2 \kappa}$. Note this is precisely the LCS parameter $\xi$ defined for Type IIB in \ref{fb-subsec: Type IIB ingredients}.

In terms of the vacua equations, one can give a more explicit expression for the tadpole condition in this setup. One begins with the topological quantity
\be
N_{\rm flux} =  \bar{e} m - \bar{e}_i m^i + \bar{m}_a \hat{m}^a = \bar{\rho} \tilde{\rho} - \bar{\rho}_i \tilde{\rho}^i +  \bar{\rho}_a^\prime \hat{\rho}^{a}\, ,
\label{Ft-NfluxrhoIIB}
\ee
which at vacua can be expressed as
\begin{eqn}
N_{\rm flux} &\stackrel{\rm vac}{=} \frac{t^0 \kappa}{6} \left(\tilde{\rho}^2 +  \frac{(\tilde{\rho}^0)^2}{(t^0)^2} + \frac{2}{3} g_{ab}^\kappa  \tilde{\rho}^a\tilde{\rho}^b  +
\frac{3}{2\kappa^2} g^{ab}_\kappa
\rho_a^\prime \rho_b^\prime \right)
\, ,
\label{Ft-NfluxvacIIB}
\end{eqn}
where we have used the conditions \eqref{Ft-eq:MinkIIB} and therefore neglected the effect of $K^{(3)}$. This approximation is justified if we aim to obtain the restriction on the fluxes  that arise in the different weak coupling, large complex structure limits $\kappa, t^0 \to \infty$, as done in section \ref{Ft-sec:tadpole}. As in there (see also \cite[appendix D]{Blanco-Pillado:2020wjn}),  we must set $\tilde{\rho} = 0$ when $t^0, \kappa/6 > \sqrt{N_{\rm flux}}$ in order to find vacua, and therefore in this regime $\tilde{\rho}^0 = m^0$, $\tilde{\rho}^a = m^a$. The remaining fluxes will then be constrained depending on the different limits that we take, which we can classify in a slightly more explicit manner as compared to the general case.

Indeed, let us  consider a scaling of the form $t^0 \sim \kappa^r \to \infty$, with $r \in \R$. If $r \geq 1$ then $t^0 \kappa g_{ab}^\kappa$ will diverge, and we will have to set $m^a = \tilde{\rho}^a$ to zero. We will also have that $t^0  g^{ab}_\kappa/\kappa$ diverges, and so $m_a = \bar{\rho}_a^\prime$ must vanish as well.  We then recover a simplified flux lattice such that $\vec{q}^{\, t} = (0,m^0,0, \hat{m}^a, 0, \bar{e}_a, \bar{e}_0, \bar{e})$, and the tadpole is given by $N_{\rm flux} = - m^0\bar{e}_0$. Alternatively, if $r<1$ the $m^0 = \tilde{\rho}^0$ must be set to zero and, generically, the same applies for $m^a = \tilde{\rho}^a$. The question is then whether $\hat{m}^a = \hat{\rho}^a$ and $m_a = \bar{\rho}_a^\prime$ must vanish or not. In fact, to have a non-trivial tadpole we need that $N_{\rm flux} = \sum_a \hat{m}^a m_a \neq 0$, and one can convince oneself that this is only possible if $t^0$ scales like $\kappa g_{aa}^\kappa$, for at least some $a$. All these are cases in which $r<1$ and $\vec{q}^{\, t} = (0,0,0, \hat{m}^a, m_a, \bar{e}_a, \bar{e}_0, \bar{e})$, which we will consider as another subset of vacua. Finally, one can check that this classification is unchanged if we add to \eqref{Ft-NfluxvacIIB} the corrections that arise from imposing \eqref{Ft-eq:MinkcorrIIB}. Let us now analyze the moduli stabilization of both classes of vacua:

\subsubsection*{IIB1: $\vec{q}^{\, t} = (0,0,0, \hat{m}^a, {m}_a, \bar{e}_a, \bar{e}_0, \bar{e})$}

This case falls into the generic class of vacua discussed in section \ref{Ft-sec:moduli}. We have that the vacua equations \eqref{Ft-eq:MinkcorrIIB} reduce to 
\begin{subequations}
 \label{Ft-eq:MinkcorrIIB1}
\begin{align}
\label{Ft-eq:MinkcorrIIB1 0}
    \bar \rho &=  -\frac{3\eps}{4}   \kappa_a \hat m^a \,,\\
       \label{Ft-eq:MinkcorrIIB1 i}
    \bar \rho_i  &=0 \,, \qquad i= 0, a\, ,\\
    {m}_b - \frac{2}{3} \frac{\kappa}{t^0} g_{bc}^\kappa \hat m^c & = \eps \frac{3}{4t^0}  \frac{\kappa_b}{\kappa}  \kappa_a \hat m^a \, ,
     \label{Ft-eq:MinkcorrIIB1 mu}
\end{align}
\end{subequations}
to first order in $\eps$. We may now apply the general discussion in section \ref{Ft-sec:moduli} to set bounds on the saxion vevs. Using \eqref{Ft-eq:MinkcorrIIB1 mu} we find that
\be
{m}_a = A\kappa_a + C_a + \cO(\eps) \, , \qquad \hat{m}^a = 2At^0t^a + C^a + \cO(\eps)\, ,
\label{Ft-solIIB1mu}
\ee
where $C_at^a=0$, $C^a\kappa_a=0$ and $C^a\kappa_{ab}=-t^0C_b$. We therefore obtain the inequality 
\be
N_{\rm flux} = 2A^2 t^0 \kappa + C_a C^a+\cO(\eps) \geq 2A^2 t^0 \kappa +\cO(\eps)\, ,
\ee
while from \eqref{Ft-eq:MinkcorrIIB1 0} one can see that
\be
\label{Ft-eq: IIB1 A}
A = -\frac{4}{9} \frac{\bar{\rho}}{t^0 K^{(3)}}  \, . 
\ee
From here we find the following bound for the complex structure saxions,
\be
\frac{\kappa}{t^0} < N_{\rm flux}^{2p+1} d^{2-4p} (K^{(3)})^2 \, , 
\label{Ft-boundIIB1}
\ee
where $p \leq h^{2,1}(C_6)+1$ is bounded by the number of complex structure plus dilaton fields, and $d ={\rm g.c.d} (\{\hat{m}^a, {m}_a\})$. Here we have used a reasoning similar to the one below \eqref{Ft-Meq} to arrive to the inequality $N_{\rm flux}^p \bar{\rho} \gtrsim d^{2p-1}$. Finally, notice that taking into account that in this scheme $t^0 \sim \kappa g_{aa}^\kappa$, we end up with a bound for the saxions which is, again, roughly of the form \eqref{Ft-boundgen}.

To obtain a more concrete scheme one may consider that the matrix $M_{ij}\equiv \hat{m}^\mu\zeta_{\mu,ij}$, introduced below \eqref{Ft-barrhoeqs}, is invertible. In the type IIB limit and with our particular choice of fluxes this matrix is given by
\begin{equation}
 M= 
\begin{pmatrix}
0 & m_a \\ m_b & S_{ab} \\
\end{pmatrix} \, , \qquad S_{ab}\equiv \kappa_{abc}\hat{m}^{c}\, .
\end{equation}
For simplicity, we work under the hypothesis that $S_{ab}$ is invertible. If that is the case, the inverse matrix  $M^{ij}$ has the form
\begin{equation}
 M^{-1}= \mathcal{H}^{-1}
\begin{pmatrix}
-1 & S^{ac}m_c \\ S^{bc}m_{c} & SS^{ab}-S^{ac}S^{bd}m_cm_d \\
\end{pmatrix} \, , \qquad \mathcal{H} \equiv S^{ab}m_am_b\, ,
\label{ft-eq: M-1 type IIB1}
\end{equation}
where $S^{ab}$ is the inverse of $S_{ab}$. Notice that for $M$ to be invertible we have to further ensure $\mathcal{H} \ne 0$. If this last condition is not satisfied, the kernel of $M$ is given by 
\begin{equation}
    \text{ker}(M) = \langle (1, -S^{ab} m_b)^t \rangle\,,
\end{equation}
such that we have a flat direction along $T^i = (\tau, -\tau S^{ab} m_b)^t$. Given the identifications of table \ref{Ft-table: type IIB F-theory dictionary} this precisely reproduces  the flat direction found in \cite{Demirtas:2019sip} for  $S_{ab}$ invertible but $\mathcal{H} =0$.

Using these results we can achieve stabilization of all the moduli of the system. Starting with the axions, from \eqref{Ft-eq:MinkcorrIIB1 i} we have 
\begin{align}
    b^0&=-\mathcal{H}^{-1}S^{ab}\bar{e}_am_b+\bar{e}_0S^{-1}\, ,\\
    b^a&=-\bar{e}_0 \mathcal{H}^{-1}S^{ab}m_b-\bar{e}_bS^{ab}+\mathcal{H}^{-1} S^{ac}m_c S^{bd}m_d\bar{e}_b=-S^{ab}(\bar{e}_b+b^0m_b)\, .
\end{align}

Regarding the saxions,  the expression \eqref{Ft-eq:MinkcorrIIB1 mu} at leading order provides us with a system of $h^{2,1}(C_6)$ independent equations of order $4$ in the set of $h^{2,1}(C_6)$ $\{t^a\}$. Hence, we can use it to express all the saxions in terms of the saxionic direction $t^0$. We can then  substitute our results in \eqref{Ft-eq:MinkcorrIIB1 0} and employ the first order corrections in $\epsilon$ to stabilize the remaining direction $t^0$. Note that we are able to ignore the corrections in \eqref{Ft-eq:MinkcorrIIB1 mu} because the first leading contribution of the saxions in \eqref{Ft-eq:MinkcorrIIB1 0} is already linear in the parameter $\epsilon$.

Looking to the shape of $M^{ij}$  we observe a very straightforward flux choice for which the matrix $M_{ij}$ is invertible, and which is related to the Ansatz taken in \cite{Blanco-Pillado:2020hbw}. Indeed, let us consider that $\hat{m}^a\neq0$ $\forall a$ and take $C^a=0$ (which means $C_a=0$). Then $\hat{m}^a \propto t^a$ and $S_{ab}\propto\kappa_{ab}$, as in \cite{Blanco-Pillado:2020hbw}. Moreover the ratio $t^a/t^0$ is easily fixed at leading order, since \eqref{Ft-solIIB1mu} gives
\begin{equation}
    A=\frac{1}{4r (t^0)^2}\,, \qquad \frac{t^a}{t^0}=2\hat{m}^a r\, ,
    \label{ft-eq: behaviour with Ca=0}
\end{equation}
with $r=\frac{m_a\hat{m}^a}{\kappa_{abc}\hat{m}^a\hat{m}^b\hat{m}^c}$. Working now with \eqref{Ft-eq:MinkcorrIIB1 i} we have
\begin{align}
    b^0&=\frac{1}{r^2S_{ab}\hat{m}^a\hat{m}^b}\left(\bar{e}_0-r\hat{m}^a\bar{e}_a\right)\, ,\\
    b^a&=-S^{ab}\bar{e}_b-r\hat{m}^ab^0\, .
\end{align}
Finally, \eqref{Ft-eq: IIB1 A} determines the vev for the saxion $t^0$.

Note that in this particular setup the total tadpole $N_{\rm flux} = \sum_a {m}_a\hat{m}^a$ is a sum of positive terms and so it exceeds in value to $h^{2,1}(C_6)$. As pointed out in \cite{Bena:2020xrh} this kind of behaviour leads to a significant tension between tadpole cancellation and full moduli stabilization for a large number of moduli. From our perspective, this would favour vacua where $C^a \neq 0$. In that case, one should apply \eqref{Ft-originTC} to see whether $N_{\rm flux}$ is bounded from below by $h^{2,1}(C_6)$ or not. We will consider this family of solutions in greater detail during the following chapter.

\subsubsection*{IIB2: $\vec{q}^{\, t} = (0,m^0,0, \hat{m}^a, 0, \bar{e}_a, \bar{e}_0, \bar{e})$}

This case is dual, via mirror symmetry, to the type IIA non-supersymmetric Minkowski flux vacua constructed in \cite{Palti:2008mg} and analyzed from the viewpoint of the bilinear potential \eqref{Ft-bilinear} in \cite{Escobar:2018rna}. As shown in there, in this case one can solve for the vev of each field in terms of the flux vacua and the correction $\eps$. One starts with the following vacua equations 
\begin{subequations}
 \label{Ft-eq:MinkcorrIIB2}
\begin{align}
    \bar \rho  &=  0 \,,\\
    t^0\bar{e}_0 +\frac{1}{6}\kappa m^0 &= \eps\, \frac{\kappa}{6} \frac{1+ 4\eps}{2 - \eps} m^0 \,,\\
    \bar \rho_a  &= \eps\, \frac{3}{2}  \frac{\kappa_a}{2 - \eps} m^0  \,,\\
   \hat \rho^a & = 0 \,,
\end{align}
\end{subequations}
which at first order in $\eps$ are equivalent to \eqref{Ft-eq:MinkcorrIIB}, restricted to this choice of fluxes.
Borrowing the results from \cite[section 4.1]{Escobar:2018rna} and adapting them to our notation we obtain the solution
\begin{eqn}
b^0 & = - \frac{1}{3\bar e_0(m^0)^2}\left(\kappa_{abc}\hat{m}^{a}\hat{m}^{b}\hat{m}^{c}-3e_a\hat{m}^{a}m^0\right)-\frac{\bar{e}}{\bar e_0}\, , \\
b^a & = - \frac{\hat{m}^a}{m^0}\, ,
\label{Ft-axionsIIB2}
\end{eqn}
for the axions and 
\begin{eqn}
t^0 & = - \frac{1}{6}\frac{m^0}{\bar{e}_0} \kappa \left(1\kappa - \eps \frac{1 + 4\eps}{2- \eps}\right) 
\, , \\
\kappa_a & = \frac{2- \eps }{3 (m^0)^2 \eps }  \left(2m^0 \bar{e}_a -\kappa_{abc} \hat{m}^{b}\hat{m}^{c} - 2m^0 \bar{e}_a \right)\, ,
\label{saxionsIIB2}
\end{eqn}
for the saxions. Note that the $\kappa_a$ are determined implicitly, and that acceptable vacua correspond to saxion vevs within $\{t^a > 0 | \epsilon \ll 1\}$, which imposes a constraint on the flux quanta.\footnote{Explicit solutions to the equations for $\kappa_a$ have been proposed in \cite{Palti:2008mg}, assuming  homogeneous vevs for all $t^a$. Additionally, these equations are similar to those determining the K\"ahler moduli vevs in type IIA AdS$_4$ CY  orientifolds \cite{DeWolfe:2005uu,Marchesano:2019hfb}, and so explicit solutions for such a setup will translate into Minkowski vacua in this context.}

Notice that $t^0 \sim \kappa/6$, as could have been guessed from the leading order equation  \eqref{Ft-eq:MinkIIB 00} and the fact that $\bar{\rho}_0 = \bar{e}_0$. Also
\be
\frac{\kappa}{\kappa_a}  \lesssim (m^0)^2 |K^{(3)}| < N_{\rm flux}^2 d^{-2} |K^{(3)}|  \, , 
\label{Ft-boundIIB2}
\ee
with $d = {\rm g.c.d.} (m^0,\bar{e}_0)$. This results in an upper bound on the value of the complex structure saxions which is roughly of the form \eqref{Ft-boundgen} with $p =\frac{3}{2}$, even if the moduli stabilization scheme under discussion is different from the one in section \ref{Ft-sec:moduli}. Note also that this bound is consistent with the regime in which $\eps \ll 1$, whenever $(m^0)^2 |K^{(3)}|$ is moderately larger than 1.  

This class of vacua stands in tension to the Tadpole Conjecture proposed in \cite{Bena:2020xrh}, since the flux contribution to the tadpole $N_{\rm flux} = - m^0\bar{e}_0$ does not openly depend on the number of complex structure moduli. There seems to be no incompatibility between achieving full moduli stabilization and having an $N_{\rm flux}$ that it is bounded. A key property for this to happen is the fact that most of the RR flux quanta that implement moduli fixing do not contribute to the tadpole because they do not pair up with $\bar{e}_0$ in the intersection matrix.  What is true is that all complex structure saxions $t^a$ are stabilized only when the effect of the correction $K^{(3)}$ is taken into account \cite{Palti:2008mg} which suggests that, in this case, decoupling the expression for $N_{\rm flux}$ from the number of complex structure moduli comes at the cost of having several light fields. In the next section we will generalize this scheme to genuine F-theory setups. We will see that most of the features of the type IIB case will be realized except for the bound \eqref{Ft-boundIIB2}, which may or may not be present.

\section{The linear scenario}
\label{Ft-s:linear}

The moduli stabilization scheme \textbf{IIB2} for type IIB orientifolds provides a class of compactifications in which the flux contribution to the D3-brane tadpole $N_{\rm flux}$ is independent from the number of complex structure moduli $h^{3,1}(Y_8)$, and nevertheless one can achieve full moduli stabilization. Therefore it is quite simple to stabilize all complex structure moduli and at the same time satisfy the bound $24 N_{\rm flux} \leq \chi(Y_8)$, a scenario whose realization has been recently doubted \cite{Braun:2020jrx,Bena:2020xrh,Bena:2021wyr}, see also \cite{Betzler:2019kon}. In the following we would like to generalize the key features of scheme \textbf{IIB2} to more general F-theory compactifications, providing a wider family of solutions in tension with the Tadpole Conjecture of \cite{Bena:2020xrh}. 

We will dub this more general setup the {\em linear scenario}, because the key ingredient will be a four-fold $Y_8$ such that at least one complex structure saxion $t_L$ only appears linearly on $\CK = \frac{3}{2}e^{-K_{\rm cs}}$ and in the superpotential. This means that $\CK$ takes the form
\begin{eqn}\label{Ft-cKlin}
 \CK = 4 \CK_L t_L + f \, ,
\end{eqn}
with $\CK_L \equiv  \cK_{L abc}t^a t^b t^c$, and $f \equiv f(t^a)$ a function independent of $t_L$ and homogeneous of degree four on the remaining saxions $t^a$. This kind of K\"ahler potential is found when the mirror four-fold $X_8$ is a  smooth three-fold fibration over $\mathbb{P}^1$,\footnote{Note that in order for $t_L$ to appear only linearly in $\cK$ the mirror $X_8$ needs to have a nef effective divisor $D_L$ such that $D_L^2=0$. The normal bundle $\mathcal{O}_{X_8}(D_L)|_{D_L}$ of $D_L$ is then trivial and by adjunction it  follows that $c_1(D_L)$ vanishes. This is satisfied whenever $X_8$ is a fibration of a CY three-fold, $K3\times \mathbb{T}^2$ or an abelian variety over $\mathbb{P}^1$, in which case $D_L$ corresponds to the class of the generic fibre. See \cite{Lee:2019wij} for a related discussion for CY three-folds.}  see section \ref{Ft-sec:counter} for an explicit example. In this case the leading saxion-dependent matrix \eqref{Ft-ZAB} is 
\be
\label{Ft-ZABlin}
 Z  = \frac{1}{2{\cal V}_3^2}
 \begin{pmatrix}
\frac{\cK}{24} & &  &  & & & -1 \\
& \frac{\cK}{6} g_{LL} & \frac{\cK}{6} g_{LL} \varepsilon_a  & &  &  1 &   \\ 
& \frac{\cK}{6} g_{LL} \varepsilon_a & \frac{\cK}{6} g_{ab}   &  & \delta^a_b &  & \\
 & & &   g_{\mu\nu} -\eta_{\mu\nu} & & &  \\
& & \delta^b_a   & &   \frac{6}{\cK}\tilde{g}^{ab} &   -\frac{6}{\cK}\tilde{g}^{ab}\varepsilon_a &  \\
& 1 &   & &  -\frac{6}{\cK}\tilde{g}^{ab}\varepsilon_a & \frac{6}{\cK}g^{-1}_{LL} + \frac{6}{\cK}\tilde{g}^{ab}\varepsilon_a \varepsilon_b&  \\
-1  & & &    & & &  \frac{24}{\cK} \\
\end{pmatrix} \, ,
\ee

where
\begin{eqn}
\frac{\cK}{6} g_{LL} = \frac{1}{6}\frac{\cK_L}{t_L\left(1 +\frac{1}{4}\frac{f}{\cK_Lt_L} \right)}\, , \qquad \varepsilon_a = \p_a \left( \frac{f}{4\CK_L}\right) ,
\end{eqn}
and $\tilde{g}^{ac}g_{cb} = \delta^a_b$. We now consider a limit which takes one or several of the saxions $t^a$ to infinity such that
 \begin{align}\label{Ft-limitL}
    t_L \sim \cK_L \to \infty \, .
\end{align}
We also assume that $t_L$ grows faster than any of the other saxions $t^a$, so that we realize the hierarchy $t_L \gg t^a$. Along such a limit $\cK \to \infty$ and $\cK g_{ab} \to \infty$. This implies that in order to find vacua we need to set  $\tilde{\rho} = \tilde{\rho}^a =0$, which translates into the flux constraint $m = m^a =0$. We also have that
\begin{eqn}
\frac{\cK}{6} g_{LL} \stackrel{\eqref{Ft-limitL}}{\to} \frac{1}{6}\frac{\cK_L}{t_L}\, ,
\label{Ft-limgLL}
\end{eqn}
and one may find vacua with $m^L \neq 0$ in this regime. Finally, one describes the fluxes $\hat{m}^\mu$ by constructing the set of four-forms $\sigma_\mu$ in the mirror four-fold $X_8$. As mentioned, we assume that $X_8$ is a three-fold fibration over $\mathbb{P}^1$. Due to this fibration structure, a basis of holomorphic four-cycles on $X_8$ can be generated from the K\"ahler cone generators $D_a$ of the fibre $X_3$ 
\begin{align}\label{Ft-linearscenarioH1}
   H_a = D_a . D_L \,,  
\end{align}
as well as by fibering the Mori cone generators $\mathcal{C}^a$ of $X_3$ over the base $\mathbb{P}^1$
\begin{align}\label{Ft-linearscenarioH2}
    H_{\hat a} = \mathcal{C}^a \rightarrow \mathbb{P}^1 \,. 
\end{align}
This last set of basis elements is related to the holomorphic four-cycles $\gamma_{ij}=D_i . D_j$ as 
\begin{align}
    [\gamma_{ab}] = \cK_{Labc}\delta^{c\hat c} [H_{\hat c}]\,. 
\end{align}
The integral basis of four-form classes $[\sigma_\mu]$ is then $\{[\sigma_\mu]\}=\{[H_a],[H_{\hat a}]\}$, and so the four-cycle index splits as $\mu = \{a, \hat{a}\}$, like in the type IIB case, and then \eqref{Ft-ZABlin} takes the form \eqref{Ft-blockZ}. The intersection matrix $\eta_{\mu\nu}$ is given by 
\begin{align}
    \eta_{a \hat b}=\delta_{a \hat b}\,,\qquad \eta_{a b}=0\,,
\end{align}
plus $\eta_{\hat a \hat b}$ in general non-vanishing and quite involved (see \eqref{Ft-intersectionexampleIII} for its form in an explicit example). The form of $g_{\mu\nu}$ will then in general be quite complicated, but given the non-vanishing entries of the intersection matrix $\eta_{\mu\nu}$, setting $\hat{m}^{\hat a}=0$ leads to $\eta_{\mu\nu} \hat{m}^\mu\hat{m}^\nu =0$ and the only contribution to the tadpole is
\be
N_{\rm flux} = - m^L\bar{e}_L\, , 
\ee
which, just as in the \textbf{IIB2} scheme, is independent from $h^{3,1}(Y_8)$. We then recover the flux vector
\begin{align}\label{Ft-eq:fluxchoiceL}
    \vec q^{\, t} = (0, m^L, 0, \hat m^a,0,   \bar e_a, \bar e_L,  \bar e)\,,
\end{align}
and we find that $m^L$ has the same role as $m^0$ in the \textbf{IIB2} setup of section \ref{Ft-sec:moduliIIB}. It moreover follows that the flux-axion polynomials \eqref{Ft-corrhos} reduce to 
\begin{align}
    \nonumber \bar \rho &= \bar e +e_Lb^L+\bar e_a b^a + \cK_{L abc}\left(\frac{1}{2}\hat m^a b^b b^c+\frac{1}{6} m^L   b^a b^b b^c\right) \,, \\
    \nonumber \bar \rho_a &= \bar e_a +  \cK_{Labc}\left(\hat m^b b^c + \frac{1}{2} m^L  b^bb^c\right)\,,\\
    \label{Ft-rhoAexample} \bar \rho_L  &= \bar e_L  \,,\qquad \bar \rho_a^\prime = 0 \,,\qquad \hat \rho^a = \hat m^a + m^L  b^a \,,\\
    \nonumber\tilde \rho^a &=0 \,,\qquad \tilde \rho^L  = m^L \,,\qquad \tilde\rho=0\, ,
\end{align}
which we recognise as the flux-axion polynomials in the \textbf{IIB2} setup upon the identifying $\cK_{Labc}$ with $\kappa_{abc}$. The leading-order vacua equations read
\begin{subequations}
    \label{Ft-vaclin}
\begin{align}
\bar{\rho}  &= 0\, ,\\
\label{Ft-vaclin1}
\bar{e}_L + \frac{\cK}{6}g_{LL} m^L  & = 0 \, ,\\
\label{Ft-vaclin2}
\bar{\rho}_a - \varepsilon_a  \bar{e}_L & =  0\, ,\\
\hat{\rho}^a  & =  0\, ,
 \end{align}
\end{subequations}
and can be solved like for the \textbf{IIB2} scheme. Indeed, the first and fourth equations fix the vev for the axions as
\begin{eqn}
b^L & = - \frac{1}{3\bar e_L(m^L)^2}\left(\cK_{L abc}\hat{m}^{a}\hat{m}^{b}\hat{m}^{c}-3e_a\hat{m}^{a}m^L\right)-\frac{\bar{e}}{\bar e_L}\, , \\
b^a & = - \frac{\hat{m}^a}{m^L}\, ,
\label{Ft-axionsL}
\end{eqn}
and the remaining ones the vev for the saxions. In particular we find that $\cK g_{LL}/6 \simeq \cK_L/6t_L$ must lie in the range $(N_{\rm flux}^{-1}, N_{\rm flux})$, and that 
\begin{equation}
    t^L=-\frac{1}{6}\frac{\mathcal{K}_Lm_L}{e_L}-\frac{f}{4\mathcal{K}_{L}}\, ,
\end{equation}
\begin{eqn}
\varepsilon_a  = \frac{m^L \bar{e}_a -\oh \cK_{Labc} \hat{m}^{b}\hat{m}^{c}}{ m^L\bar{e}_L}  = - \frac{N_a}{N_{\rm flux}} \quad \implies \quad N_{\rm flux} |\varepsilon_a| \gtrsim 1\, .
\label{Ft-Nfluxineqlin}
\end{eqn}
Here we have defined $N_a \equiv  m^L \bar{e}_a - \frac{1}{2} \cK_{L abc} \hat m^b \hat m^c$ as a monodromy-invariant flux combination in the present setup, more precisely the analogue of the third invariant listed in appendix \ref{Ft-ap:invariants}. We also obtain the inequality
\begin{eqn}
\frac{\cK}{6}g_{LL} |\varepsilon_a| \gtrsim N_{\rm flux}^{-2}\, .
\label{Ft-Nfluxineqlin2}
\end{eqn}

Notice that in the present setup the leading vacua equations  in principle suffice to find a set of vacua with full moduli fixing, unlike in the \textbf{IIB2} scheme. This is due to the fact that $\varepsilon_a$ appears at leading order. Nevertheless, further corrections will also contribute to the above equations, and in some cases they are needed to understand the implications of the inequality \eqref{Ft-Nfluxineqlin}. We can read off such corrections from \eqref{Ft-apeq:vacuumrho_i}, focusing on those  that involve $K^{(3)}_L$, which are the leading ones. To leading order  in $\epsilon_L \equiv \frac{3K^{(3)}_L}{2\CK_L}$ one can also extract them from \eqref{Ft-eq:Minkcorrrhoi}, obtaining that in \eqref{Ft-vaclin2} now we have 
\begin{eqn}
 \varepsilon_a  =\p_a \left( \frac{f}{4\CK_L}\right) - 6\, \eps_L\cK_L\frac{\cK_a}{g_{LL}\cK^2}  \stackrel{\eqref{Ft-limitL}}{\to} g_a^\infty - \frac{27}{4} K^{(3)}_L  \frac{t_L}{\cK_L} \frac{\cK_{L a}}{\cK_L} \, .
\end{eqn}
Here we have defined $ g_a^\infty$ as the asymptotic behaviour of $\p_a \left( \frac{f}{4\CK_L}\right)$ along the limit \eqref{Ft-limitL}. Notice that the second term asymptotes as $ \frac{\cK_{L a}}{\cK_L} \to 0$, and so the qualitative behaviour of the system depends on the functional behaviour of $g_a^\infty$. We have two possibilities:
\begin{itemize}
 
    \item[-] If $g_a^\infty \to 0$ for some $a$, then \eqref{Ft-Nfluxineqlin} will set an upper bound on this limit. If moreover the $K_L^{(3)}$ correction dominates over $g_a$, then  the bound will be similar to \eqref{Ft-boundIIB2}. Indeed, from \eqref{Ft-Nfluxineqlin2} we then obtain
  \begin{eqn}
 \frac{9}{8} |K^{(3)}_L|  \frac{\cK_{L a}}{\cK_L} \gtrsim N_{\rm flux}^{-2} \, .
\end{eqn}

    \item[-] If for all $a$, $g_a^\infty$ tends to a finite number bigger than $N_{\rm flux}^{-1}$, then \eqref{Ft-Nfluxineqlin} is automatically satisfied and no bound is imposed on the saxion vevs in order to find vacua in this region. This is for instance the case of the overall rescaling  $t^a  \to \lambda t^a, \lambda \to \infty$, since due to the homogeneity of $f$ and $\CK_L$ all the $g_a^\infty$ tend to quotients of intersection numbers. Therefore for sufficiently large values of $N_{\rm flux}$ \eqref{Ft-Nfluxineqlin} becomes trivial. Notice that in this case the monodromy-invariant flux combination $N_a$ only scans a finite number of values along the limit, and so the set of inequivalent flux vacua in this regime should be finite.
    
\end{itemize}
A potential third possibility would be that $g_a \to \infty, \forall a$, which would also imply that the bound \eqref{Ft-Nfluxineqlin} is automatically satisfied and that all possible values of $N_a$ are scanned along the limit, yielding an infinity of flux vacua. However, this scenario is not realized here. To see this, we note that the limit \eqref{Ft-limitL} requires that we have to blow up (some of) the saxions $t^a$, possibly at different rates. Now, there is always (at least) one saxion $t^*$ that grows the fastest. Due to the fibrations structure of the mirror $X_8$ the terms appearing in $f$ are determined by the intersection numbers $\cK_{Labc}$ of the fibre $X_3$ and the details of the twist of $X_3$ over $\mathbb{P}^1$. As a consequence in the limit \eqref{Ft-limitL} we can estimate 
\begin{align}
    f \lesssim t^* \cK_L\,.  
\end{align}
With this information, we can now evaluate the component $|g_*^\infty|$  as 
\begin{align}
    4|g_*^\infty|=\left|\partial_* \left(\frac{f}{\cK_L}\right)\right| \leq \left|\frac{\partial_* f}{\cK_L} \right| + \left|\frac{f \partial_* \cK_L}{\cK_L^2}\right| \lesssim \left|\frac{\cK_L}{\cK_L} \right|+ \left|\frac{t^* \cK_L \partial_* \cK_L}{\cK_L^2}\right| \,.
\end{align}
We can further use $t^* \partial_* \cK_L \lesssim \mathcal{O}(\cK_L)$ to see that the second term on the RHS is finite in the limit \eqref{Ft-limitL}. Given that the first term on the RHS is $\mathcal{O}(1)$ we find that at least $|g_*^\infty|$ can never diverge along \eqref{Ft-limitL}, and so the possibility $g_a\rightarrow \infty$, $\forall a$ cannot be realized. 

\subsubsection*{Beyond large complex structure}

As we have seen, the linear scenario is quite natural in the context of F-theory four-fold compactifications at large complex structure, and one may construct several explicit examples like the one discussed in section \ref{Ft-sec:counter}.  A natural question is then if the same setup can be realized along other limits of infinite distance within the complex structure field space. To address this question let us extract the key features and the underlying geometric picture that lies behind the linear scenario, in order to connect with the results of \cite{Grimm:2019ixq}, where techniques were developed to address the features of flux potentials along general infinite distance limits.

For this, notice that the leading-order saxion-dependent matrix \eqref{Ft-ZABlin} is of the form
\be
\label{Ft-ZABTC}
2{\cal V}_3^2 Z + \chi_0 = 
 \begin{pmatrix}
H & &  &  & & &  \\
& M & M \varepsilon_a  & &  &   &  \\ 
& M \varepsilon_a & H_{ab}   &  &  &  & \\
 & & &   g_{\mu\nu}  & & &  \\
& &    & &   H^{ab} &   -H^{ab}\varepsilon_a &  \\
&  &  & &  -H^{ab}\varepsilon_a & M^{-1} + H^{ab}\varepsilon_a \varepsilon_b&  \\
 & & &    & & & H^{-1} \\
\end{pmatrix} \, ,
\ee
where $\chi_0$ is defined in \eqref{Ft-eq:chi0}, and $H^{ac}H_{cb} = \delta^a_b$. From the results of appendix \ref{Ft-ap:georho}, we can interpret this matrix as the Hodge star action on the basis of four-forms $ \{ \tilde\a, \tilde\a_i, \tilde\sigma_\mu, \tilde\b^i, \tilde\b\}$ in which the component of $G_4$ are the flux-axion polynomials $\rho_A$, see eq.\eqref{Ft-ap:G4rho}. In \eqref{Ft-ZABTC} this action is block-diagonal, which is a general feature of the large complex structure regime, cf.\eqref{Ft-ZAB}. In fact, it follows from the results of \cite{Grimm:2019ixq} that the Hodge star action is approximately block-diagonal in any complex structure region in the vicinity of an infinite distance point.

Now, it is also a general result of \cite{Grimm:2019ixq} that as we approach an asymptotic region in complex structure field space, the different blocks in the Hodge star action behave differently. Some of them tend to infinity and some of them tend to zero, while the rest remain of finite order. In the linear scenario we have that $H, H_{ab} \to \infty$, $H^{-1}, H^{ab} \to 0$, and $M$ remains of order one. To find vacua one then needs to set $\tilde{\rho} = \tilde{\rho}^a =0$, which implies the flux constraint $m = m^a =0$. Finally, it is reasonable to assume that  it is consistent with the vacua equations to set to zero some of the fluxes $\hat{m}^\mu$, in such a way that $\eta_{\mu\nu} \hat{m}^\mu\hat{m}^\nu =0$. The only contribution to the tadpole is then
\be
N_{\rm flux} = - m^L\bar{e}_L\, , 
\ee
which is independent of $h^{3,1}(Y_8)$. This leads to a vector of flux-axion polynomials of the form $\vec{\rho}^{\, t} = \left(0, m^L, 0, \hat{\rho}^a, 0,  \bar{\rho}_a, \bar{e}_L, \bar{\rho} \right)$, from where the equations of motion follow
\begin{eqn}
\bar \rho = 0\, ,\qquad \hat{\rho}^a   =  0\, , \qquad \bar{\rho}_a = \varepsilon_a  \bar{e}_L \, , \qquad \bar{e}_L + M m^L  & = 0\, .
\label{Ft-vacuagen}
\end{eqn}
From here one obtains that $M \in (N_{\rm flux}^{-1}, N_{\rm flux})$, and the inequalities
\begin{eqn}
N_{\rm flux} |\varepsilon_a| \gtrsim 1\ \implies \ M|\varepsilon_a| \gtrsim N_{\rm flux}^{-2}\, .
\label{Ft-Nfluxineqgen}
\end{eqn}
The relevance of these bounds depends on the asymptotic behaviour of the $\varepsilon_a$ along each limit. By the results of \cite{Grimm:2019ixq} one would expect that $\varepsilon_a$ either tends to zero, increasing the number of blocks in which the Hodge star action is divided, or it remains finite. If all $\varepsilon_a$ tend to zero, then we recover a bound for the saxion vevs, just as in the \textbf{IIB2} scheme of section \ref{Ft-sec:moduliIIB}. If they do not, there is a priori no bound for the saxion vevs, but the values that the monodromy-invariant flux bilinear $N_a$ can take is limited, and so should be the number of inequivalent flux vacua. 

As we depart from the large complex structure region, some of the entries of \eqref{Ft-ZABTC} will stop being zero, and the above block-diagonal structure will be further broken. A clear example of this is the effect of $K^{(3)}$ corrections in the \textbf{IIB2} scheme, that besides generating a non-vanishing $\varepsilon_a$, induce additional non-vanishing off-diagonal entries in \eqref{Ft-ZABTC}. However, in that case such additional corrections do not deform significantly the set of vacua equations \eqref{Ft-vacuagen}, as can be appreciated from \eqref{Ft-eq:MinkcorrIIB2}. As a result, this moduli stabilization scheme can be taken to be valid on a large region of complex structure field space. Whether this last feature is also present along limits outside of the large complex structure regime is yet to be seen, although the robustness of the equations in the \textbf{IIB2} setup suggests that this could well be the case.


\section{Examples}
\label{Ft-s:examples}

As stressed in section \ref{Ft-s:potential}, the most subtle part of the flux potential is the piece related to the four-forms $\sigma^\mu$, whose basis is not known in general. Exceptions to this  are four-folds $Y_8$ whose mirror dual $X_8$ is a smooth fibration, of which the setups in sections \ref{Ft-s:IIB} and \ref{Ft-s:linear} are particular subcases. In this section we provide explicit constructions that illustrate our previous results, by considering two types of fibrations for $X_8$. In section \ref{Ft-sec:elliptic} we apply our framework to the case in which $X_8$ is an elliptic fibration, which is a natural generalization of the type IIB case. In section  \ref{Ft-sec:twofield} we study a concrete two-field model of this setup, and show how the bounds for the saxion vevs obtained in section \ref{Ft-sec:moduli} are realized in practice. Section \ref{Ft-sec:counter} considers a four-fold $X_8$ that is a fibration of a Calabi--Yau three-fold over $\mathbb{P}^1$, yielding a concrete realisation of the linear scenario of section \ref{Ft-s:linear}.

\subsection{Elliptically fibered mirror}
\label{Ft-sec:elliptic}

A natural generalization of the type IIB limit is given by Calabi--Yau four-folds $Y_8$ whose mirror $X_8$ is a smooth, elliptically fibered four-fold with a section. In this case all the topological invariants of $X_8$ are determined by the three-fold base $B_6$ and so, as pointed out in \cite{Cota:2017aal}, one has explicit control over the set of four-forms $\sigma^\mu$. In our language, this allows us to determine the intersection numbers $ \zeta_{\mu, ij}$ explicitly, specify the form of the flux potential, and to carry out our analysis with the same degree of detail as in the type IIB limit. 

To see how this works, let us construct explicitly a basis of holomorphic $2p$-cycles classes in the mirror four-fold $X_8$, as done in the type IIB case. On the three-fold base $B_6$ of $X_8$, a basis of holomorphic $2p$-cycles is given by the point class $\mathcal{O}_{pt}$, the generators of the Mori cone  $[\mathcal{C}'^a]$, $a = 1, \dots, h^{1,1}(B_6) = h^{1,1}(X_8)-1$, the divisors classes $[D_a']$ that generate the K\"ahler cone, and the class of $B_6$. The relevant topological invariants for us will be the triple intersection numbers and the first Chern class of $B_6$:
\be
\kappa_{abc} = [D_a']\cdot [D_b']\cdot [D_c']\, , \quad \text{and} \quad c_1(B_6) = c_1^a  [D_a']\, .
\ee
We embed the holomorphic cycles of $B_6$ into $X_8$ by using the projection of the fibration $\pi$ and the divisor class of the section $[E]$. In particular, the  Mori cone of $X_8$ is generated by
\be
[\mathcal{C}^a] = [E . \pi^{-1} (\mathcal{C}'^a)] \, , \qquad [\mathcal{C}^0]\, ,
\ee
with $[\mathcal{C}^0]$ the class of the fibre. The K\"ahler cone is generated by the dual basis of divisor classes 
\be\label{Ft-kahlerconeelliptic}
[D_a] =\pi^*[D_a']\, , \qquad [D_0] = [E] + \pi^* c_1(B_6)\, .
\ee
Similarly to the type IIB case, we can construct a set of holomorphic four-cycle classes as 
\be
[\gamma_{ij}]  = [D_i . D_j] \, , \quad i = \{0, a\}\, .
\label{Ft-gammafib}
\ee
Again, all holomorphic four-cycle classes can be generated from linear combinations of $[\gamma_{ij}]$, but \eqref{Ft-gamma} does not form a basis because it is not a linearly independent set. For the case at hand, one can construct such a basis from
\be
  [H_a] = [D_0 . \pi^{-1}(D_a)] \,, \qquad [H_{\hat{a}}] = \pi^{*}[\mathcal{C}^a] \, ,
  \label{Ft-4formbasisfib}
\ee
which reduces to \eqref{Ft-4formbasis} when the fibration is trivial. This is a different choice of basis compared to the one taken in \cite{Cota:2017aal}, but more convenient for our purposes. The integral basis of four-form classes $[\sigma_\mu]$ that correspond to the period \eqref{Ft-period4} is then given by $\{[\sigma_\mu]\} = \{[H_a], [H_{\hat{a}}]\}$, and so $\mu = \{a, \hat{a}\}$, with $a, \hat{a} = 1, \dots, h^{1,1}(B_6)$. Notice that the number of elements of the basis \eqref{Ft-gammafib} is $2h^{1,1}(X_8) -2$, smaller than the $\half h^{1,1}(X_8) (h^{1,1}(X_8)+1)$ elements in \eqref{Ft-gammafib}. The tensor $\zeta^\mu_{ij}$ connecting both sets of four cycles as
\be
[\gamma_{ij}] = \zeta^\mu_{ij} [\sigma_\mu] = \zeta^a_{ij} [H_a] + \zeta^{\hat{a}}_{ij} [H_{\hat{a}}]\, ,
\ee
is specified by
\begin{align}
\zeta^a_{0b} = \zeta^a_{b0}  = \delta_{ab}\, , \qquad   \zeta_{a,bc}\equiv \zeta^{\hat{a}}_{bc}\eta_{\hat{a}a}= \kappa_{abc}\, ,\qquad \zeta_{00}^a = c_1^a \,,
    \label{Ft-Koszulfib}
\end{align}
with all remaining components vanishing. This clearly reduces to \eqref{Ft-Koszul} for $c_1^a =0$, and one can check that it satisfies the relation \eqref{Ft-interrel}. The intersection matrix for the basis \eqref{Ft-4formbasisfib} is given by 
\begin{align}\label{Ft-etafibration}
    \eta_{\hat{a}\hat{b}}=0\,,\qquad \eta_{a\hat{b}} = \delta_{a\hat{b}} \,,\qquad \eta_{ab}=  \kappa_{abc}c_1^c \equiv c_{ab} \,,
\end{align}
and so applying  \eqref{Ft-interrel} we recover
\begin{eqn}
    \cK_{0abc}&= \kappa_{abc}\,,\qquad \qquad\qquad \cK_{00ab}= \kappa_{abc} c_1^c\equiv c_{ab} \,,\\
    \cK_{000a}&= \kappa_{abc} c_1^b c_1^c\equiv c_a \,,\qquad \cK_{0000}=\kappa_{abc}c_1^a c_1^b c_1^c\equiv c\, ,
    \label{Ft-c1rel}
\end{eqn}
which indeed are  the quadruple intersection numbers of the elliptically fibered four-fold $X_8$. Furthermore, for a $X_8$ a smooth Weierstrass model the Euler characteristic $\chi(X_8)$ can be calculated from the adjunction formula as 
\begin{align}
    \chi(X_8) = \int_{B_6} \left[12 c_1(B_6)\wedge c_2(B_6) + 360 \, c_1(B_6)\wedge c_1(B_6) \wedge c_1(B_6) \right]\, ,
\end{align}
which also gives the Euler characteristic for the mirror $Y_8$. In the mirror four-fold $Y_8$ \eqref{Ft-Koszulfib} translates, via \eqref{Ft-hatm}, into the following dictionary for the set of $G_4$-flux quanta 
\be
{m}_a \equiv \delta_{a\hat{b}} \hat{m}^{\hat{b}} = \oh  \kappa_{abc} m^{bc}\, , \qquad \hat{m}^a = m^{0a} + \oh c_1^a m^{00}\, ,
\label{Ft-hatmfib}
\ee
which are the generalization of the type IIB fluxes $m_a$, $\hat{m}^a$ to the present case and  the actual $G_4$-flux quanta,\footnote{Because the mirror manifold $X_8$ is a smooth elliptic fibration the quantization condition for the $G_4$ flux \cite{Witten:1996md} is trivial, in the sense that $[G_4]$ must be an integer class \cite{Collinucci:2010gz}. In the present setup this implies that $m_a, \hat{m}^a \in \Z$. In fact, all flux quanta in \eqref{Ft-G4} should be integers when $X_8$ is a smooth elliptically fibered four-fold.} while $m^{ij}$ should be seen as auxiliary quanta. The superpotential then reads
\begin{align}
    W^{\rm corr}=&\, \bar{e}+\bar{e}_i T^{i}+\frac{1}{2}\kappa_{abc}\bar{m}^cT^bT^c+T^0 c_{ab}\bar{m}^a T^b+\frac{1}{2} (T^0)^2c_a\bar{m}^a+T^0T^a\bar{m}_a+\frac{1}{2}(T^0)^2 c_1^a\bar{m}_a\nonumber\\
    &+\frac{1}{6}m^0\kappa_{abc}T^aT^bT^c+\frac{1}{2}T^0\kappa_{abc}m^aT^bT^c+\frac{1}{2}m^0T^0c_{ab}T^aT^b+\frac{1}{2}(T^0)^2c_{ab}m^aT^b\nonumber\\
    &+\frac{1}{2}m^0 (T^0)^2c_{a}T^a+\frac{1}{6}(T^0)^3c_{a}m^a+\frac{1}{6}m^0(T^0)^3c  \nonumber\\
    &+\frac{m}{24}\left(c(T^0)^4+4(T^0)^3c_a T^a+6(T^0)^2c_{ab}T^aT^b+4T^0\kappa_{abc}T^aT^bT^c\right) \nonumber \\
    & -iK_{i}^{(3)}(m^i+mT^i)\, ,
\end{align}
where we have included the polynomial corrections of section \ref{Ft-sec:poly}, and in particular the flux redefinition \eqref{Ft-eq:fluxshift}. Similarly, the corrected K\"ahler potential is 
\begin{equation}
    K_{\rm cs}^{\rm corr}=-\log\left(\frac{2}{3}(4t^0\kappa+6(t^0)^2\kappa_ac^a_1+4(t^0)^3\kappa_{ab}c^a_1c^b_1+(t^0)^4c)+4K_i^{(3)}t^i\right)\, .
\end{equation}

It then follows from our general analysis that we recover a scalar potential of the form \eqref{Ft-bilinear}, where the flux-axion polynomials are given by 
\begin{align}
 \nonumber \bar{\rho} =&\, \bar{e} + \bar{e}_0 b^0+  \bar{e}_ab^a+ \bar{m}_a\left(b^a+c_1^a b^0\right) b^0 + \frac{1}{2}  \kappa_{abc} \hat{m}^{a} b^bb^c + \frac{1}{2} \kappa_{abc} \hat m^a c_1^bb^0 (b^c+ c_1^cb^0) \\
 \nonumber &+\frac{1}{6}\kappa_{abc}\left(3 m^ab^bb^cb^0 +3m^a c_1^b (b^0)^2 b^c+m^a c_1^b c_1^c (b^0)^3 +m^0(b^a+c_1^ab^0)(b^b+c_1^bb^0)(b^c+c_1^c b^0)\right)\\ 
 \nonumber&+\frac{m}{24} \kappa_{abc}\left(4b^ab^bb^c b^0+6b^a b^b c_1^c(b^0)^2 +4 b^ac_1^b c_1^c (b^0)^3 + c_1^a c_1^bc_1^c (b^0)^4\right)\,,\\
\nonumber \bar{\rho}_0=& \,\bar{e}_0+\bar{m}_a(b^a +c_1^ab^0)+\kappa_{abc}\hat m^ac_1^b(b^c+c_1^cb^0) +\frac{1}{2}\kappa_{abc} \left(m^a+c_1^am^0\right)(b^b+c_1^b b^0)(b^c+c_1^cb^0) \\
\nonumber &+\frac{m}{6}(b^a+c_1^a b^0)(b^b+c_1^b b^0)(b^c+c_1^cb^0) \,,\\
\nonumber \bar{\rho}_a=& \,\bar{e}_a + \bar{m}_ab^0 +  \kappa_{abc} \hat{m}^{b}\left(b^c + c_1^cb^0\right)  + \kappa_{abc} \left(m^b b^0 \left(b^c + \frac{1}{2} c_1^c b^0\right) +\frac{1}{2}m^0 (b^b+b^0c_1^b)(b^c+b^0c_1^c)\right)\\
 &+ \frac{m}{6}\kappa_{abc}\left( 3b^b b^c b^0+3(b^0)^2 c_1^b b^c +(b^0)^3 c_1^b c_1^c\right)\,,\\
\nonumber \bar{\rho}_a^\prime=& \,\bar{m}_a +  \kappa_{abc} \left(m^b b^c + \frac{1}{2} m b^b b^c\right) \,,\\
\nonumber \bar{\rho}^{a}=&\, \bar{m}^{a} + m^a b^0 +m^0 b^a+ c_1^a m^0 b^0 + m \left(b^0 b^a +\frac{1}{2} c_1^a (b^0)^2\right)  \,,\\
\nonumber \tilde \rho^a=&\, m^a + m b^a \,, \\
\nonumber \tilde \rho^0=&\,m^0 + mb^0 \,, \\
\nonumber \tilde \rho=&\,m\, ,
\end{align}
and the saxion-dependent matrix reads, in the limit where the corrections $K^{(3)}_i$ can be neglected
\begin{align}
\label{Ft-ZABfib}
Z^{AB} =
\frac{e^K\cK}{3}
\begin{pmatrix}
\frac{\cK}{24} & & & & & -1 \\
&  \frac{\cK}{6} g_{ij} & & &  \d^i_j &  \\
& &   \tilde{B}_a{}^c\tilde{A}_{cd}\tilde{B}^d{}_b  & \tilde{B}_a{}^b & & \\
& & \tilde{B}^b{}_a & \tilde{A}^{ab} &  & \\
&  \d^i_j & & & \frac{6}{\cK} g^{ij} &  \\
-1 & & & & & \frac{24}{\cK} \\
\end{pmatrix} \, ,
\end{align}
with $\vec{\rho}^{\, t} = \left(\tilde{\rho},  \tilde{\rho}^i,   \bar{\rho}^{a},  \bar{\rho}_a^\prime, \bar{\rho}_i,    \bar\rho   \right)$. Here we have separated the tensor $g_{\mu\nu} -\eta_{\mu\nu}=  \frac{6}{\cK}  g^{ij}_{P}\zeta_{\mu i}\zeta_{\nu j}$   in \eqref{Ft-ZAB} into four blocks, reflecting the  splitting $\mu = \{a,\hat{a}\}$.  In particular, the matrices that appear in \eqref{Ft-ZABfib} are related to the metric $g_{\mu\nu}$ defined below \eqref{Ft-ZABdiag} by
\begin{equation}
   \tilde{A}^{ab}\delta_{a\hat{c}}\delta_{b\hat{d}}= {g}_{\hat{c}\hat{d}}\, , \qquad \tilde{B}_a{}^b\delta_{b\hat{c}}= {g}_{a\hat{c}} - {\delta}_{a\hat{c}}\, ,
   \label{Ft-eq: gmunu fib}
\end{equation}
and their explicit form is
\begin{align}
\label{Ft-tildeA}
  \tilde{A}^{ab}=&-2\left[\cK^{00}\left(t^at^b+t^0(t^ac_1^b+t^bc_1^a)+(t^0)^2c_1^ac_1^b\right)+\cK^{0a}t^0(t^b+t^0c_1^b)+\cK^{0b}t^0(t^a+t^0c_1^a)\right.\nonumber\\
    &\left.+\cK^{ab}(t^0)^2\right]+\frac{2(t^0)^2}{\cK}\left[4t^at^b+2t^0(t^ac_1^b+t^bc_1^a)+(t^0)^2c_1^ac_1^b\right]\, , \\
   \tilde{B}^b{}_a =&\   -2\left[\cK^{00}\left(\kappa_{ac}c_1^ct^b+t^0(c_at^b+\kappa_{ac}c_1^cc_1^b)+(t^0)^2c_ac_1^b\right)+\cK^{0b}t^0(\kappa_{ac}c_1^c+c_at^0)\right.\nonumber\\
    &\left.+\cK^{0c}\left(\kappa_{ac}t^b+t^0(\kappa_{ac}c_1^b+c_{ac}t^b)+(t^0)^2c_{ac}c_1^b\right)+\cK^{bc}t^0(\kappa_{ac}+t^0c_{ac})\right]\nonumber\\
    &+\frac{2t^0}{\cK}\left[2\kappa_a t^b+(t^0)\left(4\kappa_{ac}c_1^ct^b+\kappa_ac_1^b\right)+(t^0)^2\left(2\kappa_{ac}c_1^cc_1^b+2c_at^b\right)+(t^0)^3c_ac_1^b\right]\, .
  \end{align}
Finally, $\tilde{A}_{ac}\tilde{A}^{cb} = \delta_a^b$, from where the structure \eqref{Ft-blockZ} is manifest.

\subsubsection*{Moduli stabilization}

Let us now write down the Minkowski vacuum equations for the case at hand, and study to what extent the results from the Type IIB orientifold limit generalize to this class of compactifications. In this setup the on-shell conditions \eqref{Ft-eq:Mink} become
\begin{subequations}
    \label{Ft-eq:Mink elliptic compact}
\begin{empheq}[box=\widefbox]{align}
 \bar\rho &= \frac{1}{24}\mathcal{K}\tilde{\rho}   \\
  \bar{\rho}_i&=-\frac{1}{6}\mathcal{K}g_{ij}\tilde{\rho}^j    \\
  \bar \rho_a'&= \Gamma_{ab} \bar \rho^b  \label{Ft-eq:Mink elliptic compact rho mu}
\end{empheq}
\end{subequations}
where we have defined $\Gamma_{ab} \equiv - \tilde{A}_{ac}\tilde{B}^c{}_b$. An explicit expression for this matrix is given in appendix \ref{Ft-ap:elliptic}, from where one can see that for vanishing $c_1^a$, $\Gamma_{ab}  \to \frac{2}{3} \frac{\kappa}{t^0} g_{ab}^\kappa$, and  we recover \eqref{Ft-eq:MinkIIB}.  Using the vacuum equations we can rewrite the  flux contribution to the tadpole as
\begin{align}
\label{Ft-eq: elliptic fibered flux}
    N_{\rm flux} = \bar{\rho} \tilde{\rho} - \bar{\rho}_i \tilde{\rho}^i + \frac{1}{2} \eta_{\mu\nu} \bar{\rho}^{\mu} \bar{\rho}^{\nu}
    \stackrel{\rm vac}{=} \frac{\mathcal{K}}{24}\left(\tilde{\rho}^2+4g_{ij}\tilde{\rho}^i\tilde{\rho}^j\right)+\frac{1}{2}\left(c_{ab} + \Gamma_{ab} + \Gamma_{ba} \right) \bar{\rho}^a\bar{\rho}^b\, ,
\end{align}
where the last term is positive definite by construction, as it equals $\oh g_{\mu\nu}\bar{\rho}^{\mu} \bar{\rho}^{\nu}$.  Hence, as in the type IIB case, in order not to overshoot the D3-brane tadpole we have to set $\tilde \rho=m=0$ and we further set $\tilde \rho^a=m^a=0$. However, unlike in the type IIB case the saxion $t^0$ now enters with a fourth power in $\cK$. Thus, based on our general discussion in section \ref{Ft-s:vacua}, we also need to demand $\tilde \rho^0=m^0=0$ to find vacua that do not violate the tadpole constraint at large complex structure.   

As before, including the corrections $K_i^{(3)}$ will modify the vacuum equations. At linear order in these corrections we have \eqref{Ft-eq:Minkcorr} adapted to this setup, which reads:
\begin{subequations}
\label{Ft-eq:Minkcorr elliptic}
\begin{align}
&\bar\rho- \frac{1}{24}\mathcal{K}\tilde{\rho}  =  - \frac{3}{8}\eps_it^i \left[ \frac{\cK}{18}\tilde{\rho} +  \varpi \right] \, , \\
& \bar{\rho}_i+\frac{1}{6}\mathcal{K}g_{ij}\tilde{\rho}^j   =   \frac{1}{3}  \CK_i  \left( \eps_j -  \epsilon_k t^k \frac{ \cK_j}{\cK} \right)\tilde{\rho}^j   - \frac{1}{6}\epsilon_i \cK_j \tilde \rho^j   \, , \\
 &\bar{\rho}_a'- \Gamma_{ab} \bar{\rho}^b  = \frac{E_a}{4t^0} \left[\frac{\cK}{2}\tilde{\rho}+ \varpi\right] \, ,
 \end{align}
 \end{subequations}
where we have defined $\varpi = (2t^at^0+(t^0)^2c_1^a)\bar{\rho}_a'+(\kappa_a+2\kappa_{ab}c_1^bt^0+c_a(t^0)^2)\bar{\rho}^a$ and
\begin{equation}
    E_a=\left[\epsilon_b-\frac{\cK_b}{(\cK-2\cK_0 t^0)}\left(2\epsilon_c t^c - \epsilon_i t^i\right)\right]\left[\delta_a^b-\frac{\cK_a c_1^b t^0}{\cK-2\cK_0 t^0+\cK_a c_1^a t^0}\right] \, .
\end{equation}
Let us now turn to the restricted flux scenario $m=m^i=0$ which yields $\tilde{\rho}=\tilde{\rho}^i=0$, $\bar{\rho}^a=\hat{m}^a$ and $\bar{\rho}_a'= {m}_a$. In this case \eqref{Ft-eq:Minkcorr elliptic} reduces to
\begin{subequations}
\label{Ft-eq: minkcorr elliptic simplify}
\begin{align}
 \bar\rho &=- \frac{3}{8}\eps_it^i \, \varpi \, , \label{Ft-eq: minkcorr elliptic simplify bar rho}\\
  \bar{\rho}_i& =0\, ,\\
 m_a - \Gamma_{ab}  \hat{m}^b &= \frac{1}{4} \left(\eps_a -\frac{\cK_a(\epsilon_ct^c-\epsilon_0t^0+t^0\epsilon_bc_1^b)}{\cK-2\cK_0t^0+\cK_ac_1^at^0}\right)   \varpi \, .
 \label{Ft-eq: minkcorr elliptic simplify rho mu}
 \end{align}
\end{subequations}
In order to stabilize all complex structure fields, we need to choose the flux quanta $(m_a, \hat m^a)$ such that the matrix $M$ defined in \eqref{Ft-Meq} is invertible. In the present case of a smooth elliptic fibration the matrix $M$ is given by 
\begin{align}\label{Ft-Mellipticfibration}
 M = \left(\begin{matrix} M_{00} & M_{0a} \\ M_{a0} & M_{ab} \end{matrix} \right) =  \left(\begin{matrix} c_1^a m_a + c_{a}\hat m^a & m_a + c_{ab}\hat m^b \\ m_a + c_{ab}\hat m^b  & \kappa_{abc} \hat m^b \end{matrix} \right) \,. 
\end{align}
To see whether this matrix is invertible, let us define the matrices $S_{ab} \equiv \kappa_{abc} \hat m^b$ and
\begin{align}
    \tilde S_{ab} = S_{ab} - \frac{\left(m_a + c_{ac}\hat m^c\right) \left(m_b + c_{bd}\hat m^d\right)}{c_1^c m_c +c_c\hat m^c}\,.
\end{align}
Now the block-matrix \eqref{Ft-Mellipticfibration} is invertible if one of the two is fulfilled 
\begin{eqn}\label{Ft-invertibleMelliptic}
    a)&:\qquad S_{ab}\; \;\text{invertible} \qquad\qquad \text{and} \qquad m_a S^{ab} m_b + c_1^a m_a \ne 0 \,,\\
    b)&: \qquad c_1^c m_c +c_c\hat m^c \ne 0 \,,\;\qquad \text{and} \qquad \tilde S_{ab}\; \;\text{invertible}\,. 
\end{eqn}

The solution \eqref{Ft-splitmu} now reads 
\begin{equation}
    m_a = A\kappa_a + C_a + \cO(\eps_i) \, , \qquad \hat{m}^a = A\left(2t^at^0+(t^0)^2c_1^a\right) + C^a + \cO(\eps_i)\, , \label{Ft-eq: ansatz elliptic fib}
\end{equation}
with the coefficients $C_a$ and $C^a$ satisfying
\begin{equation}
    C_at^0=-(\kappa_{ab}+c_{ab}t^0)C^b\, , \qquad (c^a\kappa_{ab}t^0+\kappa_b)C^b=0\, .
\end{equation}
Then \eqref{Ft-eq: minkcorr elliptic simplify bar rho} allows us to recover \eqref{Ft-astiA}
\begin{equation}
    A=-\frac{4\bar{\rho}}{9 K^{(3)}_i t^i}\, .
\end{equation}
In addition, \eqref{Ft-eq: minkcorr elliptic simplify rho mu} simplifies to
\begin{equation}
    m_a- \Gamma_{ab} \hat{m}^b=\frac{A}{4}\left(\eps_a -\frac{\cK_a(\epsilon_ct^c-\epsilon_0t^0+t^0\epsilon_bc_1^b)}{\cK-2\cK_0t^0+\cK_ac_1^at^0}\right)\, ,
\end{equation}
and \eqref{Ft-eq: elliptic fibered flux} becomes
\begin{equation}
    N_{\rm flux}=\frac{1}{2}A^2\cK-\frac{1}{t^0}(\kappa_{ab}+\frac{1}{2}c_{ab}t^0)C^aC^b+\mathcal{O}(\epsilon_i)\geq \frac{1}{2}A^2\cK +\mathcal{O}(\epsilon_i)\, .
\end{equation}
At this point we may apply the reasoning below \eqref{Ft-Meq} to obtain the inequality $N^p_{\rm flux}\bar{\rho}\gtrsim d^{2p-1}$ with $d=\textrm{gcd}(m_a,\hat{m}^a)$ and $p\leq h^{(3,1)}$. Hence we conclude that
\begin{equation}
\cK <   N_{\rm flux}^{2p+1}d^{2-4p}  (K_i^{(3)}t^i)^2 \, .
\end{equation}

\subsection{A two-field model}
\label{Ft-sec:twofield}

As a concrete example of a Calabi--Yau four-fold $Y_8$ for which the mirror $X_8$ is elliptically fibered, let us consider $X_8$ to be the degree 24 hypersurface in $\mathbb{P}^{5}_{(1,1,1,1,8,12)}$. This manifold has been studied in the context of moduli stabilization for instance in \cite{Cota:2017aal}. This hypersurface can be viewed as an elliptic fibration over $\mathbb{P}^3$ with intersection polynomial 
\begin{align}
    I(X_8) = 64 D_0^4 + 16 D_0^3 D_1 + 4 D_0^2 D_1^2 + D_0D_1^3\,,
\end{align}
where $D_0$ is the K\"ahler cone divisor associated to the zero section $E$, $[D_1]=\pi^*[H]$ the pull back of the hyperplane class in $\mathbb{P}^3$ and $c_1(\mathbb{P}^3)=4H$. For this four-fold we have the following basis of four-cycles 
\begin{align}
    [H_1] = [D_0 . D_1]\,,\qquad [H^1]= \pi^*[\mathcal{C}^1]\,,
\end{align}
with $\mathcal{C}^1$ the single Mori cone generator of $\mathbb{P}^3$. The non-vanishing components of the tensor $\zeta^\mu_{ij}$ as in \eqref{Ft-Koszulfib} are thus given by
\begin{align}
    \zeta^{1}_{01} = 1 \,,\qquad \zeta_{1,11}= \kappa_{111}=1\,,\qquad \zeta^1_{00}=c_1^1=4\,,
\end{align}
and the intersection matrix $\eta$ reduces to 
\begin{align}
    \eta^{11}=0\,,\qquad \eta^1_1=1\,,\qquad \eta_{11}=4\,.
\end{align}
Furthermore, the corrections $K^{(3)}_i$ for this example are given by 
\begin{align}
    K^{(3)}_0 = -3860\frac{\zeta(3)}{(2\pi)^3}\,,\qquad K^{(3)}_1=  -960\frac{\zeta(3)}{(2\pi)^3}\,. 
\end{align}
Finally, the Euler number of $X_{24}$ and its mirror $Y_8=X_{24}^*$ is $\chi(X_{24})=\chi(X_{24}^*)= 23328$. With this preparation we can now look at flux vacua for F-theory on  $X_{24}^*$ in the large complex structure regime. To find vacua for large values of the saxions $t^0, t^1$ we restrict to the  flux vector 
\begin{align}
    \vec{q}^t = (0,0,0, \hat m^1, m_1, \bar e_1, \bar e_0, \bar e)\,. 
\end{align}
The vacuum equations $\bar \rho_i=0$ then translate to 
\begin{align}
    \bar e_0 + m_1\left(b^1 +4b^0\right) +4\hat m^1\left(b^1 +4b^0\right)  =0 \,,\qquad \bar e_1 +\hat m^1\left(b^1+4b^0\right) +m_1 b^0=0\,, 
\end{align}
such that the matrix $M$ in \eqref{Ft-Mellipticfibration} is given by 
\begin{align}
    M = \left(\begin{matrix} 4m_1 +16\hat m^1 & m_1 + 4\hat m^1 \\ m_1 +4\hat m^1 & \hat m^1  \end{matrix}\right)\,,
\end{align}
which is invertible provided $m_1+4\hat m^1 \ne 0$ and $m_1\ne 0$. In case this is fulfilled we obtain
\begin{align}
    b^0&=\frac{\hat m^1 \bar e_0}{m_1 \left(4\hat m^1 +m_1\right)}-\frac{\bar e_1}{m_1}\,,\\
    b^1&=-4\frac{\hat m^1 \bar e_0}{m_1 \left(4\hat m^1 +m_1\right)} -\frac{\bar e_0}{4\hat m^1+m_1}+\frac{4\bar e_1}{m_1}\,. 
\end{align}
From here we can deduce 
\begin{align}\label{Ft-barrhoX24}
\bar \rho = \frac{2\left(4\hat m^1+m_1\right)m_1 \bar e - m_1 \bar e_1 \bar e_0 + \hat m^1\bar e_0\left(\bar e_0 -4\bar e_1\right)+3\bar e_0\bar e_1\left(4\hat m^1+m_1\right)}{2\left(4\hat m^1+m_1\right)m_1} \,, 
\end{align}
for which the numerator is a combination of integer fluxes and thus at least of $\mathcal{O}(1)$ if non-vanishing. 
We can further use \eqref{Ft-eq:Mink elliptic compact rho mu} to solve the vacuum equations for $\rho^\mu$ at leading order. For our particular two-modulus case we have
\begin{align}
    m_1 = \Gamma_{11} \hat m^1\,,
\end{align}
with
\begin{align}
    \Gamma_{11} = \frac{(t^1)^4+ 4t^0(t^1)^3}{2(t^1)^3t^0+12(t^1)^2 (t^0)^2+ 16 (t^0)^3t^1}\,. 
\end{align}
The corrected equations of motion for $\bar \rho$ now give
\begin{align}\label{Ft-eqofmbarrhoX24}
\frac{2\left(4\hat m^1+m_1\right)m_1 \bar e - m_1 \bar e_1 \bar e_0 + \hat m^1\bar e_0\left(\bar e_0 -4\bar e_1\right)+3\bar e_0\bar e_1\left(4\hat m^1+m_1\right)}{2\left(4\hat m^1+m_1\right)m_1} = -\frac{3}{8}\epsilon_i t^i \zeta_\mu m^\mu\,,
\end{align}
with 
\begin{align}\label{Ft-zetamummuX24}
    \zeta_\mu m^\mu = \left[(t^1)^2 + 4t^1 t^0 +16 (t^0)^2\right]\hat m^1 + \left(t^1 t^0 +4(t^0)^2\right) \Gamma_{11} \hat m^1\, .
\end{align}
Furthermore, in this model the contribution to the tadpole is then given by 
\begin{align}\label{Ft-NfluxX24}
        N_\text{flux}=\hat m^1m_1 + 4(\hat m^1)^2= \left(\Gamma_{11}+4\right)(\hat m^1)^2\,.
\end{align}
We now want to find the bound on $\cK$ and $t^i$ for which we expect solutions to the vacuum equations similar to \eqref{Ft-eq: K3 saxion bound}. To that end, let us distinguish three different cases depending on the hierarchy between $t^1$ and $t^0$: 
\begin{itemize}
    \item[$i)$] {\em Vacua with the hierarchy $t^1\gg t^0$}. In this case we can approximate
    \begin{align}
        m_1 = \left[\frac{t^1}{2t^0}\left(1+ 4 \frac{t^0}{t^1} + \mathcal{O}\left(\frac{t^0}{t^1}\right)^2 \right)\right]\hat m^1= \left(\frac{t^1}{2t^0}+2\right)\hat m^1+ \mathcal{O}\left(\frac{t^0}{t^1}\right)\,. 
    \end{align}
    Thus in order to have the required hierarchy we need $m_1 \gg \hat m^1$ such that the contribution to the tadpole goes essentially as $N_\text{flux}\gtrsim \Gamma_{11}$. From \eqref{Ft-barrhoX24} we then find
    \begin{align}\label{Ft-NfluxbarrhoX24i}
        \qquad N_\text{flux}^2 \bar \rho\gtrsim 1\,,
    \end{align} 
    i.e. we would expect \eqref{Ft-eq: K3 saxion bound} to hold for $p=2$. The RHS of \eqref{Ft-eqofmbarrhoX24} to leading order is then given by
    \begin{align}
        -\frac{9}{16} \frac{K_0^{(3)} t^0 + K_1^{(3)}t^1}{t^0 (t^1)^3} \left((t^1)^2 \hat m^1+t^1 t^0 m_1\right) = \frac{27}{32} \frac{K_1^{(3)}}{t^0} \hat m^1+\mathcal{O}\left(\frac{1}{t^1}\right) \,. 
    \end{align}
    Using the bound \eqref{Ft-NfluxbarrhoX24i} we can derive 
    \begin{align}
        t^0 \lesssim \left(4\hat m^1+m_1\right)\left(m_1 \hat m^1\right)|K_1^{(3)}| = \Gamma_{11}^2(\hat m^1)^3 |K_1^{(3)}|\lesssim |K_1^{(3)}| N_\text{flux}^{2}\,.
    \end{align}
    Accordingly, $t^1$ is bounded by
    \begin{align}
        t^1\sim \frac{m^1}{\hat m^1} t^0 \lesssim \left(4\hat m^1+m_1\right)\left(m_1 m_1\right)|K_1^{(3)}|\lesssim N_\text{flux}^{3}|K_1^{(3)}|\,.
    \end{align}
    Combining the bound for $t^0$ and $t^1$ we find 
    \begin{align}
         \cK \lesssim \left(K^{(3)}_i t^i\right)^2 N_\text{flux}^5\,,
    \end{align}
    in accordance with \eqref{Ft-eq: K3 saxion bound} for $p=2$. 
    \item[$ii)$] {\em Vacua with both saxions  of the same order, i.e. $t^0/t^1=\gamma$ with $\gamma\sim \mathcal{O}(1)$}. In this case
    \begin{align}
         m_1 = \left(\gamma^{-1} \frac{1+4\gamma }{2+12\gamma+16\gamma^2}\right)\hat m^1\,,
    \end{align}
    such that in order for $\gamma$ to be $\mathcal{O}(1)$ we also need $\hat m^1$ and $m_1$ to be of the same order. Accordingly, from \eqref{Ft-barrhoX24} and \eqref{Ft-NfluxX24} we find 
    \begin{align}
        \qquad N_\text{flux} \bar \rho\gtrsim 1\,,
    \end{align}
    such that we expect the bound \eqref{Ft-eq: K3 saxion bound} with $p=1$. We can now set a bound on the overall saxion $t^1$. Using \eqref{Ft-zetamummuX24} we have that 
    \begin{align}
        -\frac{3}{8}\epsilon_i t^i \zeta_\mu m^\mu \sim \frac{1}{t^1} \hat m^1 f(\gamma)\,,
    \end{align}
    with $f$ a function of $\gamma$. From here, we derive the bound 
    \begin{align}
        t^1 \lesssim \hat m^1 \left(4\hat m^1+m_1\right)m_1 |K_1^{(3)} + \gamma K_0^{(3)}| \lesssim N_\text{flux}^{3/2} |K_1^{(3)} + \gamma K_0^{(3)}|\,, 
    \end{align}
    and similar for $t^0$. Combining the scaling of $t^0$ and $t^1$ we find the bound 
    \begin{align}
         \cK \lesssim \left(K^{(3)}_i t^i\right)^2 N_\text{flux}^3\,,
    \end{align}
    in accordance with \eqref{Ft-eq: K3 saxion bound} with $p=1$. 
    \item[$iii)$] {\em Vacua with the hierarchy $t^0\gg t^1$}. Here we find 
    \begin{align}
        m_1 = \left[ \frac{1}{4}\left(\frac{t^1}{t^0}\right)^2 + \mathcal{O}\left(\frac{t^1}{t^0}\right)^3 \right] \hat m^1   \,,
    \end{align}
    such that we need to impose $\hat m^1\gg m_1$ to achieve the required hierarchy. In view of \eqref{Ft-NfluxX24} and \eqref{Ft-barrhoX24} we then find the bound 
    \begin{align}\label{Ft-NfluxbarrhoX24iii}
       \qquad N_\text{flux}^{1/2}\bar \rho \gtrsim 1\,,
    \end{align}
    which should lead to \eqref{Ft-eq: K3 saxion bound} with $p=1/2$. In this regime we have that 
    \begin{align}
        -\frac{3}{8} \epsilon_i t^i \zeta_\mu m^\mu \sim -K_0^{(3)}\frac{\hat m^1}{t^0}\,.
    \end{align}
    From here we can then derive the bounds
    \begin{align}
        t^0 \lesssim 4 (\hat m^1)^2 m_1 |K_0^{(3)}| \lesssim  N_\text{flux} |K_0^{(3)}|\,, \qquad t^1\lesssim N_\text{flux}^{1/2} |K_0^{(3)}|\,.
    \end{align}
    Putting things together we then find 
    \begin{align}
         \cK \lesssim \left(K^{(3)}_i t^i\right)^2 N_\text{flux}^2\,,
    \end{align}
    in accordance with \eqref{Ft-eq: K3 saxion bound} for $p=1/2$.
\end{itemize}
\subsubsection*{Type IIB limit}
F-theory compactified on the four-fold $X_{24}^*$ can be viewed as the F-theory lift of type IIB compactified on the mirror quintic which has a single complex structure modulus $T^1$. The intersection number and Euler characteristic of the mirror, i.e. in the quintic itself, are 
\begin{align}
    \kappa_{111}=1\,,\qquad \chi_E=-200\,.  
\end{align}
The main difference to the case of $X_{24}^*$ discussed before is that now $t^0$ only appears linearly in the K\"ahler potential. In this case, the set of vacuum equations simplifies considerably. For instance at the classical level \eqref{Ft-eq:MinkIIB mu} reduces to
\begin{align}
     \bar{\rho}'_1 = \frac{1}{2} \frac{t^1}{t^0} \hat \rho^1 \,.
\end{align}
Focusing on the restricted flux case $\vec{q}^{\, t} = (0,0,0, \hat{m}^a, \bar{m}_a, \bar{e}_a, \bar{e}_0, \bar{e})$ this translates into 
\begin{align}
    \frac{t^1}{t^0} = \frac{2m_1}{\hat m^1}\,. 
\end{align}
In this case the equation for $\rho_i$ read 
\begin{align}
    \bar \rho_0 = \bar e_0 + m_1 b^1 =0 \,, \qquad \bar \rho_1 = \bar e_1 + m_1 b^0 + \hat m^1 b^1 =0\,,
\end{align}
which are solved by 
\begin{align}
    b^1 = -\frac{\bar e_0}{m_1} \,,\qquad b^0 = -\frac{\bar e_1}{m_1} + \frac{\hat m^1}{m_1^2} \bar e_0\,. 
\end{align}
This can be inserted into $\bar \rho$ to find 
\begin{align}
    \bar \rho = \bar e +\frac{1}{2} \bar e_i b^i = \frac{1}{m_1^2} \left((m_1)^2\bar e - \bar e_1\bar e_0 m_1 +\frac{1}{2} \bar e_0^2 \hat m^1\right)\,. 
\end{align}
We can now give an estimate for the range where the moduli $t^0$, $t^1$ can be fixed, based on \eqref{Ft-eq:MinkcorrIIB1 0}: \begin{align}
     (m_1)^2\bar e - \bar e_1\bar e_0 m_1 +\frac{1}{2}\bar e_0^2 \hat m^1 = \frac{9}{8} K^{(3)} \frac{\hat m^1}{t^1} (m_1)^2.
\end{align}
As in the case of $X_{24}^*$ we distinguish three cases: 
\begin{itemize}
    \item[$i)$] {\em At the vacuum we have the hierarchy $t^1\gg t^0$}. In this case we have $m_1 \gg \hat m^1$ such that the flux contribution to the tadpole is determined by $m_1$. As a consequence
    \begin{align}
        t^1 \lesssim |K^{(3)}|\, m_1 \left(m_1 \hat m^1\right) < |K^{(3)}| N^{2}_\text{flux}\,,
    \end{align}
    and accordingly 
    \begin{align}
        \frac{\kappa}{t^0} \lesssim (K^{(3)})^2 N_\text{flux}^5\,, 
    \end{align}
    which agrees with \eqref{Ft-boundIIB1} for $p=2$. 
    \item[$ii)$] {\em At the vacuum $t^0\sim t^1$}.  For this we need $m_1\sim \hat m^1$. In this case we find 
    \begin{align}
        t^1 \lesssim |K^{(3)}|\, m_1 \left(m_1 \hat m^1\right) < |K^{(3)}| N^{3/2}_\text{flux}\,,
    \end{align}
    where we used $N_\text{flux}^{1/2} \gtrsim \hat m^1 \sim m_1$. Hence 
    \begin{align}
        \frac{\kappa}{t^0}\lesssim (K^{(3)})^2 N_\text{flux}^3\,, 
    \end{align}
    which corresponds to \eqref{Ft-boundIIB1} for $p=1$.
    \item[$iii)$] {\em At the vacuum we have the hierarchy $t^0\gg t^1$}. In this case we have $N_\text{flux}\gtrsim \hat m^1\gg m_1$ such that our bound becomes 
    \begin{align}
        t^1 \lesssim |K^{(3)}|\, m_1 \left(m_1 \hat m^1\right) < |K^{(3)}| N_\text{flux}\,,
    \end{align} 
    and 
    \begin{align}
        \frac{\kappa}{t^0}\lesssim (K^{(3)})^2 N_\text{flux}\,, 
    \end{align}
    reproducing \eqref{Ft-boundIIB1} for $p=1/2$. 
\end{itemize}
We see that compared to the $X_{24}^*$ discussion the bound on $t^1$ in the case $i)$ is stronger whereas it is the same in case $ii)$ and even less constraining in case $iii)$. In the present example we further have 
\begin{align}
    |K^{(3)}| = \frac{\zeta(3)}{8\pi^3} |\chi_E| \simeq 1\,, 
\end{align}
such that $|K^{(3)}|\, m_1 \left(m_1 \hat m^1\right)$ can be made moderately larger than 1 to ensure that we are always in the regime where the perturbations $\epsilon \ll 1$.  

\subsection{A realisation of the linear scenario}
\label{Ft-sec:counter}

In section \ref{Ft-s:linear} we discussed a linear scenario that resembles certain features of the \textbf{IIB2} scheme  and in particular allows for full moduli stabilization for a flux choice with only one contribution to the D3-brane tadpole $N_\text{flux}$. In the following we would like to give an explicit example of an F-theory construction that realizes this linear scenario. In our concrete model the number of complex structure moduli is four, but as discussed in section \ref{Ft-s:linear} the construction can be easily generalized to an arbitrary $h^{3,1}(Y_8)$. 

As pointed out in section \ref{Ft-s:linear} we can realize the linear scenario in case the mirror manifold $X_8$ admits a fibration of a Calabi--Yau three-fold $X_6$ over a $\mathbb{P}^1$. As the example in this section, we take the mirror manifold $X_8$ to be a triple fibration $\mathbb{T}^2 \rightarrow \mathbb{P}^1 \rightarrow \mathbb{P}^1 \rightarrow \mathbb{P}^1$, which can either be seen as an elliptic fibration over a base $B_6=\mathbb{P}^1\rightarrow \mathbb{F}_2$ or as a fibration of a Calabi--Yau $X_6=\mathbb{T}^2 \rightarrow \mathbb{F}_1$ over $\mathbb{P}^1$. Here, $\mathbb{F}_n$ the $n$-th Hirzebruch surfaces. Such a manifold can be constructed using toric methods -- the toric data for this manifold is given e.g. in \cite{Mayr:1996sh}. For this model we have four generators of the K\"ahler cone $D_0, \, D_1, \,  D_2$ and $D_L$ with intersection polynomial 
\begin{eqn}\label{Ft-intersectionpolyIII}
    I(Y_8) =& \left(8D_0^3+D_0 D_1 D_2 + D_0D_2^2 + 2 D_0^2 D_1 + 3 D_0^2 D_2\right)D_L + 6 D_0^2 D_2 D_1 + 2D_0 D_2 D_1^2 \\
    &+ 2 D_0D_2^2D_1 + 16 D_0^3 D_1 + 2 D_0 D_2^3 +4D_0^2 D_1^2 + 6 D_0^2 D_2^2 + 18 D_0^3 D_2 +52 D_0^4\,.
\end{eqn}
We can identify $D_0$ as the K\"ahler cone generator related to the zero section of the elliptic fibration as in \eqref{Ft-kahlerconeelliptic}. Furthermore $D_L$, satisfying $D_L. D_L=0$, denotes the class of the generic Calabi--Yau three-fold fibre $X_6$ and $D_1$ and $D_2$ are the divisors dual to the curves inside the base $\mathbb{F}_1$ of $X_6$. From \eqref{Ft-intersectionpolyIII} we can read off
\begin{eqn}
    \cK =& \left[8(t^0)^3+t^0t^1t^2 + t^0(t^2)^2 + 2 (t^0)^2 t^1 + 3 (t^0)^2 t^2\right]t_L + 6 (t^0)^2 t^2t^1 + 2t^0t^2(t^1)^2 \\
    &+ 2 t^0(t^2)^2t^1 + 16 (t^0)^3 t^1 + 2 t^0(t^2)^3 +4(t^0)^2 (t^1)^2 + 6 (t^0)^2 (t^2)^2 + 18 (t^0)^3 t^2 +52 (t^0)^4\,.
\end{eqn}
In the following we will use the indices $a,b,\dots$ to refer to $i=0,1,2$ and $\alpha,\beta, \dots$ to refer just to $i=1,2$. The first Chern class of the base $B_6$ is given by 
\begin{align}
    c_1(\mathbb{F}_1\rightarrow \mathbb{P}^1) = 2D_2 + D_1\,, 
\end{align}
and the corrections $K_i^{(3)}$ can be read off from \cite{Mayr:1996sh}
\begin{align}\label{Ft-c_3ExampleIII}
    c_3(Y_8) D_i = -3136 D_0-960 D_1 -1080D_2 - 480 D_L\,. 
\end{align}
Since $X_8$ can be seen as an elliptic fibration, a basis of four-cycles is given as in \eqref{Ft-4formbasisfib}. However, here we choose a different basis of four-cycles that is better suited for the study of the linear scenario that is given by \eqref{Ft-linearscenarioH1} and \eqref{Ft-linearscenarioH2}. The first set of four-cycles is given by divisors of the generic three-fold fibre $X_6$: 
\begin{align}
    H_0= D_0.D_L \,,\qquad  H_{\alpha} = D_\alpha . D_L\, , \qquad \alpha = 1,2\,. 
\end{align}
The second set of four-cycles is obtained by fibering the Mori-cone generators $\mathcal{C}^\alpha$ of $X_6$ over the base $\mathbb{P}^1$. The so-obtained four-cycles $H$ satisfy 
\begin{align}
    D_\alpha.D_\beta = \lambda_{\alpha \beta} H_{\hat 0}\,,\;\, D_0.D_\alpha = \lambda_{\alpha\beta}\left(\delta^{\beta \hat{\beta}}H_{\hat \beta} +c_1^\beta H_{\hat 0}\right)\,,\;\, D_0.D_0 =\lambda_{\alpha\beta }c_1^\alpha \left(\delta^{\beta \hat \beta} H_{\hat \beta} +c_1^\beta H_{\hat 0}\right)\,,
\end{align}
where $\lambda_{\alpha\beta}=\cK_{L 0 \alpha \beta}$ is the intersection on the two-fold base of $X_6$. From here we can read off the non-vanishing components of the $\zeta$ tensor 
\begin{eqn}
    \zeta^0_{L 0}&=1\,,\qquad \zeta^\alpha_{L\beta} =\delta^\alpha_\beta \,,\qquad \zeta^{\hat 0}_{\alpha \beta}=\lambda_{\alpha \beta}\,,\qquad \zeta^{\hat \alpha}_{\beta 0}=\delta^{\hat \alpha \alpha} \lambda_{\alpha \beta}\,,\\
    \zeta^{\hat \alpha}_{00}&=\delta^{\hat \alpha \alpha} \lambda_{\alpha \beta}\,,\qquad \zeta^{\hat 0}_{\alpha 0}=\lambda_{\alpha\beta}c_1^\beta\,,\qquad \zeta^{\hat 0}_{00}=\lambda_{\alpha \beta}c_1^\alpha c_1^\beta. 
\end{eqn}
The non-vanishing components of the intersection matrix $\eta_{\mu \nu}$ in the four-cycle sector are
\begin{eqn}\label{Ft-intersectionexampleIII}
     \eta_{a\hat b} &= \delta_{a\hat b} \,,\qquad \eta_{\hat 0\hat \alpha}=\delta_{\hat \alpha \alpha} \lambda^{\alpha \gamma} \lambda^{\delta \rho} D_0 D_\gamma D_\delta D_\rho\\
     \qquad \eta_{\hat \alpha \hat \beta} &= \delta_{\hat \alpha \alpha} \delta_{\hat \beta \beta}\left[\lambda^{\alpha \gamma}\lambda^{\beta \delta} D_0^2 D_\gamma D_\delta-(c_1^\alpha \lambda^{\beta \delta}+c_1^\beta \lambda^{\alpha \delta})\lambda^{\gamma\rho} D_0 D_\delta D_\gamma D_\rho\right]\,. 
\end{eqn}
In the following, we use the notation $ m_a = \delta_{a\hat a} \hat m^{\hat{a}}$ for the fluxes associated to $H_{\hat a}$. With this information, we can now look for solutions to the vacuum equations. We are interested in vacua that realize the linear scenario of section \ref{Ft-s:linear} and hence look at the limit \eqref{Ft-limitL}, which in the present case 
 can be viewed as some sort of Sen's limit. As before, to find vacua in the region probed by this limit we must set $\tilde \rho=m=0$, and since $\cK g_{ab}$ will generically diverge we also set $\tilde\rho^a=m^a=0$ in order not to violate the tadpole constraint. However, we can have $m^L\ne 0$ since \eqref{Ft-limgLL} is finite. If we further set $\bar \rho_a'=m_a=0$ by \eqref{Ft-intersectionexampleIII} we have a single pair of fluxes contributing to the D3-brane tadpole as $ N_\text{flux} = - m^L \bar{e}_L$. This results  in  the following restricted flux vector \eqref{Ft-eq:fluxchoiceL}:
\begin{align}\label{Ft-eq:fluxchoiceexampleIII}
    \vec q^{\, t} = (0, m^L, 0,0, \hat m^a,0,   \bar e_\alpha, \bar e_0, \bar e_L,
    \bar e)\,,
\end{align}
and the following flux-axion polynomials: 
\begin{align}
    \nonumber \bar \rho &= \bar e +\bar e_i b^i + \frac{ \cK_{L 0 \alpha \beta} }{2}\left(\hat m^0 (b^\alpha+c_1^\alpha b^0)(b^\beta +c_1^\beta b^0)+\hat m^\alpha b^0 (2b^\beta +c_1^\beta b^0)\right)+\frac{1}{6} \cK_{L abc} m^L b^a b^b b^c \,, \\
    \nonumber \bar \rho_0 &= \bar e_0 +  \cK_{L 0 \alpha \beta}\left(\hat m^\alpha (b^\beta+c_1^\beta b^0)+ c_1^\alpha \hat m^0(b^\beta +c_1^\beta b^0)\right) + \frac{1}{2} \cK_{L 0ab} m^L  b^0 b^a b^b\,,\\
    \nonumber \bar \rho_\alpha &= \bar e_\alpha + \hat m^0 \cK_{L 0 \alpha \beta}\left(b^\beta +c_1^\beta b^0 \right) + \hat m^\beta \cK_{L 0 \alpha \beta}b^0  + \frac{1}{2} \cK_{\alpha L ab} m^L b^\beta b^a b^b\,, \\
    \bar \rho_L  &= \bar e_L  \,,\\
    \nonumber\bar \rho_a^\prime &= 0 \,,\\
    \nonumber\hat \rho^a &= \hat m^a + m^L  b^a \,,\\
    \nonumber\tilde \rho^a &=0 \,, \\ \nonumber \tilde \rho^L  &= m^L \,, \\ \nonumber \tilde\rho&=0\, ,
\end{align}
where we used that the intersection numbers $\cK_{L abc}$ are related to $\lambda_{\alpha \beta}$ and $c_1^\alpha$ via
\begin{align}\label{Ft-kappaIII}
    \cK_{L0\alpha \beta} = \lambda_{\alpha \beta}\,,\qquad \cK_{L00\alpha}= \lambda_{\alpha \beta}c_1^\beta \,,\qquad \cK_{L000}= \lambda_{\alpha \beta}c_1^\alpha c_1^\beta\, .
\end{align}
One can check that the polynomials in \eqref{Ft-rhoAexample} correspond to those obtained in the general linear scenario in section \ref{Ft-s:linear}. The axions are stabilized as in \eqref{Ft-axionsL}, and also the stabilization of the saxions works as in the general case. For concreteness, let us focus on the overall rescaling
\begin{align}\label{Ft-limitlambda}
   t^a = v^a \lambda \,,\qquad v^a\sim \mathcal{O}(1)\,,\qquad \lambda\rightarrow \infty\,. 
\end{align}
together with $t_L  \sim \lambda^3 \to \infty$. We can thus write $\cK= 4 t_L   \kappa(v)\lambda^3 + f(v) \lambda^4$ and $\cK_a =3 t_L \kappa_a(v)\lambda^2 + f_a(v)\lambda^3$. The values for the parameters $v^a$ can then be inferred  from the equation of motion for $\bar \rho_a$ as in \eqref{Ft-vaclin2} where $\varepsilon_a$ in the present example is given by 
\begin{align}
    \varepsilon_a = \frac{g_{L a}}{g_{LL}}   - 6\eps_L\cK_L\frac{\cK_a}{g_{L L}\cK^2}\quad \stackrel{\eqref{Ft-limitL}}{\to} \quad \frac{\kappa(v) f_a(v) -\frac{3}{4} \kappa_a(v) f(v)}{\kappa(v)^2}- \frac{27}{4} K^{(3)}_L \frac{\kappa_a(v)}{\kappa(v)^2} \frac{t_L}{\lambda^4}\,. 
\end{align}
Then the equation of motion \eqref{Ft-Nfluxineqlin}  fixes the $v^a$.  Since by assumption the $v^a$ are of order one, we also find $\varepsilon_a\sim \mathcal{O}(1)$, $\forall a$, such that the bound $N_\text{flux} |\varepsilon_a| \ge 1$
is trivially satisfied. As a result there is no upper bound for the value of $\lambda$, in accordance with the general discussion in section \ref{Ft-s:linear}. Still, since $N_a$ is a monodromy-invariant flux combination there should only be  a finite number of inequivalent vacua along the limit. Finally, the ratio $t_L/\lambda^3$ is fixed by \eqref{Ft-vaclin1}: 
\begin{align}
     \bar e_L = -\frac{\cK}{6} g_{LL}  m^L \rightarrow -\frac{\lambda^3}{t_L} \frac{m^L}{6} \ \implies \ \frac{t_L}{\lambda^3} \lesssim N_\text{flux}\,. 
\end{align}
We thus conclude that the present example indeed captures all the key features discussed for the general linear scenario in section \ref{Ft-s:linear}.

\section{Summary}
\label{Ft-s:conclu}

In this chapter we analyzed flux potentials and their vacua for F-theory compactifications on smooth elliptically fibered Calabi--Yau four-folds. We restricted our analysis to the regime of moderate to large complex structure,  where the complex structure moduli split into an axionic and a saxionic component and the periods of the holomorphic four-form $\Omega$ can be well approximated by polynomial expressions, neglecting exponentially suppressed terms. 
In this regime we provided an explicit expression for the scalar potential that allows for a systematic study of its vacua. To arrive at this result, we used that  the periods of the four-fold in the large complex structure regime are captured, through homological mirror symmetry, by the central charges of B-branes wrapping the holomorphic $2p$-cycles in the mirror four-fold. This strategy was promoted in \cite{CaboBizet:2014ovf,Cota:2017aal} to calculate the Gukov-Vafa-Witten superpotential. 

Since in our limit of consideration exponential corrections to the periods can be ignored, the resulting axionic shift symmetry allows us to separate the scalar potential into a saxion-dependent matrix $Z^{AB}$ and a set of flux-axion polynomials $\rho_A$ that depend on the axions and the $G_4$-flux quanta, in a similar manner to the type IIA compactifications studied in chapter \ref{ch: systematics}. This structure is in fact a general feature of the scalar potential close to generic large distance singularities, as argued in \cite{Grimm:2019ixq}. In terms of the $\rho_A$ the vacua conditions, i.e. the self-duality constraint for the $G_4$-flux, take the particularly simple form \eqref{Ft-eq:Mink} and can be analyzed systematically. Using this form of the self-duality condition allowed us to directly compute the flux contribution to the D3-brane tadpole $N_{\rm flux}$ in terms of the $\rho_A$ on-shell values.  

Our analysis shows that for generic Calabi--Yau four-folds we have to restrict the choice of fluxes in order not to violate tadpole cancellation parametrically. This led us to consider the generic flux choice \eqref{Ft-truncflux}. In fact this constraint on the possible fluxes can be viewed as a generalization of the result of \cite{Brodie:2015kza,Marsh:2015zoa}, where it was shown that in 4d type IIB/F-theory compactifications switching on the flux associated to the top period is inconsistent with tadpole cancellation and moduli stabilization at large complex structure. 

As it turns out, our generic choice of fluxes compatible with the tadpole cancellation is too constrained in order for the leading vacua equations to stabilize all complex structure fields. In particular, the analysis of the set of leading order vacua equations revealed that at least one saxionic direction necessarily remains flat. This problem is circumvented when polynomial corrections to the periods are included. While most of these polynomial corrections can be treated as a re-definition of the flux quanta, the correction $K^{(3)}_i$, that is related to the third Chern class of the mirror four-fold, has important consequences for the vacua equations as it gives a correction to the action of the Hodge $\star$ operator on $Y_8$. Including this correction allows us to generically stabilize all the complex structure fields. Still, to achieve full moduli stabilization the fluxes need to be chosen in such a way that the matrix $M$ appearing in \eqref{Ft-Meq} is invertible. Invertibility of this matrix should be read as a constraint on the fluxes $\hat{m}^\mu$ contributing to the tadpole $N_{\rm flux}$. In the light of the recent conjecture put forward in \cite{Bena:2020xrh, Bena:2021wyr} it would be very interesting to translate this constraint into a precise relation between $N_{\rm flux}$ and the number of fields that need to be stabilized, which a priori could exist for this particular family of vacua.  

In any event, we observed that in this generic flux scenario the regime for the saxion vevs in which we can find vacua without violating tadpole cancellation is bounded from above by $|K^{(3)}| N_\text{flux}^{p+\oh}$.\footnote{We stress that even taking into account this upper bound, we can find vacua consistent with our approximation of neglecting exponentially suppressed terms, since the saxion vevs are still allowed to be moderately large depending on the precise value of $K^{(3)}$. } As discussed in section \ref{Ft-s:vacua}, the exponent $p$ is bounded by the number of complex structure fields in the system, and the upper bound on the saxionic vevs can be understood as arising due to the full stabilization of the complex structure moduli by means of perturbatively suppressed terms. This bound on the saxion vevs nicely parallels the prediction for the total number of flux vacua based on statistical methods \cite{Ashok:2003gk,Denef:2004ze, Denef:2004cf}. Indeed, it was found that the number of vacua in type IIB flux compactifications grows like $N_{\rm flux}^{Q/2}$, with $Q$ the number of flux quanta. Since in type IIB the number of flux quanta is twice the number of complex structure plus dilaton fields, our bound on the saxion vevs is indeed in line with the expected number of flux vacua in type IIB. It would be interesting to make this link more precise, also by adding the D7-brane flux contribution as in \cite{Gomis:2005wc}.

Reducing our general F-theory setup to type IIB, we connected with several existing results in the literature. We realized that the flux choice made in \cite{Blanco-Pillado:2020hbw} is one of the simplest that guarantee that the matrix $M$ is invertible, implying that all complex structure moduli and the dilaton are fixed. In our scheme, the mass spectrum clearly depends on the correction $K^{(3)}$, as one of the fields is only stabilized when they are taken into account. This is also consistent with the results of \cite{Blanco-Pillado:2020wjn,Blanco-Pillado:2020hbw}, since the parameter $\xi$ that controls their mass spectrum is a simple function of $K^{(3)}$ (we will explore the type IIB limit in more detail in the following chapter). Furthermore,  we also showed that in one particular case in which the matrix $M$ is not invertible, we recover the residual flat direction found in \cite{Demirtas:2019sip} for the same flux choices. In that reference it was shown that this flat direction can be stabilized by including non-perturbative corrections, possibly yielding to an exponentially small superpotential. Our analysis of section \ref{Ft-sec:moduli} provides an F-theory generalization of both of these type IIB constrained flux scenarios, and we expect them to display similar features, see e.g. \cite{Honma:2021klo}. In particular, notice that the vacuum obtained in \cite{Demirtas:2019sip} after including exponential corrections is located at $\mathcal{O}(1)$ values for the saxionic fields. This is analogous to our observation that the small corrections which yield full stabilization of all complex structure fields set an upper bound for the regime in which we expect to find vacua. Based on our analysis presented in this chapter, it would be interesting to investigate whether also in general F-theory models non-perturbative corrections can lift the perturbatively flat direction of the potential when $M$ is not invertible. 

Besides the class of vacua associated to the flux choice \eqref{Ft-truncflux}, which is present in generic F-theory models, we found a second class of vacua arising for a different pattern of flux quanta when at least one of the complex structure fields only enters linearly in $e^{-K}$ and the superpotential. In this case there exists a region in field space where we can fix all complex structure moduli with the flux choice \eqref{Ft-eq:fluxchoiceL}, without violating the tadpole constraint. Most importantly, for this flux choice there is only a pair of  flux quanta that contribute to the tadpole. As we argued in section \ref{Ft-s:linear}, in the linear scenario the full moduli stabilization can be achieved provided the matrix $Z^{AB}$ entering the scalar potential has enough off-diagonal components. In the type IIB limit these off-diagonal components are again related to the $K^{(3)}$ correction and reproduce the mirror dual of the Minkowski vacua studied in \cite{Escobar:2018rna}. However, as discussed in section \ref{Ft-s:linear} in the generic F-theory setup we do not necessarily need to rely on the $K^{(3)}$ corrections, and full moduli stabilization can be already achieved just on the level of the classical contributions to the periods of the four-fold. Notice that in this case the off-diagonal terms of $Z^{AB}$ are not necessarily suppressed in the large complex structure limit. As a consequence there is in general no bound on the value of the saxion vevs for which we can find these kind of vacua. Still, as argued in section \ref{Ft-s:linear}, we expect the number of vacua in this class to be finite. This follows from inequivalent vacua being characterized by a monodromy-invariant integer which can only take values in a finite range.


\ifSubfilesClassLoaded{%
\bibliography{biblio}%
}{}

\end{document}

\graphicspath{{Images/Type_IIB_at_LCS}}

\ifSubfilesClassLoaded{%
\tableofcontents
}{}

\setcounter{chapter}{8}
\chapter{Analytics of type IIB flux vacua and their mass spectra}
\label{ch: typeIIB}

In the previous chapter we considered the technicalities of moduli stabilization of F-theory at the large complex structure regime including polynomial corrections. The easiest case one could study in this framework is type IIB Calabi--Yau  (CY) orientifolds with three-form fluxes. But, as we already observed in section \ref{Ft-s:IIB}, despite its relative simplicity, in practice it is hard to achieve analytical control when describing this setup. In particular, as soon as there are several complex structure moduli stabilized by fluxes, the analytic description of the set of vacua is typically lost, except in some special cases where the use of discrete isometry groups allows for a consistent reduction of the complex structure sector \cite{Kachru:2003aw,Balasubramanian:2005zx,Conlon:2005ki,
Westphal:2006tn,Giryavets:2003vd,Giryavets:2004zr,DeWolfe:2004ns,Denef:2004dm,Louis:2012nb,Cicoli:2013cha} possibly down to a single field \cite{Klemm:1992tx,Doran:2007jw,Candelas:2017ive,Braun:2011hd,Batyrev:2008rp,Doran:2005gu,Candelas:2019llw,Joshi:2019nzi,Grimm:2019ixq,Blanco-Pillado:2020wjn}. The same statement applies to the mass spectrum of the fields that are stabilized by fluxes, which depends on the scalar potential and the vacuum expectation values (vevs) of the fields. These two ingredients, vevs and mass spectra, are crucial in order to implement full moduli stabilization, and therefore to develop an overall picture of the ensemble of vacua and to extract its phenomenological features. 

The aim of this chapter is to expand upon the content of section \ref{Ft-s:IIB} and improve the current state of affairs, by providing a class of type IIB flux configurations where the vevs and mass spectrum in the axio-dilaton and complex structure sector can be described analytically.\footnote{In most type IIB CY schemes that implement full moduli stabilization, the flux-induced vevs and masses are independent of the K\"ahler moduli stabilization details, and can therefore be seen as properties of the final vacuum. In this chapter we will not discuss K\"ahler moduli stabilization, and we will dub as {\em flux vacua} those vevs in the axio-dilaton and complex structure sector that solve their equations of motion at tree-level in 4d Minkowski.} This analytic description is independent of the number of complex structure fields, and the key ingredient to implement it is a simplified description of the Calabi--Yau holomorphic three-form periods in some asymptotic region. We focus on the region of Large Complex Structure (LCS), where such periods can be expressed as polynomials of the complex structure fields, up to exponential terms that can be neglected. It is precisely in this region where recent progress in describing the flux-induced mass spectrum \cite{Blanco-Pillado:2020wjn,Blanco-Pillado:2020hbw} and the flux potential \cite{Marchesano:2021gyv} analytically and for an arbitrary number of fields has been made, so it is a particularly promising regime to look at. In this work we show how these two different set of results are connected to each other, and how they can be merged into a single framework that leads to a more detailed analysis of such flux vacua. 

As we previously discussed, in order to find vacua in the LCS limit some flux quanta must be set to zero in order to satisfy the tadpole constraints, which in the context of Type IIB led to the introduction of two different families of flux configurations, dubbe IIB1 and IIB2. In this chapter we will focus on the former and show that it  corresponds to a set of compactifications in which the flux-induced superpotential is quadratic in the axio-dilaton and complex structure fields. It follows from here that such set of flux vacua splits into three distinct classes, that can be classified according to the nature of the field directions that are unfixed by fluxes.\footnote{More precisely, these are flat directions at the approximation level in which all polynomial corrections to the leading behaviour of the periods are included, while exponential corrections are neglected.} In the first class, in which supersymmetry is broken in the K\"ahler sector, all fields in the complex structure/axio-dilaton sector are stabilized. Moreover, the simplest choice of fluxes leads to the {\em no-scale aligned} vacua of \cite{Blanco-Pillado:2020wjn}. In this case, one can describe the field vevs in terms of quadratic and cubic equations, and apply the techniques of \cite{Blanco-Pillado:2020wjn} to obtain the flux-induced mass spectrum analytically, for an arbitrary number of complex structure moduli. The second class also breaks supersymmetry in the K\"ahler sector, but now contains one or more axion-like fields that are flat directions of the flux potential. Finally, in the third class, vacua are fully supersymmetric and, remarkably, they always contain some complexified flat directions.

These results can be compared to other strategies in the literature employed to analyze the same setup. For instance, one may compute the flux-induced mass spectrum by first extracting the Hessian from the analytic expression for the scalar potential provided in chatper \ref{ch: Ftheory}. While this analysis is in general quite involved, one can see that for the axionic sector of the IIB1 scenario one obtains a perfect match with our analytic expressions. A different, more direct method is to perform a numerical analysis of the flux vacua solutions and their mass spectra. When applying this approach to the IIB1 scenario the result is two-fold: On the one hand, it shows that the analytical control inside the IIB1 setup allows to very efficiently find flux configurations yielding consistent vacua. On the other hand, various features of the numerical vacua are shown to precisely match the analytical results developed in this thesis, supporting the robustness of the analysis.

The chapter is organized as follows: In section \ref{IIB-sec:Bilinear}, we provide a coarse-grained classification of vacua that can arise from a quadratic superpotential and uncover the supersymmetric and the two non-supersymmetric families mentioned above. We detail here what is the \emph{IIB1 scenario} for which, precisely, the superpotential takes a bilinear form. In section \ref{IIB-sec:nonsusy}, we explore the non-supersymmetric vacua highlighted in the generic classification in more detail. We focus on a specific branch of vacua by assuming an Ansatz for the saxions, where, upon further refinement to two cases, we can express analytically the vacuum expectation values of the axio-dilaton  and all complex structure moduli. We prove here that one of these two cases falls into the \emph{no-scale aligned} class described in \cite{Blanco-Pillado:2020wjn}, so that we are able to determine their complete tree-level mass spectra analytically. Details about the computation of these masses are presented in appendices \ref{IIB-sec:NSA} and \ref{IIB-ap:spot}. In section \ref{IIB-sec:susy}, we briefly investigate the supersymmetric family exhibited from the generic classification. In section \ref{IIB-sec:numerics}, we numerically generate and analyze an ensemble of IIB1 vacua that fits into the \emph{no-scale aligned} branch in a toy two-parameter model. We end up  with some conclusions and prospects in section \ref{IIB-sec:Conclusion}.



\section{Vacua from a quadratic superpotential}
\label{IIB-sec:Bilinear}

In this section, we consider Type IIB compactification on a Calabi-Yau orientifold $Y_6$ at large complex structure, so we can use all the results developed in section \ref{fb-subsec: Type IIB ingredients}. In this context  we present a generic classification of the flux vacua arising from  from superpotentials that take a generic bilinear form, i.e., that are of the following kind:
\begin{equation}
W=\oh\vec{Z}^t M \vec{Z} + \vec{L} \cdot \vec{Z} +Q\ ,
\label{IIB-eq:W_bilinear}
\end{equation}
where $\vec{Z}\equiv (\tau, \vec{z})$ and where the $(h^{2,1}+1)$-dimensional matrix $M$, the vector $\vec L$ and the scalar $Q$ are real flux-dependent quantities. Note that the matrix $M$ is symmetric by construction. 
As we will see in section \ref{IIB-sec:IIB1_def}, the \emph{IIB1 scenario} that is of interest in this paper is precisely designed to get a quadratic structure from the superpotential \eqref{IIB-eq:Wfull}.
In the rest of the chapter, we will apply the general formulas derived here in more detail and push the analytical developments. Note that generically the superpotential is cubic in the complex structure/axio-dilaton sector, as shown in section \ref{fb-sec: typeIIB compactafications}.

Let us denote the covariant derivatives with respect to $\tau$ and $z^i$ with $i\in h^{2,1}(Y_6)$ in a vector notation $\vec D\equiv (D_\tau,D_i)$. Likewise, we package the first derivatives of the Kähler potential within the vector $\vec\partial K\equiv(K_\tau,K_i)$, which is purely imaginary and axion-independent (see eqs.~\eqref{IIB-eq:Kt} and \eqref{IIB-eq:Ki}). The vacuum equations then take the form
\begin{equation}
\vec D W=0\ \Longleftrightarrow\  M\vec Z +\vec L+ (\vec\partial K) W=0\ .
\label{IIB-eq:vacuum_eqs_bilinear}
\end{equation}
The superpotential at vacua enjoys a reality property. Indeed, decomposing $\vec Z\equiv B+i\vec T$ into eq.~\eqref{IIB-eq:W_bilinear} yields
\begin{equation}
\Im(W)=\vec B^t M\vec T+\vec L\cdot\vec T\ .
\end{equation}
On the other hand, and thanks to this expression for $\Im(W)$, the real part of \eqref{IIB-eq:vacuum_eqs_bilinear} contracted with $\vec T$ gives
\begin{equation}
\Im(W)\left(1+i\vec T\cdot\vec\partial K\right)=-\frac{4+\xi}{2(1+\xi)}\Im(W)=0\ .
\end{equation}
Here, we made use of eq.~\eqref{IIB-eq:Ki} and the definition of the LCS parameter $\xi$ introduced in eq.~\eqref{IIB-eq:def_xi} to express $\vec T\cdot\vec\partial K$. Since $\xi$ cannot be equal to $-4$, as explained below \eqref{IIB-eq:def_xi}, we deduce that $\Im(W)$ vanishes at vacua so that the superpotential is real on-shell.
With this result at hand, the vacuum equations \eqref{IIB-eq:vacuum_eqs_bilinear} split into
\begin{align}
M\vec B&=-\vec L\ ,\label{IIB-eq:B_gen}\\
M\vec T&=i(\vec\partial K) W\ ,\label{IIB-eq:T_gen}
\end{align}
which in particular imply that $\vec{L}$ should be in the image of the matrix $M$ in order to find a vacuum solution, which is a non-trivial requirement on the flux quanta when $M$ is not invertible. For this reason it is natural to discuss separately those cases in which the matrix $M$ is regular and when it is not. In both cases, 
using \eqref{IIB-eq:B_gen}, we can write the superpotential at vacua like
\begin{equation}
\label{IIB-eq:Wvac_int}
W=-\oh\vec T^tM\vec T+Q'\ ,    
\end{equation}
where $Q'$ is a flux-dependent quantity defined by
\begin{equation}
\label{IIB-eq:Q'}
Q'\equiv Q - \oh\vec L^t M^+ \vec{L}\ ,
\end{equation}
and $M^+$ is the generalized inverse of $M$, whose explicit expression we give below. Then, from \eqref{IIB-eq:T_gen} and \eqref{IIB-eq:Wvac_int} we deduce that
\begin{equation}
\label{IIB-eq:Wvac}
W=\frac{Q'}{1-\frac{i}{2}\vec T\cdot\vec\partial K}=\frac{4}{3}\frac{1+\xi}{\xi}Q'=-\frac{2}{3}e^{-K_{\rm cs}}\frac{Q'}{\Im\kappa_0}\ .    
\end{equation}
Therefore, when approaching the LCS point at $\xi=0$, the superpotential diverges. Also, notice that supersymmetric vacua are only possible if $Q'=0$.

\subsection[When \texorpdfstring{$M$}{M} is invertible]{\bm When \texorpdfstring{$M$}{M} is invertible}
\label{IIB-sec:Minvertible}
 When  $M$ has an inverse then $M^+ = M^{-1}$, and so eq.~\eqref{IIB-eq:B_gen} stabilizes all the axions at 
 \begin{equation}
 \vec B=-M^{-1}\vec L \ .\label{IIB-eq:B_nonsusy}
 \end{equation}
 On the other hand, eq.~\eqref{IIB-eq:T_gen} is implicit on the saxions since $\vec\partial K$ and $W$ depend on $\vec T$. This is summed up in the following expression for $\vec Z$:
\begin{equation}
\vec Z=-M^{-1}\left(\vec L+(\vec\partial K) W\right)\ .
\label{IIB-eq:Z}
\end{equation}
 The superpotential at vacua reads as \eqref{IIB-eq:Wvac} with  $Q'$ given by
\begin{equation}
Q'=Q-\oh\vec L^tM^{-1}\vec L\ .
\end{equation}
As noted above, supersymmetric vacua only arise if $Q'=0$. But with $M$ invertible this would imply that $\vec T=\vec 0$ due to \eqref{IIB-eq:T_gen}. Supersymmetric vacua are thus forbidden when $M$ is regular.

\subsection[When \texorpdfstring{$M$}{M} is singular]{\bm When \texorpdfstring{$M$}{M} is singular}
\label{IIB-sec:Mnoninvertible}

As mentioned earlier, eq.~\eqref{IIB-eq:B_gen} tells us that $\vec L$ lies in the image of $M$ since $\vec{L}=M(-\vec B)$. As a consequence, the field directions inside the kernel of $M$ do not enter the superpotential. Thus, in the LCS approximation, the axionic directions that correspond to $\ker(M)$ do not enter the scalar potential at all, implying a number of flat directions. To describe the number of these flat directions one must distinguish between supersymmetric and non-supersymmmetric vacua:
    \begin{itemize}
        \item When $W\neq 0$, which corresponds to flux choices such that $Q' \neq 0$, we have that $\textrm{rank}(M)$ of the axions are stabilized, while $h^{2,1}+1-\textrm{rank}(M)$ constraints on the flux quanta must be satisfied in order for vacua to exist.
        To see this, we can diagonalize the matrix $M$ to a matrix $D\equiv\text{diag}(\lambda_0,\dots,\lambda_{r-1},0,\dots,0)$ with $\lambda_0,\dots,\lambda_{r-1}$ representing the $r\equiv\textrm{rank}(M)$ non-zero eigenvalues of the matrix, and where there are as many zeroes as the dimension of the kernel. We write the similarity transformation with a matrix $N$ like
        \begin{equation}
        M= N^tDN\quad\text{ and }\quad N^t=N^{-1}\ .
        \end{equation}
        Defining $\vec B'\equiv N\vec B$ and $\vec L'\equiv N\vec L$, the axionic system of equations \eqref{IIB-eq:B_gen} becomes
        \begin{equation}
        D\vec B'=-\vec L'\ .
        \end{equation}
        We now split the $h^{2,1}+1$ indices $\{0,i\}$ like $\alpha\in\{0,\dots,r-1\}$ and $\beta\in\{r,\dots,h^{2,1}\}$ to get the following vacuum expectation values and constraints:
        \begin{equation}
        b'^\alpha=-\frac{\vec L'^\alpha}{\lambda_\alpha}\quad\text{ and }\quad \vec L'^\beta=0\ .
        \end{equation}

        The superpotential at vacua \eqref{IIB-eq:Wvac} involves the quantity $Q'$ which again is flux-dependent-only and reads
        \begin{equation}
        \label{IIB-eq:Q'sing}
        Q'=\oh \vec L' \cdot\vec B '+Q=-\oh \sum_\alpha\frac{(\vec L'^\alpha)^2}{\lambda_\alpha}+Q = - \oh \vec{L}^t M^+ \vec{L}+ Q \ ,
        \end{equation}
        where $M^+= N^t D^+ N$ and $D^+ \equiv\text{diag}(\lambda_0^{-1},\dots,\lambda_{r-1}^{-1},0,\dots,0)$.
        As for the saxions, they satisfy the non-linear implicit relation \eqref{IIB-eq:T_gen}, where the superpotential $W$ takes the saxion-dependent form \eqref{IIB-eq:Wvac}. Since all axions enter in this condition, one generically expects that its solution stabilizes all of them. 
        \item When $W=0$, we read from \eqref{IIB-eq:vacuum_eqs_bilinear} that the vacuum solutions are
        \begin{equation}
        \label{IIB-eq:sol_susy}
        \vec Z=\vec B+\ker\, (M)\ ,
        \end{equation}
        and so only $\textrm{rank}(M)$ complex moduli are stabilized. As in the previous case, the same $h^{2,1}+1-\textrm{rank}(M)$ constraints on the flux quanta should hold. Moreover, $Q'=0$ provides one additional constraint on the fluxes. In total, we expect the fluxes to satisfy $h^{2,1}+2-\textrm{rank} (M)$ relations in order to fall into this supersymmetric class of vacua.
    \end{itemize}

\subsection{The IIB1 family}
\label{IIB-sec:IIB1_def}

In this subsection, we introduce the \emph{IIB1 scenario} described in \ref{Ft-s:IIB}, which in the language of Type IIB (see table \ref{Ft-table: type IIB F-theory dictionary}) is characterized by putting the following flux quanta to zero:
\begin{equation}
\label{IIB-eq:def_IIB1}
\text{IIB1 flux configuration:}\quad f_A^0=0\ ,\  h_A^0=0\ \text{ and }\ h_A^i=0\ ,\ i\in\{1,\dots,h^{2,1}\}\ . 
\end{equation}

We can motivate the interest on this Ansatz by looking at its effects on the type IIB superpotential \eqref{IIB-eq:Wfull}. The choice $f_A^0 = h_A^0 = 0$, i.e. $N_A^0 = 0$, has important consequences. We see that it removes the ``pure complex structure'' cubic, highest-order term $z^iz^jz^k$, from the superpotential. This ends up being quite a non-trivial effect, since it leads to solutions arbitrarily close to the LCS point, as opposed to the $N_A^0 \neq 0$ case \cite{Sousa:2014qza,Marsh:2015zoa,Brodie:2015kza}. In one-parameter models, this choice of fluxes has been proven to lead to completely different mass spectra than in the generic $N_A^0 \neq 0$ case, along with its own statistical ensembles of vacua \cite{Blanco-Pillado:2020wjn}. Following a similar reasoning as to the statements above, we remark that with the additional choice $h_A^i=0$ we get $N_A^i = f_A^i$, which removes the mixed (complex structure and axio-dilaton) cubic term $z^i z^j \tau$ from the superpotential, and only leaves a quadratic one on $z^i z^j $. 

Thus, the IIB1 flux choice ensures that the superpotential takes the bilinear form \eqref{IIB-eq:W_bilinear} with $\vec{Z}^t = (\tau, \vec{z}^t)$ and the following flux-dependent quantities:
\begin{equation}
\label{IIB-eq:MLQ}
M\equiv\begin{pmatrix}
0 & -\vec h^{B\, t}\\
-\vec h^{B} & S_{ij}
\end{pmatrix},
\quad \vec L\equiv(-h_0^B,f_i^B+a_{ij}f_A^j)\ ,\quad Q\equiv f_0^B-\hat{a}_if_A^i\ ,
\end{equation}
and where the matrix $S$ is defined as $S_{ij}\equiv\kappa_{ijk}f_A^k$. We further write $\vec L\equiv (L_0,L_i)$ so that $L_0\equiv -h_0^B$ and $L_i\equiv f_i^B+a_{ij}f_A^j$. Note that in the following sections, we will focus on flux configurations for which the matrix $S$ is invertible. When it is the case, the invertibility of $M$ is determined by the value of $\det(M)/\det(S)\equiv\cH= h_i^BS^{ij}h_j^B$.

In the previous chapter we expressed the vacuum equations descending from the F-theory ones and wrote them at first order in the LCS parameter $\xi$.
In the following, we will generalize this analysis and extend it to the full LCS region, i.e. for arbitrary $\xi$, by applying the generic results of the present section. We consider the non-supersymmetric (section \ref{IIB-sec:nonsusy}) and supersymmetric (section \ref{IIB-sec:susy}) vacua highlighted above and, in both cases, fully analytical relations for the axions and saxions vacuum locations are displayed. In the non-supersymmetric case, the analytical control over the saxions comes at the cost of restricting to a particular branch of solutions that we know is not unique thanks to numerics. Moreover, yet in a further subclass, we are able to express the vacuum expectation values with formulas that are exact in $\xi$ and we are able to uncover the scalar mass spectrum analytically.


\section{Non-supersymmetric vacua}
\label{IIB-sec:nonsusy}

We study here the non-supersymmetric flux vacua exhibited in the previous section, that can arise both with $M$ invertible or singular. We recall that the vacuum equations reduce to \eqref{IIB-eq:B_gen} and \eqref{IIB-eq:T_gen}, where the superpotential at vacua takes the form \eqref{IIB-eq:Wvac}. We thus have
\begin{align}
M\vec B&=-\vec L\ ,\label{IIB-eq:B_nonsusybis}\\
M\vec T&=-\frac{2}{3}ie^{-K_{\rm cs}}\frac{Q'}{\Im\kappa_0}(\vec\partial K)\ .\label{IIB-eq:T_nonsusy}
\end{align}
We first focus on the saxionic system which can be recast as
\bea
- 3 h^B_i t^i t^0 &=& e^{-K_{\rm cs}} \frac{Q'}{\Im \kappa_0} =  \frac{4}{3} \frac{Q'}{\Im \kappa_0} \kappa_{ijk}t^it^kt^k -2Q'\ ,\label{IIB-eq:Q/k}\\
- h^B_i t^0 + S_{ij} t^j  &=&  \frac{4}{3} \frac{Q'}{\Im \kappa_0} \kappa_{ijk}t^jt^k\ ,
\label{IIB-saxions}
\eea
from which it seems natural to define the following rescaled variables:
\be
x^0 \equiv  \frac{4}{3} \frac{Q'}{\Im \kappa_0} t^0\ , \qquad x^i \equiv  \frac{4}{3} \frac{Q'}{\Im \kappa_0} t^i \ . 
\ee

In terms of these rescaled variables, the above equations read
\bea
\label{IIB-dilatonx}
- 3 h^B_i x^i x^0 &=&   \kappa_{ijk}x^ix^kx^k - \S\alpha\ ,\\
- h^B_i x^0 + S_{ij} x^j  & = &\kappa_{ijk}x^jx^k\ ,
\label{IIB-saxionsx}
\eea
where
\be
\label{IIB-eq:def_alpha}
\alpha \equiv \frac{2^5 Q'^3 }{ 3^2 (\Im \kappa_0)^2\S}\ , \qquad \S \equiv \kappa_{ijk} f_A^if_A^jf_A^k\ .
\ee
Notice that eq.~\eqref{IIB-saxionsx} only depends on triple intersection numbers and fluxes bounded by the D3-brane tadpole. Therefore, one expects $x^A \sim {\cal O}(N_{\rm flux}^{1/2})$, with $A\in\{0,i\}$ and $\Nf=-f_A^ih_i^B$. To generate larger values for the saxions $t^A$, one may consider flux choices such that
\be
\frac{Q'}{\Im \kappa_0} \ll 1\ .
\ee
When it is the case,
\be
1 \gg |\alpha| \simeq |\xi|\ ,
\ee
so vacua satisfying this condition may be compatible with a large complex structure regime description.

The system of equations \eqref{IIB-dilatonx} and \eqref{IIB-saxionsx} is rather involved as it is, so we will propose an Ansatz to make analytical progress, that we will further refine into two cases in which we are able to obtain concrete results. Our working assumption will be that the matrix $S$ is invertible, and we will oftentimes also assume that $\S \neq 0$, in order to define $\alpha$ as above. To build the Ansatz we take inspiration from the analysis performed in section \ref{Ft-s:IIB} and perform a decomposition of the flux quanta $f^i_A$ and $h_i^B$ in terms of saxion vevs  as follows
\begin{equation}
    f^i_A = A t^i + C^i\ , \qquad h_i^B = B \kappa_{ijk}t^jt^k + C_i\ ,
    \label{IIB-decomp}
\end{equation}
with $C^i \kappa_{ijk}t^jt^k = C_it^i =0$. This fully general decomposition was helpful in the study of the equations of motion, which required the relations $A=t^0 B$ and $-C_i t^0=\kappa_{ij}C^j$. However, in order to provide concrete expressions for the vacuum expectation values of the moduli including first order polynomial corrections, in \eqref{ft-eq: behaviour with Ca=0} we restricted the flux space to the case $C_i=C^i=0$ and linearized the equations in $\xi$. We now aim to extend this Ansatz and to consider the effect of polynomial corrections at all orders. To do so we turn on the vector $C^i$ but demand a concrete relation with the flux quanta. We thus propose the Ansatz
\be
\label{IIB-eq:ansatz}
t^i \equiv \hat{t} f_A^i + \tilde{t} S^{ij} h_i^B  \quad\implies\quad x^i \equiv \hat{x} f_A^i + \tilde{x} S^{ij} h_i^B\ .
\ee
The vacua equations then read
\bea
\label{IIB-dilaton2}
3 x^0 \left( N_{\rm flux} \hat{x} - {\cH}\tilde{x}\right) & = & \hat{x}^3 \S - 3N_{\rm flux} \hat{x}^2 \tilde{t} + 3{\cH} \hat{x} \tilde{x}^2+ \tilde{x}^3  \kappa^{\cH} - \S \a\ ,\\
h^B_i (\tilde{x} -x^0) + S_i \hat{x}   & = & \hat{x}^2 S_i + 2\hat{x}\tilde{x} h_i^B  + \tilde{x}^2 \kappa^{\cH}_i\ ,
\label{IIB-saxions2}
\eea
where we have defined
\be
\kappa^{\cH}_i \equiv \kappa_{ijk}S^{jl}S^{km}   h_l^B h_m^B\ , \qquad \kappa^{\cH} \equiv \kappa_{ijk} S^{il} S^{jm}S^{kn}   h_l^B h_m^B  h_n^B\ ,
\ee
and recall that $\cH\equiv\det(M)/\det(S)=h_i^BS^{ij}h_j^B$. Upon contracting \eqref{IIB-saxions2} with $f_A^i$ and with $S^{ij}h_j^B$, and plugging back into \eqref{IIB-saxions2}, we obtain a consistency flux condition that reads
\be
\left(N_{\rm flux}^2 - \S {\cH} \right)  \kappa^{\cH}_i  + \left(\S \kappa^{\cH}  + {\cH} N_{\rm flux} \right) h^B_i + \left( \kappa^{\cH} N_{\rm flux}  +  {\cH}^2 \right)S_i =0\ ,
\label{IIB-flux1}
\ee
where $S_i\equiv \kappa_{ijk}f_A^jf_A^k$.

As evoked above, progressing without further refining the branch under consideration seems very involved. However, we notice that the constraint \eqref{IIB-flux1} is compatible with the relation $\cH=0$, which will define our first subclass of interest developed in section \ref{IIB-sec:S0}. This case falls into the kind of non-supersymmetric vacua described in section \ref{IIB-sec:Mnoninvertible} where the matrix $M$ is singular. The other subclass to be studied in the sequel assumes the Ansatz \eqref{IIB-eq:ansatz} with the simplification $\tilde t=0$, and will be discussed in section \ref{IIB-sec:simpler}

\subsection[A subcase with \texorpdfstring{$M$}{M} singular]{\bm A subcase with $M$ singular}
\label{IIB-sec:S0}

In this subsection, we push the analytics sketched above with the further flux condition
\begin{equation}
\cH=h_i^BS^{ij}h_j^B=0\ .    
\end{equation}
In this case, the matrix $M$ has a one-dimensional kernel generated by $\langle (1,S^{ij}h_j^B)\rangle$. From the generic discussion of section \ref{IIB-sec:Mnoninvertible}, we then expect one constraint to arise from the axionic system \eqref{IIB-eq:B_nonsusybis} as well as one flat direction. More precisely, we have
\begin{align}
h_i^BS^{ij}L_j=h_0^B\quad\text{ and }\quad b^i=-S^{ij}L_j+b^0S^{ij}h_j^B\ .
\end{align}

The saxionic system given by eqs.~\eqref{IIB-dilaton2} and \eqref{IIB-saxions2} reduces to the following one when $\cH=0$:
\bea
3 N_{\rm flux} \hat{x} x^0& = & \hat{x}^3 \S - 3N_{\rm flux} \hat{x}^2 \tilde{t}+ \tilde{x}^3  \kappa^{\cH} - \S \a\ ,\\
h^B_i (\tilde{x} -x^0) + S_i \hat{x}   & = & \hat{x}^2 S_i + 2\hat{x}\tilde{x} h_i^B  + \tilde{x}^2 \kappa^{\cH}_i\ ,
\eea
and the flux condition \eqref{IIB-flux1} becomes\footnote{Notice that this condition is automatically satisfied for models with two complex structure moduli where $\cH =0$, because then the vector in \eqref{IIB-fluxcond} is always orthogonal to $f_A^i$ and $S^{ij}h^B_j$. }
\be
N_{\rm flux}^2 \kappa^{\cH}_i  + \S \kappa^{\cH}h^B_i + \kappa^{\cH} N_{\rm flux}S_i =0\ .
\label{IIB-fluxcond}
\ee
One can manipulate the system of equations to arrive at an expression giving $\tilde x$ as a function of $\hat x$, a relation giving $x^0$ as a function of $\hat x$ and $\tilde x$ and an equation involving only $\hat x$. Indeed we have\footnote{These expressions assume $\kappa^\cH\neq0$ and $\S\neq 0$. If not, we find $\hat x=1$, $\tilde x=-x^0$ and one saxion is left unstabilized. When $\kappa^\cH=0$ and $\S\neq 0$, the flux relation $\alpha=1$ should also be satisfied.}
\begin{align}
&\tilde x^2=\frac{\Nf}{\kappa^\cH}\hat x(\hat x-1)\ ,\label{IIB-eq:xtilde}\\
&x^0=\frac{\S\hat x(\hat x-1)}{\Nf}+\tilde x-2\hat x\tilde x\ ,\label{IIB-eq:x0}\\
&\left(2\hat x^3-3\hat x^2+\alpha\right)^2=16\frac{\Nf^3}{\S^2\kappa^\cH}\hat x^3(\hat x-1)^3\ .\label{IIB-eq:sixth}
\end{align}

The last equation involving only $\hat x$ is polynomial of sixth order. To proceed, we can neglect $\alpha$ to find approximate solutions valid close to the LCS point. The polynomial then becomes only of third order and can be written like
\be
 \hat{x}^3 - 3  \hat{x}^2  + 3\frac{\beta - 3/4}{\beta-1}  \hat{x} - \frac{\beta}{\beta-1} \simeq 0\ , \quad\text{ with }\quad \beta \equiv 4 \frac{N_{\rm flux}^3}{\S^2\kappa^{\cH}}\ .
\ee 
This cubic equation admits three roots, either one real and two complex or three reals. If we label them $\hat x_0$, $\hat x_1$ and $\hat x_2$, they are given by
\begin{equation}
\hat x_k=1 + \frac{j^k\gamma}{2} -\frac{1}{2j^k\gamma(\beta-1)}\ ,\quad k\in\{0,1,2\}\quad \text{ and }\quad j\equiv \frac{-1+i\sqrt{3}}{2}\ ,\label{IIB-eq:xk}
\end{equation}
and where $\gamma$ is such that
\begin{equation}
\gamma^3\equiv \frac{1}{\beta-1}\left(1+\sqrt{\frac{\beta}{\beta-1}}\right)\ .
\end{equation}
Note that we cannot determine in full generality which of these solutions correspond to the real ones. With a solution for $\hat x$, eq.~\eqref{IIB-eq:xtilde} allows to compute $\tilde x$ so that we can deduce $x^i$ from the Ansatz. On the other hand, eq.~\eqref{IIB-eq:x0} allows to compute $x^0$. From the definitions of the rescaled variables, one can then deduce the vacuum expectation values of the saxions $t^0$ and $t^i$.

We can refine this approximate solution, valid near the LCS point, by using a pertubative approach. Indeed, if we denote the above approximate solution $\hat x^{(0)}$, we can write
\begin{equation}
\hat x=\hat x^{(0)}+\delta\hat x\ ,\label{IIB-eq:xkrefined}
\end{equation}
with $\delta\hat x\sim\mathcal{O}(\alpha)\ll 1$. Plugging this into the full equation \eqref{IIB-eq:sixth} and restricting to first order in $\alpha$ yields
\begin{equation}
\delta\hat x=\frac{(2\hat x^{(0)}-3)\S^2\kappa^\cH \alpha }{6(\hat x^{(0)}-1)\left[4\Nf^3(\hat x^{(0)}-1)(2\hat x^{(0)}-1)-\S^2\kappa^\cH\hat x^{(0)}(2\hat x^{(0)}-3)\right]} + {\cal O}(\alpha^2)\ . 
\end{equation}
One can plug this refined value of $\hat{x}$ into \eqref{IIB-eq:sixth}, and again linearize the equation to obtain its value to the next order in $\alpha$. The procedure can be repeated to provide an analytic expression up to any order in $\alpha$.

\subsection{A simpler Ansatz for full analyticity}
\label{IIB-sec:simpler}

Another very interesting subclass of vacua arises when one considers a particular restriction of the Ansatz proposed in \eqref{IIB-eq:ansatz}. This restriction consists in assuming $\tilde t=0$, so that we are left with
\begin{equation}
\label{IIB-eq:that}
t^i\equiv \hat t f_A^i\quad\implies\quad x^i \equiv \hat{x} f_A^i\ .
\end{equation}
For reasons that will be clearer later, we call this branch of vacua the \emph{no-scale aligned} branch. The vacuum equations for this branch reduce to
\bea
\label{IIB-dilaton1}
3 N_{\rm flux} \hat{x} x^0 &=&   \hat{x}^3 \S - \S \a\ ,\\
- h^B_i x^0 + S_{i} \hat{x}  & = & S_i \hat{x}^2\ .
\label{IIB-saxions1}
\eea
Contracting \eqref{IIB-saxions1} with $f_A^i$ we obtain 
\be
\S (\hat{x}^2 -\hat{x}) = N_{\rm flux} x^0\ ,
\label{IIB-saxions1c}
\ee
so we deduce that $\S \neq 0$. Plugging this equation back into \eqref{IIB-saxions1}, we obtain a condition for the flux vector $\vec{h}_B$:
\be
 h^B_i = - N_{\rm flux}\frac{S_i}{\S} \quad \Longrightarrow \quad h_i^B = -  \hat{h}^B \frac{S_i}{q}\ ,
\label{IIB-hBicond}
\ee
where $\hat{h}^B \in \mathbb{Z}$ and $q \equiv \gcd (S_i)$. This flux relation can be thought of as a simpler version of \eqref{IIB-flux1} for this particular Ansatz. It is worth noting that in the language of chapter \ref{ch: Ftheory},  \eqref{IIB-eq:that} and \eqref{IIB-hBicond} correspond to the choice $C_i$, $C^i=0$ using the decomposition \eqref{Ft-solIIB1mu}. If we  assume that the matrix $S$ is invertible, the above relation implies that $\cH \equiv h^B_iS^{ij} h^B_j \neq 0$, and so $M$ is regular. We are thus in the generic case described in section \ref{IIB-sec:Minvertible}. In the sequel, we will solve the axionic and saxionic systems of equations.

\subsubsection{Moduli stabilization}

\paragraph{Axions:}

The axions are stabilized at $\vec B=-M^{-1}\vec L$. The inverse of the matrix $M$ defined in eq.~\eqref{IIB-eq:MLQ} cannot be expressed in full generality but it can under the assumption that the matrix $S$ is invertible.\footnote{And in this case we saw above that $\cH\neq0$.} When it is the case, we have \eqref{ft-eq: M-1 type IIB1}
\begin{equation}
\label{IIB-eq:Minv}
M^{-1}=\frac{1}{\cH}\begin{pmatrix}
-1 & -S^{jk}h_k^B\\
-S^{ik}h_k^B & \cH S^{ij}-S^{ik}S^{jl}h_k^Bh_l^B
\end{pmatrix}.
\end{equation}
This yields
\begin{align}
\begin{split}
\label{IIB-eq:bs}
b^0 &= \frac{h_i^B S^{ij} L_j - h_0^B}{\cH}\ ,\\
b^i &= S^{ij} \left( b^0 h_j^B - L_j \right) \ .
\end{split}
\end{align}
Note that the quantity $Q'$ in this case is given by
\begin{align}
	Q' = f_0^B - f_A^i \hat{a}_i + \frac{(h_i^B S^{ij} L_j - h_0^B)^2}{2 h_i^B S^{ij} h_j^B} - \frac{1}{2} L_i S^{ij} L_j\ .
	\label{IIB-eq:rho}
\end{align}

\paragraph{Saxions:}

For the saxions, the relation \eqref{IIB-saxions1c} allows to solve for $\hat x$ as a function of $x^0$. We find
\be
\hat{x} = \frac{1}{2} \left( 1 \pm \sqrt{1 + 4 \frac{N_{\rm flux}}{\S} x^0}  \right)\ .
\label{IIB-premaster}
\ee
We now plug \eqref{IIB-dilaton1} into this expression, to obtain
\be
2\hat{x} = 1 \pm \sqrt{ 1 + \frac{4}{3}\left(\hat{x}^2-\frac{\a}{\hat{x}}\right)}\ ,
\ee
which yields the following cubic equation:
\be
2\hat{x}^3 -3\hat{x}^2 + \a = 0\ .
\label{IIB-cubic0}
\ee

The discriminant $\Delta$ of the cubic can be expressed simply as a function of $\alpha$ like
\begin{equation}
\Delta=4\alpha(\alpha-1)\ .
\end{equation}
When $\alpha<0$ or $\alpha>1$, the discrimant is positive and there is a single real root given by
\begin{equation}
\label{IIB-sol}
\hat x=\oh\left(1+\Gamma+\frac{1}{\Gamma}\right)\quad\text{ where }\quad \Gamma^3\equiv 1-2\left(\alpha+\sqrt{\alpha(\alpha-1)}\right)\ . 
\end{equation}
When $\alpha\in[0,1]$, the discriminant is negative and there are three real roots. The formula above is still valid to describe one of them if one defines the square and cubic roots as principal values. A (unique or not) solution for $\hat t$ is thus always given by
\begin{equation}
\label{IIB-eq:solthat}
\hat t=\frac{3}{8}\frac{\Im\kappa_0}{Q'}\left(1+\Gamma+\frac{1}{\Gamma}\right)\ .
\end{equation}
We will show below that this expression for $\hat t$ with roots defined as principal values always gives the unique physical solution. With this exact expression for $\hat t$ at hand, we can use \eqref{IIB-premaster} to isolate $t^0$. With the help of eq.~\eqref{IIB-eq:Q/k} that we repeat here
\begin{equation}
\label{IIB-eq:Q/kbis}
3\Nf\hat tt^0=e^{-K_{\rm cs}}\frac{Q'}{\Im\kappa_0}\ ,
\end{equation}
we arrive at
\begin{align}
\label{IIB-eq:t0t}
    t^0 = 
    \frac{q}{\hat h^B}
    \frac{2 \S \hat t^3-3\Im\kappa_0}{4\S\hat t^3+3\Im\kappa_0} \ \hat{t}\ .
\end{align}
Using \eqref{IIB-eq:solthat}, we can express a useful relation between the LCS parameter $\xi$ and the quantity $\alpha$:
\begin{align}
    \label{IIB-eq:flux_const_away}
    &\frac{\xi}{(\xi-2)^3}=\frac{\alpha}{27}\ .
\end{align}

\paragraph{Physical solutions:}

Let us now take a more detailed look at the physical solutions depending on the sign of $\alpha$. From eq.~\eqref{IIB-eq:Q/kbis} above, we see that the sign of $\hat t$ is the same as that of the ratio $Q'/\Im\kappa_0$. We thus have:
\begin{itemize}
    \item When $\alpha<0$, then if $\hat t>0$ we deduce $Q'<0$ from the definition of $\alpha$ and thus $\Im\kappa_0<0$ from \eqref{IIB-eq:Q/kbis}. If $\hat t<0$ we deduce $Q'>0$ from the definition of $\alpha$ and still $\Im\kappa_0<0$ from \eqref{IIB-eq:Q/kbis}. Thus, $\alpha$ negative corresponds exclusively to models with a negative $\Im\kappa_0$. For those models, we mentioned in section \ref{fb-subsec: Type IIB ingredients} that $\xi$ should be in the range $[0,1/2]$ for the Kähler metric to be well-defined with positive eigenvalues. By solving $\xi<1/2$, we can deduce a lower bound that the solution $\hat x$ of the cubic should satisfy. We find
    \begin{equation}
    \hat x>2^{1/3}|\alpha|^{1/3}\ .
    \end{equation}
    Equivalently, \eqref{IIB-eq:flux_const_away} yields $\alpha>-4$.
    \item When $\alpha>0$, same arguments lead to conclude that no matter what the sign of $\hat t$ is, $Q'$ has the same and $\Im\kappa_0$ is positive. For those models, we should have $\xi\in[-1,0]$. Solving $\xi>-1$, we find
    \begin{equation}
    \hat x>|\alpha|^{1/3}\ ,
    \end{equation}
    and equivalently, \eqref{IIB-eq:flux_const_away} yields $\alpha<-1$.
\end{itemize}

Figure \ref{IIB-roots} shows the values of the roots of the cubic equation \eqref{IIB-cubic0} as a function of $\alpha$ as well as the bounds derived above. We observe that for $\alpha<0$, the Kähler cone bound is violated when $\alpha<-4$ and when $\alpha>1$, there is no physical solution as expected. When $0<\alpha<1$, we observe that only one root is compatible with the Kähler cone condition. Moreover, it turns out that this is the one that can be expressed like \eqref{IIB-sol} with the proper principal value definitions of the roots.
\begin{figure}[!h]
\centering
\includegraphics[scale=0.45]{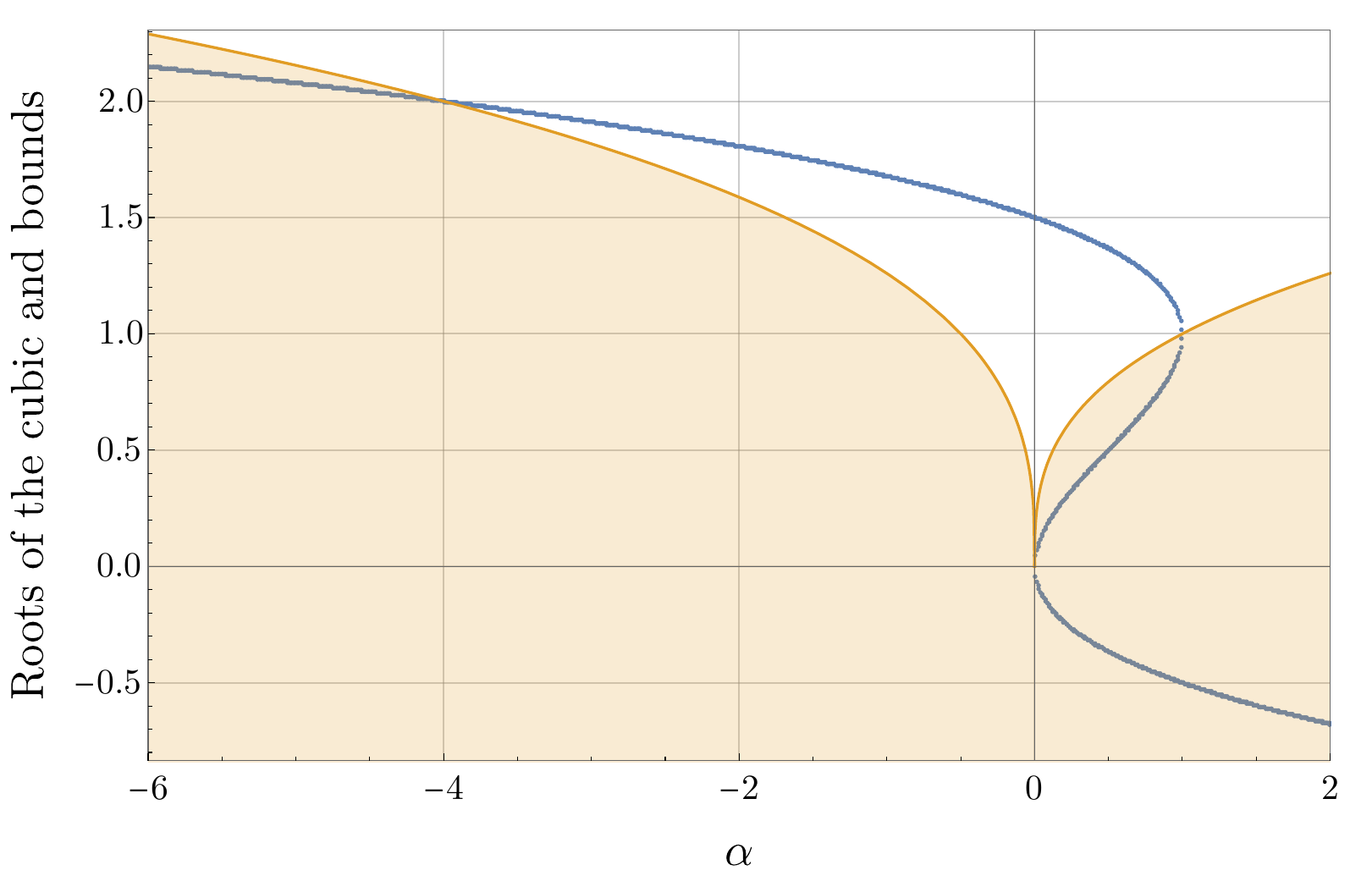}
\caption{The roots of the cubic \eqref{IIB-cubic0} with respect to the parameter $\alpha$.}
\label{IIB-roots}
\end{figure}

Apart from the full analytical expressions for the moduli vacuum expectation values, valid at arbitrary $\xi$, the simpler Ansatz under consideration here also allows to uncover the scalar mass spectrum. Computing these masses is the purpose of the next subsection.

\subsubsection{Mass spectrum}
\label{IIB-sec:masses}

To uncover the mass spectrum, we make use of the symplectic decomposition of the flux vector introduced in \cite{Denef:2004ze}, which reads
\begin{equation}
    N = \sqrt{4\pi} e^{K_{\rm cs}} \left( -i W \bar{\Pi} + 2 t^0 D_{\bar{\tau}} D_{\bar{j}} \bar{W} K^{\bar{j}i} D_i \Pi \right)\ .
    \label{IIB-eq:N_structure}
\end{equation}
Inserting the flux constraints of the IIB1 setup $f_A^0=h_A^0=h_A^i=0$ inside the above expression yields two relations
\begin{equation}
W = -2i t^0 D_{\bar{\tau} \bar{j}} \bar{W} K^{\bar{j}i} K_i\ \quad\text{ and }\quad f_A^i= 2 e^{K_{\rm cs}} \left( t^0 D_{\bar{\tau} \bar{j}} \bar{W} K^{\bar{j}i} - t^i W \right)\ ,
\end{equation}
from which we deduce
\begin{equation}
\label{IIB-eq:DtW_Int}
D_{\tau i} W = \frac{1}{2t^0} K_{i\bar{j}} \left( e^{-K_{\rm cs}} f_A^j + 2 t^j \bar{W} \right)\ .
\end{equation}
Now we can make use of the proportionality relations \eqref{IIB-eq:that} that defines the Ansatz to replace $f_A^j$ in the above formula and factor a term $K_{i\bar j}t^j$. From eqs.~\eqref{IIB-eq:Ki} and \eqref{IIB-eq:Kij}, this factor reads
\begin{align}
    K_{i\bar{j}} t^j = - 2 \mathring{\kappa}_{ijk} t^j t^k + 4 \mathring{\kappa}_{imn} \mathring{\kappa}_{jpq} t^m t^n t^j t^p t^q = i \left( 1 - 2 \mathring{\kappa} \right) K_i\ ,
\end{align}
where we have defined $\mathring{\kappa} \equiv e^{K_{\rm cs}} \kappa_{ijk} t^i t^j t^k$. Plugging this result back into eq.~\eqref{IIB-eq:DtW_Int} yields
\begin{equation}
D_{\tau i} W = \frac{ i \left( 1 - 2 \mathring{\kappa} \right)}{2t^0} \left( 2 \bar{W} -  \frac{e^{-K_{\rm cs}}}{\hat t}\right)K_i\ .
\end{equation}

These steps show that under the IIB1 flux configuration and for our branch of solution of interest, the two-derivative of the superpotential with respect to the axio-dilaton and some complex structure field is proportional to the first derivative of the Kähler potential with respect to this latter modulus. As such, the IIB1 scenario fullfills the prerequisite for the derivation of the \emph{no-scale aligned} mass spectrum, introduced in \cite{Blanco-Pillado:2020hbw} and reviewed in appendix~\ref{IIB-sec:NSA}. The tree-level mass spectrum is thus given by \eqref{IIB-eq:nsa_mass_spectrum_app}, which we repeat here:

\begin{equation}
	\frac{\mu^2_{\pm \lambda}}{m_{3/2}^2} = 
	\left\lbrace
	\begin{array}{ll}
		\left( 1 \pm \sqrt{\frac{1 -2 \xi}{3}} \hat{m} (\xi) \right)^2  & \lambda = 0  \\
		\left( 1 \pm \sqrt{\frac{1 -2 \xi}{3}} (\hat{m} (\xi))^{-1} \right)^2  & \lambda = 1 \\
		\left( 1 \pm \frac{1+\xi}{3} \right)^2 & \lambda = 2, \ldots, h^{2,1}
	\end{array}
	\right.
	\label{IIB-eq:nsa_mass_spectrum}
\end{equation}
where we have defined the quantities
\begin{align}
\begin{split}
&\hat{m} (\xi) \equiv \frac{1}{\sqrt{2}} \left( 2 + \kappa (\xi)^2 - \kappa (\xi) \sqrt{4 + \kappa (\xi)^2} \right)^{1/2}\ ,\\
&\kappa (\xi)\equiv 2 (1+ \xi)^2 / \sqrt{3(1-2\xi)^3}\ .
\end{split}    
\end{align}
The evolution of this normalized mass spectrum is displayed in fig.~\ref{IIB-analytical_mass_spectrum}.
\begin{figure}[!t]
\centering
\includegraphics[scale=0.45]{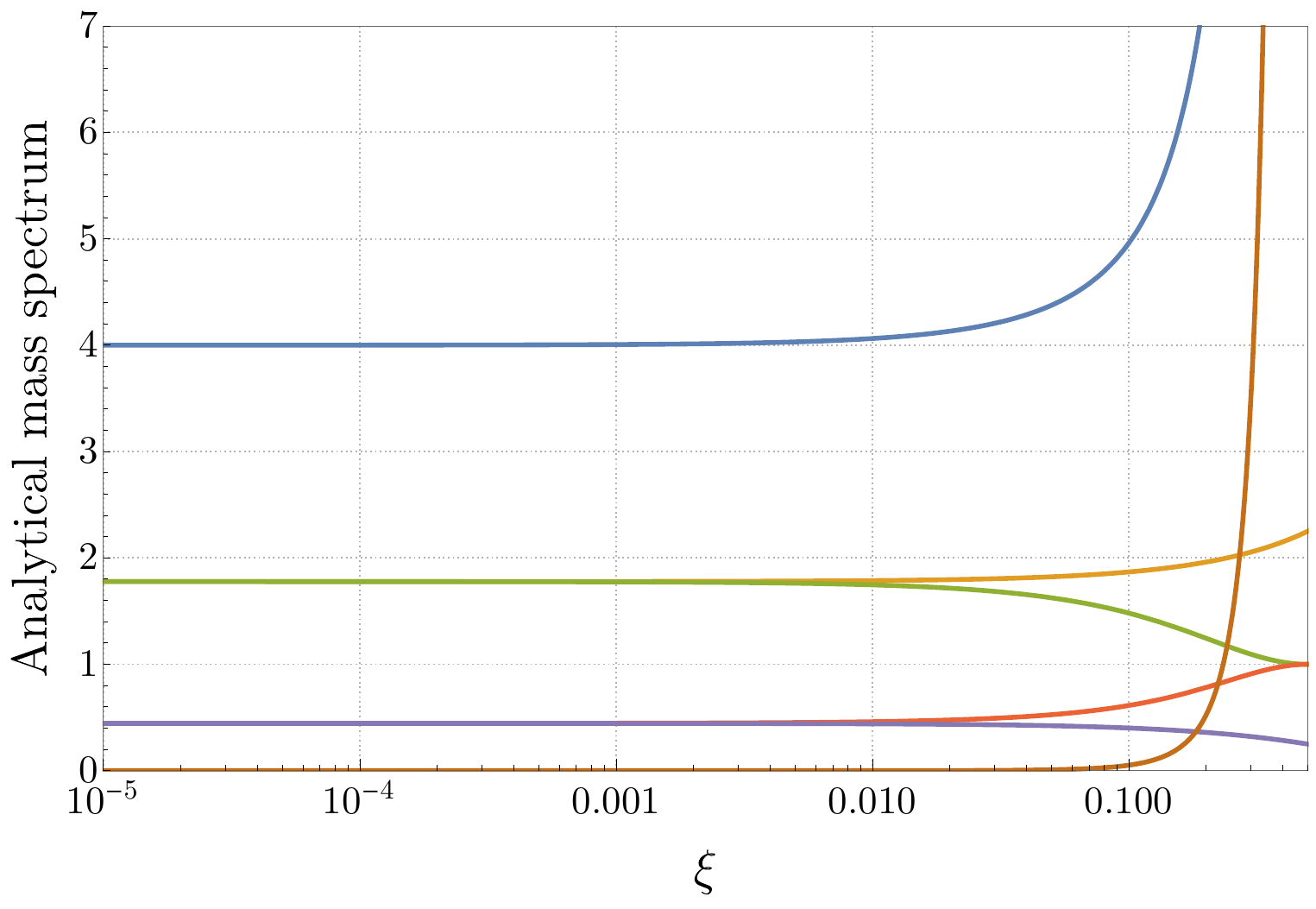}
\caption{Evolution of the scalar mass spectrum \eqref{IIB-eq:nsa_mass_spectrum} with respect to the LCS parameter $\xi$. The correspondence between the curves and the labels $\pm\lambda$ of the different modes is as follows: Blue curve is $+1$; Orange curve is $+\lambda$, $\lambda=2,\dots,h^{2,1}$; Green curve is $+0$; Red curve is $-0$; Purple curve is $-\lambda$, $\lambda=2,\dots,h^{2,1}$ and Brown curve is $-1$.}
\label{IIB-analytical_mass_spectrum}
\end{figure}
Expanded around the LCS point at $\xi=0$, the spectrum reads
\begin{equation}
	\frac{\mu^2_{\pm \lambda}}{m_{3/2}^2} = 
	\left\lbrace
	\begin{array}{ll}
		\frac{16}{9}+\mathcal O(\xi)\ ,\  \frac{4}{9}+\mathcal O(\xi)& \lambda = 0  \\[5pt]
		4+\mathcal O(\xi)\ ,\ \frac{9}{4}\xi^2+\mathcal O(\xi^3)  & \lambda = 1 \\[5pt]
		\frac{16}{9}+\mathcal O(\xi)\ ,\  \frac{4}{9}+\mathcal O(\xi) & \lambda = 2, \ldots, h^{2,1}
	\end{array}
	\right.
\end{equation}
Notice that the mode labeled by $-1$ becomes rapidly massless as $\xi\to 0$, as can also be seen from fig.~\ref{IIB-analytical_mass_spectrum}. This is also true in Planck units, since the gravitino mass dependence on $\xi$ is given by
\begin{equation}
m_{3/2}^2=\frac{3}{2\V^2}\frac{\S\hat h^B}{q(2-\xi)} M_{\rm P}^2=\frac{3}{2\V^2}\frac{\Nf}{2-\xi}M_{\rm P}^2\ . 
\end{equation}
This nicely matches the expectations put forward in the previous chapter, where we found that given the choice of fluxes \eqref{IIB-hBicond},  polynomial corrections are required to stabilize all moduli, and that otherwise a field is left unstabilized. It is thus natural to identify such a field with the lightest mode of the spectrum, whose mass goes proportional to $\xi$ as we approach the LCS point. 

All these results are verified by appendix \ref{IIB-ap:spot}, which develops a  different approach to the computation of the mass spectrum. This method works directly with the scalar potential \eqref{Ft-scalarpot} particularized to the Type IIB case using the expressions in section \ref{Ft-s:IIB},  from where the Hessian can be obtained. One can see that in terms of the Hessian, the axion-like fields and their saxionic partners are decoupled. Therefore, by analyzing one of these two sets, it enables us to distinguish between axions and saxions in \eqref{IIB-eq:nsa_mass_spectrum}. In particular,  appendix \ref{IIB-ap:spot} works out explicit analytic expressions for the axionic masses of the no-scale aligned branch, obtaining a perfect match with half of the spectrum in \eqref{IIB-eq:nsa_mass_spectrum}. One can then check that the lightest field of \eqref{IIB-eq:nsa_mass_spectrum} is not one of the axion-like fields and that it instead belongs to the saxionic sector.

\subsubsection{Generating flux vacua}
\label{IIB-sec:generate_vacua}

In the previous paragraphs we have studied how a choice of fluxes which satisfies
\begin{align}
	f_A^0 = h_A^0 = h_A^i = 0\ , \quad h_i^B = - N_{\rm flux} \frac{S_i}{\S} \ , 
\end{align} 
admits an analytical solution for the real and imaginary parts of the axio-dilaton and all of the complex structure moduli, as long as the rest of the fluxes satisfy the constraints outlined above. Indeed, given such a choice of fluxes, one may compute the axionic  components using eqs.~\eqref{IIB-eq:bs}. On the other hand, we have seen that given the Ansatz $t^i \equiv \hat{t} f_A^i$ for the complex structure saxions, one may use eq.~\eqref{IIB-eq:solthat} to compute $\hat{t}$ and, finally, use \eqref{IIB-eq:t0t} to determine the value of $t^0$. As a consequence, the search for flux vacua in the branch we have described here can be completely automatized.

Note that once $f_A^i$ and $h_i^B$ are fixed, one is free to choose $f_0^B$, $h_0^B$ and $f_i^B$ without changing the D3-tadpole. Thanks to the relation \eqref{IIB-eq:flux_const_away}, the definition of $\alpha$ \eqref{IIB-eq:def_alpha} and the definition of $Q'$ in \eqref{IIB-eq:rho}, these flux quanta may be easily tuned to generate vacua at the desired distance from the LCS point. This procedure has been explicitly carried out in the two-parameter example explored in section \ref{IIB-sec:numerics}.

In particular, this can also be useful to easily generate tuples of fluxes which yield vacua close to the LCS point, where exponentially suppressed 
corrections to the tree-level prepotential  may be neglected. From \eqref{IIB-eq:flux_const_away}, we find that vacua close to the LCS point where $|\xi| \ll 1$ satisfy
\begin{align}
	\label{IIB-eq:LCS_xi_flux}
	\xi \approx -  \frac{2^8}{3^5 (\Im \kappa_0)^2 \S} \, Q'^3 \ ,
\end{align}
where we recall that $\S\equiv \kappa_{ijk} f_A^i f_A^j f_A^k$ and $Q'$ has been defined in \eqref{IIB-eq:rho}. Thus, we need $Q'$ to be small and negative. An easy way to satisfy such a condition is by choosing $f_i^B = - f_A^j a_{ij}$, so that $L_i = 0$. In that case, $Q'$ is simplified to
\begin{align}
	f_i^B = - f_A^j a_{ij} \ \Longrightarrow \ Q' = f_0^B - f_A^i \hat{a}_i + \frac{1}{2\S} \left( \frac{\S h_0^B}{N_{\rm flux}} \right)^2 .
\end{align}
Thus, having chosen $f_A^i$ and $N_{\rm flux}$, we can easily generate pairs of $f_0^B$ and $h_0^B$ which yield vacua with small $\xi$. 

\section{Supersymmetric vacua}
\label{IIB-sec:susy}

We now turn our attention to supersymmetric vacua which, as already mentioned, always contain a number of complex flat directions at the level of approximation to which we are working. One important feature of these vacua is that the flux quanta need to satisfy a series of constraints, in agreement with recent results in the literature. While obtaining the vevs for the stabilized fields is  straightforward, working out the mass spectra for these vacua turns out to be more involved than in the no scale aligned case.  

\subsection{Moduli stabilization and flat directions}

Here we describe the supersymmetric class of vacua defined in section \ref{IIB-sec:Mnoninvertible}. As already said there and similarly to the case above, the requirement that $M\vec B=-\vec L$ generates $h^{2,1}+1-\textrm{rank} (M)$ constraints that the fluxes must satisfy to fall into this case. The solutions for the moduli are expressed like $\vec Z=\vec B+\ker (M)$ such that there are $h^{2,1}+1-\textrm{rank} (M)$ complex flat directions, and the additional requirement $W=0$ at vacua provides one more constraint on fluxes. If we put this back into the vacuum equations \eqref{IIB-eq:T_gen}, we obtain a simple linear system of equations where axions and saxions are decoupled:
\begin{align}
\begin{split}
 M\vec B&=-\vec L\ ,\\
M\vec T&=0\ .
\end{split}
\end{align}
The equation regarding the saxions can be further decomposed in the following relations
\begin{equation}
\label{IIB-eq:sijtj}
    h_i^B t^i=0\ ,\qquad S_{ij}t^j=h_i^Bt^0\ .
\end{equation}
Remembering now the decomposition discussed in \eqref{IIB-decomp}, we observe that supersymmetric vacua require $A=B=0$ and $C_i$, $C^i\neq0$, which contrasts with the set of non-supersymmetric solutions described by \eqref{IIB-eq:that}.

In order to make analytical progress, let us study again the subclass when the matrix $S$ possesses an inverse denoted $S^{ij}$ in components. When this is the case, the rank of $M$ is at least $h^{2,1}$ and for $M$ not to be invertible, it cannot be more than that. The non-invertibility of $M$ translates into the requirement
\begin{equation}
\label{IIB-eq:S0}
\cH=h_i^BS^{ij}h_j^B=0\ .    
\end{equation}
When solving $M\vec B=-\vec L$, as expected we derive one constraint and one axion is left unstabilized (this is the same situation as in section \ref{IIB-sec:S0}):
\begin{align}
&h_i^BS^{ij}L_j=h_0^B\ ,\label{IIB-eq:cons_X}\\
&b^i=-S^{ij}L_j+b^0S^{ij}h_j^B\ .
\end{align}
Besides, the kernel of $M$ is one dimensional and given by
\begin{equation}
\ker (M)=\langle (1,S^{ij}h_j^B)\rangle\ .
\end{equation}
We thus have
\begin{equation}
\vec Z=\vec B+\ker (M)\quad\Longleftrightarrow\quad
\left\{
\begin{matrix*}[l]
\: \tau=b^0+\lambda\\[4pt]
z^i=b^i+\lambda S^{ij}h_j^B
\end{matrix*} 
\right. ,
\end{equation}
where $\lambda$ is some complex number that we can fix using the first equation of the system:
\begin{equation}
\Re (\lambda)=0\quad\text{ and }\quad \Im (\lambda)=t^0\ .    
\end{equation}
The second set of equations then gives expressions for $t^i$ with $t^0$ as a free parameter.\footnote{Note that here we applied naively the generic relation of section \ref{IIB-sec:Bilinear} but we could have expressed $t^i$ easily from eq.~\eqref{IIB-eq:sijtj}.}  Summarizing, we have
\begin{align}
b^i&=-S^{ij}L_j+b^0S^{ij}h_j^B\ ,\label{IIB-eq:bi_susy}\\[5pt]
t^i&=S^{ij}h_j^Bt^0\ .
\end{align}
These relations define the two real flat directions that we expected from the general analysis.

One last constraint arising from the requirement of a vanishing superpotential is to be uncovered. Demanding $Q'=0$ from eq.~\eqref{IIB-eq:Q'} yields
\begin{equation}
\label{IIB-eq:cons_susy}
f_0-\hat{a}_if_A^i-\oh L_iS^{ij}L_j=0\ .
\end{equation}

Following similar arguments to the ones presented in section \ref{IIB-sec:generate_vacua}, a straightforward choice of fluxes which satisfies all the above conditions, eqs.~\eqref{IIB-eq:S0}, \eqref{IIB-eq:cons_X} and \eqref{IIB-eq:cons_susy}, is based on picking $f_A^i$ and $f^B_i$ such that
\begin{align}
	\hat{a}_i f_A^i \in \mathbb{Z}\ , \quad a_{ij} f_A^j \in \mathbb{Z}\ , \quad f_i^B = - a_{ij} f_A^j \ .
\end{align}
This automatically implies
\begin{align}
	f_0^B = \hat{a}_i f_A^i\ , \quad h_0^B = 0\ ,
\end{align}
so all that is left to do is to find $h_i^B$ such that
\begin{align}
	h_i^B  S^{ij} h_j^B = 0 \ .
	\label{IIB-eq:hhS}
\end{align}

Notice that the flux constraints \eqref{IIB-eq:S0} and \eqref{IIB-eq:cons_susy} agree with the tree-level conditions exposed in \cite{Demirtas:2019sip,Demirtas:2020ffz} where the authors further consider exponentially suppressed corrections in order to generate small flux superpotentials. The complex flat direction we found here when $S$ is invertible also seems to generalize the supersymmetric vacua uncovered in \cite{Cicoli:2022vny} to arbitrary Calabi--Yau geometries.

\subsection{Towards the mass spectrum}

In this section we push the computation of the mass spectrum for the supersymmetric vacua as far as we can. In the end, however, we will not be able to express it analytically in full generality like for the non-supersymmetric vacua with the simple saxionic Ansatz. It is still interesting to understand what prevents us from doing so.

As we proved in the section above, the supersymmetric vacua satisfy
\begin{align}
	t^i = v^i t^0 \ , \quad v^i \equiv S^{ij} h_j^B\ ,\quad\text{ with } \quad h_i^B v^i = h_i^B S^{ij} h_j^B = 0\ .
\end{align}
We will follow the same logic as in the derivation of the mass spectrum for no-scale aligned vacua presented in appendix~\ref{IIB-sec:NSA}. This means we want to simplify the Kähler metric as best as we can, in order to obtain the simplest form possible for the matrix $Z_{AB}\equiv e^{K/2}D_AD_BW$ where the indices $A$, $B$ run into $\{\tau,z^i\}$. As reviewed in appendix \ref{IIB-sec:NSA} and shown in \cite{Sousa:2014qza}, the scalar masses $\mu_{\pm\lambda}$, $\lambda=0,\dots,h^{2,1}$ are simply given in the supersymmetric case by the fermion masses $m_{\lambda}$:
\begin{equation}
\mu_{\pm\lambda}=m_{\lambda}\ ,
\end{equation}
which correspond to the eigenvalues of the matrix $Z$.

To start orthonormalizing the Kähler metric \eqref{IIB-eq:Kij}, we can introduced two vielbeins inspired by the two preferred directions of the supersymmetric vacua: $t^i=t^0v^i$ and $f_A^i$. Notice, as we will explicitly see shortly, that in the non-supersymmetric branch studied earlier, these two vectors are aligned, which implies the alignment of $D_iD_\tau W$ with $K_i$ and hence the ``no-scale aligned'' property of the vacua, which enabled us to uncover the mass spectrum. We thus define the two vielbeins $e_1^i$ and $e_2^i$ like
\begin{equation}
    e_1^i\equiv \frac{t^i}{x}\quad\text{ and }\quad e_2^i\equiv \frac{f_A^i}{y}\ ,
\end{equation}
where $x$ and $y$ are normalization factors that can be straightforwardly expressed like
\begin{equation}
    x=\frac{\sqrt{3(2-\xi)}}{2(1+\xi)}\ ,\quad y=\sqrt{2t^0\Nf e^{K_{\rm cs}}}\ ,
\end{equation}
with $\Nf=-f_A^ih_i^B$. These two vielbeins are indeed orthogonal since we can show that
\begin{equation}
e_1^iK_{ij}e_2^j\propto h_i^BS^{ij}h_j^B=0\ .
\end{equation}
Plugging the vielbeins into the Kähler metric \eqref{IIB-eq:Kij}, we can obtain expressions for the rescaled Yukawa couplings $\mathring{\kappa}_{abc}$ involving the  direction $1$ similar to \eqref{IIB-eq:kappas_circle_app}:
\begin{align}
    \label{IIB-eq:kappas_circle}
    \mathring{\kappa}_{111} = \frac{2 (1+\xi)^2}{\sqrt{3(1-2\xi)^3}}\ ,\quad \mathring{\kappa}_{a' 11} = 0\ ,\quad \mathring{\kappa}_{a' b' 1} = \frac{-(1+\xi)}{\sqrt{3 (1 - 2 \xi)}} \delta_{a'b'}\ ,
\end{align}
where the prime indices run from $2$ onwards.

With this, we are now ready to see the special role played by these two directions: Direction $1$ is aligned with the no-scale direction while direction $2$ is aligned with $Z_{0a}$. Indeed, making use of \eqref{IIB-eq:Ki} and the symplectic decomposition of the flux vector \eqref{IIB-eq:N_structure} we find
\begin{equation}
    K_a=e_a^iK_i=2ix^2\mathring\kappa_{a11}\propto\delta_a^1\quad\text{ and }\quad Z_{0a}=ye^{K/2-K_{\rm cs}}\delta_a^2 \ .
\end{equation}
Finally, using eq.~\eqref{IIB-eq:Denef}, the expression for $Z_{ab}$ is
\begin{equation}
    Z_{ab}=-iye^{K/2-K_{\rm cs}}\mathring\kappa_{ab2}\ .
\end{equation}
Precisely because directions $1$ and $2$ are not aligned, we lack information to characterize the rescaled Yukawa couplings $\mathring\kappa_{ab2}$ and the only matrix elements we have control of are
\begin{equation}
\begin{aligned}
	Z_{0a} &= ye^{K/2-K_{\rm cs}}\delta_a^2\ ,\qquad\quad  &&Z_{11}=0\ , \\[7pt]
	Z_{1a} &= \frac{iy}{2x} e^{K/2-K_{\rm cs}} \delta_a^2\ , &&Z_{22} =- i e^{K/2} y^{-2} \S\ ,
\end{aligned}
\end{equation}
while the elements $Z_{2\tilde a}$ and $Z_{\tilde a\tilde b}$ are unknown for $\tilde a$, $\tilde b$ running from 3 onwards. The canonically normalized fermion mass matrix then reads
\begin{align}
	Z = \left(
	\begin{array}{ccc|c}
		0 & 0 & Z_{02} & 0  \\
		0 & 0 & Z_{12} & 0  \\
		Z_{02} & Z_{12} & Z_{22} & Z_{2\tilde a}  \\ \hline
		0 & 0 & Z_{2\tilde a} & Z_{\tilde a \tilde b} 
	\end{array}
	\right) .
	\label{IIB-eq:zab_w0}
\end{align}

Remember that the scalar masses correspond to the fermion ones, only doubled. The mass matrix \eqref{IIB-eq:zab_w0} cannot be diagonalized in full generality but it is easy to see that it features a massless mode, which thus translates into two massless directions in the scalar potential. This matches the expectations of the previous subsection.

\section{A numerical set of vacua in a two-parameter model}
\label{IIB-sec:numerics}

The goal of this section is to provide a numerical cross-check of the analytical results exposed in the previous section for the non-supersymmetric class of vacua following the \emph{no-scale aligned} branch with $t^i\propto f_A^i$. To this end, we generate an ensemble of IIB1 flux vacua in a two-parameter model by solving the vacuum equations numerically and then check various properties of these vacua. The model in question is the one arising from a symmetric point in the moduli space of the Calabi--Yau hypersurface $\mathbb{CP}^4_{[1,1,1,6,9]}$. We will first see how the analytical control of the IIB1 scenario enables us to generate a large number of vacua in the LCS regime very efficiently and we then show the perfect agreement between the features of these numerical vacua and the expectations from the analytics presented in section \ref{IIB-sec:simpler}. 

\subsection{Generating flux tuples}

The first step to generate a numerical ensemble of vacua is to create a set of flux tuples meant to be run through in search for solutions of the vacuum equations. In order to reduce a bit the number of parameters, we consider the following restriction on the flux quanta $f_A^i$, $i=1,2$:
\begin{equation}
    f_A^1=f_A^2\equiv\hat f_A\ .
\end{equation}
If we trust our Ansatz \eqref{IIB-eq:that}, this means that at the vacua we will have $t^1=t^2$.

We want flux configurations that do not overshoot the tadpole D3-charge bound $Q_{\text D3}$. With an O7-plane/D7-brane configuration identical to the one used in \cite{Demirtas:2019sip} and described in \cite{Louis:2012nb}, the induced D3-charge is restricted to satisfy $Q_{\text D3}\leq 138$. The flux contribution to the tadpole $\Nf$ depends only on $\hat f_A$ and $\hat h^B$ and thus we first generate a set of tuples for these flux quanta subject to the tadpole constraint. More precisely, we consider all flux entries in the range $[-6,6]$ and produce $14$ configurations satisfying the tadpole bound.

The fluxes remaining to be fixed at this point are $f_0^B$, $f_1^B$, $f_2^B$ and $h_0^B$. For the sake of efficiency, instead of generating a random set of tuples for them, we make use of our analytical expectations derived in section \ref{IIB-sec:simpler}. This is done by expressing the flux-dependent quantity $\alpha$ defined in \eqref{IIB-eq:def_alpha} in terms of the unfixed flux quanta and by ensuring a choice of the latter such that $\alpha$ lies in the range $[-4,0]$. Since we want to cross-check our $\xi$-dependent analytics, we can do more than that and produce flux tuples that we expect to span the whole allowed range for $\xi$. To this end, we subdivide the $\alpha$ range $[-4,0]$ into $200$ pieces and try to find fluxes ${f_0^B,f_1^B,f_2^B,h_0^B}$ to fall into each piece, for each of the $14$ configurations ${\hat f_A,\hat h^B}$ previously generated. This results into a set of $2650$ full flux configurations that will use in the next subsection.\footnote{Note that all these steps are very easy and quick to implement so that a much bigger set of flux configurations could be generated effortlessly.}

\subsection{Vacua analysis}

We numerically implemented the vacuum equations and searched for solutions for each flux configuration of our ensemble. The two-parameter model is characterized by the following topological quantities that fully define the prepotential \eqref{IIB-eq:full_prepotential} (neglecting exponentially suppressed corrections):
\begin{equation}
\begin{aligned}
&\kappa_{111}=9\ ,\quad&&\kappa_{112}=3\ ,\quad &&\kappa_{122}=1\ ,\quad\kappa_{222}=0\ ,\\
&\kappa_{11}=-\frac{9}{2}\ ,\quad&&\kappa_{22}=0\ ,\quad&&\kappa_{12}=-\frac{3}{2}\ ,\\
&\kappa_1=\frac{17}{4}\ ,\quad&&\kappa_2=\frac{3}{2}\ ,\quad&&\kappa_0=-540\frac{\zeta(3)}{(2i\pi)^3}\ .
\end{aligned}
\end{equation}
As expected from our careful choice of fluxes guided by the analytics, each flux tuple yields a consistent vacuum inside the Kähler cone. The vacua are displayed in the $(t^1,t^0)$-plane in fig.~\ref{IIB-t1t0_fit}.

A first analytical relation that we can check is eq.~\eqref{IIB-eq:t0t}. In the case at hand with $f_A^1=f_A^2=\hat f_A$, we have $q=(\hat f_A)^2$ and $\S=21(\hat f_A)^3$. The relation then becomes 
\begin{equation}
\label{IIB-eq:t0t_num}
t^0 =-\frac{\hat f_A}{\hat h^B}\frac{14 (t^1)^3-\Im\kappa_0}{28(t^1)^3+\Im\kappa_0} \ t^1\ .
\end{equation}
The comparison between this analytical formula and the data of our ensemble of vacua is displayed in fig.~\ref{IIB-t1t0_fit}. We observe a perfect match between the two.

\begin{figure}[!ht]
\centering
\includegraphics[scale=0.4]{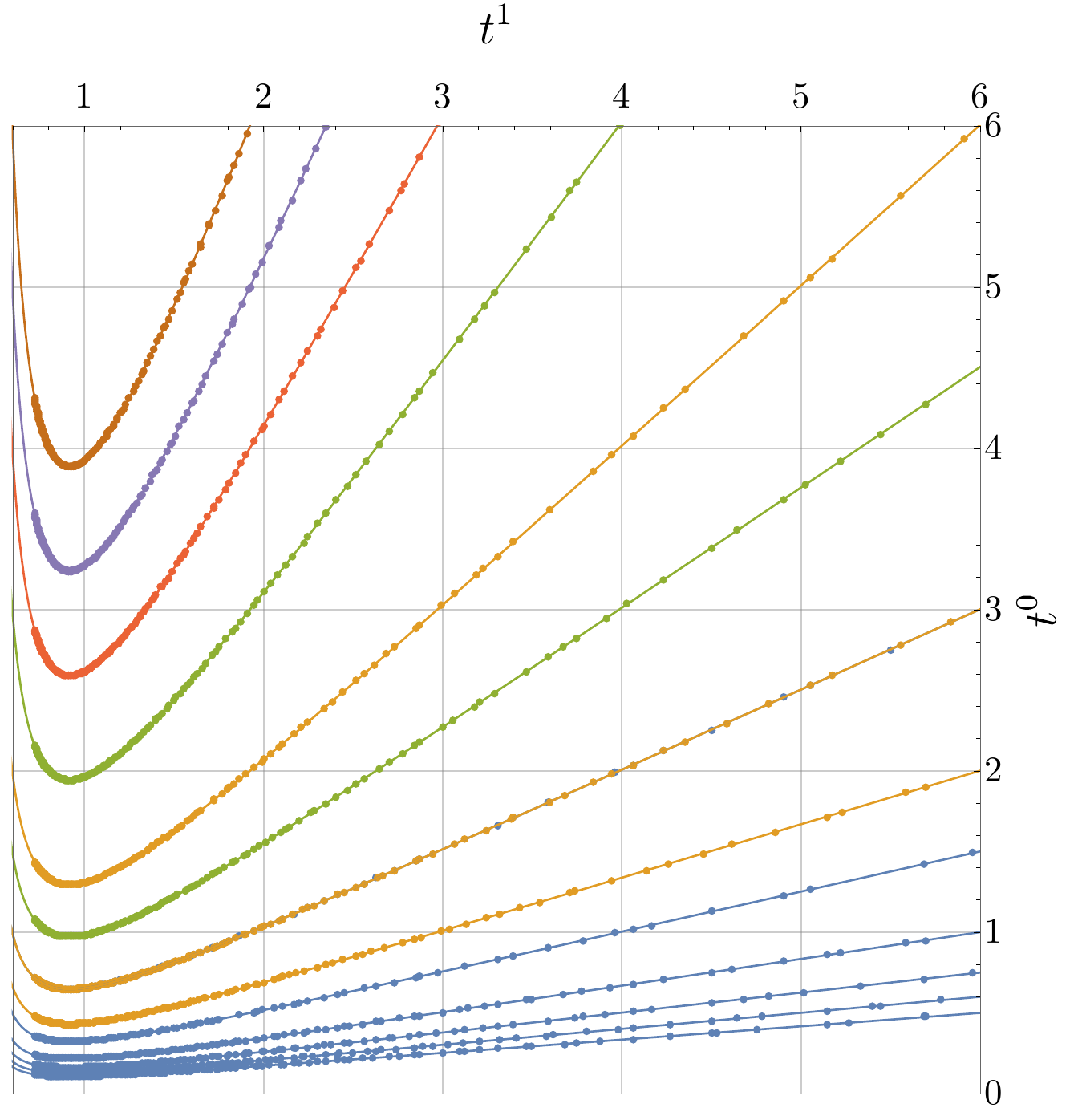}
\caption{This plot shows the locations of the numerically generated IIB1 vacua in the $(t^1,t^0)$-plane. The vacua are depicted with different colors corresponding to different values of $\hat f_A$, with different branches corresponding to different values for $\hat h^B$, present in the ensemble. For a given color, the expression \eqref{IIB-eq:t0t_num} is displayed on top of the numerical data. We observe a perfect agreement. More precisely, the colors correspond to the following fluxes: Blue: $\hat f_A=1,\,\hat h^B=1,\dots,6$; Orange: $\hat f_A=2,\,\hat h^B=1,\dots,3$; Green: $\hat f_A=3,\,\hat h^B=1,2$; Red: $\hat f_A=4,\,\hat h^B=1$; Purple: $\hat f_A=5,\,\hat h^B=1$; and Brown: $\hat f_A=6,\,\hat h^B=1$. As explained in section \ref{IIB-sec:instantons}, vacua with $\xi<0.17$ \ie with $t^1,\ t^2\gtrsim 1$ are expected to be safe under instanton corrections as the relative changes induced by the corrections on the moduli space and other quantities are small.}
\label{IIB-t1t0_fit}
\end{figure}

Another non-trivial result we can check is the relation between $\xi$ and the quantity $\alpha$ (see eqs.~\eqref{IIB-eq:def_alpha} and \eqref{IIB-eq:flux_const_away}). Figure \ref{IIB-xi_alpha} shows a nice fit of the data by the analytical expression.

\begin{figure}[!ht]
\centering
\includegraphics[scale=0.48]{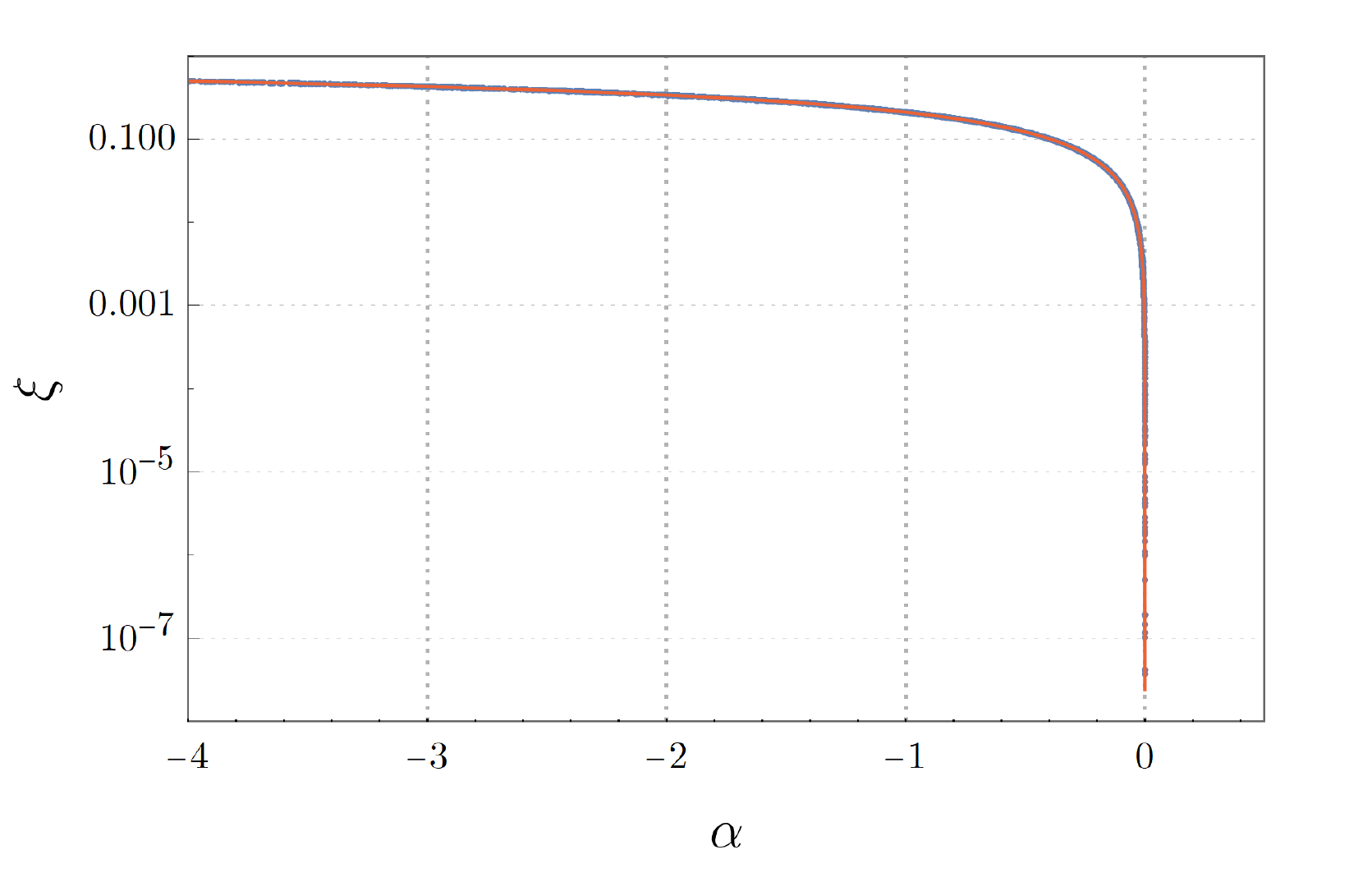}
\caption{This plots shows the values of $\xi$ against $\alpha$ for the numerical vacua of our ensemble. The relation \eqref{IIB-eq:flux_const_away} is plotted in red and fits perfectly the data points.}
\label{IIB-xi_alpha}
\end{figure}

One last important result to be checked is the mass spectrum in the vacua. We have shown in section \ref{IIB-sec:simpler} that the vacua under consideration fall into the definition of the \emph{no-scale aligned} setup whose mass spectrum normalized by the gravitino mass $m_{3/2}$ is given as a function of $\xi$ by eq.~\eqref{IIB-eq:nsa_mass_spectrum}. The canonically normalized masses, numerically computed for each vacuum, are displayed in fig.~\ref{IIB-spectrum}. We again observe that the numerical results perfectly match the analytical expectations displayed in fig.~\ref{IIB-analytical_mass_spectrum} in section \ref{IIB-sec:masses}.

\begin{figure}[!ht]
\centering
\includegraphics[scale=0.5]{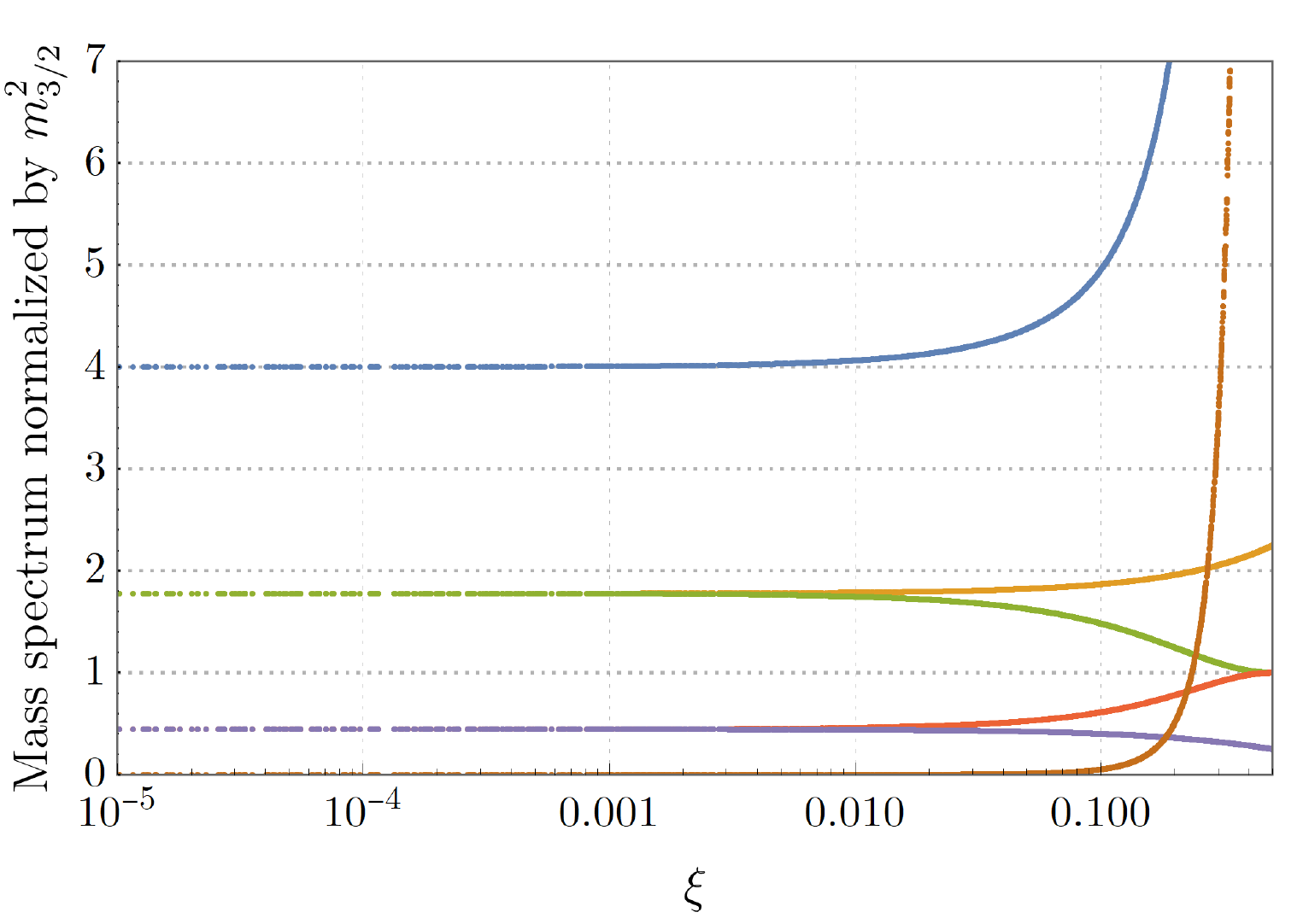}
\caption{This plot shows the squared masses, normalized by the gravitino mass squared, numerically obtained in the set of vacua. They precisely reproduce the analytical behaviour \eqref{IIB-eq:nsa_mass_spectrum} displayed in fig.~\ref{IIB-analytical_mass_spectrum}.}
\label{IIB-spectrum}
\end{figure}

\subsection{Exponential corrections}
\label{IIB-sec:instantons}

Of course we expect exponential corrections in the prepotential \eqref{IIB-eq:full_prepotential} to become more and more relevant as the LCS parameter $\xi$ goes away from the LCS point and gets closer to the boundary at $\xi=1/2$. In specific examples and following \cite{Blanco-Pillado:2020wjn,Blanco-Pillado:2020hbw}, we can evaluate the effect of the exponentially suppressed corrections by computing their relative effects on the geometry of the moduli space and other physical quantities.

For the $\mathbb{CP}^4_{[1,1,1,6,9]}$ hypersurface, the dominant exponential corrections are expressed like \cite{Blanco-Pillado:2020hbw,Candelas:1994hw}
\begin{equation}
\F_{\rm inst}=-\frac{135}{2\pi^3}ie^{2i\pi z^1}-\frac{3}{8\pi^3}ie^{2i\pi z^2}\ ,
\end{equation}
and we can use them to numerically compute the relative errors induced on the Kähler metric, the gravitino mass $m_{3/2}$ and $e^{K_{\rm cs}}$. Note that this definition for the validity of the perturbative result is rather conservative and much more stringent than only requiring the non-perturbative part of the prepotential to be dominated by the perturbative one. We find that vacuum expectation values for $t^1=t^2$ slightly above $1$ are enough to guarantee the stability of the perturbative vacua since all the relative corrections are smaller than a small threshold of $5\%$. In terms of the LCS parameter, $\xi<0.17$ ensures robustness of the perturbative results.


\section{Summary}
\label{IIB-sec:Conclusion}

In this chapter, we investigated the specific type IIB family of flux vacua at large complex structure introduced in chapter \ref{ch: Ftheory} and called \emph{IIB1 scenario}. Arising as a type IIB limit from an F-theory construction, the vacuum equations were studied there at first order in the LCS parameter $\xi$ defined in \eqref{IIB-eq:def_xi}, i.e., not too far from the LCS point. The analysis of the current chapter extends these results by exploring in more detail different classes of vacua allowed by the IIB1 setup, and by pushing their analytical resolutions (computation of the complex structure and axio-dilaton vevs as well as mass spectra) as far as possible.

The IIB1 choice of fluxes ensures that all cubic terms disappear from the flux-induced superpotential such that it is simply quadratic in the axio-dilaton and complex structure fields. A very generic and coarse-grained classification of vacua arising from such a quadratic structure reveals the existence of one supersymmetric family and two non-supersymmetric ones, depending on the definiteness or not of the bilinear form involved in the superpotential. More precisely, a regular bilinear structure forbids supersymmetric vacua while a singular one allows vacua that are either supersymmetric or not. In any of these cases, the vacuum equations nicely split into two separate systems: A very simple one involving only the axions (thanks to the independence on the axions of the superpotential at vacua), and a more involved one relating the saxions. Moduli stabilization can then be studied separately for these two sets of fields.

We then explored the three classes mentioned above further in detail. The supersymmetric vacua are described by very simple vacuum equations thanks to the vanishing of the superpotential on-shell. Restricting to fluxes such that the matrix $S$, with $S_{ij}\equiv \kappa_{ijk}f_A^k$ involving the triple intersection numbers of the mirror manifold, is invertible (a recurring assumption in this paper), we saw that the supersymmetric vacua feature one complex flat direction and are similar to those used in \cite{Demirtas:2019sip,Demirtas:2020ffz} to achieve small superpotentials. They also generalize the supersymmetric models studied in \cite{Cicoli:2022vny} to arbitrary Calabi--Yau compactifications.  For these supersymmetric vacua, we addressed the computation of the  scalar masses, and it seems that further analytical progress in obtaining the mass spectrum for models with $h^{2,1}>2$ requires more definite knowledge of the model under study.

The two non-supersymmetric classes highlighted above differ if the bilinear form involved in the superpotential is degenerate or not. The effect of a non-trivial kernel is to generate one flux constraint and one flat direction for each dimension of the kernel of the bilinear form. As a particular case, when the matrix $M$ representing the form is invertible, all axions are stabilized. Whether $M$ is regular or not, the saxionic system of equations is highly non-linear and generically stabilizes all fields. As a counterpart, it is trickier to handle. To make analytical progress, we proposed an Ansatz \eqref{IIB-eq:ansatz} for the saxions and studied the subsequent vacuum equations. This led us to consider two further refined branches where we could provide analytic expressions for all the vevs of the axio-dilaton and complex structure fields, and even express analytically the scalar mass spectrum for one of these branches.

The first branch is a subcase where the matrix $M$ is singular with a specific uni-dimensional kernel. One axionic direction is thus left as a flat direction. The saxionic vacuum equations produce a sixth order polynomial relation, from which we can express the saxion vevs. The polynomial can be analytically solved using a perturbative expansion in the LCS parameter $\xi$. The second branch is uncovered when assuming a simpler sub-Ansatz \eqref{IIB-eq:that} for the saxions. It is shown to be allowed only when $M$ is regular, so that all axions are fixed. The saxionic system yields a manageable cubic polynomial such that the vevs can be fully expressed within the LCS region. Moreover, we showed that this branch falls into the \emph{no-scale aligned} family studied in \cite{Blanco-Pillado:2020wjn,Blanco-Pillado:2020hbw}, for which the scalar mass spectrum can be fully expressed analytically in terms of the LCS parameter. As already observed in the previous chapter and expected from the necessity of incorporating polynomial corrections to stabilize all moduli in this context, these kind of mass spectra feature a mode becoming lighter  as one gets closer to the LCS point.

We checked numerically the validity  of our  approximations in the non-supersymmetric \emph{no-scale aligned} branch, and in particular the accuracy of the mass spectrum. We did this by investigating a small ensemble of IIB1 vacua in this branch, generated numerically. We worked with the two-parameter model coming from a symmetric point in the moduli space of the Calabi--Yau hypersurface $\mathbb{CP}^4_{[1,1,1,6,9]}$. In addition to providing a solid cross-check of the analytics derived in the paper, the numerical analysis shows that the IIB1 scenario gives a setup where we can very efficiently generate vacua numerically at (almost) arbitrary distance of the LCS point desired.

Thus we conclude that the simple Ansatz presented in section \ref{IIB-sec:simpler} allows for complete analytical control over both the distance to the LCS point and the vevs of all complex structure moduli and the axio-dilaton. As such, this setup can be extremely useful to consider further corrections to the tree-level solutions, either by the inclusion of stringy corrections which would render more accurate solutions, or by the inclusion of exponentially suppressed corrections to the prepotential. An interesting line of work in this sense can be the stabilization of the Kähler sector through different means, either through racetrack potentials \cite{Kachru:2003aw,Balasubramanian:2005zx} or by more generic mechanisms 
\cite{AbdusSalam:2020ywo}.

The analytics derived in this chapter hold for models with an arbitrary number of complex structure moduli at large complex structure. However, one should keep in mind that when the number of moduli is large, the flux-induced contribution to the D3-brane tadpole may go out of control as proposed by the Tadpole Conjecture \cite{Bena:2020xrh,Bena:2021wyr}. In the setup of our simple Ansatz, it is worth noticing that our estimates for the flux-induced tadpole $\Nf$ are in the same footing as the solutions discussed in \cite{Plauschinn:2021hkp,Lust:2021xds,Grimm:2021ckh,Grana:2022dfw}. This is because, on the one hand, the Ansatz forces the flux quanta $f_A^i$ to be non-zero and to have a same common sign for the saxionic vevs to be well-defined. On the other hand, the constraint \eqref{IIB-hBicond} on the fluxes $h_i^B$ also imposes these quanta to be non-zero and have the same sign, such that $\Nf$ is a generically a sum of $h^{2,1}$ positive terms. As a consequence, the tadpole contribution indeed grows with the number of moduli in this context. However, we cannot say much more in this sense for the more involved Ansatz \eqref{IIB-eq:ansatz} where flux quanta are less restricted or even for solutions outside this generic Ansatz.

We should also point out that in our numerical analysis we are using a model where effectively only two moduli play the game thanks to a consistent truncation, and thus, small tadpoles can be achieved there without too much tinkering. This is also in line with \cite{Lust:2022mhk}, where a similar reasoning is applied to F-theory compactifications built at loci of discrete symmetry groups of the moduli space. Even though the tadpole conjecture is generically very sound, it is also true that such symmetric models may allow for non-generic solutions where the tadpole is small. This idea was further explored and tested in the context of F-theory in \cite{Braun:2023pzd}. For the cases considered in that reference, complex structure moduli stabilization by fluxes that have low tadpole charge could only be realized at special points in moduli space associated to large gauge symmetries.

\ifSubfilesClassLoaded{%
\bibliography{biblio}%
}{}

\end{document}


\part[\textcolor{Teja}{Conclusions}]{\scshape \textcolor{Teja}{\huge Conclusions}}
\label{part: conclusions}

\fancyhf{}
\renewcommand{\chaptermark}[1]{\markboth{#1}{}}
\renewcommand{\sectionmark}[1]{\markright{#1}}
\fancyhf[EHL]{\textit{\thechapter. \nouppercase{\leftmark}}}
\fancyhf[OHR]{}
\fancyhf[EFC,OFC]{\thepage}

\graphicspath{{Images/Type IIB at LCS}}

\ifSubfilesClassLoaded{%
\tableofcontents
}{}

\setcounter{chapter}{9}
\chapter{Conclusions and final thoughts}
\label{ch: conclusions}

Flux compactifications are one of the cornerstones in the process of building phenomenologically viable string theory models. In general, the effect of  background fluxes is two-fold. On the one hand, they generate a potential that stabilizes the moduli of the Calabi--Yau orientifold compactification, yielding families of supersymmetric and non-supersymmetric vacua. On the other hand, they generate a warp factor, a varying dilaton and deform the background away from the Calabi--Yau metric. In this thesis we have analyzed these effects in different contexts using the bilinear formalism of the 4-dimensional effective potential as a guiding thread that ties all our results together. 

After introducing in chapter \ref{ch: basics} the basic concepts of String Theory, in chapter \ref{ch: calabi-yau} we presented the geometrical tools commonly used in string compactifications and summarized the main properties of the bilinear formalism. Thus, we observed that the effective potential generated in Type IIA compactifications with NSNS and RR fluxes can be decomposed in terms of monodromy invariant flux-axion polynomials and a bilinear operator that only depends on the saxions and the internal geometry. The simplifying power of this expression was used to great effect in \cite{Marchesano:2019hfb}, allowing to classify the vacua of 4-dimensional massive type IIA into several families, one of which was supersymmetric. The latter was later uplifted to a 10-dimensional solution in \cite{Marchesano:2020qvg} going beyond the smearing approximation through a perturbative expansion in terms of the string coupling. 

The initial part of this thesis aimed to build on top of the results of the two aforementioned papers in the framework of Type IIA. 

First, in chapter \ref{ch: systematics}  we proved that the bilinear structure of the scalar potential is preserved in more general cases when geometric and non-geometric fluxes are also present. We analyzed the flux invariants that appear in type IIA Calabi--Yau, geometric and non-geometric flux compactifications and discussed their role  in determining the vacuum expectation values of the saxions at the minima of the potential.  Introducing an Ansatz motivated by the goal of finding metastable de Sitter vacua, we studied the equations of motion of Type IIA with metric fluxes. Despite our initial intentions, we found that only AdS vacua were allowed, both in SUSY and non-SUSY setups. In the process, we generalized several results from the literature, like \cite{Camara:2005dc, Dibitetto:2011gm, Hertzberg:2007wc,Flauger:2008ad}. We also discussed the stability of the non-SUSY solutions and concluded that a sizeable subset of them was perturbatively stable. Finally, we searched for scale separation in the generic branch of our Ansatz without success.

All these results demonstrate that analyzing the bilinear form of the scalar potential provides a systematic strategy to determine the vacua of this class of compactifications, overarching previous results in the literature. To obtain a clear overall picture it would be important to generalize our analysis in several directions. Regarding scale separation, it is necessary to point out the existence of another set of solutions that was not considered throughout most of the discussion of chapter \ref{ch: systematics} and which has the potential to generate scale separated AdS vacua, like those found in \cite{Cribiori:2021djm}. Thus it would be interesting to study such set of branches of vacua, found and omitted when refining the original, Ansatz and see what new properties they display. It would also be useful to employ other on-shell F-term Ansatz beyond  \eqref{sys-solsfmax} that also guarantee vacua metastability. Indeed, our analysis of the Hessian shows that, for certain geometric flux compactifications, perturbative stability occurs for a very large region of the parameter space of our F-term Ansatz, and it would be important to determine how general this result is. Lastly, a natural extension of our results would be to  consider in more detail the role of the D-term potential and the non-geometric fluxes as elements capable of generating de Sitter vacua.

Second, in chapter \ref{ch: bionic} we turned our attention to another interesting aspect of the analysis performed by \cite{Marchesano:2019hfb}: the existence of perturbatively stable non-SUSY DGKT-like vacua.  Turning off metric fluxes, we focused on one of  the non-SUSY branches, characterized by the relation $G^4_{\cancel{SUSY}}=-G^4_{SUSY}$ (non-susy A1-S1 branch in table \ref{cy-table: summary ads vacua}). Taking inspiration from the uplift of the supersymmetric solution performed in \cite{Marchesano:2020qvg}, we provided an analogous description for this non-SUSY branch and compared both uplifts. We observed that in the smearing approximation configurations with D4 or D8 domain walls had always $Q\leq T$, both for SUSY and non-SUSY solutions, and thus decays are at most marginal. However, going beyond the smearing approximation, we found that bound states of D8s  wrapping the internal manifold with D6s wrapping internal 3-cycles can have $Q>T$ in the case of non-SUSY backgrounds. These kind of objects, called BIons, can therefore yield perturbative instabilities that satisfy the Swampland conjectures. Their behaviour was later studied in explicit examples using toroidal geometries in chapter \ref{ch: membranes}. There, we derived a general formula to compute the corrections to the smearing approximation in a general toroidal orbifold and observed a possible tension with the Weak Gravity Conjecture depending on the particular distribution of the D6s along the inner manifold. These problems have been recently addressed in \cite{Marchesano:2022rpr}. In this work, the authors used our results (presented in \cite{Marchesano:2021ycx, Casas:2022mnz}) as a stepping stone to build more exotic bound states of branes which are threaded by non-diluted worldvolume fluxes. In those new bound states, additional terms dominate over the bionic corrections, providing configurations with $Q>T$ as predicted by the WGC.

This line of research has created many paths that lead in different directions. On the one hand, there are other families of non-supersymmetric $AdS_4$ vacua (dubbed S2 in table \ref{cy-table: summary ads vacua}) whose uplift and non-perturbative stability still need to be studied. On the other hand, the stability of non-SUSY AdS remains an open question in configurations with D4 domain walls or D8 domain walls where no D6 branes are present.

The second part of the thesis aimed to expand the bilinear formulation of the scalar potential to include F-theory and Type IIB and explore the vacua structure of the 4-dimensional effective theories that result from flux compactifications in such setups. We started in chapter \ref{ch: Fintro} with a review of these two theories, where we emphasized the usefulness of F-theory as a framework to understand string compactifications and their vacua due to its connections through dualities to most String Theory constructions. Then, in chapter \ref{ch: Ftheory}, we aimed to improve the analytical control over F-theory flux potentials and their vacua in order to better address questions regarding the consistency of these solutions. To do so, we considered Calabi-Yau four-folds which are elliptic fibrations over a three-fold base and work with the standard Gukov-Vafa-Witten superpotential in the Large Complex Structure region. Through the use of homological mirror symmetry  we found that the bilinear structure of the scalar potential in type IIA also arises in the complex structure sector of F-theory, a behaviour that is preserved even when we include polynomial corrections (hence extending our analysis to regions where the complex structure saxions are only moderately large, so that the instanton corrections can still be neglected). Expanding to linear order in the polynomial corrections, we expressed the equations of motion of the complex structure moduli in a compact way  and discussed the restrictions that the D3 brane tadpole constraint imposes on the potential choices of flux vacua. This led us to find two main families of solutions. In the first, more generic one, moduli stabilization can be achieved once the polynomial corrections are included  and the saxionic vacuum expectation values are bounded by the choice of fluxes. In the second one, that arises when one of the saxions only appears linearly in the expression for the volume of the inner mirror manifold (and hence named the linear scenario), moduli stabilization can be achieved at leading order, the vacuum expectation values are unbounded and there is a single contribution to the tadpole. Therefore, the latter was in tension with the Tadpole Conjecture \cite{Bena:2020xrh}. This conflict was addressed in \cite{Lust:2021xds, Plauschinn:2021hkp, Grimm:2021ckh}, where it was argued using the statistical data of \cite{Demirtas:2018akl} that the pair of fluxes that contributes to the tapdole cannot be chosen arbitrarily but need to scale with the number of moduli in order to remain in the Large Complex structure regime, making the result compatible with the conjecture. 

Lastly, in chapter \ref{ch: typeIIB} we particularized our F-theory results to Type IIB and found that, in the context of the first family of flux quanta mentioned above, the superpotential becomes quadratic in the axio-dilaton and complex structure fields. In terms of this new bilinear form, we could work with the equations of motion at all orders in polynomial corrections and we were able to classify the possible solutions into three classes: a generic non-supersymmetric branch of solutions and two more constraint branches (one supersymmetric and the other non-supersymmetric) associated to the cases in which the bilinear became singular.   This allowed for a clear discussion on the structure of vacua  and, choosing a particular Ansatz for the generic non-SUSY case, enabled us to provide the explicit mass spectrum in terms of the flux quanta. Meanwhile, we also proved that SUSY solutions always display flat directions at this level of approximation.  Through that analysis, we were thus able to highlight the role that polynomial corrections play in moduli stabilization.

The analysis presented in these two chapters offers some avenues to follow in the future: first of all it would be interesting to calculate the precise mass spectra for the F-theory as we did for the Type IIB case to further verify the hierarchy between the masses of the fields in the complex structure sector. Second, one could try to generalize our explicit expressions to other asymptotic regions in the four-fold complex structure moduli space, as classified in \cite{Grimm:2019ixq}.  As an initial step, one could consider infinite distance limits that involve intersections with divisors corresponding to conifold-like singularities, like it has been recently considered in \cite{Demirtas:2020ffz, Alvarez-Garcia:2020pxd} for Type IIB models. It would also be important to address the stabilization of the Kähler sector, which would likely require considering instanton corrections. 

Finally, regarding the Tadpole Conjecture, one could try to extend the proof found in \cite{Grana:2022dfw} beyond the strict asymptotic limits to include polynomial corrections as the ones we considered.  This regime was tested in \cite{Coudarchet:2023mmm} for a Type IIB toroidal orbifold example and it was found that the bound predicted by the tadpole conjecture was saturated for flux configurations that displayed several symmetries.  The relation with points of high symmetry  was also explored in \cite{Lust:2022mhk}, where examples were found in tension with the conjecture in the bulk of the moduli space of F-theory. It would therefore be very interesting to see whether the Tadpole Conjecture breaks when we move away from the strict asymptotic limit and how. Even if it is not satisfied in general, this conjecture provides a powerful insight on the efficacy and limitations of moduli stabilization techniques. It also points to a trend that shows that, contrary to naive expectations, the discretum of type IIB/F-theory flux vacua should be dominated by Calabi–Yau manifolds with a small complex structure sector.

This thesis has improved our understanding of the moduli stabilization process and the vacua structure of 4-dimensional effective theories arising from flux compactifications in different String regimes. Throughout this journey, we have been able to test several Swampland conjectures,  providing useful insights into the properties and limits of the String Landscape. We conclude by restating that flux compactifications constitute an extremely valuable framework in which to study the consequences and predictions of String Theory. Their analysis brings us closer to achieving a description that reflects the nature of quantum gravity in the observable Universe and we hope to keep contributing to the growth of the subject in the future.

\newpage

\chapter*{\vspace{-3.5cm}Conclusiones y comentarios finales}
\vspace{-0.5 cm}

Las compactificaciones con flujos son una de las piedras angulares en el proceso de construcción de modelos de Teoría de Cuerdas fenomenológicamente viables. En general, el efecto de los flujos es doble. Por un lado, generan un potencial que estabiliza los módulos de compatificaciones sobre orientifolds de variedades Calabi--Yau, dando lugar a familias de vacíos supersimétricos y no supersimétricos. Por otro lado, generan un factor de acoplo no trivial entre el espacio interno y externo, un dilatón variable y deforman la geometría alejándola de la métrica de Calabi--Yau. En esta tesis hemos analizado dichos efectos en diferentes contextos, utilizando el formalismo bilineal del potencial efectivo en 4 dimensiones como el hilo conductor que une todos nuestros resultados. 

Tras introducir en el capítulo \ref{ch: basics} los conceptos básicos de la Teoría de Cuerdas, en el capítulo \ref{ch: calabi-yau} presentamos las herramientas geométricas comúnmente utilizadas en las compactificaciones de cuerdas y resumimos las principales propiedades del formalismo bilineal. De este modo, observamos que el potencial efectivo generado en compactificaciones de Tipo IIA con NSNS y flujos RR puede descomponerse en términos de polinomios invariantes bajo monodromías con los axiones y los cuantos de flujo como variables y un operador bilineal que sólo depende de los saxiones y de la geometría interna. El poder simplificador de esta expresión se utilizó con gran efecto en \cite{Marchesano:2019hfb} permitiendo clasificar los vacíos de la teoría tipo IIA masiva en 4 dimensiones en varias familias, una de las cuales era supersimétrica. Esta última fue posteriormente elevada a una solución de 10 dimensiones en \cite{Marchesano:2020qvg}, mejorando la aproximación de \textit{smearing} mediante una expansión perturbativa en términos de la constante de acoplamiento.

La parte inicial de esta tesis tuvo como objetivo ampliar los resultados de estos dos trabajos en el marco de la teoría Tipo IIA. 

En primer lugar, en el capítulo \ref{ch: systematics} demostramos que la estructura bilineal del potencial escalar se conserva en casos más generales cuando también están presentes flujos geométricos y no geométricos. Analizamos las combinaciones invariantes de flujos que aparecen en las compactificaciones sobre variedades Calabi--Yau, incluyendo los casos con flujos geométricos y no geométricos, y discutimos el papel que estos invariantes desempeñan en la determinación de los valores esperados de los saxiones en los mínimos del potencial.  Introduciendo un Ansatz motivado por el objetivo de encontrar vacíos de Sitter metaestables, estudiamos las ecuaciones de movimiento de Tipo IIA con flujos métricos y descubrimos que, a pesar de nuestras intenciones iniciales, sólo es posible obtener vacíos AdS, tanto en configuraciones SUSY como no-SUSY. En el proceso, generalizamos varios resultados de la literatura como \cite{Camara:2005dc, Dibitetto:2011gm, Hertzberg:2007wc,Flauger:2008ad}. También discutimos la estabilidad de las soluciones no-SUSY y concluimos que un subconjunto considerable de ellas era perturbativamente estable. Por último, buscamos sin éxito  separación de escalas en la rama genérica de nuestro Ansatz.

Todos estos resultados demuestran que el análisis de la forma bilineal del potencial escalar proporciona una estrategia sistemática para determinar los vacíos de esta clase de compactificaciones, ampliando resultados anteriores en la literatura. Huelga decir que, para obtener una imagen global clara, sería importante generalizar nuestro análisis en varias direcciones. En cuanto a la separación de escalas, es importante señalar la existencia de otra familia de soluciones asociada a nuestro Ansatz que no fue considerada en la mayor parte de la discusión del capítulo \ref{ch: systematics} y que tiene el potencial de generar vacíos AdS con separación de escalas, como los encontradas en \cite{Cribiori:2021djm}. Por tanto, sería interesante analizar este otro conjunto de ramas de vacíos, encontrado y omitido al refinar el Ansatz original, y ver qué nuevas propiedades presenta. También sería útil buscar otros Ansatzs distintos a \eqref{sys-solsfmax} para los F-terms on-shell  que también garanticen la metaestabilidad de los vacíos. De hecho, nuestro análisis del hessiano muestra que, para ciertas compactificaciones con flujos métricos, la estabilidad perturbativa tiene lugar para una región muy grande del espacio de parámetros de nuestro F-term Ansatz, y sería importante determinar cómo de general es este resultado. Por último, una extensión natural de nuestros resultados sería considerar con más detalle el papel del D-potencial y los flujos no geométricos como elementos potencialmente capaces de generar vacíos de Sitter.

En segundo lugar, en el capítulo \ref{ch: bionic} dirigimos nuestra atención a otro aspecto interesante del análisis realizado por \cite{Marchesano:2019hfb}: la existencia de vacíos perturbativamente estables no supersimétricos del tipo DGKT.  Ignorando los flujos métricos, nos centramos en una de las ramas no-SUSY, caracterizada por la relación $G^4_{\cancel{SUSY}}=-G^4_{SUSY}$ (rama no-SUSY A1-S1 en la tabla \ref{cy-table: summary ads vacua}). Inspirándonos en la elevación de la solución supersimétrica realizada en \cite{Marchesano:2020qvg}, proporcionamos una descripción análoga para esta rama no-SUSY y comparamos ambos resultados. Observamos que en la aproximación \textit{smearing} las configuraciones con paredes de dominio D4 o D8 tienen siempre $Q\leq T$, tanto para las soluciones SUSY como para las no-SUSY, y por tanto los decaimientos son a lo sumo marginales. Sin embargo, yendo más allá de la aproximación \textit{smearing} encontramos que estados ligados de D8s envolviendo el espacio interno junto con D6s envolviendo 3-ciclos internos pueden tener $Q>T$ en el caso de vacíos no-SUSY. Este tipo de objetos, denominados BIones, pueden producir inestabilidades perturbativas que satisfagan las conjeturas de la Ciénaga. Su comportamiento se estudió posteriormente en ejemplos explícitos utilizando geometrías toroidales en el capítulo \ref{ch: membranes}. De este modo, derivamos una fórmula general para calcular las correcciones a la aproximación  \textit{smearing} en un orbifold toroidal genérico y observamos una posible tensión con la Conjetura de la Gravedad Débil dependiendo de la distribución particular de las D6s a lo largo de la varieadad interna. Estos problemas se han tratado recientemente en \cite{Marchesano:2022rpr}. En dicho artículo, los autores utilizaron nuestros resultados (presentados en \cite{Marchesano:2021ycx, Casas:2022mnz}) como punto de partida para construir estados ligados  más exóticos constituidos por branas recorridas por flujos de \textit{worldvolume} no diluidos. En estos nuevos estados ligados dominan términos adicionales sobre las correcciones biónicas, proporcionando configuraciones con $Q>T$, como predice la WGC.

Esta línea de investigación ha creado muchos caminos que se abren en distintas direcciones. Por un lado, existen otras familias de vacíos $AdS_4$ no supersimétrios (denominadas S2 en la tabla \ref{cy-table: summary ads vacua}) cuyo levantamiento a 10 dimensiones y estabilidad no-perturbativa aún no han sido estudiados. Por otro lado, la estabilidad de AdS no-SUSY sigue siendo una cuestión abierta en configuraciones con paredes de dominio D4 o paredes de dominio D8 en las que no hay presencia de branas D6.

El objetivo de la segunda parte de la tesis fue extender la formulación bilineal del potencial escalar más allá del Tipo IIA y a través de ella considerar la estructura de los vacíos de las teorías efectivas de 4 dimensiones que surgen de la compactificación con flujos en teoría F y en Tipo IIB. En el capítulo \ref{ch: Fintro} comenzamos  con una revisión de estas dos teorías, donde enfatizamos la utilidad de la teoría F como marco en el que entender  compactificaciones y sus vacíos gracias a sus conexiones a través de dualidades con la mayoría de las construcciones de la Teoría de Cuerdas. Posteriormente, en el capítulo \ref{ch: Ftheory}, nos propusimos mejorar el control analítico sobre los potenciales con flujos de la teoría F y sus vacíos con el fin de abordar mejor las cuestiones relativas a la consistencia de estas soluciones. Para ello, consideramos variedades Calabi-Yau de 8 dimensiones que fuesen fibraciones elípticas sobre una base 6-dimensional Calabi-Yau y trabajamos con el superpotencial  de Gukov-Vafa-Witten estándar en la región de Gran Estructura Compleja. Mediante el uso de la simetría especular homológica encontramos que la estructura bilineal del potencial escalar en la teoría tipo IIA también surge en el sector de estructura compleja de la F-teoría, un comportamiento que se conserva incluso cuando incluimos correcciones polinómicas (extendiendo así nuestro análisis a regiones donde los saxiones de estructura compleja son sólo moderadamente grandes, de modo que las correcciones instantónicas aún pueden despreciarse). Expandiendo hasta el orden lineal en las correcciones polinómicas, expresamos las ecuaciones de movimiento de los módulos de estructura compleja de una forma compacta y discutimos las restricciones que la relación \textit{Tadpole} de la brana D3 impone a las posibles elecciones de cuantos de flujo. Esto nos llevó a encontrar dos familias principales de soluciones. En la primera, más genérica, la estabilización de los módulos puede lograrse una vez que se incluyen las correcciones polinómicas y los valores de vacío esperados para el sector saxiónico están acotados por la elección de los flujos. En la segunda familia, que surge cuando uno de los saxiones sólo aparece linealmente en la expresión para el volumen de la variedad especular (y de ahí que se denomine escenario lineal), la estabilización de los módulos puede lograrse a orden cero, los valores de vacío esperados no están acotados y existe una única contribución al \textit{tadpole}. Por lo tanto, este último escenario se halla en tensión con la Conjetura \textit{Tadpole} \cite{Bena:2020xrh}. Este conflicto se abordó en \cite{Lust:2021xds, Plauschinn:2021hkp, Grimm:2021ckh}, donde se argumentó usando los datos estadísticos de \cite{Demirtas:2018akl} que la pareja de flujos que contribuye al \textit{Tadpole} no puede elegirse arbitrariamente, sino que necesita escalar con el número de módulos para permanecer en el régimen de Gran Estructura Compleja, haciendo que el resultado sea compatible con la conjetura. 

Finalmente, en el capítulo \ref{ch: typeIIB} particularizamos nuestros resultados de teoría F a la teoría tipo IIB y descubrimos que, en el contexto de la primera familia de cuantos de flujo mencionada anteriormente, el superpotencial se vuelve cuadrático en el axiodilatón y en los campos de estructura compleja. En términos de esta nueva forma bilineal, pudimos trabajar con las ecuaciones de movimiento a todos los órdenes en las correcciones polinómicas y pudimos clasificar las posibles soluciones en tres clases: una rama de soluciones genérica no supersimétrica y otras dos ramas restringidas (una supersimétrica y otra no supersimétrica) asociadas a los casos en los que la forma bilineal se volvía singular.   Esto permitió una discusión clara sobre la estructura de los vacíos y, eligiendo un Ansatz particular para el caso genérico no-SUSY, nos permitió proporcionar el espectro de masas explícito en términos de los cuantos de flujo. Por otro lado, también demostramos que las soluciones SUSY siempre muestran direcciones planas a este nivel de aproximación.  A través de ese análisis, pudimos así poner de relieve el papel que desempeñan las correcciones polinómicas en la estabilización de los módulos.

El análisis presentado en estos dos capítulos ofrece varios caminos a seguir en el futuro: en primer lugar, sería interesante calcular los espectros de masas precisos para la teoría F, como hicimos para el caso de la Tipo IIB, con el fin de verificar en un contexto más general la jerarquía entre las masas de los campos en el sector de estructura compleja. En segundo lugar, se podría intentar extender nuestras expresiones explícitas a otras regiones asintóticas en el espacio de moduli de estructura compleja de variedades Calabi-Yau de 8 dimensiones, clasificadas en \cite{Grimm:2019ixq}.  Como paso inicial se podrían considerar límites de distancia infinita que impliquen intersecciones con divisores correspondientes a singularidades tipo conifold, como se ha considerado recientemente en \cite{Demirtas:2020ffz, Alvarez-Garcia:2020pxd} para modelos Tipo IIB. También sería importante abordar la estabilización del sector de Kähler, lo que probablemente requeriría considerar correcciones instantónicas. 

Por último, con respecto a la conjetura \textit{Tadpole}, se podría intentar extender la demostración hallada en \cite{Grana:2022dfw} más allá de los límites asintóticos estrictos para incluir correcciones polinómicas como las que hemos considerado.  Este régimen se testó en \cite{Coudarchet:2023mmm} para un ejemplo de orbifold toroidal de Tipo IIB y se encontró que el límite predicho por la conjetura \textit{Tadpole} estaba saturado para configuraciones de flujos que presentaban varias simetrías.  La relación con puntos de alta simetría también se exploró en \cite{Lust:2022mhk}, donde se encontraron ejemplos en tensión con la conjetura en el interior del espacio de moduli de la teoría F. Por lo tanto, sería muy interesante ver si la Conjetura \textit{Tadpole} se rompe cuando nos alejamos del límite asintótico estricto y cómo. Incluso si no se satisface en general, esta conjetura proporciona una poderosa visión sobre la eficacia y los límites de las técnicas de estabilización de módulos. También apunta a una tendencia que muestra que, contrariamente a la intuición, el espacio discreto de vacíos con flujos de las teorías tipo IIB/F debería estar dominado por variedades de Calabi-Yau con un  sector pequeño de estructura compleja.

En esta tesis hemos mejorado nuestra comprensión del proceso de estabilización de los módulos y de la estructura de los vacíos de teorías efectivas de 4 dimensiones que surgen de compactificaciones con flujos en diferentes regímenes de Teoría de Cuerdas. A lo largo de este viaje, hemos sido capaces de contrastar varias conjeturas de la Ciénaga, proporcionando ideas útiles sobre las propiedades y los límites del Paisaje de Cuerdas. Concluimos reafirmando que las compactificaciones con flujos constituyen un marco extremadamente valioso para estudiar las consecuencias y predicciones de la Teoría de Cuerdas. Su análisis nos acerca a la consecución de una descripción que refleje la naturaleza de la gravedad cuántica en el Universo observable y esperamos seguir contribuyendo al desarrollo del campo en el futuro.

\ifSubfilesClassLoaded{%
\bibliography{biblio}%
}{}

\end{document}


\part[\textcolor{Teja}{Appendices}]{\scshape \textcolor{Teja}{\huge Appendices}}
\label{part: Appendix}

\fancyhf{}
\renewcommand{\chaptermark}[1]{\markboth{#1}{}}
\renewcommand{\sectionmark}[1]{\markright{#1}}
\fancyhf[EHL]{\textit{\thechapter. \nouppercase{\leftmark}}}
\fancyhf[OHR]{\textit{\thesection. \nouppercase{\rightmark}}}
\fancyhf[EFC,OFC]{\thepage}

\appendix

\ifSubfilesClassLoaded{%
\tableofcontents
}{}

\chapter{Notation and Conventions}
\label{ch: ap conventions}

In this appendix, we will detail the conventions for the basic differential geometry operations used in the thesis. 

Let $M$ be a manifold of dimension $m$ and $\alpha_r$ and $r$-form over that manifold. In a basis $\{dx^i\}$ of $T^*M$, $\alpha_r$ admits the expansion
\begin{equation}
    \alpha_r= \frac{1}{r!}\alpha_{1,\dots,r}dx^1\wedge\dots \wedge dx^r\,.
\end{equation}

\textbf{Interior product}

The interior product is an operation associated to a section  of the tangent bundle  of the manifold, $X$, that acts over forms as follows
\begin{equation}
    \iota_X: \Omega^s(M)\rightarrow  \Omega^{s-1}(M)\,.
\end{equation}
More specifically, let $X$ be a vector field and $\omega \in \Omega^s(M)$. The action of the interior product is defined as 
\begin{equation}
    \iota_X \omega(x_1,\dots ,x_{s-1})\equiv \omega(X,x_1,\dots, x_{s-1})=\frac{1}{(s-1)!}X^\mu \omega_{\mu,\nu_1,\dots ,\nu_{s-1}}dx^{\nu_1}\wedge\dots\wedge dx^{\nu_{s-1}}\,.
\end{equation}
This interior product can be easily extended to act as a general product between antisymmetric tensors and forms. Let $A$ be an antisymmetric $(r,0)$-tensor. Then $A$ admits a decomposition of the form $A=1/r! A^{\mu_1\dots\mu_r}\partial_{x_{\mu_1}}\wedge\dots\wedge\partial_{x_{\mu_r}}$ and we define its product with a s-form $\omega$ as 
\begin{equation}
    \iota_A \omega=\begin{cases}
        \frac{1}{r!}A^{\mu_1\dots \mu_r}\iota_{x_{\mu_r}}(\dots\iota_{x_{\mu_1}}(\omega)\dots)\phantom{\partial_{x_{s+1}}\wedge \dots \wedge \partial_{x_{r}}}\, \qquad  {\rm if \, } r\leq s\,\\
        \frac{1}{r!} A^{\mu_1\dots \mu_r}\iota_{x_{\mu_s}}(\dots\iota_{x_{\mu_1}}(\omega)\dots)\partial_{x_{s+1}}\wedge \dots \wedge \partial_{x_{r}}\,\qquad {\rm if \, } r> s\,
    \end{cases}
\end{equation}

Finally, the interior product induces a product inside the space of forms. For $r<s\in \mathbb{N}$, this product is given by
\begin{equation}
    \begin{array}{ccccc}
         \cdot: \Omega^r & \times & \Omega^s&\rightarrow& \Omega^{s-r}\\
         \hspace{0.3cm} \alpha &\cdot &\omega &\mapsto & \iota_{\hat{\alpha}}\omega
    \end{array}
    \label{eq: interior product of forms}
\end{equation}
where $\hat{\alpha}$ is the dual $(r,0)$-tensor to the r-form $\alpha$. If the manifold is endowed with a metric $g$, the duality can be made explicit through
\begin{equation}
    \hat{\alpha}=\frac{1}{r!}\alpha_{\mu_1\dots \mu_r}g^{\mu_1\nu_1}\dots g^{\mu_r\nu_r}\frac{\partial}{\partial x^{\nu_1}}\wedge\dots \wedge \frac{\partial}{\partial x^{\nu_r}}\,.
\end{equation}

\textbf{Hodge star}

If the manifold has a metric $g$, we can define the Hodge star operator in $m$ dimensions acting on a $r$ form $\alpha_r$ by
\begin{equation}
    \star_m\alpha_r=\frac{1}{r!(m-r)!}\sqrt{|g|}\epsilon_{i_1\dots i_m}\alpha^{i_{m-r+1}\dots i_m}dx^{i_1}\wedge\dots \wedge dx^{i_{m-r}}\,.
\end{equation}
This induces an inner product of $r$ forms as follows
\begin{equation}
(\alpha,\beta)\equiv \sigma \int_M \star \alpha\wedge \beta =\int_M (\alpha\cdot \beta)\textrm{dvol}_m\,,
\label{ap.conv-eq: hodge product}
\end{equation}
where $\sigma=\pm 1$ depending on the conventions relating the volume form and the Hodge star action and
\begin{equation}
    \alpha\cdot\beta\equiv \frac{1}{r!}\alpha_{i_1\dots i_r}\beta^{i_1\dots i_r}\,.
\end{equation}
For the particular case $\alpha=\beta$, we denote $(\alpha,\alpha)\equiv |\alpha|^2$.

It is important to note that, for the sake of adhering to standard conventions of type IIA and F-theory/Type IIB, we have made changes in our choice of volume form and calibrations between parts II and part III of this thesis. The most important relations to consider are summarized in table \ref{ap.conv-table: calibrations}.

\begin{table}[htbp]
\centering
\begin{tabular}{|c|c|}
\hline
Part II    & Part III    \\ \hline
$\sigma=1$ & $\sigma=-1$ \\
$e^{-J_c}$ & $e^{J_c}$   \\ \hline
\end{tabular}
\caption{Different conventions for the Hodge dual - volume relation  \eqref{ap.conv-eq: hodge product} and the calibrations (see appendix \ref{ch: ap complex geometry}) between the two main blocks of the thesis. }
\label{ap.conv-table: calibrations}
\end{table}

\ifSubfilesClassLoaded{%
\bibliography{biblio}%
}{}

\end{document}

\ifSubfilesClassLoaded{%
\tableofcontents
}{}

\chapter{Complex Geometry}
\label{ch: ap complex geometry}

In this appendix, we will provide a useful mathematical framework for analyzing the characteristics of the compact space: group structures and generalized complex geometry. For a deeper take on the subject we refer the reader to \cite{Koerber:2010bx,Tomasiello:2022dwe}.

\section{Structure Groups}

Given a compact d-dimensional manifold $X$, it is always possible to construct the tangent bundle $TX$ which associates to each point $p\in X$ the tangent vector space at that point $T_p X$. The local description around a point is given by a direct product of the form $X\times T_p X$ (trivialization of the fibration). However, such a decomposition is not valid to describe the global structure. To account for that, transition functions must be defined to build the tangent bundle. These functions establish how the fiber transforms between two patches $U_\alpha$ and $U_\beta$ of the base manifold $X$. Once the tangent bundle is built, one can take a section of this bundle, consisting of a map that assigns an element of the tangent space to every point in $X$.  The set of sections will be denoted by $\Gamma$ and the space of p-forms (space sections of antisymmetric products of the cotangent bundle) by $\Omega^p(X,\mathbb{R})$.

 Let us consider  two patches of the manifold, $U_\alpha, U_\beta \subset X$, with non-empty intersection. For a given point $p\in U_\alpha\cap U_\beta$ there are two descriptions (local trivializations) of the fibration depending on the patch. Each one assigns a different basis to the tangent vector space, i.e. $(p,e_a)$ and $(p,e_a')$. The transition function between both local trivializations at $p$, $t_{\beta\alpha}(p)$, acts on the basis of the tangent space as $e'_a = e_b (t_{\beta\alpha})^b_a \,.$

The set of all transition functions $t_{\alpha\beta}$ between local trivializations at a point $p$ forms a group called structure group. In the most general case, this group is the general linear group $GL(d,\mathbb{R})$. Depending on the particular topological properties of $X$  (like the existence of a globally defined object), the structure group could be reduced to a proper subgroup $G\subset GL(d,\mathbb{R})$ by appropriately choosing the basis of the tangent space at each local trivialization. In that case, the manifold $X$ is said to have a G-structure. 

A basic example is a Riemannian manifold with a globally defined metric. The metric can be used to fix the length of the basis elements at each patch, reducing $GL(d,\mathbb{R})$ to $O(d,\mathbb{R})$. If the manifold is orientable, it can be further reduced to $SO(d)$.

We will focus on two particular G-structures, their relation and refinement: the almost complex structure and the pre-symplectic structure. They will become the fundamental pieces of our compactified space.

\subsection{Almost complex structure}

It is characterized by the existence of a globally defined tensor map
\begin{equation}
    I:TX\rightarrow TX\,,
\end{equation}
that respects the bundle structure and verifies that $I^2=-\mathbb{I}$. Such a map can only be constructed when the dimension of the manifold is even, in which case the structure group is reduced to $GL(d/2, \mathbb{C})$. At a given point $n\in X$, the action of $I$ over $T_p X$ splits the vector space into two  eigenspaces with eigenvalues $+i$ and $-i$. This decomposition is generally dependent on the point $p$. The space of p-forms $\Omega^n(X,\mathbb{R})$ can be refined by distinguishing if the entries of the cotangent space are associated with positive of negative eigenspaces of the complex structure operator, which leads to defining the subspaces $\Omega^{q,r}(X,\mathbb{R})$ with $q+r=n$. 

At a given point, the subspace with positive eigenvalue $L$ is generated by $d/2$ independent 1-forms $\theta^a$ which define a local section on the bundle of $d/2$-forms
\begin{equation}
    \Omega=\theta^1\wedge \dots \wedge \theta^{d/2}\,.
\end{equation}
The basis $\theta^a$ is determined by $I$ up to $GL(d/2,\mathbb{C})$ transformations, which means that $\Omega$ can differ by a complex function from one point of the manifold to another. If $\Omega$ is required to be a global form non-vanishing everywhere, the structure group is further simplified to $SL(d/2,\mathbb{C})$.

In the case in which one can introduce a  basis of holomorphic coordinates $z^a$ on $X$ such that the eigenspace $L$ is spanned at any point by $\{\partial/\partial z^a|a=1,\dots, d/2\}$ and the transformations relating the coordinates between patches are holomorphic, the manifold is said to have a \textit{complex structure}.\footnote{More formally a complex structure is an almost structure group that is integrable, which means that the action of the Lie bracket over two vector fields is closed in the set of sections.}  Then the exterior derivative satisfies the simple property $d\phi\in \Omega^{p+1,q}(X)\oplus \Omega^{p,q+1}(X)$ with $\phi\in \Omega^{p,q}(X)$ and can thus be decomposed into the standard Dolbeault operators $\partial,\bar{\partial}$ \cite{nakahara2003geometry}.

Note that a complex structure does not imply the existence of a globally defined $d/2$-form. The map between overlapping local patches can change $\Omega$ up to a complex factor. Reversely, the existence of a globally defined non-vanishing everywhere $d/2$-form $\Omega$ defines an almost complex structure through $L=\{v\in TX| \iota_v \bar{\Omega}=0\}$ with $\iota_v$ the interior product defined in appendix \ref{ch: ap conventions}, but depending on the behaviour of $d\Omega$ might not generate a full complex structure.

\subsection{Pre-symplectic structure}

It is characterized by a globally defined non-degenerate 2-form $J$, i.e. $J^{d/2}\neq0$. Its existence reduces the structure group to $\rm{Sp}(d,\mathbb{R})$. When $dJ=0$, the structure is called symplectic.\footnote{Formally, it is again equivalent to demanding integrability.}

\subsection{Hermitian metric and \texorpdfstring{$SU(d/2)$}{SU(d/2)}-structure}

Suppose that in the same manifold there is a pre-symplectic and an almost complex structure, so there is a global pre-symplectic form $J\in\Omega^2(X,\mathbb{R})$ and we can locally construct a $d/2$-form $\Omega \in \Omega^{d/2,0}(X)$. Then if the symplectic form satisfies the following compatibility condition with respect to the almost complex structure associated to $\Omega$
\begin{equation}
    I^i_k J_{ij}I^j_l=J_{kl} \Leftrightarrow J\in\Omega^{1,1}(X,\mathbb{R})\,,
    \label{ap.geo-eq: compatibility condition hermitian metric}
\end{equation}
the manifold $X$ admits a hermitian metric defined by
\begin{equation}
    g_{ij}=J_{ik}I_j{}^{k}\,,
    \label{ap.geo-eq: metric from kahler form}
\end{equation}
and the structure group becomes $U(d/2)$. Finally, if the almost complex structure form $\Omega$ is globally defined, decomposable and non-degenerate everywhere (i.e. $\Omega\wedge\bar{\Omega}\neq0$),  the structure group simplifies to $SU(d/2)$. Spaces with $SU(d/2)$-structure and generalizations of thereof will be the focus of the remaining of this section.

In a $SU(d/2)$ structure the compatibility condition \eqref{ap.geo-eq: compatibility condition hermitian metric} propagates to the symplectic and complex structure form, imposing the following relations
\begin{subequations}
    \begin{gather}
        J\wedge \Omega=0\,,\\
        d\rm{vol}_d=\frac{(-1)^{d/2}}{6}J^{d/2}=\left(\frac{i}{2}\right)^{d/2}\Omega\wedge \bar{\Omega}\,.
    \end{gather}
    \label{ap.geo-eq: normalization J Omega}
\end{subequations}

\subsection{Torsion classes and manifold classification}
\label{ap.geo-subsec: torsion classes}

Now, let us consider a 6-dimensional Riemannian manifold with a $SU(3)$-structure. The torsion tensor $T$ can be understood as an element of $\Omega^1(X)\otimes \Lambda^2(X)$, with $\Lambda^2(X)$ the space of 2-dimensional symmetric tensors on $X$.

\begin{tcolorbox}[breakable, enhanced,  colback=uam!10!white, colframe=uam!85!black, title= Holonomy and Torsion]
\begin{small}

The associated hermitian metric defines a connection on the manifold that enables to build a section out of a single point $(p_0,v_0)$ of the tangent bundle through the process of parallel transport. Considering a chart of the manifold $(U,\varphi)$ and an element of the tangent bundle $(p,v)$, a new element of the tangent space can be assigned at each point along a curve  $c(t)$ passing through $p$  by solving the following differential equation
\begin{equation}
   \nabla_c v=0\Rightarrow  \frac{d v^i}{dt}+\Gamma^{i}_{jk}\frac{dx^j (c(t))}{dt}v^k=0\,,
\end{equation}
where $\nabla_c$ represents the covariant derivative along the curve $c$, $x^\nu$ are the coordinates of the chart and we impose the initial conditions $x^\nu(c(t=0))=\varphi(p_0)$, $v^\mu(t=0)=v^\mu_0$. The coefficients $\Gamma^{i}_{jk}$ specify the connection and the compatibility with the metric requires 
\begin{equation}
    \partial_k g_{ij} -\Gamma^l_{ki}g_{lj}-\Gamma^l_{kj}g_{il}=0\,.
\end{equation}
When a vector is moved through parallel transport along a closed curved back to the initial point, it will be transformed via the above equation. The new vector $v'$ will be related to the original vector $v$  through a transformation $v'=Av$. The set of all possible transformations $A$ associated with all closed curves crossing $p$ forms a group known as \textit{holonomy group}. The fact that the connection satisfies the compatibility condition with the metric restricts this group to be $SO(d)$ or a subgroup thereof. 

Two fundamental quantities study how the parallelly transported vectors behave, providing a great insight about the geometry of the manifold: the torsion and the curvature. We will focus on the torsion, which can be defined as
\begin{equation}
\begin{aligned}
    T: \Omega^1(X)\otimes \Omega^1(X)& & \rightarrow   & \hspace{0.2cm}\Omega^1(X)\, \\
        (Y,Z)\hspace{0.6cm} & & \rightarrow & \hspace{0.2cm} \nabla_Y Z\nabla_Z Y-[Y,Z]\,,
\end{aligned}
\end{equation}
where in this case $\nabla_Y$ represents the covariant derivative along the tangent curve to the dual vector of $Y$. From its definition, it can be concluded that the torsion measures the failure to close the parallelogram built from the small displacement vectors   and their images through  parallel transport along the curves induced by each other.

\end{small}
\end{tcolorbox}

Using that $\Lambda^2$ is isomorphic to the Lie algebra $\mathit{so}(6)$ \cite{Chiossi:2002tw} and splitting $\mathit{so}(6)$ as the sum of its $\mathit{su}(3)$ subalgebra and its orthogonal complement, we have $T\in \Omega^1\otimes (\mathit{su}(3)\oplus\mathit{su}(3)^\perp)$. Taking advantage of the $SU(3)$ structure, we can restrict the action of the torsion to $SU(3)$ invariant forms, which gives rise to the intrinsic torsion $T_0$. Using known properties of $SU(3)$ representation theory it is possible to conclude \cite{Grana:2005jc}
\begin{equation}
\begin{aligned}
        T_0\in \Omega_1\otimes\mathit{su}(3)^\perp & = (\mathbf{3}\oplus \bar{\mathbf{3}})\otimes (\mathbf{1}+\mathbf{3}+\bar{\mathbf{3}})\\
        & \begin{array}{cccccccc}       
        =& (\mathbf{1}\oplus\mathbf{1})&\oplus&(\mathbf{8}\oplus\mathbf{8})&\oplus&(\mathbf{6}\oplus\bar{\mathbf{6}})&\oplus &2(\mathbf{3}\oplus\bar{\mathbf{3}})\,.\\
          & W_1   & & W_2 & & W_3 & & W_4,W_5 
        \end{array} 
\end{aligned}
\end{equation}
Therefore, the intrinsic torsion can be described through five torsion classes $W_1,\dots, W_5$ that have a simple interpretation in terms of representations of the $SU(3)$ group. $W_1$ is a complex scalar, $W_2$ is complex $(1,1)$ form, $W_3$ is a real $(1,2)+(2,1)$-form, $W_4$ is a real one form and $W_5$ is a complex $(1,0)$-form. In addition, $W_2$ and $W_3$ satisfy a primitivity condition
\begin{equation}
    W_2\wedge \omega\wedge \omega=W_3\wedge \omega=0\,.
\end{equation}

The torsion classes can be used to systematically classify the different geometries associated with a $SU(3)$ structure depending on the behaviour of the external derivatives of the pre-symplectic and almost complex structure forms. In general, they decompose as follows in terms of $SU(3)$ representations
\begin{subequations}
  \begin{align}
    dJ & =-\frac{3}{2}\Im(\bar{W}_1\Omega)+W_4\wedge J+W_3\,,\\
    d\Omega & = W_1J^2+W_2\wedge J+\bar{W}_5\wedge\Omega \,.
\end{align}  
\label{ap.geo-eq: torsion classes}
\end{subequations}
When $W_1=W_2=0$ the manifold has a complex structure; otherwise $d\Omega$ would not be $(3,1)$-form, breaking the characteristic behaviour of the exterior derivative in complex manifolds. A symplectic manifold requires $dJ=0$ and hence $W_1=W_3=W_4=0$. When a manifold is both complex and symplectic, it is called Kähler and therefore has $W_5$ as the only non-vanishing torsion class. The symplectic form of a Kähler manifold is commonly known as Kähler form and its holonomy group is slightly bigger than the structure group ($U(3)$). When $W_5$ also vanishes, the Kähler manifold has trivial torsion and $SU(3)$ holonomy. This particular subclass of Kähler manifolds are named Calabi-Yau manifolds. In table \ref{ap.geo-table: torsion classes} we present a classification of some of the most important manifolds with $SU(3)$ structure. 

\begin{table}[htp]
\begin{center}
\def\arraystretch{1.5}
\begin{tabular}{c|c}
Vanishing Torsion Classes & Manifold      \\ \hline
$W_1,W_2$                 & Complex       \\
$W_1,W_3,W_4$             & Symplectic    \\
$\re(W_1),\re(W_2), W_3,W_4$             & Half-flat    \\
$W_2,W_3,W_4,W_5$         & Nearly Kähler \\
$W_1,W_2,W_3,W_4$         & Kähler        \\
$W_1,W_2,W_3,W_4,W_5$     & Calabi-Yau   
\end{tabular}
\caption{Classification of $SU(3)$-structure manifolds in terms of torsion classes. }
\label{ap.geo-table: torsion classes}
\end{center}
\end{table}

It is worth noting that the categories used to classify the $SU(3)$ structure manifolds are generally not fully contained inside that set. It is clear, for example, that there can be complex and symplectic manifolds without $SU(3)$ structure. The torsion classes  identify the intersection between those structures and the $SU(3)$ one. More subtle is the Kähler manifold, which, as an independent definition, only requires $U(3)$-structure and a closed pre-symplectic form. Thus, the assumption of a decomposable globally defined non-degenerate 3-form $\Omega$  can be relaxed to the demand of an almost complex structure compatible with the pre-symplectic form. This more general manifold can still be described in terms of the $SU(3)$-classification by considering $\Omega$ as a bundle-valued form\footnote{A global $d/2$ form $\Omega$ is a section of the canonical bundle $\Lambda^{d/2,0}T^*X$ (antisymmetric product of $d/2$ copies of the positive eigenspace associated to the almost complex structures $I$ acting on the cotangent space). A never vanishing section exists if and only if the bundle is trivial or, equivalently, if the first Chern class vanishes. If the bundle is not topologically trivial, $\Omega$ becomes a twisted $d/2$-form where $W_5$ plays the role of a connection ($(d-W_5\wedge)\Omega=0$). } built as a section of $\Lambda^{d/2,0}T^*X \otimes (\Lambda^{d/2,0}T^*X)^{-1}$ whose covariant derivative vanishes \cite{Tomasiello:2022dwe}.

\begin{tcolorbox}[breakable, enhanced, colback=uam!10!white, colframe=uam!85!black, title=Kähler potential]
\begin{small}

A Kähler manifold is endowed with a complex structure, making it possible to define a global basis of holomorphic and antiholomorphic variables ($z^i,\bar{z}^{\bar{i}})$ that diagonalize the almost complex structure operator $I$. Using this basis in the compatibility condition  \eqref{ap.geo-eq: compatibility condition hermitian metric} and \eqref{ap.geo-eq: metric from kahler form}, we can write the pre-symplectic (Kähler) form as
\begin{equation}
    J=ig_{i\bar{j}}dz^{i}\wedge d\bar{z}^{\bar{j}}\,.
\end{equation}
Given that a Kähler manifold requires $dJ=0$, we have
\begin{equation}
    dJ=i\partial_l g_{i\bar{j}} dz^l\wedge dz^i\wedge d\bar{z}^{\bar{j}}+i\partial_{\bar{\lambda}}g_{\mu\bar{\nu}}d\bar{z}^{\bar{l}}\wedge dz^{i}\wedge d\bar{z}^{\bar{j}}\,,
\end{equation}
which means that 
\begin{equation}
    \frac{\partial g_{i\bar{j}}}{\partial z^l}=\frac{\partial g_{l\bar{j}}}{\partial z^i}\,,\qquad \frac{\partial g_{i\bar{j}}}{\partial \bar{z}^{\bar{l}}}=\frac{\partial g_{i\bar{l}}}{\partial \bar{z}^{\bar{j}}}\,.
\end{equation}
Thus we conclude that the metric is locally expressed as the second derivative of a scalar function
\begin{equation}
    g_{i\bar{j}}= \frac{\partial^2 K}{\partial z^i\partial \bar{z}^j}\,.
\end{equation}
Such locally defined function is known as Kähler potential.
\end{small}
\end{tcolorbox}

\section{Spinors and polyforms}

\subsection{Spinors and gamma matrices}

Spinors are elements of the fundamental representation of the Clifford algebra $Cl$. The simplest way to introduce them is by first considering the matrix representation of that algebra, the gamma matrices, and its relation with the spin group $Spin(d)$. The gamma matrices are thus defined imposing the Clifford algebra
\begin{equation}
    \{\gamma_i,\gamma_j\}=g_{ij}\mathbb{I}\,.
\end{equation}
They can be used as building blocks to construct other matrices through antisymmetric products
\begin{equation}
    \gamma_{n_1\dots n_k}\equiv \gamma_{[n_1}\dots \gamma_{n_k]}\,.
\end{equation}
It is not difficult to see that the matrices $-1/2\gamma_{ij}$ verify the algebra $\mathit{so}(d)$, providing an alternative representation $\rho_S$. In this context, spinors are introduced as the elements of the vector space in which the representation $\rho_S$ operates. Under an infinitesimal transformation that acts on a vector as $\delta v^i=\lambda^i{}_j v^j$, it acts on a spinor with $\delta \psi= \rho_s(\lambda)\psi=-\frac{1}{2}\lambda_{ij}\gamma^{ij}$, where the upper indices are raised by the metric. As it is well known, spinors are not rigorously a representation of the group $SO(d)$. An additional minus sign arises on spinorial states when performing a $2\pi$ rotation. Even though they share the same algebra, the associated group $Spin(d)$ is a larger group that constitutes the universal cover of $SO(d)$.

Spinors can be  classified in regards to their image under the action of the matrix $\gamma=(-1)^{d/2}\gamma^1\cdots \gamma^d$. Spinors that behave as eigenstates of $\gamma$ are called Weyl spinors and are divided into two complementary sets that constitute irreducible representations of $\rho_S$. If $\gamma\psi=\psi$, $\psi$ has positive chirality and it is denoted by $\psi_+$. If instead $\gamma\psi=-\psi$, it is said to have negative chirality and represented by $\psi_-$. The representation $\rho_S$ is the direct sum of the positive and negative chirality representations.

Of particular interest are pure spinors, defined as those annihilated by a maximal number ($d/2$) of independent gamma matrices. For the case $d=6$, this condition is easily verified since all Weyl spinors are pure \cite{Tomasiello:2022dwe}.

An $SU(d/2)$ structure can be simply characterized in terms of a metric and a globally defined pure spinor $\eta$. The associated pre-symplectic and globally defined almost complex structure forms are built as follows
\begin{equation}
    J_{ij}=i\eta^{\dagger}\gamma_{ij}\eta\,,\qquad \Omega_{i_1\dots i_{d/2}}=(\eta^{c})^\dagger \gamma_{i_1\dots i_{d/2}}\eta\,,
    \label{ap.geo-eq: J and Omega from spinors}
\end{equation}
where $\eta^{c}$ is the complex conjugate of $\eta$. In $d=6$, the chirality operator $\gamma$ changes sign under complex conjugation and therefore $\eta$ and $\eta^{c}$ have different chirality and are usually denoted by $\eta_+$ and $\eta_-$ respectively.

\subsection{Polyforms and Clifford map}
\label{ap.geo-subsec: polyforms}

A polyform is a formal sum of forms of different dimensions $\alpha=\alpha_1+\dots +\alpha_n$ with $\alpha_i\in \Omega^{k_i}(X)$. The set of polyforms has a group structure and is denoted by $\Omega^\bullet\equiv \oplus_{k=0}^d \Omega^k(X)$.

It is often useful to describe spinor operators in terms of forms. One can do so thanks to the Clifford map, which establishes an isomorphism between polyforms and operators acting on spinors, i.e. a correspondence $\Omega^\bullet\leftrightarrow Cl$ given by \cite{Tomasiello:2022dwe, Koerber:2010bx}
\begin{equation}
    \alpha_k=\frac{1}{k!}\alpha_{m_1\dots m_k}dx^{m_1}\wedge\dots\wedge dx^{m_k}\leftrightarrow \slashed{\alpha}=\frac{1}{k!}\alpha_{m_1\dots m_k}\gamma^{m_1\dots m_k}\,.
\end{equation}
Since they have two spinor indices, the matrices $\gamma^{m_1\dots m_k}$ and their linear functions are usually called bispinors.  

The exterior algebra $\Omega^\bullet$ has the advantage that the product of its elements is much simpler than that of $Cl$. For example, using the definition of the symmetric product and the Clifford algebra, the product of two gamma matrices gives $\gamma^m\gamma^n=\gamma^{mn}+g^{mn}$. Generalizing this result to other products, we can map the left action of gamma matrices on bispinors to the action of a polyform
\begin{equation}
    \overrightarrow{\gamma}^m \leftrightarrow dx^m + g^{mn}\iota_n\,,
\end{equation}
where we denote by $\iota_n$ the interior product with $dx^n$ and the arrow emphasizes that the identification is only valid when the gamma matrix acts on the left.

The action on the right can be derived using the anticommutation relations of gamma matrices and bispinors, obtaining
\begin{equation}
    \overleftarrow{\gamma}^m=(dx^m\wedge - g^{mn}\iota_n)(-1)^{\rm{deg}}\,,
\end{equation}
where $\rm{deg}$ is the operator that counts the degree of the form it is acted upon.

In turn, the Clifford map enables to introduce a double Clifford algebra on the space of forms through the external and internal products. They satisfy $\{dx^i\wedge ,dx^j\wedge \}=\{\iota_i,\iota_j\}=0$ and $\{dx^i\wedge,\iota_j\}=\delta^i_j$. Then, it is possible to define the generalized gamma matrices
\begin{equation}
    \Gamma_A=\{dx^1\wedge, \dots, dx^d,\iota_1,\dots,\iota_d\}\,.
    \label{ap.geo-eq: gamma double clifford}
\end{equation}
They satisfy a Clifford algebra relation with metric 
\begin{equation}
    \mathcal{I}=\left(\begin{array}{cc}
        0 & \mathbb{I}_d \\
        \mathbb{I}_d & 0
    \end{array}\right)\,,
    \label{ap.geo-eq: clifford metric}
\end{equation}
that has signature $(d,d)$. Thus the matrices $\Gamma$ generate $Cl(d,d)$.

In this framework and following the parallelism between forms and spinors, a pure form is defined as one that is annihilated by half of the gamma matrices $\Gamma$. To find the most general expression of a pure form, we should demand it to be annihilated by operator combinations of the form $\iota_i+b_{ij}dx^j\wedge$. Grouping the coefficients $b_{ij}$ into a two form $b$, it is then possible to show  that any pure form can be obtained as a component of a polyform of the following kind \cite{Tomasiello:2022dwe}
\begin{equation}
    \Phi=\alpha_1\wedge \dots \wedge \alpha_k \wedge e^{b}\,,
    \label{ap.geo-eq: pure form expansion}
\end{equation}
where $\alpha_i$ are one-forms and $b$ is a complex two-form.

Given two pure spinors $\eta$ and $\eta^{c}$, the tensor product $\eta\otimes (\eta^{c})^\dagger$ is a bispinor that can be identified through the Clifford map with a pure form. Through careful use of the previously explained properties and the traces of gamma matrices, the bispinor can be expanded in the space of forms as
\begin{equation}
    \eta\otimes\eta^{c}=\sum_{k=0}^d \frac{1}{2^{d/2}k!}((\eta^{c})^\dagger \gamma_{n_k\dots n_1}\eta) dx^{n_1}\wedge \dots \wedge dx^{n_k}\,.
\end{equation}
The above relation is known as Fierz identity.

Returning to the $SU(d/2)$ structure, recall we have a pure spinor $\eta$ and its complex conjugate $\eta^c$, which were used to construct the almost complex structure and pre-symplectic structure. One can build two useful pure forms from these two spinors and relate them to the $SU(d/2)$ forms using \eqref{ap.geo-eq: J and Omega from spinors} and the Fierz identity
\begin{equation}
    \begin{aligned}
    \Phi&\equiv 2^{d/2} \eta\otimes \eta^\dagger =e^{-iJ}\,,\\
    \tilde{\Phi}&\equiv  2^{d/2} \eta\otimes (\eta^{c})^\dagger =\Omega\,.
    \end{aligned}
    \label{ap.geo-eq: pure forms of J and Omega}
\end{equation}

\begin{tcolorbox}[breakable, enhanced,  colback=uam!10!white, colframe=uam!85!black, title=Calibrations]
\small
    An almost calibration form $\phi$ is an k-form on a Riemannian manifold $M$ such that in every point $p\in M$ and every k-dimensional subspace $T$ of the tangent space $T_pM$ 
        \begin{equation}
            \sqrt{\det g|_T}\geq \phi|_T\,,
            \label{ap.geo-eq: calibration inequality}
        \end{equation}
    where $g|_T$ is the restriction of the metric of $M$ to the subspace $T$ and $\phi|_T$ is the component of $\phi$ along the space $T$. The above inequality must always be saturated in some subspace $T$ at every $p\in M$.
        
   A submanifold $\Sigma$ is calibrated if in every point $p\in\Sigma$ the bound \eqref{ap.geo-eq: calibration inequality} is saturated.
   When $\phi$ is closed $(d\phi=0)$, the almost calibration is promoted to a calibration. If that is the case, a calibrated submanifold minimizes the volume within its homology class.

    Using the properties of the gamma matrix and the Clifford map, it can be shown (\cite{Tomasiello:2022dwe}) that the pure forms of the $SU(3)$ structure \eqref{ap.geo-eq: pure forms of J and Omega} calibrate any closed cycle. 
    \begin{itemize}
        \item Even cycles are calibrated by $\Re(e^{i\theta}\Phi)=\Re(e^{i\theta}e^{-iJ})$ with $\theta$ arbitrary.
        \item Odd cycles are calibrated by $\Re(e^{i\theta}\tilde{\Phi})=\Re(e^{i\theta}\Omega)$ where again $\theta$ is arbitrary.
    \end{itemize}
    They are calibrations when the pure forms are closed, and hence when the manifold is a Calabi-Yau.
\end{tcolorbox}

\section{Generalized complex structure}
\label{ap.geo-sec: generalized complex structure}

\subsection{Generalities}

In order to study compactifications with non-trivial flux backgrounds, it will be useful to use the formalism of generalized complex geometry, which was originally introduced in \cite{Hitchin:2003cxu,Gualtieri:2003dx}. It extends the results from the previous section to broader scenarios  by working simultaneously with the tangent and cotangent spaces in the generalized tangent bundle $TX\oplus T^*X$. The motivation to work on this space is based on the observation that the set of vectors and one-forms is naturally equipped with a metric of signature $(d,d)$ and generates a double Clifford algebra through \eqref{ap.geo-eq: gamma double clifford}. Thus, a section of the generalized tangent bundle $\mathbb{X}=(x,\chi)\in TX\oplus T^*X$ acts on a polyform $\Phi$ as 
\begin{equation}
    \mathbb{X}\cdot \Phi=\iota_x \Phi+\chi\wedge\Phi \,,
\end{equation}
and, as a consequence, the generalized tangent space is always naturally endowed with a metric $\mathcal{I}$ with signature $(d,d)$ .

The notion of G-structure can be extended to this new context. A generalized G-structure reduces the group of transition functions on the generalized tangent bundle $TX\oplus T^*X$. In particular, the presence of a metric with signature $(d,d)$ already restricts the structure group from $GL(2d,\mathbb{R})$ to $O(d,d)$. Furthermore, it is possible to see that the volume form  defined by the product described above does not depend on the choice  of orientation, which  reduces the generalized structure group to $SO(d,d)$. 

\begin{tcolorbox}[breakable, enhanced,  colback=uam!10!white, colframe=uam!85!black, title={$SO(d,d)$ generators}]
    \small
    
    The structure group $SO(d,d)$ is generated by elements of the form \cite{Koerber:2010bx}
    \begin{equation}
        \mathcal{O}_A=\left(\begin{array}{cc}
            A & 0 \\
            0 &  (A^T)^{-1}
        \end{array}\right)\,\qquad \mathcal{O}_B=\left(\begin{array}{cc}
            \mathbb{I}_d & 0 \\
             -B & \mathbb{I}_d 
        \end{array}\right)\,,\qquad \mathcal{O}_{\beta}=\left(\begin{array}{cc}
            \mathbb{I}_d & \beta  \\
            0 & \mathbb{I}_d 
        \end{array}\right)\,,
        \label{ap.geo-eq: b and beta generators}
    \end{equation}
    where $A\in GL(d,\mathbb{R})$ is the standard structure group of the tangent bundle, $B$ is a 2-form and $\beta$ is an antisymmetric 2-vector. The action of the generators $\mathcal{O}_B$ (B-transform) and $\mathcal{O}_\beta$ ($\beta$-transformation) over a generalized tangent vector $\mathbb{X}=x+\xi$ is
    \begin{equation}
        \mathcal{O}_B\mathbb{X}=x+(\xi-\iota_X B)\,,\qquad  \mathcal{O}_\beta \mathbb{X}=(x+\iota_\beta \xi)+\xi\,.
    \end{equation}
\end{tcolorbox}

In this context, we can consider a generalized almost complex structure, defined as the real map
\begin{equation}
    \mathcal{J}: TX\oplus T^*X\rightarrow TX\oplus T^*X \quad \rm{with} \quad \mathcal{J}^2=-\mathbb{I}_{2d}\,.
\end{equation}
As expected, given a manifold with an almost complex structure $I$, it is straightforward to construct a generalized almost complex structure taking
\begin{equation}
    \mathcal{J}_I=\left(\begin{array}{cc}
        -I & 0 \\
         0 & I^T
    \end{array}\right)\,.
\end{equation}
One of the advantages of the new formalism resides in the fact that a manifold with a pre-symplectic structure characterized by a pre-symplectic form $J$ can be naturally provided with an almost complex structure as well by taking
\begin{equation}
    \mathcal{J}_J=\left(\begin{array}{cc}
        0 & -J^{-1}  \\
        -J & 0
    \end{array}\right)\,.
\end{equation}
The existence of several distinct generalized almost complex structures over the same manifold leads to considering how they can interact. Motivated by the above example and the compatibility condition \eqref{ap.geo-eq: compatibility condition hermitian metric} required to define a hermitian metric, two generalized almost complex structures $\mathcal{J}_1,\mathcal{J}_2$ are said to be compatible if $[\mathcal{J}_1,\mathcal{J}_2]=0$ and 
\begin{equation}
    \mathcal{G}=\mathcal{I}\mathcal{J}_1\mathcal{J}_2\,,
\end{equation}
is a positive-definite metric in $TX\oplus T^*X$.

Given a non-degenerate\footnote{A pure for is non-degenerate if $\Phi\wedge\bar{\Phi}\neq0$.} pure form, one can associate a generalized almost complex structure through the following identification
\begin{equation}
    \mathbb{X}\cdot \Phi=0\Leftrightarrow \mathcal{I}\mathbb{X}=i\mathbb{X}\,,
\end{equation}
for any section  $\mathbb{X}$ of the generalized tangent bundle. Therefore, the existence of a pure form with such characteristics induces a generalized almost complex structure and reduces the structure group to $SU(d,d)$.

Elaborating on this notion, two pure forms constitute a compatible pure pair if their corresponding induced generalized almost complex structures are compatible and if they have equal norm. A compatible pair reduces the structure group of the generalized tangent bundle to $SU(d/2)\times SU(d/2)$ \cite{Koerber:2010bx,Tomasiello:2022dwe}.

To end this review, we consider the most general way to represent a compatible pair of pure spinors. Recalling that a generic pure form can be written like \eqref{ap.geo-eq: pure form expansion} and demanding initially $b=0$, it is possible to prove that two pure forms are compatible if they can be written in terms of two pure spinors $\eta^1,\eta^2$ as
\begin{equation}
    \Phi_1=2^{d/2} \eta^1\otimes(\eta^2)^\dagger\,,\qquad \Phi_2= 2^{d/2}\eta^1\otimes (\eta^{2c})^\dagger\,,
\end{equation}
with $\eta^{2c}$ the complex conjugate spinor of $\eta^{2}$. A detailed explanation of these results can be found in \cite{Tomasiello:2022dwe} and it relies on the properties of generalized almost complex structure, the Clifford map and the Fierz identity.  To account for the additional degree of freedom associated with the $b$ two-form, one must allow a possible B-transform \eqref{ap.geo-eq: b and beta generators} of the pure forms above. The most general compatible pair is then
\begin{equation}
    \Phi_1=N e^{B}\wedge \eta^1\otimes(\eta^2)^\dagger\,,\qquad \Phi_2=N e^{B}\wedge \eta^1\otimes(\eta^{2c})^\dagger\,,
\end{equation}
with $N$ a normalization factor. When $\eta^1\propto\eta^2$ and $B=0$ we recover the standard $SU(3)$-structure with hermitian metric $g$ and the generalized metric $\mathcal{G}$ is just
\begin{equation}
    \mathcal{G}=\left(\begin{array}{cc}
         g & 0 \\
         0 & g^{-1}
    \end{array}\right)\,.
\end{equation}
In the generic case
\begin{equation}
    \mathcal{G}=\left(\begin{array}{cc}
        1 & B \\
        0 & 1
    \end{array}\right)\left(\begin{array}{cc}
         g & 0 \\
         0 & g^{-1}
    \end{array}\right)\left(\begin{array}{cc}
        1 & 0 \\
        -B & 1
    \end{array}\right)\,.
\end{equation}
It is then possible to conclude that a pair of compatible pure forms are uniquely determined by a hermitian metric $g_{mn}$, a 2-form $B_{mn}$ and a warping factor $e^A$ (through the normalization function $N$). Therefore, they provide an excellent tool to describe compactification in spaces with no trivial NSNS backgrounds.

\subsection{\texorpdfstring{$SU(3)\times SU(3)$}{SU(3)xSU(3)} generalized structure}

In a compact 3-fold with non-trivial holonomy we can have up to two independent globally defined spinors $\eta^1,\eta^2$ (and the conjugates). The $SU(3)\times SU(3)$ generalized structure allows to describe all possible combinations, including the $SU(3)$-structure case, in a unified framework. Let us briefly discuss how it works. For an in-depth look, we refer the reader to \cite{Grana:2005sn, Halmagyi:2009xun, Saracco:2012wc, Tomasiello:2022dwe}.
  
We have seen that each of these two spinors defines a SU(3) structure. Furthermore, we can construct the pure forms
\begin{equation}
    \Phi_+\equiv \eta_+^1\otimes (\eta^2_+)^\dagger\,,\qquad \Phi_-=\eta^1_+\otimes(\eta^2_-)^\dagger\,,
 \end{equation}
which in turn define a $SU(3)\times SU(3)$ structure. A careful analysis of the chirality of the different independent 6-dimensional spinors allows  to write the following relation \cite{Tomasiello:2022dwe}
\begin{equation}
    \eta^2_+=a\eta^1_+\frac{1}{2}v^m\gamma_m \eta^1_-\,,
\end{equation}
with $a$ a scalar and $v$ a 1-form. In addition, we can parametrize the inner product of the two spinors in terms of an angle and a complex phase
\begin{equation}
    \eta^{2\dagger}_+\eta^1_+\equiv N \cos\psi e^{i\theta}\,,
\end{equation}
with $\psi$ and $\theta$ parameters varying through the manifold. Then, it is possible to see that the SU(3)-structure forms associated to $\eta^1_+$ and $\eta^2_+$ can be expressed as
\begin{equation}
    \Omega^a=v\wedge \omega^a\,, \qquad J^a=j^a+\frac{i}{2}v\wedge \bar{v}\,,
\end{equation}
where $j^a$ is a real 2-form  and $\omega^a$ is a complex 2-form. They are not independent of each other: $j^a$ and $\omega^a$ are functions of a single pair $j,\omega$ as follows
\begin{equation}
    \left(\begin{array}{c}
         j^{1,2}  \\
         \Im \omega^{1,2} 
    \end{array}\right)=\left(\begin{array}{cc}
        \cos(\psi) & \mp \sin\psi \\
         \pm \sin \psi & \cos\psi 
    \end{array}\right)\left(\begin{array}{c}
         j  \\
         \im\omega 
    \end{array}\right)\,,\qquad \Re\omega^{1,2}=\Re\omega\,.
\end{equation}
They also satisfy
\begin{equation}
\begin{gathered}
     \iota_v j=\iota_{\bar{v}}j= 0 \,,\qquad \iota_v\omega=\iota_{\bar v}\omega=0\,,\\
     j\wedge\omega=\omega\wedge \omega=0\,,\qquad \omega\wedge\bar{\omega}=2j^2\,.
\end{gathered}
\end{equation}
Expanding, making use of the Clifford map and choosing an appropriate normalization that takes into account the warp factor of the metric, it is possible to write
\begin{equation}
    \Phi_+=e^{3A-\phi}e^{i\theta}\cos\psi e^{-iJ_\psi-\tan\psi{\rm Re}\omega}\,,\qquad \Phi_-=e^{3A-\phi}\cos\psi v\wedge e^{i\omega_\psi -\tan\psi {\rm Re} \omega}\,,
    \label{ap.geo-eq: pure forms SU3xSU3}
\end{equation}
with
\begin{equation}
    J_\psi\equiv \frac{j}{\cos\psi}+\frac{i}{2}v\wedge \bar{v}\,, \qquad \omega_\psi \equiv \frac{1}{\sin\psi}\left(\im \omega- i\cos\psi\re\omega\right)\,.
\end{equation}
When $\psi=0$, we recover a $SU(3)$ structure and when $\psi=\pi/2$ we obtain a $SU(2)$ structure.

\ifSubfilesClassLoaded{%
\bibliography{biblio}%
}{}

\end{document}

\ifSubfilesClassLoaded{%
\tableofcontents
}{}

\chapter[Geometric and Non-geometric Fluxes and Vacua]{Geometric and Non-geometric\\ Fluxes and Vacua}
\label{ch: ap systematic}

\section{Fluxes and axion polynomials}
\label{ap.sys-sec: conv}

In type IIA orientifold compactifications, geometric and non-geometric fluxes are defined in terms of their action on the basis of $p$-forms of table \ref{cy-table: harmonic basis}, that correspond to the harmonic representatives of $p$-form cohomology classes of a would-be Calabi--Yau manifold $X_6$. In this framework, and following the conventions in  \cite{Ihl:2007ah}, the action of the different NS fluxes on each $p$-form is determined as 

\bea
\label{sys-eq:fluxActions0}
& & \hskip-1.3cm \ell_s H \wedge {\bf 1} = -h_\mu \beta^\mu , , \qquad \qquad \  \ \ell_s  H \wedge \alpha_\mu  =   h_\mu   \Phi_6 \nonumber\, ,\\
& &  \hskip-1.3cm \ell_s f \triangleleft \om_a = -f_{a \mu}\, \beta^\mu \, , \qquad \qquad \,\ell_s  f \triangleleft \varpi_\alpha = \hat{f}_{\alpha}{}^\mu\, \alpha_\mu \, , \nonumber\\
& &  \hskip-1.4cm \ell_s  f \triangleleft \alpha_\mu = -f_{a \mu} \, \tilde\om^a\, ,  \qquad \qquad \, \ell_s f \triangleleft \beta^\mu = -\, \hat f_{\alpha}{}^\mu \, \tilde\varpi^\alpha \,,\nonumber\\
& & \\
& & \hskip-1.3cm \ell_s Q \triangleright \tilde\om^a = -Q^{a}{}_{\mu}\, \beta^\mu\, , \qquad \quad \ \ \ell_s  Q \triangleright  \tilde\varpi^\alpha =  \hat{Q}^{\alpha \mu}\, \alpha_\mu \, , \nonumber\\
& & \hskip-1.3cm\ell_s  Q \triangleright \alpha_\mu = \, Q^{a}{}_{\mu} \, \om_a\, ,  \qquad \qquad \ell_s  Q \triangleright \beta^\mu =  \hat{Q}^{\alpha\, \mu} \, \varpi_\alpha\,, \nonumber\\
& & \hskip-1.3cm \ell_s  R \bullet {\Phi_6} = R_\mu\, \beta^\mu \, , \qquad \qquad \ \ \ell_s  R \bullet \alpha_\mu =  R_\mu   {\bf 1} \,, \nonumber
\eea
where $\Phi_6$ is the normalized volume form $\frac{1}{\ell_s^6}\int_{X_6}\Phi_6=1$ and we also have that $H\wedge \beta^\mu  = R \bullet \beta^\mu  = 0$. The NS flux quanta are $h_\mu, f_{a\, \mu},  \hat f_{\a}{}^\mu, Q^a{}_\mu,  \hat Q^{\a\, \mu}, \hat  R_\mu \in \bZ$. This specifies the action of the twisted differential operator \eqref{cy-eq: twistedD} on each $p$-form, and in particular the superpotential \eqref{cy-eq: general NS superpotential expanded} and the RR potential transformation \eqref{cy-eq: C3change} leading to the D-term potential.

\subsubsection*{Axionic flux orbits and the $P$-matrices}

From the superpotential it is easy to read the gauge-invariant flux-axion polynomials \eqref{sys-RRrhos} and \eqref{sys-NSrhos}. Then, as in the Calabi--Yau case \cite{Herraez:2018vae}, one can check that all the remaining entries of $\rho_{\cal A}$ can be generated by taking derivatives of the {\it master polynomial} $\rho_0$. Indeed, in our more general case one finds that
\bea
\label{sys-eq:drho0}
& & \frac{\partial\rho_0}{\partial b^a} = \rho_a\, , \quad \frac{\partial\rho_0}{{\partial b^a}{\partial b^b}} = {\cal K}_{abc} \, \tilde\rho^c\, , \quad \frac{\partial\rho_0}{{\partial b^a}{\partial b^b}{\partial b^c}} = {\cal K}_{abc} \, \tilde\rho\,, \quad \frac{\partial\rho_0}{\partial \xi^\mu} = \rho_\mu\,  , \\
& & \frac{\partial\rho_0}{{\partial b^a}{\partial \xi^\mu}} = \rho_{a\mu}\, , \quad \frac{\partial\rho_0}{{\partial b^a}{\partial b^b}{\partial \xi^\mu}} = {\cal K}_{abc} \, \tilde\rho^c{}_\mu\, , \quad \frac{\partial\rho_0}{{\partial b^a}{\partial b^b}{\partial b^c}{\partial \xi^\mu}} = {\cal K}_{abc} \, \tilde\rho_\mu \, ,\nonumber
\eea
while all the other derivatives vanish. Just like in \cite{Herraez:2018vae}, one can understand these relations from the fact that the matrix ${\cal R}$ in relating quantized and gauge invariant fluxes can be written as
\be
{\cal R}\equiv e^{b^a P_a + \xi^K P_K} \,,
\ee
with $P_a$ and $P_\mu$ nilpotent matrices.  Indeed, given \eqref{sys-eq:invRmat} one can check that
\begin{eqnarray}
\label{sys-eq:P-matrices}
& & \hskip-0cm P_a = \begin{bmatrix}
0 & \vec\delta_a^t & 0 & 0 & 0 & 0 & 0 & 0 \\
0 & 0 & {\cal K}_{abc} & 0 & 0 & 0 & 0 & 0 \\
0 & 0 & 0 & \vec\delta_a & 0 & 0 & 0 & 0 \\
0 & 0 & 0 & 0 & 0 & 0 & 0 & 0 \\
0 & 0 & 0 & 0 & 0 & \vec\delta_a^t  \, \delta_\mu^\nu & 0 & 0 \\
0 & 0 & 0 & 0 & 0 & 0 & {\cal K}_{abc} \, \delta_\mu^\nu & 0 \\
0 & 0 & 0 & 0 & 0 & 0 & 0 & \vec\delta_a \, \delta_\mu^\nu \\
0 & 0 & 0 & 0 & 0 & 0 & 0 & 0 \\
\end{bmatrix}\, ,
\end{eqnarray}
and
\begin{eqnarray}
& & P_\mu = \begin{bmatrix}
0 & 0 & 0 & 0 & \vec\delta_\mu^t & 0 & 0 & 0 \\
0 & 0 & 0 & 0 & 0 &  \vec\delta_a^t  \, \delta_\mu^\nu & 0 & 0 \\
0 & 0 & 0 & 0 & 0 & 0 &  \vec\delta_a  \, \delta_\mu^\nu & 0 \\
0 & 0 & 0 & 0 & 0 & 0 & 0 &  \vec\delta_\mu^t \\
0 & 0 & 0 & 0 & 0 & 0 & 0 & 0 \\
0 & 0 & 0 & 0 & 0 & 0 & 0 & 0 \\
0 & 0 & 0 & 0 & 0 & 0 & 0 &  0 \\
0 & 0 & 0 & 0 & 0 & 0 & 0 & 0 \\
\end{bmatrix}\, .
\end{eqnarray}

\subsubsection*{Constraints from Bianchi identities}

On compactifications with geometric and non-geometric fluxes, one important set of consistency constraints are the flux Bianchi identities. In our setup, these can be obtained by imposing that the twisted differential ${\cal D}$ in \eqref{cy-eq: twistedD} satisfies the idempotency constraint ${\cal D}^2 = 0$ when applied on the $p$-form basis of table \ref{cy-table: harmonic basis}  \cite{Grana:2006hr,Ihl:2007ah,Robbins:2007yv}. Applying the definitions \eqref{sys-eq:fluxActions0}, one obtains\footnote{Compared to \cite{Robbins:2007yv}, in our setup the flux components $h^\mu$, $R^\mu$, $f_a{}^\mu, Q^{a\mu}, \hat{f}_{\alpha \mu}$ and $\hat{Q}^\alpha{}_\mu$ are projected out. }
\bea
\label{sys-eq: bianchids2}
& & h_\mu\, \hat{f}_\alpha{}^\mu = 0\, , \quad h_\mu\, \hat{Q}^{\alpha \mu} = 0\, , \quad f_{a\mu}\, \hat{f}_\alpha{}^\mu = 0\, , \quad f_{a\mu}\, \hat{Q}^{\alpha \mu} =0\, , \nonumber\\
& & R_\mu\, \hat{Q}^{\alpha \mu} = 0\, , \quad R_\mu \, \hat{f}_\alpha{}^\mu = 0\, , \quad Q^a{}_\mu \, \hat{Q}^{\alpha \mu} = 0\, , \quad \hat{f}_\alpha{}^{\mu} \,Q^a{}_\mu=0\, ,\\
& & \hat{f}_\alpha{}^{[\mu}\, \hat{Q}^{\alpha \nu]} = 0\, , \quad h_{[\mu} \, R_{\nu]} + f_{a[\mu}\, Q^a{}_{\nu]} = 0\,. \nonumber
\eea

\section{Curvature and sGoldstino masses} \label{ap.sys: curvature} In this appendix we will show  that the directions \eqref{sys-partialmax} minimize respectively $R_{a\bar c d\bar d}g^ag^bg^cg^d$ and $R_{\mu\hat\nu\rho\hat\sigma}g^{\mu}g^{\hat\rho}g^{\nu}g^{\hat\sigma}$. To do so we will follow closely \cite{Covi:2008ea,Covi:2008zu}.

\subsubsection*{Curvature}
\label{ap.sys: curvature1}
Before talking about the extrema conditions, there are some relations that must be introduced. Consider a K\"ahler  potential depending on some set of complex chiral fields $\phi^A$ obeying a no-scale type condition:
\begin{align}
\label{sys-noscale}
    K^A K_A=p\, ,
\end{align}
where $K_A=\nabla_A K$, $K^A=G^{A\bar B} K_{\bar B}$ and $G_{A\bar B}=\partial_A\partial_{\bar B} K$.
Taking the derivative with respect to $\nabla_B$ in \eqref{sys-noscale} one obtains:
\begin{align}
K_B+K^A\nabla_B K_A=0\, ,
\end{align}
and deriving now with respect to $\nabla_{ C}$ we find:
\begin{align}
2\nabla_CK_B+K^A\nabla_C\nabla_B K_A=0\, . \label{sys-hola}
\end{align}
Equation \eqref{sys-hola} can be contracted with $K^CK^{\bar D}$ and $K^{\bar D}$  to obtain respectively
\begin{align}
R_{C\bar{D} M\bar{N}}K^CK^MK^{\bar{N}}K^{\bar D}&=\textcolor{black}{2p}\, ,&		R_{C\bar{D} M\bar{N}}K^MK^{\bar{N}}K^{\bar D}&=\textcolor{black}{2}K_C \label{sys-curva}\, .
\end{align}
We will need these two last relations to study the extrema of $R_{A\bar B C\bar D}g^Ag^{\bar B}g^Cg^{\bar D}$

\subsubsection*{sGoldstino masses}
\label{ap.sys: curvature2}
As discussed in section \ref{sys-ss:fterms},
the relevant parameter to compute the sGoldstino masses is
\be
\hat{\sig} = \frac{2}{3}-R_{A\bar B C \bar D} f^{A} f^{\bar B} f^{C} f^{\bar D}\, ,
\label{sys-app:sigma}
\ee
which we are interested in maximise. In this sense, it was shown in \cite{Covi:2008zu} that the extrema of \eqref{sys-app:sigma} are  given  by the $f_{0A}$ satisfying the implicit relation: 
\begin{align}
    f_{0A}=\frac{R_{A\bar B C\bar D}f^{\bar B }_0f^{C}_0f^{\bar D }_0}{R_{A\bar B C\bar D}f_0^{A }f^{\bar B }_0f^{C}_0f^{\bar D }_0}\, .
    \label{sys-app:ex}
\end{align}
Using the results above it is now straightforward to see that $f_{0A}=e^{i\alpha}\frac{K_A}{\sqrt{p}}$, $\alpha\in \mathds{R}$ are solutions of \eqref{sys-app:ex} and therefore extrema of \eqref{sys-app:sigma}.

\subsubsection*{Type IIA on a CY$_3$} \label{ap.sys: curvature3}
The moduli space metric of IIA on a CY$_3$ orientifold is described  from the K\"ahler  potential:
\begin{align}
    K=K_K+K_Q\, ,
\end{align}
where the subindex $K$ refers to the K\"ahler sector whereas we use $Q$ for the complex sector. All the relations discussed above can be applied independently to $K_K$ with $p=3$ and to $K_Q$ with $p=4$. In particular, this shows that \eqref{sys-partialmax} extremize respectively $R_{a\bar c d\bar d}g^ag^bg^cg^d$ and $R_{\mu\hat\nu\rho\hat\sigma}g^{\mu}g^{\hat\rho}g^{\nu}g^{\hat\sigma}$. Regarding the character of the points one can show that they are minima by doing small perturbations around these directions.

If one just considered the  K\"ahler sector or the complex sector (meaning taking $K_Q=0$ in the first case and $K_T=0$ in the second case) this would be the end of the story. Nevertheless, since in general we want to have both contributions, there are some subtleties one has to take into account. The point is that now  $R_{A\bar B C\bar D}g^Ag^Bg^Cg^D$ does not have just ``one" contribution but two independent contributions:
\be
R_{A\bar B C\bar D}g^Ag^Bg^Cg^D=R_{a\bar c d\bar d}g^ag^bg^cg^d+R_{\mu\hat\nu\rho\hat\sigma}g^{\mu}g^{\hat\rho}g^{\nu}g^{\hat\sigma}\, ,
\label{sys-app:sigma2}
\ee
and the novelty is that a new extremum appears :
\begin{align}
    f_0^A=\frac{1}{\sqrt{7}}\left\{K_a,e^{i\alpha}K_\mu\right\}
    \label{sys-app:max}
\end{align}
with $\alpha\in\mathds{R}$, which is precisely the one discussed below \eqref{sys-solsfmax}. Doing again a small perturbation around the points, it can be shown that now both $f_0^A=\left\{e^{i\alpha}\frac{K_a}{\sqrt{3}},0\right\}$ and $f_0^A=\left\{0,e^{i\alpha}\frac{K_\mu}{\sqrt{4}}\right\}$ are saddle points of \eqref{sys-app:sigma2} whereas \eqref{sys-app:max} is a  minimum.

\section{Analysis of the Hessian}
\label{ap.sys: Hessian}

In this appendix we will compute the Hessian of the scalar potential and study its properties. We will first focus on the F-term potential, whose complexity will require a detailed analysis and the use of a simplified version of our Ansatz. Once the associated Hessian matrix has been found, we will evaluate the result in both the SUSY and the non-SUSY branches independently, in order to obtain information regarding their stability. Finally, we will briefly discuss the general behaviour of the D-term potential Hessian matrix. We will work in Planck units, so $\kappa_4=1$.

\subsection*{F-term Potential}
\label{ap.sys: Hessianf}
Starting from \eqref{sys-eq:potentialgeom} and evaluating the second derivatives along the vacuum equations we obtain:
\bes
\begin{align}
    e^{-K}\frac{\partial^2 V_F}{\partial \xi^\sigma \partial \xi^\lambda}|_{\text{vac}}=&8\rho_\lambda\rho_\sigma+2g^{ab}\rho_{a\sigma}\rho_{b\lambda}\, ,\\
    e^{-K}\frac{\partial^2 V_F}{\partial\xi^\sigma \partial b^a}|_{\text{vac}}=&8\rho_\sigma \rho_a+8\rho_0\rho_{a\sigma}+2g^{bc}\mathcal{K}_{abd}\rho_{c\sigma}\tilde{\rho}^d\, ,\\
    e^{-K}\frac{\partial^2 V_F}{\partial \xi^\lambda \partial u^\sigma}|_{\text{vac}}=&0\, ,\\
    e^{-K}\frac{\partial^2 V_F}{\partial \xi^\sigma\partial t^a}|_{\text{vac}}=&2\partial_a g^{bc}\rho_{b\sigma}\rho_c\, ,\\
    e^{-K}\frac{\partial^2 V_F}{\partial b^a \partial b^b}|_{\text{vac}}=&8\rho_a\rho_b+8\rho_0\mathcal{K}_{abc}\tilde{\rho}^c+2g^{cd}\mathcal{K}_{ace}\mathcal{K}_{bdf}\tilde{\rho}^e\tilde{\rho}^f+2g^{cd}\mathcal{K}_{abc}\rho_d\tilde{\rho}+\frac{8\mathcal{K}^2}{9}g_{ab}\tilde{\rho}^2\nonumber\\
    &+2c^{\mu\nu}\rho_{a\mu}\rho_{b\nu}\, ,\\
    e^{-K}\frac{\partial^2 V_F}{\partial u^\sigma \partial b^a}|_{\text{vac}}=&2\partial_\sigma c^{\mu\nu}\rho_{a\mu}\rho_\nu\, ,\\
    e^{-K}\frac{\partial^2 V_F}{\partial b^a\partial t^b}|_{\text{vac}}=&2\partial_b g^{cd}\mathcal{K}_{ace}\rho_d\tilde{\rho}^e+\left(\frac{16\mathcal{K}}{3}\mathcal{K}_b g_{ac}+\frac{8\mathcal{K}^2}{9}\partial_b g_{ac}\right)\tilde{\rho}^c\tilde{\rho}\, ,\\
    \frac{\partial^2 V_F}{\partial u^\sigma\partial u^\lambda}|_{\text{vac}}=&V_F\partial_\sigma \partial_\lambda K-V_F\partial_\sigma K\partial_\lambda K\nonumber\\
    &+e^{K}\left[\partial_\sigma \partial_\lambda c^{\mu\nu}\rho_\mu\rho_\nu +t^at^b(\partial_\lambda\partial_\sigma c^{\mu\nu}\rho_{a\mu}\rho_{b\nu}-8\rho_{a\sigma}\rho_{b\lambda})+2g^{ab}\rho_{a\sigma}\rho_{b\lambda}\right]\, ,\\
    \frac{\partial^2 V_F}{\partial t^a\partial u^\sigma}|_{\text{vac}}=&V_F\partial_\sigma \partial_a K-V_F\partial_\sigma K \partial_a K+e^{K}\left[-4\mathcal{K}_a\tilde{\rho}^b\rho_{b\sigma}+4\mathcal{K}_a\tilde{\rho}\rho_\sigma\right.\nonumber\\
    &\left.-8\rho_{a\sigma}\rho_{b\mu}u^\mu t^b-8\rho_{b\sigma}\rho_{a\mu}u^\mu t^b+2\partial_\sigma c^{\mu\nu}\rho_{a\mu}\rho_{b\nu}t^b+2\partial_ag^{bc}\rho_{b\mu} u^\mu\rho_{c\sigma}\right]\, ,\\
     \frac{\partial^2 V_F}{\partial t^a\partial t^b}|_{\text{vac}}=&V_F\partial_a \partial_b K-V_F\partial_a K \partial_b K+e^{K}\left[\partial_a\partial_b g^{cd}\rho_c\rho_d+2\mathcal{K}_a\mathcal{K}_b\tilde{\rho}^2\right.\nonumber\\
     &\left.+\left(8\mathcal{K}_a\mathcal{K}_b g_{cd}+\frac{16\mathcal{K}}{3}\mathcal{K}_{ab}g_{cd}+\frac{8\mathcal{K}}{3}\mathcal{K}_a\partial_{b}g_{cd}+\frac{8\mathcal{K}}{3}\mathcal{K}_b\partial_{a}g_{cd}+\frac{4\mathcal{K}^2}{9}\partial_a\partial_bg_{cd}\right)\tilde{\rho}^c\tilde{\rho}^d\right.\nonumber\\
     &\left.+\frac{4\mathcal{K}}{3}\mathcal{K}_{ab}\tilde{\rho}^2-8\mathcal{K}_{ab}\tilde{\rho}^c\rho_{c\nu}u^\nu+8\mathcal{K}_{ab}\tilde{\rho}\rho_\nu u^\nu+2\tilde{c}^{\mu\nu}\rho_{a\mu}\rho_{b\nu}+\partial_a\partial_b g^{cd}\rho_{c\mu}\rho_{d\nu}u^\mu u^\nu\right]\, .
\end{align}
\ees
If we now introduce the ansatz \eqref{sys-Ansatz} and make use of the decomposition of the metric in its primitive and non primitive parts -see  \eqref{sys-eq: primitive metric}- we are left with:
\bes
\label{sys-eq: partial second derivatives anst general}
\begin{align}
\label{sys-ap1}
    e^{-K}\frac{\partial^2 V_F}{\partial \xi^\sigma \partial \xi^\lambda}|_{\text{vac}}=&(8E^2 +\frac{1}{6}F^2)\mathcal{K}^2\partial_\lambda K\partial_\sigma K +2g_{P}^{ab}\rho_{a\sigma}\rho_{b\lambda}\, ,\\
    e^{-K}\frac{\partial^2 V_F}{\partial \xi^\sigma \partial b^a}|_{\text{vac}}=&(8BE-\frac{4}{3}CF)\mathcal{K}^2\partial_aK\partial_\sigma K +(8A-\frac{4}{3}C)\mathcal{K}\rho_{a\sigma}\, ,\\
    e^{-K}\frac{\partial^2 V_F}{\partial \xi^\lambda \partial u^\sigma}|_{\text{vac}}=&0\, ,\\
    e^{-K}\frac{\partial^2 V_F}{\partial \xi^\sigma \partial t^a}|_{\text{vac}}=&-16B\mathcal{K}\rho_{a\sigma}\, ,\\
     e^{-K}\frac{\partial^2 V_F}{\partial b^a\partial b^b}|_{\text{vac}}=&2c^{\mu\nu}_{P}\rho_{a\mu}\rho_{b\nu}+(8B^2+\frac{4}{9}C^2+\frac{2}{9}D^2+\frac{2}{9}F^2)\mathcal{K}^2\partial_a K\partial_b K\nonumber\\
     &+(8AC-8BD-\frac{4}{3}C^2-\frac{4}{3}D^2)\mathcal{K}\mathcal{K}_{ab}\, ,\\
    e^{-K}\frac{\partial^2 V_F}{\partial b^a\partial u^\sigma}|_{\text{vac}}=&-16E\mathcal{K}\rho_{a\sigma}\, ,\\
     e^{-K}\frac{\partial^2 V_F}{\partial b^a \partial t^b}|_{\text{vac}}=&(-16BC+\frac{8}{3}CD)\mathcal{K}\mathcal{K}_{ab}\, ,\\
     e^{-K}\frac{\partial^2 V_F}{\partial u^\sigma\partial u^\lambda}|_{\text{vac}}=&(8E^2+\frac{F^2}{6})\mathcal{K}^2\partial_\sigma K\partial_\lambda K-\frac{G_{\mu\nu}}{G}(16E^2-\frac{1}{3}F^2-\frac{4}{3}DE+\frac{1}{3}CF)\mathcal{K}^2\nonumber\\
     &+2g^{ab}_{P}\rho_{a\sigma}\rho_{b\lambda} \label{sys-eq: uu}\, ,\\
    e^{-K}\frac{\partial^2 V_F}{\partial u^\sigma\partial t^a}|_{\text{vac}}=&(-8E^2+\frac{1}{6}F^2)\mathcal{K}^2\partial_a K\partial_\sigma K-\frac{4}{3}F\mathcal{K}\rho_{a\sigma}\, ,\\
     e^{-K}\frac{\partial^2 V_F}{\partial t^a \partial t^b}|_{\text{vac}}=&(8B^2+\frac{4}{9}C^2+\frac{2}{9}D^2+\frac{2}{9}F^2)\mathcal{K}^2\partial_aK \partial_b K+ (-96B^2-\frac{8}{3}C^2+\frac{4}{3}F^2)\mathcal{K}\mathcal{K}_{ab}\nonumber\\
     &+2c^{\mu\nu}_P\rho_{a\mu}\rho_{b\nu}\label{sys-apf}\, ;
\end{align}
\ees
where we have used the following relations
\begin{align}
    \partial_b g_{ac}t^c=-2g_{ab}\, ,\\
    \partial_\sigma\partial_\lambda c^{\mu\nu}\partial_\mu K \partial_\nu K=32 c_{\mu\nu}\, ,\\
    \partial_a\partial_b g^{cd}\partial_c K \partial_d K=32 g_{ab}\, ,\\
    \partial_a\partial_b g_{cd}t^ct^d=6g_{ab}\, .
\end{align}
Unfortunately, it is not possible to provide a general description of the stability using the results above. As discussed in section \ref{sys-sec: stabalidity}, for an arbitrary $\rho_{a\mu}$ one needs to know explicitly the internal metric. Only if we restrict ourselves to the case in which $\rho_{a\mu}$ has rank one are we able to derive a universal analysis. Therefore, from now on we will set
\begin{align}
\rho_{a\mu}&=-\frac{F}{12}\mathcal{K}\partial_aK_T \partial_\mu K_Q \label{sys-eq: geom r1}\, .
\end{align}
Plugging this expression back into \eqref{sys-eq: partial second derivatives anst general} the on-shell second derivatives of the potential are finally reduced to:
\bes
\begin{align}
    e^{-K}\frac{\partial^2 V_F}{\partial \xi^\sigma\partial \xi^\lambda}|_{\text{vac}}=&(8E^2+\frac{1}{6}F^2)\mathcal{K}^2\partial_\sigma K\partial_\lambda K\, ,\\
    e^{-K}\frac{\partial^2 V_F}{\partial \xi^\sigma\partial b^a}|_{\text{vac}}=&(8EB-\frac{2}{3}AF-\frac{2}{9}CF)\mathcal{K}^2\partial_\sigma K\partial_a K\, ,\\
    e^{-K}\frac{\partial^2 V_F}{\partial \xi^\sigma \partial u^\lambda}|_{\text{vac}}=&0\, ,\\
    e^{-K}\frac{\partial^2 V_F}{\partial \xi^\sigma\partial t^a}|_{\text{vac}}=&\frac{4}{3}BF\mathcal{K}^2\partial_a K\partial_\sigma K\, ,\\
    e^{-K}\frac{\partial^2 V_F}{\partial b^a\partial b^b}|_{\text{vac}}=&(8B^2+\frac{4}{9}C^2+\frac{2}{9}D^2+\frac{2}{9}F^2)\mathcal{K}^2\partial_a K\partial_b K\nonumber\\
    &+(8AC-8BD-\frac{4}{3}C^2-\frac{4}{3}D^2)\mathcal{K}\mathcal{K}_{ab}\, ,\\
    e^{-K}\frac{\partial^2 V_F}{\partial u^\sigma \partial b^a}|_{\text{vac}}=&\frac{4}{3}EF\mathcal{K}^2\partial_aK\partial_\sigma K\, ,\\
    e^{-K}\frac{\partial^2 V_F}{\partial b^a \partial t^b}|_{\text{vac}}=&(-16BC+\frac{8}{3}CD)\mathcal{K}\mathcal{K}_{ab}\, ,\\
    e^{-K}\frac{\partial^2 V_F}{\partial u^\sigma\partial u^\lambda}|_{\text{vac}}=&(8E^2+\frac{1}{6}F^2)\mathcal{K}^2\partial_\sigma K\partial_\lambda K-\frac{G_{\mu\nu}}{G}(16E^2-\frac{1}{3}F^2-\frac{4}{3}DE+\frac{1}{3}CF)\mathcal{K}^2\, ,\\
    e^{-K}\frac{\partial^2 V_F}{\partial u^\sigma\partial t^a}|_{\text{vac}}=&(-8E^2+\frac{5}{18}F^2)\mathcal{K}^2\partial_\sigma K\partial_a K\, ,\\
    e^{-K}\frac{\partial^2 V_F}{\partial t^a\partial t^b}|_{\text{vac}}=&(8A^2+16B^2+\frac{2}{9}C^2+32E^2-\frac{8}{9}F^2)\mathcal{K}^2\partial_aK\partial_b K\nonumber\, ,\\
    &+(-96B^2-\frac{8}{3}C^2+\frac{4}{3}F^2)\mathcal{K}\mathcal{K}_{ab}\, .
\end{align}
\ees
In order to make the computations manageable, we follow the same procedure as in \cite{Marchesano:2019hfb} and consider a basis of canonically normalized fields by performing the following change of basis:
\begin{align}
   \left(\xi^\mu, b^a\right)\rightarrow \left(\hat{\xi}, \hat{b}, \xi^{\hat{\mu}}, b^{\hat{a}}\right)&\, , &     \left( u^\mu, t^a\right)\rightarrow \left(\hat{u}, \hat{t}, u^{\hat{\mu}}, t^{\hat{a}}\right)& \, ,
\end{align}
where $\left\{ \hat{b}, \hat{t}\right\}$ $\left(\left\{ \hat{\xi},\hat{u}\right\}\right)$ are unit vectors along the subspace corresponding to $g_{ab}^{NP}|_{\text{vac}}$ $\left(c_{\mu\nu}^{NP}|_{\text{vac}}\right)$ and  $\left\{b^{\hat{a}}, t^{\hat{a}}\right\}$ $\left(\left\{ \xi^{\hat{\mu}},u^{\hat{\mu}}\right\}\right)$\footnote{Notice that $\hat a=1,\dots,h^{1,1}_--1$; $\hat \mu=1,\dots,h^{2,1}$} correspond analogously to vectors of unit norm with respect to $g_{ab}^{P}|_{\text{vac}}$ $\left(c_{\mu\nu}^{P}|_{\text{vac}}\right)$. We can then rearrange the Hessian $\hat H$ in a $8\times 8$ matrix with basis $(\hat{\xi}, \hat{b}, \xi^{\hat{\mu}}, b^{\hat{a}},\hat{u}, \hat{t}, u^{\hat{\mu}}, t^{\hat{a}})$ so that it reads
\begin{equation}
    \hat{H}_F=e^K\mathcal{K}^2F^2\begin{pmatrix}\frac{384{E_F}^{2}+8}{3} & H_{12} & 0 & 0 & 0 & \frac{32B}{\sqrt{3}} & 0 & 0\cr 
    H_{12} & H_{22} & 0 & 0 & \frac{32E_F}{\sqrt{3}} & H_{26} & 0 & 0\cr 
    0 & 0 & 0 & 0 & 0 & 0 & 0 & 0\cr 0 & 0 & 0 & H_{44} & 0 & 0 & 0 & H_{48}\cr 
    0 & \frac{32E_F}{\sqrt{3}} & 0 & 0 & H_{55} & H_{56} & 0 & 0\cr 
    \frac{32B_F}{\sqrt{3}} & H_{26} & 0 & 0 & H_{56} & H_{66} & 0 & 0\cr 
    0 & 0 & 0 & 0 & 0 & 0 & H_{77} & 0\cr 
    0 & 0 & 0 & H_{48} & 0 & 0 & 0 & H_{88}\end{pmatrix}\, ,
    \label{sys-eq: hessian}
\end{equation}
where we have defined:
\begin{align}
    H_{22}=&\frac{8{D_F}^{2}-96B_FD_F+32{C_F}^{2}+96A_FC_F+864{B_F}^{2}+24}{9}\, ,\\
    H_{44}=&\frac{8{D_F}^{2}+48B_FD_F+8{C_F}^{2}-48A_FC_F}{9}\, ,\\
    H_{55}=&-\frac{192{E_F}^{2}-48D_FE_F+12C_F-20}{3}\, ,\\
    H_{66}=&\frac{3456{E_F}^{2}-8{C_F}^{2}+576{B_F}^{2}+864{A_F}^{2}-80}{9}\, ,\\
    H_{77}=&\frac{192{E_F}^{2}-16D_FE_F+4C_F-4}{3}\, ,\\
    H_{88}=&\frac{16{C_F}^{2}+576{B_F}^{2}-8}{9}\, ,\\
    H_{12}=&8\sqrt{3}\left(8B_FE_F-\frac{2C_F}{9}-\frac{2A_F}{3}\right)\\
    H_{26}=& \frac{32C_FD_F-192B_FC_F}{9}\, ,\\
    H_{48}=& -\frac{16C_FD_F-96B_FC_F}{9}\, ,\\
    H_{56}=& 8\sqrt{3}\left( \frac{5}{18}-8{E_F}^{2}\right)\, .
\end{align}
Note that \eqref{sys-eq: hessian} defines a symmetric matrix whose components are determined once we chose a vacuum. In other words, given an extremum of the potential, one just needs to plug the correspondent $\left\{A_F,B_C,C_F,D_F\right\}$ into \eqref{sys-eq: hessian} to analyze its perturbative stability. The physical masses of the moduli will be given by $1/2$  of the eigenvalues of the Hessian.

Once the  explicit form of Hessian has been introduced, we are ready to discuss the spectrum of the two branches obtained in the main text. This will be done in detail below.

\subsubsection*{SUSY light spectrum}
\label{sys-sap:SUSYH}
We consider now the Hessian of the F-term potential associated to the supersymemtric branch of solutions. As explained in sections \ref{sys-branchvacu} and \ref{sys-sec: summary} this solution is characterized by
\begin{align}
\label{sys-sussysol}
    A_F&=-3/8\, ,   &    B_F&=-3E_F/2\, ,   &    C_F&=1/4\, ,   &    D_F&=15E_F\, .
\end{align}
Then, one just has to plug \eqref{sys-sussysol} into \eqref{sys-eq: hessian}, diagonalize and divide by $1/2$ to obtain the corresponding mass spectrum. The result is:
\begin{equation}
\label{sys-susymass}
   m^2= F^2e^K\mathcal{K}^2\left\{0,-\frac{1}{2}(1+16E_F^2),-\frac{1}{18}+56E_F^2\pm 
    \frac{1}{3}\sqrt{1+160E_F^2+2304E_F^4},\lambda_5,\lambda_6,\lambda_7,\lambda_8\right\}\, ,
\end{equation}
where the  $\lambda_i$ are the four roots of
\begin{align}
   0=&-160380+18662400E_F^2+62547240960E_F^4+2721784135680E_F^6+29797731532800E_F^8\nonumber\\
   &+(-19971-33191568E_F^2-4174924032E_F^4-74992988160E_F^6)18\lambda\nonumber\\
   &+(4483+1392480E_F^2+55800576E_F^4)\left(18\lambda\right)^2+(-133-13392E_F^2)\left(18\lambda\right)^3+\left(18\lambda\right)^4
    \label{sys-eq: implicit}\, .
\end{align}
In order to discuss the stability, we must compare \eqref{sys-susymass} to the BF bound, which for this case takes the value:
\begin{equation}
    m_{BF}^2=\frac{3}{4}V|_{\text{vac}}=-(\frac{9}{16}+9E_F^2)e^K\mathcal{K}^2 F^2\, .
\end{equation}
It is straightforward to see that the first non-zero eigenvalue can be rewritten as:
\begin{align}
  m_2^2= -\frac{1}{2}(1+16E_F^2)=\frac{8}{9}m_{BF}^2\, .
\end{align}
Regarding the other masses, although they can also be written as functions of $m_{BF}$ their expressions are not that illuminating. In this sense, one can check that the third eigenvalue is always positive, whereas  $m_4^2$ has a  negative region -respecting the the BF bound- for $|E_F|\lesssim 0.1$. Finally, the dependence of the four remaining eigenvalues with $E_F$, conveyed as implicit solutions of \eqref{sys-eq: implicit}, has to be studied numerically. One finds that only one of them enters in a negative region -again above $m_{BF}^2$- for $|E_F|\lesssim 0.04$.

We conclude that the SUSY vacuum may have up to three tachyons, though only one is preserved for $|E_F| \gtrsim 0.1$. None of them violates the BF bound, as it is expected for this class of vacua. To finish this part of the appendix, let us also write the tachyonic directions:
\begin{itemize}
    \item $m_2^2$. Direction: $u^{\hat{\mu}}$.\footnote{For the complex axions, the direction $\xi^{\hat\mu}$ is the one with zero eigenvalue.}
    \item $m_4^2$.  Direction: linear combination of $b^{\hat{a}}$ and $t^{\hat{a}}$.
    \item $m_5^2=F^2e^K\mathcal{K}^2\lambda_5$ (lowest solution of \eqref{sys-eq: implicit}). Direction: combination of all non primitive directions, i.e. $\hat{\xi}$, $\hat{b}$, $\hat{u}$ and $\hat{t}$.
\end{itemize} 

\subsubsection*{Non-SUSY branch}
\label{sys-sap:nonSUSYH}

We end this section of the appendix by analysing the Hessian of the F-term potential associated with the non-SUSY solutions. As it was studied in detail in the main text, this branch has to be defined implicitly in terms of the $A_F$ and $C_F$ solving equation \eqref{sys-eq: megaeqF} (check table \ref{sys-vacuresul} and figure \ref{sys-fig: generalsol} for details). In consequence, trying to explore the stable regions analytically is, in practice, impossible, and things must be computed numerically. What we have done is to extract  the physical $A_F$ and $C_F$ satisfying \eqref{sys-eq: megaeqF}, plug them into \eqref{sys-eq: hessian} -$B_F$, $D_F$ and $E_F$ are determined once $A_F$ and $C_F$ are chosen- and study the mass spectrum. Despite the numerical approach, results can be obtained easily.  

After performing a complete analysis, we conclude that a single mode is responsible for the stability of the solution. In other words, seven out of the eight masses respect the BF bound at every point of the Non-SUSY branch. Therefore, the behaviour of the aforementioned mode is precisely the one which determines the unstable region (red points) in figure \ref{sys-fig: excludsol}. For the sake of completeness, let us write it explicitly:
\begin{align}
\label{sys-tachmass}
  m^2=&-F^2e^K\mathcal{K}^2\left[9(12A_F^2-1)((2A_F+C_F)(6A_F+C_F)-1)\right]^{-1}\left[-9+7776A_F^6+5184A_F^5C_F\right.\nonumber\\
    &+4A_FC_F(2+C_F)(C_F^2-5C_F+9)+1296A_F^4(C_F^2-2)+144A_F^3C_F(C_F^2+C_F-9)\nonumber\\
    &\left.-C_F(C_F-2)(C_F^2+6C_F-1)+6A_F^2(C_F^4+8C_F^3-46C_F^2+4C_F+45)\right]\, .
\end{align}
As it happened in the SUSY case for the mode with mass $\frac{8}{9}m_{BF}^2$, the direction of the mode with mass \eqref{sys-tachmass} is given by  $u^{\hat{\mu}}$. It is worth to point out that we are not saying that the other modes do not yield tachyons, but they are always above the $BF$ bound.  As discussed below figure \ref{sys-fig: excludsol}, these other tachyons are localized close to the regions where $m^2$ defined in  \eqref{sys-tachmass} violates the BF bound.

\subsection*{D-term potential}

We perform a similar analysis with the D-terms. Starting from \eqref{sys-eq: D-potentialgeom} and evaluating the second derivatives along the vacuum equations, we obtain that the only non-vanishing second partial derivatives of the potential $V_D$ are
\begin{align}
     \frac{\partial^2 V_D}{\partial u^\mu \partial u^\nu}=&\frac{3}{\mathcal{K}}\partial_\mu c_{\nu\sigma}\partial_\lambda K \tilde{g}^{\alpha\beta}\hat{\rho}^\sigma_\alpha\hat{\rho}^\lambda_\beta+\frac{12}{\mathcal{K}}c_{\mu\sigma}c_{\nu\lambda} \tilde{g}^{\alpha\beta}\hat{\rho}^\sigma_\alpha\hat{\rho}^\lambda_\beta\, ,\\
   \frac{\partial^2 V_D}{\partial u^\mu \partial t^a}=&\frac{3}{\mathcal{K}}c_{\mu\sigma}\partial_\lambda K\partial_a \tilde{g}^{\alpha\beta}\hat{\rho}^\sigma_\alpha\hat{\rho}^\lambda_\beta\,-\frac{9\mathcal{K}_a}{\mathcal{K}^2}c_{\mu\sigma}\partial_\lambda K \tilde{g}^{\alpha\beta}\hat{\rho}^\sigma_\alpha\hat{\rho}^\lambda_\beta ,\\
    \frac{\partial^2 V_D}{\partial t^a \partial t^b}=&(\partial_\sigma K\partial_\lambda K\hat{\rho}^\sigma_\alpha\hat{\rho}^\lambda_\beta)\left(\frac{3}{8\mathcal{K}} \partial_a\partial_b g^{\alpha\beta}\,-\frac{9\mathcal{K}_a}{8\mathcal{K}^2} \partial_b g^{\alpha\beta}\,\right.\nonumber\\
    &\left.-\frac{9\mathcal{K}_b}{8\mathcal{K}^2} \partial_a g^{\alpha\beta}\,+\frac{27\mathcal{K}_a\mathcal{K}_b}{4\mathcal{K}^3}\partial_\sigma K\partial_\lambda K  g^{\alpha\beta}\,-\frac{9\mathcal{K}_{ab}}{4\mathcal{K}^2}  g^{\alpha\beta}\,\right) .
\end{align}
If we now take into consideration the ansatz \eqref{sys-Ansatz} together with the Bianchi identity $f_{a\mu}\hat{f}_\alpha^\mu=0$, we have that, on-shell, $\partial_\mu K \hat{\rho}^\mu_\alpha=0$. Hence the saxionic sector of the D-term Hessian becomes
\begin{align}
    \p_A\p_B V_D=\left(\begin{matrix}\frac{12}{\mathcal{K}}c_{\mu\sigma}c_{\nu\lambda} \tilde{g}^{\alpha\beta}\hat{\rho}^\sigma_\alpha\hat{\rho}^\lambda_\beta    &   0\\
    0   &   0
    \end{matrix}\right)\, ,
    \label{sys-dhessian}
\end{align}
which is clearly positive-semidefinite for any choice of the geometric fluxes.

\ifSubfilesClassLoaded{%
\bibliography{biblio}%
}{}

\end{document}

\ifSubfilesClassLoaded{%
\tableofcontents
}{}

\chapter{10d Analysis Tools of Type IIA}
\label{ch: ap 10 Analysis Tools}

\section{10d equations of motion}
\label{bio-ap:10deom}

In this appendix we will discuss how the SUSY \eqref{cy-eq: solutionflux}  and the non-SUSY backgrounds \eqref{bio-solutionfluxnnosusy} solve the 10d equations of motion, introduced in section \ref{cy-subsec: democratic formulation}. We will limit ourselves to emphasize the contrast between the supersymmetric and non-supersymmetric cases.

\subsubsection*{Supersymmetric case}

In our approximation, the internal part of the first equation is
\begin{align}
\label{bio-ap1eq1}
0&=\mathrm{d}\left(e^{4A}\star_{\rm CY} \hat{G}_{2}\right)+e^{4A}H \wedge \star_{\rm CY} \hat{G}_{4}+\CO\left(g_s\right)=0+\CO\left(g_s\right)\, ,
\end{align}
where we have used that $\hat{G}_2$ is known up to $\CO\left(g_s\right)$ -see  \eqref{cy-eq: G2sol}. Since the natural scaling of a  $p$-form is $g_s^{-p/3}$, the total error we are making in solving  this equation is  $\CO(g_s^{8/3})$.

The internal part of second equation, at our level of approximation, reads
\begin{align}
0=d\left(e^{4A}\star_{\rm CY} \hat{G}_4\right)=4g_se^{4A}\hat{G}_0d\varphi\wedge J_{\text{CY}}+\frac{e^{4A}}{g_s}d\star_{\rm CY} \left(J_{\rm CY}\wedge d\im v\right)+\CO(g_s^2)\, ,
\label{bio-eqg4}
\end{align}
it is more or less straightforward to check that
\begin{align}
\label{bio-auxg4}
\frac{1}{g_s} d\star_{\rm CY} \left(J_{\rm CY}\wedge d\im v\right)=4\hat{G}_0g_s\star_{\rm CY}\left( J_{\rm CY}\wedge d_c\varphi\right)=-4\hat{G}_0g_s J_{\rm CY}\wedge d\varphi\, ,
\end{align}
which cancels out the first term of  \eqref{bio-eqg4} and satisfies the equation up to order $ \CO(g_s^{3})$ compared to the natural scaling of a three-form.

It remains to check equation \eqref{bio-problem}, which is the most cumbersome. We will go term by term and write just the internal parts to make the computation clearer. At the level of approximation that we are working the second term in the r.h.s  is 
\begin{align}
e^{4A}\star_6 \hat{G}_4\wedge \hat{G}_2=\frac{12}{5}\hat{G}_0 d\varphi\wedge\im\Omega_{\rm CY}-\frac{3}{5}\hat{G}_0\star_{\rm CY}\hat{G}_2+\CO(g_s)\, ,
\end{align}
while the first term reads
\begin{align}\nonumber
d\left(e^{-2\phi+4A}\star _6 H\right)& =d\left[ e^{-2\phi+4A}\star_6\left(\frac{2}{5}\hat{G}_0g_se^{-A}\re\Omega-\frac{1}{2}d\re\left(\bar{v}\cdot\Omega_{\rm CY}\right)\right)\right]+\CO(g_s)\\
& = \frac{2 }{5}\hat{G}_0g_sd\left(e^{-2\phi+3A}\im\Omega\right)-\frac{e^{-2\phi+4A}}{2}d\star_6d\re\left(\bar{v}\cdot\Omega_{\rm CY}\right)+\CO(g_s)\, .
\label{bio-ap:divided}
\end{align}
The first contribution to \eqref{bio-ap:divided} can then be rewritten as
\begin{align}
\frac{2 }{5}\hat{G}_0g_sd\left(e^{-2\phi+3A}\im\Omega\right)=\frac{2\hat{G}_0}{5}\left(4d\varphi\im\Omega_{\rm CY}-\star_{\rm CY}\hat{G}_2\right)+\CO(g_s)\, ,
\end{align}
whereas for the second contribution, a long calculation shows that
\begin{align}
-\frac{e^{-2\phi+4A}}{2}d\star_6d\re\left(\bar{v}\cdot\Omega_{\rm CY}\right)=-4\hat{G}_0 d\varphi\wedge\im\Omega_{\rm CY}+\CO(g_s)\, .
\end{align}
Finally, putting everything together \eqref{bio-problem} reduces to 
\begin{align}
0=\left(\frac{12}{5}+\frac{8}{5}-4\right)e^{4A}\hat{G}_0d\varphi\wedge\im\Omega_{\rm CY}+\left(-\frac{2}{5}-\frac{3}{5}+1\right)e^{4A}\hat{G}_0\star_{\rm CY}\hat{G}_2+\CO(g_s)\, ,
\end{align}
which,  as a $4$-form equation, we are solving it with an error  $\CO(g_s^{7/3})$.

\subsubsection*{Non-supersymmetric case}

In the non-SUSY solution, only the fields $H$ and $\hat{G}_4$ change, so it is enough to check the equations involving these quantities.

Let us start by the Bianchi identities, which we ignored in the previous section. To start with we can look at
\begin{align}
\label{bio-bianH}
d \hat{G}_4= \hat{G}_2\wedge H\, .
\end{align}
The changes in $\hat{G}_4^{\text{non-SUSY}}$ appear in the harmonic and the closed parts, so the LHS is the same as the  $\hat{G}_4^{\text{SUSY}}$. The changes in $H^{\text{non-SUSY}}$ are of order $\CO(g_s^{7/3})$, giving a  contribution beyond the order at which \eqref{bio-bianH} is being solved: we can ignore them and recover the RHS of the SUSY solution as well. The other BIs which could be sensitive to the non-SUSY novelties are  $d\hat{G}_2$ and $dH$. For both of them, the changes appear beyond the order of approximation in which they are being solved, so we can just neglect them.

Regarding the equations of motion, for $\hat{G}_4$  the internal part now reads
\begin{align}
d\left(e^{4A}\star_{\rm CY} \hat{G}_4\right)=-\frac{24}{5} g_se^{4A}\hat{G}_0d\varphi\wedge J_{\text{CY}}-\frac{6e^{4A}}{5g_s}d\star_{\rm CY} \left(J_{\rm CY}\wedge d\im v\right)+\CO(g_s^2)=0+\CO(g_s^2)\, ,
\end{align}
where we have used \eqref{bio-auxg4}. As in the SUSY case, it is solved at total order $\CO(g_s^3)$. 

Finally, the equation for $H$ is again the most tedious. Following the reasoning of the previous section, we will directly write each of the contributions to the internal part. On the one side
\begin{align}
e^{4A}\star_6 \hat{G}_4\wedge \hat{G}_2=-\frac{12}{5}\hat{G}_0 d\varphi\wedge\im\Omega_{\rm CY}+\frac{3}{5}\hat{G}_0\star_{\rm CY}\hat{G}_2+\CO(g_s)\, ,
\end{align}
on the other side
\begin{align}
d\left(e^{-2\phi+4A}\star _6 H\right)=\frac{12}{5}\hat{G}_0d\varphi\wedge\im\Omega_{\rm CY}-\frac{8}{5}\hat{G}_0\star_{\rm CY}\hat{G}_2+\mathcal{O}(g_s)\, ,
\end{align}
and \eqref{bio-problem} reduces to
\begin{align}
0=\left(-\frac{12}{5}+\frac{12}{5}\right)e^{4A}\hat{G}_0d\varphi\wedge\im\Omega_{\rm CY}+\left(-\frac{8}{5}+\frac{3}{5}+1\right)e^{4A}\hat{G}_0\star_{\rm CY}\hat{G}_2+\CO(g_s)\, ,
\end{align}
which is again solved  at total order $\CO(g_s^{7/3})$.

\subsubsection*{Einstein and dilaton equations}

To show how our expressions satisfy these two last constraints (specified in \eqref{cy-eq: dilaton eom} and \eqref{cy-eq: Einstein eom}), we will use the results derived in \cite{Junghans:2020acz}, focusing again on the changes introduced by the non-SUSY case. At leading order the equations evaluated for the non-SUSY solution coincide with the equations evaluated in the SUSY background, so they are satisfied in the first case provided they are solved in the second case -as it happens-. When the changes come into play, they do it at least at order $|F_4|^2\sim e^{-2\phi}|H|^2\sim \CO(g_s^{4/3})$. Nevertheless, to solve the equations at this order, we need to consider terms in $e^{A}$ and $e^{-\phi}$ which are beyond our approximation. 
  In other words, the modifications introduced in the non-SUSY case are seen by the Einstein and dilaton equations at the next order in the expansion.


\section{DBI computation}
\label{bio-ap:dbi}

The BIonic D8-brane system of section \ref{bio-s:bion} is defined by the profile \eqref{bio-BIonrel} for the transverse D8-brane position. In this appendix we check that this relation fulfils the basic requirement of a BPS condition, in the sense that it linearizes the DBI action of the D8-brane, at least at the level of approximation at which we work in the main text. 

The DBI action of a D8-brane wrapping $X_6$ is given by 
\be
 S_{\rm DBI}^{\rm D8} = - \frac{2\pi}{\ell_s^9} \int dt dx^1 dx^2 \int_{X_6} d^6 \xi   e^{3A - \phi} e^{\frac{3Z}{R}} \sqrt{\det \left(g_{ab} + \p_aZ\p_bZ +\cF_{ab} \right) } \, ,
 \label{bio-ap:DBID8}
\ee
where the D8-brane transverse position $Z$ is seen as a function on $X_6$. For BPS configurations the integrand simplifies, in the sense that the square root linearizes and corresponds to integrating a six-form over $X_6$. To see how this happens for the BIon configuration, let us use the matrix determinant lemma to rewrite things as
\be
\det \left(g_{ab} + \p_aZ\p_bZ +\cF_{ab} \right)  =  \det g \, \det \left(\II + g^{-1} \cF\right) \left( 1 + \p Z \cdot (g+\cF)^{-1} \cdot \p Z\right)\, .
\label{bio-inidet}
\ee
Then using that $\cF$ is antisymmetric one can deduce that
\begin{align}
 \det \left(\II + g^{-1} \cF\right) = 1 - \frac{t_2}{2} + \frac{t_2^2}{8} -\frac{t_4}{4} + \frac{\det  \cF}{\det g}\, ,
\end{align}
where  $t_k = \Tr\, g^{-1} \cF^k$. Using in addition the Woodbury matrix identity we obtain 
\be
\p Z \cdot (g+\cF)^{-1} \cdot \p Z = \p Z \cdot \sum_{k=0}^\infty \left(g^{-1} \cF\right)^{2k} g^{-1} \cdot \p Z\, .
\ee

One may then combine all these expressions to compute \eqref{bio-inidet}. Recall however that our unsmeared background description is only accurate below $\cO(g_s^2)$ corrections in the $g_s$ expansion. As pointed out in  \cite{Junghans:2020acz,Marchesano:2020qvg} a flux of the form \eqref{bio-cfsol} is suppressed as $\cO(g_s^{3/2})$ compared to a harmonic two-form and, because of \eqref{bio-BIonrel}, the same suppression holds for $\p Z$. This means that we are only interested in terms up to quadratic order in the worldvolume flux or $\p Z$ in the DBI action, or equivalently up to quartic order in \eqref{bio-inidet}. That is, we are interested in computing the following terms
\be
\left(1 - \oh \Tr \tilde\cF^2 \right) \left(1 + (\p Z)^2\right)   + \frac{1}{8} \left( \Tr \tilde\cF^2\right)^2 -\frac{1}{4}  \Tr \tilde\cF^4 - \left( \p Z\cdot  \tilde \cF \right)^2 \, ,
\ee
where $\tilde \cF \equiv g_{\rm CY}^{-1} \cF$, and $(\p Z)^2 = g_{\rm CY}^{ab} \p_a Z \p_b Z$, etc. To proceed we split the worldvolume flux as in section \ref{bio-ss:bionnosusy}
\be
\tilde \cF_1 \equiv g^{-1}  \cF^{(1,1)} \, , \qquad \tilde \cF_2 \equiv g^{-1} \cF^{(2,0)+(0,2)} \, ,
\ee
assuming that $\cF^{(1,1)}$ is primitive, and use the following identity
\be
\Tr \tilde\cF^4 = \frac{1}{4} \left(\Tr \tilde \cF^2 \right)^2 +  \left(\Tr \tilde \cF_1^2 \right)\left(\Tr \tilde \cF_2^2 \right) + 4 \Tr \left([\tilde \cF_1, \tilde \cF_2]^2\right) \, ,
\ee
to arrive to
\be
\left( 1 - \frac{1}{4} \Tr \tilde\cF^2 + \oh (\p Z)^2 \right)^2 -  \frac{1}{4}\left((\p Z)^2  \Tr \tilde\cF^2 +  \Tr \tilde \cF_1^2 \, \Tr \tilde \cF_2^2 +  (\p Z)^4 \right)  -\Tr \left([\tilde \cF_1, \tilde \cF_2]^2\right) - \left( \p Z\cdot  \tilde \cF \right)^2\, .
\ee
Finally, one can see that \eqref{bio-BIonrel} and primitivity imply that
\be
 (\p Z)^2 = -  \Tr \tilde\cF_2^2\, , \qquad \left( \p Z\cdot  \tilde \cF \right)^2 = \left( \p Z\cdot  \tilde \cF_1 \right)^2 = - \Tr \left([\tilde \cF_1, \tilde \cF_2]^2\right)\, ,
\ee
and so we are left with
\be
\left( 1 - \frac{1}{4} \Tr \tilde\cF^2 + \oh (\p Z)^2 \right)^2 = \left( 1 - \frac{1}{4} \Tr \left(\tilde\cF_1^2  - \tilde\cF_2^2\right) + (\p Z)^2 \right)^2\, .
\ee
When plugged into \eqref{bio-ap:DBID8} this translates into
\be
 S_{\rm DBI}^{\rm D8} = - \frac{2\pi}{\ell_s^9} \int dt dx^1 dx^2  g_s^{-1} e^{\frac{3z_0}{R}} \int_{X_6} \left[-\frac{1}{6}J_{\rm CY}^3 + \oh J_{\rm CY} \wedge \cF^2 + \star_{\rm CY} dZ \wedge dZ + \cO(g_s^{4/3}) \right]
 \label{bio-ap:DBID8fin}
\ee
where we used that in our approximation $\cF_1 \equiv \cF^{(1,1)}$ is a primitive (1,1)-from, and as a result $-\oh \Tr \tilde \cF_1^2 d{\rm vol}_{X_6} = \star_{\rm CY} \cF_1 \wedge \cF_1 = J_{\rm CY} \wedge \cF_1 \wedge \cF_1$. Finally, we have expanded $e^{3A-\phi} = g_s^{-1} + \cO(g_s)$ and $e^{\frac{3Z}{R}} = e^{\frac{3z_0}{R}}\left(1 - \frac{12 \ell_s}{|m|R}  \varphi \right) + \cO(g_s^{8/3})$, and used that $\int_{X_6} \varphi = 0$.


\section{BIonic strings and SU(4) instantons}
\label{bio-ap:IIBion}

The BIonic solution found in section \ref{bio-s:bion} is not unique of type IIA flux compactifications. It can also be found when one wraps a D7-brane on the whole internal manifold of type IIB warped Calabi--Yau compactifications with background three-form fluxes. The advantage of this type IIB setup compared to the type IIA one considered in the main text is two-fold: {\it i)} we know the exact 10d background and  {\it ii)} we can directly connect it to the Abelian $SU(4)$ instanton solutions that define Donaldson--Thomas theory \cite{Donaldson:1996kp}.

\subsubsection*{IIB BIonic strings}

Let us consider a type IIB warped Calabi--Yau compactification, namely a metric background of the form
\be
ds^2 = e^{2A}ds^2_{\pr^{1,3}} + e^{-2A} ds^2_{X_6}\, ,
\ee
where $X_6$ is endowed with a Calabi--Yau metric. On top of it we can add background fluxes $H$ and $F_3$ which are quantized harmonic three-forms of $X_6$ sourcing the warp factor. Let us consider the case in which $\ell_s^{-2} [H]$ is Poincar\'e dual to a three-cycle class with a special Lagrangian representative $\Pi$ calibrated by $\im \Omega_{\rm CY}$. That is:
\be
 \ell_s^{-2} [H] = {\rm P.D.} [\Pi]  = \ell_s^{-3} \d (\Pi)\, ,
\ee
where $\d (\Pi)$ is the bump delta-function of $X_6$ with support in $\Pi$. 

We now wrap a D7-brane on the internal six-dimensional space, as in \cite[section 6]{Evslin:2007ti}, and extended along $(t,x^1,0,0)$. The Freed--Witten anomaly induced by the $H$-flux can be cured by a D5-brane wrapping $-\Pi$, extended along $(t, x^1,0, x^3>0)$ and ending on the D7-brane. This configuration describes a 4d string to which a 4d membrane is attached. Microscopically this is due to the Freed--Witten anomaly. Macroscopically it as a result of the type IIB axion $C_0$ gaining an F-term axion-monodromy potential generated by the internal $H$-flux \cite{BerasaluceGonzalez:2012zn,Marchesano:2014mla,Blumenhagen:2014gta}. 

The Bianchi identity for the D7-brane worldvolume flux reads
\be
d\cF =  H - \ell_s^{-1}  \d (\Pi)\, ,
\label{bio-ap:BIF}
\ee
and finding its solution works as in \cite[section 5]{Marchesano:2020qvg}, see also \cite[section 3.4]{Hitchin:1999fh}. We have that
\be
\ell_s^{-1}\cF = d^{\dag}_{\rm CY} K = - J_{\rm CY} \cdot d \left( \hat\varphi \im \Omega_{\rm CY} - \star_{\rm CY} K\right) \, ,
\ee
up to a harmonic piece. Here the function $\hat{\varphi}$ satisfies $\int_{X_6} \hat{\varphi} =0$ and 
\be
\Delta_{\rm CY}  \hat \varphi = \left(\frac{{\cal V}_{\Pi}}{{\cal V}_{\rm CY}} - \delta^{(3)}_{\Pi}\right) \, , \qquad \delta_{\Pi}^{(3)} = \star_{\rm CY} \left[ \im \Omega \wedge \d (\Pi)\right]\, ,
\ee
while the three-form current $K$ is defined as in \eqref{cy-eq: formK} with the replacement $\varphi \to \hat{\varphi}/4$. The main difference with respect to the type IIA solution is that this one is exact. The 10d BPS configurations is therefore described by a BIon solution with profile
\be
\star_{\rm CY} dX^3 =  \im \Omega_{\rm CY} \wedge \cF\, ,
\ee
from where we deduce that $X^3 = -  \ell_s \hat{\varphi}$. This would correspond to a DBI action such that
\bea\nonumber
S_\text{DBI}^{\rm D7} &= & - \frac{2\pi}{\ell_s^9} \int dt dx^1 dx^2  g_s^{-1}  \int_{X_6} e^{2A} \sqrt{\det \left(g_{ab} + e^{2A} \p_a X^3 \p_b X^3  + \mathcal{F}_{ab}\right)} \\ 
& =&- \frac{2\pi}{\ell_s^9} \int dt dx^1 dx^2  g_s^{-1} \int_{X_6}  - \frac{e^{-4A}}{6}J^3_{\rm CY} + \frac{1}{2} \cF \wedge \cF \wedge J_{\rm CY}  +  \star_{\rm CY} dX^3 \wedge  dX^3\, ,
\label{bio-DBI}
\eea
as would follow from the results of \cite{Evslin:2007ti}. 

Besides being an exact solution, the D7-brane setup has the interesting feature that the transverse space to the D7 is given by $\pr\times S^1$. As a result one is able to relate the D7 BIon system to a gauge configuration that is defined on $\pr\times S^1 \times X_6$. The natural object where such a gauge theory is defined is a  D9-brane dual to the BIonic D7-brane. As we will now discuss, this construction leads us directly to the setup where Donaldson--Thomas theory is defined. 

\subsubsection*{The Donaldson--Thomas setup}

In a Calabi--Yau four-fold $X_8$ we can define a complex star operator $*$ that maps a $(0,q)$-form $\alpha$ to a $(0,4-q)$-form $* \a$ such that 
\be
\alpha \wedge * \alpha =\frac{1}{4} |\alpha|^2 \bar{\Omega}
\ee
where $\Omega$ is the holomorphic four-form of $X_8$, normalized such that $\Omega \wedge \bar{\Omega} = 16 \, d{\rm vol}_{X_8}$. It turns out that $*$ maps $(0,2)$-forms to $(0,2)$-forms, and that $*^2 =1$. One can then define two eigenspaces of $(0,2)$-forms such that $* \alpha_\pm = \pm \a_\pm$. In particular, one may take the $(0,2)$-component of a real non-Abelian gauge flux $F$ on $X_8$ and demand that $* F^{0,2} = - F^{0,2}$, or in other words that $F^{0,2}_+ =0$. This is one of the conditions of Donaldson--Thomas $SU(4)$ instanton equations  \cite{Donaldson:1996kp}, that read
\begin{subequations}
\label{bio-DT}
\begin{align}
\label{bio-DT1}
       F_+^{0,2}  & = 0\, ,\\
       F \wedge J^3 & = 0 \, .
        \label{bio-DT2}
\end{align}
\end{subequations}

To connect with the more familiar Hodge star operator $\star$, one can use that, when acting on $(0,q)$-forms, $\bar{\star} = \frac{1}{4}  \Omega \wedge *$ \cite{Baulieu:1997jx}. Therefore we deduce that
\be
\star F^{0,2}_\pm = \pm \frac{1}{4} \bar \Omega \wedge F^{2,0}_\pm\, .
\label{bio-DTH}
\ee
From here we deduce that $ F_\pm^{0,2} =0$ is equivalent to
\begin{subequations}
\label{bio-DTre}
\begin{align}
\label{bio-DTrere}
\star \re F^{0,2} & = \pm\frac{1}{4} \re \Omega \wedge F\, , \\
\star  \im F^{0,2} & = \mp \frac{1}{4}\im \Omega \wedge F \ \implies \ F \wedge F \wedge \im \Omega = 0\, .
\label{bio-DTreim}
\end{align}
\end{subequations}
and also implies
\be
\Tr \left(\re F^{0,2} \wedge \star \re F^{0,2}\right) = \frac{1}{4} \Tr \left(\re F^{0,2}_+ \wedge \re F^{0,2}_+  - \re F^{0,2}_- \wedge \re F^{0,2}_- \right) \wedge \re \Omega\, .
\label{bio-splitFpm}
\ee

\subsubsection*{The dictionary}

To connect with the D7 BIon configuration, we consider the Donaldson--Thomas equations for an Abelian gauge theory in the following Calabi--Yau background
\be
  \pr  \times S^1 \times X_6\, ,
\label{bio-DTsetup}
\ee
with complex coordinates $\{ \omega =  x + i\theta, z^1, z^2, z^3\}$ and holomorphic four-form
\be
\Omega_4 = \left(dx + i d\theta\right) \wedge \Omega_3 \, .
\ee
We now consider a gauge field strength of the form 
\be
\cF = \cF_{X_6} + \cF_{\rm Bion}\, ,
\label{bio-fluxDT}
\ee
where $\cF_{X_6}$ is a two-form on $X_6$ and 
\be
\cF_{\rm Bion} = F_{xi} \, dx \wedge dz^i + {\rm c.c.}
\ee
so that there is no component of the flux along $d\theta$, and as a result $\cF^4 = 0$. 

The dictionary with the D7 BIon configuration can then be done by simple dimensional reduction along $\pr \times S^1$. After that, we recover a gauge theory on $X_6$ with gauge field strength $\cF_{X_6}$ and a non-trivial profile for the transverse position field $X$, seen as a function on $X_6$
\be
\p X = - F_{xi} dz^i\, .
\ee
Notice that
\be
\cF_{\rm BIon} =  dZ \wedge dx =  \frac{1}{2}\left( \p X + \bar{\p} X \right) \wedge \left( d\omega + d\bar{\omega}\right)  \implies \cF_{\rm BIon}^{0,2} = - \frac{1}{2} d\bar{\omega} \wedge  \bar{\p} Z\, .
\ee
Therefore to satisfy \eqref{bio-DT1} we need to impose 
\be
d\bar{\omega} \wedge  \bar{\p} X = - \oh \star_4 \left(\bar{\Omega}_4 \wedge \cF_{X_6}\right) \implies   \bar{\p} X  = \frac{i}{2}  \star_{X_6}  \left( \bar{\Omega}_3 \wedge \cF_{X_6}\right)\, ,
\ee
from where we deduce the following relations
\bea
\label{bio-BIon1}
\star_{X_6} dZ & = &   \im \Omega_3 \wedge \cF_{X_6} \, ,\\
\star_{X_6} d^c Z  & = &   \re \Omega_3 \wedge \cF_{X_6}\, .
\label{bio-BIon2}
\eea
Eq.\eqref{bio-BIon1} corresponds to the BIon equation of section \ref{bio-s:bion}, while \eqref{bio-BIon2} looks like a new, independent equation. In principle we would expect that it is also satisfied by the BIon solution, and so it would be interesting to understand its implications. Notice that we can translate \eqref{bio-DT1} into the condition 
\be
\cF_{X_6}^{0,2} = \frac{1}{8} \star_4 \left( \bar{\Omega}_4 \wedge \p X \wedge d\omega\right) \implies \cF_{X_6}^{0,2} = -\frac{i}{4} \star_3  \left(\bar{\Omega}_3 \wedge \p X \right) \, ,
\ee
which in turn implies
\bea
\label{bio-BIon3}
\re \cF_{X_6}^{2,0}  & = & - \frac{1}{4} \star_3  \left(dX \wedge\im \Omega_3 \right) = \frac{1}{4} \star_3  d\left(\hat{\varphi} \im \Omega_3 \right) \, ,\\
\im \cF_{X_6}^{2,0}  & = & - \frac{1}{4} \star_3  \left(dX \wedge\re \Omega_3 \right) = \frac{1}{4} \star_3  d \left(\hat{\varphi} \re \Omega_3 \right) \, .
\label{bio-BIon4}
\eea
Eq.\eqref{bio-BIon3} corresponds to \eqref{bio-cF2} adapted to this setup, while \eqref{bio-BIon4} is equivalent to \eqref{bio-BIon2}. Finally, imposing \eqref{bio-DT2} amounts to require that $\cF_{X_6}$ is primitive, as the BIon solution fulfils. 

The relation between the solutions to the Bianchi identity of the form \eqref{bio-ap:BIF} and the Abelian SU(4) instanton equations of \cite{Donaldson:1996kp} was already pointed out in \cite[section 3.4]{Hitchin:1999fh}. We find it quite amusing that a BIonic D7-brane and the corresponding worldvolume flux on a D9-brane give a neat physical realisation of this correspondence. It would be interesting to understand if this description has any implications for the theory of invariants developed in \cite{Donaldson:1996kp}.


\section{Moduli stabilization in \texorpdfstring{$T^6/(\IZ_2 \times \IZ_2)$}{T6/Z2xZ2}}
\label{mem-ap:Z2xZ2}

In this appendix we consider the moduli stabilization of the K\"ahler sector in the $T^6/(\IZ_2 \times \IZ_2)$ orientifold background with $(h^{1,1}, h^{2,1})_{\rm orb} = (51,3)$. As in \cite{DeWolfe:2005uu,Ihl:2006pp}, we look for vacua where the twisted two- and four-cycles are blown up due to the presence of background four-form fluxes. As pointed out in \cite{Marchesano:2019hfb}, for the class of type IIA flux vacua analyzed in the main text the K\"ahler moduli stabilization conditions amount to
\be
 {\cal K}_a = - \epsilon \frac{10}{3m} \hat{e}_a\, , \qquad \hat{e}_a \coloneqq e_a - \oh \frac{{\cal K}_{abc}m^bm^c}{m} - \oh \cK_{aab} m^b + mK_a^{(2)} \, ,
 \label{mem-Kahler}
\ee
where
\be
\CK_a = - \int_{X_6} J_{\rm CY} \wedge J_{\rm CY} \wedge \omega_a\, , \quad e_a = - \frac{1}{\ell_s^5} \int_{X_6} \bar{G}_4  \wedge \omega_a \in \IZ\, , \quad m^a =   \frac{1}{\ell_s^5} \int_{X_6} \bar{G}_2 \wedge \tilde{\omega}^a  \in \IZ \, ,
\ee
and $\eps=\pm 1$ distinguishes between supersymmetric and non-supersymmetric vacua, as in eq.\eqref{mem-intflux}. The connection with this set of equations can be made by taking into account the dependence of $G_4$ on $\bar{G}_4$, $\bar{G}_2$, $G_0$ and the B-field axions, something that it is usually done in the smearing approximation. In any event, in the following we will consider compactifications where $m^a =0$, so that these subtleties disappear and \eqref{mem-Kahler} simplifies. 

To look for solutions to this equation we need to compute the quantity $\oh\CK_a$, that in our conventions measures the volume of holomorphic four-cycles or divisors. For this we need to parametrize the K\"ahler form in terms of such divisors, including the exceptional ones, and compute their triple intersection numbers. This exercise was done in \cite{Denef:2005mm} for the above orbifold background $T^6/(\IZ_2 \times \IZ_2)$ with a type IIB orientifold projection that leads to O3- and O7-planes. Notice that the orientifold projection that we are interested in is different, as it leads to type IIA O6-planes. Therefore, we will take the approach of \cite{Marchesano:2019hfb} and solve \eqref{mem-Kahler} for the unorientifolded orbifold geometry $T^6/(\IZ_2 \times \IZ_2)$. Then, following the remarks in section \ref{mem-ss:fluxquant}, we will demand that $e_a \in 2\IZ$ for the four-form flux quanta defined in the covering space  $T^6/(\IZ_2 \times \IZ_2)$. The necessary topological data for this case can be extracted from the results of \cite{Lust:2006zh,Reffert:2006du}.

The K\"ahler form for the blown-up orbifold $T^6/(\IZ_2 \times \IZ_2)$ reads
\be
J = r_i R_i - t_{1\a, 2\b} E_{1\a, 2\b} - t_{2\b,3\g} E_{2\b,3\g} - t_{3\g,1\a} E_{3\g,1\a} \, , 
\ee
where $\a, \b, \g$ run over the four fixed points of a given $T^2$. Also, $R_i$ and $E_A \equiv E_{i\a, j\b}$ correspond to divisors that satisfy the linear equivalence relation  \cite{Lust:2006zh,Reffert:2006du}
\be
R_1 \simeq 2D_{1\a} + \sum_\b E_{1\a, 2\b} + \sum_\g E_{3\g,1\a} \qquad \forall \a\, ,
\label{mem-Ris}
\ee
that differs by a factor of 2 compared to \cite[eq.(6.2)]{Denef:2005mm}, due to the lack of orientifold action.  Similar relations hold for $R_{2}$ and $R_3$. With these conventions and assuming the symmetric resolution of \cite{Denef:2005mm} one finds that the intersection form is given by 
\be
\begin{aligned}
{\cal I}=&\, 2R_1R_2R_3-2\Big(\sum_{\alpha\beta}E^2_{1\alpha,2\beta}R_3+\ldots\Big)+4\Big(\sum_{\alpha\beta}E^3_{1\alpha,2\beta}+\ldots\Big)\\
&-\Big[\sum_{\alpha\beta\gamma}E_{1\alpha,2\beta}(E^2_{2\beta,3\gamma}+E^2_{3\gamma,1\alpha})+\ldots\Big]+\sum_{\alpha\beta\gamma}E_{1\alpha,2\beta}E_{2\beta,3\gamma}E_{3\gamma,1\alpha}
\end{aligned}
\ee
where $\ldots$ are $(1,2,3)$ cyclically permuted terms. This matches the results of \cite[section B.19.4]{Reffert:2006du}. 

With this intersection form one can compute the quantity ${\cal K}_a$ for each divisor $R_i$ and $E_{i\a, j\b}$. For simplicity we assume that all twisted moduli and untwisted moduli are equal among them:
\be
r_i = r\, , \qquad t_A \equiv t_{i\a, j\b} = t\, .
\ee
One then obtains that
\be
{\cal K}_i = 4r^2 - 32 t^2\, \qquad {\cal K}_A = 4rt - 12t^2\, .
\ee
A sensible flux Ansatz to solve \eqref{mem-Kahler} is $e_i = e$ and $e_A \equiv  e_{i\a, j\b} = f$, with $e, f \in 2\IZ$. Equation \eqref{mem-Kahler} then reduces to
\be
4r^2 - 32 t^2 = -\epsilon \frac{10}{3m} e \, , \qquad 4rt - 12t^2 = -\epsilon \frac{10}{3m} f\, .
\ee
Since four-form flux quanta are not constrained by tadpoles, it is easy to choose values for $e,f$ such that ${\cal K}_i$ and ${\cal K}_A$ are positive and $r \gg t$. Let us parametrize a solution as $r = xt$, with $x \gg 1$. For supersymmetric vacua $(\epsilon=-1)$ we obtain
\be
10 e = 3m (4x^2-32) t^2\, , \qquad 10 f = 3m (4x-12) t^2\, ,
\ee
and so 
\be
e = \frac{x^2 -8}{x-3} f\, .
\ee
It is thus simple to find reasonable solutions by taking $x \in \mathbb{N}$, like for instance $x = 10$,  $f=126m$ and $t = \sqrt{15}$. For non-supersymmetric vacua one should only flip the sign of the fluxes. 

What is important, though, is that the values for $r$ and $t$ correspond to the interior of the K\"ahler cone. From \cite[eq.(6.11)]{Denef:2005mm} this amounts to require that $r > 4t > 0$. This is satisfied as long as $t>0$ and $x> 4$, which is in general quite easy to achieve. 

\section{Curvature corrections in \texorpdfstring{$T^6/(\IZ_2 \times \IZ_2)$}{T6/Z2xZ2}}
\label{mem-ap:Z2xZ2curv}

In order to check the WGC for 4d membranes one needs to compute the curvature correction $ \Delta_{\rm D8}^{\rm curv}$. In this appendix we perform its computation for the case of $X_6 = T^6/(\IZ_2 \times \IZ_2)$, again assuming the symmetric resolution of \cite{Denef:2005mm}.\footnote{We would like to thank T.~Courdarchet and R.~Savelli for important discussions regarding this computation.}  For this, we use the result of this reference that claims that the divisors $D_{i \a}$ that appear in \eqref{mem-Ris} have the topology of $\P^1 \times \P^1$, and the exceptional divisors $E_{i\a, j\b}$ that of $\P^1 \times \P^1$ with four blown-up points. Using toric geometry techniques one can compute the intrinsic topological data for each of these divisors. The results are shown in table \ref{mem-table:divisors}, where we applied the relations
\be
c_2(X_6) . S = \chi(S) - S^3\, , \qquad {\rm and} \qquad 12\chi(\cO_S) = \chi(S)\ + S^3 \, .
\ee
\renewcommand{\arraystretch}{0.9}
\begin{table}[H]
$$
\begin{array}{|c|c|c|c|c|}
\hline S & K_S^2 & \chi(S)  & \chi(\cO_S) & c_2(X_6) . S  \\
\hline 
D_{i\a}  & 8 & 4 & 1 & -4 \\
\hline
E_{i\a, j\b} & 4 & 8 & 1 & 4\\
\hline
\end{array}
$$
\caption{Topological data of divisors on $T^6/(\mathbb{Z}_2\times \mathbb{Z}_2)$.}
\label{mem-table:divisors}
\end{table}

With these results it is easy to see that $c_2(X_6) . R_{i\a} = 24$, from where we obtain
\be
\frac{1}{24} c_2(X_6) . J = \sum_i r_i - \frac{1}{6} \sum_{\a,\b\g} \left( t_{1\a, 2\b} +  t_{2\b,3\g} +  t_{3\g,1\a}\right)\, .
\label{mem-c2J}
\ee
Going to the orbifold limit $t_{i \a, j\b} \to 0$, one recovers \eqref{mem-D8curvz2xz2} by using the dictionary  $T^i_{\rm D4} = e^{K/2} t^i = 2 e^{K/2} r_i$ that can be deduced from \eqref{mem-Ris}.

\ifSubfilesClassLoaded{%
\bibliography{biblio}%
}{}

\end{document}

\ifSubfilesClassLoaded{%
\tableofcontents
}{}

\chapter{Technical Aspects of F-theory vacua}

\section{Geometric interpretation of the \texorpdfstring{$\rho_A$}{pA}}
\label{Ft-ap:georho}

In this appendix we provide a geometric interpretation of the flux-axion polynomials $\rho_A$, introduced in section \ref{Ft-s:potential} to describe the scalar potential in regions of large complex structure, as well as of the saxion-dependent matrix $Z^{AB}$ that appears in \eqref{Ft-ZAB}. While our discussion is restricted to the large complex structure region, our reasoning can be easily generalized to other limits in which approximate axionic shift symmetries appear in the moduli space metric.

To understand the flux-axion polynomials geometrically, one may first realize that they can be seen as the components of the  flux $G_4$ in a particular basis of four-forms. More precisely we have that
\begin{align}
    G_4 = \bar \rho \tilde \alpha - \bar \rho^i \tilde \alpha_i + \bar{\rho}^\mu \tilde \sigma_\mu - \rho_i \tilde \beta^i + \rho \tilde \beta\, ,
    \label{Ft-ap:G4rho}
\end{align}
where we have defined
\begin{eqn}\label{Ft-apeq:tildealpha}
    \tilde \alpha &= \alpha + b^i \alpha_i +\frac{1}{2} b^i b^j \zeta^\mu_{ij} \sigma_\nu^Y +\frac{1}{6} \cK_{ijkl}b^ib^j b^k \beta^l +\frac{1}{24} \cK_{ijkl}b^ib^j b^k b^l\beta\,,\\
    \tilde \alpha_i &=\alpha_i + \zeta^{\mu}_{ij}b^j \sigma_\mu^Y + \frac{1}{2} \cK_{ijkl}b^j b^k \beta^l + \frac{1}{6} \cK_{ijkl}b^j b^k b^l \beta \,,\\
    \tilde \sigma_\mu &=\sigma_\mu^Y + \zeta_{\mu kl} b^k \beta^l +\frac{1}{2}\zeta_{\mu kl} b^kb^l \beta\,,\\
    \tilde \beta^i &= \beta^i + b^i \beta\,,\\
    \tilde \beta &= \beta \,. 
\end{eqn} 
The geometric interpretation of the $\rho$'s then boils down to the geometric significance of this tilded set of four-forms, in comparison with the basis of integer four-forms $\{ \a, \a_i, \sigma_\mu^Y, \b^i, \b\}$, that span the horizontal subspace  $H_H^4(Y_8)$. As one can check, two key properties properties of this new basis are that:
\begin{itemize}

\item[{\it i)}] It has the same intersection numbers as the initial basis $\{ \a, \a_i, \sigma_\mu^Y, \b^i, \b\}$.

\item[{\it ii)}] Their elements are invariant under monodromies around the large complex structure point.  

\end{itemize}

The first property can be easily checked by direct computation, and it implies the tadpole identity \eqref{Ft-Nfluxrho}. The second one follows from the characterisation of the large complex structure monodromies as \eqref{Ft-monoexp}, given that the monodromy generators $\hat{P}_i$ also specify the change of basis $\{ \a, \a_i, \sigma_\mu^Y, \b^i, \b\} \to \{ \tilde\a, \tilde\a_i, \tilde\sigma_\mu, \tilde\b^i, \tilde\b\}$.  Combined, these two properties also allows us to relate the saxion-dependent matrix $Z^{AB}$ with  the action of the Hodge star operator on the basis $\{ \tilde\a, \tilde\a_i, \tilde\sigma_\mu, \tilde\b^i, \tilde\b\}$.

Indeed, this tilded basis is particularly suitable to express monodromy-invariant quantities like the holomorphic four-form $\Omega$ and its derivatives. To simplify the discussion, let us ignore the contribution of the corrections $K_i^{(3)}$ to the expression of $\Omega$. That is, we consider the expression \eqref{Ft-Omega}, from where we find
\begin{align}\label{Ft-eqap:Omega}
    \Omega & = \tilde \alpha + i t^i \tilde \alpha_i - \frac{1}{2}\zeta^\mu \tilde \sigma_\mu - \frac{i}{6}\cK_i \tilde \beta^i + \frac{\cK}{24} \tilde \beta \,, \\
    \label{Ft-eqap:derOmega}
    D_i \Omega & =   \tilde \alpha_i + i \zeta^\mu_i \tilde \sigma_\mu - \frac{1}{2}\cK_{ik} \tilde \beta^k - \frac{i}{6}\cK_i \tilde \beta +\frac{2 i \cK_i}{\cK}\left[\tilde \alpha + i t^i\tilde \alpha_i-\frac{1}{2}\zeta^\mu \tilde \sigma_\mu- \frac{i}{6}\cK_i \tilde \beta^i+\frac{\cK}{24}\tilde \beta \right]\,, \\
    \label{Ft-eqap:derderOmega}
     D_iD_j \Omega & = \zeta_{ij}^\mu \tilde\sigma_\mu + i \cK_{ijk}\tilde \beta^k -\frac{1}{2} \cK_{ij} \tilde \beta -\left(g_{ij} + \frac{4 \cK_i \cK_j}{\cK^2}\right) \Omega + \frac{2i\cK_i}{\cK}\partial_{T^j} \Omega + \frac{2i\cK_j}{\cK}\partial_{T^i} \Omega\\
     \nonumber &+\left(\frac{2i \cK_{ij} t^k}{\cK} -\frac{2i}{\cK}\left(\delta_i^k\cK_j + \delta_j^k \cK_i) \right)+i \cK^{kl}\cK_{ijl}\right) D_k\Omega \, .
\end{align}
We now use the fact that the Hodge star operator has a simple action on each of these four-forms
\be
\star \Omega = \Omega\, ,\quad \star D_i \Omega = - D_i\Omega\, , \quad \star D_iD_j \Omega = D_i D_j\Omega\, ,
\ee
and in particular that 
\be
\frac{1}{3}t^it^j D_iD_j \Omega + \Omega = i t^i \tilde \alpha_i - \frac{2}{3} \zeta^\mu \tilde\sigma_\mu - \frac{i}{6} \cK_{i} \tilde \beta^i
\ee
is self dual. From the real part of this expression we obtain that 
\begin{align}\label{Ft-eqap:starsigma1}
    \star \left(\zeta^\mu \tilde \sigma_\mu\right)=\zeta^\mu \tilde \sigma_\mu \,,
\end{align}
and from its imaginary part that
\begin{align}
    \star \left(t^i \tilde \alpha_i -\frac{\cK_i}{6} \tilde \beta^i \right)&= t^i \tilde \alpha_i -\frac{\cK_i}{6} \tilde \beta^i\,.\label{Ft-eqap:alphaistar0}
\end{align}
In addition, using that $\star\Omega = \Omega$ and the above relations we obtain
\begin{align}
    \star \left(\tilde \alpha + \frac{\cK}{24} \tilde \beta\right) &= \tilde \alpha  + \frac{\cK}{24} \tilde \beta\, .
\end{align}
Moreover, from $\star D_i \Omega = - D_i\Omega$ we obtain the following two conditions
\begin{align}\label{Ft-eqap:alphaistar1}
     \star\left(\tilde \alpha_i -\frac{\cK_{ik}}{2}\tilde \beta^k\right) & = - \tilde \alpha_i + \frac{\cK_{ik}}{2}\tilde \beta^k -\frac{2}{3} \frac{\cK_i\cK_k}{\cK} \tilde \beta^k + \frac{4\cK_i}{\cK}\tilde \alpha_k t^k\,,\\
\star \left(\zeta^\mu_{i}\tilde \sigma_\mu-\frac{1}{6}\cK_i \tilde \beta \right) & =  - \zeta^{\mu}_i\tilde \sigma_\mu+ \frac{1}{6} \cK_i \tilde \beta -\frac{4\cK_i}{\cK} \left(\tilde \alpha -\frac{1}{2} \zeta^\mu \tilde \sigma_\mu +\frac{\cK}{24} \tilde \beta\right)\,,
\end{align}
where we used \eqref{Ft-eqap:alphaistar0}. Taking this into account as well as the above relations, one finds that the action of the Hodge star operator on the basis $\{ \tilde\a, \tilde\a_i, \tilde\sigma_\mu, \tilde\b^i, \tilde\b\}$ must be given by
\begin{eqn}\label{Ft-eqap:starrest}
    \star \tilde \alpha &= \frac{\cK}{24} \tilde \beta \,,\qquad\qquad  \star \tilde \beta= \frac{24}{\cK} \tilde \alpha\,,\\
    \star \tilde \alpha_i& = -\frac{1}{6} \cK g_{ij} \tilde \beta^i \,,\qquad  \star \tilde \beta^i = - \frac{6}{\cK}g^{ij} \tilde \alpha_j\,,
\end{eqn}
together with \eqref{Ft-eqap:starsigma1} and
\begin{align}\label{Ft-eqap:starsigma2}
    \star \left(\zeta^\mu_i -\frac{\cK_i}{\cK}\zeta^\mu \right)\tilde\sigma_\mu= -\left( \zeta^\mu_i - \frac{\cK_i}{\cK}\zeta^\mu\right) \tilde \sigma_\mu \,.
\end{align}

It is now easy to identify the action of the Hodge star with the diagonal entries of the saxion-dependent matrix \eqref{Ft-ZAB}. More precisely, we have that the matrix $2{\cal V}_3^2 Z + \chi_0$ defined in there corresponds to the entries of the standard four-form metric
\be
G^{AB} = \int_{Y^4} \omega^A \wedge \star \omega^B\, ,
\label{Ft-Hmetric}
\ee
with $\{\omega^A\} = \{ \tilde\a, \tilde\a_i, \tilde\sigma_\mu, \tilde\b^i, \tilde\b\}$, computed to the same level of approximation. In fact, to fully show this statement one must verify that
\begin{align}
    g_{\mu \nu} = \int_{Y_8} \tilde \sigma_\mu \wedge \star \tilde \sigma_\nu\,,
    \label{Ft-ap:sigmetric}
\end{align}
with $g_{\mu\nu}$ as defined below \eqref{Ft-ZABdiag}. This is easy to argue from the results above. For this, let us perform the decomposition
\be
\rho^\mu \tilde{\sigma}_\mu = \left(A\zeta^\mu + B^\mu +C^\mu\right) \tilde{\sigma}_\mu\, ,
\ee
with components such that
\begin{align}
    B^\mu=  \left(\zeta_i^\mu-\frac{\cK_i}{\cK} \zeta^\mu \right)\xi^i\,, \qquad \zeta_{\mu i}C^\mu=0\ \forall i\, ,
    \label{Ft-ap:sigmadec}
\end{align}
for some arbitrary vector $\xi^i$. This splitting is directly related to the decomposition introduced in \eqref{Ft-splitmu}, to which one can give a geometric meaning in terms of self-duality properties. Indeed, it follows from \eqref{Ft-eqap:starsigma1} and \eqref{Ft-eqap:starsigma2} that the first and second components are Hodge self-dual and anti-self-dual, respectively, and it is easy to convince oneself (either using mirror symmetry or \eqref{Ft-eqap:derderOmega}) that $\star C^\mu \tilde\sigma_\mu = C^\mu \tilde\sigma_\mu$. Putting all these together we have 
\begin{align}
\rho^\mu \rho^\nu \int_{Y_8} \tilde \sigma_\mu \wedge \star \tilde \sigma_\nu & = \rho^\mu \eta_{\mu\nu}\rho^\nu  - 2 B^\mu \eta_{\mu\nu} B^\nu = \rho^\mu \eta_{\mu \nu}\rho^\nu - 2 \xi^i \left(\cK_{ij} - \frac{\cK_i \cK_j}{\cK} \right) \xi^j \nonumber \\
& = \rho^\mu \eta_{\mu \nu}\rho^\nu-2 \rho^\mu \left(\cK^{ij} - \cK^{-1} t^i t^j \right)\zeta_{\mu i}\zeta_{\nu j} \rho^\nu\,,
\end{align}
where we have used that $ \xi^i (\cK_{ij} - \frac{\cK_i \cK_j}{\cK})= (\zeta_{\mu j} -\frac{\cK_j}{\cK} \zeta_\mu)\rho^\mu$, and so \eqref{Ft-ap:sigmetric} follows.

Notice that our results imply a prescription to construct the flux-axion polynomials $\rho_A$, without the knowledge of \eqref{Ft-apeq:tildealpha}, and that one can apply it to any other field space region with approximate axionic symmetries. Indeed, given a real integral basis of horizontal four-forms $\{\omega^A\}$ one may construct an alternative basis $\{\tilde \omega^A\}$ from axion-independent linear combinations of the real and imaginary parts of $\Omega$, $D_i\Omega$ and $D_iD_j\Omega$, so that the elements of the new basis are automatically monodromy-invariant. One must moreover choose the new basis such that $\chi^{AB} \equiv \int \tilde \omega^A \wedge\tilde \omega^B = \int \omega^A \wedge \omega^B$. We then define the flux-axion polynomials $\rho_A$ as the coefficients of the four-form flux in this basis, and the saxion-dependent matrix in terms of its Hodge and intersection products:
\be
G_4 = \rho_A \tilde{\omega}^A\, , \qquad Z^{AB} = \frac{1}{2{\cal V}_3^2}  \left(G^{AB} - \chi^{AB}\right)\, ,
\ee
with $G^{AB}$ defined as in \eqref{Ft-Hmetric}.


\section{Curvature corrections on four-folds}
\label{Ft-ap:curvature}

In this appendix we cover several technical details regarding the polynomial corrections discussed in section \ref{Ft-sec:poly}. In \ref{Ft-sap:corrper} we elaborate on the computation of the corrections to the  periods and the intersection matrix, both seen as curvature corrections in the dual Calabi--Yau four-fold $X_8$. In \ref{Ft-sap:corrka} we provide an alternative derivation of the corrected K\"ahler potential \eqref{Ft-Kcscorr}. In \ref{Ft-sap:corrFterm} we provide the flux potential including all the polynomial corrections. In \ref{Ft-sap:corrvac} we focus on the corrections to the F-terms, which we use to provide the corrected vacuum equations. 

\subsection{Corrected periods and intersection matrix}
\label{Ft-sap:corrper}

Section \ref{Ft-sec:poly} discusses the polynomial corrections to the four-fold periods in the large complex structure regime. These can be obtained via mirror symmetry from the central charges of B-branes wrapped on holomorphic $(2p)$-cycles in the mirror four-fold $X_8$. In the large volume regime the leading polynomial form of the central charge of a $(2p)$-brane that corresponds to a complex $\mathcal{E}$ is given by 
\begin{align}\label{Ft-eqap:centralcharge}
    Z(\mathcal{E})= \int_{X_8} e^J \Gamma_\mathbb{C}(X_8) \lambda\left(\text{ch}(\mathcal{E})\right) \,,
\end{align}
where $J$ is the complexified K\"ahler class. The Calabi--Yau $n$-fold complex $\Gamma$-class is given by 
\begin{align}\label{Ft-eqap:gammaclass}
    \Gamma_\mathbb{C} (X_n)= \sqrt{\text{Td}(X_n)} \exp(i\Lambda_{X_n})\,,
\end{align}
with $\text{Td}(X_n)$ the Todd class of $X_n$ and
\begin{align}
    \Lambda_{X_n} = \frac{\zeta(3)}{(2\pi)^3}c_3 + \frac{\zeta(5)}{(2\pi)^5}\left(c_5-c_2c_3\right) + \dots 
\end{align}
To evaluate these central charges one needs a basis of $(2p)$-branes, which we take as type IIA D$(2p)$-branes on a four-fold $X_8$. For $p\ne2$ such a basis was constructed in \cite{Gerhardus:2016iot}: the D8-brane wrapped on $X_8$ is associated with the structure sheaf $\mathcal{O}_{X_8}$ with Chern character $\text{ch}(\mathcal{O}_{X_8}) = 1$. 
A basis of D6-branes is given by the sheaves $\mathcal{O}_{D_i}$ with $D_i$ the generators of the K\"ahler cone. For these sheaves the Chern character is given by 
\begin{align}
    \text{ch}(\mathcal{O}_{D_i})=D_i -\frac{1}{2} D_i^2 + \frac{1}{6} D_i^3 - \frac{1}{24} D_i^4\,. 
\end{align}
A basis for D2-branes is obtained from the Mori cone generators $C^i$ via $\mathcal{C}^i= \iota_!\mathcal{O}_{C^i}\left(K^{1/2}_{C^i}\right)$ for which the Chern character is simply
\begin{align}
    \text{ch}(\mathcal{C}^i)= C^i \,. 
\end{align}
Finally, as shown in \cite{Cota:2017aal} in many cases a basis of D4-branes can be constructed from the intersection of two divisors $D_i . D_j$. The Chern character of the associated sheaf $\mathcal{O}_{D_i. D_j}$ is then 
\begin{align}
    \text{ch}(\mathcal{O}_{D_i . D_j}) = D_i . D_j -\frac{1}{2} D_i.D_j. \left(D_i + D_j\right) + \frac{1}{12}D_i. D_j \left(2 D_i^2 + 3 D_iD_j + 2 D_j^2\right)\,.
\end{align}
Using these expressions for the Chern characters, the central charges in \eqref{Ft-eqap:centralcharge} can be explicitly evaluated yielding the periods \eqref{Ft-eq:corrper}. Let us stress that these expressions for the central charges are valid in the large volume regime. Away from these limits in principle exponential corrections need to be taken into account that do not necessarily converge in the entire classical K\"ahler cone. In order to ensure that we are in the regime of validity of the polynomial approximation to the central charges we impose that the classical contribution to the central charges of $8$-, $6$- and $4$-branes is suitably large. We will in particular assume that the curvature corrections due to $c_i(X_8)$ are small compared to the leading polynomial expression in \eqref{Ft-eq:corrper}. As an example of what this constraint entails, let us consider the central charge of a D6-brane on a divisor $D_i$ that satisfies $D_i.D_i.D_i.D_j=0$, $\forall D_j$.   In the limit of large $t^i$ we find that
\begin{eqn}
    Z(\mathcal{O}_{D_i}) = - \frac{1}{6} \cK_{iijk} T^i T^j T^k - \frac{1}{24} T^i \int c_2 \wedge D_i\wedge D_i + \dots = -T^i\left(\frac{1}{6} \cK_{iijk} T^j T^k + K_{ii}^{(2)}\right) +\dots
\end{eqn}
Since the term in the brackets is constant for large values of $t^i$ we see that for the curvature correction to be subleading we need to impose
\begin{align}\label{Ft-eqap:constraintK2}
     \frac{1}{6} \cK_{iijk} t^j t^k > |K_{ii}^{(2)}|\,, 
\end{align} 
which is a condition on the other saxions. For a related discussion of the role of the second Chern class for the validity of the perturbative expansion in  type IIA on CY three-folds, see \cite{Lee:2019wij}.

Besides the periods, to extract the form of the flux potential we also need the corrected intersection matrix $\chi$ associated to the integer basis of $2p$-cycles on the Calabi--Yau four-fold $X_8$. As reviewed in the main text, this intersection matrix is given by the open string index
\begin{align}\label{Ft-eqapp:chi}
    \chi(\mathcal{E}, \mathcal{F}) = \int_{X_8} \text{Td}(X_8) \lambda(\text{ch} \mathcal{E}) (\text{ch}\mathcal{F})\,,
\end{align}
where the Todd class is given by \eqref{Ft-eq: Todd} and $\mathcal{E}$ and $\mathcal{F}$ are complexes corresponding to the branes wrapped on the $2p$-cycles. 
Using the Chern characters of the associated complexes reviewed above, we can calculate the intersection matrix to be
\begin{align}
    \chi = \left( \begin{matrix} \frac{1}{720} \int 3c_2^2 -c_4 & -K_{ii}^{(2)} -\frac{1}{24} \cK_{iiii} & \chi(\cO_{D_i. D_j}, \cO_Y)&0 &1 \\ 
    -K_{kk}^{(2)} -\frac{1}{24} \cK_{kkkk} &\chi(\cO_{D_k}, \cO_{D_i})  & -\frac{1}{2} \cK_{kkij} +\frac{1}{2}(\cK_{kiij}+\cK_{kijj})&-\delta_k^i &0\\
    \chi(\cO_{D_k. D_l}, \cO_Y) & -\frac{1}{2} \cK_{iikl} + \frac{1}{2} (\cK_{klli} + \cK_{kkli}) & \cK_{klij}&0&0\\
    0&-\delta_i^k &0&0&0 \\ 1 &0&0&0&0\end{matrix} \right) \,,
\end{align}
where 
\begin{align}
         \chi(\cO_{D_i. D_j}, \cO_Y)=\frac{1}{12} (2\cK_{iiij} + 3\cK_{iijj}+ 2 \cK_{ijjj} )+2 K^{(2)}_{ij}\,,
     \\
     \chi(\cO_{D_i}, \cO_{D_k})=-2 K_{ik}^{(2)} + \frac{1}{4} \cK_{iikk} -\frac{1}{6} (\cK_{iiik}+ \cK_{ikkk})\,. 
\end{align}
This matrix can now be rewritten as a product of three matrices 
\begin{align}
    \chi = \hat \Lambda^t \hat \chi_0 \hat \Lambda\,, 
\end{align}
where 
\begin{align}
\label{Ft-hatLambda}
    \hat \Lambda = \left(\begin{matrix} 1&0&0&0&0 \\ 0&\delta_i^j & 0&0&0 \\ \frac{1}{24} c_2^{jl} & - \frac{1}{2}\delta^{j}_{i} \delta^{l}_{i} & \delta^j_i \delta^l_k & 0&0 \\ 0 & \frac{1}{6}\cK_{jjji} +K_{ji}^{(2)} & -\frac{1}{2}\left(\cK_{jiik} +\cK_{jikk}\right) &\delta_j^i&0 \\ 
    K^{(0)}& -\frac{1}{24}\cK_{iiii} - \half K_{ii}^{(2)} & \lambda_{ik} &0&1 \end{matrix}\right)\,,
\end{align}
with $\lambda_{ik}= \frac{1}{12} \left(2\cK_{iiik} + 3\cK_{iikk} + 2\cK_{ikkk}\right) + K_{ik}^{(2)}$, and we have 
\begin{align}
    \hat \chi_0 = \left(\begin{matrix} 0&0&0&0&1 \\ 0&0&0&-\delta_j^i &0 \\ 0&0&\cK_{ijkl} &0&0\\ 0&-\delta_i^j &0&0&0 \\ 1&0&0&0&0   \end{matrix}\right)\,.
\end{align}

As emphasized in the main text, to describe the potential in terms of physical fluxes we need to rewrite $\hat \chi_0$ so that it describes an intersection on the actual basis of four-cycles $\sigma_\mu$. We can do this by defining
\begin{align}
    \hat \chi_0 = \Theta^t \chi_0 \Theta \,,
\end{align}
with 
\begin{align}
    \Theta = \left(\begin{matrix}1&0&0&0&0 \\0& \delta_i^j &0&0&0 \\ 0&0& \zeta_{ij}^\mu &0&0\\ 0&0&0&\delta_j^i &0 \\ 0&0&0&0&1 \end{matrix}\right)\,. 
    \label{Ft-Theta}
\end{align}
Then, by defining $\Lambda = \Theta \hat \Lambda$ we arrive at the expression \eqref{Ft-eq:Lambda} and
\begin{align}
    \chi=\Lambda^t \chi_0 \Lambda\,,
\end{align}
with $\chi_0$ given in \eqref{Ft-eq:chi0}. 

\subsection{Corrections to the K\"ahler potential}
\label{Ft-sap:corrka}

In the main text, we derived the polynomial corrections to the Kähler potential \eqref{Ft-Kcs} via the correction to the periods of $\Omega$ and the intersection numbers. We noted that the resulting K\"ahler potential \eqref{Ft-Kcscorr} remains of the classical form up to a term proportional to the third Chern class of the mirror. In the following we will review a more direct way to arrive at the same result, based on the results of \cite{Halverson:2013qca}. 

In \cite{Halverson:2013qca} the Kähler potential on the complexified Kähler moduli space of general Calabi--Yau $n$-fold $X_n$ was argued to be of the form 
\begin{align}\label{Ft-eqapp:KahlerXn}
 e^{-K}= \int_{X_n} \exp\left(2i\sum\limits_{i=1}^{h^{(1,1)}(X_n)} t^i D_i\right)\left(\frac{\hat\Gamma_{\mathbb{C}}(X_n)}{\bar{\hat\Gamma}_{\mathbb{C}}(X_n)}\right)+ \mathcal{O}(e^{2\pi i T})\,,
\end{align}
based on calculating the perturbative corrections to the $S^2$ partition function of the associated gauged linear sigma model. Here $t^i = \text{Im}\,T^i$ is the saxionic part of the complexified Kähler moduli of $X_n$ and $\hat \Gamma_\mathbb{C} (X_n)$ is the complex $\Gamma$-class \eqref{Ft-eqap:gammaclass} that also appears in the calculation of the central charges \eqref{Ft-eqap:centralcharge}. Since the Todd class is real, its contribution to the Kähler potential drops out and we are left only with contributions from the term $\exp(i\Lambda_{X_n})$. For Calabi--Yau four-folds there is only one term in $\Lambda_{X_8}$ proportional to the third Chern class, indicating that only the third Chern class gives a correction to the K\"ahler potential. Evaluating \eqref{Ft-eqapp:KahlerXn} for a four-fold thus yields 
\begin{align}
    e^{-K}= \frac{2}{3} \cK_{ijkl}t^i t^j t^k t^l - \frac{4\zeta(3)}{(2\pi)^3} \int_{X_8} c_3(X_8) . D_i t^i = \frac{2}{3} \cK_{ijkl}t^it^j t^k t^l + 4 K_i^{(3)} t^i\,,
\end{align}
up to exponentially-suppressed corrections, with $K_i^{(3)}$ defined as in \eqref{Ft-K23}. This polynomial structure was previously conjectured in \cite{Honma:2013hma}, and one can easily check that it agrees with \eqref{Ft-Kcscorr}. 

\subsection{Corrected F-term potential}
\label{Ft-sap:corrFterm}

To compute the F-term potential we use the standard Cremmer el al. formula \cite{Cremmer:1982en}

\begin{equation}
    e^{-K}V_F=g^{m\bar{n}}D_m W D_{\bar{n}}\bar{W}-3|W|^2\, ,
\end{equation}
where $D_m W=\partial_{T^m}W+(\partial_{T^m}K) W$, $g^{m\bar{n}}$ is the inverse field space metric and $m, n$ run over all moduli. Ignoring corrections to the K\"ahler sector of the compactification we recover the standard cancellation of no-scale structure models and the above expression simplifies to
\begin{align}
    e^{-K}V_F&=g^{i\bar{j}}D_{i}W D_{\bar{j}} \bar{W}\\
    &=g^{i\bar{j}}\left[\Re W_i \Re W_{\bar{j}} +\Im W_i \Im W_{\bar{j}}+\left((\Re W)^2+\Im(W)^2\right)K_i\bar{K}_{\bar{j}}+K_i W\bar{W}_{\bar{j}}+\bar{K}_{\bar{j}}W_i \bar{W}\right]\, ,\nonumber
\end{align}
where $W_i=\partial_i W$ and now $i,j$ only run over complex structure moduli.  We proceed to consider the version of the superpotential \eqref{Ft-eq:supo and der corr} and the Kähler potential \eqref{Ft-Kcscorr} that include the polynomial corrections:
\begin{align}
     W&=\, \bar{\rho}_0+i\bar{\rho}_it^i-\frac{1}{4}\mathcal{K}_{ij}\bar{\rho}^{ij}-i \left(\frac{1}{6}\mathcal{K}_i + K_i^{(3)}\right)\tilde{\rho}^i + \left(\frac{\mathcal{K}}{24} +  K_i^{(3)}t^i \right)\tilde{\rho} \, ,
\end{align}
\be
K_{\rm cs}=-\log\left(\frac{2}{3}\mathcal{K}_{ijkl}t^it^jt^kt^l + 4 K_i^{(3)} t^i \right)\, ,
\ee
where the $\rho$'s are given by \eqref{Ft-corrhos}. From the Kähler potential we can derive the corrected version of the metric of the complex structure moduli space. We have
\begin{align}\label{Ft-apeq:KTi}
    K_{T^i}&\equiv \partial_{T^i} K_{\rm cs}=\frac{i\left(2\mathcal{K}_i+3K^{(3)}_i\right)}{\mathcal{K}+6K^{(3)}_kt^k}=\frac{i}{2\cK}\frac{4\cK_i+\cK\epsilon_i}{1+\epsilon_k t^k}\, ,\\
    g_{ij}&\equiv \partial_{T^i}\partial_{\bar{T^j}}  K_{\rm cs}\nonumber\\
    &=\frac{1}{\left(1+\epsilon_k t^k\right)^2}\left[\frac{4\cK_i\cK_j}{\cK^2}-\frac{3\cK_{ij}}{\cK}+\frac{1}{\cK}\left(\cK_i\epsilon_j+\cK_j\epsilon_i-3\cK_{ij}\epsilon_kt^k\right)+\frac{1}{4}\epsilon_i\epsilon_j\right]\, ,
\end{align}
where we have defined $\epsilon_i\equiv 6K^{(3)}_i/\cK$. The inverse metric can be computed as a series in powers in $\epsilon_i$, whose first terms are given by
\begin{align}
    g^{ij}=&(1+\epsilon_kt^k)^2 \left[\frac{4}{3}t^it^j-\frac{1}{3}\cK\cK^{ij}+\frac{\epsilon_k\cK}{3}\left(\cK^{ij} t^k+\cK^{ik} t^j+\cK^{jk}t^i\right)\right.\nonumber\\
    &\left.+\left[\frac{\cK^2}{12}\cK^{ik}\cK^{jl}-\frac{\cK}{3}\left(\cK^{ij} t^k t^l+t^it^j\cK^{kl}\right)+\frac{4}{3}t^it^j t^kt^l\right]\epsilon_k\epsilon_l+\cO(\epsilon_k^3)\right]\, .
\end{align}
Working with the inverse metric in its full extension would be extremely cumbersome. We take a different approach with the final aim of obtaining an expression for the scalar potential where the uncorrected part can be easily identified. To do so we make use of the following relation:
\begin{equation}
    g^{ik}K_{T^{i}}=2it^i-\frac{3i}{2}\frac{\tilde{\epsilon}^i}{(1+\epsilon)^2}\, ,
\end{equation}
with $\epsilon=\epsilon_it^i$ and $\tilde{\epsilon}^i=g^{ij}(\epsilon_j-4\epsilon\cK_j/\cK)$. Then $V_F$ becomes
\begin{align}
    e^{-K}V_F=&g^{ij}(\Re W_i\Re W_j+\Im W_i\Im W_j)+4\Re W(\Re W+t^i\Im W_i)\nonumber\\
    &+4\Im W(\Im W -t^i\Re W_i)+\frac{3\tilde{\epsilon}^i}{(1+\epsilon)^2}\left(\Im W \Re W_i -\Re W\Im W_i\right)\nonumber\\
    &-\left((\Re W)^2+(\Im W)^2\right)L\, ,
\end{align}
where
\begin{equation}
    L=\frac{3\epsilon}{1+\epsilon}+\frac{3\tilde{\epsilon}^i}{4(1+\epsilon)^3\cK}(4\cK_i+\cK\epsilon_i)\, .
\end{equation}
Substituting the superpotential and its derivatives in terms of the flux polynomials and denoting the uncorrected metric and its inverse \eqref{Ft-metric} by $g^0_{ij}$ and $g_0^{ij}$, respectively, we arrive to
\begin{align}
   e^{-K}V_F=&\; 4\left(\bar{\rho}-\frac{\mathcal{K}}{24}\tilde{\rho}\right)^2+g_0^{ij}\left(\rho_i+\frac{\mathcal{K}}{6}g^0_{ik}\tilde{\rho}^k\right)\left(\rho_j+\frac{\mathcal{K}}{6}g^0_{jl}\tilde{\rho}^l\right)+
    (g^{ij}-t^it^j)\zeta_{\mu i}\zeta_{\nu j} \bar{\rho}^\mu  \bar{\rho}^\nu\nonumber\\
    &+\frac{1}{36}\left(g^{ij}-g_0^{ij}\right)\cK_i\cK_j\tilde{\rho}^2+\frac{1}{36}g^{ij}(2\cK_i\cK\epsilon_j+\cK^2\epsilon_i\epsilon_j)\tilde{\rho}^2-\frac{1}{12}\epsilon\cK^2\tilde{\rho}^2+\frac{2\epsilon}{3}\cK\bar{\rho}\tilde{\rho}\nonumber\\
    &+\frac{1}{3}\left(g_0^{ij}-g^{ij}\right)\zeta_{\mu i}\cK_j \bar{\rho}^\mu\tilde{\rho}-\frac{\cK}{3}g^{ij}\zeta_{\mu i}\epsilon_j\bar{\rho}^\mu\tilde{\rho}+\frac{\epsilon}{3}\cK\zeta_\mu\bar{\rho}^\mu\tilde{\rho}+(g^{ij}-g_0^{ij})\bar{\rho}_i\bar{\rho}_j\nonumber\\
    &-\left(g^{ij}-g^{ij}_0\right)\cK_{jk}\bar{\rho}_i\tilde{\rho}^k+\frac{1}{4}\left(g^{ij}-g^{ij}_0\right)\cK_{ik}\cK_{jl}\tilde{\rho}^k\tilde{\rho}^l-\frac{2\cK}{3}\epsilon_j\tilde{\rho}^j\bar{\rho}_it^i-\frac{\cK}{9}\cK_i\epsilon_j\tilde{\rho}^i\tilde{\rho}^j+\frac{\cK^2}{9}(\epsilon_i\tilde{\rho}^i)^2\nonumber\\
    &+\frac{3\tilde{\epsilon}^i}{(1+\epsilon)^2}\left[\bar{\rho}_i\bar{\rho}_k t^k-\frac{1}{6}(\cK_k\delta^j_i+3\cK_{ik}t^j)\bar{\rho}_j\tilde{\rho}^k-\frac{1}{6}\cK\epsilon_k\tilde{\rho}^k\bar{\rho}_i+\frac{1}{12}(\cK_j+\cK\epsilon_j)\cK_{ik}\tilde{\rho}^j\tilde{\rho}^k\right.\nonumber\\
    &-\bar{\rho}\zeta_{\mu i}\bar{\rho}^\mu+\frac{1}{6}(\cK_i+\cK\epsilon_i)\tilde{\rho}\bar{\rho}+\frac{1}{2}\zeta_\mu\bar{\rho}^\mu\zeta_{\nu i}\bar{\rho}^\nu-\frac{1}{12}(\cK_i+\cK\epsilon_i)\tilde{\rho}\zeta_\mu\bar{\rho}^\mu\nonumber\\
    &\left.-\left(\frac{1}{24}+\frac{\epsilon}{6}\right)\cK \tilde{\rho}\zeta_{\mu i}\bar{\rho}^\mu+\frac{\cK}{6}\left(\frac{1}{24}+\frac{\epsilon}{6}\right)(\cK_i+\cK\epsilon_i)\tilde{\rho}^2\right]\nonumber\\
    &-L\left[\bar{\rho}^2+\frac{1}{4}(\zeta_\mu\bar{\rho}^\mu)^2+\left(\frac{1}{24}+\frac{\epsilon}{6}\right)^2\cK^2\tilde{\rho}^2-\zeta_\mu\bar{\rho}^\mu\bar{\rho}+\left(\frac{1}{12}+\frac{\epsilon}{3}\right)\cK\tilde{\rho}\bar{\rho}\right.\nonumber\\
    &\left.-\zeta_\mu\bar{\rho}^\mu\left(\frac{1}{24}+\frac{\epsilon}{6}\right)\cK\tilde{\rho}+(\bar{\rho}_it^i)^2+\frac{1}{36}[(\cK_i+\cK\epsilon_i)\tilde{\rho}^i]^2-\frac{1}{3}\bar{\rho}_it^i(\cK_j+\cK\epsilon_j)\tilde{\rho}^j\right]\, . \label{Ft-scalarpotcorr}
\end{align}
One can then see that in the limit $\eps_i \to 0$ we recover the leading form of the potential \eqref{Ft-scalarpot} from the first line of this expression. Notice that as expected all terms are quadratic on the flux-axion polynomials $\rho_A$, and so one has a potential of the form \eqref{Ft-bilinear}. The expression for the matrix $Z$ is, however, much more complicated than \eqref{Ft-ZAB}, with several new non-vanishing entries that destroy its block-diagonal structure. 

We can use the result in \eqref{Ft-scalarpotcorr} to generalize \eqref{Ft-ZAB} to account for the presence of linear order corrections in $\epsilon_i$. The new matrix will be given by
\begin{eqn}
Z = Z_0 + \eps_k Z^k + \cO(\eps_k^2) \, ,
\end{eqn}
where $Z_0$ is the uncorrected matrix from \eqref{Ft-ZAB} and $Z^k$ is given by
\begin{align}
&2{\cal V}_3^2 Z^{k}=\begin{pmatrix}
\frac{\cK t^k}{48} & &-\frac{t^k}{2}\zeta_\mu+\frac{\cK}{8}\cK^{ik}\zeta_{\mu i} & &  -t^k\\
& \mathcal{X}^k_{ij} & &  \mathcal{Y}^{ki}_j& &  \\
-\frac{t^k}{2}\zeta_\nu+\frac{\cK}{8}\cK^{ik}\zeta_{\nu i}& & \mathcal{M}^{k}_{\mu\nu} & & 3\cK^{ik}\zeta_{\nu i}+\frac{6t^k}{\cK}\zeta_\nu  \\
&  \mathcal{Y}^{kj}_i& &  \mathcal{Z}^{kij} &  \\
-t^k & &3\cK^{ik}\zeta_{\mu i}+\frac{6t^k}{\cK}\zeta_\mu & &  \frac{36t^k}{\cK} \\
\end{pmatrix} \, ,
\end{align}
where we have arranged the flux-axion polynomials in a vector of the form
$\vec{\rho}^{\, t} = \left(\tilde{\rho}, \tilde{\rho}^i,   \bar{\rho}^{\mu}, \bar{\rho}_i,   \bar{\rho}   \right)$ and we have defined
\begin{align}
    \mathcal{X}_{ij}^k\equiv&\frac{t^k}{\cK}\cK_i\cK_j-\frac{t^k}{2}\cK_{ij}-\frac{1}{12}\delta_i^k\cK_j-\frac{1}{12}\delta_j^k\cK_i\, ,\\
    \mathcal{Y}_j^{ki}\equiv& \frac{2t^k}{\cK}t^i\cK_j+t^k\delta^i_j-\frac{1}{2}\cK_j\cK^{ik}-\frac{3t^i}{2}\delta_j^k\, ,\\
    \mathcal{Z}^{kij}\equiv&\frac{4t^k}{\cK}t^it^j-2\cK^{ij}t^k-\cK^{ik}t^j-\cK^{jk}t^i\, ,\\
    \mathcal{M}_{\mu\nu}^k\equiv& \frac{t^k}{\cK}\zeta_\mu\zeta_\nu-2t^k\cK^{ij}\zeta_{\mu i}\zeta_{\nu j}+\frac{1}{2}\cK^{ik}\zeta_{\mu i}\zeta_{\nu}+\frac{1}{2}\cK^{ik}\zeta_{\nu i}\zeta_{\mu}\, .
\end{align}
In general, given the complicated form of the potential, it is easier to characterize the corrected vacuum equations in terms of the corrected F-terms, as we now turn to discuss.

\subsection{Corrected vacuum equations}
\label{Ft-sap:corrvac}

The polynomial corrections to the superpotential \eqref{Ft-supofinal} and K\"ahler potential \eqref{Ft-Kcscorr} modify the  on-shell conditions \eqref{Ft-eq:Mink} at leading order. In the following we would like to compute such a modification which, as pointed out in the main text, essentially depends on $K_i^{(3)}$. Using 
 \eqref{Ft-eq:supo and der corr} and \eqref{Ft-apeq:KTi} we find the F-term condition $D_i W=0$ to be equivalent to 
\begin{eqn}\label{Ft-apeq:DiW=0}
        &\left(\cK + 6 K_j^{(3)} t^j\right)\left[\bar \rho_i +i \zeta_{\mu,i}\bar \rho^{\mu} -\frac{1}{2} \cK_{ij} \tilde \rho^j -\frac{i}{6} \cK_i \tilde \rho- i K_i^{(3)} \tilde \rho\right]\\
    &=-2i \left(\cK_i +\frac{3}{2} K_i^{(3)}\right)\left[\bar \rho + i\bar \rho_j t^j - \frac{1}{2}\zeta_\mu \bar \rho^{\mu} -\frac{i}{6}\cK_j \tilde \rho^j -i K_j^{(3)} \tilde \rho^j+\frac{\cK}{24}\tilde \rho + K_j^{(3)} t^j \tilde \rho \right]\,. 
\end{eqn}
Contracting this expression with $t^i$ yields 
\begin{eqn}\label{Ft-apeq:tiDiW=0}
    &\left(\cK + 6 K_i^{(3)} t^i\right)\left[\bar \rho_jt^j +i \zeta_\mu \bar \rho^\mu -\frac{1}{2} \cK_j \tilde \rho^j -\frac{i}{6} \cK \tilde \rho-  i K_j^{(3)}t^j \tilde \rho\right]\\
    &=-2i \left(\cK +\frac{3}{2} K_i^{(3)}t^i\right)\left[\bar \rho + i\bar \rho_j t^j - \frac{1}{2}\zeta_\mu \bar \rho^\mu -\frac{i}{6}\cK_j \tilde \rho^j -i K_j^{(3)} \tilde \rho^j+\frac{\cK}{24}\tilde \rho + K_j^{(3)} t^j \tilde \rho \right]\,.
\end{eqn}
We now split this equation into real and imaginary part. The real part gives 
\begin{align}
   \left(1 +\half \epsilon_it^i\right) \bar \rho_i t^i = -\left(1 +\frac{5}{2} \epsilon_i t^i\right) \frac{\cK_j\tilde \rho^j }{6} + \left(4 + \epsilon_i t^i\right)\frac{\cK\epsilon_j  \tilde \rho^j}{12}\,, 
\end{align}
and the imaginary part 
\begin{align}\label{Ft-apeq:vacuumrho}
   (1 + \frac{1}{4} \epsilon_it^i) \bar \rho = \frac{\cK \tilde \rho}{24} - \frac{\cK \epsilon_i t^i\tilde \rho}{96}  - \frac{3  \epsilon_i t^i \zeta_\mu \bar \rho^{\mu}}{8} +\frac{\cK(\epsilon_i t^i)^2 \tilde \rho}{24}\, ,
\end{align}
where again $\epsilon_i \equiv 6 K_i^{(3)}/\cK$. Inserting the above expressions back into \eqref{Ft-apeq:DiW=0} we obtain 
\begin{align}\label{Ft-apeq:vacuumrho_i}
    \bar \rho_i = \frac{1}{2} \cK_{ij}\tilde \rho^j -\frac{2\left(\frac{\cK_i}{\cK} +\frac{1}{4} \epsilon_i \right)}{1+ \half \epsilon_j t^j - \half (\epsilon_k t^k)^2}\left[\frac{1}{3} \cK_j \tilde \rho^j + \frac{1}{3} \epsilon_k t ^k \cK_j \tilde \rho^j - \frac{1}{6} \cK \epsilon_j \tilde \rho^j \left(1 + \epsilon_k t ^k \right)\right]\,,
\end{align}
and 
\begin{eqn}\label{Ft-apeq:vacuumrhomu}
    \left(\zeta_{\mu, i} - \frac{\cK_i}{\cK} \zeta_\mu \right) \bar \rho^\mu =& -\frac{1}{8}\left(1+\epsilon_k t^k\right)\left( \cK_i \epsilon_k t^k - \epsilon_i \cK \right)\tilde \rho - \frac{5}{4}\epsilon_k t^k \zeta_{\mu, i}\hat \rho^\mu \\
    &+ \left(\frac{\cK_i}{\cK}\epsilon_k t^k +\frac{1}{4} \epsilon_i\right) \zeta_\mu \bar \rho^\mu  + \frac{1}{4} \left(\epsilon_i \epsilon_k t^k \zeta_\mu - (\epsilon_k t^k )^2 \zeta_{\mu, i}\right) \bar \rho^\mu\,.
\end{eqn}
As expected, in the limit $\epsilon_i \to 0$ equations \eqref{Ft-apeq:vacuumrho}, \eqref{Ft-apeq:vacuumrho_i} and \eqref{Ft-apeq:vacuumrhomu} reduce to the classical vacuum equations \eqref{Ft-eq:Mink}. To capture the leading effect of the corrections, we can also expand to linear order in $\epsilon_i$ to find 
\begin{subequations}\label{Ft-apeq:vaclinear}
\begin{align}
   \bar \rho- \frac{1}{24}\cK \tilde \rho  &= - \frac{1}{48} \epsilon_i t^i \left[\cK \tilde \rho + 18 \zeta_\mu \bar \rho^\mu \right] + \cO(\eps_i^2) \,,\\
    \bar \rho_i + \frac{1}{6} \cK g_{ij} \tilde \rho^j &=  -\frac{1}{6} \epsilon_i \cK_j \tilde \rho^j -\frac{1}{3} \cK_i\left(\epsilon_j t^j \frac{\cK_k}{\cK} - \epsilon_k\right) \tilde \rho^k + \cO(\eps_i^2)  \,,\\
    \left(\zeta_{\mu, i} - \frac{\cK_i}{\cK} \zeta_\mu \right)\bar \rho^\mu &= \frac{1}{8} \left(\eps_i - \epsilon_k t^k \frac{\cK_i }{\cK} \right)   \left( \cK \tilde \rho +  2 \zeta_\mu \bar\rho^{\mu}\right)+ \cO(\eps_i^2) \, ,
\end{align}
\end{subequations}
which gives \eqref{Ft-eq:Minkcorr} in the main text. If we further impose the condition for supersymmetric vacua $W=0$ we get the additional constraints
\begin{subequations}
\begin{align}
    \bar \rho_i t^i &= \frac{1}{4} \left(\cK \epsilon_i \tilde \rho^i -\epsilon_i t^i \cK_j \tilde \rho^j \right) + \cO(\eps_i^2) \,,\\
    \zeta_\mu \bar{\rho}^\mu &=\frac{\cK}{6}\left(1 + \eps_it^i \right) \tilde{\rho} + \cO(\eps_i^2) \, .
\end{align}
\end{subequations}
where we have also made use of the linearized equations \eqref{Ft-apeq:vaclinear}.


\section{Flux invariants and moduli fixing}
\label{Ft-ap:invariants}

The introduction of the flux-axion polynomials $\rho_{A}$ is a powerful technique that allows for the study of moduli stabilization in a clear and systematic way. Since the flux polynomials depend on the axions $b^i$, fixing the moduli amounts to solve the system of algebraic equations in the saxions $t^i$ and the flux polynomials $\rho_{A}$ that arises from the vanishing derivatives of the scalar potential with respect to the set of moduli.

As discussed in section \ref{sys-ss:invariants}, using the $\rho_{A}$ as a stepping stone to stabilize the $b^i$ may lead to some questions regarding whether it is actually possible to accomplish this task, since in most examples the number of polynomials will exceed the rank of the system of equations. The solution to this problem comes through the fact that the $\rho_{A}$ are not a set of fully independent variables. There are many constraints that arise from their definition and they can be expressed by the set of combinations of flux polynomials which are invariant under shifts of the axions. More precisely, we look for invariant multilinear combinations of $\rho_{A}$ under the transformation $\vec\rho \rightarrow R(b) \vec\rho$, with $R(b)$ given by \eqref{Ft-eq: axion matrix}. After some algebra we find that these invariants are
\begin{subequations}
\label{Ft-eq: invariants}
\begin{align}
    \tilde{\rho}^2\rho_i-\tilde{\rho}\zeta_{\mu,ij}\tilde{\rho}^j\bar{\rho}^\mu+\frac{1}{3}\mathcal{K}_{ijkl}\tilde{\rho}^j\tilde{\rho}^k\tilde{\rho}^l & = m^2e_i - m \zeta_{\mu,ij} m^j \bar{m}^\mu + \frac{1}{3} \mathcal{K}_{ijkl} m^jm^km^l  \, ,\\
    \bar{\rho} \tilde{\rho} - \bar{\rho}_i \tilde{\rho}^i + \frac{1}{2} \eta_{\mu\nu} \bar{\rho}^{\mu} \bar{\rho}^{\nu} & = \bar{e} m - \bar{e}_i m^i + \frac{1}{2} \eta_{\mu\nu} \bar{m}^\mu\bar{m}^\nu\, , \label{Ft-eq: invariants tadpole} \\
    \bar{\rho}^\mu\tilde{\rho}-\frac{1}{2}\zeta^\mu_{ij}\tilde{\rho}^i\tilde{\rho}^j & = \bar{m}^\mu m -\frac{1}{2}\zeta^\mu_{ij} m^im^j \, ,\\
    \tilde{\rho} & = m\, .
\end{align}
\end{subequations}

Looking at \eqref{Ft-eq:Mink}, we could think the system is composed of $2h^{(3,1)}+1$ linearly independent equations but note that the last family of equations has an additional constraint, since $\left(  \cK \zeta_{\mu i}- \mathcal{K}_i\zeta_{\mu}\right)t^i $ is trivially zero.  Therefore we actually have $2h^{(3,1)}$ equations in the variables $\{t^i, \rho_{A}\}$, which amount to  $3h^{(3,1)}+h^{(2,2)}+2$ unknowns. If it were not for the invariants this would imply that we have an extremely unconstrained system. However, the existence of invariant combinations of axion polynomials greatly reduces the number of degrees of freedom. From \eqref{Ft-eq: invariants} we see that we have  $2+h^{(3,1)}+h^{(2,2)}$ constraints. Consequently, the $\rho$'s move in an orbit of dimension $h^{(3,1)}$ which is just enough to fix all the axions using half of the vacua equations. The remaining $h^{(3,1)}$ vacua equations can be used to fix the saxions $t^a$.

 Notice that, by construction, the multilinear combinations of flux quanta in the rhs of \eqref{Ft-eq: invariants} are invariant under the monodromies ${\cal T}_i$ around the complex structure point, see  \eqref{Ft-monoexp}. This implies that they label flux-inequivalent vacua, and therefore that the saxion vevs should only depend on such invariants, simply because the value of the invariants $\rho_A$ in the vacuum also must depend on them.  Finally, in some specific scenarios where some flux quanta vanish, like in sections \ref{Ft-sec:moduli},  \ref{Ft-sec:moduliIIB} and \ref{Ft-s:linear}, 
the flux-axion polynomials will simplify and some other combinations of fluxes may play the role of those in  \eqref{Ft-eq: invariants}. For instance, only   \eqref{Ft-eq: invariants tadpole} remains non-vanishing in the moduli stabilization scheme of section \ref{Ft-sec:moduli}, but other invariants like $\bar{m}^\mu$ appear in this case.


\section{Vacua equations for elliptic fibered mirrors}
\label{Ft-ap:elliptic}

In this appendix we analyze the vacua equations for the particular case in which the mirror manifold $X_8$ is elliptically fibered, as considered in section \ref{Ft-sec:elliptic}. In particular we want to provide an explicit expression for $\Gamma_{ab} \equiv - \tilde{A}_{ac}\tilde{B}^c{}_b$ in \eqref{Ft-eq:Mink elliptic compact rho mu}. While one could simply compute the inverse of \eqref{Ft-tildeA} and apply the definition, in the following we would like to obtain an expression for $\Gamma$ directly from  \eqref{Ft-eq:Mink mu}, in the same spirit as in \eqref{Ft-vacua mu strategy}. This strategy should be useful in cases where $X_8$ is not a fibration, and so the index splitting $\mu = \{ a, \hat{a}\}$ does not occur. Then, in general $g_{\mu\nu} -\eta_{\mu\nu}$ will be a singular matrix, and we cannot have an expression of the form \eqref{Ft-ZABfib}, because $\tilde{A}$ does not have an inverse.

To proceed one may expand the vacua equations  \eqref{Ft-eq:Mink mu} in the basis \eqref{Ft-4formbasisfib}. This is equivalent to consider the equations
\begin{align}
    \zeta^b_0\left[ \tilde{B}_b{}^c\tilde{A}_{cd}\tilde{B}^d{}_e  \bar{\rho}^e + \tilde{B}_b{}^c \bar{\rho}_c'\right]=0 \, ,\\
     \zeta^b_a \left[ \tilde{B}_b{}^c\tilde{A}_{cd}\tilde{B}^d{}_e  \bar{\rho}^e + \tilde{B}_b{}^c \bar{\rho}_c'\right] +\zeta_{ab} \left[\tilde{B}^b{}_c\bar{\rho}^c+ \tilde{A}^{bc}\bar{\rho}_c'\right]=0\, ,
\end{align}
which are in turn equivalent to \eqref{Ft-eq:Mink elliptic compact rho mu}. Expanding \eqref{Ft-eq:Mink mu} using \eqref{Ft-Koszulfib} and after some algebra we obtain:
\begin{subequations}\label{Ft-eqap:mink a}
\begin{align}
    \cK \left(t^a + c_1^a t^0\right)(\bar \rho_a + c_{ab}\bar \rho^b) &=\cK_0\left[t^0(2t^c + t^0 c_1^c)(\bar \rho_c + c_{cb} \bar \rho^c) + \kappa_b \bar \rho^b \right]\,, \\
    \cK\left(\kappa_{ab} \bar \rho^b + t^0 (\bar \rho_a + c_{ab} \bar \rho^b)\right)&= \cK_a \left[ t^0(2 t^b + t^0 c_1^b) (\bar \rho_b + c_{bc} \bar \rho^c) +\kappa_b \bar \rho^b\right] \, ,
\end{align}
\end{subequations}
which can be simplified with the following change of basis
\begin{align}
    \varrho_a = \bar \rho_a + c_{ab}\bar \rho^b\,,\qquad \varrho^a = \bar \rho^a \, ,
\end{align}
in terms of which \eqref{Ft-eqap:mink a} read 
\begin{subequations}\label{Ft-eqap:varmink a}
\begin{align}
    \cK \left(t^a + c_1^a t^0\right)\varrho_a &=\cK_0\left[t^0(2t^c + t^0 c_1^c)\varrho_c+ \kappa_b \varrho^b \right]\,, \label{Ft-eqap:varrho0}\\
    \cK\left(\kappa_{ab} \varrho^b + t^0 \varrho_a \right)&= \cK_a \left[ t^0(2 t^b + t^0 c_1^b) \varrho_b+\kappa_b \varrho^b\right] \,.\label{Ft-eqap:varrhoa}
\end{align}
\end{subequations}

Note that there is some redundancy among this set of equations, inherited from the fact that the contraction of \eqref{Ft-eq:Mink mu} with $t^i$ vanishes identically. To extract the information contained in \eqref{Ft-eqap:varrhoa} that is independent of \eqref{Ft-eqap:varrho0} we introduce  two projection operators 
\begin{align}
    \left(\mathbb{P}_p\right)^a_b = \delta^a_b - \frac{\cK_a t^b}{\cK - \cK_0t^0} \,,\qquad  \left(\mathbb{P}_{np}\right)^a_b=\frac{\cK_a t^b}{\cK - \cK_0t^0}\,. 
\end{align}
Then applying $\mathbb{P}_p$ to \eqref{Ft-eqap:varrhoa} we obtain 
\begin{align}
    t^0\left(\varrho_{a}-\frac{\cK_a}{\cK -\cK_0t^0} t^c\varrho_{c}\right)=\frac{\cK_a}{\cK - \cK_0t^0}\kappa_{b} \varrho^{b} 
- \kappa_{ab}\varrho^{b} \,,
\end{align}
which is solved by 
\begin{align}\label{Ft-ansatzPp}
    \varrho_{a}=\mathcal{G}_{ab} \varrho^{b}\, , \quad \text{with} \quad  \mathcal{G}_{ab} \equiv \left[\frac{\cK_a v_b}{\cK-\cK_0t^0} + \frac{1}{t^0}\left(\frac{\cK_a}{\cK-\cK_0t^0} \kappa_{b} -\kappa_{ab} \right)\right]\,,
\end{align}
where $v_b$ is a vector that still needs to be determined. Projecting \eqref{Ft-eqap:varrhoa} with 
 $\mathbb{P}_{np}$ is equivalent to \eqref{Ft-eqap:varrho0}, which can be rewritten as 
\begin{align}
    (\cK -\cK_0t^0)\left(t^a \varrho_{a} + t^0 c_1^a\varrho_{0a} \right)=\cK_0\left(t^0 t^a\varrho_{a}+\kappa_b \varrho^{b}\right)\,.
\end{align}
From this equation we can determine $v_b$ to be 
\begin{align}
    v_b= \frac{\left(\cK -\cK_0 t^0 \right)c_1^c \kappa_{cb} + (\cK_0-c_1^c \cK_c) \kappa_b}{\cK-2\cK_0t^0 +t^0c_1^a \cK_a}\,, 
\end{align}
such that the matrix $\Gamma_{ab} =  \mathcal{G}_{ab} - c_{ab}$ is given by 
\begin{align}
    \Gamma_{ab} =&\, \frac{1}{\cK - \cK_0 t^0}\left[ \frac{ \left(\cK -\cK_0 t^0 \right)c_1^c \kappa_{cb} \cK_a+ (\cK_0-c_1^c \cK_c) \cK_a \kappa_b}{\cK-2\cK_0t^0 +t^0c_1^a \cK_a} + \frac{1}{t^0}\cK_a\kappa_{b}   \right] -\frac{1}{t^0} \kappa_{ab} - c_{ab}\\
    =&\, \frac{\cK_a(\kappa_b+t^0\kappa_{bc}c_1^c)}{t^0(\cK-2\cK_0 t^0+t^0c_1^a\cK_a)}-\frac{1}{t^0}\kappa_{ab} - c_{ab}\, .
\end{align}
Finally, we may rewrite $\Gamma_{ab}$ in terms of base quantities by expanding it in $t^0 c_1^b$. The result is:
\begin{eqn}\label{Ft-eqap:expansionAab}
     t^0\left(2\kappa+3c_1^c\kappa_c t^0 + c_1^c c_1^d\kappa_{cd}(t^0)^2 \right)& \Gamma_{ab} =3 \kappa_a \kappa_b  -2 \kappa  \kappa_{ab}  \\ 
     &+ t^0\left(3\kappa_{ac}  \kappa_{b}  + 3 \kappa_{a}  \kappa_{bc}  - 3\kappa_{ab} \kappa_c  - 2 \kappa \kappa_{abc}  \right)c_1^c\\ 
    & +(t^0)^2\left(3\kappa_{ac} \kappa_{bd} -\kappa_{cd} \kappa_{ab}  + \kappa_{acd} \kappa_b -3 \kappa_{abc}  \kappa_d  \right)c_1^cc_1^d \\
     &+  (t^0)^3\left(\kappa_{acd} \kappa_{be}  - \kappa_{cd} \kappa_{abe} \right)c_1^c c_1^d c_1^e\, .
\end{eqn}


\ifSubfilesClassLoaded{%
\bibliography{biblio}%
}{}

\end{document}

\ifSubfilesClassLoaded{%
\tableofcontents
}{}

\chapter{Type IIB Mass spectrum}

\section{Mass spectrum of no-scale aligned vacua}
\label{IIB-sec:NSA}

The no-scale aligned vacua described in \cite{Blanco-Pillado:2020wjn} are defined by the following relation between two-derivatives of the superpotential and one-derivative of the Kähler potential as well as two constraints on flux quanta:
\begin{align}
	D_{\tau} D_i W \propto K_i\quad\text{ and }\quad f_A^0=h_A^0=0\ .
	\label{IIB-eq:nsa_def}
\end{align}
These vacua feature an analytical mass spectrum expressed solely in terms of the LCS parameter. In this section, we present the key steps of its derivation.

One of the main difficulties to obtain the mass spectrum in generic points of field-space is the fact that one has to compute eigenvalues with respect to the field space metric $K_{i\bar{j}}$. In order to overcome this difficulty, it is customary to introduce real vielbein $e_i^a$ which render the metric to a canonical form \cite{Blanco-Pillado:2020wjn,Marsh:2015zoa}, such that
\begin{align}
	K_{i\bar{j}} = e_i^a \delta_{ab} e_{\bar{j}}^b \ , \quad \delta_{ab} = e_a^i K_{i\bar{j}} e_b^{\bar{j}} \ .
\end{align} 
In what follows, we will reserve letters $i,j,\ldots$ to refer to curved indices in field space, while $a,b,\ldots$ will label flat indices. 

From the block-diagonal form of the metric in the axio-dilaton and complex structure sectors, we can easily see that $e_0^{\tau} = -2 t^0$. On the other hand, since we are free to choose the first vielbein to diagonalize the metric, we will pick 
\begin{align}
	e_1^i \equiv \frac{t^i}{x}\ ,
\end{align}
where $x$ is a normalization factor. Plugging this into the field-space metric \eqref{IIB-eq:Kij}, we get
\begin{align}
  \delta_{ab} = e_a^i K_{i\bar{j}} e^j_b = - 2 x \mathring{\kappa}_{ab1} + 4 x^4 \mathring{\kappa}_{a11} \mathring{\kappa}_{b11}\ .
\end{align}
Using the definition of the LCS parameter \eqref{IIB-eq:def_xi} and the previous equation, we can obtain several identities:
\begin{align}
    \label{IIB-eq:kappas_circle_app}
	x = \frac{\sqrt{3 (1 -2 \xi)}}{2(1 + \xi)}\ , \ \mathring{\kappa}_{111} = \frac{2 (1+\xi)^2}{\sqrt{3(1-2\xi)^3}}\ , \ \mathring{\kappa}_{a' 11} = 0\ , \ \mathring{\kappa}_{a' b' 1} = \frac{-(1+\xi)}{\sqrt{3 (1 - 2 \xi)}} \delta_{a'b'}\ ,
\end{align}
where the prime indices $a',b'$ run from 2 onwards. 

The ``1'' direction in the vielbein turns out to have special significance. Contracting its corresponding vielbein with $K_i$ given in eq.~\eqref{IIB-eq:Ki}, we find
\begin{align}
	Z_{0a}\propto K_a = e_a^i K_i = 2 i x^2 \mathring{\kappa}_{a11} \propto \delta_a^1\ ,
\end{align}
where the matrix $Z$ is defined below. Thus, the introduction of the vielbein into our problem not only simplifies expressions involving the field-space metric or its inverse, but it also aligns the so-called no-scale direction $K_i$ with the 1-direction. 

We are now prepared to tackle the computation of the mass spectrum. In order to do this, we will proceed in the lines of \cite{Blanco-Pillado:2020wjn}. As explicitly proven in that work, the mass spectrum can be neatly written as \cite{Sousa:2014qza}
\begin{equation}
	\mu^2_{\pm \lambda} = (m_{3/2} \pm m_{\lambda})^2\ ,
	\label{IIB-eq:mass_rel}
\end{equation}
where $\lambda = 0, \ldots, h^{2,1}$ and we have defined the gravitino mass $m_{3/2} \equiv e^{K/2} |W|$ as well as the fermion masses $m_{\lambda}$. The easiest way to obtain the latter ones is through the diagonalization of the following matrix\footnote{The first metric factor must be introduced due to the kinetic term of the scalar fields being potentially non-canonical.}
\begin{align}
	(Z^{\dag} Z)^A_B \equiv K^{A\bar{C}} \bar{Z}_{\bar{C}\bar{D}} K^{\bar{D}E} Z_{EB}\ , 
\end{align}
where $Z_{AB} \equiv e^{K/2} D_A D_B W$ and the indices $A,B,\dots$ run into $\{\tau,z^i\}$. Thus, eigenvalues of $Z^{\dag}Z$ will yield the masses  $m^2_{\lambda}$. 

In order to compute these values we will employ several simplifying schemes. First of all, it is easy to check that $Z_{\tau \tau} = 0$ at supersymmetric vacua described by the tree-level LCS prepotential. Another useful identity is \cite{Denef:2004ze,Candelas:1990pi}
\begin{equation}
\label{IIB-eq:Denef}
	Z_{ij} = - (\tau - \bar{\tau}) e^{K_{\rm cs}} \kappa_{ijk} K^{k\bar{l}} \bar{Z}_{\bar{\tau} \bar{l}}\ .
\end{equation}
This identity can be easily rewritten in terms of the vielbein introduced above:
\begin{align}
	Z_{ab} = i \mathring{\kappa}_{abc} \delta^{cd} \bar{Z}_{0d}\ .
	\label{IIB-eq:zab_id}
\end{align}
On the other hand, since the vacua we are studying have the no-scale-aligned property \eqref{IIB-eq:nsa_def}, we have that $Z_{0a} \propto \delta_a^1$ and therefore,
\begin{align}
	Z_{ab} = i \mathring{\kappa}_{ab1} \bar{Z}_{01} \ . 
\end{align}
Note that we have a closed expression in terms of $\xi$ for all the required $\mathring{\kappa}_{ab1}$ that will appear when constructing $Z$. Using eq.~\eqref{IIB-eq:kappas_circle_app}, the matrix reads
\begin{equation}
Z_{AB}\!=\!\begin{pmatrix}
0 & Z_{01} & 0\\
Z_{01} & i\mathring{\kappa}_{111}\bar Z_{01} & 0\\
0 & 0 & i\mathring{\kappa}_{a'b'1}\bar Z_{01}
\end{pmatrix}
\!=\!\begin{pmatrix}
0 & Z_{01} & 0\\
Z_{01} & \frac{2i (1+\xi)^2}{\sqrt{3(1-2\xi)^3}}\bar Z_{01} & 0\\
0 & 0 & \frac{-i(1+\xi)}{\sqrt{3(1-2\xi)}}\delta_{a'b'}\bar Z_{01}
\end{pmatrix}.
\end{equation}

The diagonalization of $Z^\dagger Z$ gives the following eigenvalues:
\begin{equation}
	m_\lambda^2 = 
	\left\lbrace
	\begin{array}{ll}
		 \hat{m} (\xi)^2|Z_{01}|^2  & \lambda = 0  \\[5pt]
		(\hat{m} (\xi))^{-2}|Z_{01}|^2  & \lambda = 1 \\[5pt]
		\frac{(1+\xi)^2}{3(1-2\xi)}|Z_{01}|^2 & \lambda = 2, \ldots, h^{2,1}
	\end{array}
	\right.
	\label{IIB-eq:fermion_masses}
\end{equation}
where we have defined the quantities
\begin{align}
\begin{split}
&\hat{m} (\xi) \equiv \frac{1}{\sqrt{2}} \left( 2 + \kappa (\xi)^2 - \kappa (\xi) \sqrt{4 + \kappa (\xi)^2} \right)^{1/2}\ ,\\
&\kappa (\xi)\equiv\mathring{\kappa}_{111}= 2 (1+ \xi)^2 / \sqrt{3(1-2\xi)^3}\ .
\end{split}    
\end{align}
In order to deal with the dependency on $|Z_{01}|$, we use the other defining feature of no-scale aligned vacua, namely $f_A^0 = h_A^0 = 0$. According to the decomposition of the flux vector given in eq.~\eqref{IIB-eq:N_structure} together with the form of the period vector \eqref{IIB-eq:period}, this choice of fluxes leads to
\begin{align}
	W = -2i t^0 D_{\bar{\tau} \bar{j}} \bar{W} K^{\bar{j}i} K_i = i D_{0a} W \delta^{ab} K_b  \ \Rightarrow \ &
	& m_{3/2} = \sqrt{\frac{3}{1 - 2 \xi}} |Z_{01}| .
	\label{IIB-eq:m32_id}
\end{align}
Therefore, when plugging the eigenvalues $m_{\lambda}^2$ into eq.~\eqref{IIB-eq:mass_rel}, we can factorize an $m_{3/2}^2$ factor and obtain the scalar masss spectrum at no-scale-aligned vacua:
\begin{equation}
    \text{NSA mass spectrum:}\quad
	\frac{\mu^2_{\pm \lambda}}{m_{3/2}^2} = 
	\left\lbrace
	\begin{array}{ll}
		\left( 1 \pm \sqrt{\frac{1 -2 \xi}{3}} \hat{m} (\xi) \right)^2  & \lambda = 0  \\
		\left( 1 \pm \sqrt{\frac{1 -2 \xi}{3}} (\hat{m} (\xi))^{-1} \right)^2  & \lambda = 1 \\
		\left( 1 \pm \frac{1+\xi}{3} \right)^2 & \lambda = 2, \ldots, h^{2,1}
	\end{array}
	\right.
	\label{IIB-eq:nsa_mass_spectrum_app}
\end{equation}

\renewcommand{\theequation}{B.\arabic{equation}}

\section{Scalar potential and mass matrix}
\label{IIB-ap:spot}

In this section we present a detailed derivation of the scalar potential that describes the  IIB1 scenario and use it to directly compute the Hessian of the axionic sector, hence providing an alternative way to obtain the associated mass spectrum. 

\subsection{Metric tensor}

In the main text we found the vacuum equations using the no-scale structure of type IIB and working only with the superpotential. This procedure proved to be a powerful simplifying tool. However we now wish to go back to the results of chapter \ref{ch: Ftheory} and write the scalar potential for the  IIB1 scenario with corrections to all orders. The first step in this process is to revisit the Kähler potential and analyze the moduli space metric in more detail. From \eqref{IIB-eq: complex Kahler potential} we have
\begin{equation}
     K_{\text{cs}} = - \log \left( \frac{4}{3} \kappa_{ijk} t^i t^j t^k (1+\xi) \right)\ .
\end{equation}
Taking partial derivatives with respect to the dilaton and complex structure moduli we find
 \begin{align}
    K_{\tau \tau}=&\ \frac{1}{4(t^0)^2}\ ,\\
    K_{\tau i}=&\ 0\ ,\\
    K_{ij}=&\ \frac{9}{4}\frac{\kappa_i\kappa_j}{\kappa^2(1+\xi)^2}-\frac{3}{2}\frac{\kappa_{ij}}{\kappa(1+\xi)}=K^{\rm o}_{ij}\frac{\kappa}{\kappa(1+\xi)}-\frac{9}{4}\frac{\kappa_i\kappa_j}{\kappa^2(1+\xi)^2}\xi\ ,
\end{align}  
with $\kappa_{ij}\equiv\kappa_{ijk} t^k$, $\kappa_i\equiv\kappa_{ij}t^j$ and $\kappa\equiv\kappa_i t^i$. Finally we denote by $K^{\rm o}_{ij}$ the leading order metric, that is, the metric in the limit $\xi\rightarrow 0$.

Using the last expression for $K_{ij}$, it is straightforward to obtain its inverse in terms of the inverse of the leading order metric:
\begin{equation}
    K^{ij}=K_{\rm o}^{ij}(1+\xi)+\frac{4\xi(1+\xi)}{1-2\xi}t^it^j\ .
\end{equation}
Following the same line of reasoning as in appendix \ref{Ft-sap:corrFterm} we also compute
\begin{equation}
    K^{ij}K_{j}=2it^i\frac{1+\xi}{1-2\xi}\ .
\end{equation}
Finally, note that the metric leading order metric splits in its primitive and non primitive components as
\begin{align}
\begin{split}
    K_{ij}^{\rm oNP}&=\frac{3}{4}\frac{\kappa_i\kappa_j}{\kappa^2}\ ,\\
    K_{ij}^{\rm oP}&=\frac{3}{2}\frac{\kappa_i\kappa_j}{\kappa^2}-\frac{3}{2}\frac{\kappa_{ij}}{\kappa}\ .
\end{split}
\end{align}
In particular they satisfy $K_{ij}^{\rm oP}t^j=0$ and $K^{ij}_{\rm oP}\kappa_j=0$. We can replicate this split for the full metric to find
\begin{align}
\begin{split}
    K_{ij}^{\rm NP}&=K_{ij}^{\rm oNP}\frac{1-2\xi}{(1+\xi)^2}\ ,\\
    K_{ij}^{\rm P}&=K_{ij}^{\rm oP}\frac{1}{1+\xi}\ .
    \label{IIB-eq: primitive and no primitive corrected}
\end{split}
\end{align} 
\subsection{Scalar potential}

The scalar potential of the type IIB1 scenario can be derived following the same steps as the computation performed in appendix \ref{Ft-sap:corrFterm}. We start with the standard Cremmer et al. formula \cite{Cremmer:1982en} for the F-term potential in F-theory
\begin{equation}
    e^{-K}V_F=K^{m\bar{n}}D_m W D_{\bar{n}}\bar{W}-3|W|^2\ ,
\end{equation}
where $D_m W=\partial_{m}W+(\partial_{m}K) W$, $K^{m\bar{n}}$ is the inverse field space metric and $m, n$ run over all moduli. Ignoring corrections to the K\"ahler sector of the compactification we recover the standard cancellation of no-scale structure models and the above expression simplifies to
\begin{align}
    e^{-K}V_F=&\ K^{A\bar{B}}D_{A}W D_{\bar{B}} \bar{W}\nonumber\\
    =&\ K^{A\bar{B}}\left[\Re W_A \Re W_{\bar{B}} +\Im W_A \Im W_{\bar{B}}+\left((\Re W)^2+\Im(W)^2\right)K_A\bar{K}_{\bar{B}}\right.\nonumber\\
    &\left.+K_A W\bar{W}_{\bar{B}}+\bar{K}_{\bar{B}}W_A \bar{W}\right]\ ,
\end{align}
with $W_A\equiv\partial_A W$ and now $A,B\in\{0,i\}$ only run over the dilaton and complex structure moduli.

Using our knowledge of the metric and its properties, the above expression can be expanded to
\begin{align}
    e^{-K}V_F=&\ \frac{4-2\epsilon}{1-2\epsilon}\left((\Re W)^2+(\Im W)^2\right)+K^{ij}(\Re W_i\Re W_j+ \Im W_i\Im W_j)\nonumber\\
    &+4t^i \frac{1+\epsilon}{1-2\epsilon}[\Re W \Im W_i- \Im W \Re W_i]+4(t^0)^2[(\Re W_0)^2+(\Im W_0)^2]\nonumber\\
    &+4t^0[\Re W\Im W_0-\Im W\Re W_0]\ .
\end{align}
 We proceed to consider the version of the superpotential and the Kähler potential  described in the main text in eq.~\eqref{IIB-eq:W_bilinear}:
\begin{equation}
W=\oh\vec{Z}^t M \vec{Z} + \vec{L} \cdot \vec{Z} +Q\ ,
\end{equation}
Splitting the real and imaginary parts we see that
\begin{align}
\begin{split}
        \Re W=&\ \frac{1}{2}\vec{B}M\vec{B}+\vec{L}\cdot\vec{B}+Q-\frac{1}{2}\vec{T}M\vec{T}=\rho-\frac{1}{2}\kappa_i f_A^i+t^0t^ih_i^B\ ,\\
        \Im W=&\ \vec{T}\cdot(M\vec{B}+\vec{L})=\rho_A t^A\ ,
\end{split}        
\end{align}
where we have defined
\begin{align}
\begin{split}
        \rho&\equiv\frac{1}{2}\vec{B}M\vec{B}+\vec{L}\cdot\vec{B}+Q\ ,\\
        \rho_A&\equiv M_{AB}b^B+L_A\ .
\end{split}
\end{align}
Similarly, the real and imaginary parts of the partial derivatives of the superpotential can be written as follows:
\begin{eqnarray}
                \Re W_0=&\ \bar{\rho}_0\ , \qquad & \Im W_0=-t^ih_i^B\ ,\nonumber\\
        \Re W_i=&\ \bar{\rho}_i\ , \qquad & \Im W_i= -t^0h_i^B+ \kappa_{ij}f_A^j\ .
\end{eqnarray}
Substituting, expanding, rearranging and using the expressions found in the previous section we conclude that
\begin{align}
    V_Fe^{-K}=&\ 4\rho^2+4(\rho_0t^0)^2+(1+\xi)\left(K_{\rm o}^{ij}\rho_i\rho_j+(t^0)^2 K_{\rm o}^{ij} h_i^B h_j^B+\frac{4}{9}K_{ij}^{\rm o} f^i_Af^j_A+\frac{4}{3}t^0\kappa h_i^B f^i_A\right)\nonumber\\
    &+\frac{\xi}{1-2\xi}\bigg[6\rho^2+6\rho\kappa_if_A^i+\frac{1}{2}(\kappa_if_A^i)^2+6(\rho_0 t^0)^2-2(\rho_it^i)^2\!-2(t^0h_i^Bt^i)^2\\
    &+2\xi\left[2(\rho_it^i)^2+2(t^it^0h_i^B)^2+(\kappa_i f^i_A)^2\right]\bigg]\ ,\nonumber
\end{align}
which at leading order recovers the result \eqref{Ft-ZABIIB} in the IIB1 scenario.

\subsection{Hessian}

Now that we have the potential, we can compute the second derivatives. We focus only on the simpler axionic directions. For that mission, the following relations prove to be very useful.
\begin{equation}
\begin{aligned}
&\frac{\partial \rho}{\partial b^0}=\rho_0\ , &&\frac{\partial \rho}{\partial b^i}=\rho_i\ , &&\frac{\partial \rho_0}{\partial b^0}=0\ ,\\ 
&\frac{\partial \rho_0}{\partial b^i}=-h_i^B\ ,\qquad\qquad &&\frac{\partial\rho_i}{\partial b^0}=-h_i^B\ ,\qquad\qquad  &&\frac{\partial \rho_i}{\partial b^j}=\kappa_{ijk}f^k_A\ .
\end{aligned}
\end{equation}
Thanks to them we obtain that the first derivatives can be written as
    \begin{align}
        \frac{\partial V}{\partial b^0}e^{-K}=&\ 8\rho\rho_0-(1+\xi)(2K^{ij}_{\rm o}h_i^B\rho_j)\nonumber\\
        &+\frac{3\xi}{1-2\xi}(4\rho\rho_0+2\rho_0\kappa_i f_A^i+\frac{4}{3}h^B_it^i\rho_j t^j -\frac{8}{3}\xi h^B_i t^i\rho_jt^j)\ ,\\
        \frac{\partial V}{\partial b^i}e^{-K}=& \ 8\rho\rho_i-8\rho_0t^0h_i^B t^0+(1+\xi)(2K_{\rm o}^{jk}\kappa_{ijl}f_A^l\rho_k)\nonumber\\
        &+\frac{3\xi}{1-2\xi}\left[4\rho\rho_i+2\rho_i\kappa_jf_A^j-4h_i^B\rho_0(t^0)^2-\frac{4}{3}\rho_jt^j\kappa_{ik}f_A^k+\frac{8}{3}\xi \rho_jt^j\kappa_{ik}f_A^k\right]\ .\nonumber
    \end{align}
We proceed with the second derivatives. From the above expressions we can already see that axions and saxions are decoupled in the vacuum. Noting that  that the $\rho$'s do not depend on the saxions and that the equation of motion \eqref{IIB-eq:B_gen} implies $\rho_A=0$, it is easy to see that the cross terms involving derivatives of saxions and axions vanish. Therefore, the saxionic and axionic mass matrices are decoupled. Focusing on the pure axionic sector we find
    \begin{align}
        \frac{\partial^2 V}{(\partial b^0)^2}e^{-K}=&\ 8\rho_0^2+(1+\xi)(2K_{\rm o}^{ij}h_i^Bh_j^B)+\frac{3\xi}{1-2\xi}\left(4\rho_0^2-\frac{4}{3}(h_i^Bt^i)^2+\frac{8}{3}\xi (h_i^Bt^i)^2\right)\ ,\nonumber\\
        \frac{\partial^2 V}{\partial b^i\partial b^0}e^{-K}=&\ 8\rho_i\rho_0-8\rho h_i^B-(1+\xi)2K_{\rm o}^{jk}\kappa_{ijl}f_A^l h_k^B\\
        &+\frac{3\xi}{1-2\xi}(4\rho_i\rho_0-4\rho h_i^B-2h_i^B\kappa_jf_A^j+\frac{4}{3}h_j^Bt^j\kappa_{ik}f_A^k-\frac{8}{3}\xi h_j^Bt^j \kappa_{ik}f_A^k)\ ,\nonumber\\
         \frac{\partial^2 V}{\partial b^i\partial b^j}e^{-K}=&\ 8\rho_i\rho_j+8\rho\kappa_{ijk}f_A^k+8h_i^Bh_j^B(t^0)^2+(1+\xi)(2K^{kl}_{\rm o}\kappa_{ikm}\kappa_{jln}f_A^mf_A^n)\nonumber\\
         &+\frac{3\xi}{1-2\xi}(4\rho_i\rho_j+4\rho\kappa_{ijk}f_A^k+2\kappa_{ijk}f_A^j\kappa_l f_A^l+4h_i^Bh_j^B(t^0)^2-\frac{4}{3}\kappa_{ik}f_A^k\kappa_{jl}f_A^l\nonumber\\
         &+\frac{8}{3}\xi \kappa_{ik}\kappa_{jl}f_A^kf_A^l)\ .\nonumber
    \end{align}
To evaluate the Hessian in the vacuum, we introduce the equations of motion and restrict ourselves to the Ansatz considered in the main text \eqref{IIB-eq:that}. Hence, from now on the results will be only valid in a particular subbranch of the non-supersymmetric vacua with $M$ regular. The relation for the axions demands $\rho_A=0$ while the Ansatz \eqref{IIB-eq:that} in combination with the equations of motion of the saxions \eqref{IIB-hBicond} implies 
\begin{equation}
    f_A^i= \frac{t^i}{\hat{t}}\ ,\qquad  h_i^B=-\frac{q^{-1}\hat{h}^B}{\hat{t}^2}\kappa_i\ .
\end{equation}
For the sake of convenience we rewrite the last two relations in terms of the coefficients of the decomposition introduced in \eqref{IIB-decomp}. Then, with the help of \eqref{IIB-eq:t0t} we have the simple relations
\begin{equation}
    f_A^i=-t^0 r_\xi B\ , \qquad h_i^B= B\kappa_i\ ,
\end{equation}
where $B=-A/(t^0 r_\xi)$, $A=1/\hat{t}$ and we have defined
\begin{equation}
    r_\xi\equiv \frac{2-\xi}{1+\xi}\ .
\end{equation}
Finally, when the axionic equations of motion are satisfied, eq.~\eqref{IIB-eq:B_gen} means $\rho=Q'$ and using \eqref{IIB-eq:flux_const_away} and the above definitions we can derive the following equation
\begin{equation}
    \rho=\frac{3}{2}\frac{\xi}{\xi+1}\kappa t^0 B\ .
\end{equation}
Putting all together, we conclude that the Hessian evaluated in the branch \eqref{IIB-eq:that} takes the form
    \begin{align}
    \label{IIB-eq: axionic derivatives + ansatz}
        \frac{\partial^2 V}{(\partial b^0)^2}e^{-K}=&\ \frac{4}{3}B^2(2-\xi)\kappa^2\ ,\nonumber\\
        \frac{\partial^2 V}{\partial b^i\partial b^0}e^{-K}=&\ \frac{4B^2\kappa t^0 (2-\xi)^2}{3(1+\xi)}\kappa_i \ ,\\
         \frac{\partial^2 V}{\partial b^i\partial b^j}e^{-K}=&\ \frac{4 B^2 \left(2 \xi^4-16 \xi^3+30 \xi^2-19 \xi+14\right) (t^0)^2}{3 (\xi+1)^2 (2 \xi-1)}\kappa_i\kappa_j+\frac{8 B^2 (\xi-2)^2 k^2 (t^0)^2}{9 (\xi+1)} K_{ij}^0\ .\nonumber
    \end{align}
The last step is to write the Hessian for canonically normalized fields. We separate the dilaton and the non-primitive directions by considering an orthogonal basis of the form $\mathcal{B}\equiv\{e_0,e_1, e_\alpha\}$ where the elements are chosen such that $e_0^0K_{00}e_0^0=1$, $e_1^i K_{ij}^{\rm NP} e_1^j=1$ and  $e_\alpha^i K_{ij}^{\rm oP}e_\alpha^j=1$ $\forall\alpha$, with $K_{ij}^{\rm P} e^i_1=0=K_{ij}^{\rm NP} e^j_{\alpha}$. To make the basis explicit we make use of \eqref{IIB-eq: primitive and no primitive corrected}. We have
    \begin{align}
    \begin{split}
        e_0=&\ \{2t^0,0,\dots, 0\}\ ,\\
        e_1^a=&\ \frac{2}{\sqrt{3}}\frac{1+\xi}{\sqrt{1-2\xi}}t^a\ .
    \end{split}
    \end{align}
Note that since $K_{ij}^{\rm P} e^i_\alpha e^j_\beta=\delta_{\alpha \beta}$, then $K_{ij}^{\rm oP} e^i_\alpha e^j_\beta=\delta_{\alpha \beta} (1+\xi)$. Projecting the  Hessian \eqref{IIB-eq: axionic derivatives + ansatz} along the directions of our canonically normalized basis, we obtain the final following form:
\begin{equation}
    H=e^K B^2\kappa^2 (t^0)^2\left(\begin{array}{ccc}
        \frac{16}{3}(2 - \xi) & \frac{16(2-\xi)^2}{3\sqrt{3-6\xi}} & 0 \\
        \frac{16(2-\xi)^2}{3\sqrt{3-6\xi}} & \frac{16 \left(2 \xi^4-16 \xi^3+30 \xi^2-19 \xi+14\right) }{9 (1-2 \xi)^2} & 0 \\
        0 & 0 & \frac{8}{9}(2-\xi)^2
    \end{array}\right)\, .
\end{equation}
Ignoring the global factors, this matrix has the following eigenvalues:
\begin{align}
\begin{split}
    \lambda_1=&\ \frac{16 (\xi-2)}{9 (2 \xi-1)^3} \left(2 \xi^4-25 \xi^3+30 \xi^2-19 \xi+5\right.\\
    &\left.+\sqrt{1-2 \xi} \sqrt{-(\xi-2)^3 \left(2 \xi^4-37 \xi^3+30 \xi^2-10 \xi+2\right)}\right)\ ,\\
    \lambda_2=&\ \frac{16 (\xi-2)}{9 (2 \xi-1)^3} \left(2 \xi^4-25 \xi^3+30 \xi^2-19 \xi+5\right.\\
    &\left.-\sqrt{1-2 \xi} \sqrt{-(\xi-2)^3 \left(2 \xi^4-37 \xi^3+30 \xi^2-10 \xi+2\right)}\right)\ ,\\
    \lambda_3=&\ \frac{8}{9} (\xi-2)^2\ .
\end{split}
\end{align} 
The first two eigenvalues have multiplicity one whereas the last one has multiplicity $h^{2,1}-1$. Adding the factors and remembering  that the mass spectrum gets and additional factor $1/2$, the masses will be given by
\begin{equation}
    m_i^2=\frac{1}{2}e^K A^2 \kappa^2 (t^0)^2 \lambda_i\, M_{\rm P}^2\ .
\end{equation}
To compare with the results found in \eqref{IIB-eq:nsa_mass_spectrum_app}, we expand the exponential of the Kähler potential
\begin{equation}
    e^K=\frac{1}{\mathcal{V}^2}\frac{1}{2t^0}\frac{1}{\frac{4}{3}\kappa(1+\xi)}\ ,
\end{equation}
and the gravitino mass
\begin{equation}
m_{3/2}^2=\frac{3}{2\mathcal{V}^2}\frac{\Nf}{2-\xi} M_{\rm P}^2  =  \frac{3}{2\mathcal{V}^2}\frac{B^2 t^0 \kappa }{1+\xi} M_{\rm P}^2 \, .
\end{equation}
Putting all together we conclude that the  eigenvalues coincide with the results in \eqref{IIB-eq:nsa_mass_spectrum_app}. This calculation has the advantage that it enables us to distinguish the axionic and saxionic masses. The axionic ones under consideration here then correspond to the following choices of signs in \eqref{IIB-eq:nsa_mass_spectrum_app}
\begin{align}
\begin{split}
    m_1^2= &\ m_{3/2}^2 \left( 1 + \sqrt{\frac{1 -2 \xi}{3}} (\hat{m} (\xi))^{-1} \right)^2\ ,\\
    m_2^2 = &\ m_{3/2}^2 \left( 1 - \sqrt{\frac{1 -2 \xi}{3}} \hat{m} (\xi) \right)^2\ ,\\
    m_3^2 =& \ m_{3/2}^2 \left(1-\frac{1+\xi}{3}\right)^2\ .
\end{split}
\end{align}

\ifSubfilesClassLoaded{%
\bibliography{biblio}%
}{}

\end{document}


\backmatter
\addcontentsline{toc}{chapter}{Bibliography}
\small
\singlespacing
\bibliography{biblio}

\providecommand{\href}[2]{#2}\begingroup\raggedright\begin{thebibliography}{100}

\bibitem{Marchesano:2020uqz}
F.~Marchesano, D.~Prieto, J.~Quirant and P.~Shukla, \emph{{Systematics of Type
  IIA moduli stabilisation}},
  \href{https://doi.org/10.1007/JHEP11(2020)113}{\emph{JHEP} {\bfseries 11}
  (2020) 113} [\href{https://arxiv.org/abs/2007.00672}{{\ttfamily
  2007.00672}}].

\bibitem{Marchesano:2021gyv}
F.~Marchesano, D.~Prieto and M.~Wiesner, \emph{{F-theory flux vacua at large
  complex structure}},
  \href{https://doi.org/10.1007/JHEP08(2021)077}{\emph{JHEP} {\bfseries 08}
  (2021) 077} [\href{https://arxiv.org/abs/2105.09326}{{\ttfamily
  2105.09326}}].

\bibitem{Marchesano:2021ycx}
F.~Marchesano, D.~Prieto and J.~Quirant, \emph{{BIonic membranes and AdS
  instabilities}}, \href{https://doi.org/10.1007/JHEP07(2022)118}{\emph{JHEP}
  {\bfseries 07} (2022) 118}
  [\href{https://arxiv.org/abs/2110.11370}{{\ttfamily 2110.11370}}].

\bibitem{Casas:2022mnz}
G.F.~Casas, F.~Marchesano and D.~Prieto, \emph{{Membranes in AdS$_{4}$
  orientifold vacua and their Weak Gravity Conjecture}},
  \href{https://doi.org/10.1007/JHEP09(2022)034}{\emph{JHEP} {\bfseries 09}
  (2022) 034} [\href{https://arxiv.org/abs/2204.11892}{{\ttfamily
  2204.11892}}].

\bibitem{Coudarchet:2022fcl}
T.~Coudarchet, F.~Marchesano, D.~Prieto and M.A.~Urkiola, \emph{{Analytics of
  type IIB flux vacua and their mass spectra}},
  \href{https://doi.org/10.1007/JHEP01(2023)152}{\emph{JHEP} {\bfseries 01}
  (2023) 152} [\href{https://arxiv.org/abs/2212.02533}{{\ttfamily
  2212.02533}}].

\bibitem{Coudarchet:2023mmm}
T.~Coudarchet, F.~Marchesano, D.~Prieto and M.A.~Urkiola, \emph{{Symmetric
  fluxes and small tadpoles}},
  \href{https://doi.org/10.1007/JHEP08(2023)016}{\emph{JHEP} {\bfseries 08}
  (2023) 16} [\href{https://arxiv.org/abs/2304.04789}{{\ttfamily 2304.04789}}].

\bibitem{systematicsfollowup}
R.~Carrasco, T.~Coudarchet, F.~Marchesano and D.~Prieto, \emph{{New families of
  scale separated vacua}},
  \href{https://doi.org/10.1007/JHEP11(2023)094}{\emph{JHEP} {\bfseries 11}
  (2023) 094} [\href{https://arxiv.org/abs/2309.00043}{{\ttfamily
  2309.00043}}].

\bibitem{Einstein:1916vd}
A.~Einstein, \emph{{The foundation of the general theory of relativity.}},
  \href{https://doi.org/10.1002/andp.19163540702}{\emph{Annalen Phys.}
  {\bfseries 49} (1916) 769}.

\bibitem{LIGOScientific:2016aoc}
{\scshape LIGO Scientific, Virgo} collaboration, \emph{{Observation of
  Gravitational Waves from a Binary Black Hole Merger}},
  \href{https://doi.org/10.1103/PhysRevLett.116.061102}{\emph{Phys. Rev. Lett.}
  {\bfseries 116} (2016) 061102}
  [\href{https://arxiv.org/abs/1602.03837}{{\ttfamily 1602.03837}}].

\bibitem{Einstein:1916cc}
A.~Einstein, \emph{{Approximative Integration of the Field Equations of
  Gravitation}}, {\emph{Sitzungsber. Preuss. Akad. Wiss. Berlin (Math. Phys. )}
  {\bfseries 1916} (1916) 688}.

\bibitem{CMS:2012qbp}
{\scshape CMS} collaboration, \emph{{Observation of a New Boson at a Mass of
  125 GeV with the CMS Experiment at the LHC}},
  \href{https://doi.org/10.1016/j.physletb.2012.08.021}{\emph{Phys. Lett. B}
  {\bfseries 716} (2012) 30} [\href{https://arxiv.org/abs/1207.7235}{{\ttfamily
  1207.7235}}].

\bibitem{ATLAS:2012yve}
{\scshape ATLAS} collaboration, \emph{{Observation of a new particle in the
  search for the Standard Model Higgs boson with the ATLAS detector at the
  LHC}}, \href{https://doi.org/10.1016/j.physletb.2012.08.020}{\emph{Phys.
  Lett. B} {\bfseries 716} (2012) 1}
  [\href{https://arxiv.org/abs/1207.7214}{{\ttfamily 1207.7214}}].

\bibitem{Parker:2018vye}
R.H.~Parker, C.~Yu, W.~Zhong, B.~Estey and H.~M\"uller, \emph{{Measurement of
  the fine-structure constant as a test of the Standard Model}},
  \href{https://doi.org/10.1126/science.aap7706}{\emph{Science} {\bfseries 360}
  (2018) 191} [\href{https://arxiv.org/abs/1812.04130}{{\ttfamily
  1812.04130}}].

\bibitem{Feynman:1963ax}
R.P.~Feynman, \emph{{Quantum theory of gravitation}}, {\emph{Acta Phys. Polon.}
  {\bfseries 24} (1963) 697}.

\bibitem{tHooft:1974toh}
G.~'t~Hooft and M.J.G.~Veltman, \emph{{One loop divergencies in the theory of
  gravitation}}, {\emph{Ann. Inst. H. Poincare Phys. Theor. A} {\bfseries 20}
  (1974) 69}.

\bibitem{Goroff:1985sz}
M.H.~Goroff and A.~Sagnotti, \emph{{QUANTUM GRAVITY AT TWO LOOPS}},
  \href{https://doi.org/10.1016/0370-2693(85)91470-4}{\emph{Phys. Lett. B}
  {\bfseries 160} (1985) 81}.

\bibitem{Witten:1995ex}
E.~Witten, \emph{{String theory dynamics in various dimensions}},
  \href{https://doi.org/10.1016/0550-3213(95)00158-O}{\emph{Nucl. Phys.}
  {\bfseries B443} (1995) 85}
  [\href{https://arxiv.org/abs/hep-th/9503124}{{\ttfamily hep-th/9503124}}].

\bibitem{Dai:1989ua}
J.~Dai, R.G.~Leigh and J.~Polchinski, \emph{{New Connections Between String
  Theories}}, \href{https://doi.org/10.1142/S0217732389002331}{\emph{Mod. Phys.
  Lett. A} {\bfseries 4} (1989) 2073}.

\bibitem{Polchinski:1995mt}
J.~Polchinski, \emph{{Dirichlet Branes and Ramond-Ramond charges}},
  \href{https://doi.org/10.1103/PhysRevLett.75.4724}{\emph{Phys. Rev. Lett.}
  {\bfseries 75} (1995) 4724}
  [\href{https://arxiv.org/abs/hep-th/9510017}{{\ttfamily hep-th/9510017}}].

\bibitem{Taylor:2015xtz}
W.~Taylor and Y.-N.~Wang, \emph{{The F-theory geometry with most flux vacua}},
  \href{https://doi.org/10.1007/JHEP12(2015)164}{\emph{JHEP} {\bfseries 12}
  (2015) 164} [\href{https://arxiv.org/abs/1511.03209}{{\ttfamily
  1511.03209}}].

\bibitem{Vafa:2005ui}
C.~Vafa, \emph{{The String landscape and the swampland}},
  \href{https://arxiv.org/abs/hep-th/0509212}{{\ttfamily hep-th/0509212}}.

\bibitem{Marchesano:2019hfb}
F.~Marchesano and J.~Quirant, \emph{{A Landscape of AdS Flux Vacua}},
  \href{https://doi.org/10.1007/JHEP12(2019)110}{\emph{JHEP} {\bfseries 12}
  (2019) 110} [\href{https://arxiv.org/abs/1908.11386}{{\ttfamily
  1908.11386}}].

\bibitem{Ibanez:2012zz}
L.E.~Ibanez and A.M.~Uranga, \emph{{String theory and particle physics: An
  introduction to string phenomenology}}, Cambridge University Press (2, 2012).

\bibitem{Blumenhagen:2013fgp}
R.~Blumenhagen, D.~L{\"u}st and S.~Theisen, \emph{{Basic concepts of string
  theory}},
  \href{https://doi.org/10.1007/978-3-642-29496-9}{\emph{Springer-Verlag}
  (2013) }.

\bibitem{Becker:2006dvp}
K.~Becker, M.~Becker and J.H.~Schwarz, \emph{{String theory and M-theory: A
  modern introduction}}, Cambridge University Press (12, 2006),
  \href{https://doi.org/10.1017/CBO9780511816086}{10.1017/CBO9780511816086}.

\bibitem{Polchinski:1998rq}
J.~Polchinski, \emph{{String theory. Vol. 1: An introduction to the bosonic
  string}}, Cambridge Monographs on Mathematical Physics, Cambridge University
  Press (12, 2007),
  \href{https://doi.org/10.1017/CBO9780511816079}{10.1017/CBO9780511816079}.

\bibitem{Polchinski:1998rr}
J.~Polchinski, \emph{{String theory. Vol. 2: Superstring theory and beyond}},
  Cambridge Monographs on Mathematical Physics, Cambridge University Press (12,
  2007),
  \href{https://doi.org/10.1017/CBO9780511618123}{10.1017/CBO9780511618123}.

\bibitem{Green:2012oqa}
M.B.~Green, J.H.~Schwarz and E.~Witten, \emph{{Superstring Theory Vol. 1}:
  {25th Anniversary Edition}}, Cambridge Monographs on Mathematical Physics,
  Cambridge University Press (11, 2012),
  \href{https://doi.org/10.1017/CBO9781139248563}{10.1017/CBO9781139248563}.

\bibitem{Green:2012pqa}
M.B.~Green, J.H.~Schwarz and E.~Witten, \emph{{Superstring Theory Vol. 2}:
  {25th Anniversary Edition}}, Cambridge Monographs on Mathematical Physics,
  Cambridge University Press (11, 2012),
  \href{https://doi.org/10.1017/CBO9781139248570}{10.1017/CBO9781139248570}.

\bibitem{Tong:2009np}
D.~Tong, \emph{{String Theory}},
  \href{https://arxiv.org/abs/0908.0333}{{\ttfamily 0908.0333}}.

\bibitem{Kiritsis:1997hj}
E.~Kiritsis, \emph{{Introduction to superstring theory}}, vol.~B9 of
  \emph{Leuven notes in mathematical and theoretical physics}, Leuven U. Press,
  Leuven (1998), [\href{https://arxiv.org/abs/hep-th/9709062}{{\ttfamily
  hep-th/9709062}}].

\bibitem{Brennan:2017rbf}
T.D.~Brennan, F.~Carta and C.~Vafa, \emph{{The String Landscape, the Swampland,
  and the Missing Corner}},
  \href{https://doi.org/10.22323/1.305.0015}{\emph{PoS} {\bfseries TASI2017}
  (2017) 015} [\href{https://arxiv.org/abs/1711.00864}{{\ttfamily
  1711.00864}}].

\bibitem{Palti:2019pca}
E.~Palti, \emph{{The Swampland: Introduction and Review}},
  \href{https://doi.org/10.1002/prop.201900037}{\emph{Fortsch. Phys.}
  {\bfseries 67} (2019) 1900037}
  [\href{https://arxiv.org/abs/1903.06239}{{\ttfamily 1903.06239}}].

\bibitem{vanBeest:2021lhn}
M.~van Beest, J.~Calder\'on-Infante, D.~Mirfendereski and I.~Valenzuela,
  \emph{{Lectures on the Swampland Program in String Compactifications}},
  \href{https://doi.org/10.1016/j.physrep.2022.09.002}{\emph{Phys. Rept.}
  {\bfseries 989} (2022) 1} [\href{https://arxiv.org/abs/2102.01111}{{\ttfamily
  2102.01111}}].

\bibitem{Grana:2021zvf}
M.~Gra\~na and A.~Herr\'aez, \emph{{The Swampland Conjectures: A Bridge from
  Quantum Gravity to Particle Physics}},
  \href{https://doi.org/10.3390/universe7080273}{\emph{Universe} {\bfseries 7}
  (2021) 273} [\href{https://arxiv.org/abs/2107.00087}{{\ttfamily
  2107.00087}}].

\bibitem{Agmon:2022thq}
N.B.~Agmon, A.~Bedroya, M.J.~Kang and C.~Vafa, \emph{{Lectures on the string
  landscape and the Swampland}},
  \href{https://arxiv.org/abs/2212.06187}{{\ttfamily 2212.06187}}.

\bibitem{Polyakov:1981rd}
A.M.~Polyakov, \emph{{Quantum Geometry of Bosonic Strings}},
  \href{https://doi.org/10.1016/0370-2693(81)90743-7}{\emph{Phys. Lett. B}
  {\bfseries 103} (1981) 207}.

\bibitem{Neveu:1971rx}
A.~Neveu and J.H.~Schwarz, \emph{{Factorizable dual model of pions}},
  \href{https://doi.org/10.1016/0550-3213(71)90448-2}{\emph{Nucl. Phys. B}
  {\bfseries 31} (1971) 86}.

\bibitem{Ramond:1971gb}
P.~Ramond, \emph{{Dual Theory for Free Fermions}},
  \href{https://doi.org/10.1103/PhysRevD.3.2415}{\emph{Phys. Rev. D} {\bfseries
  3} (1971) 2415}.

\bibitem{Gliozzi:1976qd}
F.~Gliozzi, J.~Scherk and D.I.~Olive, \emph{{Supersymmetry, Supergravity
  Theories and the Dual Spinor Model}},
  \href{https://doi.org/10.1016/0550-3213(77)90206-1}{\emph{Nucl. Phys. B}
  {\bfseries 122} (1977) 253}.

\bibitem{Horowitz:1991cd}
G.T.~Horowitz and A.~Strominger, \emph{{Black strings and P-branes}},
  \href{https://doi.org/10.1016/0550-3213(91)90440-9}{\emph{Nucl. Phys. B}
  {\bfseries 360} (1991) 197}.

\bibitem{Bergshoeff:2001pv}
E.~Bergshoeff, R.~Kallosh, T.~Ortin, D.~Roest and A.~Van~Proeyen, \emph{{New
  formulations of D = 10 supersymmetry and D8 - O8 domain walls}},
  \href{https://doi.org/10.1088/0264-9381/18/17/303}{\emph{Class. Quant. Grav.}
  {\bfseries 18} (2001) 3359}
  [\href{https://arxiv.org/abs/hep-th/0103233}{{\ttfamily hep-th/0103233}}].

\bibitem{Lerche:1986cx}
W.~Lerche, D.~Lust and A.N.~Schellekens, \emph{{Chiral Four-Dimensional
  Heterotic Strings from Selfdual Lattices}},
  \href{https://doi.org/10.1016/0550-3213(87)90115-5}{\emph{Nucl. Phys. B}
  {\bfseries 287} (1987) 477}.

\bibitem{QuirantPellin:2022vyp}
J.~Quirant~Pell\'\i{}n, \emph{{Aspects of type IIA AdS4 orientifold vacua}},
  Ph.D. thesis, U. Autonoma, Madrid, 2022.
\newblock \href{https://arxiv.org/abs/2210.08466}{{\ttfamily 2210.08466}}.

\bibitem{Banks:1988yz}
T.~Banks and L.J.~Dixon, \emph{{Constraints on String Vacua with Space-Time
  Supersymmetry}},
  \href{https://doi.org/10.1016/0550-3213(88)90523-8}{\emph{Nucl.Phys.}
  {\bfseries B307} (1988) 93}.

\bibitem{Banks:2010zn}
T.~Banks and N.~Seiberg, \emph{{Symmetries and Strings in Field Theory and
  Gravity}}, \href{https://doi.org/10.1103/PhysRevD.83.084019}{\emph{Phys.Rev.}
  {\bfseries D83} (2011) 084019}
  [\href{https://arxiv.org/abs/1011.5120}{{\ttfamily 1011.5120}}].

\bibitem{Harlow:2018jwu}
D.~Harlow and H.~Ooguri, \emph{{Constraints on Symmetries from Holography}},
  \href{https://doi.org/10.1103/PhysRevLett.122.191601}{\emph{Phys. Rev. Lett.}
  {\bfseries 122} (2019) 191601}
  [\href{https://arxiv.org/abs/1810.05337}{{\ttfamily 1810.05337}}].

\bibitem{Harlow:2018tng}
D.~Harlow and H.~Ooguri, \emph{{Symmetries in quantum field theory and quantum
  gravity}}, \href{https://doi.org/10.1007/s00220-021-04040-y}{\emph{Commun.
  Math. Phys.} {\bfseries 383} (2021) 1669}
  [\href{https://arxiv.org/abs/1810.05338}{{\ttfamily 1810.05338}}].

\bibitem{McNamara:2019rup}
J.~McNamara and C.~Vafa, \emph{{Cobordism Classes and the Swampland}},
  \href{https://arxiv.org/abs/1909.10355}{{\ttfamily 1909.10355}}.

\bibitem{ArkaniHamed:2006dz}
N.~Arkani-Hamed, L.~Motl, A.~Nicolis and C.~Vafa, \emph{{The String landscape,
  black holes and gravity as the weakest force}},
  \href{https://doi.org/10.1088/1126-6708/2007/06/060}{\emph{JHEP} {\bfseries
  0706} (2007) 060} [\href{https://arxiv.org/abs/hep-th/0601001}{{\ttfamily
  hep-th/0601001}}].

\bibitem{Harlow:2022ich}
D.~Harlow, B.~Heidenreich, M.~Reece and T.~Rudelius, \emph{{Weak gravity
  conjecture}}, \href{https://doi.org/10.1103/RevModPhys.95.035003}{\emph{Rev.
  Mod. Phys.} {\bfseries 95} (2023) 035003}
  [\href{https://arxiv.org/abs/2201.08380}{{\ttfamily 2201.08380}}].

\bibitem{Coleman:1967ad}
S.R.~Coleman and J.~Mandula, \emph{{All Possible Symmetries of the S Matrix}},
  \href{https://doi.org/10.1103/PhysRev.159.1251}{\emph{Phys. Rev.} {\bfseries
  159} (1967) 1251}.

\bibitem{Ooguri:2016pdq}
H.~Ooguri and C.~Vafa, \emph{{Non-supersymmetric AdS and the Swampland}},
  \href{https://doi.org/10.4310/ATMP.2017.v21.n7.a8}{\emph{Adv. Theor. Math.
  Phys.} {\bfseries 21} (2017) 1787}
  [\href{https://arxiv.org/abs/1610.01533}{{\ttfamily 1610.01533}}].

\bibitem{Freivogel:2016qwc}
B.~Freivogel and M.~Kleban, \emph{{Vacua Morghulis}},
  \href{https://arxiv.org/abs/1610.04564}{{\ttfamily 1610.04564}}.

\bibitem{Ooguri:2006in}
H.~Ooguri and C.~Vafa, \emph{{On the Geometry of the String Landscape and the
  Swampland}},
  \href{https://doi.org/10.1016/j.nuclphysb.2006.10.033}{\emph{Nucl. Phys.}
  {\bfseries B766} (2007) 21}
  [\href{https://arxiv.org/abs/hep-th/0605264}{{\ttfamily hep-th/0605264}}].

\bibitem{Lust:2019zwm}
D.~L\"ust, E.~Palti and C.~Vafa, \emph{{AdS and the Swampland}},
  \href{https://doi.org/10.1016/j.physletb.2019.134867}{\emph{Phys. Lett. B}
  {\bfseries 797} (2019) 134867}
  [\href{https://arxiv.org/abs/1906.05225}{{\ttfamily 1906.05225}}].

\bibitem{Duff:1986hr}
M.J.~Duff, B.E.W.~Nilsson and C.N.~Pope, \emph{{Kaluza-Klein Supergravity}},
  \href{https://doi.org/10.1016/0370-1573(86)90163-8}{\emph{Phys. Rept.}
  {\bfseries 130} (1986) 1}.

\bibitem{Douglas:2006es}
M.R.~Douglas and S.~Kachru, \emph{{Flux compactification}},
  \href{https://doi.org/10.1103/RevModPhys.79.733}{\emph{Rev. Mod. Phys.}
  {\bfseries 79} (2007) 733}
  [\href{https://arxiv.org/abs/hep-th/0610102}{{\ttfamily hep-th/0610102}}].

\bibitem{Tsimpis:2012tu}
D.~Tsimpis, \emph{{Supersymmetric AdS vacua and separation of scales}},
  \href{https://doi.org/10.1007/JHEP08(2012)142}{\emph{JHEP} {\bfseries 08}
  (2012) 142} [\href{https://arxiv.org/abs/1206.5900}{{\ttfamily 1206.5900}}].

\bibitem{Gautason:2015tig}
F.F.~Gautason, M.~Schillo, T.~Van~Riet and M.~Williams, \emph{{Remarks on scale
  separation in flux vacua}},
  \href{https://doi.org/10.1007/JHEP03(2016)061}{\emph{JHEP} {\bfseries 03}
  (2016) 061} [\href{https://arxiv.org/abs/1512.00457}{{\ttfamily
  1512.00457}}].

\bibitem{Font:2019uva}
A.~Font, A.~Herráez and L.E.~Ibáñez, \emph{{On scale separation in type II
  AdS flux vacua}}, \href{https://doi.org/10.1007/JHEP03(2020)013}{\emph{JHEP}
  {\bfseries 03} (2020) 013}
  [\href{https://arxiv.org/abs/1912.03317}{{\ttfamily 1912.03317}}].

\bibitem{Montero:2022ghl}
M.~Montero, M.~Rocek and C.~Vafa, \emph{{Pure supersymmetric AdS and the
  Swampland}}, \href{https://doi.org/10.1007/JHEP01(2023)094}{\emph{JHEP}
  {\bfseries 01} (2023) 094}
  [\href{https://arxiv.org/abs/2212.01697}{{\ttfamily 2212.01697}}].

\bibitem{Buratti:2020kda}
G.~Buratti, J.~Calderon, A.~Mininno and A.M.~Uranga, \emph{{Discrete
  Symmetries, Weak Coupling Conjecture and Scale Separation in AdS Vacua}},
  \href{https://doi.org/10.1007/JHEP06(2020)083}{\emph{JHEP} {\bfseries 06}
  (2020) 083} [\href{https://arxiv.org/abs/2003.09740}{{\ttfamily
  2003.09740}}].

\bibitem{Conlon:2021cjk}
J.P.~Conlon, S.~Ning and F.~Revello, \emph{{Exploring the holographic
  Swampland}}, \href{https://doi.org/10.1007/JHEP04(2022)117}{\emph{JHEP}
  {\bfseries 04} (2022) 117}
  [\href{https://arxiv.org/abs/2110.06245}{{\ttfamily 2110.06245}}].

\bibitem{Apers:2022tfm}
F.~Apers, J.P.~Conlon, S.~Ning and F.~Revello, \emph{{Integer conformal
  dimensions for type IIa flux vacua}},
  \href{https://doi.org/10.1103/PhysRevD.105.106029}{\emph{Phys. Rev. D}
  {\bfseries 105} (2022) 106029}
  [\href{https://arxiv.org/abs/2202.09330}{{\ttfamily 2202.09330}}].

\bibitem{Apers:2022zjx}
F.~Apers, M.~Montero, T.~Van~Riet and T.~Wrase, \emph{{Comments on classical
  AdS flux vacua with scale separation}},
  \href{https://doi.org/10.1007/JHEP05(2022)167}{\emph{JHEP} {\bfseries 05}
  (2022) 167} [\href{https://arxiv.org/abs/2202.00682}{{\ttfamily
  2202.00682}}].

\bibitem{Danielsson:2018ztv}
U.H.~Danielsson and T.~Van~Riet, \emph{{What if string theory has no de Sitter
  vacua?}}, \href{https://doi.org/10.1142/S0218271818300070}{\emph{Int. J. Mod.
  Phys.} {\bfseries D27} (2018) 1830007}
  [\href{https://arxiv.org/abs/1804.01120}{{\ttfamily 1804.01120}}].

\bibitem{Andriot:2019wrs}
D.~Andriot, \emph{{Open problems on classical de Sitter solutions}},
  \href{https://doi.org/10.1002/prop.201900026}{\emph{Fortsch. Phys.}
  {\bfseries 67} (2019) 1900026}
  [\href{https://arxiv.org/abs/1902.10093}{{\ttfamily 1902.10093}}].

\bibitem{Obied:2018sgi}
G.~Obied, H.~Ooguri, L.~Spodyneiko and C.~Vafa, \emph{{De Sitter Space and the
  Swampland}},  \href{https://arxiv.org/abs/1806.08362}{{\ttfamily
  1806.08362}}.

\bibitem{Ooguri:2018wrx}
H.~Ooguri, E.~Palti, G.~Shiu and C.~Vafa, \emph{{Distance and de Sitter
  Conjectures on the Swampland}},
  \href{https://doi.org/10.1016/j.physletb.2018.11.018}{\emph{Phys. Lett. B}
  {\bfseries 788} (2019) 180}
  [\href{https://arxiv.org/abs/1810.05506}{{\ttfamily 1810.05506}}].

\bibitem{Garg:2018reu}
S.K.~Garg and C.~Krishnan, \emph{{Bounds on Slow Roll and the de Sitter
  Swampland}}, \href{https://doi.org/10.1007/JHEP11(2019)075}{\emph{JHEP}
  {\bfseries 11} (2019) 075}
  [\href{https://arxiv.org/abs/1807.05193}{{\ttfamily 1807.05193}}].

\bibitem{Junghans:2018gdb}
D.~Junghans, \emph{{Weakly Coupled de Sitter Vacua with Fluxes and the
  Swampland}}, \href{https://doi.org/10.1007/JHEP03(2019)150}{\emph{JHEP}
  {\bfseries 03} (2019) 150}
  [\href{https://arxiv.org/abs/1811.06990}{{\ttfamily 1811.06990}}].

\bibitem{Banlaki:2018ayh}
A.~Banlaki, A.~Chowdhury, C.~Roupec and T.~Wrase, \emph{{Scaling limits of dS
  vacua and the swampland}},
  \href{https://doi.org/10.1007/JHEP03(2019)065}{\emph{JHEP} {\bfseries 03}
  (2019) 065} [\href{https://arxiv.org/abs/1811.07880}{{\ttfamily
  1811.07880}}].

\bibitem{Grimm:2019ixq}
T.W.~Grimm, C.~Li and I.~Valenzuela, \emph{{Asymptotic Flux Compactifications
  and the Swampland}},
  \href{https://doi.org/10.1007/JHEP06(2020)009}{\emph{JHEP} {\bfseries 06}
  (2020) 009} [\href{https://arxiv.org/abs/1910.09549}{{\ttfamily
  1910.09549}}].

\bibitem{Andriot:2022way}
D.~Andriot, L.~Horer and P.~Marconnet, \emph{{Charting the landscape of (anti-)
  de Sitter and Minkowski solutions of 10d supergravities}},
  \href{https://doi.org/10.1007/JHEP06(2022)131}{\emph{JHEP} {\bfseries 06}
  (2022) 131} [\href{https://arxiv.org/abs/2201.04152}{{\ttfamily
  2201.04152}}].

\bibitem{Kachru:2003aw}
S.~Kachru, R.~Kallosh, A.D.~Linde and S.P.~Trivedi, \emph{{De Sitter vacua in
  string theory}},
  \href{https://doi.org/10.1103/PhysRevD.68.046005}{\emph{Phys. Rev.}
  {\bfseries D68} (2003) 046005}
  [\href{https://arxiv.org/abs/hep-th/0301240}{{\ttfamily hep-th/0301240}}].

\bibitem{Balasubramanian:2005zx}
V.~Balasubramanian, P.~Berglund, J.P.~Conlon and F.~Quevedo, \emph{{Systematics
  of moduli stabilisation in Calabi-Yau flux compactifications}},
  \href{https://doi.org/10.1088/1126-6708/2005/03/007}{\emph{JHEP} {\bfseries
  03} (2005) 007} [\href{https://arxiv.org/abs/hep-th/0502058}{{\ttfamily
  hep-th/0502058}}].

\bibitem{Conlon:2005ki}
J.P.~Conlon, F.~Quevedo and K.~Suruliz, \emph{{Large-volume flux
  compactifications: Moduli spectrum and D3/D7 soft supersymmetry breaking}},
  \href{https://doi.org/10.1088/1126-6708/2005/08/007}{\emph{JHEP} {\bfseries
  08} (2005) 007} [\href{https://arxiv.org/abs/hep-th/0505076}{{\ttfamily
  hep-th/0505076}}].

\bibitem{Bedroya:2019tba}
A.~Bedroya, R.~Brandenberger, M.~Loverde and C.~Vafa, \emph{{Trans-Planckian
  Censorship and Inflationary Cosmology}},
  \href{https://doi.org/10.1103/PhysRevD.101.103502}{\emph{Phys. Rev. D}
  {\bfseries 101} (2020) 103502}
  [\href{https://arxiv.org/abs/1909.11106}{{\ttfamily 1909.11106}}].

\bibitem{Calderon-Infante:2022nxb}
J.~Calder\'on-Infante, I.~Ruiz and I.~Valenzuela, \emph{{Asymptotic accelerated
  expansion in string theory and the Swampland}},
  \href{https://doi.org/10.1007/JHEP06(2023)129}{\emph{JHEP} {\bfseries 06}
  (2023) 129} [\href{https://arxiv.org/abs/2209.11821}{{\ttfamily
  2209.11821}}].

\bibitem{Grimm:2004ua}
T.W.~Grimm and J.~Louis, \emph{{The Effective action of type IIA Calabi-Yau
  orientifolds}},
  \href{https://doi.org/10.1016/j.nuclphysb.2005.04.007}{\emph{Nucl. Phys. B}
  {\bfseries 718} (2005) 153}
  [\href{https://arxiv.org/abs/hep-th/0412277}{{\ttfamily hep-th/0412277}}].

\bibitem{Grimm:2004uq}
T.W.~Grimm and J.~Louis, \emph{{The Effective action of N = 1 Calabi-Yau
  orientifolds}},
  \href{https://doi.org/10.1016/j.nuclphysb.2004.08.005}{\emph{Nucl.Phys.}
  {\bfseries B699} (2004) 387}
  [\href{https://arxiv.org/abs/hep-th/0403067}{{\ttfamily hep-th/0403067}}].

\bibitem{Grimm:2005fa}
T.W.~Grimm, \emph{{The Effective action of type II Calabi-Yau orientifolds}},
  \href{https://doi.org/10.1002/prop.200510253}{\emph{Fortsch. Phys.}
  {\bfseries 53} (2005) 1179}
  [\href{https://arxiv.org/abs/hep-th/0507153}{{\ttfamily hep-th/0507153}}].

\bibitem{Grana:2005jc}
M.~Grana, \emph{{Flux compactifications in string theory: A Comprehensive
  review}}, \href{https://doi.org/10.1016/j.physrep.2005.10.008}{\emph{Phys.
  Rept.} {\bfseries 423} (2006) 91}
  [\href{https://arxiv.org/abs/hep-th/0509003}{{\ttfamily hep-th/0509003}}].

\bibitem{Blumenhagen:2006ci}
R.~Blumenhagen, B.~K{\"o}rs, D.~L{\"u}st and S.~Stieberger,
  \emph{{Four-dimensional String Compactifications with D-Branes, Orientifolds
  and Fluxes}},
  \href{https://doi.org/10.1016/j.physrep.2007.04.003}{\emph{Phys.Rept.}
  {\bfseries 445} (2007) 1}
  [\href{https://arxiv.org/abs/hep-th/0610327}{{\ttfamily hep-th/0610327}}].

\bibitem{Denef:2007pq}
F.~Denef, M.R.~Douglas and S.~Kachru, \emph{{Physics of String Flux
  Compactifications}},
  \href{https://doi.org/10.1146/annurev.nucl.57.090506.123042}{\emph{Ann. Rev.
  Nucl. Part. Sci.} {\bfseries 57} (2007) 119}
  [\href{https://arxiv.org/abs/hep-th/0701050}{{\ttfamily hep-th/0701050}}].

\bibitem{Koerber:2010bx}
P.~Koerber, \emph{{Lectures on Generalized Complex Geometry for Physicists}},
  \href{https://doi.org/10.1002/prop.201000083}{\emph{Fortsch. Phys.}
  {\bfseries 59} (2011) 169} [\href{https://arxiv.org/abs/1006.1536}{{\ttfamily
  1006.1536}}].

\bibitem{Tomasiello:2022dwe}
A.~Tomasiello, \emph{{Geometry of String Theory Compactifications}}, Cambridge
  University Press (1, 2022),
  \href{https://doi.org/10.1017/9781108635745}{10.1017/9781108635745}.

\bibitem{Lust:2004ig}
D.~Lust and D.~Tsimpis, \emph{{Supersymmetric AdS(4) compactifications of IIA
  supergravity}},
  \href{https://doi.org/10.1088/1126-6708/2005/02/027}{\emph{JHEP} {\bfseries
  02} (2005) 027} [\href{https://arxiv.org/abs/hep-th/0412250}{{\ttfamily
  hep-th/0412250}}].

\bibitem{Acharya:2006ne}
B.S.~Acharya, F.~Benini and R.~Valandro, \emph{{Fixing moduli in exact type IIA
  flux vacua}},
  \href{https://doi.org/10.1088/1126-6708/2007/02/018}{\emph{JHEP} {\bfseries
  02} (2007) 018} [\href{https://arxiv.org/abs/hep-th/0607223}{{\ttfamily
  hep-th/0607223}}].

\bibitem{Marchesano:2020qvg}
F.~Marchesano, E.~Palti, J.~Quirant and A.~Tomasiello, \emph{{On supersymmetric
  AdS$_{4}$ orientifold vacua}},
  \href{https://doi.org/10.1007/JHEP08(2020)087}{\emph{JHEP} {\bfseries 08}
  (2020) 087} [\href{https://arxiv.org/abs/2003.13578}{{\ttfamily
  2003.13578}}].

\bibitem{Junghans:2020acz}
D.~Junghans, \emph{{O-Plane Backreaction and Scale Separation in Type IIA Flux
  Vacua}}, \href{https://doi.org/10.1002/prop.202000040}{\emph{Fortsch. Phys.}
  {\bfseries 68} (2020) 2000040}
  [\href{https://arxiv.org/abs/2003.06274}{{\ttfamily 2003.06274}}].

\bibitem{Bielleman:2015ina}
S.~Bielleman, L.E.~Ibanez and I.~Valenzuela, \emph{{Minkowski 3-forms, Flux
  String Vacua, Axion Stability and Naturalness}},
  \href{https://doi.org/10.1007/JHEP12(2015)119}{\emph{JHEP} {\bfseries 12}
  (2015) 119} [\href{https://arxiv.org/abs/1507.06793}{{\ttfamily
  1507.06793}}].

\bibitem{Herraez:2018vae}
A.~Herraez, L.E.~Ibanez, F.~Marchesano and G.~Zoccarato, \emph{{The Type IIA
  Flux Potential, 4-forms and Freed-Witten anomalies}},
  \href{https://doi.org/10.1007/JHEP09(2018)018}{\emph{JHEP} {\bfseries 09}
  (2018) 018} [\href{https://arxiv.org/abs/1802.05771}{{\ttfamily
  1802.05771}}].

\bibitem{yau1978ricci}
S.-T.~Yau, \emph{On the ricci curvature of a compact k{\"a}hler manifold and
  the complex monge-amp{\'e}re equation, i}, {\emph{Communications on pure and
  applied mathematics} {\bfseries 31} (1978) 339}.

\bibitem{Hubsch:1992nu}
T.~Hubsch, \emph{{Calabi-Yau manifolds: A Bestiary for physicists}}, World
  Scientific, Singapore (1994).

\bibitem{tian1987smoothness}
G.~Tian, \emph{Smoothness of the universal deformation space of compact
  calabi-yau manifolds and its peterson-weil metric},  in \emph{Mathematical
  aspects of string theory}, pp.~629--646, World Scientific (1987).

\bibitem{Candelas:1990pi}
P.~Candelas and X.~de~la Ossa, \emph{{Moduli Space of {Calabi-Yau} Manifolds}},
  \href{https://doi.org/10.1016/0550-3213(91)90122-E}{\emph{Nucl.Phys.}
  {\bfseries B355} (1991) 455}.

\bibitem{Candelas:1985en}
P.~Candelas, G.T.~Horowitz, A.~Strominger and E.~Witten, \emph{{Vacuum
  configurations for superstrings}},
  \href{https://doi.org/10.1016/0550-3213(85)90602-9}{\emph{Nucl. Phys. B}
  {\bfseries 258} (1985) 46}.

\bibitem{Dixon:1985jw}
L.J.~Dixon, J.A.~Harvey, C.~Vafa and E.~Witten, \emph{{Strings on Orbifolds}},
  \href{https://doi.org/10.1016/0550-3213(85)90593-0}{\emph{Nucl. Phys.}
  {\bfseries B261} (1985) 678}.

\bibitem{Vafa:1994rv}
C.~Vafa and E.~Witten, \emph{{On orbifolds with discrete torsion}},
  \href{https://doi.org/10.1016/0393-0440(94)00048-9}{\emph{J. Geom. Phys.}
  {\bfseries 15} (1995) 189}
  [\href{https://arxiv.org/abs/hep-th/9409188}{{\ttfamily hep-th/9409188}}].

\bibitem{Aspinwall:1994ev}
P.S.~Aspinwall, \emph{{Resolution of orbifold singularities in string theory}},
  {\emph{AMS/IP Stud. Adv. Math.} {\bfseries 1} (1996) 355}
  [\href{https://arxiv.org/abs/hep-th/9403123}{{\ttfamily hep-th/9403123}}].

\bibitem{Lust:2006zh}
D.~Lust, S.~Reffert, E.~Scheidegger and S.~Stieberger, \emph{{Resolved Toroidal
  Orbifolds and their Orientifolds}},
  \href{https://doi.org/10.4310/ATMP.2008.v12.n1.a2}{\emph{Adv. Theor. Math.
  Phys.} {\bfseries 12} (2008) 67}
  [\href{https://arxiv.org/abs/hep-th/0609014}{{\ttfamily hep-th/0609014}}].

\bibitem{Reffert:2007im}
S.~Reffert, \emph{{The Geometer's Toolkit to String Compactifications}},  in
  \emph{{Conference on String and M Theory Approaches to Particle Physics and
  Cosmology}}, 6, 2007 [\href{https://arxiv.org/abs/0706.1310}{{\ttfamily
  0706.1310}}].

\bibitem{Angelantonj:1996uy}
C.~Angelantonj, M.~Bianchi, G.~Pradisi, A.~Sagnotti and Y.S.~Stanev,
  \emph{{Chiral asymmetry in four-dimensional open string vacua}},
  \href{https://doi.org/10.1016/0370-2693(96)00869-6}{\emph{Phys. Lett. B}
  {\bfseries 385} (1996) 96}
  [\href{https://arxiv.org/abs/hep-th/9606169}{{\ttfamily hep-th/9606169}}].

\bibitem{Berkooz:1996dw}
M.~Berkooz and R.G.~Leigh, \emph{{A D = 4 N=1 orbifold of type I strings}},
  \href{https://doi.org/10.1016/S0550-3213(96)00543-3}{\emph{Nucl.Phys.}
  {\bfseries B483} (1997) 187}
  [\href{https://arxiv.org/abs/hep-th/9605049}{{\ttfamily hep-th/9605049}}].

\bibitem{Aldazabal:1998mr}
G.~Aldazabal, A.~Font, L.E.~Ibanez and G.~Violero, \emph{{D = 4, N=1, type IIB
  orientifolds}},
  \href{https://doi.org/10.1016/S0550-3213(98)00666-X}{\emph{Nucl. Phys. B}
  {\bfseries 536} (1998) 29}
  [\href{https://arxiv.org/abs/hep-th/9804026}{{\ttfamily hep-th/9804026}}].

\bibitem{Cvetic:2001nr}
M.~Cvetic, G.~Shiu and A.M.~Uranga, \emph{{Chiral four-dimensional N=1
  supersymmetric type 2A orientifolds from intersecting D6 branes}},
  \href{https://doi.org/10.1016/S0550-3213(01)00427-8}{\emph{Nucl. Phys. B}
  {\bfseries 615} (2001) 3}
  [\href{https://arxiv.org/abs/hep-th/0107166}{{\ttfamily hep-th/0107166}}].

\bibitem{Acharya:2002ag}
B.S.~Acharya, M.~Aganagic, K.~Hori and C.~Vafa, \emph{{Orientifolds, mirror
  symmetry and superpotentials}},
  \href{https://arxiv.org/abs/hep-th/0202208}{{\ttfamily hep-th/0202208}}.

\bibitem{Hitchin:1999fh}
N.J.~Hitchin, \emph{{Lectures on special Lagrangian submanifolds}},
  {\emph{AMS/IP Stud. Adv. Math.} {\bfseries 23} (2001) 151}
  [\href{https://arxiv.org/abs/math/9907034}{{\ttfamily math/9907034}}].

\bibitem{Brunner:2003zm}
I.~Brunner and K.~Hori, \emph{{Orientifolds and mirror symmetry}},
  \href{https://doi.org/10.1088/1126-6708/2004/11/005}{\emph{JHEP} {\bfseries
  11} (2004) 005} [\href{https://arxiv.org/abs/hep-th/0303135}{{\ttfamily
  hep-th/0303135}}].

\bibitem{Lust:2008zd}
D.~Lust, F.~Marchesano, L.~Martucci and D.~Tsimpis, \emph{{Generalized
  non-supersymmetric flux vacua}},
  \href{https://doi.org/10.1088/1126-6708/2008/11/021}{\emph{JHEP} {\bfseries
  11} (2008) 021} [\href{https://arxiv.org/abs/0807.4540}{{\ttfamily
  0807.4540}}].

\bibitem{Maldacena:2000mw}
J.M.~Maldacena and C.~Nunez, \emph{{Supergravity description of field theories
  on curved manifolds and a no go theorem}},
  \href{https://doi.org/10.1142/S0217751X01003935,
  10.1142/S0217751X01003937}{\emph{Int. J. Mod. Phys.} {\bfseries A16} (2001)
  822} [\href{https://arxiv.org/abs/hep-th/0007018}{{\ttfamily
  hep-th/0007018}}].

\bibitem{koerber2007ten}
P.~Koerber and L.~Martucci, \emph{{From ten to four and back again: how to
  generalize the geometry}},
  \href{https://doi.org/10.1088/1126-6708/2007/08/059}{\emph{JHEP} {\bfseries
  08} (2007) 059} [\href{https://arxiv.org/abs/0707.1038}{{\ttfamily
  0707.1038}}].

\bibitem{Koerber:2007jb}
P.~Koerber and L.~Martucci, \emph{{D-branes on AdS flux compactifications}},
  \href{https://doi.org/10.1088/1126-6708/2008/01/047}{\emph{JHEP} {\bfseries
  01} (2008) 047} [\href{https://arxiv.org/abs/0710.5530}{{\ttfamily
  0710.5530}}].

\bibitem{Koerber:2007hd}
P.~Koerber and D.~Tsimpis, \emph{{Supersymmetric sources, integrability and
  generalized-structure compactifications}},
  \href{https://doi.org/10.1088/1126-6708/2007/08/082}{\emph{JHEP} {\bfseries
  08} (2007) 082} [\href{https://arxiv.org/abs/0706.1244}{{\ttfamily
  0706.1244}}].

\bibitem{Marolf:2000cb}
D.~Marolf, \emph{{Chern-Simons terms and the three notions of charge}},  in
  \emph{{Quantization, gauge theory, and strings. Proceedings, International
  Conference dedicated to the memory of Professor Efim Fradkin, Moscow, Russia,
  June 5-10, 2000. Vol. 1+2}}, pp.~312--320, 2000
  [\href{https://arxiv.org/abs/hep-th/0006117}{{\ttfamily hep-th/0006117}}].

\bibitem{DeWolfe:2005uu}
O.~DeWolfe, A.~Giryavets, S.~Kachru and W.~Taylor, \emph{{Type IIA moduli
  stabilization}},
  \href{https://doi.org/10.1088/1126-6708/2005/07/066}{\emph{JHEP} {\bfseries
  07} (2005) 066} [\href{https://arxiv.org/abs/hep-th/0505160}{{\ttfamily
  hep-th/0505160}}].

\bibitem{Saracco:2012wc}
F.~Saracco and A.~Tomasiello, \emph{{Localized O6-plane solutions with Romans
  mass}}, \href{https://doi.org/10.1007/JHEP07(2012)077}{\emph{JHEP} {\bfseries
  07} (2012) 077} [\href{https://arxiv.org/abs/1201.5378}{{\ttfamily
  1201.5378}}].

\bibitem{DeLuca:2021mcj}
G.B.~De~Luca and A.~Tomasiello, \emph{{Leaps and bounds towards scale
  separation}}, \href{https://doi.org/10.1007/JHEP12(2021)086}{\emph{JHEP}
  {\bfseries 12} (2021) 086}
  [\href{https://arxiv.org/abs/2104.12773}{{\ttfamily 2104.12773}}].

\bibitem{Villadoro:2005cu}
G.~Villadoro and F.~Zwirner, \emph{{N=1 effective potential from dual type-IIA
  D6/O6 orientifolds with general fluxes}},
  \href{https://doi.org/10.1088/1126-6708/2005/06/047}{\emph{JHEP} {\bfseries
  06} (2005) 047} [\href{https://arxiv.org/abs/hep-th/0503169}{{\ttfamily
  hep-th/0503169}}].

\bibitem{Camara:2005dc}
P.G.~Camara, A.~Font and L.E.~Ibanez, \emph{{Fluxes, moduli fixing and
  MSSM-like vacua in a simple IIA orientifold}},
  \href{https://doi.org/10.1088/1126-6708/2005/09/013}{\emph{JHEP} {\bfseries
  09} (2005) 013} [\href{https://arxiv.org/abs/hep-th/0506066}{{\ttfamily
  hep-th/0506066}}].

\bibitem{Shelton:2005cf}
J.~Shelton, W.~Taylor and B.~Wecht, \emph{{Nongeometric flux
  compactifications}},
  \href{https://doi.org/10.1088/1126-6708/2005/10/085}{\emph{JHEP} {\bfseries
  0510} (2005) 085} [\href{https://arxiv.org/abs/hep-th/0508133}{{\ttfamily
  hep-th/0508133}}].

\bibitem{Aldazabal:2006up}
G.~Aldazabal, P.G.~Camara, A.~Font and L.~Ibanez, \emph{{More dual fluxes and
  moduli fixing}},
  \href{https://doi.org/10.1088/1126-6708/2006/05/070}{\emph{JHEP} {\bfseries
  0605} (2006) 070} [\href{https://arxiv.org/abs/hep-th/0602089}{{\ttfamily
  hep-th/0602089}}].

\bibitem{Shelton:2006fd}
J.~Shelton, W.~Taylor and B.~Wecht, \emph{{Generalized Flux Vacua}},
  \href{https://doi.org/10.1088/1126-6708/2007/02/095}{\emph{JHEP} {\bfseries
  02} (2007) 095} [\href{https://arxiv.org/abs/hep-th/0607015}{{\ttfamily
  hep-th/0607015}}].

\bibitem{Micu:2007rd}
A.~Micu, E.~Palti and G.~Tasinato, \emph{{Towards Minkowski Vacua in Type II
  String Compactifications}},
  \href{https://doi.org/10.1088/1126-6708/2007/03/104}{\emph{JHEP} {\bfseries
  03} (2007) 104} [\href{https://arxiv.org/abs/hep-th/0701173}{{\ttfamily
  hep-th/0701173}}].

\bibitem{Ihl:2007ah}
M.~Ihl, D.~Robbins and T.~Wrase, \emph{{Toroidal orientifolds in IIA with
  general NS-NS fluxes}},
  \href{https://doi.org/10.1088/1126-6708/2007/08/043}{\emph{JHEP} {\bfseries
  0708} (2007) 043} [\href{https://arxiv.org/abs/0705.3410}{{\ttfamily
  0705.3410}}].

\bibitem{Wecht:2007wu}
B.~Wecht, \emph{{Lectures on Nongeometric Flux Compactifications}},
  \href{https://doi.org/10.1088/0264-9381/24/21/S03}{\emph{Class. Quant. Grav.}
  {\bfseries 24} (2007) S773}
  [\href{https://arxiv.org/abs/0708.3984}{{\ttfamily 0708.3984}}].

\bibitem{Robbins:2007yv}
D.~Robbins and T.~Wrase, \emph{{D-terms from generalized NS-NS fluxes in type
  II}}, \href{https://doi.org/10.1088/1126-6708/2007/12/058}{\emph{JHEP}
  {\bfseries 0712} (2007) 058}
  [\href{https://arxiv.org/abs/0709.2186}{{\ttfamily 0709.2186}}].

\bibitem{Kachru:2002sk}
S.~Kachru, M.B.~Schulz, P.K.~Tripathy and S.P.~Trivedi, \emph{{New
  supersymmetric string compactifications}},
  \href{https://doi.org/10.1088/1126-6708/2003/03/061}{\emph{JHEP} {\bfseries
  03} (2003) 061} [\href{https://arxiv.org/abs/hep-th/0211182}{{\ttfamily
  hep-th/0211182}}].

\bibitem{Plauschinn:2018wbo}
E.~Plauschinn, \emph{{Non-geometric backgrounds in string theory}},
  \href{https://doi.org/10.1016/j.physrep.2018.12.002}{\emph{Phys. Rept.}
  {\bfseries 798} (2019) 1} [\href{https://arxiv.org/abs/1811.11203}{{\ttfamily
  1811.11203}}].

\bibitem{Buscher:1987sk}
T.H.~Buscher, \emph{{A Symmetry of the String Background Field Equations}},
  \href{https://doi.org/10.1016/0370-2693(87)90769-6}{\emph{Phys. Lett. B}
  {\bfseries 194} (1987) 59}.

\bibitem{Gukov:1999ya}
S.~Gukov, C.~Vafa and E.~Witten, \emph{{CFT's from Calabi-Yau four folds}},
  \href{https://doi.org/10.1016/S0550-3213(00)00373-4}{\emph{Nucl. Phys. B}
  {\bfseries 584} (2000) 69}
  [\href{https://arxiv.org/abs/hep-th/9906070}{{\ttfamily hep-th/9906070}}].

\bibitem{Taylor:1999ii}
T.R.~Taylor and C.~Vafa, \emph{{R R flux on Calabi-Yau and partial
  supersymmetry breaking}},
  \href{https://doi.org/10.1016/S0370-2693(00)00005-8}{\emph{Phys. Lett. B}
  {\bfseries 474} (2000) 130}
  [\href{https://arxiv.org/abs/hep-th/9912152}{{\ttfamily hep-th/9912152}}].

\bibitem{Louis:2004xi}
J.~Louis and W.~Schulgin, \emph{{Massive tensor multiplets in N=1
  supersymmetry}}, \href{https://doi.org/10.1002/prop.200410193}{\emph{Fortsch.
  Phys.} {\bfseries 53} (2005) 235}
  [\href{https://arxiv.org/abs/hep-th/0410149}{{\ttfamily hep-th/0410149}}].

\bibitem{Blumenhagen:2015lta}
R.~Blumenhagen, A.~Font and E.~Plauschinn, \emph{{Relating double field theory
  to the scalar potential of N = 2 gauged supergravity}},
  \href{https://doi.org/10.1007/JHEP12(2015)122}{\emph{JHEP} {\bfseries 12}
  (2015) 122} [\href{https://arxiv.org/abs/1507.08059}{{\ttfamily
  1507.08059}}].

\bibitem{Carta:2016ynn}
F.~Carta, F.~Marchesano, W.~Staessens and G.~Zoccarato, \emph{{Open string
  multi-branched and K\"ahler potentials}},
  \href{https://doi.org/10.1007/JHEP09(2016)062}{\emph{JHEP} {\bfseries 09}
  (2016) 062} [\href{https://arxiv.org/abs/1606.00508}{{\ttfamily
  1606.00508}}].

\bibitem{Escobar:2018rna}
D.~Escobar, F.~Marchesano and W.~Staessens, \emph{{Type IIA flux vacua and
  $\alpha'$-corrections}},
  \href{https://doi.org/10.1007/JHEP06(2019)129}{\emph{JHEP} {\bfseries 06}
  (2019) 129} [\href{https://arxiv.org/abs/1812.08735}{{\ttfamily
  1812.08735}}].

\bibitem{Escobar:2018tiu}
D.~Escobar, F.~Marchesano and W.~Staessens, \emph{{Type IIA Flux Vacua with
  Mobile D6-branes}},
  \href{https://doi.org/10.1007/JHEP01(2019)096}{\emph{JHEP} {\bfseries 01}
  (2019) 096} [\href{https://arxiv.org/abs/1811.09282}{{\ttfamily
  1811.09282}}].

\bibitem{Bousso:2000xa}
R.~Bousso and J.~Polchinski, \emph{{Quantization of four form fluxes and
  dynamical neutralization of the cosmological constant}},
  \href{https://doi.org/10.1088/1126-6708/2000/06/006}{\emph{JHEP} {\bfseries
  06} (2000) 006} [\href{https://arxiv.org/abs/hep-th/0004134}{{\ttfamily
  hep-th/0004134}}].

\bibitem{Brown:1987dd}
J.D.~Brown and C.~Teitelboim, \emph{{Dynamical Neutralization of the
  Cosmological Constant}},
  \href{https://doi.org/10.1016/0370-2693(87)91190-7}{\emph{Phys. Lett. B}
  {\bfseries 195} (1987) 177}.

\bibitem{Brown:1988kg}
J.D.~Brown and C.~Teitelboim, \emph{{Neutralization of the Cosmological
  Constant by Membrane Creation}},
  \href{https://doi.org/10.1016/0550-3213(88)90559-7}{\emph{Nucl. Phys. B}
  {\bfseries 297} (1988) 787}.

\bibitem{Derendinger:2004jn}
J.-P.~Derendinger, C.~Kounnas, P.M.~Petropoulos and F.~Zwirner,
  \emph{{Superpotentials in IIA compactifications with general fluxes}},
  \href{https://doi.org/10.1016/j.nuclphysb.2005.02.038}{\emph{Nucl. Phys.}
  {\bfseries B715} (2005) 211}
  [\href{https://arxiv.org/abs/hep-th/0411276}{{\ttfamily hep-th/0411276}}].

\bibitem{Becker:2007zj}
K.~Becker, M.~Becker and J.H.~Schwarz, \emph{{String theory and M-theory: A
  modern introduction}}, Cambridge University Press (12, 2006).

\bibitem{Gao:2017gxk}
X.~Gao, P.~Shukla and R.~Sun, \emph{{Symplectic formulation of the type IIA
  nongeometric scalar potential}},
  \href{https://doi.org/10.1103/PhysRevD.98.046009}{\emph{Phys. Rev.}
  {\bfseries D98} (2018) 046009}
  [\href{https://arxiv.org/abs/1712.07310}{{\ttfamily 1712.07310}}].

\bibitem{House:2005yc}
T.~House and E.~Palti, \emph{{Effective action of (massive) IIA on manifolds
  with SU(3) structure}},
  \href{https://doi.org/10.1103/PhysRevD.72.026004}{\emph{Phys. Rev.}
  {\bfseries D72} (2005) 026004}
  [\href{https://arxiv.org/abs/hep-th/0505177}{{\ttfamily hep-th/0505177}}].

\bibitem{Grana:2006kf}
M.~Grana, R.~Minasian, M.~Petrini and A.~Tomasiello, \emph{{A Scan for new N=1
  vacua on twisted tori}},
  \href{https://doi.org/10.1088/1126-6708/2007/05/031}{\emph{JHEP} {\bfseries
  05} (2007) 031} [\href{https://arxiv.org/abs/hep-th/0609124}{{\ttfamily
  hep-th/0609124}}].

\bibitem{Aldazabal:2007sn}
G.~Aldazabal and A.~Font, \emph{{A Second look at N=1 supersymmetric AdS(4)
  vacua of type IIA supergravity}},
  \href{https://doi.org/10.1088/1126-6708/2008/02/086}{\emph{JHEP} {\bfseries
  02} (2008) 086} [\href{https://arxiv.org/abs/0712.1021}{{\ttfamily
  0712.1021}}].

\bibitem{Koerber:2008rx}
P.~Koerber, D.~Lust and D.~Tsimpis, \emph{{Type IIA AdS(4) compactifications on
  cosets, interpolations and domain walls}},
  \href{https://doi.org/10.1088/1126-6708/2008/07/017}{\emph{JHEP} {\bfseries
  07} (2008) 017} [\href{https://arxiv.org/abs/0804.0614}{{\ttfamily
  0804.0614}}].

\bibitem{Caviezel:2008ik}
C.~Caviezel, P.~Koerber, S.~Kors, D.~Lust, D.~Tsimpis and M.~Zagermann,
  \emph{{The Effective theory of type IIA AdS(4) compactifications on
  nilmanifolds and cosets}},
  \href{https://doi.org/10.1088/0264-9381/26/2/025014}{\emph{Class. Quant.
  Grav.} {\bfseries 26} (2009) 025014}
  [\href{https://arxiv.org/abs/0806.3458}{{\ttfamily 0806.3458}}].

\bibitem{Koerber:2010rn}
P.~Koerber and S.~Kors, \emph{{A landscape of non-supersymmetric AdS vacua on
  coset manifolds}},
  \href{https://doi.org/10.1103/PhysRevD.81.105006}{\emph{Phys. Rev.}
  {\bfseries D81} (2010) 105006}
  [\href{https://arxiv.org/abs/1001.0003}{{\ttfamily 1001.0003}}].

\bibitem{Hertzberg:2007wc}
M.P.~Hertzberg, S.~Kachru, W.~Taylor and M.~Tegmark, \emph{{Inflationary
  Constraints on Type IIA String Theory}},
  \href{https://doi.org/10.1088/1126-6708/2007/12/095}{\emph{JHEP} {\bfseries
  12} (2007) 095} [\href{https://arxiv.org/abs/0711.2512}{{\ttfamily
  0711.2512}}].

\bibitem{Haque:2008jz}
S.S.~Haque, G.~Shiu, B.~Underwood and T.~Van~Riet, \emph{{Minimal simple de
  Sitter solutions}},
  \href{https://doi.org/10.1103/PhysRevD.79.086005}{\emph{Phys. Rev.}
  {\bfseries D79} (2009) 086005}
  [\href{https://arxiv.org/abs/0810.5328}{{\ttfamily 0810.5328}}].

\bibitem{Caviezel:2008tf}
C.~Caviezel, P.~Koerber, S.~Kors, D.~Lust, T.~Wrase and M.~Zagermann, \emph{{On
  the Cosmology of Type IIA Compactifications on SU(3)-structure Manifolds}},
  \href{https://doi.org/10.1088/1126-6708/2009/04/010}{\emph{JHEP} {\bfseries
  04} (2009) 010} [\href{https://arxiv.org/abs/0812.3551}{{\ttfamily
  0812.3551}}].

\bibitem{Flauger:2008ad}
R.~Flauger, S.~Paban, D.~Robbins and T.~Wrase, \emph{{Searching for slow-roll
  moduli inflation in massive type IIA supergravity with metric fluxes}},
  \href{https://doi.org/10.1103/PhysRevD.79.086011}{\emph{Phys. Rev.}
  {\bfseries D79} (2009) 086011}
  [\href{https://arxiv.org/abs/0812.3886}{{\ttfamily 0812.3886}}].

\bibitem{Danielsson:2009ff}
U.H.~Danielsson, S.S.~Haque, G.~Shiu and T.~Van~Riet, \emph{{Towards Classical
  de Sitter Solutions in String Theory}},
  \href{https://doi.org/10.1088/1126-6708/2009/09/114}{\emph{JHEP} {\bfseries
  09} (2009) 114} [\href{https://arxiv.org/abs/0907.2041}{{\ttfamily
  0907.2041}}].

\bibitem{Danielsson:2010bc}
U.H.~Danielsson, P.~Koerber and T.~Van~Riet, \emph{{Universal de Sitter
  solutions at tree-level}},
  \href{https://doi.org/10.1007/JHEP05(2010)090}{\emph{JHEP} {\bfseries 05}
  (2010) 090} [\href{https://arxiv.org/abs/1003.3590}{{\ttfamily 1003.3590}}].

\bibitem{Danielsson:2011au}
U.H.~Danielsson, S.S.~Haque, P.~Koerber, G.~Shiu, T.~Van~Riet and T.~Wrase,
  \emph{{De Sitter hunting in a classical landscape}},
  \href{https://doi.org/10.1002/prop.201100047}{\emph{Fortsch. Phys.}
  {\bfseries 59} (2011) 897} [\href{https://arxiv.org/abs/1103.4858}{{\ttfamily
  1103.4858}}].

\bibitem{Andriot:2018wzk}
D.~Andriot, \emph{{On the de Sitter swampland criterion}},
  \href{https://doi.org/10.1016/j.physletb.2018.09.022}{\emph{Phys. Lett.}
  {\bfseries B785} (2018) 570}
  [\href{https://arxiv.org/abs/1806.10999}{{\ttfamily 1806.10999}}].

\bibitem{GomezReino:2006dk}
M.~Gomez-Reino and C.A.~Scrucca, \emph{{Locally stable non-supersymmetric
  Minkowski vacua in supergravity}},
  \href{https://doi.org/10.1088/1126-6708/2006/05/015}{\emph{JHEP} {\bfseries
  05} (2006) 015} [\href{https://arxiv.org/abs/hep-th/0602246}{{\ttfamily
  hep-th/0602246}}].

\bibitem{GomezReino:2006wv}
M.~Gomez-Reino and C.A.~Scrucca, \emph{{Constraints for the existence of flat
  and stable non-supersymmetric vacua in supergravity}},
  \href{https://doi.org/10.1088/1126-6708/2006/09/008}{\emph{JHEP} {\bfseries
  09} (2006) 008} [\href{https://arxiv.org/abs/hep-th/0606273}{{\ttfamily
  hep-th/0606273}}].

\bibitem{GomezReino:2007qi}
M.~Gomez-Reino and C.A.~Scrucca, \emph{{Metastable supergravity vacua with F
  and D supersymmetry breaking}},
  \href{https://doi.org/10.1088/1126-6708/2007/08/091}{\emph{JHEP} {\bfseries
  08} (2007) 091} [\href{https://arxiv.org/abs/0706.2785}{{\ttfamily
  0706.2785}}].

\bibitem{Covi:2008ea}
L.~Covi, M.~Gomez-Reino, C.~Gross, J.~Louis, G.A.~Palma and C.A.~Scrucca,
  \emph{{de Sitter vacua in no-scale supergravities and Calabi-Yau string
  models}}, \href{https://doi.org/10.1088/1126-6708/2008/06/057}{\emph{JHEP}
  {\bfseries 06} (2008) 057} [\href{https://arxiv.org/abs/0804.1073}{{\ttfamily
  0804.1073}}].

\bibitem{Covi:2008zu}
L.~Covi, M.~Gomez-Reino, C.~Gross, G.A.~Palma and C.A.~Scrucca,
  \emph{{Constructing de Sitter vacua in no-scale string models without
  uplifting}}, \href{https://doi.org/10.1088/1126-6708/2009/03/146}{\emph{JHEP}
  {\bfseries 03} (2009) 146} [\href{https://arxiv.org/abs/0812.3864}{{\ttfamily
  0812.3864}}].

\bibitem{Blumenhagen:2013hva}
R.~Blumenhagen, X.~Gao, D.~Herschmann and P.~Shukla, \emph{{Dimensional
  Oxidation of Non-geometric Fluxes in Type II Orientifolds}},
  \href{https://doi.org/10.1007/JHEP10(2013)201}{\emph{JHEP} {\bfseries 1310}
  (2013) 201} [\href{https://arxiv.org/abs/1306.2761}{{\ttfamily 1306.2761}}].

\bibitem{Shukla:2019akv}
P.~Shukla, \emph{{Rigid nongeometric orientifolds and the swampland}},
  \href{https://doi.org/10.1103/PhysRevD.103.086010}{\emph{Phys. Rev. D}
  {\bfseries 103} (2021) 086010}
  [\href{https://arxiv.org/abs/1909.10993}{{\ttfamily 1909.10993}}].

\bibitem{Gao:2018ayp}
X.~Gao, P.~Shukla and R.~Sun, \emph{{On Missing Bianchi Identities in
  Cohomology Formulation}},
  \href{https://doi.org/10.1140/epjc/s10052-019-7291-5}{\emph{Eur. Phys. J.}
  {\bfseries C79} (2019) 781}
  [\href{https://arxiv.org/abs/1805.05748}{{\ttfamily 1805.05748}}].

\bibitem{deCarlos:2009fq}
B.~de~Carlos, A.~Guarino and J.M.~Moreno, \emph{{Flux moduli stabilisation,
  Supergravity algebras and no-go theorems}},
  \href{https://doi.org/10.1007/JHEP01(2010)012}{\emph{JHEP} {\bfseries 01}
  (2010) 012} [\href{https://arxiv.org/abs/0907.5580}{{\ttfamily 0907.5580}}].

\bibitem{Shukla:2019dqd}
P.~Shukla, \emph{{$T$ -dualizing de Sitter no-go scenarios}},
  \href{https://doi.org/10.1103/PhysRevD.102.026014}{\emph{Phys. Rev. D}
  {\bfseries 102} (2020) 026014}
  [\href{https://arxiv.org/abs/1909.08630}{{\ttfamily 1909.08630}}].

\bibitem{Cribiori:2021djm}
N.~Cribiori, D.~Junghans, V.~Van~Hemelryck, T.~Van~Riet and T.~Wrase,
  \emph{{Scale-separated AdS4 vacua of IIA orientifolds and M-theory}},
  \href{https://doi.org/10.1103/PhysRevD.104.126014}{\emph{Phys. Rev. D}
  {\bfseries 104} (2021) 126014}
  [\href{https://arxiv.org/abs/2107.00019}{{\ttfamily 2107.00019}}].

\bibitem{Dibitetto:2011gm}
G.~Dibitetto, A.~Guarino and D.~Roest, \emph{{Charting the landscape of N=4
  flux compactifications}},
  \href{https://doi.org/10.1007/JHEP03(2011)137}{\emph{JHEP} {\bfseries 03}
  (2011) 137} [\href{https://arxiv.org/abs/1102.0239}{{\ttfamily 1102.0239}}].

\bibitem{Breitenlohner:1982bm}
P.~Breitenlohner and D.Z.~Freedman, \emph{{Positive Energy in anti-De Sitter
  Backgrounds and Gauged Extended Supergravity}},
  \href{https://doi.org/10.1016/0370-2693(82)90643-8}{\emph{Phys. Lett.}
  {\bfseries 115B} (1982) 197}.

\bibitem{Conlon:2006tq}
J.P.~Conlon, \emph{{The QCD axion and moduli stabilisation}},
  \href{https://doi.org/10.1088/1126-6708/2006/05/078}{\emph{JHEP} {\bfseries
  0605} (2006) 078} [\href{https://arxiv.org/abs/hep-th/0602233}{{\ttfamily
  hep-th/0602233}}].

\bibitem{bhatia2013matrix}
R.~Bhatia, \emph{Matrix analysis}, vol.~169, Springer Science \& Business Media
  (2013).

\bibitem{Camara:2011jg}
P.G.~Camara, L.E.~Ibanez and F.~Marchesano, \emph{{RR photons}},
  \href{https://doi.org/10.1007/JHEP09(2011)110}{\emph{JHEP} {\bfseries 1109}
  (2011) 110} [\href{https://arxiv.org/abs/1106.0060}{{\ttfamily 1106.0060}}].

\bibitem{Behrndt:2004km}
K.~Behrndt and M.~Cvetic, \emph{{General N = 1 supersymmetric flux vacua of
  (massive) type IIA string theory}},
  \href{https://doi.org/10.1103/PhysRevLett.95.021601}{\emph{Phys. Rev. Lett.}
  {\bfseries 95} (2005) 021601}
  [\href{https://arxiv.org/abs/hep-th/0403049}{{\ttfamily hep-th/0403049}}].

\bibitem{BEDULLI20071125}
L.~Bedulli and L.~Vezzoni, \emph{The ricci tensor of su(3)-manifolds},
  \href{https://doi.org/https://doi.org/10.1016/j.geomphys.2006.09.007}{\emph{Journal
  of Geometry and Physics} {\bfseries 57} (2007) 1125 }.

\bibitem{Ali:2006gd}
T.~Ali and G.B.~Cleaver, \emph{{The Ricci curvature of half-flat manifolds}},
  \href{https://doi.org/10.1088/1126-6708/2007/05/009}{\emph{JHEP} {\bfseries
  05} (2007) 009} [\href{https://arxiv.org/abs/hep-th/0612171}{{\ttfamily
  hep-th/0612171}}].

\bibitem{Silverstein:2007ac}
E.~Silverstein, \emph{{Simple de Sitter Solutions}},
  \href{https://doi.org/10.1103/PhysRevD.77.106006}{\emph{Phys. Rev.}
  {\bfseries D77} (2008) 106006}
  [\href{https://arxiv.org/abs/0712.1196}{{\ttfamily 0712.1196}}].

\bibitem{Shukla:2019wfo}
P.~Shukla, \emph{{Dictionary for the type II nongeometric flux
  compactifications}},
  \href{https://doi.org/10.1103/PhysRevD.103.086009}{\emph{Phys. Rev. D}
  {\bfseries 103} (2021) 086009}
  [\href{https://arxiv.org/abs/1909.07391}{{\ttfamily 1909.07391}}].

\bibitem{Andriot:2018ept}
D.~Andriot, \emph{{New constraints on classical de Sitter: flirting with the
  swampland}}, \href{https://doi.org/10.1002/prop.201800103}{\emph{Fortsch.
  Phys.} {\bfseries 67} (2019) 1800103}
  [\href{https://arxiv.org/abs/1807.09698}{{\ttfamily 1807.09698}}].

\bibitem{Maldacena:1998uz}
J.M.~Maldacena, J.~Michelson and A.~Strominger, \emph{{Anti-de Sitter
  fragmentation}},
  \href{https://doi.org/10.1088/1126-6708/1999/02/011}{\emph{JHEP} {\bfseries
  02} (1999) 011} [\href{https://arxiv.org/abs/hep-th/9812073}{{\ttfamily
  hep-th/9812073}}].

\bibitem{Gaiotto:2009mv}
D.~Gaiotto and A.~Tomasiello, \emph{{The gauge dual of Romans mass}},
  \href{https://doi.org/10.1007/JHEP01(2010)015}{\emph{JHEP} {\bfseries 01}
  (2010) 015} [\href{https://arxiv.org/abs/0901.0969}{{\ttfamily 0901.0969}}].

\bibitem{Antonelli:2019nar}
R.~Antonelli and I.~Basile, \emph{{Brane annihilation in non-supersymmetric
  strings}}, \href{https://doi.org/10.1007/JHEP11(2019)021}{\emph{JHEP}
  {\bfseries 11} (2019) 021}
  [\href{https://arxiv.org/abs/1908.04352}{{\ttfamily 1908.04352}}].

\bibitem{Apruzzi:2019ecr}
F.~Apruzzi, G.~Bruno De~Luca, A.~Gnecchi, G.~Lo~Monaco and A.~Tomasiello,
  \emph{{On AdS$_{7}$ stability}},
  \href{https://doi.org/10.1007/JHEP07(2020)033}{\emph{JHEP} {\bfseries 07}
  (2020) 033} [\href{https://arxiv.org/abs/1912.13491}{{\ttfamily
  1912.13491}}].

\bibitem{Bena:2020xxb}
I.~Bena, K.~Pilch and N.P.~Warner, \emph{{Brane-Jet Instabilities}},
  \href{https://doi.org/10.1007/JHEP10(2020)091}{\emph{JHEP} {\bfseries 10}
  (2020) 091} [\href{https://arxiv.org/abs/2003.02851}{{\ttfamily
  2003.02851}}].

\bibitem{Suh:2020rma}
M.~Suh, \emph{{The non-SUSY AdS$_{6}$ and AdS$_{7}$ fixed points are brane-jet
  unstable}}, \href{https://doi.org/10.1007/JHEP10(2020)010}{\emph{JHEP}
  {\bfseries 10} (2020) 010}
  [\href{https://arxiv.org/abs/2004.06823}{{\ttfamily 2004.06823}}].

\bibitem{Guarino:2020jwv}
A.~Guarino, J.~Tarrio and O.~Varela, \emph{{Brane-jet stability of
  non-supersymmetric AdS vacua}},
  \href{https://doi.org/10.1007/JHEP09(2020)110}{\emph{JHEP} {\bfseries 09}
  (2020) 110} [\href{https://arxiv.org/abs/2005.07072}{{\ttfamily
  2005.07072}}].

\bibitem{Guarino:2020flh}
A.~Guarino, E.~Malek and H.~Samtleben, \emph{{Stable Nonsupersymmetric
  Anti\textendash{}de Sitter Vacua of Massive IIA Supergravity}},
  \href{https://doi.org/10.1103/PhysRevLett.126.061601}{\emph{Phys. Rev. Lett.}
  {\bfseries 126} (2021) 061601}
  [\href{https://arxiv.org/abs/2011.06600}{{\ttfamily 2011.06600}}].

\bibitem{Basile:2021vxh}
I.~Basile, \emph{{Supersymmetry Breaking and Stability in String Vacua: brane
  dynamics, bubbles and the swampland}},
  \href{https://doi.org/10.1007/s40766-021-00024-9}{\emph{Riv. Nuovo Cim.}
  {\bfseries 1} (2021) 98} [\href{https://arxiv.org/abs/2107.02814}{{\ttfamily
  2107.02814}}].

\bibitem{Apruzzi:2021nle}
F.~Apruzzi, G.~Bruno De~Luca, G.~Lo~Monaco and C.F.~Uhlemann,
  \emph{{Non-supersymmetric AdS$_6$ and the swampland}},
  \href{https://arxiv.org/abs/2110.03003}{{\ttfamily 2110.03003}}.

\bibitem{Bomans:2021ara}
P.~Bomans, D.~Cassani, G.~Dibitetto and N.~Petri, \emph{{Bubble instability of
  mIIA on $\mathrm{AdS}_4\times S^6$}},
  \href{https://doi.org/10.21468/SciPostPhys.12.3.099}{\emph{SciPost Phys.}
  {\bfseries 12} (2022) 099}
  [\href{https://arxiv.org/abs/2110.08276}{{\ttfamily 2110.08276}}].

\bibitem{Narayan:2010em}
P.~Narayan and S.P.~Trivedi, \emph{{On The Stability Of Non-Supersymmetric AdS
  Vacua}}, \href{https://doi.org/10.1007/JHEP07(2010)089}{\emph{JHEP}
  {\bfseries 07} (2010) 089} [\href{https://arxiv.org/abs/1002.4498}{{\ttfamily
  1002.4498}}].

\bibitem{Lanza:2019xxg}
S.~Lanza, F.~Marchesano, L.~Martucci and D.~Sorokin, \emph{{How many fluxes fit
  in an EFT?}}, \href{https://doi.org/10.1007/JHEP10(2019)110}{\emph{JHEP}
  {\bfseries 10} (2019) 110}
  [\href{https://arxiv.org/abs/1907.11256}{{\ttfamily 1907.11256}}].

\bibitem{Lanza:2020qmt}
S.~Lanza, F.~Marchesano, L.~Martucci and I.~Valenzuela, \emph{{Swampland
  Conjectures for Strings and Membranes}},
  \href{https://doi.org/10.1007/JHEP02(2021)006}{\emph{JHEP} {\bfseries 02}
  (2021) 006} [\href{https://arxiv.org/abs/2006.15154}{{\ttfamily
  2006.15154}}].

\bibitem{Aharony:2008wz}
O.~Aharony, Y.E.~Antebi and M.~Berkooz, \emph{{On the Conformal Field Theory
  Duals of type IIA AdS(4) Flux Compactifications}},
  \href{https://doi.org/10.1088/1126-6708/2008/02/093}{\emph{JHEP} {\bfseries
  02} (2008) 093} [\href{https://arxiv.org/abs/0801.3326}{{\ttfamily
  0801.3326}}].

\bibitem{Font:2006na}
A.~Font, L.E.~Ibanez and F.~Marchesano, \emph{{Coisotropic D8-branes and
  model-building}},
  \href{https://doi.org/10.1088/1126-6708/2006/09/080}{\emph{JHEP} {\bfseries
  09} (2006) 080} [\href{https://arxiv.org/abs/hep-th/0607219}{{\ttfamily
  hep-th/0607219}}].

\bibitem{Marchesano:2014iea}
F.~Marchesano, D.~Regalado and G.~Zoccarato, \emph{{On D-brane moduli
  stabilisation}}, \href{https://doi.org/10.1007/JHEP11(2014)097}{\emph{JHEP}
  {\bfseries 11} (2014) 097} [\href{https://arxiv.org/abs/1410.0209}{{\ttfamily
  1410.0209}}].

\bibitem{Palti:2008mg}
E.~Palti, G.~Tasinato and J.~Ward, \emph{{WEAKLY-coupled IIA Flux
  Compactifications}},
  \href{https://doi.org/10.1088/1126-6708/2008/06/084}{\emph{JHEP} {\bfseries
  06} (2008) 084} [\href{https://arxiv.org/abs/0804.1248}{{\ttfamily
  0804.1248}}].

\bibitem{Miyaoka1987}
Y.~Miyaoka, \emph{The chern class and kodaira dimension of a minimal variety.},
  {\emph{Adv. Stud. Pure Math.} {\bfseries 10} (1987) 449}.

\bibitem{Gibbons:1997xz}
G.W.~Gibbons, \emph{{Born-Infeld particles and Dirichlet p-branes}},
  \href{https://doi.org/10.1016/S0550-3213(97)00795-5}{\emph{Nucl. Phys. B}
  {\bfseries 514} (1998) 603}
  [\href{https://arxiv.org/abs/hep-th/9709027}{{\ttfamily hep-th/9709027}}].

\bibitem{Evslin:2007ti}
J.~Evslin and L.~Martucci, \emph{{D-brane networks in flux vacua, generalized
  cycles and calibrations}},
  \href{https://doi.org/10.1088/1126-6708/2007/07/040}{\emph{JHEP} {\bfseries
  07} (2007) 040} [\href{https://arxiv.org/abs/hep-th/0703129}{{\ttfamily
  hep-th/0703129}}].

\bibitem{Donaldson:1996kp}
S.K.~Donaldson and R.P.~Thomas, \emph{{Gauge theory in higher dimensions}},  in
  \emph{{Conference on Geometric Issues in Foundations of Science in honor of
  Sir Roger Penrose's 65th Birthday}}, pp.~31--47, 6, 1996.

\bibitem{Held:2010az}
J.~Held, D.~Lust, F.~Marchesano and L.~Martucci, \emph{{DWSB in heterotic flux
  compactifications}},
  \href{https://doi.org/10.1007/JHEP06(2010)090}{\emph{JHEP} {\bfseries 06}
  (2010) 090} [\href{https://arxiv.org/abs/1004.0867}{{\ttfamily 1004.0867}}].

\bibitem{Banks:2006hg}
T.~Banks and K.~van~den Broek, \emph{{Massive IIA flux compactifications and
  U-dualities}},
  \href{https://doi.org/10.1088/1126-6708/2007/03/068}{\emph{JHEP} {\bfseries
  03} (2007) 068} [\href{https://arxiv.org/abs/hep-th/0611185}{{\ttfamily
  hep-th/0611185}}].

\bibitem{Blumenhagen:2005mu}
R.~Blumenhagen, M.~Cvetic, P.~Langacker and G.~Shiu, \emph{{Toward realistic
  intersecting D-brane models}},
  \href{https://doi.org/10.1146/annurev.nucl.55.090704.151541}{\emph{Ann. Rev.
  Nucl. Part. Sci.} {\bfseries 55} (2005) 71}
  [\href{https://arxiv.org/abs/hep-th/0502005}{{\ttfamily hep-th/0502005}}].

\bibitem{Marchesano:2007de}
F.~Marchesano, \emph{{Progress in D-brane model building}},
  \href{https://doi.org/10.1002/prop.200610381}{\emph{Fortsch. Phys.}
  {\bfseries 55} (2007) 491}
  [\href{https://arxiv.org/abs/hep-th/0702094}{{\ttfamily hep-th/0702094}}].

\bibitem{Hitchin:2010qz}
N.~Hitchin, \emph{{Lectures on generalized geometry}},
  \href{https://arxiv.org/abs/1008.0973}{{\ttfamily 1008.0973}}.

\bibitem{Gomis:2005wc}
J.~Gomis, F.~Marchesano and D.~Mateos, \emph{{An Open string landscape}},
  \href{https://doi.org/10.1088/1126-6708/2005/11/021}{\emph{JHEP} {\bfseries
  11} (2005) 021} [\href{https://arxiv.org/abs/hep-th/0506179}{{\ttfamily
  hep-th/0506179}}].

\bibitem{Mininno:2020sdb}
A.~Mininno and A.M.~Uranga, \emph{{Dynamical tadpoles and Weak Gravity
  Constraints}}, \href{https://doi.org/10.1007/JHEP05(2021)177}{\emph{JHEP}
  {\bfseries 05} (2021) 177}
  [\href{https://arxiv.org/abs/2011.00051}{{\ttfamily 2011.00051}}].

\bibitem{Maldacena:2001xj}
J.M.~Maldacena, G.W.~Moore and N.~Seiberg, \emph{{D-brane instantons and K
  theory charges}},
  \href{https://doi.org/10.1088/1126-6708/2001/11/062}{\emph{JHEP} {\bfseries
  11} (2001) 062} [\href{https://arxiv.org/abs/hep-th/0108100}{{\ttfamily
  hep-th/0108100}}].

\bibitem{Marchesano:2006ns}
F.~Marchesano, \emph{{D6-branes and torsion}},
  \href{https://doi.org/10.1088/1126-6708/2006/05/019}{\emph{JHEP} {\bfseries
  05} (2006) 019} [\href{https://arxiv.org/abs/hep-th/0603210}{{\ttfamily
  hep-th/0603210}}].

\bibitem{Frey:2002hf}
A.R.~Frey and J.~Polchinski, \emph{{N=3 warped compactifications}},
  \href{https://doi.org/10.1103/PhysRevD.65.126009}{\emph{Phys. Rev. D}
  {\bfseries 65} (2002) 126009}
  [\href{https://arxiv.org/abs/hep-th/0201029}{{\ttfamily hep-th/0201029}}].

\bibitem{Blumenhagen:2003vr}
R.~Blumenhagen, D.~Lust and T.R.~Taylor, \emph{{Moduli stabilization in chiral
  type IIB orientifold models with fluxes}},
  \href{https://doi.org/10.1016/S0550-3213(03)00392-4}{\emph{Nucl. Phys. B}
  {\bfseries 663} (2003) 319}
  [\href{https://arxiv.org/abs/hep-th/0303016}{{\ttfamily hep-th/0303016}}].

\bibitem{Cascales:2003zp}
J.F.G.~Cascales and A.M.~Uranga, \emph{{Chiral 4d string vacua with D branes
  and NSNS and RR fluxes}},
  \href{https://doi.org/10.1088/1126-6708/2003/05/011}{\emph{JHEP} {\bfseries
  05} (2003) 011} [\href{https://arxiv.org/abs/hep-th/0303024}{{\ttfamily
  hep-th/0303024}}].

\bibitem{Font:1988mk}
A.~Font, L.E.~Ibanez and F.~Quevedo, \emph{{$Z(N$) X $Z$(m) Orbifolds and
  Discrete Torsion}},
  \href{https://doi.org/10.1016/0370-2693(89)90864-2}{\emph{Phys. Lett. B}
  {\bfseries 217} (1989) 272}.

\bibitem{Angelantonj:1999ms}
C.~Angelantonj, I.~Antoniadis, G.~D'Appollonio, E.~Dudas and A.~Sagnotti,
  \emph{{Type I vacua with brane supersymmetry breaking}},
  \href{https://doi.org/10.1016/S0550-3213(00)00052-3}{\emph{Nucl. Phys. B}
  {\bfseries 572} (2000) 36}
  [\href{https://arxiv.org/abs/hep-th/9911081}{{\ttfamily hep-th/9911081}}].

\bibitem{Marchesano:2004xz}
F.~Marchesano and G.~Shiu, \emph{{Building MSSM flux vacua}},
  \href{https://doi.org/10.1088/1126-6708/2004/11/041}{\emph{JHEP} {\bfseries
  11} (2004) 041} [\href{https://arxiv.org/abs/hep-th/0409132}{{\ttfamily
  hep-th/0409132}}].

\bibitem{Blumenhagen:2005tn}
R.~Blumenhagen, M.~Cvetic, F.~Marchesano and G.~Shiu, \emph{{Chiral D-brane
  models with frozen open string moduli}},
  \href{https://doi.org/10.1088/1126-6708/2005/03/050}{\emph{JHEP} {\bfseries
  03} (2005) 050} [\href{https://arxiv.org/abs/hep-th/0502095}{{\ttfamily
  hep-th/0502095}}].

\bibitem{Douglas:1998xa}
M.R.~Douglas, \emph{{D-branes and discrete torsion}},
  \href{https://arxiv.org/abs/hep-th/9807235}{{\ttfamily hep-th/9807235}}.

\bibitem{Gomis:2000ej}
J.~Gomis, \emph{{D-branes on orbifolds with discrete torsion and topological
  obstruction}},
  \href{https://doi.org/10.1088/1126-6708/2000/05/006}{\emph{JHEP} {\bfseries
  05} (2000) 006} [\href{https://arxiv.org/abs/hep-th/0001200}{{\ttfamily
  hep-th/0001200}}].

\bibitem{Denef:2005mm}
F.~Denef, M.R.~Douglas, B.~Florea, A.~Grassi and S.~Kachru, \emph{{Fixing all
  moduli in a simple f-theory compactification}},
  \href{https://doi.org/10.4310/ATMP.2005.v9.n6.a1}{\emph{Adv. Theor. Math.
  Phys.} {\bfseries 9} (2005) 861}
  [\href{https://arxiv.org/abs/hep-th/0503124}{{\ttfamily hep-th/0503124}}].

\bibitem{Herraez:2020tih}
A.~Herraez, \emph{{A Note on Membrane Interactions and the Scalar potential}},
  \href{https://doi.org/10.1007/JHEP10(2020)009}{\emph{JHEP} {\bfseries 10}
  (2020) 009} [\href{https://arxiv.org/abs/2006.01160}{{\ttfamily
  2006.01160}}].

\bibitem{Cheung:2014vva}
C.~Cheung and G.N.~Remmen, \emph{{Naturalness and the Weak Gravity
  Conjecture}},
  \href{https://doi.org/10.1103/PhysRevLett.113.051601}{\emph{Phys. Rev. Lett.}
  {\bfseries 113} (2014) 051601}
  [\href{https://arxiv.org/abs/1402.2287}{{\ttfamily 1402.2287}}].

\bibitem{Marchesano:2022rpr}
F.~Marchesano, J.~Quirant and M.~Zatti, \emph{{New instabilities for
  non-supersymmetric AdS$_{4}$ orientifold vacua}},
  \href{https://doi.org/10.1007/JHEP10(2022)026}{\emph{JHEP} {\bfseries 10}
  (2022) 026} [\href{https://arxiv.org/abs/2207.14285}{{\ttfamily
  2207.14285}}].

\bibitem{blumenhagen2000supersymmetric}
R.~Blumenhagen, L.~G{\"o}rlich and B.~Kors, \emph{{Supersymmetric 4D
  orientifolds of type IIA with D6-branes at angles}},
  \href{https://doi.org/10.1088/1126-6708/2000/01/040}{\emph{JHEP} {\bfseries
  01} (2000) 040} [\href{https://arxiv.org/abs/hep-th/9912204}{{\ttfamily
  hep-th/9912204}}].

\bibitem{blumenhagen2003supersymmetric}
R.~Blumenhagen, L.~G{\"o}rlich and T.~Ott, \emph{{Supersymmetric Intersecting
  Branes on the Type IIA T6/Z4 Orientifold}},
  \href{https://doi.org/10.1088/1126-6708/2003/01/021}{\emph{JHEP} {\bfseries
  08} (2003) 021} [\href{https://arxiv.org/abs/hep-th/0211059}{{\ttfamily
  hep-th/0211059}}].

\bibitem{Ihl:2006pp}
M.~Ihl and T.~Wrase, \emph{{Towards a Realistic Type IIA T**6/Z(4) Orientifold
  Model with Background Fluxes. Part 1. Moduli Stabilization}},
  \href{https://doi.org/10.1088/1126-6708/2006/07/027}{\emph{JHEP} {\bfseries
  07} (2006) 027} [\href{https://arxiv.org/abs/hep-th/0604087}{{\ttfamily
  hep-th/0604087}}].

\bibitem{Forste:2001gb}
S.~Forste, G.~Honecker and R.~Schreyer, \emph{{Orientifolds with branes at
  angles}}, \href{https://doi.org/10.1088/1126-6708/2001/06/004}{\emph{JHEP}
  {\bfseries 06} (2001) 004}
  [\href{https://arxiv.org/abs/hep-th/0105208}{{\ttfamily hep-th/0105208}}].

\bibitem{Reffert:2006du}
S.~Reffert, \emph{{Toroidal Orbifolds: Resolutions, Orientifolds and
  Applications in String Phenomenology}}, Ph.D. thesis, Munich U., 2006.
\newblock \href{https://arxiv.org/abs/hep-th/0609040}{{\ttfamily
  hep-th/0609040}}.

\bibitem{Strominger:1985ku}
A.~Strominger, \emph{{TOPOLOGY OF SUPERSTRING COMPACTIFICATION}},  in
  \emph{{Workshop on Unified String Theories Santa Barbara, California, July
  29-August 16, 1985}}, 1985.

\bibitem{lust2007moduli}
D.~L{\"u}st, S.~Reffert, W.~Schulgin and S.~Stieberger, \emph{{Moduli
  stabilization in type IIB orientifolds (I)}},
  \href{https://doi.org/10.1016/j.nuclphysb.2006.12.018}{\emph{Nuclear Physics
  B} {\bfseries 766} (2007) 68}
  [\href{https://arxiv.org/abs/hep-th/0506090}{{\ttfamily hep-th/0506090}}].

\bibitem{forste2001supersymmetric}
S.~F{\"o}rste, G.~Honecker and R.~Schreyer, \emph{{Supersymmetric ZN$\times$ ZM
  orientifolds in 4D with D-branes at angles}},
  \href{https://doi.org/10.1016/S0550-3213(00)00616-7}{\emph{Nuclear Physics B}
  {\bfseries 593} (2001) 127}
  [\href{https://arxiv.org/abs/hep-th/0008250}{{\ttfamily hep-th/0008250}}].

\bibitem{Vafa:1996xn}
C.~Vafa, \emph{{Evidence for F theory}},
  \href{https://doi.org/10.1016/0550-3213(96)00172-1}{\emph{Nucl. Phys. B}
  {\bfseries 469} (1996) 403}
  [\href{https://arxiv.org/abs/hep-th/9602022}{{\ttfamily hep-th/9602022}}].

\bibitem{Denef:2008wq}
F.~Denef, \emph{{Les Houches Lectures on Constructing String Vacua}},
  {\emph{Les Houches} {\bfseries 87} (2008) 483}
  [\href{https://arxiv.org/abs/0803.1194}{{\ttfamily 0803.1194}}].

\bibitem{Weigand:2010wm}
T.~Weigand, \emph{{Lectures on F-theory compactifications and model building}},
  \href{https://doi.org/10.1088/0264-9381/27/21/214004}{\emph{Class. Quant.
  Grav.} {\bfseries 27} (2010) 214004}
  [\href{https://arxiv.org/abs/1009.3497}{{\ttfamily 1009.3497}}].

\bibitem{Weigand:2018rez}
T.~Weigand, \emph{{F-theory}}, {\emph{PoS} {\bfseries TASI2017} (2018) 016}
  [\href{https://arxiv.org/abs/1806.01854}{{\ttfamily 1806.01854}}].

\bibitem{Wiesner:2021pgd}
M.~Wiesner, \emph{{Quantum corrections and the Swampland}}, Ph.D. thesis,
  Madrid, Autonoma U., 2021.

\bibitem{Ellis:1983sf}
J.R.~Ellis, A.B.~Lahanas, D.V.~Nanopoulos and K.~Tamvakis, \emph{{No-Scale
  Supersymmetric Standard Model}},
  \href{https://doi.org/10.1016/0370-2693(84)91378-9}{\emph{Phys. Lett. B}
  {\bfseries 134} (1984) 429}.

\bibitem{Michelson:1996pn}
J.~Michelson, \emph{{Compactifications of type IIB strings to four-dimensions
  with nontrivial classical potential}},
  \href{https://doi.org/10.1016/S0550-3213(97)00184-3}{\emph{Nucl. Phys. B}
  {\bfseries 495} (1997) 127}
  [\href{https://arxiv.org/abs/hep-th/9610151}{{\ttfamily hep-th/9610151}}].

\bibitem{Giddings:2001yu}
S.B.~Giddings, S.~Kachru and J.~Polchinski, \emph{{Hierarchies from fluxes in
  string compactifications}},
  \href{https://doi.org/10.1103/PhysRevD.66.106006}{\emph{Phys. Rev. D}
  {\bfseries 66} (2002) 106006}
  [\href{https://arxiv.org/abs/hep-th/0105097}{{\ttfamily hep-th/0105097}}].

\bibitem{Hori:2003ic}
K.~Hori, S.~Katz, A.~Klemm, R.~Pandharipande, R.~Thomas, C.~Vafa et~al.,
  \emph{{Mirror symmetry}}, vol.~1 of \emph{Clay mathematics monographs}, AMS,
  Providence, USA (2003).

\bibitem{Kontsevich:1994dn}
M.~Kontsevich, \emph{{Homological Algebra of Mirror Symmetry}},
  \href{https://arxiv.org/abs/alg-geom/9411018}{{\ttfamily alg-geom/9411018}}.

\bibitem{Mayr:1996sh}
P.~Mayr, \emph{{Mirror symmetry, N=1 superpotentials and tensionless strings on
  Calabi-Yau four folds}},
  \href{https://doi.org/10.1016/S0550-3213(97)00196-X}{\emph{Nucl. Phys.}
  {\bfseries B494} (1997) 489}
  [\href{https://arxiv.org/abs/hep-th/9610162}{{\ttfamily hep-th/9610162}}].

\bibitem{Douglas:2000gi}
M.R.~Douglas, \emph{{D-branes, categories and N=1 supersymmetry}},
  \href{https://doi.org/10.1063/1.1374448}{\emph{J. Math. Phys.} {\bfseries 42}
  (2001) 2818} [\href{https://arxiv.org/abs/hep-th/0011017}{{\ttfamily
  hep-th/0011017}}].

\bibitem{Cota:2019cjx}
C.F.~Cota, A.~Klemm and T.~Schimannek, \emph{{Topological strings on genus one
  fibered Calabi-Yau 3-folds and string dualities}},
  \href{https://doi.org/10.1007/JHEP11(2019)170}{\emph{JHEP} {\bfseries 11}
  (2019) 170} [\href{https://arxiv.org/abs/1910.01988}{{\ttfamily
  1910.01988}}].

\bibitem{huybrechts2005complex}
D.~Huybrechts, \emph{Complex geometry: an introduction}, vol.~78, Springer
  (2005).

\bibitem{iritani2009integral}
H.~Iritani, \emph{An integral structure in quantum cohomology and mirror
  symmetry for toric orbifolds}, {\emph{Advances in Mathematics} {\bfseries
  222} (2009) 1016} [\href{https://arxiv.org/abs/0903.1463}{{\ttfamily
  0903.1463}}].

\bibitem{Cicoli:2013cha}
M.~Cicoli, D.~Klevers, S.~Krippendorf, C.~Mayrhofer, F.~Quevedo and
  R.~Valandro, \emph{{Explicit de Sitter Flux Vacua for Global String Models
  with Chiral Matter}},
  \href{https://doi.org/10.1007/JHEP05(2014)001}{\emph{JHEP} {\bfseries 05}
  (2014) 001} [\href{https://arxiv.org/abs/1312.0014}{{\ttfamily 1312.0014}}].

\bibitem{Mayr:2000as}
P.~Mayr, \emph{{Phases of supersymmetric D-branes on Kahler manifolds and the
  McKay correspondence}},
  \href{https://doi.org/10.1088/1126-6708/2001/01/018}{\emph{JHEP} {\bfseries
  01} (2001) 018} [\href{https://arxiv.org/abs/hep-th/0010223}{{\ttfamily
  hep-th/0010223}}].

\bibitem{Blanco-Pillado:2020wjn}
J.J.~Blanco-Pillado, K.~Sousa, M.A.~Urkiola and J.M.~Wachter, \emph{{Towards a
  complete mass spectrum of type-IIB flux vacua at large complex structure}},
  \href{https://doi.org/10.1007/JHEP04(2021)149}{\emph{JHEP} {\bfseries 04}
  (2021) 149} [\href{https://arxiv.org/abs/2007.10381}{{\ttfamily
  2007.10381}}].

\bibitem{Greene:1989ya}
B.R.~Greene, A.D.~Shapere, C.~Vafa and S.-T.~Yau, \emph{{Stringy Cosmic Strings
  and Noncompact Calabi-Yau Manifolds}},
  \href{https://doi.org/10.1016/0550-3213(90)90248-C}{\emph{Nucl. Phys. B}
  {\bfseries 337} (1990) 1}.

\bibitem{Bianchi:2011qh}
M.~Bianchi, A.~Collinucci and L.~Martucci, \emph{{Magnetized E3-brane
  instantons in F-theory}},
  \href{https://doi.org/10.1007/JHEP12(2011)045}{\emph{JHEP} {\bfseries 12}
  (2011) 045} [\href{https://arxiv.org/abs/1107.3732}{{\ttfamily 1107.3732}}].

\bibitem{Grimm:2010ks}
T.W.~Grimm, \emph{{The N=1 effective action of F-theory compactifications}},
  \href{https://doi.org/10.1016/j.nuclphysb.2010.11.018}{\emph{Nucl. Phys. B}
  {\bfseries 845} (2011) 48} [\href{https://arxiv.org/abs/1008.4133}{{\ttfamily
  1008.4133}}].

\bibitem{Witten:1996md}
E.~Witten, \emph{{On flux quantization in M theory and the effective action}},
  \href{https://doi.org/10.1016/S0393-0440(96)00042-3}{\emph{J. Geom. Phys.}
  {\bfseries 22} (1997) 1}
  [\href{https://arxiv.org/abs/hep-th/9609122}{{\ttfamily hep-th/9609122}}].

\bibitem{Becker:1996gj}
K.~Becker and M.~Becker, \emph{{M theory on eight manifolds}},
  \href{https://doi.org/10.1016/0550-3213(96)00367-7}{\emph{Nucl. Phys.}
  {\bfseries B477} (1996) 155}
  [\href{https://arxiv.org/abs/hep-th/9605053}{{\ttfamily hep-th/9605053}}].

\bibitem{Haack:2001jz}
M.~Haack and J.~Louis, \emph{{M theory compactified on Calabi-Yau fourfolds
  with background flux}},
  \href{https://doi.org/10.1016/S0370-2693(01)00464-6}{\emph{Phys. Lett. B}
  {\bfseries 507} (2001) 296}
  [\href{https://arxiv.org/abs/hep-th/0103068}{{\ttfamily hep-th/0103068}}].

\bibitem{Cota:2017aal}
C.F.~Cota, A.~Klemm and T.~Schimannek, \emph{{Modular Amplitudes and
  Flux-Superpotentials on elliptic Calabi-Yau fourfolds}},
  \href{https://doi.org/10.1007/JHEP01(2018)086}{\emph{JHEP} {\bfseries 01}
  (2018) 086} [\href{https://arxiv.org/abs/1709.02820}{{\ttfamily
  1709.02820}}].

\bibitem{Dasgupta:1999ss}
K.~Dasgupta, G.~Rajesh and S.~Sethi, \emph{{M theory, orientifolds and G -
  flux}}, \href{https://doi.org/10.1088/1126-6708/1999/08/023}{\emph{JHEP}
  {\bfseries 08} (1999) 023}
  [\href{https://arxiv.org/abs/hep-th/9908088}{{\ttfamily hep-th/9908088}}].

\bibitem{Sethi:1996es}
S.~Sethi, C.~Vafa and E.~Witten, \emph{{Constraints on low dimensional string
  compactifications}},
  \href{https://doi.org/10.1016/S0550-3213(96)00483-X}{\emph{Nucl. Phys. B}
  {\bfseries 480} (1996) 213}
  [\href{https://arxiv.org/abs/hep-th/9606122}{{\ttfamily hep-th/9606122}}].

\bibitem{Bena:2020xrh}
I.~Bena, J.~Blab\"ack, M.~Gra\~na and S.~L\"ust, \emph{{The tadpole problem}},
  \href{https://doi.org/10.1007/JHEP11(2021)223}{\emph{JHEP} {\bfseries 11}
  (2021) 223} [\href{https://arxiv.org/abs/2010.10519}{{\ttfamily
  2010.10519}}].

\bibitem{Bena:2021wyr}
I.~Bena, J.~Blab\"ack, M.~Gra\~na and S.~L\"ust, \emph{{Algorithmically Solving
  the Tadpole Problem}},
  \href{https://doi.org/10.1007/s00006-021-01189-6}{\emph{Adv. Appl. Clifford
  Algebras} {\bfseries 32} (2022) 7}
  [\href{https://arxiv.org/abs/2103.03250}{{\ttfamily 2103.03250}}].

\bibitem{Braun:2023pzd}
A.P.~Braun, B.~Fraiman, M.~Gra\~na, S.~L\"ust and H.~Parra~de Freitas,
  \emph{{Tadpoles and gauge symmetries}},
  \href{https://doi.org/10.1007/JHEP08(2023)134}{\emph{JHEP} {\bfseries 08}
  (2023) 134} [\href{https://arxiv.org/abs/2304.06751}{{\ttfamily
  2304.06751}}].

\bibitem{Grana:2022dfw}
M.~Gra\~na, T.W.~Grimm, D.~van~de Heisteeg, A.~Herraez and E.~Plauschinn,
  \emph{{The tadpole conjecture in asymptotic limits}},
  \href{https://doi.org/10.1007/JHEP08(2022)237}{\emph{JHEP} {\bfseries 08}
  (2022) 237} [\href{https://arxiv.org/abs/2204.05331}{{\ttfamily
  2204.05331}}].

\bibitem{Lust:2022mhk}
S.~L\"ust and M.~Wiesner, \emph{{The tadpole conjecture in the interior of
  moduli space}}, \href{https://doi.org/10.1007/JHEP12(2023)029}{\emph{JHEP}
  {\bfseries 12} (2023) 029}
  [\href{https://arxiv.org/abs/2211.05128}{{\ttfamily 2211.05128}}].

\bibitem{Quevedo:2014xia}
F.~Quevedo, \emph{{Local String Models and Moduli Stabilisation}},
  \href{https://doi.org/10.1142/S0217732315300049}{\emph{Mod. Phys. Lett. A}
  {\bfseries 30} (2015) 1530004}
  [\href{https://arxiv.org/abs/1404.5151}{{\ttfamily 1404.5151}}].

\bibitem{Baumann:2014nda}
D.~Baumann and L.~McAllister, \emph{{Inflation and String Theory}}, Cambridge
  Monographs on Mathematical Physics, Cambridge University Press (5, 2015),
  \href{https://doi.org/10.1017/CBO9781316105733}{10.1017/CBO9781316105733},
  [\href{https://arxiv.org/abs/1404.2601}{{\ttfamily 1404.2601}}].

\bibitem{Braun:2020jrx}
A.P.~Braun and R.~Valandro, \emph{{$G_{4}$ flux, algebraic cycles and complex
  structure moduli stabilization}},
  \href{https://doi.org/10.1007/JHEP01(2021)207}{\emph{JHEP} {\bfseries 01}
  (2021) 207} [\href{https://arxiv.org/abs/2009.11873}{{\ttfamily
  2009.11873}}].

\bibitem{Grimm:2020cda}
T.W.~Grimm, \emph{{Moduli space holography and the finiteness of flux vacua}},
  \href{https://doi.org/10.1007/JHEP10(2021)153}{\emph{JHEP} {\bfseries 10}
  (2021) 153} [\href{https://arxiv.org/abs/2010.15838}{{\ttfamily
  2010.15838}}].

\bibitem{Becker:1997cp}
K.~Becker and M.~Becker, \emph{{On graviton scattering amplitudes in M
  theory}}, \href{https://doi.org/10.1103/PhysRevD.57.6464}{\emph{Phys. Rev. D}
  {\bfseries 57} (1998) 6464}
  [\href{https://arxiv.org/abs/hep-th/9712238}{{\ttfamily hep-th/9712238}}].

\bibitem{Strominger:1990pd}
A.~Strominger, \emph{{SPECIAL GEOMETRY}},
  \href{https://doi.org/10.1007/BF02096559}{\emph{Commun. Math. Phys.}
  {\bfseries 133} (1990) 163}.

\bibitem{Greene:1993vm}
B.R.~Greene, D.R.~Morrison and M.R.~Plesser, \emph{{Mirror manifolds in higher
  dimension}}, \href{https://doi.org/10.1007/BF02101657}{\emph{Commun. Math.
  Phys.} {\bfseries 173} (1995) 559}
  [\href{https://arxiv.org/abs/hep-th/9402119}{{\ttfamily hep-th/9402119}}].

\bibitem{Braun:2014xka}
A.P.~Braun and T.~Watari, \emph{{The Vertical, the Horizontal and the Rest:
  anatomy of the middle cohomology of Calabi-Yau fourfolds and F-theory
  applications}}, \href{https://doi.org/10.1007/JHEP01(2015)047}{\emph{JHEP}
  {\bfseries 01} (2015) 047} [\href{https://arxiv.org/abs/1408.6167}{{\ttfamily
  1408.6167}}].

\bibitem{CaboBizet:2014ovf}
N.~Cabo~Bizet, A.~Klemm and D.~Vieira~Lopes, \emph{{Landscaping with fluxes and
  the E8 Yukawa Point in F-theory}},
  \href{https://arxiv.org/abs/1404.7645}{{\ttfamily 1404.7645}}.

\bibitem{Farakos:2017jme}
F.~Farakos, S.~Lanza, L.~Martucci and D.~Sorokin, \emph{{Three-forms in
  Supergravity and Flux Compactifications}},
  \href{https://doi.org/10.1140/epjc/s10052-017-5185-y}{\emph{Eur. Phys. J. C}
  {\bfseries 77} (2017) 602}
  [\href{https://arxiv.org/abs/1706.09422}{{\ttfamily 1706.09422}}].

\bibitem{Bandos:2018gjp}
I.~Bandos, F.~Farakos, S.~Lanza, L.~Martucci and D.~Sorokin,
  \emph{{Three-forms, dualities and membranes in four-dimensional
  supergravity}}, \href{https://doi.org/10.1007/JHEP07(2018)028}{\emph{JHEP}
  {\bfseries 07} (2018) 028}
  [\href{https://arxiv.org/abs/1803.01405}{{\ttfamily 1803.01405}}].

\bibitem{Valenzuela:2016yny}
I.~Valenzuela, \emph{{Backreaction Issues in Axion Monodromy and Minkowski
  4-forms}}, \href{https://doi.org/10.1007/JHEP06(2017)098}{\emph{JHEP}
  {\bfseries 06} (2017) 098}
  [\href{https://arxiv.org/abs/1611.00394}{{\ttfamily 1611.00394}}].

\bibitem{Grimm:2020ouv}
T.W.~Grimm and C.~Li, \emph{{Universal axion backreaction in flux
  compactifications}},
  \href{https://doi.org/10.1007/JHEP06(2021)067}{\emph{JHEP} {\bfseries 06}
  (2021) 067} [\href{https://arxiv.org/abs/2012.08272}{{\ttfamily
  2012.08272}}].

\bibitem{Cremmer:1982en}
E.~Cremmer, S.~Ferrara, L.~Girardello and A.~Van~Proeyen, \emph{{Yang-Mills
  Theories with Local Supersymmetry: Lagrangian, Transformation Laws and
  SuperHiggs Effect}},
  \href{https://doi.org/10.1016/0550-3213(83)90679-X}{\emph{Nucl. Phys. B}
  {\bfseries 212} (1983) 413}.

\bibitem{Marsh:2015zoa}
M.C.D.~Marsh and K.~Sousa, \emph{{Universal Properties of Type IIB and F-theory
  Flux Compactifications at Large Complex Structure}},
  \href{https://doi.org/10.1007/JHEP03(2016)064}{\emph{JHEP} {\bfseries 03}
  (2016) 064} [\href{https://arxiv.org/abs/1512.08549}{{\ttfamily
  1512.08549}}].

\bibitem{Gerhardus:2016iot}
A.~Gerhardus and H.~Jockers, \emph{{Quantum periods of Calabi\textendash{}Yau
  fourfolds}},
  \href{https://doi.org/10.1016/j.nuclphysb.2016.09.021}{\emph{Nucl. Phys. B}
  {\bfseries 913} (2016) 425}
  [\href{https://arxiv.org/abs/1604.05325}{{\ttfamily 1604.05325}}].

\bibitem{Honma:2017uzn}
Y.~Honma and H.~Otsuka, \emph{{On the Flux Vacua in F-theory
  Compactifications}},
  \href{https://doi.org/10.1016/j.physletb.2017.09.062}{\emph{Phys. Lett. B}
  {\bfseries 774} (2017) 225}
  [\href{https://arxiv.org/abs/1706.09417}{{\ttfamily 1706.09417}}].

\bibitem{Demirtas:2019sip}
M.~Demirtas, M.~Kim, L.~Mcallister and J.~Moritz, \emph{{Vacua with Small Flux
  Superpotential}},
  \href{https://doi.org/10.1103/PhysRevLett.124.211603}{\emph{Phys. Rev. Lett.}
  {\bfseries 124} (2020) 211603}
  [\href{https://arxiv.org/abs/1912.10047}{{\ttfamily 1912.10047}}].

\bibitem{Blanco-Pillado:2020hbw}
J.J.~Blanco-Pillado, K.~Sousa, M.A.~Urkiola and J.M.~Wachter, \emph{{Universal
  Class of Type-IIB Flux Vacua with Analytic Mass Spectrum}},
  \href{https://doi.org/10.1103/PhysRevD.103.106006}{\emph{Phys. Rev. D}
  {\bfseries 103} (2021) 106006}
  [\href{https://arxiv.org/abs/2011.13953}{{\ttfamily 2011.13953}}].

\bibitem{Betzler:2019kon}
P.~Betzler and E.~Plauschinn, \emph{{Type IIB flux vacua and tadpole
  cancellation}}, \href{https://doi.org/10.1002/prop.201900065}{\emph{Fortsch.
  Phys.} {\bfseries 67} (2019) 1900065}
  [\href{https://arxiv.org/abs/1905.08823}{{\ttfamily 1905.08823}}].

\bibitem{Lee:2019wij}
S.-J.~Lee, W.~Lerche and T.~Weigand, \emph{{Emergent strings from infinite
  distance limits}}, \href{https://doi.org/10.1007/JHEP02(2022)190}{\emph{JHEP}
  {\bfseries 02} (2022) 190}
  [\href{https://arxiv.org/abs/1910.01135}{{\ttfamily 1910.01135}}].

\bibitem{Collinucci:2010gz}
A.~Collinucci and R.~Savelli, \emph{{On Flux Quantization in F-Theory}},
  \href{https://doi.org/10.1007/JHEP02(2012)015}{\emph{JHEP} {\bfseries 02}
  (2012) 015} [\href{https://arxiv.org/abs/1011.6388}{{\ttfamily 1011.6388}}].

\bibitem{Brodie:2015kza}
C.~Brodie and M.C.D.~Marsh, \emph{{The Spectra of Type IIB Flux
  Compactifications at Large Complex Structure}},
  \href{https://doi.org/10.1007/JHEP01(2016)037}{\emph{JHEP} {\bfseries 01}
  (2016) 037} [\href{https://arxiv.org/abs/1509.06761}{{\ttfamily
  1509.06761}}].

\bibitem{Ashok:2003gk}
S.~Ashok and M.R.~Douglas, \emph{{Counting flux vacua}},
  \href{https://doi.org/10.1088/1126-6708/2004/01/060}{\emph{JHEP} {\bfseries
  01} (2004) 060} [\href{https://arxiv.org/abs/hep-th/0307049}{{\ttfamily
  hep-th/0307049}}].

\bibitem{Denef:2004ze}
F.~Denef and M.R.~Douglas, \emph{{Distributions of flux vacua}},
  \href{https://doi.org/10.1088/1126-6708/2004/05/072}{\emph{JHEP} {\bfseries
  05} (2004) 072} [\href{https://arxiv.org/abs/hep-th/0404116}{{\ttfamily
  hep-th/0404116}}].

\bibitem{Denef:2004cf}
F.~Denef and M.R.~Douglas, \emph{{Distributions of nonsupersymmetric flux
  vacua}}, \href{https://doi.org/10.1088/1126-6708/2005/03/061}{\emph{JHEP}
  {\bfseries 03} (2005) 061}
  [\href{https://arxiv.org/abs/hep-th/0411183}{{\ttfamily hep-th/0411183}}].

\bibitem{Honma:2021klo}
Y.~Honma and H.~Otsuka, \emph{{Small flux superpotential in F-theory
  compactifications}},
  \href{https://doi.org/10.1103/PhysRevD.103.126022}{\emph{Phys. Rev. D}
  {\bfseries 103} (2021) 126022}
  [\href{https://arxiv.org/abs/2103.03003}{{\ttfamily 2103.03003}}].

\bibitem{Westphal:2006tn}
A.~Westphal, \emph{{de Sitter string vacua from Kahler uplifting}},
  \href{https://doi.org/10.1088/1126-6708/2007/03/102}{\emph{JHEP} {\bfseries
  03} (2007) 102} [\href{https://arxiv.org/abs/hep-th/0611332}{{\ttfamily
  hep-th/0611332}}].

\bibitem{Giryavets:2003vd}
A.~Giryavets, S.~Kachru, P.K.~Tripathy and S.P.~Trivedi, \emph{{Flux
  compactifications on Calabi-Yau threefolds}},
  \href{https://doi.org/10.1088/1126-6708/2004/04/003}{\emph{JHEP} {\bfseries
  04} (2004) 003} [\href{https://arxiv.org/abs/hep-th/0312104}{{\ttfamily
  hep-th/0312104}}].

\bibitem{Giryavets:2004zr}
A.~Giryavets, S.~Kachru and P.K.~Tripathy, \emph{{On the taxonomy of flux
  vacua}}, \href{https://doi.org/10.1088/1126-6708/2004/08/002}{\emph{JHEP}
  {\bfseries 08} (2004) 002}
  [\href{https://arxiv.org/abs/hep-th/0404243}{{\ttfamily hep-th/0404243}}].

\bibitem{DeWolfe:2004ns}
O.~DeWolfe, A.~Giryavets, S.~Kachru and W.~Taylor, \emph{{Enumerating flux
  vacua with enhanced symmetries}},
  \href{https://doi.org/10.1088/1126-6708/2005/02/037}{\emph{JHEP} {\bfseries
  02} (2005) 037} [\href{https://arxiv.org/abs/hep-th/0411061}{{\ttfamily
  hep-th/0411061}}].

\bibitem{Denef:2004dm}
F.~Denef, M.R.~Douglas and B.~Florea, \emph{{Building a better racetrack}},
  \href{https://doi.org/10.1088/1126-6708/2004/06/034}{\emph{JHEP} {\bfseries
  06} (2004) 034} [\href{https://arxiv.org/abs/hep-th/0404257}{{\ttfamily
  hep-th/0404257}}].

\bibitem{Louis:2012nb}
J.~Louis, M.~Rummel, R.~Valandro and A.~Westphal, \emph{{Building an explicit
  de Sitter}}, \href{https://doi.org/10.1007/JHEP10(2012)163}{\emph{JHEP}
  {\bfseries 10} (2012) 163} [\href{https://arxiv.org/abs/1208.3208}{{\ttfamily
  1208.3208}}].

\bibitem{Klemm:1992tx}
A.~Klemm and S.~Theisen, \emph{{Considerations of one modulus Calabi-Yau
  compactifications: Picard-Fuchs equations, Kahler potentials and mirror
  maps}}, \href{https://doi.org/10.1016/0550-3213(93)90289-2}{\emph{Nucl. Phys.
  B} {\bfseries 389} (1993) 153}
  [\href{https://arxiv.org/abs/hep-th/9205041}{{\ttfamily hep-th/9205041}}].

\bibitem{Doran:2007jw}
C.~Doran, B.~Greene and S.~Judes, \emph{{Families of quintic Calabi-Yau 3-folds
  with discrete symmetries}},
  \href{https://doi.org/10.1007/s00220-008-0473-x}{\emph{Commun. Math. Phys.}
  {\bfseries 280} (2008) 675}
  [\href{https://arxiv.org/abs/hep-th/0701206}{{\ttfamily hep-th/0701206}}].

\bibitem{Candelas:2017ive}
P.~Candelas and C.~Mishra, \emph{{Highly Symmetric Quintic Quotients}},
  \href{https://doi.org/10.1002/prop.201800017}{\emph{Fortsch. Phys.}
  {\bfseries 66} (2018) 1800017}
  [\href{https://arxiv.org/abs/1709.01081}{{\ttfamily 1709.01081}}].

\bibitem{Braun:2011hd}
V.~Braun, \emph{{The 24-Cell and Calabi-Yau Threefolds with Hodge Numbers
  (1,1)}}, \href{https://doi.org/10.1007/JHEP05(2012)101}{\emph{JHEP}
  {\bfseries 05} (2012) 101} [\href{https://arxiv.org/abs/1102.4880}{{\ttfamily
  1102.4880}}].

\bibitem{Batyrev:2008rp}
V.~Batyrev and M.~Kreuzer, \emph{{Constructing new Calabi-Yau 3-folds and their
  mirrors via conifold transitions}},
  \href{https://doi.org/10.4310/ATMP.2010.v14.n3.a3}{\emph{Adv. Theor. Math.
  Phys.} {\bfseries 14} (2010) 879}
  [\href{https://arxiv.org/abs/0802.3376}{{\ttfamily 0802.3376}}].

\bibitem{Doran:2005gu}
C.F.~Doran and J.W.~Morgan, \emph{{Mirror symmetry and integral variations of
  Hodge structure underlying one parameter families of Calabi-Yau threefolds}},
   in \emph{{Workshop on Calabi-Yau Varieties and Mirror Symmetry}},
  pp.~517--537, 5, 2005 [\href{https://arxiv.org/abs/math/0505272}{{\ttfamily
  math/0505272}}].

\bibitem{Candelas:2019llw}
P.~Candelas, X.~de~la Ossa, M.~Elmi and D.~Van~Straten, \emph{{A One Parameter
  Family of Calabi-Yau Manifolds with Attractor Points of Rank Two}},
  \href{https://doi.org/10.1007/JHEP10(2020)202}{\emph{JHEP} {\bfseries 10}
  (2020) 202} [\href{https://arxiv.org/abs/1912.06146}{{\ttfamily
  1912.06146}}].

\bibitem{Joshi:2019nzi}
A.~Joshi and A.~Klemm, \emph{{Swampland Distance Conjecture for One-Parameter
  Calabi-Yau Threefolds}},
  \href{https://doi.org/10.1007/JHEP08(2019)086}{\emph{JHEP} {\bfseries 08}
  (2019) 086} [\href{https://arxiv.org/abs/1903.00596}{{\ttfamily
  1903.00596}}].

\bibitem{Sousa:2014qza}
K.~Sousa and P.~Ortiz, \emph{{Perturbative Stability along the Supersymmetric
  Directions of the Landscape}},
  \href{https://doi.org/10.1088/1475-7516/2015/02/017}{\emph{JCAP} {\bfseries
  02} (2015) 017} [\href{https://arxiv.org/abs/1408.6521}{{\ttfamily
  1408.6521}}].

\bibitem{Demirtas:2020ffz}
M.~Demirtas, M.~Kim, L.~McAllister and J.~Moritz, \emph{{Conifold Vacua with
  Small Flux Superpotential}},
  \href{https://doi.org/10.1002/prop.202000085}{\emph{Fortsch. Phys.}
  {\bfseries 68} (2020) 2000085}
  [\href{https://arxiv.org/abs/2009.03312}{{\ttfamily 2009.03312}}].

\bibitem{Cicoli:2022vny}
M.~Cicoli, M.~Licheri, R.~Mahanta and A.~Maharana, \emph{{Flux vacua with
  approximate flat directions}},
  \href{https://doi.org/10.1007/JHEP10(2022)086}{\emph{JHEP} {\bfseries 10}
  (2022) 086} [\href{https://arxiv.org/abs/2209.02720}{{\ttfamily
  2209.02720}}].

\bibitem{Candelas:1994hw}
P.~Candelas, A.~Font, S.H.~Katz and D.R.~Morrison, \emph{{Mirror symmetry for
  two parameter models. 2.}},
  \href{https://doi.org/10.1016/0550-3213(94)90155-4}{\emph{Nucl. Phys. B}
  {\bfseries 429} (1994) 626}
  [\href{https://arxiv.org/abs/hep-th/9403187}{{\ttfamily hep-th/9403187}}].

\bibitem{AbdusSalam:2020ywo}
S.~AbdusSalam, S.~Abel, M.~Cicoli, F.~Quevedo and P.~Shukla, \emph{{A
  systematic approach to K\"ahler moduli stabilisation}},
  \href{https://doi.org/10.1007/JHEP08(2020)047}{\emph{JHEP} {\bfseries 08}
  (2020) 047} [\href{https://arxiv.org/abs/2005.11329}{{\ttfamily
  2005.11329}}].

\bibitem{Plauschinn:2021hkp}
E.~Plauschinn, \emph{{The tadpole conjecture at large complex-structure}},
  \href{https://doi.org/10.1007/JHEP02(2022)206}{\emph{JHEP} {\bfseries 02}
  (2022) 206} [\href{https://arxiv.org/abs/2109.00029}{{\ttfamily
  2109.00029}}].

\bibitem{Lust:2021xds}
S.~L\"ust, \emph{{Large complex structure flux vacua of IIB and the Tadpole
  Conjecture}},  \href{https://arxiv.org/abs/2109.05033}{{\ttfamily
  2109.05033}}.

\bibitem{Grimm:2021ckh}
T.W.~Grimm, E.~Plauschinn and D.~van~de Heisteeg, \emph{{Moduli stabilization
  in asymptotic flux compactifications}},
  \href{https://doi.org/10.1007/JHEP03(2022)117}{\emph{JHEP} {\bfseries 03}
  (2022) 117} [\href{https://arxiv.org/abs/2110.05511}{{\ttfamily
  2110.05511}}].

\bibitem{Demirtas:2018akl}
M.~Demirtas, C.~Long, L.~McAllister and M.~Stillman, \emph{{The Kreuzer-Skarke
  Axiverse}}, \href{https://doi.org/10.1007/JHEP04(2020)138}{\emph{JHEP}
  {\bfseries 04} (2020) 138}
  [\href{https://arxiv.org/abs/1808.01282}{{\ttfamily 1808.01282}}].

\bibitem{Alvarez-Garcia:2020pxd}
R.~\'Alvarez-Garc\'\i{}a, R.~Blumenhagen, M.~Brinkmann and L.~Schlechter,
  \emph{{Small Flux Superpotentials for Type IIB Flux Vacua Close to a
  Conifold}}, \href{https://doi.org/10.1002/prop.202000088}{\emph{Fortsch.
  Phys.} {\bfseries 68} (2020) 2000088}
  [\href{https://arxiv.org/abs/2009.03325}{{\ttfamily 2009.03325}}].

\bibitem{nakahara2003geometry}
M.~Nakahara, \emph{Geometry, topology and physics}, CRC press (2003).

\bibitem{Chiossi:2002tw}
S.~Chiossi and S.~Salamon, \emph{{The Intrinsic torsion of SU(3) and G(2)
  structures}},  in \emph{{International Conference on Differential Geometry
  held in honor of the 60th Birthday of A.M. Naveira Valencia, Spain, May 8-14,
  2001}}, 2002 [\href{https://arxiv.org/abs/math/0202282}{{\ttfamily
  math/0202282}}].

\bibitem{Hitchin:2003cxu}
N.~Hitchin, \emph{{Generalized Calabi-Yau manifolds}},
  \href{https://doi.org/10.1093/qjmath/54.3.281}{\emph{Quart. J. Math.}
  {\bfseries 54} (2003) 281}
  [\href{https://arxiv.org/abs/math/0209099}{{\ttfamily math/0209099}}].

\bibitem{Gualtieri:2003dx}
M.~Gualtieri, \emph{{Generalized complex geometry}}, Ph.D. thesis, Oxford U.,
  2003.
\newblock \href{https://arxiv.org/abs/math/0401221}{{\ttfamily math/0401221}}.

\bibitem{Grana:2005sn}
M.~Grana, R.~Minasian, M.~Petrini and A.~Tomasiello, \emph{{Generalized
  structures of N=1 vacua}},
  \href{https://doi.org/10.1088/1126-6708/2005/11/020}{\emph{JHEP} {\bfseries
  11} (2005) 020} [\href{https://arxiv.org/abs/hep-th/0505212}{{\ttfamily
  hep-th/0505212}}].

\bibitem{Halmagyi:2009xun}
N.~Halmagyi and A.~Tomasiello, \emph{{Generalized Kaehler Potentials from
  Supergravity}},
  \href{https://doi.org/10.1007/s00220-009-0881-6}{\emph{Commun. Math. Phys.}
  {\bfseries 291} (2009) 1} [\href{https://arxiv.org/abs/0708.1032}{{\ttfamily
  0708.1032}}].

\bibitem{Grana:2006hr}
M.~Grana, J.~Louis and D.~Waldram, \emph{{SU(3) x SU(3) compactification and
  mirror duals of magnetic fluxes}},
  \href{https://doi.org/10.1088/1126-6708/2007/04/101}{\emph{JHEP} {\bfseries
  04} (2007) 101} [\href{https://arxiv.org/abs/hep-th/0612237}{{\ttfamily
  hep-th/0612237}}].

\bibitem{BerasaluceGonzalez:2012zn}
M.~Berasaluce-Gonzalez, P.G.~Camara, F.~Marchesano and A.M.~Uranga, \emph{{Zp
  charged branes in flux compactifications}},
  \href{https://doi.org/10.1007/JHEP04(2013)138}{\emph{JHEP} {\bfseries 04}
  (2013) 138} [\href{https://arxiv.org/abs/1211.5317}{{\ttfamily 1211.5317}}].

\bibitem{Marchesano:2014mla}
F.~Marchesano, G.~Shiu and A.M.~Uranga, \emph{{F-term Axion Monodromy
  Inflation}}, \href{https://doi.org/10.1007/JHEP09(2014)184}{\emph{JHEP}
  {\bfseries 09} (2014) 184} [\href{https://arxiv.org/abs/1404.3040}{{\ttfamily
  1404.3040}}].

\bibitem{Blumenhagen:2014gta}
R.~Blumenhagen and E.~Plauschinn, \emph{{Towards Universal Axion Inflation and
  Reheating in String Theory}},
  \href{https://doi.org/10.1016/j.physletb.2014.08.007}{\emph{Phys. Lett. B}
  {\bfseries 736} (2014) 482}
  [\href{https://arxiv.org/abs/1404.3542}{{\ttfamily 1404.3542}}].

\bibitem{Baulieu:1997jx}
L.~Baulieu, H.~Kanno and I.M.~Singer, \emph{{Special quantum field theories in
  eight-dimensions and other dimensions}},
  \href{https://doi.org/10.1007/s002200050353}{\emph{Commun. Math. Phys.}
  {\bfseries 194} (1998) 149}
  [\href{https://arxiv.org/abs/hep-th/9704167}{{\ttfamily hep-th/9704167}}].

\bibitem{Halverson:2013qca}
J.~Halverson, H.~Jockers, J.M.~Lapan and D.R.~Morrison, \emph{{Perturbative
  Corrections to Kaehler Moduli Spaces}},
  \href{https://doi.org/10.1007/s00220-014-2157-z}{\emph{Commun. Math. Phys.}
  {\bfseries 333} (2015) 1563}
  [\href{https://arxiv.org/abs/1308.2157}{{\ttfamily 1308.2157}}].

\bibitem{Honma:2013hma}
Y.~Honma and M.~Manabe, \emph{{Exact Kahler Potential for Calabi-Yau
  Fourfolds}}, \href{https://doi.org/10.1007/JHEP05(2013)102}{\emph{JHEP}
  {\bfseries 05} (2013) 102} [\href{https://arxiv.org/abs/1302.3760}{{\ttfamily
  1302.3760}}].

\end{thebibliography}\endgroup

\end{document}